\documentclass[11.5pt,a4paper]{book}
\usepackage[utf8]{inputenc}
\usepackage[spanish, catalan, english]{babel} 
\usepackage[toc,page]{appendix}
\usepackage{amsmath}
\usepackage{mathtools}
\usepackage{amsfonts}
\usepackage{amssymb}
\usepackage{graphicx}
\usepackage{cite}
\usepackage{float}
\usepackage{eso-pic}
\usepackage{soul}

\usepackage{url}

\usepackage{breakurl}

\usepackage[dvipsnames]{xcolor}
\usepackage[
    colorlinks = true,
    linkcolor = Blue,
    citecolor = BrickRed,
    urlcolor = ForestGreen,
    pdfprintscaling = None,
    pdfstartview = FitH,
    breaklinks = true
]{hyperref}

\usepackage{subfigure}
\usepackage{geometry}
\usepackage[skip=10pt,font=small]{caption}
\newtheorem{thm}{Theorem}
\newtheorem{lemma}{Lemma}
\newtheorem{proposition}{Proposition}
\usepackage{fancyhdr}
\usepackage{setspace}
\usepackage{titlesec}
\usepackage{lmodern} 

\titlespacing*{\chapter}
  {0pt}    
  {-50pt}    
  {20pt}   

\usepackage{layout}
\usepackage{pdfpages}

\newcommand{\ChapterNumber}{\ifnum\value{chapter}>0 \thechapter.\fi}
\newcommand{\SectionNumber}{\ChapterNumber\arabic{section}.}
\newcommand{\SubsectionNumber}{\SectionNumber\arabic{subsection}.}

\titleformat{\chapter}[display]
  {\normalfont\Huge\bfseries\centering}  
  {\rule{\textwidth}{1pt}}             
  {0mm}                                  
  {\ChapterNumber~}                         
  [\vspace{-4mm}\rule{\textwidth}{1pt}]  

\titleformat{\section}
  {\normalfont\Large\bfseries}
  {\SectionNumber}{0.4em}{}

\titleformat{\subsection}
  {\normalfont\large\bfseries}
  {\SubsectionNumber}{0.4em}{}
  
\begin{document}

\sloppy

\includepdf[pages=-]{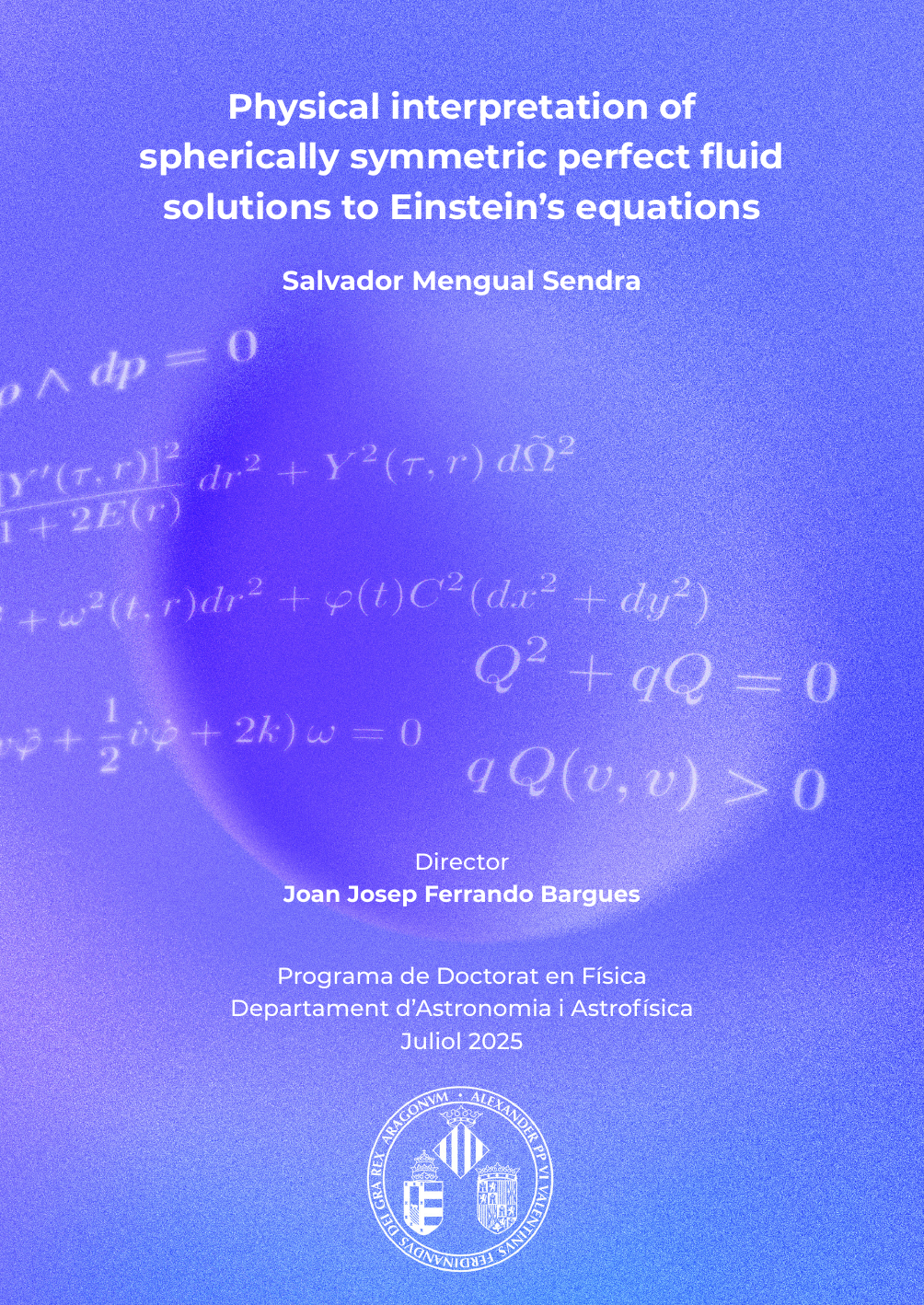}

\newpage
\thispagestyle{empty}
\null

\includepdf[pages=-]{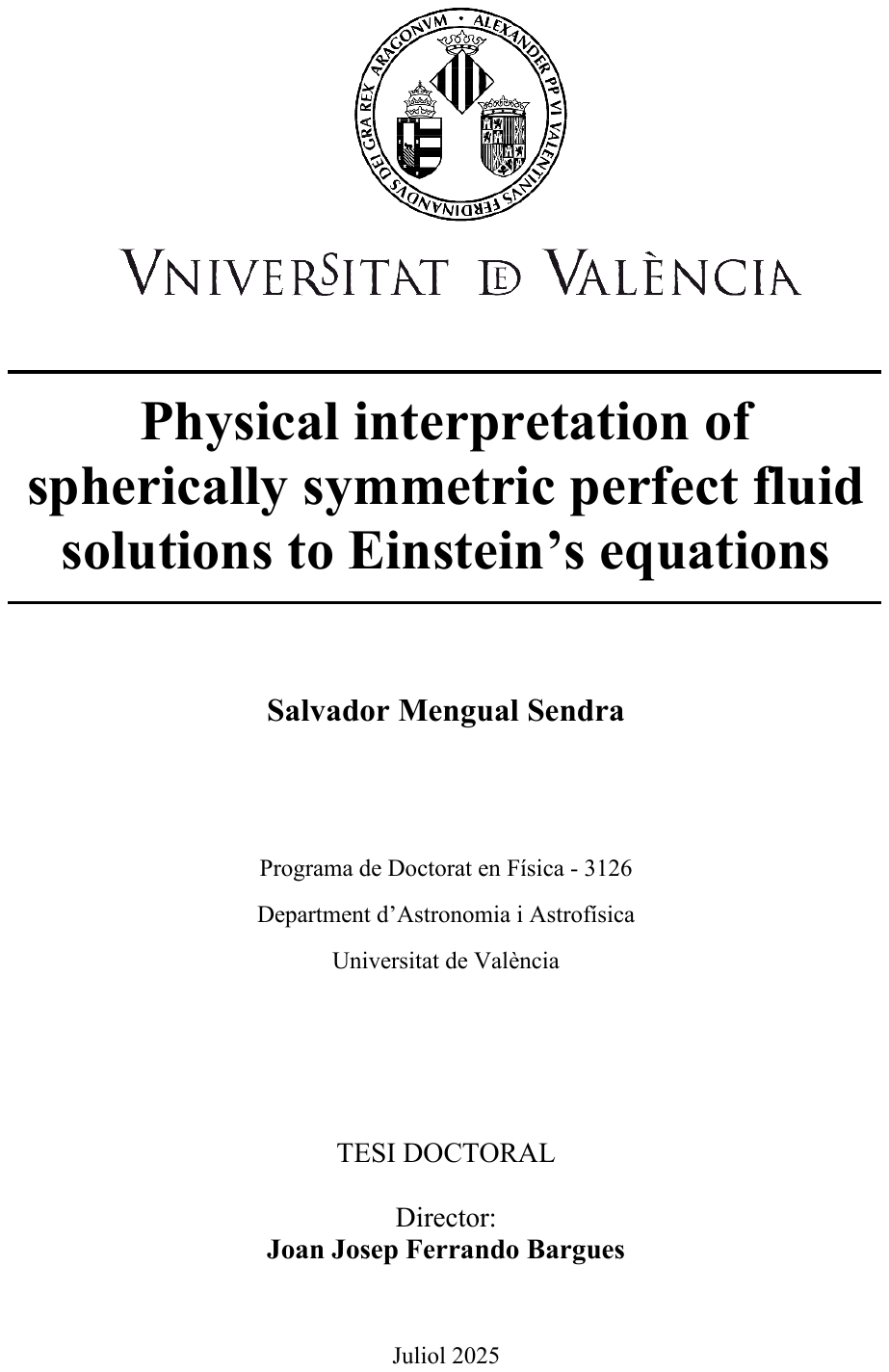}

\newpage
\thispagestyle{empty}
\null

\newpage
\thispagestyle{empty}

\begin{flushleft}
\begin{minipage}{0.5\textwidth}
\vspace{1cm}
“This is insanity!" \\
“No, this is scholarship!” \\[1em]
\raggedright ― Shallan Davar, \\ \textit{Words of Radiance} by Brandon Sanderson
\end{minipage}
\end{flushleft}

\newpage
\thispagestyle{empty}
\null

\begin{otherlanguage}{catalan}

\newpage

\pagenumbering{Roman}
\section*{Agraïments}
Per algun motiu, he posposat escriure els agraïments fins que no m'ha quedat més remei. Mentre escric aquestes línies, i potser aquest era el motiu inconscient de la meua reticència, m'adone que, tot i haver acabat ja d'escriure la tesi, m'està resultant més difícil trobar les paraules per a agrair, i fer justícia, a totes les persones que, d'una manera o d'una altra, m'han ajudat a arribar fins ací. Però ho vull intentar. \\ \\
He de començar per Joan, perquè sense ell aquesta tesi no hauria sigut ni remotament possible. Però és que, a més, has fet el procés d'allò més fàcil de dur, fins i tot divertit. Gràcies per la teua proximitat, per la teua comprensió i pel teu suport incondicional. \\ \\
També vull donar les gràcies a tot el Departament d'Astronomia i Astrofísica, perquè la familiaritat que m'heu fet sentir no es troba en cap altre departament. Cal fer una menció especial al personal de secretaria, Manel, Arancha i Ana, que a banda de ser un dels principals responsables d'aquest sentiment d'acollida, també fan un treball impecable amb el pitjor malson d'un científic: la burocràcia. \\ \\
Als companys de despatx de la primera planta: Ignasi, Adam (a qui no cal que li traduïsca estes línies) i l'\textit{alto príncipe}, Óscar; i als de la quarta planta: Kiara, Marco i Yannik. També a la resta del \textit{comando dinaor}: María, Isac, Mónica i Pau. Gràcies a tots pels moments surrealistes (a l'ascensor i fora d'ell) que fan el dia a dia més amé. \\ \\
I would also like to thank all members of the Thesis Committee for accepting such a by no means negligible task. Many thanks to Juan Antonio Morales Lladosa, Alfonso García-Parrado Gómez-Lobo, Alena Pravdová, Susana Planelles Mira, Raül Vera Jiménez and Vojt\v{e}ch Pravda. \\ \\
A Alfonso, además, debo agradecerle su hospitalidad durante mi estancia en Córdoba, así como su esfuerzo por transmitirme sus conocimientos sobre el funcionamiento de \textit{xAct} y \textit{Mathematica} en general. La parte III de esta tesis no sería lo que es sin su ayuda. \\ \\
I'm also grateful to Igor Khavkine, for hosting me in Prague during my stay there, for his understanding and for his patience teaching me rigorous maths. We still have a lot of work ahead. \\ \\
No puc acabar amb l'àmbit científic sense agrair a la gent de Bilbao, per acollir-me en el seu grupet cada any als EREP. \\ \\
Però no tot és el món acadèmic, i és que, des que vaig arribar a València fa ja deu anys (socorro), s'han creuat pel meu camí persones a qui també els dec molt: Javi, Sara, Paula, Joan, Saül, Salva, Rasmi, Xavi, Calata i Pedro; i, més recentment, Carolina, Edu i Noé. \\ \\
I a Vicent, no solament per la meravellosa portada que t'has currat i les exhaustives revisions estilístiques de la tesi, sinó també per acompanyar-me durant aquest procés, per aguantar les meues \textit{frikades}, animar-les i, fins i tot, formar-ne part. I per tot el que queda. \\ \\
Quina millor manera de tancar els agraïments que tornant a casa? A Joan, Júlia, David, Maria, Judith, Mar, Rosalia i Naomi, per ser insoportables, tots, i per ser casa. \\ \\
Finalment, a tota la meua família i, en especial, als meus pares, per tot. \\ \\
\begin{flushright}
Amb molta estima, \\
Salva
\end{flushright}

\newpage

\section*{Resum$\vspace{-1mm}$}
La teoria de la Relativitat General d'Albert Einstein va revolucionar la comprensió prèvia que es tenia del funcionament de la gravetat, substituint la concepció newtoniana, segons la qual aquesta és una força, per una interpretació geomètrica: la gravetat apareix com una conseqüència de la deformació de l'espaitemps deguda a la presència de matèria i energia. La relació entre aquests dos elements és descrita de manera matemàtica per les equacions de camp d'Einstein, un sistema d'equacions diferencials acoblades altament no lineals. A causa de la complexitat d'aquestes equacions, totes les famílies de solucions que es coneixen han estat obtingudes per a casos particulars imposant certes simetries, restriccions al contingut de matèria o altres condicions que en simplifiquen la integració. \\ \\[-1mm]
La mètrica de Schwarzschild fou la primera solució exacta de les \mbox{equacions d'Einstein,} obtinguda l'any 1916 sota les assumpcions d'absència de constant cosmològica, simetria esfèrica i per a buit. Poc després, es generalitzaria aquesta solució per a incorporar un camp electromagnètic i constant cosmològica. Diverses dècades després, s'inclouria també la rotació de la font i càrrega elèctrica. Més enllà de les solucions de buit, la solució de l'interior de Schwarzschild estén la solució de Schwarzschild per a tenir en compte la presència d'un fluid perfecte estàtic i amb simetria esfèrica de densitat d'energia constant. La cerca de solucions amb densitat d'energia no constant va donar lloc al desenvolupament de les equacions d'estructura este\lgem ar. \\ \\[-1mm]
Les solucions anteriors modelitzen el camp gravitatòri d'objectes compactes, però una de les primeres intencions d'Einstein a l'hora de buscar solucions de la seua teoria fou construir un model de l'Univers. En aquest context, se sol considerar que l'Univers a gran escala es pot aproximar com un medi continu. Així, en el seu primer model de l'Univers, Einstein va considerar que aquest era estàtic, homogeni i isòtrop, introduint la constant cosmològica per a contrarestar el co\lgem apse gravitatòri. Tant aquest model, com el proposat poc després per de Sitter per poder explicar els desplaçaments al roig observats, assumien que l'Univers és estàtic. Friedmann fou el primer en estudiar un model no estàtic, fins i tot abans que s'observara l'expansió de l'Univers. Després d'aquest descobriment, aquesta idea va adquirir rellevància i fou desenvolupada amb més detall, donant lloc al que avui dia es coneixen com els models de Friedmann-Lemaître-Robertson-Walker (FLRW). \\
Fins eixe moment, s'havien buscat solucions cosmològiques que modelitzaren un Univers homogeni i isòtrop, el que es coneix com el principi cosmològic. Amb el pas del temps, però, es van començar a considerar models basats en principis cosmològics menys restrictius, com les solucions homogènies però anisòtropes o aquelles en què el principi cosmològic se satisfà solament en la part espacial de la mètrica. També s'han desenvolupat solucions que abandonen la hipòtesi de l'homogeneïtat amb l'objectiu de modelitzar situacions físiques més diverses, com estructures cosmològiques o interiors este\lgem ars. \\ \\
Un conjunt de solucions (inhomogènies) que ha tingut un paper fonamental a l'hora de desenvolupar la teoria de la Relativitat General és el dels espaitemps amb simetria esfèrica, atès que molts escenaris físics importants presenten aquesta geometria. Les solucions en què la font és un fluid perfecte també han sigut considerades per molts autors, sent les solucions de FLRW l'exemple paradigmàtic. Un fluid perfecte és un fluid en què els coeficients de transport (el coeficient de conductivitat tèrmica i els coeficients de viscositat) són negligibles i la força de tensions és perpendicular a la superfície de contacte. \\ \\
Les solucions estàtiques de fluid perfecte amb simetria esfèrica constitueixen el model més simple per a l'estudi de l'estructura de l'interior este\lgem ar. El model de Lemaître-Tolman, una solució amb simetria esfèrica i pressió nu\lgem a, també s'ha emprat per a analitzar tant el co\lgem apse gravitacional d'una estrella com inhomogeneïtats cosmològiques. Així mateix, nombrosos treballs han explorat altres espaitemps amb simetria esfèrica no estàtics en què la font és un fluid perfecte, i es coneix un nombre considerable de solucions exactes d'aquest tipus. No obstant això, com ocorre en tots els camps de la física, una solució d'una equació pot no ser físicament admissible, i moltes de les solucions esmentades anteriorment no tenen una interpretació clara com a fluids físicament realistes. \\ \\
Aquest fet motiva l'objectiu d'aquesta tesi, estudiar la possible interpretació física de les solucions de fluid perfecte que admeten un grup tridimensional d'isometries G$_3$ actuant sobre òrbites bidimensionals S$_2$ (G$_3/$S$_2$), és a dir, solucions de fluid perfecte amb simetria esfèrica, plana o hiperbòlica. \\ \\ 
Les mètriques pertanyents a aquesta família de solucions es poden agrupar en els models T i els models R, segons si el gradient de la curvatura de les òrbites és tangent o no al flux del fluid. Tots els models T són geodèsics i estan inclosos dins de la família de solucions de Szekeres-Szafron de classe II. Dins dels models R, aquells que són geodèsics corresponen a les mètriques de Lemaître-Tolman, la subfamília de les solucions de Szekeres-Szafron de classe I amb un G$_3/$S$_2$. Pel que fa als models R no geodèsics, aquests inclouen la subfamília de solucions de Stephani-Barnes que admet un G$_3/$S$_2$ i altres famílies de solucions que, a diferència de les anteriors, no tenen com a límit les mètriques de FLRW. \\ \\
En la primera part d'aquesta tesi es fa un recull dels resultats previs necessaris per a donar una interpretació física a un tensor d'energia perfecte i s'expandeixen amb la nostra pròpia aportació. L'objectiu de la segona part és fer un pas més en l'estudi de la interpretació física de les solucions de fluid perfecte que admeten un G$_3/$S$_2$. En la tercera part de la tesi, es presenta una eina per al programa \textit{Mathematica} dissenyada per facilitar l'estudi de solucions exactes de les equacions d'Einstein.

	\subsubsection*{Interpretació física d'un tensor d'energia perfecte}
	El Capítol 2 recopila els fragments de treballs anteriors necessaris per a abordar el contingut de les dues primeres parts de la tesi. En primer lloc, s'hi explica com estudiar la possible interpretació física d'un tensor solució de les equacions de conservació de fluid perfecte, al qual anomenarem \textit{tensor d'energia perfecte}. S'hi posa especial èmfasi en la distinció entre les magnituds físiques del fluid que intervenen en les equacions de la hidrodinàmica, les \textit{magnituds hidrodinàmiques}, i aquelles que cal introduir per poder interpretar les solucions d'aquestes equacions com a fluids en equilibri termodinàmic local, les \textit{magnituds termodinàmiques}. Les primeres inclouen el \textit{flux del fluid}, la \textit{densitat d'energia} i la \textit{pressió} del fluid, i en descriuen l'evolució, mentre que cada conjunt de les segones, anomenat \textit{esquema termodinàmic}, proporciona una interpretació termodinàmica concreta de la solució. Aquests estan formats per la \textit{densitat de matèria}, l'\textit{energia interna específica}, l'\textit{entropia específica} i la \textit{temperatura}. Tot seguit, s'indica com determinar quan un conjunt de magnituds hidrodinàmiques admet esquemes termodinàmics associats (\textit{problema directe}) i com obtindre'ls (\textit{problema invers}). \\
	La \textit{condició sònica} dona la solució del problema directe fent ús, únicament, de magnituds hidrodinàmiques. La condició diu que l'anomenada \textit{funció indicatriu} del fluid és una funció de la densitat d'energia i de la pressió. En eixe cas, a més, aquesta representa el quadrat de la velocitat del so. Això permet determinar d'una manera molt directa quan una solució de les equacions d'Einstein de fluid perfecte admet esquemes termodinàmics associats. Les anomenarem \textit{solucions termodinàmiques}. \\ \\ 
	A banda de la condició d'equilibri termodinàmic local, cal imposar una sèrie de restriccions addicionals perquè les solucions siguen físicament admissibles. En primer lloc, cal considerar les \textit{condicions d'energia} de Pleba\'nski. Segons aquestes, per a tot observador la densitat d'energia del fluid ha de ser positiva i el caràcter causal del flux de moment ha de ser temporal o nul. En segon lloc, cal demanar que els esquemes termodinàmics tinguen una temperatura, densitat de matèria i energia interna específica positives. Finalment, per a poder desenvolupar una teoria consistent de les ones de xoc al fluid, s'han d'imposar les \textit{condicions de compressibilitat}. Aquestes requereixen que la velocitat de l'ona siga inferior a la de la llum, que hi haja un increment de l'entropia durant el xoc i que l'estat posterior a aquest quede unívocament determinat per l'estat anterior. \\ \\
	A la part final del capítol es presenten diverses ferramentes per estudiar la compatibilitat de les solucions amb algunes interpretacions concretes. D'una banda, per a que una solució siga compatible amb l'equació d'estat d'un gas ideal genèric s'ha de verificar la \textit{condició sònica ideal}. En eixe cas, s'hi mostra com obtindre l'esquema termodinàmic corresponent al gas ideal genèric, que anomenarem \textit{esquema termodinàmic ideal}. També es presenten les expressions que les condicions de realitat física prenen per a aquest cas particular. D'altra banda, per a que una solució siga compatible amb l'equació d'estat d'un gas ideal clàssic, la seua funció indicatriu ha de prendre una forma concreta. A més a més, un vector unitari temporal, geodèsic i amb expansió és el flux d'un gas ideal clàssic si, i sols si, no té vorticitat i l'expansió és homogènia. Finalment, s'enumeren les condicions que s'han de satisfer per a que una solució siga compatible amb la presència de coeficients de transport no nuls. En eixe cas, s'estaria modelitzant un fluid no perfecte experimentant una evolució que es pot descriure mitjançant el tensor d'energia d'un fluid perfecte. \\ \\
	La nostra aportació a aquest estudi queda recollida en el Capítol 3. En aquest, es pretén obtindre la condició que caracteritza la compatibilitat d'una solució amb l'equació d'estat d'un gas de Synge, un gas ultrarelativista no degenerat. Aquesta equació d'estat, que obtingué Synge a partir de la teoria cinètica microscòpica integrant el límit ultrarelativista de la funció de distribució de Maxwell-Boltzmann, queda escrita en termes de funcions de Bessel modificades de segona espècie. Així doncs, el primer que es fa és obtindre una caracterització purament hidrodinàmica d'aquesta equació d'estat a través d'una equació diferencial per a la funció indicatriu. També s'hi obtenen altres caracteritzacions no necessàriament hidrodinàmiques emprant  l'energia interna específica i el coeficient adiabàtic generalitzat. Tot això, permet explorar a continuació el comportament del gas de Synge a altes i baixes temperatures, així com establir-ne la teoria a la Rainich corresponent. \\ \\
	La presència de les funcions de Bessel modificades de segona espècie en l'equació d'estat del gas de Synge ha fet que en la literatura s'hagen buscat diferents maneres d'aproximar-la en els límits a altes o baixes temperatures. En la següent part del capítol, s'analitzen diverses d'aquestes aproximacions i s'empra l'estudi hidrodinàmic anterior per a obtindre noves aproximacions acceptables en tot el rang de temperatures. En particular, es recupera l'aproximació de Taub-Mathews i es generalitza per a qualsevol valor del coeficient adiabàtic $\gamma$. \\ \\
	Finalment, es considera l'evolució isentròpica d'un gas ideal i s'aplica aquest estudi per a obtindre l'equació de Friedmann per a un gas ideal de TM.

	\subsubsection*{Interpretació física de solucions de fluid perfecte amb simetria esfèrica}
	Al Capítol 4 s'estudia la termodinàmica dels models T. L'anàlisi comença amb l'obtenció de les expressions de les magnituds hidrodinàmiques i de la funció indicatriu, en termes de les funcions coordenades de la mètrica, juntament amb les expressions generals dels esquemes termodinàmics associats. La riquesa dels diversos esquemes termodinàmics admissibles ve donada per dues funcions arbitràries \mbox{d'una funció d'estat.} \\ \\ \\ \\
	A continuació, ens centrem en els models T ideals, és a dir, aquells que presenten les propietats hidrodinàmiques d'un gas ideal genèric. Imposant la condició sònica ideal, es determinen les condicions que caracteritzen aquesta subfamília i es particularitzen, per a aquest cas, les expressions de les magnituds obtingudes ante\mbox{rior}ment. Les expressions de les funcions mètriques i de les variables hidrodinàmiques queden en funció d'un paràmetre $\tilde{\gamma}$ i una funció arbitrària de la coordenada espacial. A més, es presenten els esquemes termodinàmics corresponents a tres models particulars: l'esquema termodinàmic ideal, el d'un model de Lima-Tiomno i el dels models amb la temperatura del límit FLRW. \\ \\
	L'anàlisi de les condicions de realitat física mostra que, per a diferents valors del paràmetre $\tilde{\gamma}$, existeixen diferents regions de l'espaitemps en què aquestes se satisfan. A més, s'hi estudia l'evolució dels models en aquestes regions. \\ \\ 
	Finalment, s'obté la família de solucions ideals que generalitza la solució de McVittie-Wiltshire-Herlt amb curvatura arbitrària i es demostra que no pot modelitzar un fluid perfecte en equilibri termodinàmic local. Així mateix, s'hi argumenta que cap model T no pot descriure un gas ideal clàssic, ja que la seua expansió no és homogènia. \\ \\
	El capítol continua amb l'estudi de les equacions de camp dels models T. En primer lloc, s'analitzen en detall i s'exposa l'algoritme d'integració de Herlt, que permet obtindre la solució del cas amb simetria esfèrica mitjançant quatre quadratures. Una revisió de l'algoritme ens permet demostrar que pot ser generalitzat per a incloure les tres simetries possibles i que, en realitat, només és necessari resoldre dues quadratures. \\ \\
	A partir d'aquest mètode, i mitjançant un canvi de variables i una elecció de la coordenada temporal adequats, es reformula l'algoritme per tal que, en el cas de simetria plana, permeta obtindre la solució sense resoldre cap quadratura. Així, s'empra aquest mètode tant per a recuperar solucions ja conegudes com per a obtindre'n una de nova. \\ \\
	Tot seguit, es proposen dos algoritmes nous per obtindre solucions amb simetria esfèrica o hiperbòlica que tampoc involucren cap quadratura. A tall d'exemple, s'obtenen algunes solucions ja conegudes i una de nova per al cas de simetria esfèrica. També es mostra que totes les solucions noves trobades inclouen dominis espaitemporals en què les condicions de realitat física se satisfan. \\ \\
	Per concloure el capítol, s'estudien les solucions de Kompaneets-Chernov-Kantowski-Sachs (KCKS), que corresponen al límit homogeni dels models T. Així, en aquesta darrera part s'exploren en detall les seues interpretacions com a límit isentròpic dels models T i com a evolució isentròpica d'altres models compatibles amb l'evolució d'un gas ideal clàssic. \\ \\
	A continuació, es passa a l'estudi dels models R. Concretament, al Capítol 5 s'analitzen des d'un punt de vista termodinàmic els models R que admeten una sincronització plana. Totes aquestes solucions són geodèsiques i, per tant, coincideixen amb les mètriques de Lemaître-Tolman (LT) espacialment planes. Per començar, s'hi fa un comentari sobre els diferents mètodes que es poden emprar per resoldre les equacions de camp per a aquest cas i, com a exemple, s'obté la solució dels models de LT plans amb pressió nu\lgem a i constant cosmològica. \\ \\ 
	Després, es duu a terme l'estudi termodinàmic general dels models R que admeten una sincronització plana: s'obtenen les expressions de les magnituds hidrodinàmiques i de la funció indicatriu, i es determinen les expressions dels esquemes termodinàmics compatibles. \\ \\ 
	Per aprofundir en la interpretació física de les solucions, cal disposar d'expressions més concretes d'aquestes magnituds. Per tant, en la següent part del capítol es consideren els models compatibles amb l'evolució d'un gas ideal genèric. La imposició de la condició sònica ideal permet obtindre l'equació que aquesta subfamília de solucions ha de verificar, una equació diferencial ordinària per a dues funcions del temps pròpi $f(\tau)$ i $g(\tau)$ i una funció d'una funció arbitrària de la coordenada espacial. Tot seguit, s'analitza la possibilitat de trobar-ne solucions mitjançant diversos procediments i s'estudia la compatibilitat de dues famílies particulars de solucions, els models de Szafron $f(\tau) = \tau^q$ amb $q \neq 1/2$ i els models amb $f(\tau) = \sqrt{\tau}$. En ambdós casos, s'analitza en detall el comportament dels models resultants particularitzant les expressions de les magnituds generals i s'obtenen les regions espaitemporals en què les condicions de realitat física se satisfan. \\
	Per acabar, s'investiguen les condicions que determinen quan els models R són consistents amb un coeficient de conductivitat tèrmica no nul. Les dues famílies de solucions ideals considerades anteriorment, per exemple, no ho són. \\ \\
	La segona part de la tesi finalitza amb el Capítol 6, dedicat a estudiar els universos de Stephani termodinàmics. Com en els capítols anteriors, el primer pas consisteix a determinar les expressions generals de les magnituds hidrodinàmiques, la funció indicatriu i els esquemes termodinàmics associats. \\ \\
	Per aprofundir en la interpretació física d'aquests models, cal introduir restriccions addicionals que tinguen cert interés físic. En primer lloc, es consideren els models compatibles amb l'equació d'estat de gas ideal genèric, els anomenats universos de Stephani ideals. Aquesta família de solucions, que depèn de cinc paràmetres, ja ha estat estudiada en el passat, per la qual cosa ací ens limitem a resumir els resultats necessaris per a dur a terme el nostre objectiu. L'estudi de les equacions de camp ens força a distingir els \textit{models regulars} i els \textit{models singulars}. En segon lloc s'analitza la compatibilitat de les solucions amb un coeficient de conductivitat tèrmica no nul, i es demostra que únicament els models de FLRW ho permeten. A continuació, s'estudien les condicions que cal imposar per tal d'aproximar un gas ideal clàssic a baixes temperatures fins a primer ordre. Finalment, es determinen les restriccions necessàries perquè les solucions tinguen un bon comportament a altes temperatures també fins a primer ordre. \\ \\
	A la següent part del capítol s'exploren els models en què un observador comòbil amb la velocitat del fluid pot mesurar radiació prova isòtropa. Aquesta condició és especialment interessant a causa de l'alt grau d'isotropia que s'observa en la radiació del fons còsmic de microones. Així doncs, es determinen les restriccions que aquesta condició imposa en la mètrica i s'obté la seua expressió i la de les magnituds hidrodinàmiques per al cas de simetria esfèrica. La mètrica de Dabrowski pertany a aquesta família de solucions, però s'hi demostra que aquesta mètrica en particular no satisfà les condicions de compressibilitat, necessàries per a poder ser interpretada com un fluid en equilibri termodinàmic local. \\ \\ \\
	En aquesta part s'estudien també algunes propietats generals de la subfamília compatible amb l'equació d'estat d'un gas ideal genèric, com ho són l'expressió de la funció indicatriu corresponent i els dominis en què les condicions d'energia i les de compressibilitat se satisfan. També s'hi obté l'equació de Friedmann generalitzada per al cas de gas ideal compatible amb un observador comòbil que mesura radiació isòtropa, que dona lloc a tres casos diferents. \\ \\
	Finalment, es tenen en compte tots els resultats anteriors per identificar els universos de Stephani que poden representar un gas ultrarelativista amb radiació isòtropa mesurada per l'observador comòbil. Aquesta família de solucions aproxima un gas de Synge a altes temperatures a primer ordre i satisfà les condicions de compressibilitat. S'exploren en detall els models singulars, per als quals s'obtenen les evolucions temporals i els perfils radials de les magnituds termodinàmiques associades. Això permet delimitar les regions de l'espaitemps en què se satisfan les condicions d'energia. Per concloure, s'estudia formalment l'equació de Friedmann generalitzada corresponent a aquesta família de solucions.
	
	\subsubsection*{Caracteritzacions IDEALs de fluid perfecte i d'espaitemps amb \textit{xAct}}
	El Capítol 7 comença plantejant el problema de, donat un sistema d'equacions (diferencials) per a dos conjunts de funcions, ${\cal D}_n(\phi_1, ..., \phi_r, \psi_1, ..., \psi_q) = 0$, trobar-ne un d'equivalent que involucre sols un dels dos conjunts, $\hat{{\cal D}}_m(\phi_1, ..., \phi_r) = 0$, que anomenarem \textit{sistema condicional}. La condició sònica, per exemple, resol aquest problema per al cas de la caracterització hidrodinámica de l'equilibri termodinàmic local. Per tant, estendrem l'ús del terme \textit{problema directe} per a fer referència a aquesta situació en general. El \textit{problema invers} consisteix aleshores a, una vegada resolt el sistema condicional, trobar les funcions $\psi_i$ restants ($i = 1, ..., q$) que resolen el sistema d'equacions inicial. \\ \\
	Aquesta manera de procedir es pot emprar per a resoldre el problema de \mbox{l'equivalència} de solucions de les equacions d'Einstein, de manera que s'obtinguen condicions que identifiquen un determinat espaitemps emprant únicament concomitants del tensor mètric. Això dona lloc a les \textit{caracteritzacions IDEALs}, és a dir caracteritzacions Intrínsiques (depenen únicament del tensor mètric), Deductives (no involucren cap procés d'inferència), Explícites (totes les expressions necessàries estan donades explícitament en funció de la mètrica) i ALgorítmiques (es poden escriure en forma de diagrama de flux amb un nombre finit de passos). \\ \\
	Aquesta filosofia es pot emprar també per a determinar algorítmicament expressions per a magnituds intrínsiques de determinats espaitemps, emprant exclusivament concomitants del tensor mètric. Aquests procediments seran anomenats \textit{determinacions IDEALs}. \\ \\
	A la segona part del capítol es presenta el concepte d'\textit{àlgebra computacional}, que fa referència a la teoria i implementació de programes d'ordinador dissenyats per a dur a terme les manipulacions i càlculs simbòlics típics en matemàtiques. Es fa un breu resum de la motivació de l'ús d'aquest tipus de programes en Relativitat General i de les diferents possibilitats de què disposen els usuaris en l'actualitat, parant especial atenció a \textit{Mathematica} i, en concret, a \textit{xAct}. \\ \\
	\textit{xAct} és un conjunt de paquets per a \textit{Mathematica} dissenyat per a la manipulació simbòlica de tensors. Incorpora funcionalitats tant per a la manipulació d'expressions generals que involucren índex abstractes com per a evaluar components tensorials en un sistema de coordenades i una mètrica concrets. Com es pot comprovar per l'àmplia literatura d'articles i tesis que n'han fet ús, es tracta d'una ferramenta molt popular i diversos autors han desenvolupat mòduls addicionals per estendre les seues funcions en direccions molt variades. \\ \\
	El capítol conclou presentant \textit{xIdeal}, el paquet per a \textit{xAct} que hem desenvolupat en co\lgem aboració amb A. García-Parrado. L'objectiu és incorporar en aquest paquet les diverses caracteritzacions i determinacions IDEALs disponibles en la literatura. A més a més, \textit{xIdeal} també disposa d'una base de dades de mètriques. L'objectiu és que l'usuari puga carregar la mètrica desitjada expressada en diferents sistemes de coordenades sense haver d'introduir-la manualment. A més a més, s'inclou una ferramenta per a que el públic puga guardar-hi les seues pròpies mètriques en cas de no estar en la base de dades. A més a més, cadascuna de les mètriques incloses va acompanyada d'un llistat de propietats conegudes al qual es pot accedir a voluntat. \\ \\ \\
	Al Capítol 8 es recullen i expliquen breument les funcions d'\textit{xIdeal} que implementen caracteritzacions i determinacions IDEALs de propietats relacionades amb fluids perfectes. Les caracteritzacions de fluid perfecte, fluid perfecte termodinàmic i gas ideal genèric estan implementades mitjançant les funcions \texttt{PerfectFluidQ}, \texttt{ThermodynamicPerfectFluidQ} i \texttt{GenericIdealGasQ} respectivament. \textit{xIdeal} també disposa de la funció \texttt{PerfectFluidVariables}, que torna les expressions de les variables hidrodinàmiques si se li proporciona com a entrada una mètrica de fluid perfecte. \\ \\
	També s'hi presenten les funcions que incorporen les caracteritzacions IDEALs dels universos de Stephani, els universos de FLRW i els espaitemps de Kustaanheimo-Qvist. Aquestes són, respectivament, \texttt{StephaniUniverseQ}, \texttt{FriedmannQ} i \texttt{KustaanheimoQvistQ}. \\ \\
	Per a concloure el capítol es presenta, de manera resumida, la caracterització IDEAL del flux d'un fluid perfecte conservat. A tal efecte, s'hi obté una classificació dels vectors unitaris temporals que es mostra en forma de taula i de diagrama de flux. Per a cada classe, es determinen les condicions necessàries i suficients que ha de verificar el vector per modelitzar el flux d'un fluid perfecte conservat i es donen les expressions de la pressió i la densitat d'energia associades. Tot i que aquesta caracterització no està incorporada en el nostre paquet encara, hi desenvolupem en detall alguns casos rellevants als quals l'hem aplicada: el cas d'un vector de Killing, un flux radial en un espaitemps amb simetria esfèrica, un fluid perfecte estacionari i amb simetria axial i, finalment, el flux d'un univers de Stephani. \\ \\
	El Capítol 9 recull la resta de funcions de què disposa ara mateix el paquet \textit{xIdeal}. En primer lloc tenim les funcions que determinen el tipus Petrov-Bel i les expressions de les direccions múltiples de Debever de la mètrica considerada, \texttt{PetrovType} i \texttt{DebeverNullDirections}. Ambdues funcions poden fer la seua funció emprant dos mètodes diferents, el primer basat en el tensor de Weyl autodual i el segon basat en la matriu de Petrov. S'hi expliquen els dos mètodes i, a continuació, s'apliquen a una mètrica de tipus Petrov II i a la mètrica de Kerr-NUT com a exemple. \\ \\ \\
	En segon lloc es presenta la determinació IDEAL de la dimensió del grup d'isometries d'una solució amb un sistema de referència invariant. El pas fonamental per a dur a terme aquesta determinació és obtindre el \textit{tensor de connexió}. S'hi explica com fer-ho per a solucions de tipus Petrov-Bel I, II i III i per a aquelles en què es coneix explícitament el sistema de referència invariant. La funció d'\textit{xIdeal} que incorpora aquest procediment s'anomena \texttt{ConnectionTensor}, i la que utilitza el tensor de connexió resultant per a determinar la dimensió del grup, \texttt{IsometryGroupDimension}. De nou, es consideren dues solucions a les que s'apliquen aquestes funcions per a exemplificar el seu funcionament: la solució homogènia i de buit de Petrov i la solució conformement plana i de radiació pura de Wills. \\ \\
	Per al cas de solucions de tipus Petrov-Bel D, N o O, la determinació del nombre d'isometries requereix una anàlisi més ampla. En aquest capítol, es fa un resum de l'estudi del cas de les solucions de tipus N de buit amb constant cosmològica: es fa una classificació d'aquestes mètriques i, per a aquelles que admeten un sistema de referència invariant, s'explica com obtindre el tensor de connexió corresponent. Per a la resta, es determina la dimensió del grup d'isometries. Aquesta recepta no està incorporada encara al nostre paquet, però sí que l'hem emprada en tres famílies de solucions per comprovar la seua eficàcia: les ones pp, les ones de Kundt i les solucions de Siklos. Aquests exemples es desenvolupen en detall. \\ \\
	La següent caracterització IDEAL que es considera en aquest capítol és la de les cosmologies espacialment homogènies. De nou, en el procés cal considerar diferents classes i, per a cadascuna d'elles, es donen les condicions necessàries i suficients perquè la solució corresponga a una cosmologia espacialment homogènia. Aquesta caracterització tampoc està automatitzada encara en forma de funció per al nostre paquet, però també s'ha aplicat a dos exemples que es detallen al final d'aquesta part del capítol: les cosmologies de tipus Bianchi I i les solucions homogènies d'Ozsváth de classes II i III. \\ \\
	Les dues darreres caracteritzacions que incorpora \textit{xIdeal} són les de les mètriques de Schwarzschild i Kerr. Aquestes es recullen al final del capítol i estan implementades a través de les funcions \texttt{StaticVacuumTypeDClassify} i \texttt{KerrSolutionQ}. 
	
	\subsubsection*{Conclusions i perspectives de futur}
	Aquesta tesi recull els principals resultats obtinguts durant el doctorat. Estructurada en tres parts, s'hi investiguen dues grans línies dins de la Relativitat General. Les dues primeres parts es dediquen a entendre millor, des d’un punt de vista físic, les solucions de fluid perfecte de les equacions d’Einstein, i la tercera es dedica a la implementació de caracteritzacions i determinacions IDEALs d'espaitemps en \textit{xIdeal}, un paquet d'\textit{xAct}. \\ \\
	Al Capítol 3 s'han expandit els coneixements previs sobre la interpretació dels tensors d’energia de fluid perfecte, mitjançant l'aplicació d'un enfocament hidrodinàmic al gas de Synge. Aquesta visió és útil per a buscar solucions prova o sistemes autogravitants que modelitzen escenaris d’altes energies, com es demostra en els capítols posteriors. L’anàlisi conceptual ha permés, a més, formular una teoria tipus Rainich per a les solucions d’Einstein-Synge. Cal destacar ací que aquests teoremes de caracterització es poden modificar lleugerament per adaptar-se a altres equacions d’estat, com ara la de Taub-Mathews. \\ \\ 
	De l'anàlisi de l'aproximació de Taub-Mathews es conclou que les diferències en el quadrat de la velocitat del so són com a màxim del 2.36$\%$. També s'ha obtingut un procediment per a obtenir altres aproximacions analítiques de l’equació d'estat de Synge, que ha demostrat que la de Taub-Mathews actua com a cas límit. Com a resultat, s’han obtingut noves aproximacions amb més precisió que aquesta. \\ \\
	En el quart capítol s’han estudiat els models T. S’hi han obtingut les expressions de les magnituds hidrodinàmiques. Del seu estudi es dedueix que aquests models poden descriure fluids en equilibri termodinàmic local, i s'han derivat les expressions generals del quadrat de la velocitat del so i dels esquemes termodinàmics associats. \\ \\
	S'han determinat, a més a més, els models T compatibles amb l’equació d'estat d’un gas ideal genèric i s’han identificat els esquemes termodinàmics associats. D'aquest treball es pot extraure que, per exemple, un model T ideal aproxima bé un gas ultrarelativista. \\ \\
	Així mateix, s’ha analitzat des d’un punt de vista termodinàmic la solució McVittie-Wiltshire-Herlt, prèviament coneguda, i s’ha demostrat que no pot ser interpretada com a fluid perfecte en l.t.e. A més, s’ha provat que cap model T no pot representar un gas ideal clàssic. \\ \\
	A continuació, s’han analitzat les equacions de camp dels models T amb \mbox{l’objectiu} de trobar la seua solució general. Revisar l’algoritme d’integració de Herlt, ens ha dut a proposar-ne una versió modificada i altres algoritmes alternatius que eviten calcular quadratures. Això ha permés donar l’expressió explícita de la solució general, amb simetria plana, esfèrica o hiperbòlica. Emprat aquest resultat hem aconseguit recuperar solucions conegudes i obtindre'n de noves. \\
	El significat físic d’aquestes solucions es pot analitzar a posteriori amb l’enfocament hidrodinàmic, tot i que és convenient poder imposar condicions físiques prèviament, com s’ha fet amb els models T ideals. Això justifica l'utilitat de disposar de diversos mètodes d'integració de les equacions. \\ \\
	Tant les solucions de KCKS com les altres dues famílies de solucions termodinàmiques de Szekeres-Szafron de classe II (els models T en són la tercera) presenten equacions de camp amb expressions anàlogues a les dels models T. Per tant, els algoritmes d'integració deduïts en aquest capítol també es poden emprar en aquestes famílies. \\ \\
	Finalment, s’ha estudiat la subfamília de KCKS dels models T i s'ha mostrat que representen evolucions isentròpiques d’un fluid l’evolució no isentròpica del qual està descrita per un model T. S’han obtingut les relacions barotròpiques corresponents, que es poden interpretar com evolucions compatibles amb l'equació d'estat de gas ideal i la $\gamma$\textit{-law} relativista. D'aquest estudi també es conclou que l’evolució de gas ideal clàssic generalitza la solució FLRW per a un gas d'aquest tipus, permetent funcions d’escala amb diferents ritmes de creixement. \\ \\
	En el capítol següent, s'han tractat els models R amb sincronització plana ortogonal. S’ha demostrat que poden descriure fluids en equilibri termodinàmic local i s’han obtingut les magnituds hidrodinàmiques i els esquemes termodinàmics associats. S'hi ha derivat l’equació diferencial que ha de satisfer una solució compatible amb l'equació d'estat d'un gas ideal genèric i s’han estudiat dues solucions particulars. En ambdós casos s’han obtingut les expressions explícites de les magnituds físiques corresponents i s’han trobat dominis espaitemporals en què es compleixen les condicions de realitat física. A més a més, s’ha provat que aquestes solucions ideals no poden modelitzar un fluid amb coeficient tèrmic no nul, però queda oberta la possibilitat que altres models R sí que puguen. \\ \\
	El segon bloc es tanca amb l’estudi dels universos de Stephani, centrant-se en aquells que permeten a un observador comòbil detectar una radiació isotropa. Això reforça el fet, conegut de fa temps, que una solució inhomogènia pot ser compatible amb l'observació de radiació isotropa i inhomogènia. Sense pretendre construir models cosmològics observacionalment ajustats, s’ha mostrat que aquestes solucions podrien servir per modelitzar inhomogeneïtats locals del nostre Univers. \\ \\
	Després d’aquests estudis detallats, podem concloure que l’enfocament hidrodinàmic introduït en la primera part de la tesi ofereix un marc eficaç per assolir el nostre objectiu, ja que permet identificar les subfamílies dels tres models considerats que representen fluids físicament realistes amb diverses propietats d’interés. Tanmateix, això representa només el primer pas cap a l’objectiu més ampli de proporcionar una interpretació física del vast conjunt de solucions exactes disponibles en la literatura. \\ \\
	A més d’aprofundir en la nostra comprensió de la teoria de la Relativitat General, la qual cosa ja suposa un assoliment important, considerem que el nostre estudi també pot resultar d’interés per a la comunitat de Relativitat Numèrica. Tot i que és cert que les solucions numèriques poden ser més adequades per modelitzar determinats escenaris físics, les solucions analítiques obtingudes amb el nostre enfocament poden servir com a catàleg d’espaitemps de referència, potencialment útils per a posar a prova codis numèrics. \\ \\
	Això no obstant, al llarg d'aquest treball hem anat deixant obertes diverses qüestions que convé recopilar per a estudiar-les en un futur.
	En primer lloc, queda pendent emprar els algoritmes d'integració dels models T per a obtindre solucions que verifiquen condicions físiques rellevants imposades \textit{a priori}. \\
	De manera semblant, caldria estudiar en detall l'equació diferencial que caracteritza els models R espacialment plans ideals per obtenir noves solucions, així com determinar les formes funcionals de la funció espacial que generen perfils d'inhomogeneïtats realistes en les que s'hi han estudiat. El mateix ocorre amb l'equació que determina els models R espacialment plans compatibles amb un coeficient tèrmic no nul. \\ \\[-2mm] 
	Dels universos de Stephani termodinàmics, queden oberts l'estudi de les condicions d'enllaç dels models singulars realistes obtinguts amb solucions de pols i el dels models regulars. També seria convenient estudiar la compatibilitat amb un observador no comòbil que mesura radiació isòtropa, que es podria aprofitar dels resultats de l'estudi cinemàtic del final del Capítol 8. Aquests, a més a més, deixen oberta la possibilitat de generalitzar el teorema de Von-Zeipel sobre velocitats estacionàries amb simetria axial. \\ \\[-2mm]
	A banda de les plantejades al llarg de la tesi, hi ha diverses qüestions sense resoldre que cal tindre en ment per al futur. Per exemple, l'anàlisi de les solucions amb un G$_3/$S$_2$ que queden pendents. O també la de solucions amb simetria axial, molt utilitzades en l'estudi de nombroses situacions astrofísiques. \\ \\[-2mm]
	Pel que fa al paquet \textit{xIdeal}, ha resultat ser una eina útil i les tretze funcions que el componen ara mateix han demostrat funcionar correctament en la majoria de casos en què han estat posades a prova. Tot i això, és evident que el paquet encara necessita millores i que hi ha moltes caracteritzacions IDEALs pendents d'implementar. En aquest sentit, el seu desenvolupament futur representa una línia de treball oberta i prometedora. \\ \\[-2mm]
	Esperem que \textit{xIdeal} esdevinga un recurs útil per a la comunitat científica que treballa amb solucions exactes de les equacions d’Einstein. Entre els seus principals avantatges hi ha la possibilitat de discernir si una solució potencialment nova és, en realitat, una de coneguda expressada en un sistema de coordenades poc habitual; d’obtenir en qüestió de segons les expressions explícites de quantitats i propietats intrínseques de l’espaitemps considerat, i de proporcionar una base de dades de mètriques amb les seues propietats més rellevants. En última instància, mitjançant l’automatització i sistematització d’aspectes clau de l’enfocament IDEAL, esperem que el nostre paquet contribuïsca a avançar la recerca en Relativitat General.

\end{otherlanguage}

\newpage

\section*{Abstract}
Einstein's equations govern the dynamics of the spacetime geometry in the theory of General Relativity. They constitute a system of highly nonlinear coupled differential equations. As a consequence, all known families of particular solutions have been derived by imposing symmetry assumptions, matter content restrictions or other simplifying conditions that enable the integration of the equations. A class of solutions that has played an essential role in developing the General Relativity theory is that of spherically symmetric spacetimes. These solutions provide a simplified yet rich framework for studying a wide range of physical phenomena. However, despite the extensive literature dedicated to them, the interest in this topic has not waned nowadays, and several open questions are currently under study. \\ \\
Nevertheless, some of these solutions have been derived without specifying an equation of state, while others have been obtained in the dust case or by prescribing a (non-physical) time-dependence of the pressure, or also by imposing particular barotropic relations to close the system of equations. While these approaches have led to a broad spectrum of exact solutions, many of them lack a clear physical interpretation. Consequently, the objective of this thesis is the study of the potential physical viability of the spherically symmetric perfect fluid solutions, a study that can also be extended to the plane and hyperbolic symmetries. \\ \\
To this end, we collect previous results on the hydrodynamic approach, which offers the necessary tools to interpret a perfect fluid energy tensor as a fluid in local thermal equilibrium. In particular, the interpretations as a generic ideal gas, as a classical ideal gas and as a fluid with non-vanishing transport coefficients are studied. We extend this work to the case of an ultrarelativistic Synge gas and we provide a method to obtain different approximations of the Synge equation of state. \\ \\ 
Next, we apply these results to analyse the interpretation of three families of perfect fluid solutions as physically admissible fluids. We start with the T-models, the family of perfect fluid solutions admitting spherical, plane or hyperbolic symmetry whose curvature has a gradient tangent to the fluid flow. We continue with the R-models (the curvature's gradient is not tangent to the fluid flow) admitting a flat synchronisation orthogonal to the fluid flow, which are geodesic. We finish with the thermodynamic Stephani universes, the non-geodesic R-models belonging to the conformally flat subfamily of the Stephani-Barnes cosmological solutions. For the three of them we obtain the general expressions of the fluid flow, the energy density, the pressure and the square of the speed of sound, given by the indicatrix function. These provide the different evolutions that the considered solutions can represent. We also obtain the general expressions of the different sets of thermodynamic quantities (thermodynamic schemes) admitted: the matter density, the specific internal energy, the specific entropy and the temperature. Each of these provides a different interpretation of the solution as a specific fluid. To analyse whether they fulfil the necessary requirements for physical reality (the Pleba\'nski energy conditions, the positivity conditions and the compressibility conditions), we first determine the subfamilies satisfying additional physical requirements. The most relevant one is the compatibility with the equation of state of a generic ideal gas. The subfamilies obtained after imposing this condition are studied in detail, investigating their evolutions, radial profiles, spacetime singularities and associated thermodynamic schemes. In all three cases, we find wide spacetime domains in which the solutions can be interpreted as perfect fluids in local thermal equilibrium with the evolution of a generic ideal gas and fulfilling all the necessary requirements for physical viability. \\ \\
In the last part of the thesis we present \textit{xIdeal}, a custom-built \textit{Mathematica} tool designed to facilitate the study of exact solutions to the Einstein equations. This package has been developed in collaboration with García-Parrado and, at the moment, includes thirteen functions that implement different IDEAL (Intrinsic, Deductive, Explicit and ALgorithmic) spacetime characterisations and determinations. It also incorporates a database of metrics with their properties. We summarise all the algorithms currently included in \textit{xIdeal} and illustrate their use through concrete examples. We also present here some IDEAL algorithms that we have developed recently and are not yet implemented. \\ \\

\begin{otherlanguage}{catalan}
\section*{Síntesi}
Les equacions d'Einstein governen la dinàmica de la geometria de l'espaitemps en la teoria de la Relativitat General. Aquestes constitueixen un sistema d’equacions diferencials altament no lineals i acoblades. Com a conseqüència, totes les famílies de solucions particulars conegudes s’han obtingut imposant hipòtesis de simetria, restriccions sobre el contingut de matèria o altres condicions que permeten integrar les equacions. Una classe de solucions que ha tingut un paper essencial en el desenvolupament de la teoria de la Relativitat General és la dels espaitemps amb simetria esfèrica. Aquestes solucions proporcionen un marc simplificat però ric per a l’estudi d’una àmplia gamma de fenòmens físics. Tanmateix, malgrat l’extensa literatura dedicada a aquest tema, l’interés no ha minvat en l’actualitat, i diverses qüestions obertes es troben actualment en estudi. \\ \\
No obstant això, algunes d’aquestes solucions s’han obtingut sense especificar una equació d’estat, mentre que d’altres han estat derivades en el cas de pols, o prescrivint una dependència temporal (no física) de la pressió, o bé imposant relacions barotròpiques particulars per a tancar el sistema d’equacions. Encara que aquests enfocaments han permés obtindre un ampli espectre de solucions exactes, moltes d’elles manquen d’una interpretació física clara. En conseqüència, l’objectiu d’aquesta tesi és l’estudi de la possible viabilitat física de les solucions de fluid perfecte amb simetria esfèrica, un estudi que també es pot estendre a les simetries plana i hiperbòlica. \\ \\
Amb aquest fi, fem un recull de resultats anteriors sobre el plantejament hidrodinàmic, que ofereix les ferramentes necessàries per interpretar un tensor d’energia de fluid perfecte com un fluid en equilibri termodinàmic local. En particular, s'estudien les interpretacions com un gas ideal genèric, un gas ideal clàssic i un fluid amb coeficients de transport no nuls. A continuació, estenem aquest estudi al cas d’un gas de Synge ultrarelativista i proporcionem un mètode per obtindre diferents aproximacions de l’equació d’estat de Synge. \\ \\
A continuació, apliquem aquests resultats per analitzar la interpretació de tres famílies de solucions de fluid perfecte com a fluids físicament admissibles. Comencem pels models T, la família de solucions de fluid perfecte que admet simetria esfèrica, plana o hiperbòlica i en què la curvatura té un gradient tangent al flux del fluid. Continuem amb els models R (on el gradient de la curvatura no és tangent al flux del fluid) que admeten una sincronització plana ortogonal al flux, els quals són geodèsics. Finalment, estudiem els universos de Stephani termodinàmics, els models R no geodèsics que pertanyen a la subfamília conformement plana de les solucions cosmològiques de Stephani-Barnes. Per a les tres famílies, obtenim les expressions generals del flux del fluid, la densitat d’energia, la pressió i el quadrat de la velocitat del so, donada per la funció indicatriu. Aquestes proporcionen les diferents evolucions que poden representar les solucions considerades. També obtenim les expressions generals dels diferents conjunts de magnituds termodinàmiques (esquemes termodinàmics) admesos: la densitat de matèria, l’energia interna específica, l’entropia específica i la temperatura. Cadascun d’aquests proporciona una interpretació diferent de la solució com un fluid determinat. Per tal d’analitzar si compleixen els requisits necessaris per a la realitat física (les condicions d’energia de Pleba\'nski, les condicions de positivitat i les condicions de compressibilitat), determinem en primer lloc les subfamílies que satisfan requisits físics addicionals. El més rellevant és la compatibilitat amb l’equació d’estat d’un gas ideal genèric. Les subfamílies obtingudes després d’imposar aquesta condició són estudiades amb detall, investigant les seues evolucions, perfils radials, singularitats espaitemporals i esquemes termodinàmics associats. En els tres casos, trobem dominis espaitemporals amplis en què les solucions poden interpretar-se com fluids perfectes en equilibri termodinàmic local amb l’evolució d’un gas ideal genèric i que compleixen tots els requisits necessaris per a la seua viabilitat física. \\ \\
En l’última part de la tesi presentem \textit{xIdeal}, una eina personalitzada desenvolupada en \textit{Mathematica} per facilitar l’estudi de solucions exactes de les equacions d’Einstein. Aquest paquet ha estat desenvolupat en co\lgem aboració amb García-Parrado i actualment inclou tretze funcions que implementen diferents caracteritzacions i determinacions IDEALs (Intrínseques, Deduïdes, Explícites i ALgorítmiques) d'espaitemps. També incorpora una base de dades de mètriques amb les seues propietats. Resumim tots els algoritmes actualment inclosos en xIdeal i n’i\lgem ustrem l’ús amb exemples concrets. A més, presentem alguns algoritmes IDEALs que hem desenvolupat recentment i que encara no estan implementats. 
\end{otherlanguage}

\begin{otherlanguage}{spanish}
\section*{Síntesis}
Las ecuaciones de Einstein rigen la dinámica de la geometría del espaciotiempo en la teoría de la Relatividad General. Estas constituyen un sistema de ecuaciones diferenciales altamente no lineales y acopladas. Como consecuencia, todas las familias conocidas de soluciones particulares se han obtenido imponiendo hipótesis de simetría, restricciones sobre el contenido de materia u otras condiciones que permiten integrar las ecuaciones. Una clase de soluciones que ha desempeñado un papel esencial en el desarrollo de la teoría de la Relatividad General es la de los espaciotiempos con simetría esférica. Estas soluciones proporcionan un marco simplificado pero rico para el estudio de una amplia gama de fenómenos físicos. Sin embargo, a pesar de la extensa literatura dedicada a este tema, el interés no ha disminuido en la actualidad, y varias cuestiones abiertas están siendo estudiadas actualmente. Por ejemplo, los análisis de la conjetura de censura cósmica han reforzado el estudio de las propiedades de las soluciones de fluido perfecto con simetría esférica. \\ \\
No obstante, algunas de estas soluciones se han obtenido sin especificar una ecuación de estado, mientras que otras han sido derivadas en el caso de polvo, o prescribiendo una dependencia temporal (no física) de la presión, o bien imponiendo relaciones barotrópicas particulares para cerrar el sistema de ecuaciones. Aunque estos enfoques han permitido obtener un amplio espectro de soluciones exactas, muchas de ellas carecen de una interpretación física clara. En consecuencia, el objetivo de esta tesis es el estudio de la posible viabilidad física de las soluciones de fluido perfecto con simetría esférica, un estudio que también puede extenderse a las simetrías plana e hiperbólica. \\ \\
Con este fin, recogemos los resultados anteriores sobre el enfoque hidrodinámico, que ofrece las herramientas necesarias para interpretar un tensor de energía de fluido perfecto como un fluido en equilibrio termodinámico local. En particular, como un gas ideal genérico, un gas ideal clásico y un fluido con coeficientes de transporte no nulos. A continuación, extendemos este estudio al caso de un gas de Synge ultrarrelativista y proporcionamos un método para obtener distintas aproximaciones de la ecuación de estado de Synge. \\ \\ \\
A continuación, aplicamos estos resultados para analizar la interpretación de tres familias de soluciones de fluido perfecto como fluidos físicamente admisibles. Comenzamos con los modelos T, la familia de soluciones de fluido perfecto que admiten simetría esférica, plana o hiperbólica y cuya curvatura tiene un gradiente tangente al flujo del fluido. Continuamos con los modelos R (el gradiente de la curvatura no es tangente al flujo del fluido) que admiten una sincronización plana ortogonal al flujo, los cuales son geodésicos. Finalmente, estudiamos los universos de Stephani termodinámicos, los modelos R no geodésicos que pertenecen a la subfamilia conformemente plana de las soluciones cosmológicas de Stephani-Barnes. Para las tres familias obtenemos las expresiones generales del flujo del fluido, la densidad de energía, la presión y el cuadrado de la velocidad del sonido, dado por la función indicatriz. Estas proporcionan las distintas evoluciones que pueden representar las soluciones consideradas. También obtenemos las expresiones generales de los diferentes conjuntos de magnitudes termodinámicas (esquemas termodinámicos) admitidos: la densidad de materia, la energía interna específica, la entropía específica y la temperatura. Cada uno de ellos proporciona una interpretación distinta de la solución como un fluido determinado. Para analizar si cumplen los requisitos necesarios para su viabilidad física (las condiciones de energía de Pleba\'nski, las condiciones de positividad y las condiciones de compresibilidad), determinamos en primer lugar las subfamilias que satisfacen requisitos físicos adicionales. El más relevante es la compatibilidad con la ecuación de estado de un gas ideal genérico. Las subfamilias obtenidas tras imponer esta condición se estudian en detalle, investigando sus evoluciones, perfiles radiales, singularidades espaciotemporales y esquemas termodinámicos asociados. En los tres casos, encontramos amplios dominios espaciotemporales en los que las soluciones pueden interpretarse como fluidos perfectos en equilibrio termodinámico local con la evolución de un gas ideal genérico y que cumplen todos los requisitos necesarios para su viabilidad física. \\ \\
En la última parte de la tesis presentamos \textit{xIdeal}, una herramienta personalizada desarrollada en \textit{Mathematica} para facilitar el estudio de soluciones exactas de las ecuaciones de Einstein. Este paquete ha sido desarrollado en colaboración con García-Parrado y actualmente incluye trece funciones que implementan distintas caracterizaciones y determinaciones IDEALes (Intrínsecas, Deducidas, Explícitas y ALgorítmicas) de espaciotiempo. También incorpora una base de datos de métricas con sus propiedades. Resumimos todos los algoritmos actualmente incluidos en \textit{xIdeal} y se ilustra su uso con ejemplos concretos. Además, presentamos algunos algoritmos IDEALes que hemos desarrollado recientemente y que aún no están implementados. 
\end{otherlanguage}

\newpage

\section*{Publications}
The contents of Chapters 3, 4, 5 and 6 of this thesis are based on the following publications, which are listed here in chronological order:
	\begin{itemize}
		\item J. J. Ferrando and S. Mengual, ``Thermodynamic approach to the T-models", \textit{Phys. Rev. D} \textbf{104} (2021) 024038
		\item J. J. Ferrando and S. Mengual, ``T-model field equations: The general solution", \textit{Phys. Rev. D} \textbf{104} (2021) 064029
		\item S. Mengual and J. J. Ferrando, ``Thermodynamic perfect fluid spheres admitting an orthogonal flat synchronization", \textit{Phys. Rev. D} \textbf{105} (2022) 124019
		\item S. Mengual, J. J. Ferrando and J. A. Sáez, ``Hydrodynamic approach to the Synge gas", \textit{Phys. Rev. D} \textbf{106} (2022) 124032
		\item S. Mengual, J. J. Ferrando and J. A. Sáez, ``Thermodynamics of the universes admitting isotropic radiation", \textit{Phys. Rev. D} \textbf{110} (2024) 044012 \\ \\
	\end{itemize}
The parts of Chapters 8 and 9 that indicate so, are based on publications in which I have contributed. In those parts, I develop in more detail my contributions and summarise the remaining necessary fragments of the works. These publications are the following:
	\begin{itemize}
		\item J. A. Sáez, S. Mengual and J. J. Ferrando, ``On the flow of perfect energy tensors", \textit{Class. Quantum Grav.} \textbf{40} (2023) 175003
		\item J. A. Sáez, S. Mengual and J. J. Ferrando, ``Dimension of the isometry group in spacetimes with an invariant frame", \textit{Class. Quantum Grav.} \textbf{40} (2023) 205020
		\item J. A. Sáez, S. Mengual and J. J. Ferrando, ``Spatially-Homogeneous Cosmologies", \textit{Class. Quantum Grav.} \textbf{41} (2024) 205013
		\item J. A. Sáez, S. Mengual and J. J. Ferrando, ``Obtaining the multiple Debever null directions", \textit{Phys. Scr.} \textbf{100} (2025) 011501
		\item J. A. Sáez, S. Mengual and J. J. Ferrando, ``Dimension of the isometry group in type N vacuum solutions: an IDEAL approach", [arXiv:2506.20427]
	\end{itemize}
All this work has also resulted in the creation of the software \textit{xIdeal}, which is explained throughout Chapters 7, 8 and 9 and can be found at:
\begin{itemize}
		\item A. García-Parrado-Gómez-Lobo and S. Mengual, ``xIdeal: IDEAL characterizations for xAct" https://github.com/wtbgagoa/xIdeal
	\end{itemize}
%




\hypersetup{
    colorlinks = true,
    linkcolor = black,
    citecolor = black,
    urlcolor = black
}
\tableofcontents
\hypersetup{
    colorlinks = true,
    linkcolor = Blue,
    citecolor = BrickRed,
    urlcolor = ForestGreen,
}

\newpage
\thispagestyle{empty}
\null

\newpage

\pagenumbering{arabic}

\chapter{Introduction} \label{chap-intro}
The theory of General Relativity, proposed by Albert Einstein in 1915 \cite{Einstein-1915}, changed our understanding of gravity by replacing the Newtonian concept of a force with a geometric interpretation: gravity arises as a consequence of the spacetime deformation caused by the presence of matter and energy. The spacetime geometry is mathematically described by the metric tensor, $g_{\mu \nu}$, and the field equations governing it are Einstein's equations \cite{Einstein-1916}:
\begin{equation} \label{Einstein-equations}
	G_{\mu \nu} + \Lambda g_{\mu \nu} = \frac{8 \pi G}{c^4} T_{\mu \nu} \, ,
\end{equation}
where $G_{\mu \nu}$ is the Einstein tensor, $\Lambda$ is the cosmological constant, $G$ is the gravitation constant, $c$ is the speed of light and $T_{\mu \nu}$ is the energy tensor describing the matter-energy content. In this thesis we will work in geometrised units such that $8 \pi G = c = 1$. Einstein's equations constitute a system of highly nonlinear coupled differential equations. As a result, all known families of particular solutions have been derived by imposing symmetry assumptions, matter content restrictions or other simplifying conditions that enable the integration of the equations. \\ \\
The Schwarzschild metric \cite{Schwarzschild} was the first exact solution to Einstein's equations. It was derived under the assumptions of absence of cosmological constant, spherical symmetry and vacuum, representing the gravitational field outside a static spherically symmetric mass. Soon after, Reissner and Nordström independently extended this solution to incorporate an electromagnetic field \cite{Reissner, Nordstrom}. Kottler, in turn, generalised the Schwarzschild metric to include a cosmological constant \cite{Kottler}. It was not until more than four decades later that Kerr obtained a solution for the case of a rotating source, the Kerr metric \cite{Kerr}, which has become crucial in the study of rotating black holes. This was soon further extended by Newman \textit{et al.}, who included both rotation and the presence of electric charge into what is now known as the Kerr-Newman metric \cite{Newman}. \\ \\ \\
In addition to vacuum solutions, efforts were made to describe gravitational fields inside matter distributions. Among the earliest models to represent such scenario was the Schwarzschild interior solution, which extends the Schwarzschild solution to account for the presence of a static and spherically symmetric perfect fluid with constant energy density \cite{Schwarzschild-interior}. The search for solutions with non-constant energy density led to the development of the Tolman-Oppenheimer-Volkoff equations \cite{Tolman-1939, Oppenheimer-Volkoff}, also known as the stellar structure equations. \\ \\
All the aforementioned solutions share the feature that they model the gravitational field of compact objects such as stars or black holes. However, one of the first things Einstein attempted when seeking physical applications of his theory was to construct a model of the Universe as a whole. In this context, it is considered that the Universe can be modelled as a continuous medium whose large-scale properties can be described by smooth physical fields. Of course, this is only an approximation. The real Universe actually exhibits a ``granular" structure: it is composed of stars that cluster into galaxies, which in turn group into galaxy clusters or other larger structures. Nevertheless, the continuous medium approximation is useful when the aim is to describe the evolution of the Universe, or of a portion of it, as a whole, rather than the dynamics of its individual constituents.\\ \\
The first relativistic model of the Universe was proposed by Einstein in 1917 \cite{Einstein-1917}. In this model, the Universe was assumed to be static, homogeneous and isotropic. To maintain a static configuration despite the attractive nature of gravity, Einstein introduced the cosmological constant $\Lambda$ as a repulsive term to counterbalance gravitational collapse. However, this model failed to predict astronomical observations, particularly the systematic redshifts observed in the spectra of extragalactic nebulae. Around the same time, De Sitter proposed an alternative model \cite{deSitter-1917}, also static, that successfully explained the observed redshifts. The downside of this model was that it could not account for the observed distribution of matter. \\ \\
Both Einstein's and de Sitter's models assumed a static Universe, which was the prevailing belief at the time. Friedmann \cite{Friedmann-1922} was the first to study, from a theoretical point of view and before the expansion of the Universe was observed, a non-static Universe model. Following the observational discovery of the Universe's expansion by Hubble in 1929, Friedmann's ideas gained renewed interest. Lemaître had already derived similar expanding solutions in 1927 working independently \cite{Lemaitre-1927}, and also proposed a physical interpretation connecting the redshifts to cosmic expansion. Subsequantly, Robertson \cite{Robertson-1929} and Walker \cite{Walker-1935} refined the mathematical formulation of these models, leading to what are now known as the Friedmann-Lemaître-Robertson-Walker (FLRW) models. They are based on the assumption that the universe is homogeneous and isotropic on large scales (a hypothesis known as the cosmological principle).\\ \\
However, over the years, models based on less restrictive cosmological principles have also been studied. On the one hand, various studies \cite{Anisotropia1, Anisotropia2, Anisotropia3} fostered interest in homogeneous but anisotropic solutions, which were first studied by Kompaneets and Chernov \cite{Kompaneets-Chernov} and later popularised, due to their symmetry properties, by Kantowski and Sachs \cite{Kantowski-Sachs}. On the other hand, there also exist models known as the Stephani universes \cite{Stephani} in which the cosmological principle is weakened and holds only at each instant of cosmological time; that is, only the spatial part of the metric exhibits maximal symmetry.\\ \\
Solutions to Einstein's field equations that abandon the assumption of homogeneity have also been developed and extensively studied. These solutions play a crucial role in advancing our understanding of more realistic and diverse scenarios. They have been applied both globally, Lemaître-Tolman \cite{Lemaitre, Tolman} solution being the most representative example (see also \cite{Bondi, Ellis-Maartens-MacCallum}), and to model local inhomogeneities and their effects, including the influence of cosmic structures on light propagation or the backreaction of inhomogeneities on cosmic expansion (see Chapter 16 of \cite{Ellis-Maartens-MacCallum}). Moreover, these models are not limited to cosmology, inhomogeneous solutions can also be used to represent stellar interiors. An important class of the inhomogeneous solutions consists of inhomogeneous cosmological models (see \cite{Inhomogeneous_Cosmological_Models} for a full classification of such solutions). A solution to the Einstein field equations is called cosmological if it can recover the FLRW solution in some limiting case. \\ \\ \\ \\
Interest in inhomogeneous cosmological models grew when several studies showed that large-scale spatial inhomogeneities of the Universe could be compatible with cosmological observations (see, for example \cite{Mustapha}). Notably, it was demonstrated that the magnitude-redshift relation with the Type Ia supernovae data can be reproduced in an inhomogeneous model without cosmological constant \cite{Celerier, Iguchi, Celerier-Krasinski}. Moreover, the structure and evolution of galaxies, clusters and voids require an analysis outside of the perturbative regime of the FLRW models. Although they are often modelled by Newtonian N-body computations, the nonlinear effects inherent in the Einstein field equations may be critical in the structure formation. Accordingly, a large number of studies have been devoted to providing exact inhomogeneous models for studying the formation of structures and for analysing the effect of the nonlinear inhomogeneities on the cosmic microwave background radiation. While some authors remain skeptical about whether inhomogeneous models can fully account for all the cosmological observations (see \cite{Ellis-Maartens-MacCallum} and references therein), several works continue investigating this subject \cite{Krasinski-Hellaby-B-C, Krasinski_14a, Krasinski_14b}, and the issue is still an open question. \\ \\
A class of (inhomogeneous) solutions to the Einstein field equations that has played an essential role in developing the General Relativity theory is that of spherically symmetric spacetimes. These solutions provide a simplified yet rich framework for studying a wide range of physical phenomena, from stellar interiors to gravitational collapse. However, despite the extensive literature dedicated to them (see, for example \cite{Krasinski-Plebanski, Kramer, Kim} and references therein), the interest in this topic has not waned nowadays, and several open issues are currently under study. For example, the analyses of the cosmic censorship conjecture have reinforced the study of the properties of the spherically symmetric perfect fluid solutions \cite{Lapi-Morales, Mosani-J}.\\ \\
A \textit{perfect fluid solution} is a solution to the Einstein field equations in which the source term is the conserved energy tensor of a perfect fluid, which we will call a \textit{perfect energy tensor}. A perfect fluid is a pascalian fluid, it has zero viscosity and an isotropic stress tensor for an observer at rest, whose transport of energy occurs only by means of mass flow, with no heat conductivity. A perfect energy tensor is completely characterised by three \textit{hydrodynamic quantities}: the \textit{fluid flow}, the \textit{energy density} and the \textit{pressure}. \\ \\
Static perfect fluid spheres are the most basic and simplest models for studying the stellar interior structure, both within Newtonian gravity and in the framework of General Relativity (see, e.g. \cite{Krasinski-Plebanski, Rezzolla}). These models assume a spherically symmetric, time-independent configuration where the gravitational attraction is balanced by the internal pressure of the fluid, leading to equilibrium solutions. In a cosmological context, the paradigmatic FLRW cosmological models are also spherically symmetric solutions, although they are not static. On the other hand, the most remarkable solution for modelling both gravitational collapse and cosmological inhomogeneities is the aforementioned Lemaître-Tolman model \cite{Lemaitre, Tolman, Krasinski-Plebanski}. This solution describes a spherically symmetric spacetime filled with dust and allows for a radially varying energy density and expansion rate. \\ \\
Other non-stationary spherically symmetric perfect fluid spacetimes have been widely considered in the literature, and enough families of exact solutions are known (see \cite{Krasinski-Plebanski, Kramer, Inhomogeneous_Cosmological_Models} and references therein). Some of these solutions have been derived without specifying an equation of state, while others have been obtained in the dust case or by prescribing a (non-physical) time-dependence of the pressure, or also by imposing particular barotropic relations to close the system of equations. While these approaches have led to a broad spectrum of exact solutions, many of them lack a clear physical interpretation. Consequently, further work is required to study the physical viability of the spherically symmetric perfect fluid solutions, a study that can also be extended to the plane and hyperbolic symmetries. \\ \\
Pleba\'nski's \cite{Plebanski_1964}	\textit{energy conditions} are widely accepted as necessary requirements for the physical plausibility of arbitrary macroscopic matter configurations in General Relatiivity. These conditions ensure, for instance, that the energy density as measured by any observer is non-negative. However, they are not sufficient to fully characterise physically realistic fluids. For a perfect fluid solution to describe the evolution of a deterministic fluid, it is necessary to add a deterministic closure. All the deterministic closures proposed in the literature for general (non-barotropic) perfect fluids are thermodynamic closures, involving additional \textit{thermodynamic quantities} such as the \textit{matter density}, the \textit{specific internal energy}, the \textit{temperature} and the \textit{specific entropy}. These variables satisfy additional laws allowing to interpret the solutions as fluids in local thermal equilibrium (l.t.e.). This thermodynamic framework is essential for distinguishing between fluids with different properties, even when they share the same energy tensor. \\ \\
In addition to the energy conditions and the l.t.e. assumption, other macroscopic physical reality conditions must be imposed to ensure that a perfect fluid solution admits a physically admissible interpretation. These conditions further constrain the range of acceptable solutions by requiring consistency with well established principles of fluid dynamics and thermodynamics. They include the compressibility conditions, which are essential for a consistent treatment of shock waves in the fluid, and the positivity of certain thermodynamic quantities.

\subsection*{Exact solutions to the Einstein equations with a perfect \mbox{fluid source}} \label{sec-intro-exact-sols}
\begin{figure}[h]
\centering
\hspace{-6mm} \includegraphics[width=0.85\textwidth]{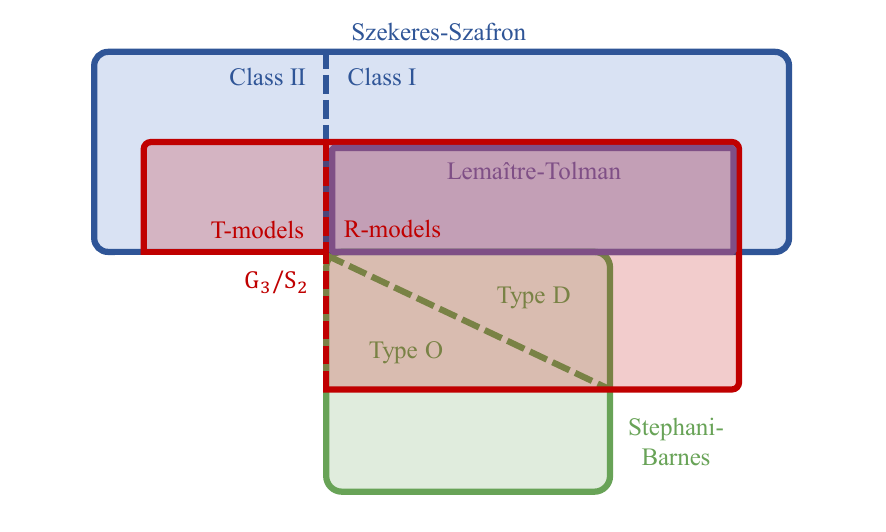}
\vspace{-3mm}
\caption{Schematic diagram representing the relations among the inhomogeneous solutions considered in this thesis.}
\label{Fig-0}
\end{figure}

	Here we present a partial classification of inhomogeneous (cosmological) solutions. In the following chapters, we will explore the potential physical interpretation of some of these families. Therefore, it is important to understand their interrelations, including their generalisations and particular subfamilies.\\ \\
	Among other inhomogeneous cosmological solutions, two families of metrics that contain a large number of such solutions as particular cases are studied in detail in \cite{Inhomogeneous_Cosmological_Models}. These two big families are the Szekeres-Szafron solutions \cite{Szekeres,Szafron} (in blue in Figure \ref{Fig-0}) and the Stephani-Barnes metrics \cite{Stephani,Barnes} (in green in Figure \ref{Fig-0}).\\ \\
	$\hspace{-2mm}$The Szekeres-Szafron family of solutions was initially derived by Szekeres \cite{Szekeres} for dust and later generalised by Szafron \cite{Szafron} for non-vanishing pressure. These solutions are geodesic and, generally, lack symmetries \cite{Inhomogeneous_Cosmological_Models}. They can be divided into two distinct subfamilies: the Szekeres-Szafron subfamily of class I and that of class II. As we will see in this thesis, to interpret a solution as a physically admissible perfect fluid model, it must be compatible with the local thermal equilibrium equations. In such case, the solution will be called thermodynamic. 
	In \cite{Krasinski-et-al}, Krasi\'nski \textit{et al.} showed that thermodynamic Szekeres-Szafron solutions of class I admit, necessarily, a three dimensional group G$_3$ of isometries acting on two-dimensional orbits S$_2$ (G$_3$/S$_2$). These solutions are the Lemaître-Tolman metrics \cite{Lemaitre,Tolman} (with pressure) and their plane and hyperbolic counterparts (in purple in Figure \ref{Fig-0}). This family of solutions is also the geodesic limit of the so called R-models, the perfect fluid solutions admitting a G$_3$/S$_2$ whose curvature has a gradient which is not tangent to the fluid flow. \\ \\
	On the other hand, there are thermodynamic Szekeres-Szafron solutions of class II without symmetries. The analysis of their thermodynamic properties and the study of the ideal gas models require distinguishing the singular and regular models \cite{C-F-S_SzSz_Singular,C-F-S_SzSz_Regular}. When the spacetime admits a G$_3$/S$_2$, the solution is a T-model, a perfect fluid solution admitting a G$_3$/S$_2$ whose curvature has a gradient which is tangent to the fluid flow. The spherical dust T-model was published in a pioneer paper by Datt \cite{Datt_1938} and rediscovered later by Ruban \cite{Ruban_1969, Ruban_1968}, and the general perfect fluid solution with a non-constant pressure was considered by Korkina and Martinenko \cite{Korkina-Martinenko_1975}. Another particular subfamily of the T-models of interest is that of the Kompaneets-Chernov-Kantowski-Sachs (KCKS) metrics \cite{Kompaneets-Chernov,Kantowski-Sachs}, which correspond to the spatially homogeneous limit.\\ \\
	Regarding the Stephani-Barnes family of solutions, it consists of all the irrotational and shear-free solutions with non-zero expansion. The only solutions in common with the Szekeres-Szafron family of solutions are the FLRW metrics \cite{Inhomogeneous_Cosmological_Models}. This family of solutions split into two subfamilies: the conformally flat ones and those of Petrov-Bel type D. The type D metrics necessarily have a G$_3$/S$_2$, while the conformally flat ones do not have symmetries in general.
	The conformally flat subfamily was completely solved by Stephani \cite{Stephani}, and are known as the Stephani Universes. Moreover, Bona and Coll \cite{Bona-Coll} showed that the thermodynamic Stephani Universes admit, necessarily, a G$_3$/S$_2$. The spherically symmetric case of the Type D subfamily was first derived by Dingle \cite{Dingle}, but the parametrisation he chose made the single Einstein field equation unworkable. The simple form that is most known nowadays was first derived by Kustaanheimo and Qvist \cite{Kustaanheimo-Qvist}. Finally, the whole family was derived by Barnes \cite{Barnes} (see also \cite{Barnes-2}), consisting of the Stephani Universes as the conformally flat subfamily, and the Kustaanheimo-Qvist class together with its plane and hyperbolically symmetric counterparts as the Type D subfamily.\\ \\ 
	In this thesis we will focus on the perfect fluid solutions admitting a G$_3$/S$_2$ (in red in Figure \ref{Fig-0}). As mentioned above, this family of solutions can be split into the T-models and the R-models, depending on whether the gradient of the orbits curvature is tangent to the fluid flow or not, respectively. Since all the T-models are geodesic (see Chapter \ref{chap-T-models}), they are completely covered by the Szekeres-Szafron solutions of class II admitting a G$_3$/S$_2$. The geodesic R-models are the Lemaître-Tolman subfamily of the class I Szekeres-Szafron family. The non-geodesic R-models include the subfamilies of Stephani-Barnes solutions admitting a G$_3$/S$_2$, namely all the Petrov type D subfamily and the thermodynamic Stephani Universes, and other families of solutions that have not been mentioned yet.\\ \\
	The R-models in the set that remains to be studied are neither geodesic nor part of the Stephani-Barnes family of solutions. This means that they do not have the FLRW metrics as a limit and, therefore, are not cosmological solutions. Indeed, the static case of the subfamily of solutions without shear and expansion of this set includes solutions such as the Schwarzschild interior (the energy density is constant) and its barotropic generalisation, which leads to the Tolman-Oppenheimer-Volkoff equation \cite{Kramer}. The shear- and expansion-free non-static subfamily can be obtained from the static ones (see \cite{Kramer}) and includes solutions modelling, for example, the interior of an incompressible sphere undergoing condensation \cite{Leibovitz-1971} and a spherically symmetric, incompressible, non-dissipative fluid following a central explosion \cite{Skripkin-1960}. However, most of these solutions have equations of state that are arguably not physically realistic, with the possible exception of the generalised Gutman-Be'spalco-Wesson solutions \cite{Herrera-PdL-1985, Kitamura-1994, Kitamura-1995b}. \\ \\
	Thus, we can see that despite the complexity of the Einstein equations, many authors have managed to overcome the difficulties involved in finding particular solutions. However, as Krasi\'nski himself states in \cite{Inhomogeneous_Cosmological_Models}, ``what is really needed is the physical and geometrical interpretation of the solutions already derived." \\ \\
	This thesis is divided into three parts. In the first part, some necessary tools for giving a perfect energy tensor a physical interpretation are provided and expanded with our own contribution. The objective of the second part, which is also the longest, is to study the potential physical interpretation of perfect fluid solutions to Einstein field equations with spherical symmetry (also plane and hyperbolic in some cases). In the third part, we present a custom-built \textit{Mathematica} tool designed with the goal of facilitating the study of exact solutions to Einstein field equations. The motivation and philosophy behind this project are explained in the first section of that part.

\part{Physical interpretation of a perfect energy tensor}

	\chapter[Conditions for physical reality]{Conditions for \\ physical reality} \label{sec-intro-physical-interpretation}
	Let us consider $(V_4, \, g)$ to be an oriented and time-oriented spacetime of signature $\left\lbrace -, +, +, + \right\rbrace$. For the metric product of two vectors, we write $(x,y) = g(x,y)$ and $x^2 = g(x,x)$. The symbols $\nabla$, $\nabla \cdot$, $\textrm{d}$, $\wedge$ and $*$ denote, respectively, the covariant derivative, the divergence operator, the exterior derivative, the exterior product and the Hodge dual operator, while $i(x)t$ stands for the interior product of a vector field $x$ and a p-tensor $t$ (see Appendix \ref{AppendixB} for further details about the notation). We shall denote with the same symbol a tensor and its associated tensors by raising and lowering indices with the metric $g$. \\ \\
	A perfect fluid is a fluid with zero viscosity and heat conductivity whose stress tensor is isotropic for a comoving observer \cite{Rezzolla}. The energetic description of the evolution of a fluid is given by its energy tensor $T$ which, for the case of a perfect fluid, is of the form:
	\begin{equation} \label{perfect-energy-tensor}
		T = (\rho + p) u \otimes u + p \, g
	\end{equation}
and fulfilling the conservative condition $\nabla \cdot T = 0$. This constraint consists of a differential system of four equations on five \textit{hydrodynamic quantities} (\textit{unit velocity} $u$, \textit{energy density} $\rho$, and \textit{pressure} $p$)
	\begin{equation} \label{energy-momentum-conservation}
		\hspace{-2cm} \textrm{C} : \qquad \qquad \qquad \textrm{d} p + u(p) u + (\rho + p) a = 0 \, , \qquad u(\rho) + (\rho + p) \theta = 0 \, ,
	\end{equation}
where $a \equiv i(u) \nabla u$ and $\theta \equiv \nabla \cdot u$ are, respectively, the acceleration and the expansion of $u$, and where $u(q)$ denotes the directional derivative, with respect to $u$, of a quantity $q$, $u(q) = u^{\alpha} \partial_{\alpha} q$. Energy tensors of the form (\ref{perfect-energy-tensor}) and satisfying the conservation equations C will be referred to as \textit{perfect energy tensors} hereafter. \\ \\
	A \textit{perfect fluid solution} to the Einstein equations is a solution to (\ref{Einstein-equations}) with an energy tensor of the form (\ref{perfect-energy-tensor}). Note that the Einstein equations imply the conservation equations C. Therefore, the energy tensor of a perfect fluid solution is a perfect energy tensor. In that case, two of the Einstein field equations become a definition of the fluid pressure $p$ and energy density $\rho$ in terms of the metric functions, while the rest become a system of differential equations that these metric functions must fulfil. However, in this part of the thesis we will focus on solutions to the conservation equations C, regardless of whether they are solutions to the Einstein equations too or not.

	\section{Local thermal equilibrium} \label{sec-lte}
\begin{figure}[t]
			\centerline{
			\parbox[c]{0.5\textwidth}{\includegraphics[width=0.49\textwidth]{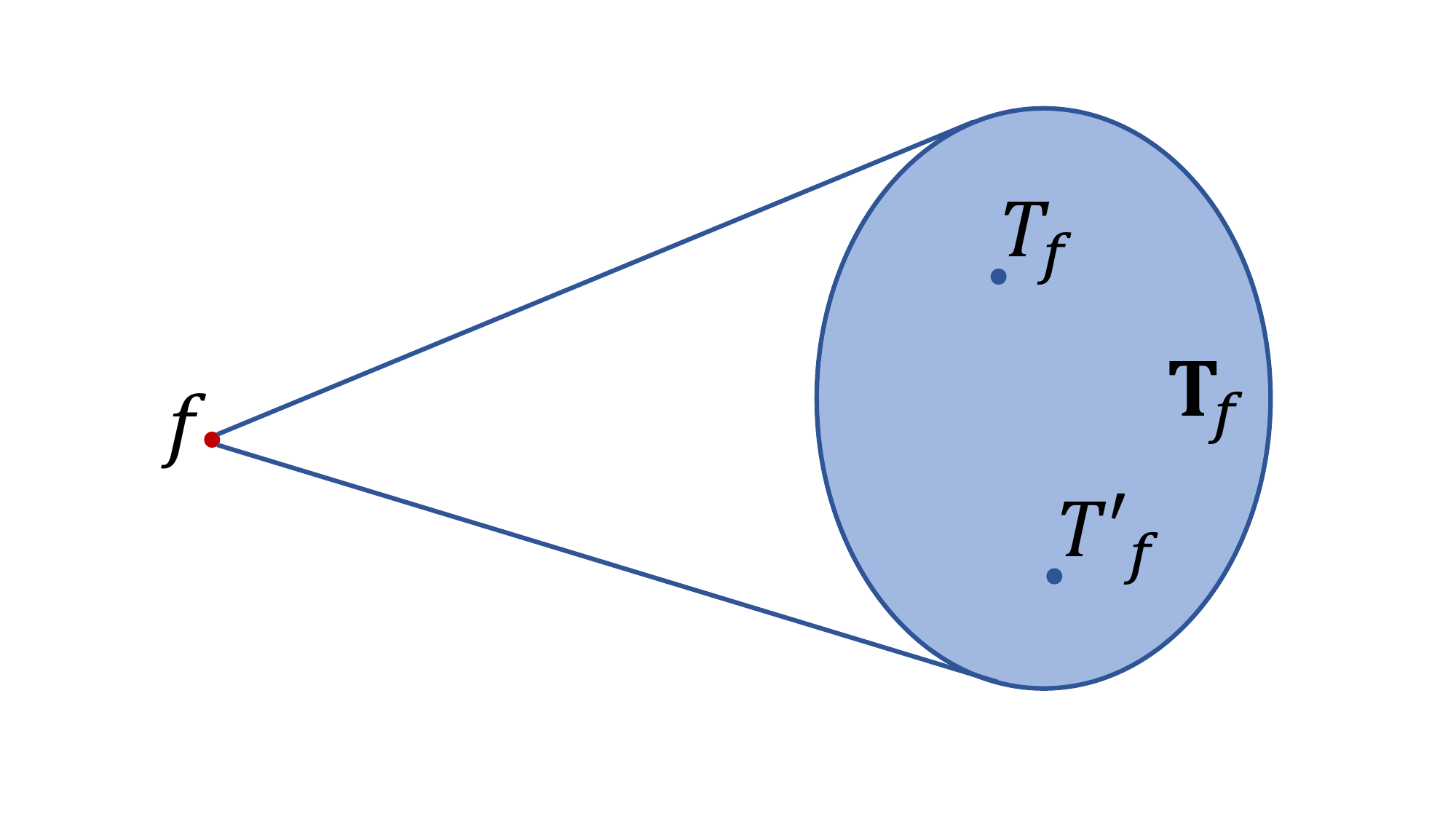}\\[-2mm] \centering{(a)}}
			\hspace{30pt}
			\parbox[c]{0.5\textwidth}{\includegraphics[width=0.49\textwidth]{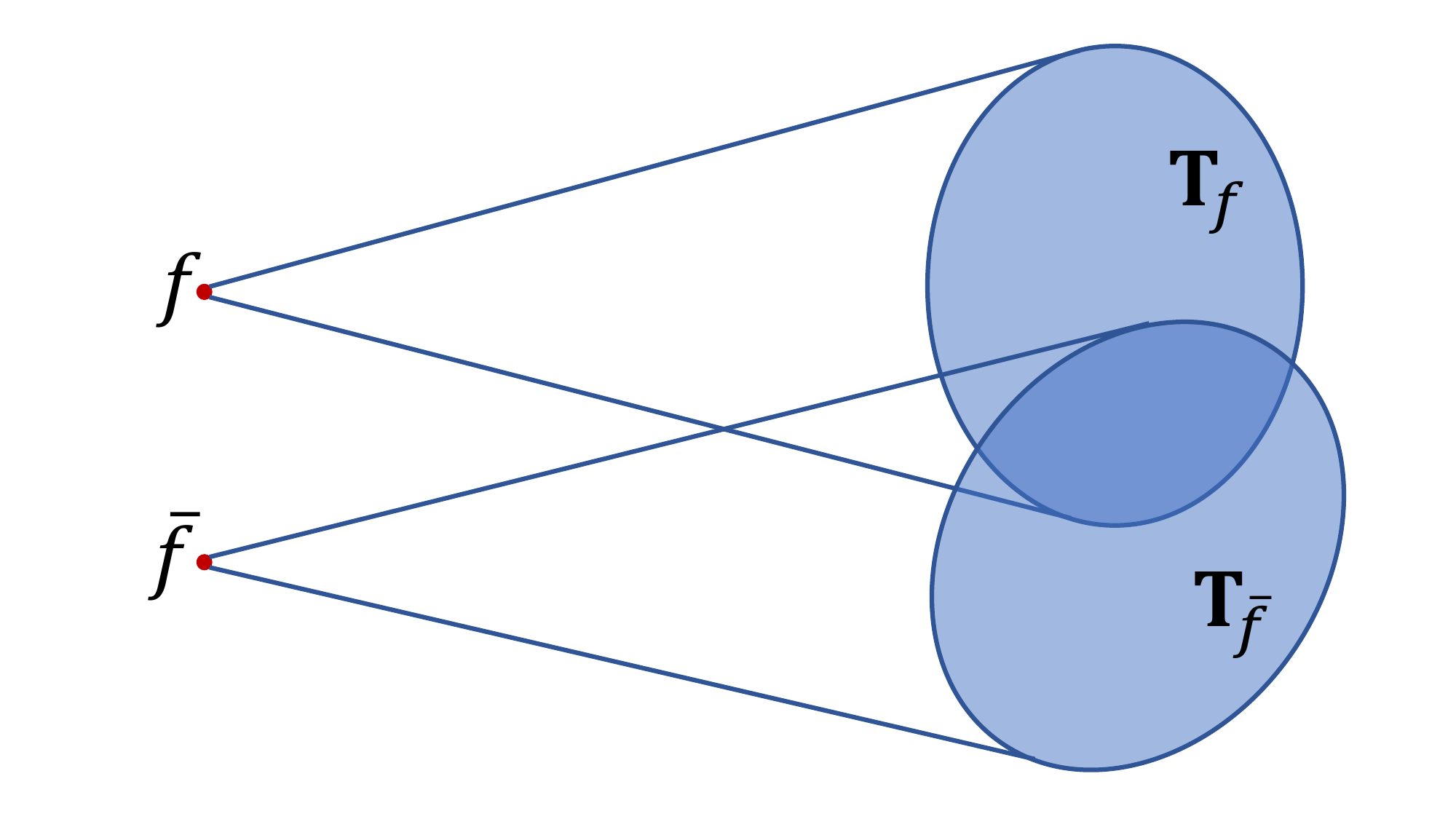}\\[-2mm] \centering{(b)}}}
			\caption{(a) $\textbf{T}_{\textit{f}}$ is the set of perfect energy tensors $T_{\textit{f}}$ corresponding to all the possible evolutions of a fluid $f$. (b) The sets $\textbf{T}_{\textit{f}}$ and $\textbf{T}_{\textit{$\bar{f}$}}$ corresponding to the possible evolutions of two different perfect fluids \textit{f} and $\textit{$\bar{f}$}$ might not be disjoint, $\textbf{T}_{\textit{f}}\cap\textbf{T}_{\textit{$\bar{f}$}}\neq\emptyset$ (reproduced from \cite{Hydro-LTE}).}
			\label{Fig-1}
		\end{figure}

\begin{figure}[t]
		\vspace{2mm}
			\centerline{
			\parbox[c]{0.5\textwidth}{\includegraphics[width=0.49\textwidth]{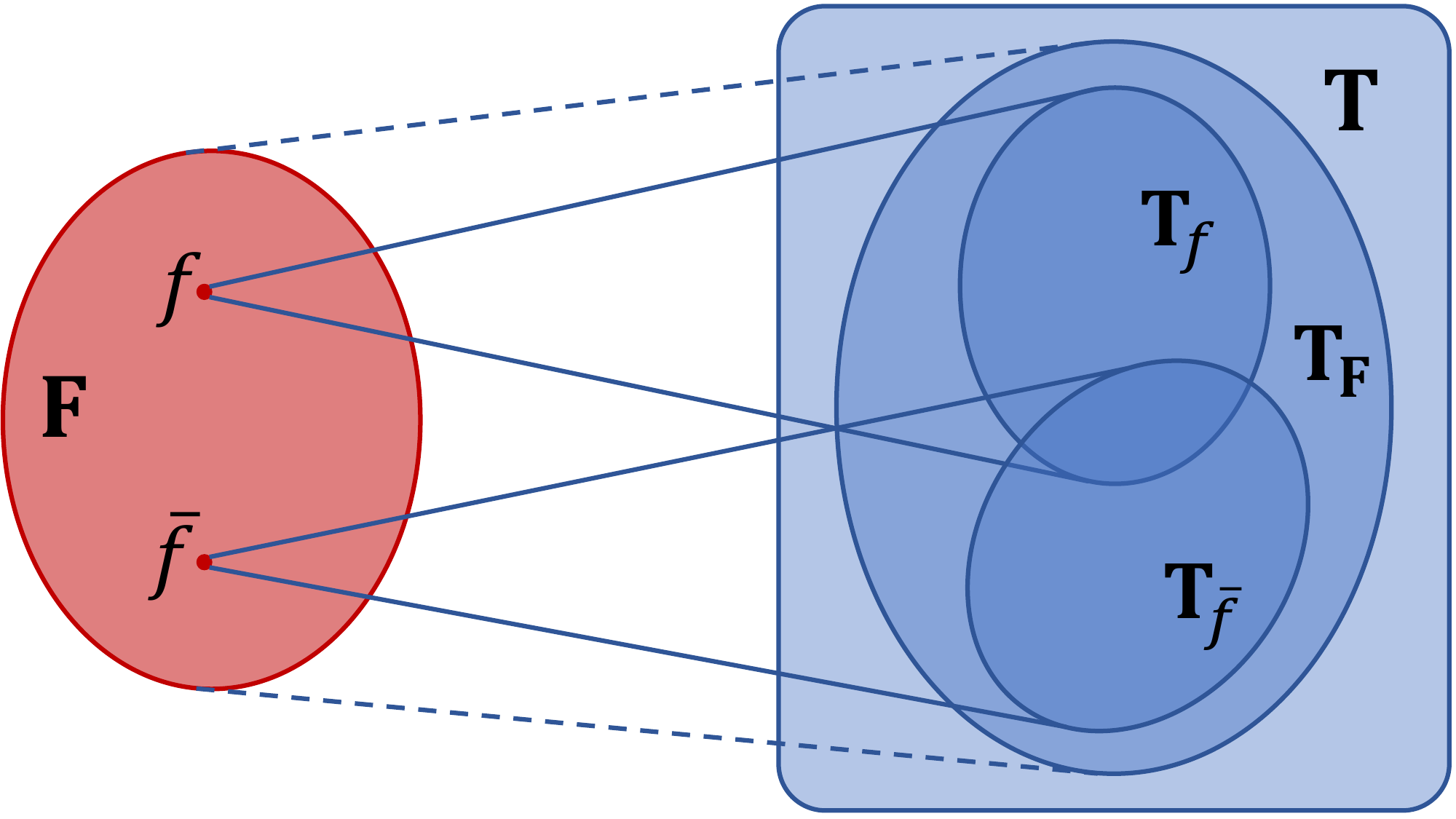}\\[2mm] \centering{(a)}}
			\hspace{30pt}
			\parbox[c]{0.5\textwidth}{\includegraphics[width=0.49\textwidth]{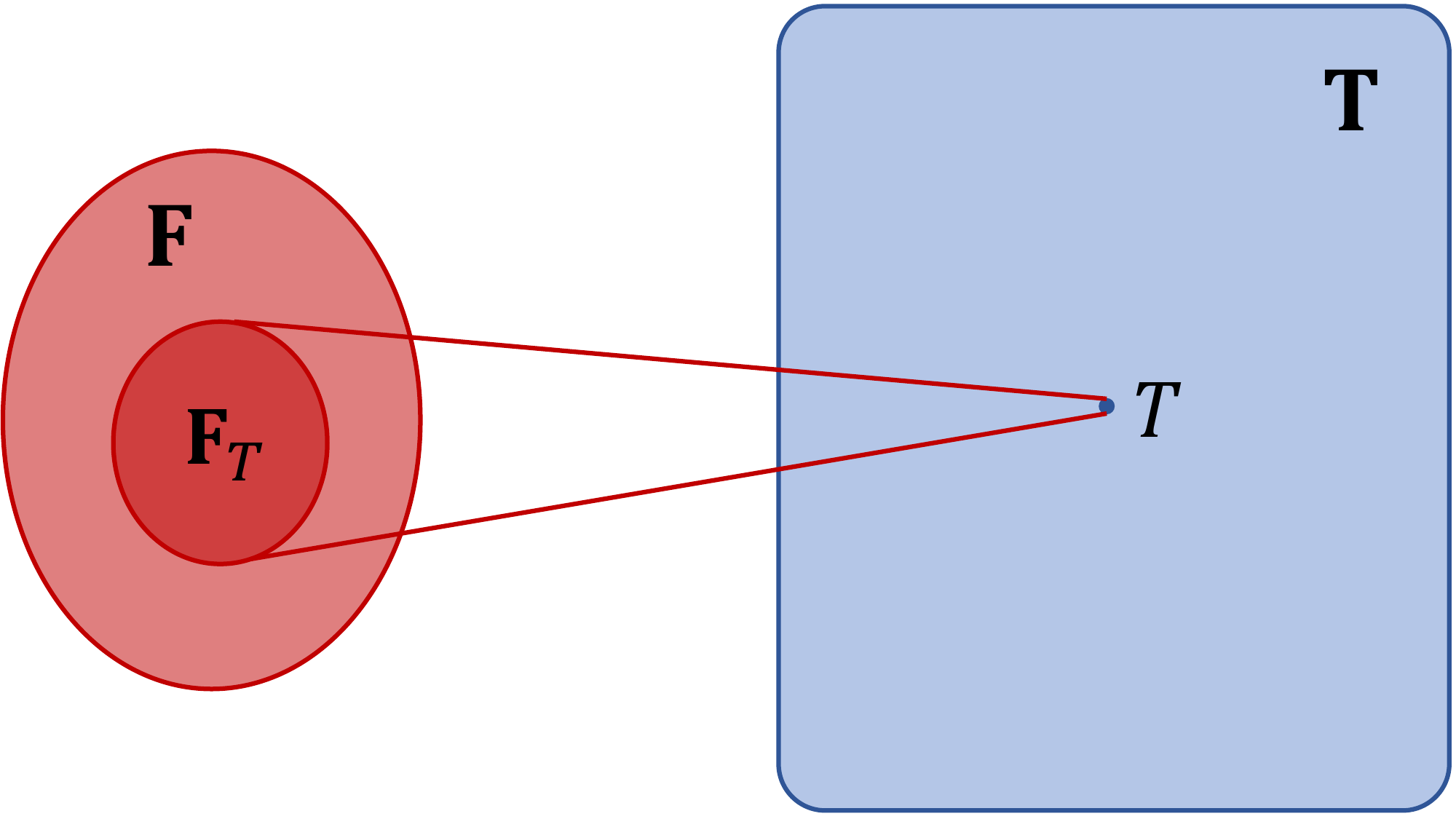}\\[2mm] \centering{(b)}}}
			\caption{(a) \textbf{F} is the set of all the perfect fluids $f$, $\textbf{T}_{\textbf{F}}$ is the set of all the perfect energy tensors corresponding to some possible evolution of a perfect fluid and \textbf{T}, the set of all the perfect energy tensors. Then $\textbf{T}_{\textbf{F}}\subset\textbf{T}$. (b) Every fluid in the subset $\textbf{F}_{\textit{T}}$ of \textbf{F} have as a possible evolution the one corresponding to \textit{T} (reproduced from \cite{Hydro-LTE}).}
			\label{Fig-2}
		\end{figure}

	As mentioned above, the conservative condition consists of a differential system of four equations for the five hydrodynamic quantities. However, from a physical perspective, more variables and equations are needed to describe a fluid. In particular, the \textit{fundamental system of relativistic hydrodynamics} consists of the energy-momentum conservation equations C and the matter conservation equation
	\begin{equation} \label{matter-conservation}
		\nabla \cdot (nu) = u(n) + n \theta = 0 \, ,
	\end{equation}
where $n$ is the \textit{matter density}. It is worth remarking that, for multi-component fluids, it is convenient to work with the rest-mass densities of the individual components. However, since we are only interested in single-component fluids here, we will consider the particle mass to be unity. \\ \\
	In any case, this fundamental system of equations is not closed from an evolution point of view (there is one fewer equation than there are variables). Generally, fluids considered in physics are deterministic, their entire evolution is uniquely determined by their configuration in a given initial hypersurface. Mathematically, this means that an additional equation must be incorporated into the fundamental system (\ref{energy-momentum-conservation}$-$\ref{matter-conservation}). Usually, this is done through an equation of state (EoS) $f(n, \rho, p) = 0$ depending only on the nature of the considered fluid \cite{Eckart, Rezzolla}. \\ \\ \\
	Moreover, adding this EoS allows one to interpret the medium as a fluid in local thermal equilibrium (l.t.e.). Indeed, in General Relativity the total energy density, $\rho$, is the sum of the rest-mass density and the internal energy density, so we have the following decomposition
	\begin{equation} \label{energy-decomposition}
		\rho = n (1 + \epsilon) \, ,
	\end{equation}
where $\epsilon$ is the \textit{specific internal energy}. Therefore, $\alpha = \textrm{d} \epsilon + p \textrm{d} \left( \frac1n \right)$ is a closed $1$-form ($\textrm{d} \alpha \wedge \alpha = 0$), which means that there exist two functions $\Theta$ and $s$ such that
	\begin{equation} \label{thermo-first-law}
		\Theta \, \textrm{d} s = \textrm{d} \epsilon + p \, \textrm{d} \left( \frac1n \right) \, .
	\end{equation}
Now, we can identify (\ref{thermo-first-law}) with the fundamental relation of the thermodynamics and interpret functions $\Theta$ and $s$ as the \textit{temperature} and the \textit{specific entropy} of the system, respectively. \\ \\
		The quantities introduced above must also satisfy additional constraints to describe a physically admissible fluid. However, before studying them, it is convenient to analyse the role of the hydrodynamic quantities $\lbrace u, \rho, p \rbrace$ when describing a fluid and the relation with the quantities introduced by the l.t.e. condition, the \textit{thermodynamic quantities} $\lbrace n, \epsilon, s, \Theta \rbrace$. This analysis was done in \cite{Hydro-LTE, C-F-S_SzSz_Regular}. In the present and the following two sections we summarise the necessary parts of such studies. \\ \\
		Every evolution of a fluid is uniquely described by an energy tensor. Thus, a set $\textbf{T}_{\textit{f}}\equiv\lbrace T_{\textit{f}}\rbrace$ of perfect energy tensors is associated to every perfect fluid $f$, corresponding to every one of its possible evolutions, as illustrated in Figure \ref{Fig-1}(a). The energetic description of every evolution of the perfect fluid $f$ thus consists in the specification of the corresponding perfect energy tensor $T_{\textit{f}}$, i.e. the specification of $u = u(x^\mu), \, \rho = \rho (x^\mu)$ and $p = p(x^\mu)$ (\textit{hydrodynamic quantities}). However, the sole energetic description of a perfect fluid is insufficient to characterise it physically since two different fluids $f$ and $\bar{f}$ can have common possible evolutions. Figure \ref{Fig-1}(b) illustrates this situation. \\ \\
		$\hspace{-2mm}$Moreover, there exist energy tensors of the form (\ref{perfect-energy-tensor}) and fulfilling (\ref{energy-momentum-conservation}) that do not correspond to any evolution of a perfect fluid. This situation is illustrated in Figure \ref{Fig-2}(a). Nevertheless, if the perfect energy tensor does correspond to the evolution of a perfect fluid, it will generically correspond to the evolution of more than one fluid, as illustrated in Figure \ref{Fig-2}(b). \\ \\
		$\hspace{-3mm}$Usually, in many physical situations one starts from a known perfect fluid $f$ and looks for energetic descriptions $T_f$ corresponding to some particular evolutions of $f$. This will be called the \textit{direct problem}. However, the \textit{inverse problem} will also be relevant here, namely that of the existence and determination of the non-empty set \textbf{F}$_T$ of all perfect fluids $f$ for which a given energy tensor $T$ is the energetic description of a particular evolution (see Section \ref{sec-IDEAL} for a general mathematical definition of the direct and inverse problems). 

		\subsection[Hydrodynamic characterisation of the local thermal equilibrium]{Hydrodynamic$\,$ characterisation$\,$ of$\,$ the$\,$ local \\ thermal$\,$ equilibrium} 
		Equations (\ref{matter-conservation}$-$\ref{thermo-first-law}) characterise the l.t.e. of a perfect energy tensor. Therefore, a perfect energy tensor fulfilling them corresponds to an evolution in l.t.e. A set of thermodynamic quantities $\lbrace n, \epsilon, s, \Theta \rbrace$ solution to these equations is called a \textit{thermodynamic scheme}, and each thermodynamic scheme corresponds to a different interpretation of the solution as a fluid. Therefore, solving the inverse problem is actually tantamount to finding the thermodynamic schemes compatible with a given perfect energy tensor. \\ \\
		But not every perfect energy tensor admits a thermodynamic scheme. In other words, not all the solutions $\lbrace u, \rho, p \rbrace$ to the conservation equations C admit an extra set of variables $\lbrace n, \epsilon, s, \Theta \rbrace$ such that the set of all of them is a solution to the whole set of equations (\ref{energy-momentum-conservation}$-$\ref{thermo-first-law}). To find which ones do, we need to solve the direct problem for the set of fluids in l.t.e. In \cite{Coll-Ferrando-termo} (see also\cite{Hydro-LTE}), the following lemma is proven:
		\begin{lemma}
			A perfect energy tensor evolves in local thermal equilibrium if, and only if, the equation
			\begin{equation} \label{specific-entropy-conservation}
				u(s) = 0
			\end{equation}
			admits solutions $s(x^\mu)$ such that $\emph{d}s \wedge \emph{d}\rho \wedge \emph{d}p = 0$.
		\end{lemma}
		This lemma solves the direct problem for the full set of fluids in l.t.e. Nevertheless, the specific entropy, $s$, is not a quantity appearing explicitly in a perfect energy tensor. To study the potential physical interpretation of perfect fluid solutions to Einstein's equations, it will be more convenient to solve the direct problem using only hydrodynamic quantities. This is also done in \cite{Coll-Ferrando-termo} (see also\cite{Hydro-LTE}): \\ \\ \\
		\begin{thm} \label{thm-hydro-LTE}
			An isoenergetic energy tensor, $u(\rho) = 0$, always evolves in l.t.e., while a non-isoenergetic one evolves in l.t.e. if, and only if, the spacetime function $\chi \equiv u(p)/u(\rho)$, called indicatrix of local thermal equilibrium, depends only on the variables $p$ and $\rho$ \\[-2mm]
			\begin{equation} \label{cond. sonica}
				\hspace{-50mm} {\rm S}: \hspace{30mm} \qquad\qquad \emph{d}\chi \wedge \emph{d}\rho \wedge \emph{d}p = 0 \, .
			\end{equation}
		\end{thm}
Now, all the quantities involved in this theorem either appear explicitly in a perfect energy tensor, hydrodynamic quantities, or can be directly obtained from them. Therefore, Theorem \ref{thm-hydro-LTE} provides a hydrodynamic characterisation of the l.t.e. Moreover, when a fluid evolves in l.t.e., then, and only then, the indicatrix $\chi$ becomes a function of state, $\chi (\rho, p)$, representing physically the square of the speed of sound $c_s$ in the fluid \cite{Hydro-LTE},
		\begin{equation} \label{indicatrix def}
			\chi (\rho,p) \equiv \frac{u(p)}{u(\rho)} = \left(\frac{\partial p}{\partial\rho} \right)_s \equiv c_s^2 \, .
		\end{equation}
For this reason, condition S is also referred to as the \textit{sonic condition}. This function of state gives all the thermodynamic information that can be given using only hydrodynamic variables.

		\subsection{Solving the inverse problem}
		After the result in Theorem \ref{thm-hydro-LTE}, the inverse problem for perfect energy tensors in l.t.e. is tantamount to the analysis of how the sonic condition S constrains the associated thermodynamic schemes. A detailed study of different cases is presented in \cite{Hydro-LTE}. However, here we will only show the result that will be needed later:
		\begin{thm} \label{Prop-general-inverse-problem}
			Let $T$ be a non-barotropic, $\textrm{d} \rho \wedge \textrm{d} p \neq 0$, and non-isobaroenergetic, $u(p) \neq 0 \neq u(\rho)$, perfect energy tensor that evolves in l.t.e. The admissible thermodynamic schemes are defined by a matter density of the form $n = \bar{n}/N(\bar{s})$ and a specific entropy of the form $s = s(\bar{s})$, where $\bar{n}(\rho, p)$ is a particular solution of {\em(\ref{matter-conservation})}, namely of
			\begin{equation} \label{problema-invers-eq-dif-n}
				n'_{\rho}+\chi(\rho,p)n'_p = \frac{n}{\rho+p}\, ,
			\end{equation}
and $N(\bar{s})$ and $s(\bar{s})$ are arbitrary functions of a particular solution $\bar{s}(\rho, p)$ of {\em(\ref{specific-entropy-conservation})}, namely of
			\begin{equation} \label{problema-invers-eq-dif-s}
				s'_{\rho}+\chi(\rho,p)s'_p = 0.
			\end{equation}
			For each pair $\left\lbrace n, s \right\rbrace$, the temperature is given by
			\begin{equation} \label{problema-invers-temperatures-p-i-rho}
				\hspace{-3mm} \Theta = \Theta^p(\rho,p) \equiv -\frac{n'_p}{n^2s'_p}(\rho + p) \quad\; \textrm{or} \quad\; \Theta = \Theta^\rho(\rho,p) \equiv \frac{1}{s'_\rho}\left[\frac{1}{n} - \frac{n'_\rho}{n^2}(\rho + p)\right],
			\end{equation}
			and the specific internal energy by
			\begin{equation} \label{problema-invers-specific-internal-energy}
				\epsilon(\rho,p) = \frac{\rho}{n(\rho,p)} - 1.
			\end{equation}
		\end{thm}
		The notation $f'_\rho\equiv(\frac{\partial f}{\partial \rho})_p$ and $f'_p\equiv(\frac{\partial f}{\partial p})_\rho$ for any function $f = f(\rho ,p)$ has been introduced in this theorem and will be used throughout the rest of the thesis.
		As a consequence of this result, we have that the richness of admissible thermodynamic schemes is given by two arbitrary functions $\left\lbrace N(\bar{s}), s(\bar{s}) \right\rbrace$. Each election of these functions determines a pair $\left\lbrace n, s \right\rbrace$ which, in turn, determines a thermodynamic scheme, and therefore an interpretation of a perfect energy tensor fulfilling the sonic condition.
		
	\section{Other necessary physical reality conditions} \label{sec-other-phys-re-conds}
	In the previous subsection, we have analysed the l.t.e. condition and we have seen that it forces us to add to the hydrodynamic quantities, the ones appearing in a perfect energy tensor, a second set of quantities, the thermodynamic quantities, necessary to give the solution a physical interpretation as a fluid. Here, we present other conditions that both sets of quantities need to verify to represent a physically admissible fluid. \\ \\
	Pleba\'nski's \cite{Plebanski_1964}	\textit{energy conditions} establish essential algebraic criteria to ensure physical plausibility. These conditions require that, for every observer, the energy density must be positive and the momentum flux must be either timelike or null. For the case of a perfect fluid, these two conditions can be expressed in terms of hydrodynamic quantities as
	\begin{equation} \label{E}
		\hspace{-52mm}{\rm E}: \hspace{37mm} \qquad\qquad -\rho < p\leqslant\rho\,.
	\end{equation}
To analyse any given perfect fluid solution, it is therefore crucial to identify the spacetime regions where these conditions are satisfied. \\
	Another fundamental requirement for thermodynamic schemes is that temperature, matter density and specific internal energy must all be positive:
		\begin{equation} \label{P}
			\hspace{-39mm} {\rm P}: \hspace{29mm} \qquad\qquad \Theta>0\, , \qquad\quad \rho > n > 0 \, .
		\end{equation}
	Finally, to develop a consistent theory of shock waves in fluids (where shock velocities remain below the speed of light, shocks are compressive supersonic waves with an associated increase in entropy, and the post-shock state is uniquely determined by the pre-shock state) it is necessary to impose the relativistic \textit{compressibility conditions} \cite{Israel_1960, Lichnerowicz_1966}. They impose the inequalities H$_1$: $(\tau'_p)_s < 0 , \, (\tau''_p)_s > 0$, and the inequality H$_2$: $(\tau'_s)_p > 0$, where the function of state $\tau = \tau(p,s)$ is the \textit{dynamic volume}, $\tau = \hat{h} / n$, $\hat{h} = h / c^2$ being the dimensionless enthalpy index and $h \equiv c^2 + \epsilon + p / n = (\rho + p) / n$ the relativistic \textit{specific enthalpy} (we explicitly include the speed of light units, $c$, here to highlight that in the classical limit, $c \, \rightarrow \, \infty$, the dynamic volume $\tau$ approaches the specific volume $1 / n$). In \cite{Coll_Ferrando_i_Saez_2020b} it is shown that, for a perfect energy tensor in l.t.e., the compressibility conditions can be expressed as
		\begin{eqnarray} 
	 		\hspace{-20mm} \textrm{H}_1: & \qquad\qquad & 0 < \chi <1 \, , \qquad (\rho + p)(\chi \chi'_p + \chi'_\rho) + 2\chi(1-\chi) > 0 \, ; \label{H1} \\[2mm]
	 \hspace{-20mm} \textrm{H}_2: & \qquad\qquad & 2 n \Theta > \frac{1}{s'_\rho} \, . \label{H2}
		\end{eqnarray}
		Note that in the general necessary macroscopic constraints C, S, E, P, H$_1$ and H$_2$ specified above, we must distinguish two types of conditions according to their nature:
	\begin{itemize}
		\item[(i)] 
			{\em Hydrodynamic constraints}: the conservation equation C, the energy conditions E, the hydrodynamic sonic condition S, and the compressibility conditions H$_1$ exclusively involve the hydrodynamic quantities $\{u, \rho, p\}$. They fully determine the hydrodynamic flow of the thermodynamic fluid in local thermal equilibrium, whether the fluid is treated as a test fluid or as the source of the gravitational field. In the latter case, they also constrain the admissible gravitational field through the Einstein equations when looking for solutions.
		\item[(ii)]
			{\em Thermodynamic constraints}: the positivity conditions P and the compressibility condition H$_2$ restrict the thermodynamic schemes $\{n, \epsilon, s, \Theta\}$ associated with a hydrodynamic flow $\{u, \rho, p\}$. They offer different physical interpretations for a given hydrodynamic perfect fluid flow. Moreover, when looking for solutions to Einstein's equations, they do not restrict the admissible gravitational field. 
	\end{itemize}
		In order to implement the above macroscopic constraints in looking for physically admissible new perfect fluid solutions and in analysing the previously known ones, we can consider the general procedure in five steps proposed in \cite{C-F-S_SzSz_Regular}:
		\begin{description}
			\item[\textbf{Step 1.}] 
				Determine the subfamily of the thermodynamic solutions by imposing the hydrodynamic sonic condition S on the solutions to the conservative \mbox{equations C}.
			\item[\textbf{Step 2.}] 
				Obtain, for this subfamily, the coordinate dependence of the hydrodynamic quantities $\{u, \rho, p\}$, and the indicatrix function $c_s^2 = \chi(\rho,p)$.
			\item[\textbf{Step 3.}] 
				Analyse, for these thermodynamic solutions, the hydrodynamic constraints for physical reality, namely, the energy conditions E and the compressibility conditions H$_1$.
			\item[\textbf{Step 4.}] 
				Obtain the thermodynamic schemes $\{n, \epsilon, s, \Theta\}$ associated with these solutions as explained in Theorem \ref{Prop-general-inverse-problem}. 
			\item[\textbf{Step 5.}] 
				Analyse, for the thermodynamic schemes $\{n, \epsilon, s, \Theta\}$ already obtained, the general thermodynamic constraints for physical reality, namely, the positivity conditions P and the compressibility condition H$_2$.
		\end{description}
		
	\section{Generic ideal gas} \label{sec-generic-ideal-gas}
	The direct and inverse problems outlined above enable us to determine whether a perfect energy tensor models the evolution of some perfect fluids. However, they do not provide information about the specific physical properties of these fluids. If we are interested in a particular family of fluids, $\textbf{G}\subset\textbf{F}$, we must solve the corresponding specific direct and inverse problems. Specifically, we need to establish a deductive criterion to identify whether a given perfect energy tensor $T$ represents the evolution of a perfect fluid within this family, as well as to determine all the perfect fluids of this family for which $T$, satisfying this criterion, follows a particular evolution. In other words, we aim to identify $\textbf{T}_{\textbf{G}}$ and, for any $\textit{T}\in\textbf{T}_{\textbf{G}}$, determine $\textbf{G}_T$. \\
	In \cite{Hydro-LTE}, such a study was performed for the paradigmatic family $\textbf{G}$ of ideal gases. A (generic) ideal gas is characterised by the EoS:
	\begin{equation} \label{eq. estat gas ideal}
		p = \tilde{k} n \Theta\, , \qquad\qquad \tilde{k} \equiv \frac{k_B}{m} \, ,
	\end{equation}
$k_B$ being Boltzmann's constant and $m$ the particle mass. As a result, the following theorems are obtained there:
	\begin{thm} \label{Theorem-ideal-direct-problem}
		The necessary and sufficient condition for a non-barotropic ($\emph{d}\rho \wedge \emph{d} p \neq 0$) and non-isoenergetic ($\dot{\rho} \neq 0$) perfect energy tensor $T = (u, \rho, p)$ to represent the l.t.e. evolution of an ideal gas is that the indicatrix function $\chi \equiv \dot{p}/\dot{\rho}$ be a function of the variable $\pi\equiv p/\rho$: \\[-2mm]
		\begin{equation} \label{cond. sonica ideal}
			\hspace{-46mm} \emph{S}^\emph{G}: \hspace{21mm} \qquad\qquad \emph{d}\chi\wedge \emph{d}\pi = 0\, ,\qquad \chi \neq \pi\, ,
		\end{equation}
which we will name the ideal sonic condition.
	\end{thm}
	\begin{thm} \label{Theorem-ideal-inverse-problem}
		A non-barotropic and non-isoenergetic perfect energy tensor satisfying {\em(\ref{cond. sonica ideal})} represents the l.t.e. evolution of the ideal gas with specific internal energy $\epsilon$, temperature $\Theta$, matter density $n$ and specific entropy $s$ given by
		\begin{equation} \label{epsilon i Theta ideals}
			\epsilon(\rho,p) = \epsilon(\pi) = e(\pi) - 1\, , \qquad \Theta(\rho,p) = \Theta(\pi) \equiv \frac{\pi}{\tilde{k}}e(\pi) \, ;
		\end{equation}
		\begin{equation} \label{n i s ideals}
			n(\rho, p) = \frac{\rho}{e(\pi)}\, , \qquad\qquad s(\rho, p) = \tilde{k}\ln\frac{f(\pi)}{\rho} \, ;
		\end{equation}
		with $e(\pi)$ and $f(\pi)$ given, respectively, by
		\begin{equation} \label{e(pi) i f(pi)}
			e(\pi) = e_0\exp \left\lbrace \int \psi(\pi) d\pi \right\rbrace\, ,\qquad f(\pi) = f_0 \exp\left\lbrace \int\phi(\pi )d\pi \right\rbrace \, ;
		\end{equation}
		where
		\begin{equation} \label{psi(pi) i phi(pi)}
			\psi(\pi) \equiv \frac{\pi}{(\chi(\pi) - \pi)(\pi+1)}\, , \qquad \phi(\pi) \equiv \frac{1}{\chi(\pi) - \pi} \, .
		\end{equation}
	\end{thm}
Theorem \ref{Theorem-ideal-direct-problem} solves the specific direct problem for the non-barotropic ideal gases, which will be our focus. Theorem \ref{Theorem-ideal-inverse-problem}, in turn, solves the corresponding specific inverse problem for a non-barotropic and non-isoenergetic \textit{ideal energy tensor} $T$, i.e. satisfying the ideal sonic condition S$^\textrm{G}$. It provides the associated \textit{ideal thermodynamic scheme}, namely the one allowing to interpret the solution as a fluid following the evolution of an ideal gas because it is an ideal gas. However, it is worth noting that this is not the only admissible thermodynamic scheme. An ideal energy tensor may, in general, admit other associated thermodynamic schemes corresponding to fluids that are not ideal gases but admit evolutions satisfying the ideal sonic condition.\\ \\
	With this, we can particularise some of the necessary physical reality conditions presented in the previous section for the case of an ideal energy tensor.
	The sonic condition, for instance, becomes in this case the ideal sonic condition S$^\textrm{G}$. The compressibility condition H$_1$ can be written as \cite{Coll_Ferrando_i_Saez_2020b}
	\begin{equation} \label{H1G}
	 	\hspace{-5mm} \textrm{H}_1^\textrm{G}: \;\qquad\qquad 0 < \chi < 1 \, , \qquad \zeta (\pi) \equiv (1 + \pi) (\chi - \pi) \chi' + 2\chi(1 - \chi) > 0 \, .
	\end{equation}
Even though there exist continuous media with a negative pressure, the EoS (\ref{eq. estat gas ideal}) together with the positivity conditions P given in (\ref{P}) imply that pressure must be positive. Therefore, energy conditions become in this case
	\begin{equation} \label{EG}
		\hspace{-48mm} \textrm{E}^\textrm{G}: \hspace{28mm} \qquad \rho > 0 \, , \qquad 0 \leqslant \pi < 1 \, .
	\end{equation}
For the case of an ideal thermodynamic scheme, the compressibility condition H$_2$ can also be rewritten as \cite{Coll_Ferrando_i_Saez_2020b}
	\begin{equation} \label{H2G}
		\hspace{-35mm} \textrm{H}_2^\textrm{G}: \hspace{16mm} \qquad\qquad \eta (\pi) \equiv (2 \pi + 1) \chi(\pi) - \pi > 0 \, . \qquad
	\end{equation}
All this considered, we can modify the procedure in the previous section as follows to study the solutions with a generic ideal gas behaviour:
	\begin{description}
			\item[\textbf{Step 1'.}] 
				Determine the subfamily of ideal gas solutions by imposing the ideal sonic condition S$^\textrm{G}$. \\[-8mm]
			\item[\textbf{Step 2'.}] 
				Obtain, for this subfamily, the hydrodynamic quantities $\{u, \rho, p\}$, and the indicatrix function $c_s^2 = \chi(\pi)$. \\[-8mm]
			\item[\textbf{Step 3'.}] 
				Analyse, for the ideal gas solutions, the hydrodynamic constraints for physical reality E$^\textrm{G}$ and H$_1^\textrm{G}$. \\[-8mm]
			\item[\textbf{Step 4'.}] 
				Obtain the thermodynamic schemes $\{n, \epsilon, s, \Theta\}$ associated with the ideal gas solutions. \\[-8mm]
			\item[\textbf{Step 5'.}] 
				Analyse, for some physically relevant thermodynamic schemes $\{n, \epsilon, s, \Theta\}$, the thermodynamic constraints for physical reality P and H$_2$.
	\end{description}  
		
	\section{Classical ideal gas} \label{sec-classic-ideal-gas}
	As a particular subfamily of the generic ideal gases, we have the classical ideal gases (CIG). They are the ideal gases, with equation of state (\ref{eq. estat gas ideal}), that also fulfil the classical dependence of the specific internal energy on the temperature 
	\begin{equation} \label{epsilon gas ideal classic}
		\epsilon = c_v \Theta. 
	\end{equation}
They are usually considered as a good approximation of an ideal gas at low temperatures. In \cite{CFS-CIG, FS-KCIG}, they are studied from a hydrodynamic and kinematic point of view, respectively, and in this subsection we summarise some results that we will need in the subsequent chapters.\\ \\
	From (\ref{eq. estat gas ideal}) and (\ref{epsilon gas ideal classic}), it follows that a CIG satisfies the \textit{classical} $\gamma$\textit{-law} equation of state
	\begin{equation} \label{eq. estat gas ideal classic}
		p = (\gamma - 1) n \epsilon, \qquad \gamma \equiv 1 + \frac{\tilde{k}}{c_v},
	\end{equation}
where $\gamma$ is the adiabatic index.
	In \cite{CFS-CIG}, the specific direct problem for the CIG is solved, leading to the following theorem:
	\begin{thm} \label{Theorem-classical-ideal-direct-problem}
		The necessary and sufficient condition for a non-barotropic ($\emph{d}\rho \wedge \emph{d} p \neq 0$) and non-isoenergetic ($\dot{\rho} \neq 0$) perfect energy tensor $T = (u, \rho, p)$ to represent the l.t.e. evolution of a CIG with adiabatic index $\gamma$ is that the indicatrix function $\chi \equiv \dot{p}/\dot{\rho}$ be of the form
		\begin{equation} \label{chi-gas-ideal-classic}
			\chi_c(\pi) = \frac{\gamma \pi}{\pi + 1} \, .
		\end{equation}
	\end{thm}
	After that, the macroscopic compressibility conditions in this case are studied and the following result is obtained \cite{CFS-CIG}: 
	\begin{proposition} \label{prop-classical-ideal-gas-compressibility-conds}
		For a classical ideal gas with $\Theta > 0$, $\rho > n > 0$, the macroscopic constraints for physical reality {\em E}$^{\mathrm{G}}$, {\em H}$^{\mathrm{G}}_1$ and {\em H}$^{\mathrm{G}}_2$ are satisfied for values of $\pi$ in a nonempty subinterval of $[0,1]$ if, and only if, the adiabatic index fulfils $\gamma > 1$. In fact, they hold in the interval
		\begin{equation} \label{cc-CIG}
			\begin{cases}  
				\; 0 < \pi < {\pi}_m \equiv \gamma - 1 \, , \quad {\rm if} \ \ 1 < \gamma \leq 2 \, , \cr
\displaystyle 
				\; 0 < \pi < \tilde{\pi}_m \equiv \frac{1}{\gamma - 1} , \quad {\rm if} \ \ \gamma \geq 2 \, . 
			\end{cases}
		\end{equation}
	\end{proposition}
	Usually (see, for example \cite{Anile}) the adiabatic index $\gamma$ is considered to be constrained by $1 < \gamma \leq 2$. The upper limit was obtained by Taub \cite{Taub} from a kinetic approach. However, the statement above shows that, under reasonable macroscopic physical requirements, it can be relaxed.\\ \\
	The isentropic evolution of a classical ideal gas is also studied in \cite{CFS-CIG}. When a non-barotropic perfect fluid has an isentropic evolution, this evolution is performed by a barotropic energy tensor. Regarding this, in \cite{CFS-CIG} they get the following:
	\begin{proposition} \label{prop-CIG-isentr}
		A perfect energy tensor $T = (u, \rho, p)$ represents the isentropic evolution of a CIG if, and only if, the following barotropic relation holds:
		\begin{equation} \label{CIG-isentropic}
			(\gamma - 1) \rho = p + B p^{\frac{1}{\gamma}} \, , \quad B = {\rm constant} \neq 0 \, .
		\end{equation}
Conversely, the barotropic evolution {\em(\ref{CIG-isentropic})} represents the isentropic evolution of a CIG with adiabatic index $\gamma$. Moreover, the matter density $n$ is given by
		\begin{equation} \label{n-CIG-isentr}
			n = \frac{B}{\gamma - 1} p^{1/\gamma} \, .
		\end{equation}
	\end{proposition}
Finally, in \cite{FS-KCIG}, Ferrando and Sáez characterise the unit velocities of the classical ideal gas solutions of the hydrodynamic equations, and the following result is shown: 
	\begin{proposition} \label{prop-kinematic-CIG}
		A geodesic and expanding timelike unit vector $u$ is the unit velocity of a classical ideal gas if, and only if, $u$ is vorticity-free and its expansion is homogeneous, that is, $u = - \textrm{d} t$ and $\theta = \theta(t)$.
	\end{proposition}

	\section[Solutions compatible with non-vanishing transport coefficients]{Solutions$\,$ compatible$\,$ with$\,$ non-vanishing \\ transport$\,$ coefficients} \label{subsec-intro-thermal-conduc-coeff}
	According to the thermodynamic theory of irreversible processes (in both the standard irreversible thermodynamics \cite{Eckart} and the extended irreversible \mbox{thermodynamics} \cite{Israel-76, Israel-St}), the transport coefficients of thermal conductivity, shear-viscosity, and bulk-viscosity appear in the constitutive equations linking dissipative fluxes (anisotropic pressures, bulk viscous pressure, and energy flux) with the kinematic coefficients of fluid flow (shear, expansion and acceleration) \cite{Rezzolla}. \\
	The perfect fluid approximation can be considered when the transport \mbox{coefficients} of a fluid vanish (or are negligible). A non-perfect fluid is a fluid with at least a non-zero transport coefficient. For this fluid, the energetic evolution is, generically, described by an energy tensor with energy flux and anisotropic pressures. However, when a non-perfect fluid admits particular evolutions in which the dissipative fluxes vanish, these evolutions are well described by a perfect energy tensor, and are usually called {\em equilibrium states} \cite{Rezzolla}. Moreover, all the thermodynamic relations of the perfect fluid hydrodynamics remain valid. Furthermore, the shear, the expansion and the acceleration of the fluid undergo strong restrictions as a consequence of the constitutive equations. Specifically: 
	\begin{itemize}
		\item[(i)]
			If the shear viscosity coefficient does not vanish, then the fluid shear vanishes. 
		\item[(ii)]
			If the bulk viscosity coefficient does not vanish, then the fluid expansion \mbox{vanishes}. 
		\item[(iii)]
			If the thermal conductivity coefficient does not vanish, then the fluid \mbox{acceleration} is constrained by the relation
		\begin{equation} \label{Fourier}
			a = - \! \perp \! \textrm{d} \ln \Theta \, , 
		\end{equation}
where $\perp$ denotes the orthogonal projection to the fluid velocity.
	\end{itemize}
	Then, under some kinematic constraints of the fluid flow, a non-perfect fluid can evolve as a perfect fluid because the dissipative fluxes can vanish, even if the transport coefficients are nonzero.
	For example, the FLRW universes can model a thermodynamic perfect fluid in isentropic evolution. Nevertheless, they could also model the evolution of a fluid with non-vanishing thermal conductivity and shear-viscosity coefficients. Indeed, in this case we have a geodesic and shear-free flow, and any homogeneous temperature is compatible with (\ref{Fourier}).
	
\chapter{Hydrodynamic approach to the Synge gas} \label{chap-Synge}
The results of the preceding chapter are based on previous works \cite{Hydro-LTE, CFS-CIG} in which both the generic ideal gas and the classical ideal gas are characterised from a hydrodynamic point of view. These results will be used in the next part of the thesis to explore when the evolutions represented by certain solutions of the Einstein equations are compatible with the corresponding equations of state. The aim of this chapter, which is based on the results of \cite{FM-Synge}, is to carry out a similar study for the Synge gas, the only relativistic fluid for which a macroscopic EoS has been obtained from microscopic kinetic theory \cite{Rezzolla}. \\ \\
In \cite{Synge}, Synge integrated the relativistic generalisation of the Maxwell-Boltzmann distribution function (also known as Maxwell-Jütner distribution \cite{Rezzolla}) over momentum space and obtained the following relations: 
\begin{equation} \label{synge}
	p = \tilde{k} n \Theta, \qquad \; \frac{\rho + p}{n} = {K_3(z) \over K_2(z)} \equiv h(z) , \quad z \equiv {1 \over \tilde{k} \Theta}, 
\end{equation}
$(\rho + p) / n = h$ being the \textit{relativistic specific enthalpy} and $K_m(z)$, the modified Bessel functions of the second kind. \\ \\
Hence, from a macroscopic point of view, Synge fluid is a specific solution of the \textit{fundamental system of relativistic hydrodynamics} defined by the following elements:
\begin{itemize}
\item[(i)]
A divergence-free perfect energy tensor, (\ref{perfect-energy-tensor}), that describes the hydrodynamic evolution of the fluid (\ref{energy-momentum-conservation}). \\ \\
\item[(ii)]
A set of {\em thermodynamic quantities} $\{n, \epsilon, s, \Theta\}$ constrained by the usual thermodynamic laws namely, the conservation of matter (\ref{matter-conservation}) and the {\em local thermal equilibrium relation} (\ref{thermo-first-law}), or, equivalently, (\ref{thermo-first-law-h}).
\item[(iii)]
The macroscopic equations of state of a relativistic non-degenerate monoatomic gas (\ref{synge}).
\end{itemize}
Note that points (i) and (ii) define the deterministic fundamental system of the perfect fluid hydrodynamics, ${\cal F} \equiv \{(\ref{energy-momentum-conservation}) (\ref{matter-conservation}) (\ref{thermo-first-law})\}$, which characterises the evolution of any perfect fluid in l.t.e. The first equation of state in (\ref{synge}) constrains the fluid to be a generic ideal gas, and the second equation in (\ref{synge}) specifies that this ideal gas is a Synge gas. \\ \\
Recall that each specific fluid has a specific expression for the function of state $c_s^2 = \chi(\rho,p)$. In particular, when the fluid is a generic ideal gas (the first equation in (\ref{synge}) holds), the indicatrix function is $\chi = \chi(\pi) \neq \pi$, $\pi \equiv p/\rho$. The first goal of this chapter is to show that the second equation of state in (\ref{synge}) imposes a first order differential equation on $\chi(\pi)$. This leads to a purely hydrodynamic characterisation of a Synge gas. \\ \\
Synge EoS (\ref{synge}) implies that the thermodynamics of the fluid needs to be formulated in terms of the modified Bessel functions. Hence, it will not have a simple analytical expression. For this reason, different approximations have been considered in the literature to study the relativistic gas.
On the one hand, some authors have used the limiting behaviour of the modified Bessel functions to obtain an approximation of the Synge EoS for high and low temperatures \cite{Thorne} (see also \cite{Rezzolla} and references therein). 
On the other hand, Taub \cite{Taub} found that the pressure $p$, energy density $\rho$ and matter density $n$ of a simple gas must satisfy a certain inequality in order to be consistent with kinetic theory, namely,
\begin{equation} \label{Taub}
	\hspace{-5mm} {\rm Ta} : \qquad \qquad \qquad \rho (\rho - 3p) \geq n^2 \, . \qquad \qquad \qquad
\end{equation}
Then, Mathews \cite{Mathews} used the corresponding equality as the EoS for a relativistic gas, and Mignone {\em et al.} \cite{Mignone, Mignone-2007} showed that it provides a reasonable approximation to the Synge EoS. From now on, we will refer to this equation as the Taub-Mathews (TM) EoS. \\
The second goal of this chapter is to analyse all these approximations and to introduce new ones. Our hydrodynamic approach (through the indicatrix function $\chi(\pi)$) enables us to study and compare the accuracy of the different approximations and to analyse whether they fulfil the usual constraints for physical reality (\ref{E}$-$\ref{H2}). These physical constraints were inferred from a macroscopic analysis. Moreover, we will also impose Taub's inequality Ta (which was inferred from the kinetic theory \cite{Taub}) on the equations of state that approximate that of a Synge gas. \\ \\
Section \ref{sec-synge} is devoted to achieving the first objective of this chapter: the purely hydrodynamic labelling of Synge gas solutions. This result is applied to analyse the behaviour of a Synge gas at low and high temperatures and to establish the Rainich theory for the Einstein-Synge solutions.
In Section \ref{sec-approximations}, we analyse equations of state that approximate the Synge EoS at low or at high temperatures, and use our hydrodynamic approach to determine acceptable approximations in the entire domain of applicability. In particular, we recover the Taub-Mathews EoS and we analyse its accuracy.
Finally, in Section \ref{sec-isentropic-evol} we consider the isentropic evolution of an ideal gas and we apply this study to obtain the Friedmann equation for a TM ideal gas.

	\section{Hydrodynamic flow of a relativistic Synge gas} \label{sec-synge}
	A Synge gas is characterised by the two equations in (\ref{synge}): the first one being the EoS of a generic ideal gas and the second one, giving an extra constraint between thermodynamic quantities. However, it is worth noting that the relation between the pair of thermodynamic quantities $(z,h)$ given in (\ref{synge}) is not the only way to characterise the Synge gas. This equation of state leads to relations between other pairs of quantities which also characterise the Synge gas. Now, the objective is to find one of these equations of state that only involves hydrodynamic quantities. \\ \\
	Recall (see Section \ref{sec-generic-ideal-gas}) that, as a consequence of the ideal gas EoS (\ref{eq. estat gas ideal}), the speed of sound is a function of the hydrodynamic quantity $\pi = p/\rho$, $c_s^2 = \chi(\pi)$. Now we show that the Synge EoS (\ref{synge}) determines $\chi(\pi)$ by imposing a specific first-order differential equation.
	
		\subsection{Hydrodynamic characterisation of the Synge gas} \label{subsec-chi-synge}
		Using that the functions $K_m(z)$ satisfy $-z^{-m}K_{m + 1}(z) = \frac{\textrm{d}}{\textrm{d}z}(z^{-m} K_m(z))$, the Synge equation (\ref{synge}) can be written as
		\begin{equation}
			h(z) = {2 \over z} - {K_2'(z) \over K_2(z)} . \label{h-K2}
		\end{equation}
The function $K_2(z)$ is a solution of the Bessel equation,
		\begin{equation}
			z^2 K_2''(z) + z K_2'(z) - (4 + z^2)K_2(z) = 0 , \label{bessel}
		\end{equation}
the only one fulfilling the boundary condition $K_2(\infty) = 0$. Then, the function $h = h(z)$ may be characterised by the following differential equation equivalent to (\ref{bessel}):
		\begin{equation}
			z[h'(z) - h^2 + 1] + 5 h = 0 , \label{bessel-h}
		\end{equation}
together with the boundary condition $h(\infty) = 1$. Moreover, this condition ensures that $h(z)$ is greater than $1$ in all the domain, as required by the definition of the relativistic specific enthalpy, $h = (\rho + p)/n$, and the positivity conditions P given in (\ref{P}). Using equation (\ref{eq. estat gas ideal}), the second equation in (\ref{epsilon i Theta ideals}) and the first one in (\ref{n i s ideals}), we can obtain the function of state that relates the quantities $(z, \pi)$: 
		\begin{equation} \label{pi-z}
			\pi = \frac{1}{z \, e} = \frac{1}{z h(z) - 1} \equiv \pi (z) \, .
		\end{equation}
Equations (\ref{bessel-h}) and (\ref{pi-z}) give us the differential equation that, together with the conditions $\pi(\infty) = 0^+$ and $\pi'(\infty) = 0$, characterise the function $\pi = \pi(z)$:
		\begin{equation} \label{bessel-pi}
			z \, \pi'(z) - (3 + z^2)\pi^2 - 2\pi + 1 = 0 \, .
		\end{equation}
In this case, the boundary conditions ensure that the function $\pi = \pi(z)$ takes values in the interval $]0, 1/3]$, a fact that is compatible with the energy conditions E$^{\rm G}$. On the other hand, by deriving the first equality in (\ref{pi-z}) with respect to $z$ and using the first equations in (\ref{e(pi) i f(pi)}$-$\ref{psi(pi) i phi(pi)}) together with (\ref{pi-z}) to eliminate $e(\pi)$ and $e'(\pi)$, we have
		\begin{equation} \label{z pi-prima}
			z \, \pi'(z) = \frac{\pi (\pi + 1)[\pi - \chi(\pi)]}{\pi [\chi(\pi) - 1] + \chi(\pi)} \, ,
		\end{equation}
which can be substituted in (\ref{bessel-pi}) to give
		\begin{equation} \label{z2-pi}
			z^2 = \frac{1}{\pi^2} - \frac{3}{\pi} + \frac{\pi}{\pi [\chi(\pi) - 1] + \chi(\pi)} - 3 \, .
		\end{equation}
Finally, deriving (\ref{z2-pi}) with respect to $z$ and using (\ref{z pi-prima}), we obtain that the indicatrix function of a Synge gas is the solution to the first-order differential equation
		\begin{subequations} \label{chi-prima}
			\begin{eqnarray} \label{chi-prima-eq}
				\chi'(\pi) &\!\!\!\! = &\!\!\!\! {\cal S}(\chi, \pi) \equiv \frac{\alpha \chi^3 + \beta \chi^2 + \gamma \chi + \delta}{\pi^3 (\pi + 1) (\pi - \chi)} \, , \\[3mm]
				\alpha &\!\!\!\! = &\!\!\!\! \alpha(\pi) \equiv 3(1 + 4\pi + 5\pi^2 + 2\pi^3) \, , \label{subeq-alpha} \\[2mm]
				\beta &\!\!\!\! = &\!\!\!\! \beta(\pi) \equiv - \pi(11 + 29\pi + 17\pi^2) \, , \label{subeq-beta} \\[2mm]
				\gamma &\!\!\!\! = &\!\!\!\! \gamma(\pi) \equiv \pi^2(13 + 17\pi) \, , \label{subeq-gamma} \\[2mm]
				\delta &\!\!\!\! = &\!\!\!\! \delta(\pi) \equiv - 5\pi^3 \, , \label{subeq-delta}
			\end{eqnarray}
		\end{subequations}
with the boundary condition $\chi'(0) = 5/3$. Then, $\chi(0) = 0$, and the solution $\chi = \chi(\pi)$ takes values in the interval $[0,1/3]$, which is compatible with the first of the compressibility conditions H$^{\rm G}$. \\ \\
		This statement certainly gives us a hydrodynamic characterisation of the Synge gas because it only uses conditions on the hydrodynamic quantities $\{u, \rho, p\}$. But it is not a deductive characterisation because the existence of a functional dependence between the quantities $\chi$ and $\pi$ does not imply that the expression of the function $\chi(\pi)$ is known, which is a necessary requirement in order to impose condition (\ref{chi-prima}). Nevertheless, a deductive characterisation easily follows:
		\begin{itemize}
\item[]
The necessary and sufficient condition for a non-isoenergetic ($\dot{\rho} \neq 0$) divergence-free energy tensor $T$ to represent the energy evolution of a Synge gas is that its hydrodynamic quantities $\{u,\rho,p\}$ fulfil the {\em Synge sonic condition}
			\begin{equation} \label{SyngeSC}
				{\rm S^{\rm S}} : \quad \qquad \ \ \textrm{d} \chi = {\cal S}(\chi, \pi) \textrm{d} \pi \, , \quad  
			\end{equation}
where ${\cal S}(\chi, \pi)$ is given in \rm(\ref{chi-prima}), and $S(0,0) = 5/3$.
		\end{itemize}
		The above result states: (i) if $\{u, \rho, p, n, s, \Theta\}$ is a solution of the fundamental system of the Synge gas hydrodynamics ${\cal F}_{\rm S} \equiv \{(\ref{energy-momentum-conservation}) (\ref{matter-conservation}) (\ref{thermo-first-law}) (\ref{eq. estat gas ideal}) (\ref{synge})\}$, then $\{u, \rho, p\}$ is a solution of the Synge hydrodynamic flow system ${\cal{H}}_{\rm S} \equiv \{(\ref{energy-momentum-conservation}) (\ref{SyngeSC})\}$, and conversely, (ii) if $\{u, \rho, p\}$ is a solution of the ideal hydrodynamic flow system ${\cal{H}}_{\rm S}$, then a solution $\{u, \rho, p, n, s, \Theta\}$ of the Synge fundamental system ${\cal F}_{\rm S}$ exists. \\ \\
		A specific ideal gas, and in particular a Synge gas, is defined by a specific indicatrix function $\chi(\pi)$, but also by the function $e = e(\pi)$. The former one is an explicit hydrodynamic quantity, $\chi = u(p)/u(\rho)$, and for that reason it is the right one to give the above hydrodynamic characterisation. Nevertheless, it can also be conceptually interesting to characterise a Synge gas in terms of the pair $(\pi, e)$. From (\ref{eq. estat gas ideal}), the second equation in (\ref{epsilon i Theta ideals}) and the first one in (\ref{n i s ideals}), we obtain
		\begin{equation}
			e(z) = h(z) - \frac{1}{z} \, .
		\end{equation}
Using this equation to write (\ref{bessel-h}) for $e = e(z)$, we get
		\begin{equation} \label{e'-1}
			z^2 e'(z) + z^2[1 - e^2(z)] + 3[1 + z e(z)] = 0 \, ,
		\end{equation}
while, from (\ref{pi-z}) we obtain
		\begin{equation} \label{e'-2}
			e'(\pi) = - \frac{\pi}{e} [1 + \pi e^2 z'(e)] = - \pi \frac{\pi e^2 + e'(z)}{e e'(z)} \, .
		\end{equation}
Then, from equations (\ref{e'-1}) and (\ref{e'-2}) we have that the function $e = e(\pi)$ of a Synge gas fulfils the boundary condition $e'(0) = 3/2$ and the first-order differential equation
		\begin{equation} \label{e'-pi}
			e'(\pi) = \frac{e(\pi)[1 + e^2(\pi)(3 \pi^2 \! + \! 3 \pi \! - \! 1)]}{ \pi [e^2(\pi)(1 \! - \!2\pi \! - \! 3\pi^2) - 1]} \, .  
		\end{equation}
		Another function of state that characterises an ideal gas is the so-called (generalised) adiabatic index $\Gamma \equiv (\partial \ln n/ \partial \ln p)_s$. For a generic ideal gas, it is related to the indicatrix function $\chi(\pi)$ by the expression \cite{Rezzolla, Krautter}
		\begin{equation} \label{Gamma}
			\Gamma(\pi) = \frac{1 + \pi}{\pi}\chi(\pi) \, .
		\end{equation}
Note that this expression shows that $\Gamma$ can be obtained from the hydrodynamic quantities $(u, \rho, p)$, and thus it can be used to identify a specific ideal gas. Equations (\ref{chi-prima}) and (\ref{Gamma}) imply that the $\Gamma = \Gamma(\pi)$ of a Synge gas is characterised by the first-order differential equation
		\begin{equation}
			\Gamma'(\pi) = \frac{(\Gamma - 1)^2[5(\pi + 1) - 3 \Gamma (2\pi + 1)]}{\pi^2(\Gamma - 1 -\pi)} \, ,
		\end{equation}
and the boundary condition $\Gamma'(0) = -5/3$.
		
		\subsection{Behaviour of the Synge gas at low and high temperatures} \label{subsec-approaches}
		In the previous subsection, we have characterised the indicatrix function $\chi(\pi)$ of a Synge gas through the differential equation (\ref{chi-prima}), which cannot be solved analytically. Now, in this subsection, we will use it to study the behaviour of $\chi(\pi)$ in the limiting cases at low and high temperatures. \\ \\
		On the one hand, the limit $\Theta = 0$ corresponds to $p = 0$, and then $\pi = 0$. On the other hand, the limit at high temperature corresponds to $z = 0$, and from expressions (\ref{h-K2}) and (\ref{pi-z}) of $h(z)$ and $\pi(z)$, to $\pi = 1/3$. Thus, the indicatrix function of a Synge gas $\chi(\pi)$ takes values in the domain $[0, \, 1/3]$. 
		Note that equation (\ref{chi-prima}) determines the value of $\chi(\pi)$ at both ends of the interval. Indeed, as commented in subsection above, we have that 
		\begin{equation} \label{chi-en-zero}
			\chi(0) = 0 \, , \quad \qquad \chi'(0) = 5/3 \, .
		\end{equation}
Then, these values and the successive derivatives of equation (\ref{chi-prima}) determine the derivatives of $\chi(\pi)$ at $\pi = 0$. This fact enables us to know the behaviour of the indicatrix function $\chi(\pi)$ of a Synge gas at low temperatures by writing its Taylor expansion around $\pi = 0$. For instance, the second, third and fourth derivatives take the values
		\begin{equation} \label{derivades-chi-zero}
			\chi''(0) = -\frac{20}{3} \, , \quad \chi'''(0) = 45 \, ,\quad \chi^{iv}(0) = 460 \, .
		\end{equation}
		Based on the behaviour of the Bessel functions close to zero, it is also possible to get that
		\begin{equation} \label{derivades-chi-1/3}
			\chi(1/3) = 1/3 \, , \qquad \chi'(1/3) = 1/2 \, ,
		\end{equation}
which is compatible with equation (\ref{chi-prima}). However, as can be seen from these values together with the derivative of equation (\ref{chi-prima}), the second derivative of $\chi(\pi)$ is not defined at $\pi = 1/3$. This is a consequence of the fact that the function $K_2(z)$ has a singular point at $z = 0$. Thus, we can only know the behaviour of the indicatrix function $\chi(\pi)$ for a Synge gas around $\pi = 1/3$ (at high temperatures) up to first order in Taylor expansion.
		
		\subsection{Rainich-like theory for the Einstein-Synge solutions} \label{subsec-Rainich}
		An Einstein-Synge solution is a solution of the Einstein field equations, (\ref{Einstein-equations}), where $T$ is a perfect energy tensor that models the energy evolution of a Synge gas. The Einstein-Synge equations involve the metric tensor $g$ and the thermodynamic quantities $\{u, \rho, p, n, s, \Theta\}$ that are constrained by the fundamental system of the Synge gas hydrodynamics ${\cal F}_{\rm S} \equiv \{(\ref{energy-momentum-conservation}) (\ref{matter-conservation}) (\ref{thermo-first-law}) (\ref{eq. estat gas ideal}) (\ref{synge})\}$. \\ \\ 
		Note that, as a consequence of the hydrodynamic characterisation obtained in the subsection above, the Einstein-Synge equations are equivalent to a system that only involves the quantities $\{g, u, \rho, p\}$: the field equations $G(g) = \kappa T$, $T \equiv \{u, \rho, p\}$ being a solution of the Synge hydrodynamic flow system ${\cal{H}}_{\rm S} \equiv \{(\ref{energy-momentum-conservation})(\ref{SyngeSC})\}$. Then, a question naturally arises: is there a system of equations involving only the metric tensor $g$ that characterises the Einstein-Synge solutions? Answering that question and obtaining those equations amounts to formulating the Rainich-like theory for the Einstein-Synge solutions. \\ \\
		The first theory that characterises the non-vacuum solutions of the Einstein field equations corresponding to a specific energy content was formulated by Rainich \cite{Rainich}. He gave the necessary and sufficient conditions for a metric tensor $g$ to be an Einstein-Maxwell solution for a regular electromagnetic field. \\ \\
		Using the algebraic characterisation of the perfect fluid energy tensor \cite{BCM-1992}, Coll and Ferrando developed in \cite{Coll-Ferrando-termo} a similar theory for the thermodynamic perfect fluid solutions of the Einstein equations, and the hydrodynamic sonic condition plays an important role in that study. Later, they also gave in \cite{Coll_Ferrando_i_Saez_2020b} the necessary and sufficient conditions for a perfect fluid solution to describe a generic ideal gas that fulfils the compressibility conditions H$_1^\textrm{G}$ and H$_2^\textrm{G}$. \\
		If we want the fluid to be a Synge gas, equation (\ref{SyngeSC}) must be added to those conditions. Nevertheless, now H$_1^\textrm{G}$ and H$_2^\textrm{G}$ identically hold, and they can be removed in the characterisation statement. Thus, if we take into account Theorems 9 and 10 in reference \cite{Coll_Ferrando_i_Saez_2020b}, we can obtain the Rainich-like theory for the Einstein-Synge solutions. Below, we present the explicit expressions of this invariant labelling. \\[2mm]
{\bf Ricci concomitants.} \textit{Consider the following scalar and tensor functions of the Ricci tensor $R$ and its derivatives:}
		\begin{eqnarray}
			t \equiv \textrm{tr} R , \qquad \qquad N \equiv R - \frac14 t \, g ,\label{fluper-definitions-1} \qquad \\[-1mm]   
			\displaystyle q \equiv - 2 \sqrt[3]{\frac{\textrm{tr} N^3}{3}} , \qquad Q \equiv N - \frac14 q \, g , \label{fluper-definitions-2} \qquad \\[-1mm]
			\label{fluper-hydro}
 			\rho = \frac14 (3 q + t) , \quad \qquad p = \frac14 (q - t) , \qquad \\[0mm]  
			\pi \equiv \frac{p}{\rho} , \quad \qquad \chi \equiv \frac{Q( \textrm{d} p , \textrm{d} \rho)}{Q( \textrm{d} \rho , \textrm{d} \rho)} . \qquad \label{Rainich-Synge}
		\end{eqnarray}
{\bf Characterisation theorem 1.} \textit{A metric is an Einstein-Synge solution in non-isoenergetic evolution if, and only if, the Ricci tensor} $R$ \textit{satisfies the invariant conditions} \\[-10mm]
		\begin{eqnarray} \label{fluper-conditions-ideal}
			Q^2 + q Q = 0 , \quad Q(v,v) > 0 , \quad -t < q \leq t , \qquad \ \\[2mm]
			Q(\textrm{d} \rho) \neq 0, \quad \textrm{d} \chi = {\cal S}(\chi, \pi) \textrm{d} \pi , \quad {\cal S}(0,0)= 5/3 , \qquad \ \label{Si(x)}
		\end{eqnarray}
\textit{where} $v$ \textit{is any timelike vector, and where} $Q$, $N$, $q$, $t$, $\rho$, $p$, $\pi$ \textit{and} $\chi$ \textit{are given in} (\ref{fluper-definitions-1}), (\ref{fluper-definitions-2}), (\ref{fluper-hydro}) \textit{and} (\ref{Rainich-Synge}) \textit{and} ${\cal S}(\chi, \pi)$ \textit{is given in} (\ref{chi-prima}).\\[2mm]
{\bf Characterisation theorem 2.} \textit{A metric is an Einstein-Synge solution in isoenergetic evolution if, and only if, the Ricci tensor} $R$ \textit{satisfies the invariant conditions given in} (\ref{fluper-conditions-ideal}) \textit{and}
		\begin{equation} \label{fluper-conditions-isoenergetic}
			Q(\textrm{d} \rho) = 0 , \qquad Q(\textrm{d} p) = 0 ,
		\end{equation}
\textit{where} $v$ \textit{is any timelike vector, and} $Q$, $N$, $q$, $t$, $\rho$ \textit{and} $p$ \textit{are given in} (\ref{fluper-definitions-1}), (\ref{fluper-definitions-2}) \textit{and} (\ref{fluper-hydro}). 

	\section{Approximations to the Synge fluid} \label{sec-approximations}
	The aim of this section is to take advantage of the results of Section \ref{sec-synge} to try to find some analytical expressions for $\chi(\pi)$ approximating that of a Synge gas, namely, the solution to the differential equation (\ref{chi-prima}).
	
		\subsection{Classical ideal gas approximation} \label{subsec-CIG-approx}
		As explained in Section \ref{sec-classic-ideal-gas} classical ideal gases are the ideal gases (EoS (\ref{eq. estat gas ideal})) fulfilling the classical $\gamma$-law equation of state (\ref{eq. estat gas ideal classic}). They are usually considered as a good approximation of an ideal gas at low temperatures, and they can equivalently be characterised as those with an indicatrix function of the form (\ref{chi-gas-ideal-classic}).
		The adiabatic index $\gamma = 5/3$ corresponds to a monoatomic gas. In this case $\chi_c(0) = 0$, $\chi_c'(0) = 5/3$, and thus it approaches a Synge gas at first order. For the sake of completeness, we analyse the constraints for physical reality for any $\gamma$. \\ \\
		In \cite{CFS-CIG}, Coll \textit{et al.} study the macroscopic compressibility conditions in this case and obtain the results summarised in Proposition \ref{prop-classical-ideal-gas-compressibility-conds} of Section \ref{sec-classic-ideal-gas}.
		Note that the interval defined in (\ref{cc-CIG}) contains the domain $[0, 1/3]$ if $4/3 \leq \gamma \leq 4$. 
		Nevertheless, it is known \cite{Taub, Rezzolla} that the relativistic kinetic theory imposes stronger restrictions. Indeed, for a generic ideal gas Taub's inequality Ta given in (\ref{Taub}) can be written as
		\begin{equation} \label{TaubG}
			\hspace{-24mm} {\rm Ta}^{\rm \! G}: \hspace{25mm} \quad \quad \ \qquad \psi \equiv e^2(\pi)(1 - 3\pi) \geq 1 \, . \qquad \qquad \quad
		\end{equation}
For the case of a classical ideal gas, this becomes
		\begin{equation} \label{TaubC}
			\hspace{-24mm} {\rm Ta}^{\rm \! C}: \hspace{25mm} \quad \quad \ \qquad \pi < \hat{\pi}_m \equiv (\gamma - 1)(5 - 3\gamma) \, . \qquad \quad   
		\end{equation}
Note that if this constraint holds for some $\pi > 0$, then necessarily $1 < \gamma < 5/3$. Moreover, the maximum value for $\hat{\pi}_m$ is $1/3$, and it is reached when $\gamma = 4/3$.
		Consequently, the classical monoatomic gas ($\gamma = 5/3$) approximates a Synge gas at first order at low temperatures and it fulfils the macroscopic constraints for physical reality H$^\textrm{G}_1$ and H$^\textrm{G}_2$ in the interval $[0, \, 2/3[$ (see Section \ref{sec-classic-ideal-gas}). However, it does not fulfil Taub's inequality at any point.
		
		\subsection{An approximation at high temperature} \label{subsec-high-approx}
		The ultrarelativistic limit of a monoatomic gas is usually obtained from the Synge EoS by taking the limit of the Bessel function $K_2(z)$ at $z = 0$ (see, for example, \cite{Rezzolla}). The thermodynamic quantities fulfil the relations
		\begin{equation} \label{s-t-e-radiacio}
			\rho = a \Theta^4 , \ \quad p = \frac13 a \Theta^4 , \ \quad S = n s_1 = \frac43 a \Theta^3 ,
		\end{equation}
where $a$ and $s_1$ are constant. 
		On the other hand, the above thermodynamic scheme can also be obtained by considering that the fluid fulfils the barotropic equation of state $\rho = 3p$, and using the usual macroscopic thermodynamic reasoning \cite{Hydro-LTE}. Now, we will see that this ultrarelativistic limit can also be obtained from our hydrodynamic approach, that is, as an approximation from the Synge indicatrix function $\chi(\pi)$. \\ \\
		Let us consider the zero-order approximation at $\pi = 1/3$, that is, $c_s^2 = \chi(\pi) = 1/3$. This indicatrix function corresponds to a specific non-barotropic ideal gas that, as can be seen by applying the expressions (\ref{e(pi) i f(pi)}) and (\ref{psi(pi) i phi(pi)}) of the ideal inverse problem, fulfils the following EoS:
		\begin{subequations} \label{e-s-high}
			\begin{eqnarray}
				e(\pi) = e_0 [(\pi + 1)^3 (1-3\pi)]^{-1/4} , \quad \label{e-pi-high} \\[2mm]
				s(\rho,p) = s_0 - \tilde{k} \ln(\rho - 3p) \, . \label{s-high} \qquad
			\end{eqnarray}
		\end{subequations}
If we impose that the ideal gas defined by equations (\ref{e-s-high}) undergoes an isentropic evolution, $s(\rho,p) = s_1$ = constant, then it fulfils the following barotropic (evolution) relation:
		\begin{equation}
			p = \frac13(\rho - \kappa) \, , \qquad \kappa = {\rm constant} \, .
		\end{equation}
Then, the behaviour of this model when $\kappa \rightarrow 0$ is that of the scheme given in equation (\ref{s-t-e-radiacio}). Thus, this usual ultrarelativistic limit to a Synge gas can be obtained as the limit of a one-parametric family of isentropic evolutions of a non-barotropic ideal gas. \\ \\ 
		A similar reasoning would allow us to interpret the so-called \textit{relativistic $\gamma$-law} models, $p = (\gamma - 1) \rho$, as the limit of a family of isentropic evolutions of the non-barotropic ideal gas defined by the indicatrix function $c_s^2 = \chi(\pi) = \gamma - 1$.
		
		\subsection{Taub-Mathews approximation} \label{subsec-TM-approx}
\begin{figure}[t]
			\centering
			\hspace{-6mm} \includegraphics[width=0.85\textwidth]{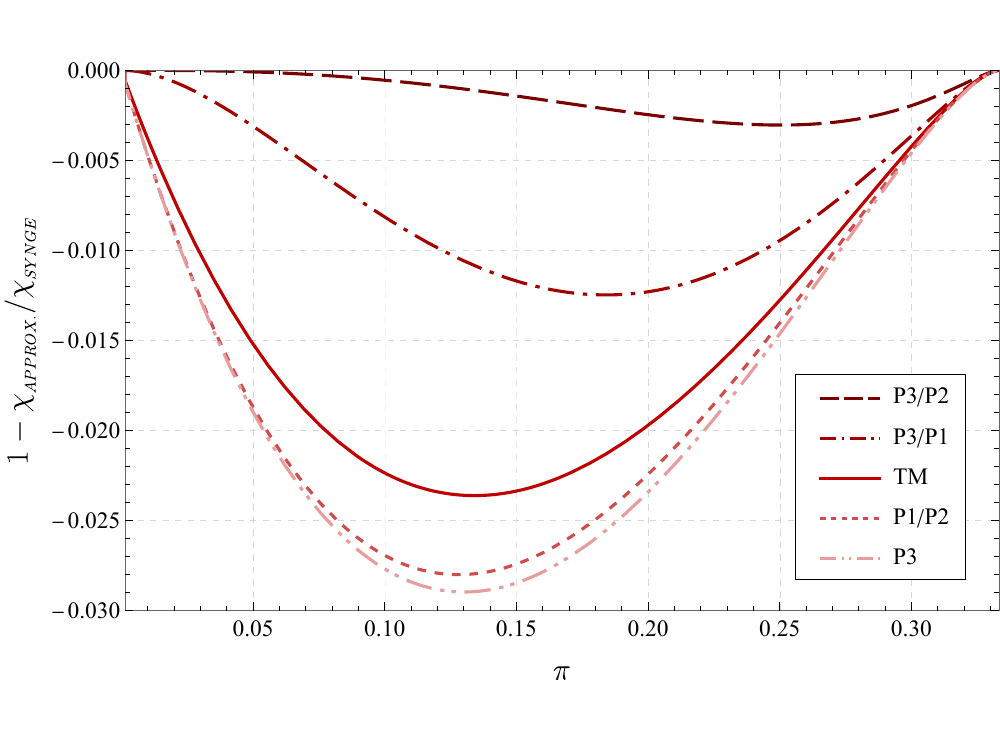}
			\vspace{-8mm}
			\caption{This plot shows the relative error with respect to the Synge EoS of the TM EoS and of the other proposed EoS. The P3/P2 and P3/P1 EoS are more accurate than the TM EoS, while P3 and P1/P2 EoS are not.}
		\label{Fig-8}
		\end{figure}

		From now on, we consider equations of state that are Pad\'e-like approximants at $\pi = 0$ and $\pi = 1/3$, which approximate the indicatrix function $\chi(\pi)$ of a Synge EoS in the entire domain $[0,1/3]$. Let us start by considering a quotient P2/P1, where Pn denotes a polynomial of degree n, for $\chi(\pi)$:
		\begin{equation} \label{ansatz-TM}
			\chi(\pi) = \frac{\pi^2 + c_1\pi + c_2}{c_3\pi + c_4} \, ,
		\end{equation}
$c_i$ being arbitrary constants with, at least, $c_3 \neq 0$. 
		Therefore, we can impose up to four conditions on the indicatrix function (\ref{ansatz-TM}). We can use the results of the previous subsection to make sure that our approximated indicatrix function behaves as that of a Synge gas up to first order, at both low and high temperatures. In other words, we can impose $\chi(0) = 0$, $\chi(1/3) = 1/3$, $\chi'(0) = 5/3$ and $\chi'(1/3) = 1/2$ on (\ref{ansatz-TM}). A straightforward calculation shows that by doing so, we get
		\begin{equation} \label{chi-TM}
			\chi(\pi) = \frac{\pi(5 - 3\pi)}{3(1 + \pi)} \, .
		\end{equation}
		In order to check whether the fluid with the above indicatrix function verifies the compressibility conditions $\rm{H}_1^{\rm{G}}$ and $\rm{H}_2^{\rm{G}}$, we simply need to substitute (\ref{chi-TM}) in (\ref{H1G}) and (\ref{H2G}), respectively. By doing so, we obtain
		\begin{eqnarray}
			\zeta = \frac{8\pi(5 - 10\pi + 9\pi^2)}{9(1 + \pi)^2} > 0 \, , \label{H1G-TM} \\
			\eta = \frac{2\pi (1 + 2\pi - 3\pi^2)}{3 (1 + \pi)} > 0 \, , \; \, \, \label{H2G-TM}
		\end{eqnarray}
if $\pi \in ]0,1/3[$. So both compressibility conditions are fulfilled. \\ \\
		Now, we can determine the ideal thermodynamic scheme by using (\ref{e(pi) i f(pi)}) and (\ref{psi(pi) i phi(pi)}), and we get
		\begin{equation} \label{e-f-TM}
			e(\pi) = \frac{1}{\sqrt{1 - 3\pi}} \, , \qquad f(\pi) = f_0 \frac{\pi^{\frac32}}{(1 - 3\pi)^2} \, ,
		\end{equation}
where we have set $e_0 = 1$ so that $\epsilon(0) = e(0) - 1 = 0$. Then, we obtain the following EoS: 
		\begin{equation} \label{TM-1}
			e^2(\pi)(1 - 3\pi) = 1 \, . \qquad \quad
		\end{equation}
Note that this EoS verifies Taub's inequality given in (\ref{TaubG}). In fact, it fulfils equality. \\ \\
		On the other hand, by inverting $e = e(\pi)$ in (\ref{e-f-TM}) and using the definition of $\pi$ and the first equation in (\ref{n i s ideals}), we get that (\ref{TM-1}) becomes
		\begin{equation} \label{TM-2}
			p = \frac13 n \left(e - \frac1e \right) \, ,
		\end{equation}
which is the equation of state proposed by Mathews \cite{Mathews} (TM EoS). Moreover, using (\ref{eq. estat gas ideal}) and the second equation in (\ref{epsilon i Theta ideals}), it can be rewritten as
		\begin{equation} \label{TM-EoS-ht}
			(h - \tilde{k} \Theta)(h - 4\tilde{k} \Theta) = 1 \, ,
		\end{equation}
which was proposed and analysed by Mignone {\em et al.} \cite{Mignone, Mignone-2007}, who showed it to be a reasonable approximation to the Synge equation of state. \\ \\
		If we compare the indicatrix function of the TM EoS with that of a Synge gas, we conclude that the relative error is less than 2.36$\%$ in the entire domain $[0, 1/3]$ (see Figure \ref{Fig-8}).

		\subsection{Generalised Taub-Mathews approximation} \label{subsec-generalized-TM-approx}
		We can try to generalise the TM EoS to an arbitrary $\gamma$ in order to obtain an analytical model for a relativistic polyatomic gas. We start from an expression of the form (\ref{ansatz-TM}) and we impose $\chi'(0) = \gamma$ instead of $\chi'(0) = \frac53$. In that case, we get the following indicatrix function (TM$\gamma$):
		\begin{equation} \label{chi-TMgamma}
			\chi(\pi) = \frac{\pi[\gamma + 3\pi (\gamma - 2)]}{1 + 3\pi (2\gamma - 3)} \, ,
		\end{equation}
which indeed reduces to (\ref{chi-TM}) for $\gamma = \frac53$. \\ \\ \\
		Now, we can easily check that $\zeta > 0$ and $\eta > 0$, and thus both \mbox{compressibility} conditions (\ref{H1G}) and (\ref{H2G}) are fulfilled for any $\gamma \geqslant 1$.
		Moreover, following the same procedure as before, we have that, in this case
		\begin{equation} \label{e-f-TMgamma}
			e(\pi) = \frac{(1 + \pi)^{\frac{5 - 3\gamma}{2(\gamma - 1)}}}{\sqrt{1 - 3\pi}} \, , \quad f(\pi) = f_0 \frac{\pi^{\frac{1}{\gamma - 1}}}{(1 - 3\pi)^2} \, ,
		\end{equation}
Then, we can obtain
		\begin{equation}
			\psi \equiv e^2(\pi)(1 - 3\pi) = (1 + \pi)^{\frac{5 - 3\gamma}{\gamma - 1}} \, ,
		\end{equation}
which implies that Taub's inequality (\ref{TaubG}) is only verified if $1 < \gamma \leqslant \frac53$. These results and all those that will be obtained in this section are summarised in Table \ref{table-6}. \\ \\ \\ \\ \\
		The first expression in (\ref{e-f-TMgamma}) can be written as
		\begin{equation} \label{TMgamma-EOS-pe}
			(n e + p)^{\frac{3\gamma - 5}{\gamma - 1}}(n e - 3p) = e^{\frac{2(\gamma - 2)}{\gamma-1}} n^{\frac{2(2\gamma - 3)}{\gamma-1}} \, ,
		\end{equation}
or, equivalently,
		\begin{equation} \label{TMgamma-EOS-ht}
			h^{\frac{3\gamma - 5}{\gamma - 1}}(h - \tilde{k} \Theta)^{\frac{2(2 - \gamma)}{\gamma - 1}}(h - 4\tilde{k} \Theta) = 1 \, .
		\end{equation}
		In \cite{Sokolov}, Sokolov \textit{et al.} proposed the simplified equation of state
		\begin{equation}
			h (h - 4\tilde{k} \Theta) = 1 \, .
		\end{equation}
It can easily be seen that this is a particular case of the TM$\gamma$ equation of state with $\gamma = 2$. Therefore, it neither reproduces the correct behaviour of the Synge EoS for low temperatures nor fulfils Taub's condition. Nevertheless, we consider it in Table \ref{table-6} for the sake of completeness.
		\begin{table*}[t]
			\begin{tabular}{cccccccl}
 				\noalign{\hrule height 1.05pt} \\[-2.5mm]
  				\hspace{-2mm} EoS$_{\textrm{\tiny{APPROX.}}}$ & $\chi(\pi)$ & $e(\pi)$ & \hspace{-4mm} $\rm{H}_1^{\rm{G}}, $  $\ \rm{H}_2^{\rm{G}}$ & $\qquad \rm{Ta^G} \qquad$ &
  				\hspace{-10mm} \\[1mm]
 				\hline \\[-2mm]
   				CIG & $\frac{\gamma \pi}{1 + \pi}$ & $\frac{\gamma - 1}{\gamma - 1 - \pi}$  & \hspace{-4mm} $4/3 < \gamma < 4$ & $\gamma = \frac43$ & \hspace{-10mm}
				\\[1.5mm]  
   				TM$\gamma$ & $\frac{\pi[\gamma + 3\pi (\gamma - 2)]}{1 + 3\pi (2\gamma - 3)}$ & $\frac{(1 + \pi)^{\frac{5 - 3\gamma}{2(\gamma - 1)}}}{\sqrt{1 - 3\pi}}$ & \hspace{-4mm} $\gamma > 1$ & $1 < \gamma \leqslant \frac53$ & \hspace{-10mm}
				\\[2mm] 
   				TM & $\frac{\pi(5 - 3\pi)}{3(1 + \pi)}$ & $\frac{1}{\sqrt{1 - 3\pi}}$ & \hspace{-4mm} $\checkmark$ & $\checkmark$ & \hspace{-10mm}
				\\[2mm] 
   				Sokolov & $\frac{2\pi}{1 + 3\pi}$ & $\frac{1}{\sqrt{(1 - 3\pi)(1 + \pi)}}$ & \hspace{-4mm} $\checkmark$ & No & \hspace{-10mm}
				\\[2mm] 
   				P3/P1 & $\frac{\pi(5 + 15\pi - 18\pi^2)}{3(1 + 5\pi)}$ & $\frac{(1 + \pi)^{\frac34}}{(1 + 3\pi)^{\frac14} \sqrt{1 - 3\pi}}$ & \hspace{-4mm} $\checkmark$ & $\checkmark$ & \hspace{-10mm}
				\\[2.0mm] 
   				P3/P2 & $\frac{5\pi(3\pi^2 + 7\pi - 4)}{57\pi^2 - 3\pi - 12}$ & $\frac{1}{(1 + \pi)^{\frac{6}{11}}(1 - \frac74 \pi)^{\frac{24}{77}}(1 - 3\pi)^{\frac{1}{2}}}$ & \hspace{-4mm} $\checkmark$ & $\checkmark$ & \hspace{-20mm}
				\\[2.8mm] 
   				P3 & $\frac16 \pi (10 - 15\pi + 9\pi^2)$ & $\frac{(\pi + 1)^{\frac{3}{14}}(1 - \frac34 \pi)^{\frac27}}{\sqrt{1 - 3\pi}}$ & \hspace{-4mm} $\checkmark$ & No & \hspace{-10mm}
				\\[1.8mm] 
   				P1/P2 & $\frac{10\pi}{3(3\pi^2 + 3\pi + 2)}$ & \ $\frac{(1 + \pi)^{\frac{3}{2}}}{(4 + 3\pi)^2 \sqrt{1 - 3\pi}}$ & \hspace{-4mm} $\checkmark$ & No & \hspace{-10mm}
				\\[-2.5mm] 
				\hspace{-20mm}  
				\\[0mm] \noalign{\hrule height 1.05pt}
			\end{tabular}
\caption{This table summarises the physical behaviour of the equations of state that approximate the Synge EoS. The two first columns show the indicatrix function $\chi(\pi)$ and the specific energy $e(\pi)$, respectively. In the third and forth columns, a $\checkmark$ indicates that the corresponding EoS fulfils the macroscopic compressibility conditions \rm{H}$_1^{\rm{G}}$, \rm{H}$_2^{\rm{G}}$ or the Taub constraint Ta$^G$ in the full interval $]0, 1/3[$. In the two first rows, the EoS depends on the adiabatic index $\gamma$; then, we indicate the values of this parameter for which these conditions hold.}
\label{table-6}
		\end{table*}

		\subsection{Other approximations} \label{subsec-other-approx}
		This subsection is devoted to finding other indicatrix functions approximating that of a relativistic Synge gas and to studying them following the same approach as in the previous subsection. \\ \\
		Firstly, we consider $\chi(\pi)$ to be the ratio of two polynomials of third and first order (P3/P1), and constrain it to match the Synge indicatrix function up to second order at low temperatures and first order at high temperatures. With that we get
		\begin{equation}
			\chi(\pi) = \frac{\pi(5 + 15\pi - 18\pi^2)}{3(1 + 5\pi)} \, . \label{chi-p3/p1}
		\end{equation}
Then, we can check that $\zeta > 0$ and $\eta > 0$, and thus both compressibility conditions (\ref{H1G}) and (\ref{H2G}) are fulfilled.
		Moreover, we can determine the specific energy by using (\ref{e(pi) i f(pi)}) and (\ref{psi(pi) i phi(pi)}). We obtain
		\begin{equation}
			e(\pi) = \frac{(1 + \pi)^{\frac34}}{(1 + 3\pi)^{\frac14} \sqrt{1 - 3\pi}} \, .
		\end{equation}
Then, we can see that the Taub constraint (\ref{TaubG}) also holds. 
		If we compare the indicatrix function (\ref{chi-p3/p1}) of the P3/P1 EoS with that of a Synge gas, we conclude that the relative error is less than 1.25$\%$, and this EoS shows better accuracy than that of TM in the entire domain $[0, 1/3]$ (see Figure \ref{Fig-8}). \\ \\
		Secondly, we take $\chi(\pi)$ to be the ratio of two polynomials of the form P3/P2. Now, we can constrain it to match the Synge indicatrix function up to third order at low temperatures and up to first order at high temperatures. In this case we obtain
		\begin{equation}
			\chi(\pi) = \frac{5\pi(3\pi^2 + 7\pi - 4)}{57\pi^2 - 3\pi - 12} \, , \qquad e(\pi) = \frac{1}{(1 + \pi)^{\frac{6}{11}}(1 - \frac74 \pi)^{\frac{24}{77}}(1 - 3\pi)^{\frac{1}{2}}} \, . \label{chi-p3/p2} \\[1mm]
		\end{equation}
Then, we can see that this P3/P2 EoS fulfils both compressibility conditions (\ref{H1G}) and (\ref{H2G}), and Taub's inequality (\ref{TaubG}).
		If we compare the indicatrix function (\ref{chi-p3/p2}) of the P3/P2 EoS with that of a Synge gas, we conclude that the relative error is less than 0.30$\%$ in the entire domain $[1, 1/3]$ (see Figure \ref{Fig-8}). \\ \\
		Finally, we look for indicatrix functions $\chi(\pi)$ that approximate the Synge EoS up to first order at both low and high temperatures. This is the case of the TM EoS, which corresponds to a ratio P2/P1. Now, we consider two new cases: a third-order polynomial P3, and a ratio of polynomials of the form P1/P2. \\ \\
		We can easily obtain the expressions for the indicatrix function $\chi(\pi)$ and the specific energy $e(\pi)$, which are listed in Table \ref{table-6}. Now, compressibility conditions (\ref{H1G}) and (\ref{H2G}) are fulfilled in both cases. Nevertheless, Taub's inequality (\ref{TaubG}) does not hold. \\ \\
		Moreover, if we compare the indicatrix function of these EoS with that of a Synge gas, we conclude that the relative error is, in the entire domain $[0, 1/3]$, less than 2.90$\%$ for the P3 EoS, and less than 2.80$\%$ for the P1/P2 EoS (see Figure \ref{Fig-8}).
		
	\section{Isentropic evolution of the Synge gas} \label{sec-isentropic-evol}
	The isentropic evolution of a non-barotropic perfect fluid is described by a barotropic energy tensor fulfilling a barotropic relation implicitly defined by $s (\rho, \, p) = constant$. Now, we analyse the isentropic evolution of a generic ideal gas and we particularise it to the Synge EoS and the TM approximation. The generalised Friedmann equation is stated for this case. 
	
		\subsection{Isentropic$\,$ evolution$\,$ of$\,$ a$\,$ generic$\,$ ideal$\,$ gas.$\,$ The \\ Taub-Mathews$\,$ case} \label{subsec-GIG-isentr}
		From the expression $s (\rho, \, p)$ of the specific entropy given in (\ref{n i s ideals}), we find that, in an isentropic evolution of a generic ideal gas, we have a barotropic relation of the form
		\begin{equation} \label{rho-isentr}
			\rho = K f(\pi) \, ,
		\end{equation}
where $K$ is an arbitrary constant. If the function $f(\pi)$ is invertible, then (\ref{rho-isentr}) gives us the following explicit barotropic relation:
		\begin{equation} \label{rel-barotropia-GIG}
			p = \varphi(\rho) \equiv \rho f^{-1}(\rho / K) \, .
		\end{equation}
By differentiating (\ref{rho-isentr}) and using (\ref{e(pi) i f(pi)}), we can also characterise the barotropic relation through a differential equation:
		\begin{equation} \label{rel-barotropia-diferencial-GIG}
			p' \equiv \varphi'(\rho) = \chi(p/\rho) \neq \pi \, .
		\end{equation}
		The isentropic evolution of a classical ideal gas with adiabatic index $\gamma$ leads to the barotropic relation (\ref{CIG-isentropic}).
		In the case of the TM approximation, the function $f(\pi)$ is given in (\ref{e-f-TM}). Thus, by substituting it in (\ref{rho-isentr}), we obtain an implicit barotropic relation:
		\begin{equation} \label{TM-barotrop}
			K^2 p^3 = \rho (\rho - 3p)^4 \, .
		\end{equation}
		For the case of the exact Synge equation of state, we do not explicitly know $f(\pi)$ nor $\chi(\pi)$. However, if we derive (\ref{rel-barotropia-diferencial-GIG}) with respect to $\rho$ and use (\ref{chi-prima}), we can obtain a second order differential equation characterising the barotropic relation $p = \varphi(\rho)$ leading to the isentropic evolution of a relativistic Synge gas: 
		\begin{equation}
			p'' \rho \pi^3 (\pi + 1) + \alpha \, p'^3 + \beta \, p'^2 + \gamma \, p' + \delta = 0 \, ,
		\end{equation}
with $\alpha$, $\beta$, $\gamma$ and $\delta$ given in (\ref{chi-prima}). \\
		
		\subsection{Generic ideal gas FLRW models. The Taub-Mathews case} \label{subsec-FLRW-GIG}
		The Friedman-Lema\^itre-Robertson-Walker universes are perfect fluid spacetimes with line element:
		\begin{equation} \label{FLRW-metric}
			\textrm{d} s^2 = -\textrm{d} \tau^2 + \frac{R^2(\tau)}{[1 + \frac14 k r^2]^2} (\textrm{d} r^2 + \textrm{d} \Omega^2) \, ,
		\end{equation}
where $\textrm{d} \Omega^2$ is a two-dimensional metric of constant curvature and $k = 0, \pm 1$. The homogeneous energy density and pressure are given, respectively, by:
		\begin{equation} \label{FLRW-rho}
			\rho = \frac{3 \dot{R}^2}{R^2} + \frac{3k}{R^2} + \Lambda \equiv \rho(R) \, ,
		\end{equation}
		\begin{equation} \label{FLRW-p}
			p = - \rho - \frac{R}{3} \partial_R \rho \equiv p(R) \, .
		\end{equation}
		What are the {\em generalised Friedmann equations} when the energy content is an ideal gas? The homogeneity of the hydrodynamic quantities $\rho$ and $p$ implies that the evolution must be isentropic. Then, as argued in the previous subsection, a barotropic relation of the form (\ref{rel-barotropia-GIG}) holds. Equation (\ref{FLRW-p}) then enables us to determine $\rho(R)$, and (\ref{FLRW-rho}) becomes a {\em Friedmann equation} for $R(\tau)$. \\ \\
		In \cite{CFS-CIG}, Coll \textit{et al.} derived this generalised Friedmann equation for the case of a classical ideal gas. We now proceed to analyse the case of a TM ideal gas. From the TM EoS (\ref{TM-2}) and the barotropic relation (\ref{TM-barotrop}), we obtain the evolution constraint
		\begin{equation} \label{n-rho}
			\rho = \rho(n) \equiv n^2(1 + \mu^2 n^{2/3}) \, ,
		\end{equation}
where $\mu \equiv 3\sqrt{3}/K$. 
		On the other hand, it is known that, for the FLRW metrics, the rest-mass density is 
		\begin{equation} \label{n-FLRW}
			n = n_0 \left(\frac{R_0}{R}\right)^3 . 
		\end{equation}
Consequently, from (\ref{n-rho}) and (\ref{n-FLRW}) we obtain the following expression for the energy density:
		\begin{equation} \label{rho-R-FLRW-TM}
			\rho (R) = \left[ n_0^2 \left( \frac{R_0}{R} \right)^6 + \rho_r^2 \left( \frac{R_0}{R} \right)^8 \right]^{1/2} ,
		\end{equation}
where $\rho_r \equiv 3\sqrt{3}\,n_0^{4/3} \! /K$. This expression and (\ref{FLRW-rho}) determine the generalised Friedmann equation for the TM-gas FLRW models. \\ \\
		It is worth noting that this generalised Friedmann equation was also obtained by de Berredo-Peixoto {\em et al.} \cite{deBerredo} following a different line of reasoning. In fact, they also use an equation of state equivalent to the TM EoS, which they compare with the exact Synge EoS. Here, we have used our hydrodynamic approach to show that the solutions to the Friedmann equation defined by (\ref{FLRW-rho}) and (\ref{rho-R-FLRW-TM}) model a TM gas evolving at constant entropy $s = k \ln[\rho_r/(3 \sqrt{3} n_0^{4/3})]$. \\ \\
	With this, we finish presenting all the theoretical tools we will need to achieve our goal of studying the potential physical plausibility of perfect fluid solutions to Einstein's equations admitting a three-dimensional group G$_3$ of isometries acting on spacelike two-dimensional orbits S$_2$ in general, or spherical symmetry in particular. Thus, in the next part of the thesis, we apply these tools to some of such families of solutions.

\addtocontents{toc}{\newpage}
\part{Physical interpretation of perfect fluid solutions with \mbox{spherical} symmetry}
		
\chapter{The T-models} \label{chap-T-models}
In this chapter, we start the study of the physical interpretation of perfect fluid solutions with a $G_3/$S$_2$ by thoroughly analysing the T-models. In comoving-synchronous coordinates, the metric of a perfect fluid solution admitting a three-dimensional group G$_3$ of isometries acting on spacelike two-dimensional orbits S$_2$ has the form \cite{Kramer}:
\begin{eqnarray} \label{metric-ss-1}
	\textrm{d}s^2 = -e^{2\nu }\textrm{d}t^2 + e^{2\lambda} \textrm{d}r^2 + Y^2 C^2 (\textrm{d}x^2 + \textrm{d}y^2), \\[3mm]
	\nu = \nu(t,r), \qquad \lambda = \lambda(t,r), \qquad Y = Y(t,r), \label{metric-ss-2} \\[1mm]
	C = C(x, y) \equiv \left[1 + \frac{k}{4}(x^2 + y^2) \right]^{-1} , \quad k = 0, \pm1 , \label{metric-ss-3}
\end{eqnarray}
where the value of $k$ distinguishes the plane, spherical and hyperbolic symmetries. \\ \\
The $r$-dependence of the functions $\nu$ and $Y$ plays an important role in the analysis of the Einstein equations for a perfect energy tensor source. Thus, usually one considers separately the cases $\nu = \nu(t)$ (geodesic motion) or/and $Y = Y(t)$ (T-models) \cite{Krasinski-Plebanski, Kramer}. On the other hand, Ruban \cite{Ruban_1969} showed that the spherically perfect fluid T-models have geodesic motion (see also \cite{Inhomogeneous_Cosmological_Models}), a result that can be extended to the plane and hyperbolic symmetries \cite{Krasinski-Plebanski}. Hereinafter, \textit{T-models} refer to the perfect fluid solutions whose metric has the form (\ref{metric-ss-1}$-$\ref{metric-ss-3}) with $\nu = \nu(t)$ and $Y = Y(t)$. \\ \\
The spherical dust T-model was published in a pioneer paper by Datt \cite{Datt_1938} and rediscovered later by Ruban \cite{Ruban_1969,Ruban_1968}, and the general perfect fluid solution with a non-constant pressure was considered by Korkina and Martinenko \cite{Korkina-Martinenko_1975}. An exhaustive list of the particular solutions presented by several authors can be found in \cite{Inhomogeneous_Cosmological_Models}, but the physical meaning of any of these solutions is doubtful. \\ \\ \\
The geometric and physical properties of the spherically symmetric dust T-models were analysed by Ruban \cite{Ruban_1968} (see also \cite{Krasinski-Plebanski}). The metric is invariant under the group of rotations but, since $Y$ depends only of $t$, the space-like 3-spaces $t = constant$ do not contain their center of symmetry. The geometry of these 3-spaces is that of a three-dimensional cylinder, that is, the direct product of a 2-sphere and an open straight line. \\ \\ 
A similar situation occurs in the flat or the hyperbolic symmetries by changing the 2-sphere by a plane or a hyperboloid. Moreover, extensive work (see \cite{Inhomogeneous_Cosmological_Models, Krasinski-Plebanski, Bonnor-ST, G-hellaby} and references therein) has been devoted to extending this study to the Szekeres solutions of class II \cite{Szekeres}, which are a generalisation with no symmetries. The solutions with non-constant pressure basically keep the geometric properties of the dust solutions \cite{Ruban_1983}, but the physical meaning of these T-models is still an open problem. \\ \\
Section \ref{Termo_T-models} is based on the results published in \cite{FM-Termo-T-models}. In it, we study the potential thermodynamic interpretations of the T-models. First, we analyse the general properties of these solutions by obtaining the hydrodynamic quantities and the hydrodynamic equation of state $c_s^2 = \chi(\rho, p)$. All the possible thermodynamic interpretations of each solution of the field equations are also presented by obtaining all the compatible thermodynamic schemes. Then, we particularise our study to the case of the T-models that are compatible with the EoS of a generic ideal gas, $p = \tilde{k} n \Theta$, to which we apply our procedure to analyse the physical reality of the solutions. \\ \\
Section \ref{T-model-general-sol} is based on \cite{FM-General-sol-T-models}, and is devoted to analysing the field equations for the T-models. First, we revisit the perfect fluid field equations for the T-models and we present some new solutions. Then, we analyse the Herlt algorithm and we show that implementing the algorithm to calculate the solution requires the realisation of two indefinite integrals. We also propose another integration algorithm to obtain the solution by quadratures. Finally, we propose new algorithms determining the general solution without quadratures and we use them to recover previously known solutions and obtaining new ones. \\ \\
Finally, in Section \ref{chap-physical interpretation KS} we study the spatially homogeneous limit of the T-models, a work that is yet unpublished. It is known \cite{Inhomogeneous_Cosmological_Models} that in this limit of the T-models, $\lambda = \lambda(t)$, they reduce to the Kompanneets-Chernov-Kantowski-Sachs (KCKS) metrics \cite{Kompaneets-Chernov, Kantowski-Sachs}. First, we study in detail the interpretation of the KCKS metrics as an isentropic evolution of a thermodynamic T-model. Then we show how to study other potential interpretations.
		
	\section{Thermodynamic approach to the T-models} \label{Termo_T-models}
	In this section, we study the potential thermodynamic interpretations of the T-models. It is worth remarking that these metrics define a subfamily of the class II Szekeres-Szafron solutions \cite{Inhomogeneous_Cosmological_Models, Szekeres, Szafron, Krasinski-Plebanski, FS-SS}, the only one remaining to be studied from a thermodynamic perspective when this work was carried out. The other thermodynamic Szekeres-Szafron solutions of class II, the singular and the regular models, where studied in \cite{C-F-S_SzSz_Singular, C-F-S_SzSz_Regular}. 

		\subsection{T-models: field equations for the metric functions} \label{sec-Tmodels}
		If we make $e^{\lambda} = \omega(t,r) > 0$, $e^{-2 \nu} = v(t) > 0$ and $Y^2 = \varphi(t) > 0$ in the metric line element (\ref{metric-ss-1}), it follows that the metric tensor of a T-model can be written as
		\begin{equation} \label{metric-T-1}
			\textrm{d}s^2 = - \frac{1}{v(t)}\textrm{d}t^2 + \omega^2(t,r) \textrm{d}r^2 + \varphi(t) C^2 (\textrm{d}x^2 + \textrm{d}y^2),
		\end{equation}
where $C$ is given in (\ref{metric-ss-3}). Moreover, from the general expressions for the field equations for the metric (\ref{metric-ss-1}) (see, for example, \cite{Krasinski-Plebanski, Kramer}), it follows that (\ref{metric-T-1}) is a perfect fluid solution if, and only if, the metric functions $v(t)$, $\omega(t,r)$ and $\varphi(t)$ meet the differential equation ($G^r{}_r = G^x{}_x$)
		\begin{equation} \label{eq-T-1}
			2 v \varphi \, \ddot{\omega} + (\dot{v} \varphi + v \dot{\varphi})\, \dot{\omega} - (v \ddot{\varphi} + \frac12 \dot{v} \dot{\varphi} + 2k) \, \omega = 0 \, ,
		\end{equation}
where a dot denotes derivative with respect to the time coordinate $t$.
		The unit velocity of the fluid $u = \sqrt{v} \, \partial_t$ is geodesic and its expansion is
		\begin{equation} \label{expansion-T-1}
			\theta = \sqrt{v} \left(\frac{\dot{\varphi}}{\varphi} + \frac{ \dot{\omega}}{\omega}\right) = \sqrt{v} \, \partial_t [\ln(\varphi \omega)] \, .
		\end{equation}
The \textit{pressure} $p$ and the \textit{energy density} $\rho$ are then given by
		\begin{eqnarray} \label{pressure-T-1}
			p = v \left[\frac14 \frac{\dot{\varphi}^2}{\varphi^2} - \frac{\ddot{\varphi}}{\varphi} - \frac12 \frac{\dot{\varphi}}{\varphi} \frac{\dot{v}}{v}\right] - \frac{k}{\varphi} \, , \\[1mm]
			\rho = v \left[\frac14 \frac{\dot{\varphi}^2}{\varphi^2} + \frac{\dot{\varphi}}{\varphi} \frac{\dot{\omega}}{\omega}\right] + \frac{k}{\varphi} \, .
\label{density-T-1}
		\end{eqnarray}
%
		Note that (\ref{eq-T-1}) is a second order linear differential equation for the function $\omega(t,r)$ when $v(t)$ and $\varphi(t)$ are given. Consequently, its general solution is of the form $\omega(t,r) = \omega_1(t) Q_1(r) + \omega_2(t) Q_2(r)$. Moreover, we can change the coordinate $r$ so that
		\begin{equation} \label{w-w1-w2}
			\omega(t,r) = \omega_1(t) + \omega_2(t) \, Q(r) \, ,
		\end{equation}
where $Q(r)$ is an arbitrary real function, and with $\omega_i(t)$ being two particular solutions to equation (\ref{eq-T-1}). \\ \\
		We have the freedom to choose the coordinate $t$ without changing the spacetime metric. Therefore, we can impose a condition on the time-dependent functions $v$, $\varphi$ and $\omega_i$ that fixes this election. For example, if we take $v(t) = 1$, then the coordinate $t$ is the proper time of the comoving observer, $t = \tau$. \\ \\
		It is worth remarking that our choice of the metric function $\varphi = Y^2$ as an unknown of the field equations leads us to equation (\ref{eq-T-1}), which is also a linear equation for $\varphi$. Then, this equation is linear for the three involved metric functions. As we will see in Section \ref{T-model-general-sol}, this fact will certainly be of interest to look for the general solution of the T-model field equations. \\ \\
		The solutions known so far have been obtained by prescribing the free metric functions in a way that allows analytical integration of field equations \cite{Krasinski-Plebanski, Kramer, Herlt}, but without any evident physical meaning. Therefore, it seems of interest to study the thermodynamic interpretation of the known solutions, as well as to obtain new solutions that meet some previously prescribed physical properties.
	
		\subsection{Thermodynamics of the T-models} \label{sec-thermo-Tmodels}
		Now, we analyse when the T-models (\ref{metric-T-1}$-$\ref{eq-T-1}) represent the evolution in local thermal equilibrium of a fluid that meets the suitable macroscopic physical constraints stated in Sections \ref{sec-lte} and \ref{sec-other-phys-re-conds}. Note that the existing symmetries imply that all the scalar invariants, and in particular the energy density $\rho$, the pressure $p$ and the indicatrix function $\chi$, depend on two functions at most. Then, out of these three quantities, only two are independent and the sonic condition S given in (\ref{cond. sonica}) identically holds. Consequently, step 1 in the procedure presented in Section \ref{sec-other-phys-re-conds} is achieved for the full set of T-models. Thus, we proceed to analyse step 2.
		
			\subsubsection{Metric and hydrodynamic quantities: unit velocity, energy density \\ and pressure}
			In the previous section, we have already given the metric and the hydrodynamic quantities of the T-models. Now, for the sake of simplicity and in order to facilitate the calculation in studying the thermodynamic properties, we choose $v = 1$. This means that the time coordinate is the proper time $\tau$ of the Lagrangian observer associated with the fluid. Then, it follows that the metric tensor of the perfect fluid T-models can be written as
			\begin{equation} \label{metric-T-2}
				ds^2 = -d\tau^2 + [\omega_1(\tau) + \omega_2(\tau) \, Q(r)]^2 dr^2 + \varphi(\tau) C^2 (dx^2 + dy^2),
			\end{equation}
where $C$ is given in (\ref{metric-ss-3}), and $\omega_i(\tau)$ are two particular solutions of the second order differential equation
			\begin{equation} \label{eq-T-2}
				2 \varphi \, \ddot{\omega} + \dot{\varphi}\, \dot{\omega} - (\ddot{\varphi} + 2k) \,\omega = 0 \, .
			\end{equation}
			The unit velocity of the fluid $u = \partial_\tau$ is geodesic and its expansion is given by
			\begin{equation} \label{expansion-T-2}
				\theta = \frac{\dot{\varphi}}{\varphi} + \frac{ \dot{\omega}}{\omega} = \partial_\tau (\ln[\varphi (\omega_1 + \omega_2 Q)]) \, . 
			\end{equation}
The pressure $p$ and the energy density $\rho$ are then given by
			\begin{eqnarray} \label{pressure-T-2}
				p = \frac14 \frac{\dot{\varphi}^2}{\varphi^2} - \frac{\ddot{\varphi}}{\varphi} - \frac{k}{\varphi} \, , \\[1mm]
				\rho = \frac14 \frac{\dot{\varphi}^2}{\varphi^2} + \frac{\dot{\varphi}}{\varphi} \, \frac{\dot{\omega}_1 + \dot{\omega}_2 Q}{\omega_1 + \omega_2 Q} + \frac{k}{\varphi} \, . \label{density-T-2}
			\end{eqnarray}
			Note that, with our choice $v(t) = 1$, the space of solutions of the T-models depends on the real functions $\{\varphi(\tau), Q(r)\}$. Moreover, the barotropic limit is achieved when $Q(r) = constant$. This leads to the spatially homogeneous limit of the T-models, the KCKS metrics \cite{Kompaneets-Chernov, Kantowski-Sachs}, which admit a group G$_4$ of isometries acting on orbits S$_3$. These barotropic models may represent an isentropic evolution of a thermodynamic fluid \cite{Hydro-LTE} (see Chapter \ref{chap-physical interpretation KS}). On the other hand, solutions in l.t.e. with constant pressure lead, necessarily, to an isobaroenergetic evolution, $\dot{\rho} = \dot{p} = 0$ \cite{Hydro-LTE}; then, the fluid expansion vanishes as a consequence of (\ref{energy-momentum-conservation}), and (\ref{expansion-T-2}) implies that $\omega$ factorises and the metric is a degenerate KCKS model. Moreover, if $\dot{\varphi} = 0$ then (\ref{pressure-T-2}) and (\ref{density-T-2}) imply that $\rho + p = 0$ and the energy conditions (\ref{E}) do not hold. Hereinafter, we will consider the T-models (\ref{metric-T-2}$-$\ref{eq-T-2}) with $Q'(r) \neq 0$, $\dot{p}(\tau) \neq 0$ and $\dot{\varphi} \neq 0$. 
			
			\subsubsection{The indicatrix function: speed of sound}
			To simplify calculations, we define the following functions:
			\begin{equation} \label{abcd}
				\sigma(\tau) \equiv \frac{\ddot{\varphi}}{\varphi} \, , \quad \beta(\tau) \equiv \frac{\dot{\varphi}^2}{\varphi^2} \, , \quad \xi(\tau) \equiv \frac{k}{\varphi} \, , \quad \Omega(\tau,r) \equiv \frac{\dot{\omega}_1 + \dot{\omega}_2 Q} { \omega_1 + \omega_2 Q}\ \, .
			\end{equation}
Then, the pressure, the energy density and the expansion take the form
			\begin{equation} \label{p-density-expansion}
				p = \frac14 \beta - \sigma - \xi , \quad \rho = \frac14 \beta + \xi \pm \sqrt{\beta} \, \Omega , \quad \theta = \pm \sqrt{\beta} + \Omega ,
			\end{equation}
			where the sign $\pm$ corresponds to the sign of $\dot{\varphi}/\varphi$. Now, we can compute the square of the speed of sound in terms of the hydrodynamic quantities $\rho$ and $p$ using the definition of the indicatrix function given in (\ref{indicatrix def}). Note that now, with the choice $v = 1$, we have $u(q) = \dot{q}$ for any scalar quantity $q$. From equation (\ref{energy-momentum-conservation}) and the expression of the expansion (\ref{p-density-expansion}), we obtain
			\begin{equation} \label{rho-punt}
				u(\rho) = \dot{\rho} = \mp\frac{1}{\sqrt{\beta}} [\rho^2 + (p + q) \rho + pq] , \quad q \equiv \frac34 \beta - \xi .
			\end{equation}
Consequently, we have that for the T-models (\ref{metric-T-2}$-$\ref{eq-T-2}), the square of the speed of sound takes the expression
			\begin{equation} \label{chi-Tmodels}
				c_s^2 = \frac{u(p)}{u(\rho)} = \chi(\rho,p) \equiv \frac{1}{{\cal A}(p) \rho^2 + {\cal B}(p) \rho + {\cal C}(p)} \, ,
			\end{equation}
where ${\cal A}$, ${\cal B}$ and ${\cal C}$ are the functions of $\tau$ (and then of $p$) given by
			\begin{equation} \label{ABCcal} 
				{\cal A}(p) \equiv \mp \frac{1}{\sqrt{\beta} \dot{p}} , \quad {\cal B}(p) \equiv {\cal A} \, (p + q) , \quad {\cal C}(p) \equiv {\cal A} \, p \,q .
			\end{equation}
			It is worth noting that the expression (\ref{chi-Tmodels}) for the square of the speed of sound is similar to those obtained for both the singular and regular models of the thermodynamic Szekeres-Szafron solutions of class II \cite{C-F-S_SzSz_Singular, C-F-S_SzSz_Regular}. This similarity is to be expected, since the T-models constitute the remaining subfamily of this class that had not yet been studied from a thermodynamic point of view. \\ \\
			The equation of state $c_s^2 = \chi(\rho,p)$ given in (\ref{chi-Tmodels}) collects all the thermodynamic information that can be expressed using exclusively hydrodynamic quantities. Note that the dependence on the variable $\rho$ is explicit, but the dependence on $p$ is implicit through the functions ${\cal A}$, ${\cal B}$ and ${\cal C}$ given in (\ref{ABCcal}). These functions only depend on $\varphi(\tau)$ and its derivatives. Thus, the explicit form of $\chi(\rho,p)$ may be obtained when a specific $\varphi(\tau)$ is given (see the following sections). \\ \\
			Once steps 1 and 2 of the procedure proposed in Section \ref{sec-other-phys-re-conds} have been achieved, we could formally impose the restrictions required in step 3 (energy and compressibility conditions H$_1$). Nevertheless, we delay this study for subclasses of solutions that fulfil complementary physical requirements, and once we have obtained the explicit form of $\chi(\rho,p)$. Now we analyse step 4 for the whole set of T-models. 

			\subsubsection{Thermodynamic scheme: Entropy, matter density and temperature} 
			In this subsection we solve the inverse problem for the T-models (\ref{metric-T-2}$-$\ref{eq-T-2}) by obtaining the full set of thermodynamic quantities: specific entropy $s$, matter density $n$ and temperature $\Theta$. The metric function $Q(r)$ plays an important role in this thermodynamic scheme. From the expressions given in (\ref{abcd}) and (\ref{p-density-expansion}) we obtain
			\begin{equation} \label{Q(prho)}
				Q = - \frac{(\rho-p) \varphi \, \omega_1 - {(\dot{\varphi} \omega_1)}^{\cdot}-2k \omega_1}{(\rho - p) \varphi \, \omega_2 - {(\dot{\varphi} \omega_2)}^{\cdot} - 2k \omega_2} \equiv Q(\rho,p) \, .
			\end{equation}
			Note that $Q = Q(\rho,p)$ is a function of state whose dependence on $\rho$ is explicit, while its dependence on $p$ is partially implicit through the functions of time $\omega_i(\tau)$ and $\varphi(\tau)$. We have that $\dot{Q} = 0$ and, consequently, $Q$ is a particular solution of $u(s) = 0$. \\
			On the other hand, from the expression (\ref{expansion-T-2}) of the expansion, it follows that $\bar{n} = [\varphi(\omega_1 + \omega_2 Q)]^{-1}$ is a particular solution of the matter conservation equation (\ref{matter-conservation}). Then, taking into account the thermodynamic study presented in Section \ref{sec-lte}, we obtain that the thermodynamic schemes associated with the T-models (\ref{metric-T-2}$-$\ref{eq-T-2}) are determined by a specific entropy $s$ and a matter density $n$ of the form
			\begin{equation} \label{s-n-Tmodels}
				s(\rho, p) = s(Q); \qquad n(\rho,p) = \frac{1}{\varphi(\omega_1 + \omega_2 Q)N(Q)},
			\end{equation}
where $s(Q)$ and $N(Q)$ are two arbitrary real functions. \\ \\
			The temperature of the thermodynamic scheme defined by each pair $\{s, n\}$ given in (\ref{s-n-Tmodels}) can be obtained from the thermodynamic relation (\ref{thermo-first-law}), which in terms of the specific enthalpy $h = (\rho + p)/n$ becomes
			\begin{equation} \label{thermo-first-law-h}
				\Theta \, \textrm{d}s = \textrm{d}h -\frac{\textrm{d}p}{n} \, .
			\end{equation}
			Expressions (\ref{p-density-expansion}) and (\ref{s-n-Tmodels}) imply that the specific enthalpy is
			\begin{eqnarray} \label{h-Tmodels}
				h = \frac{\rho+p}{n} = N(Q)[\lambda_1(\tau) + Q \lambda_2(\tau)] , \qquad \quad \\ 
				\lambda_i(\tau) \equiv \dot{\varphi} \, \dot{\omega}_i + \left[\frac{\dot{\varphi}^2}{2 \varphi} - \ddot{\varphi}\right] \omega_i = 2 Y(\dot{Y} \dot{\omega}_i - \ddot{Y} \omega_i) , \ \ \label{lambda_i}
			\end{eqnarray}
where $Y = \sqrt{\varphi}$. Then, from (\ref{thermo-first-law-h}) we have $\Theta = \left(\frac{\partial h}{\partial s}\right)_p = \frac{1}{s'(Q)} \left(\frac{\partial h}{\partial Q}\right)_t$ and, taking into account (\ref{h-Tmodels}), we obtain that for the T-models (\ref{metric-T-2}$-$\ref{eq-T-2}), the temperature $\Theta$ of the thermodynamic schemes given in (\ref{s-n-Tmodels}) takes the expression
			\begin{equation} \label{T-Tmodels}
				\Theta = \ell(Q) \lambda_1(\tau) + m(Q) \lambda_2(\tau) \equiv \Theta(\rho,p) \, , 
			\end{equation}
where $\lambda_i(\tau)$ is given in (\ref{lambda_i}) and 
			\begin{equation} \label{ell-m}
 				\ell(Q) \equiv \frac{N'(Q)}{s'(Q)} , \quad m(Q) \equiv \frac{1}{s'(Q)}[Q N'(Q) + N(Q)].
			\end{equation}
\\ \\ \\
			The final step of the procedure described in Section \ref{sec-other-phys-re-conds} consists in analysing the compatibility of the thermodynamic schemes considered above with the positivity conditions P and the compressibility condition H$_2$. This study will be efficient when we consider a specific solution and we may obtain all the thermodynamic quantities in terms of the hydrodynamic ones $\rho$ and $p$ (see the following sections).

		\subsection{T-models compatible with the equation of state of a generic$\,$ ideal$\,$ gas} \label{sec-chi-pi}
		In this section, we will analyse when the T-models and the associated thermodynamic schemes considered above are compatible with (\ref{eq. estat gas ideal}), the EoS of a generic ideal gas. In other words, here we focus on those T-models with the hydrodynamic properties of a generic ideal gas. This restriction will allow us to obtain an explicit expression of $\chi(\rho,p)$ and, consequently, facilitate the analysis of the physical reality conditions in the next subsection. Therefore, here we will follow the first steps of the procedure presented in Section \ref{sec-generic-ideal-gas}. 
		
			\subsubsection{Study of the ideal sonic condition $\textrm{S}^\textrm{G}$} 
			We must study the modified step 1 by analysing which T-models meet the ideal sonic condition S$^{\rm G}$ given in (\ref{cond. sonica ideal}). From the expression of the indicatrix function (\ref{chi-Tmodels}), it follows that (\ref{cond. sonica ideal}) is equivalent to
			\begin{equation} \label{ABCci}
				{\cal A} p^2 = c_1 \, , \qquad {\cal B} p = c_2 \, , \qquad {\cal C} = c_3 \, , \qquad c_i = constant \, .
			\end{equation}
Then, expressions (\ref{ABCcal}) lead to
			\begin{equation} \label{ideal-eq}
				c_1 = \mp \frac{p^2}{\sqrt{\beta} \dot{p}} \neq 0 \, , \qquad c_2 = \frac{c_1}{p}(p + q) \, , \qquad c_3 = \frac{c_1}{p}q \, , 
			\end{equation}
or, equivalently
			\begin{equation} \label{ideal-eq-b}
				c_2 = c_1 + c_3 \, , \qquad c_1 \, q = c_3 \, p \, , \qquad \dot{p} = \mp \frac{p^2}{c_1 \sqrt{\beta}}. 
			\end{equation}
			If $c_3 = 0$, (\ref{ideal-eq-b}) implies $q = 0$ and then $\beta = \frac34 \xi$. Since we are considering $\dot{\varphi} \neq 0$, only $k \neq 0$ is admissible in this case and function $\varphi(\tau)$ fulfils an equation of the form $\dot{\varphi}^2 = c_0^2 \varphi$. Otherwise, if $c_3 \neq 0$, from the first equation in (\ref{p-density-expansion}) and the second equation in (\ref{ideal-eq-b}), and taking into account the definitions of $\sigma$, $\beta$, $\xi$ given in (\ref{abcd}) and of $q$ given in (\ref{rho-punt}), we obtain
			\begin{equation} \label{betapdot}
				\dot{\beta} = \pm 2 \sqrt{\beta}\Big[\Big(\frac{c_3}{c_1} - 1\Big)p - \frac32 \beta \Big] , \qquad \dot{p} = \pm \frac12 \sqrt{\beta} \Big[\Big(1 - 3 \frac{c_1}{c_3}\Big)p - 3 \frac{c_1}{c_3} \beta \Big].
			\end{equation}
Then, the two expressions (\ref{ideal-eq-b}) and (\ref{betapdot}) for $\dot{p}$ lead to
			\begin{equation} \label{beta-p-2}
				2 p^2 + c_1 \Big(1-3 \frac{c_1}{c_3}\Big) p \beta - 3 \frac{c_1^2}{c_3} \beta^2 = 0 \, .
			\end{equation}
From the second expression in (\ref{ideal-eq-b}) and the definition (\ref{rho-punt}) of $q$ we obtain $p = \frac{c_1}{c_3}(\frac34 \beta -\xi)$. Then, we can substitute $p$ and equation (\ref{beta-p-2}) becomes
			\begin{equation} \label{beta-xi}
				2 \xi^2 + (3 c_1 - c_3 - 3) \beta \xi + \frac98 (1 - 2c_1 - 2c_3) \beta^2 = 0 \, .
			\end{equation}
This equation is a necessary constraint for the compatibility of the ideal sonic condition. Now we consider two cases. If $k = 0$, then $\xi = 0$, and (\ref{beta-xi}) states $2(c_1 + c_3) = 1$. Otherwise, if $k \neq 0$, for a given $\beta$, (\ref{beta-xi}) is a second degree algebraic equation for $\xi$ that must admit solution. Then, this solution is of the form $\xi = b_0 \beta$, $b_0 = constant \neq 0$. Consequently, the ideal sonic condition admits a solution if one of the two following conditions holds:
			\begin{itemize}
				\item[(i)]
					$k = 0$, and $c_2 = c_1 + c_3 = \frac12$.
				\item[(ii)]
					$k \not= 0$, and $\dot{\varphi}^2 = c_0^2 \, \varphi$, $c_0 = constant$. 
			\end{itemize}
Case (ii) leads to negative pressures and is not compatible with the generic ideal gas EoS (\ref{eq. estat gas ideal}). It will be analysed in Section \ref{sec-McVittie}. Now, we focus on case (i), the T-models with $k = 0$ which are compatible with the EoS of a generic ideal gas. From now on, we refer to them as \textit{ideal T-models}, and we carry out a detailed study of the five steps required in analysing the physical reality of these solutions. \\ \\ \\

			\subsubsection{Metric line element of the ideal T-models}
			Firstly, we achieve the first step of our procedure by completing the integration of the ideal sonic condition S$^{\rm G}$. As a consequence of the the constraints (i) for $c_i$, $c_2 = c_1 + c_3 = \frac12$, we can consider a constant $\tilde{\gamma}$ such that 
			\begin{equation} \label{ki-gamma}
				c_1 = \frac{\tilde{\gamma} - 1}{2 \tilde{\gamma}} \, , \qquad c_2 = \frac{1}{2} \, , \qquad c_3 = \frac{1}{2 \tilde{\gamma}} \, .
			\end{equation}
Then, taking into account definitions (\ref{abcd}) and that $k = 0$, equations (\ref{ideal-eq-b}) state:
			\begin{equation} \label{p-phi-ppunt}
				p = \frac34 (\tilde{\gamma} -1) \frac{\dot{\varphi}^2}{\varphi^2} \, , \qquad \dot{p} \dot{\varphi} = -\frac{2 \tilde{\gamma}}{\tilde{\gamma} - 1} \varphi p^2 \, .
			\end{equation}
This first order differential system for the functions $p(t)$ and $\varphi(t)$ can be easily integrated and we get
			\begin{equation} \label{phi-p}
				\varphi = \left[\frac32 \kappa \tilde{\gamma} (\tau - \tau_0)\right]^{\frac{4}{3 \tilde{\gamma}}}, \qquad p = 3 \kappa^2(\tilde{\gamma} - 1) \varphi^{-\frac32 \tilde{\gamma}} ,
			\end{equation}
where $\kappa$ is an arbitrary non-vanishing constant. \\ \\ 
			Note that the constant $\tau_0$ determines an origin of time and can be taken as zero. Likewise, rescaling the metric function $\varphi$ by a positive constant factor leaves the metric unchanged, as this factor can be absorbed by appropriately rescaling the coordinates $x$ and $y$ by its square root. Nevertheless, the sign of the constant $\kappa$ determines the sign of the derivative of $\varphi$, $\dot{\varphi} = 2 \kappa \varphi^{1 - \frac{3 \tilde{\gamma}}{4}}$. Consequently, $\kappa > 0$ for expanding models, and then $\tau > 0$; whereas for contracting models, $\kappa < 0$ and therefore $\tau < 0$. \\ \\
			We now determine the metric function $\omega^2 = [\omega_1(\tau) + \omega_2(\tau) Q(r)]^2$. A straightforward calculation shows that, for $k = 0$, $\omega_2 = \sqrt{\varphi}$ is a solution of equation (\ref{eq-T-2}). Moreover, $\omega_1 = \sqrt{\varphi} \alpha$ is also a solution to this equation if, and only if, $\alpha = \alpha(\tau)$ satisfies the differential equation $\dot{\alpha} = C \varphi^{-\frac32}$. Then, we can easily determine $\alpha(\tau)$ if we use the expression (\ref{phi-p}) for $\varphi(\tau)$, and then $\omega^2 = \varphi(\tau)[\alpha(\tau) + Q(r)]^2$. \\ \\ \\
			Note that, since $Q(r)$ is an arbitrary function, the metric expression is invariant if we change $\alpha$ by an additive constant and a factor (changing appropriately the function $Q$ and the coordinate $r$). Therefore, we obtain that the ideal T-models have a metric line element of the form
			\begin{equation} \label{metric-T-ideal}
				\textrm{d}s^2 = -\textrm{d}\tau^2 + \varphi(\tau)([\alpha(\tau) + Q(r)]^2 \textrm{d}r^2 + \textrm{d}x^2 + \textrm{d}y^2),
			\end{equation}
where $Q(r)$ is an arbitrary function and 
			\begin{equation} \label{phi-alpha}
				\varphi(\tau) = |\tau|^{\frac{4}{3 \tilde{\gamma}}} , \qquad \alpha(\tau) = 
					\begin{cases}
 						|\tau|^{1 - \frac{2}{\tilde{\gamma}}} , \, \qquad {\rm if} \ \ \, \tilde{\gamma} \not = 2 \, , \cr 
  						\ln |\tau| , \quad \quad \ \ \, {\rm if} \quad \tilde{\gamma} = 2 \, .
  					\end{cases}
			\end{equation}
The time coordinate takes values either in the interval $\tau > 0$ (expanding models) or in the interval $\tau < 0$ (contracting models). \\ \\[1mm]
			On the other hand, from (\ref{phi-alpha}) and expression (\ref{expansion-T-2}), the expansion of the fluid flow takes the expression
			\begin{equation} \label{expansion-T-ideal}
				\begin{array}{c}
					\displaystyle \theta = \frac{2}{\tilde{\gamma} \, \tau} \Big(1 + \frac12 \delta \Big) , \qquad \ \delta = \delta(\tau,r) \equiv \frac{\tilde{\alpha}(\tau)}{\alpha(\tau) + Q(r)} , \\[5mm] \quad \tilde{\alpha}(\tau) = 											\begin{cases} 
						(\tilde{\gamma} \! - \! 2) \alpha(\tau), \quad \, {\rm if} \ \ \tilde{\gamma} \neq 2 \ \ \cr 
						\; 2, \qquad \qquad \quad {\rm if} \ \ \tilde{\gamma} = 2 \ \
					\end{cases}
				\end{array}
			\end{equation}
where $\alpha(\tau)$ is given in (\ref{phi-alpha}). \\[-3mm]

			\subsubsection{Hydrodynamic quantities: energy density, pressure and speed of sound} 
			Now, we carry out the second step by obtaining the coordinate dependence of the hydrodynamic quantities $\rho$, $p$, and the indicatrix function $c_s^2 = \chi(\pi)$. From expressions (\ref{pressure-T-1}) and (\ref{density-T-1}), we can obtain the time dependence of the pressure and the energy density by taking $\omega_2 = \sqrt{\varphi}$ and $\omega_1 = \sqrt{\varphi}\alpha$ and making use of (\ref{phi-alpha}). \\ \\ \\ \\
			On the other hand, the indicatrix function $\chi(\pi)$ can be determined from (\ref{chi-Tmodels}) by taking into account (\ref{ABCci}) and (\ref{ki-gamma}). Then, we obtain that the pressure $p$ and the energy density $\rho$ for the ideal T-models (\ref{metric-T-ideal}$-$\ref{phi-alpha}) take the expression
			\begin{equation} \label{pressure-density-T-ideal}
				p = \frac{4(\tilde{\gamma} - 1)}{3 \tilde{\gamma}^2}\, \frac{1}{\tau^2} \, , \qquad \rho = \frac{4}{3 \tilde{\gamma}^2}\, \frac{1}{\tau^2} [1 + \delta(\tau,r)] \, ,
			\end{equation}
where $\delta(\tau,r)$ is given in (\ref{expansion-T-ideal}), and the square of the speed of sound is given by
			\begin{equation} \label{chi-T-ideal}
				c_s^2 = \chi(\pi) \equiv \frac{2 \, \tilde{\gamma} \, \pi^2}{(\pi + 1)(\pi + \tilde{\gamma} - 1)} \, , \qquad \pi \equiv \frac{p}{\rho} \, . \\ \\[2mm]
			\end{equation}

			\subsubsection{Curvature singularities and spacetime domains} 
			Expressions (\ref{expansion-T-ideal}$-$\ref{pressure-density-T-ideal}) show that the ideal T-models have a curvature singularity at $\tau = 0$ and, in the spacetime domains where $Q(r) < 0$, another one at $\alpha(\tau) + Q(r) = 0$. We briefly analyse them for the expanding models (for the contracting models the study is similar). \\ \\
			The existence of these kind of singularities has been already remarked by several authors in the homogeneous case $\omega = \omega(\tau)$ (KCKS metrics). Kantowski \cite{Kantowski} pointed out that: (i) when $\varphi(\tau_0) = 0$, the metric line element on the sphere (plane or hyperboloid) vanishes at $\tau = \tau_0$, and we have infinite energy density and pressure, and (ii) when $\omega(\tau_1) = 0$, the one-dimensional metric line element $\omega^2 \textrm{d}r^2$ vanishes at $\tau = \tau_1$, and we have infinite energy density. On the other hand, Collins \cite{Collins} showed that, under the energy conditions (\ref{E}) and the first compressibility condition in (\ref{H1}), the KCKS perfect fluid solutions are geodesically incomplete. \\ \\
			In the inhomogeneous case $\omega = \omega(\tau, r)$, we also have these curvature singularities, but the second one is not simultaneous for the comoving observer. Now, the collapsing time depends on $r$, $\tau_1 = \tau_1(r)$. 
			In our ideal T-models we have $\omega = \sqrt{\varphi(\tau)}[\alpha(\tau) + Q(r)]$ and, consequently, $\omega = 0$ when $\varphi = 0$. Thus, the full line element of the 3-spaces $\tau = constant$ vanishes at $\tau = 0$, and we have a big bang singularity. Both energy density and pressure diverge at $\tau = 0$. \\ \\ \\
			On the other hand, if $\tau_1 = \tau_1(r)$ is such that $\alpha(\tau_1) + Q(r) = 0$, the metric distance on the coordinate lines of the coordinate $r$ vanishes, and we have a singularity with a divergent energy density at $\tau = \tau_1$. \\ \\
			This analysis shows that we have two disconnected spacetime domains \mbox{defined by}
			\begin{equation}
				\begin{array}{l}
{\cal R}_0 = \{\tau > 0 ,\quad \alpha(\tau) + Q(r) < 0\} \, , \\[2mm]
{\cal R}_1 = \{\tau > 0 ,\quad \alpha(\tau) + Q(r) > 0\} \, .
				\end{array}
			\end{equation}
Note that if $Q(r) > 0 \ \ \forall r$, ${\cal R}_0 = \emptyset$. 

		\subsection{Ideal T-models: analysis of the physical reality conditions} \label{subsec-T-models-conditions}
		Now that we have an explicit expression of the hydrodynamic quantities and of $\chi(\rho,p)$, we can proceed with the analysis of the physical reality conditions. Therefore, here we will follow the last steps of the procedure presented in Section \ref{sec-generic-ideal-gas}.

			\subsubsection{Energy conditions} 
			The energy conditions E$^{\rm G}$ given in (\ref{EG}) imply $p > 0$. Then, the expression (\ref{pressure-density-T-ideal}) for the pressure means that, necessarily, $\tilde{\gamma} > 1$. Note that we have a flat FLRW limit by taking $\alpha = 0$ in the metric (\ref{metric-T-ideal}) (or, $\delta = 0$ in the expressions of the expansion and energy density). In this limit we have a barotropic evolution of the form $p = (\tilde{\gamma} - 1) \rho$. These FLRW models fulfil the energy condition (\ref{EG}) when $\tilde{\gamma} < 2$, and they are the relativistic $\gamma$-law models \cite{Assad-Lima}. The inhomogeneous models with $\tilde{\gamma} < 2$ belong to the the Szekeres-Szafron ideal singular models studied in \cite{C-F-S_SzSz_Singular}. Nevertheless, in our inhomogeneous T-models with $\tilde{\gamma} \geq 2$ there may be regions where the energy conditions meet. We will also study them here. \\ \\
			Note that $\tilde{\gamma}$ is a thermodynamic parameter that defines the EoS (\ref{chi-T-ideal}) and sets the time dependence of the metric (see (\ref{metric-T-ideal}$-$\ref{phi-alpha})). The metric also depends on an arbitrary real function $Q(r)$ which determines the inhomogeneity. If $Q = constant$, then the metric is an (homogeneous) KCKS model. \\ \\ \\
			If we denote the energy density of the FLRW limit as $\rho_F$, then we have $\rho = \rho_F(1 + \delta)$ and $p = (\tilde{\gamma} - 1) \rho_F$. Thus, the function $\delta = \delta(\tau,r)$ given in (\ref{expansion-T-ideal}) is the energy density contrast with respect to the FLRW limit. Nevertheless, note that it is not the energy density contrast with respect to a homogeneous background (the KCKS limit acquired when $Q = constant$). \\ \\
			With the notation introduced above, we have $\rho - p = \rho_F(2 - \tilde{\gamma} + \delta)$. Consequently, the solution meets the energy conditions if, and only if, $\tilde{\gamma} > 1$ and $\delta > \tilde{\gamma} - 2$. The spacetime regions where this last inequality holds strongly depend on whether $\tilde{\gamma}$ is greater than, equal to, or less than 2. The analysis of each case shows different behaviours summarised in Table \ref{table-1}. We only develop the expanding models ($\tau > 0$) in detail. The behaviour of the contracting models ($\tau < 0$) can then be obtained from the expanding ones by exchanging the future for the past. \\ \\
			\begin{table*}[t]
				\begin{tabular}{cllll}
					\noalign{\hrule height 1.05pt}
  					& \quad \ \ $\alpha(\tau)$ & \ \quad E$^{\rm G}$ & \quad \ $\tau$ &\quad \; \ $\delta$
\hspace{-10mm} \phantom{\LARGE $(\frac{A}{B})$} \\[0mm] \hline
					\hspace{-5mm} & \ \ \quad & \ \quad $Q < - \alpha $ &\quad \ $]\tau_1, \infty[ $ & \quad \; \ $> 0$, \ \quad $\infty \searrow 0$ \hspace{-20mm} \phantom{\Large $(\frac{A}{B})$}
					\\[-7mm] 
					\hspace{-3mm} \quad $\tilde{\gamma} < 2$ & \quad \ \ $\infty \searrow 0$ &   &
					\hspace{-20mm} \phantom{\large $(\frac{A}{B})^{\Big(C \Big)}$} \\[-6mm]
					\hspace{-1mm} & \ & \ \quad $Q > 0$ & \quad \ $]0, \infty[ $ & \quad \; \ $< 0$, \ \quad $(\tilde{\gamma} \! - \! 2) \nearrow 0 \ $    
\hspace{-20mm} \hspace{-20mm} \phantom{\large $(\frac{A}{B})^{\Big(C \Big)}$} 
					\\[1mm] \hline
					\hspace{-3mm} \quad $\tilde{\gamma} = 2$ & \quad \ $- \infty \nearrow \infty$ & \quad $-\alpha < Q$ & \quad \ $]\tau_1, \infty[ $ & \quad \; \ $> 0$, \ \quad $\infty \searrow 0$
					\hspace{-20mm} \phantom{\large $(\frac{A}{B})^{\big(C \big)}$} \hspace{1cm} \\[-4.3mm]
					\\[2mm] \hline
					\hspace{-3mm} \quad $\tilde{\gamma} > 2$ & \quad \ \ $0 \nearrow \infty$ & \quad $-\alpha < Q < 0$ & \quad \ $]\tau_1, \infty[ $ & \quad \; \ $> 0$, \ \quad $\infty \searrow \tilde{\gamma} \! - \! 2$
					\hspace{-20mm} \phantom{\large $(\frac{A}{B})^{\big(C \big)}$}  
					\\[2mm] \noalign{\hrule height 1.05pt}
				\end{tabular}
\caption{This table provides, for the different values of the parameter $\tilde{\gamma}$: (i) the interval where the function $\alpha(\tau)$ takes values (second column); (ii) the spacetime region ${\cal R}_1$ where the energy conditions E$^{\rm G}$ hold (third column); (iii) the time interval where E$^{\rm G}$ is kept for a given $Q(r)$ (fourth column); (iv) the sign of the energy density contrast $\delta$, and the interval where it takes values (fifth column).}
\label{table-1}
			\end{table*}
			\hspace{-4.7mm} The energy conditions E$^{\rm G}$ involve the metric functions $\alpha(\tau)$ and $Q(r)$, and they only hold in the spacetime domain ${\cal R}_1$. Moreover, the function $Q(r)$ can be chosen such that there is always a time $\tau_1$ where the energy conditions E$^{\rm G}$ hold, and then they also hold for later times. The only models with a negative $\delta$ can take place if $\tilde{\gamma} < 2$ and $Q > 0$. Then, the time coordinate covers its entire domain $\tau > 0$ (that is, region ${\cal R}_1$, since ${\cal R}_0 = \emptyset$), and $\delta$ increases from a finite negative value (at early times) to zero at later times (the model approaches the relativistic $\gamma$-law FLRW limit). Models with $\tilde{\gamma} < 2$ and a positive $\delta$ also approach the FLRW limit for later times. Models with $\tilde{\gamma} \geq 2$ have, necessarily, a positive $\delta$. When $\tilde{\gamma} = 2$ the solution approaches the shift ($\rho = p$) FLRW model at late times. If $\tilde{\gamma} > 2$, the energy density contrast decreases from large values and approaches a positive value for late times. \\ \\
			Note that, for each model, the sign of the energy density contrast does not change throughout the spacetime domain where the energy conditions E$^{\rm G}$ hold. Nevertheless, a suitable election of the function $Q(r)$ can model regions with an excess or a lack of energy density (with respect to a homogeneous KCKS background defined by a constant value of the function $Q$). \\ \\
			It is worth remarking that the expansion (\ref{expansion-T-ideal}) of our inhomogeneous model has the same sign as the FLRW limit when $\delta > -2$. This occurs for the models with a positive energy density contrast, but also for $\delta < 0$ in the domain where the energy conditions hold. This fact justifies that we speak of expanding models when $\tau > 0$, and of contracting models when $\tau < 0$.
			
			\subsubsection{Compressibility conditions $\textrm{H}^\textrm{G}_1$} 
			The compressibility conditions H$_1^{\rm G}$ were studied in \cite{C-F-S_SzSz_Singular} for the Szekeres-Szafron ideal singular models. In that case, the indicatrix function takes the form (\ref{chi-T-ideal}) with $1 < \tilde{\gamma} < 2$. The same reasoning can be applied to the ideal T-models, as it remains valid for $\tilde{\gamma} \geq 2$, and we obtain the same result as in \cite{C-F-S_SzSz_Singular}. Namely, we have that the ideal T-models (\ref{metric-T-ideal}$-$\ref{phi-alpha}) fulfil the compressibility conditions H$_1^{\rm G}$ provided that they fulfil the energy conditions E$^{\rm G}$, that is, in the domain ${\cal R}_1$. 
			
			\subsubsection{Thermodynamic schemes of the ideal T-models} 
\begin{figure}[H]
		\centerline{
		\parbox[c]{0.5\textwidth}{\hspace{-0.5cm}\includegraphics[width=0.49\textwidth]{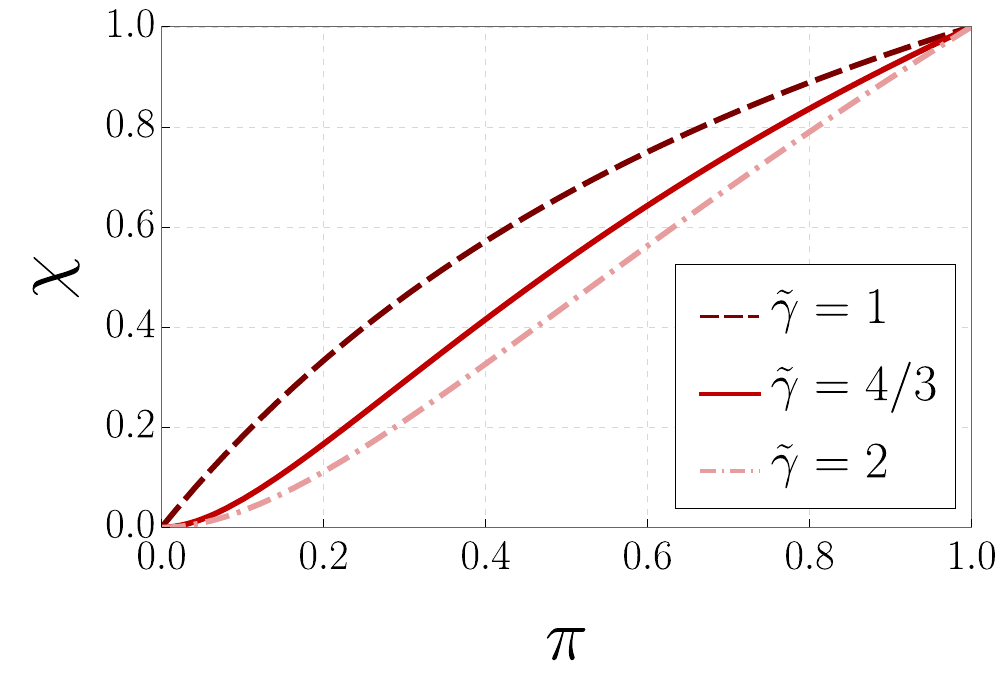}\\[-2mm] \centering{(a)}}
		\raisebox{-2pt}{\parbox[c]{0.5\textwidth}{\hspace{-0.5cm}\includegraphics[width=0.48\textwidth]{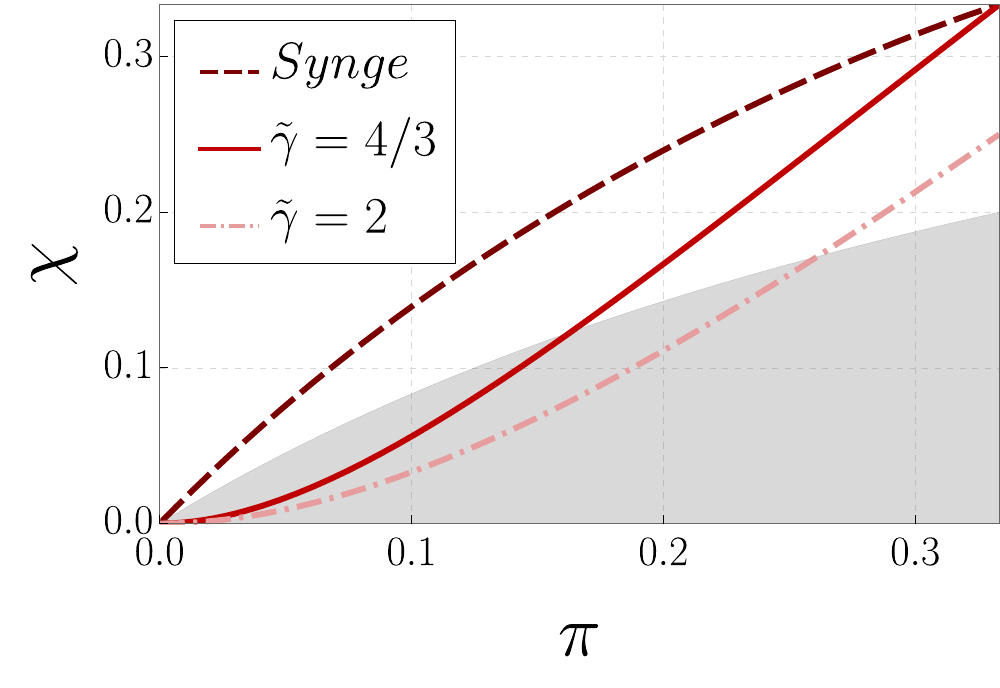}\\[-2mm] \centering{(b)}}}}
		\vspace{-2mm}
		\caption{(a) Here, we show the behaviour of the indicatrix function $\chi(\pi)$ (\ref{chi-T-ideal}) defined in the whole interval $]0,1[$ for different values of the parameter $\tilde{\gamma}$. This is the domain where the energy conditions E$^{\rm G}$ hold and, therefore, the compressibility conditions H$_1^{\rm G}$ also hold in the whole domain. (b) Here, we also show the behaviour of the square of the speed of sound but in the interval $]0,1/3[$, for the values of $\tilde{\gamma}$ that we study in detail in this section, $\tilde{\gamma} = 4/3$ and $\tilde{\gamma} = 2$, and for the Synge gas. The case $\tilde{\gamma} = 4/3$ approaches a Synge gas at high temperatures. The shaded area is forbidden by the compressibility condition H$_2^{\rm G}$. Thus, the hydrodynamic variable is limited to an interval $]\pi_m,1[$.}
	\label{Fig-01}
	\end{figure}

			In the previous two subsections we completed step 3 in analysing the physical meaning of the ideal T-models. Now we can perform step 4 by particularising the general study of the thermodynamic schemes presented at the end of Section \ref{sec-thermo-Tmodels}. Note that the thermodynamic quantities depend on the metric functions $\varphi(\tau)$, $\omega_1(\tau)$ and $\omega_2(\tau)$, and on two functions, $N(Q)$ and $s(Q)$, of the metric function $Q(r)$. The former now take the expression $\omega_1 = \sqrt{\varphi} \alpha$, $\omega_2 = \sqrt{\varphi}$, with $\varphi(\tau)$ and $\alpha(\tau)$ given in (\ref{phi-alpha}). Moreover, each choice of the latter determines a specific thermodynamic scheme with a specific entropy and a mass density given in (\ref{s-n-Tmodels}), and a temperature given in (\ref{lambda_i}$-$\ref{ell-m}). \\ \\
			Thus, we can determine the thermodynamic quantities as a function of state depending on the hydrodynamic quantities $\rho$ and $p$ if we obtain the functions $Q(\rho,p)$ and $\lambda_i(p)$ given in (\ref{Q(prho)}) and (\ref{lambda_i}). Note that $Q(\rho,p)$ can be obtained from (\ref{expansion-T-ideal}$-$\ref{pressure-density-T-ideal}). Table \ref{table-2} collects these expressions distinguishing the cases $\tilde{\gamma} \neq 2$ and $\tilde{\gamma} = 2$. \\ \\
			\begin{table*}[t]
				\begin{tabular}{cllll}
					\noalign{\hrule height 1.05pt}
					& $\quad Q(\rho,p)$ & $\quad n(\rho, p)$ & \quad $\lambda_1(p)$ & \quad $\lambda_2(p) \quad$ \hspace{-20mm} \phantom{\LARGE $(\frac{A}{B})$} \\[2mm] \hline
					\hspace{1mm} $\tilde{\gamma} \neq 2$ & $\quad \displaystyle \frac{\tilde{K}(\rho - p)}{\rho(\tilde{\gamma} - 1) - p}\, p^{\frac{2 - \tilde{\gamma}}{2 \tilde{\gamma}}}$ & $\quad \displaystyle \frac{\rho(\tilde{\gamma} - 1) - p}{K N(Q) \sqrt{p}}$ & $\quad l_1 \sqrt{p}$ & \quad $l_2 \, p^{1 - \frac{1}{\tilde{\gamma}} \; }$   
\hspace{-20mm} \phantom{\Large $\displaystyle [(\frac{A}{B})^{B}]$}  
					\\[4mm] \hline
					\hspace{1mm} $\tilde{\gamma} = 2$ & $ \quad \displaystyle \frac12 \ln(3p) + \frac{2p}{\rho - p} $ & $\quad \displaystyle \frac{\sqrt{3} \, (\rho - p)}{2 N(Q) \sqrt{p}}$ & \quad $\displaystyle  \sqrt{\frac{p}{3}} \, [2 - \ln(3p)]$ & \quad $\displaystyle \frac{2}{\sqrt{3}} \sqrt{p} \; $  
\hspace{-17.5mm} \phantom{\Large $\displaystyle [(\frac{A}{B})^{B}]$}  
					\\[4mm] \noalign{\hrule height 1.05pt}
				\end{tabular}
\caption{Thermodynamic schemes of the ideal T-models. This table offers the mass density $n(\rho,p)$ and the functions $Q(\rho,p)$ and $\lambda_i(p)$ that determine the specific entropy $s(\rho,p) = s(Q)$ and the temperature $\Theta(\rho,p) = \ell(Q) \lambda_1(\tau) + m(Q) \lambda_2(\tau)$, with $\ell(Q)$ and $m(Q)$ given in (\ref{ell-m}). The constants $\tilde{K}$, $K$, $l_1$ and $l_2$ depend on the parameter $\tilde{\gamma}$ as $\tilde{K} \equiv -(\tilde{\gamma} - 1) \hat{K}^{1 - \frac{2}{\tilde{\gamma}}}$, $K \equiv (\tilde{\gamma} - 2) \hat{K}$, $l_1\equiv 2 \hat{\gamma}$ and $l_2 \equiv \frac{4}{3 \tilde{\gamma}} \hat{K}^{\frac{1}{\tilde{\gamma}} - 1}$, where $\hat{K} \equiv \frac{2 \sqrt{\tilde{\gamma} - 1}}{\sqrt{3} \tilde{\gamma}}$.} 
\label{table-2}
			\end{table*}
			\begin{table}[t]
				\begin{tabular}{cllll}
					\noalign{\hrule height 1.05pt}
  					& $n(\rho, p)$  & $\Theta(\rho, p)$ & $s(\rho, p)$ & H$_2$ \hspace{-10mm} \phantom{\Large $(\frac{A}{B})$} 
  					\\[1mm] \hline
					\hspace{-3mm} IG & $\displaystyle \frac{{e_0}^{-1}(p \! - \! \rho)^{\frac{\tilde{\gamma}}{2 - \tilde{\gamma}}}}{\left[ \, p \! - \! \rho(\tilde{\gamma} \! - \! 1) \right]^\frac{2(\tilde{\gamma} - 1)}{2 - \tilde{\gamma}}}$ & $\displaystyle \frac{p}{\tilde{k} n(\rho, p)}$ & $\bar{s}_0 \! + \! k \ln \! \left(\! \! \frac{1}{p} \! \! \left[\frac{p - \rho(\tilde{\gamma} - 1)}{\rho - p}\right]^{\!\frac{2 \tilde{\gamma}}{2 - \tilde{\gamma}}} \! \! \right) $ & $\pi \in ] \pi_m , 1 [ $   
\hspace{-20mm}  \phantom{\Large $\displaystyle [(\frac{A}{B})^{B}]$} 
					\\[5mm] \hline
					\hspace{-3mm} LT &  $\displaystyle n_1 (\rho - p) \, p^{\frac{1}{\tilde{\gamma}} - 1}$ & $\displaystyle \Theta_1 \sqrt{p} $ & $\displaystyle s_0 \! - \! s_1 \frac{\rho(\tilde{\gamma} \! - \! 1) \! - \! p}{\rho - p} p^{\frac{-(2 - \tilde{\gamma})}{2 \tilde{\gamma}}} $ & E 
\hspace{-20mm} \phantom{\Huge $(\frac{A}{B})$}  
					\\[2.8mm] \hline
					\hspace{-2mm}FLRW & $\displaystyle n_1 \frac{\rho(\tilde{\gamma} \! - \! 1) \! - \! p}{\sqrt{p}}$ &  $ \displaystyle \Theta_1 \, p^{1 - \frac{1}{\tilde{\gamma}}}$ & $\displaystyle s_0 \! + \! s_1 \frac{\rho - p}{\rho(\tilde{\gamma} \! - \! 1) \! - \! p} p^{\frac{2 - \tilde{\gamma}}{2 \tilde{\gamma}}}$ & $\displaystyle \kappa Q \! < \! \! \frac{-2 \phi^{-\frac32 (2 - \tilde{\gamma})}}{3(2 - \tilde{\gamma})}$  
\hspace{-16.5mm} \phantom{\Huge $(\frac{A}{b})$}
					\\[4mm] \noalign{\hrule height 1.05pt}
				\end{tabular}
\caption{This table provides the explicit expression of the matter density $n$, the temperature $\Theta$ and the specific entropy $s$ in terms of the hydrodynamic quantities $\rho$ and $p$ for three specific thermodynamics, the generic ideal gas thermodynamic scheme (IG), the Lima-Tiomno model (LT), and the model with the temperature of the FLRW-limit (FLRW). The last column shows the constraints imposed by the thermodynamic compressibility condition H$_2$, where $\pi_m \equiv \frac{\tilde{\gamma} - \sqrt{17 \tilde{\gamma}^2 - 20 \tilde{\gamma} + 4}}{2(1 - 4\tilde{\gamma})}$.}
\label{table-3}
			\end{table}
			$\hspace{-4.7mm}$ Then, we have that a particular ideal T-model admits a different thermodynamic interpretation for each choice of the functions $N(Q)$ and $s(Q)$. In \cite{C-F-S_SzSz_Singular}, three thermodynamic schemes associated to the Szekeres-Szafron singular models that also apply for the ideal T-models were studied in detail for the case $\tilde{\gamma} < 2$: models with a generic ideal gas thermodynamic scheme, the Lima-Tiomno \cite{Lima-Tiomno} models, and the models with the temperature of the FLRW limit. \\ \\ 
			The first one defines a generic ideal gas and it has inhomogeneous temperature. The other two thermodynamics have homogeneous temperature, and therefore they are compatible with the current relativistic heat equations. This means that these two thermodynamic schemes model either
perfect fluids in l.t.e. or inviscid non-perfect (with non-vanishing conductivity coefficient) fluids in thermal equilibrium (see Section \ref{subsec-intro-thermal-conduc-coeff} for more details). In these three cases, step 5 of our approach has been analysed: the positivity conditions P hold, and the compressibility condition H$_2$ holds in a wide spacetime domain. All these results are summarised in Table \ref{table-3}. \\ \\[-2mm] 
			A detailed study of different thermodynamic schemes for any $\tilde{\gamma}$ falls outside the scope of this thesis. Here, we will limit ourselves to outlining some qualities of the scheme that allows us to interpret the solutions $\tilde{\gamma} = 4/3$ and $\tilde{\gamma} = 2$ as generic ideal gases in l.t.e. \\ \\[-2mm]
			\begin{table*}[t]
				\begin{tabular}{llllll} 
					\noalign{\hrule height 1.05pt}
  					& $n(\rho, p)$ & $ \; \Theta(\rho, p)$ & $ \; s(\rho, p)$ & \; H$_2$  
\hspace{-10mm} \phantom{\Large $(\frac{A}{B})$} \\[1mm] \hline
					$\tilde{\gamma} = 4/3$ & $\displaystyle \frac{(\rho - p)^{2}}{\rho - 3 p}$ & $ \; \displaystyle \frac{p}{ \tilde{k} \, n(\rho, p)}$ & $ \; \displaystyle s_0 \! + \! \tilde{k} \ln \! \left[ \frac{1}{p} \! \left[\frac{\rho - 3p}{\rho - p}\right]^{\!4} \right]$ & $ \; \displaystyle \pi \in ] \pi_1 , \frac13 [$  
\hspace{-18mm} \phantom{\Large $ \displaystyle (\frac{A}{B})^{(C)}$}  
\\[4mm] \hline
					$\tilde{\gamma} = 2$ & $\displaystyle (\rho - p) \exp\left\{\frac{2 p}{\rho-p}\right\}$ & $ \; \displaystyle \frac{p}{\tilde{k} \, n(\rho, p)}$ & $ \; \displaystyle s_0 \! - \! \tilde{k} \left[\ln p + \frac{4 p}{\rho-p}\right]$ & $ \; \pi \in ] \pi_2 , 1 [$   
\hspace{-10mm} \phantom{\Large $\displaystyle (\frac{A}{b})$} 
					\\[3.5mm] \noalign{\hrule height 1.05pt}
				\end{tabular}
\caption{This table provides, for the T-models with $\tilde{\gamma} = 4/3$ and $\tilde{\gamma} = 2$, the explicit expression of the matter density $n$, the temperature $\Theta$ and the specific entropy $s$ in terms of the hydrodynamic quantities $\rho$ and $p$ for the generic ideal gas thermodynamic scheme. Last column shows the constraints imposed by the compressibility condition H$_2$: $\pi_1 = \frac{1}{13} (\sqrt{17}-2) \approx 0.16$, and $\pi_2 = \frac{1}{7} (2\sqrt{2} - 1) \approx 0.26$.}
\label{table-4}
			\end{table*}
			In Section \ref{sec-generic-ideal-gas}, we present an algorithm obtained in \cite{Hydro-LTE} that provides all the thermodynamic quantities of the ideal gas scheme when the indicatrix function $\chi = \chi(\pi)$ is known. If we consider the expression (\ref{chi-T-ideal}) for $\chi(\pi)$ when $\tilde{\gamma} = 4/3$ or $\tilde{\gamma} = 2$ and we apply this algorithm, we obtain the thermodynamic schemes summarised in Table \ref{table-4}. On the other hand, as explained in Section \ref{sec-generic-ideal-gas}, for the ideal gas schemes the compressibility condition H$_2$ becomes H$_2^\textrm{G}$. For the $\chi(\pi)$ of the ideal gas T-models this inequality holds in an interval $]\pi_m, 1[$ as the last column in Table \ref{table-4} shows. This is also illustrated in Figure \ref{Fig-01}(b). \\
			$\hspace{-5mm}$It is worth remarking that for $\tilde{\gamma} = 4/3$ (and similarly, for any $\tilde{\gamma} < 2$) the thermodynamic quantities are not defined for $\pi = 1/3$ (similarly, for $\pi = \tilde{\gamma} - 1$). The two resulting subintervals are related with the two different cases where the energy conditions hold for $\tilde{\gamma} < 2$ (see Table \ref{table-1}). Indeed, when $Q < - \alpha$ we have $\delta > 0$ and then $\pi < \tilde{\gamma} - 1$; and when $Q > 0$ we have $\delta < 0$ and then $\pi > \tilde{\gamma} - 1$. \\ \\ 
			In Table \ref{table-4}, for the models $\tilde{\gamma} = 4/3$, we have only considered the expression of the thermodynamic quantities when the energy density contrast is positive. In this case, we have $\pi < 1/3$, and in the limit $\chi(1/3) = 1/3$, $\chi'(1/3) = 1/2$, the same values that the Synge gas (see Section \ref{sec-approximations}). This is also illustrated in Figure \ref{Fig-01}(b). Thus, this ideal gas scheme appears to be a good approximation to a relativistic gas. \\ \\
			Note that the ideal gas thermodynamic schemes considered in Table \ref{table-4} can also be obtained from the generic thermodynamic schemes in Table \ref{table-2} by considering a particular choice of the functions $s(Q)$ and $N(Q)$: for $\tilde{\gamma} = 4/3$, $s(Q) = s_0 - \frac14 \ln|Q|$ and $N(Q) = -\frac23 Q^{-2}$; whereas for $\tilde{\gamma} = 2$, $s(Q) = s_0 - \ln 3 + 2 Q $ and $N(Q) = \frac23\exp\{-Q\}$.
	
		\subsection{Analysis of the McVittie$\,$-Wiltshire$\,$-Herlt solution and \\ its generalisations} \label{sec-McVittie}
		As we will see in more detail in Section \ref{sec-integracio-1}, Herlt \cite{Herlt} proposed a method to get an inhomogeneous T-model from a known homogeneous KCKS metric. He applied it to generalise the (spherically symmetric) McVittie and Wiltshire solution \cite{McVittie}. In the canonical form (\ref{metric-T-1}), they take the time coordinate $t$ such that $Y = \sqrt{\varphi} = t$ and they look for the model with $\omega = t^m \equiv \omega_1(t)$. Then, the field equation (\ref{eq-T-1}) determines the function $v(t)$: 
		\begin{equation} \label{v-MWH}
			v(t) = \frac{1}{m^2 - 1} + C_0 \, t^{-2(m + 1)} , 
		\end{equation}
which gives the McVittie-Wiltshire solution. Then, substituting this expression in (\ref{eq-T-1}) we obtain an equation for $\omega(t)$. We know the particular solution $\omega_1(t)$, and then we can formally determine another one, $\omega_2(t)$, in terms of an integral as we will see in Section \ref{sec-integracio-1}. When $C_0 = 0$, this integral can be explicitly calculated and one obtains $\omega_2(t) = t^{-m}$. We name this specific solution the McVittie-Wiltshire-Herlt T-model (see also \cite{Kramer}). \\ \\
		We can easily recover the McVittie-Wiltshire-Herlt (MWH) solution by \mbox{working} with the proper time $\tau$ of the Lagrangian observer. When $C_0 = 0$ we have that $\textrm{d} \tau = v^{-1/2} \textrm{d} t = \sqrt{m^2-1} \, \textrm{d} t$ and, consequently, $\varphi = Y^2 = f_0 \, \tau^2$, with $f_0^{-1} = m^2 - 1 > 0$. \\ \\
		In this section we study the T-models (\ref{metric-T-2}$-$\ref{eq-T-2}) with $\varphi = f_0 \, \tau^2$. In this way, we generalise the MWH T-model to any curvature $k = 0, \pm 1$. Moreover, we analyse the macroscopic necessary condition for physical reality for these solutions. 
		
			\subsubsection{Metric line element} 
			If we take $\varphi(\tau) = f_0 \, \tau^2$, then the field equation (\ref{eq-T-2}) becomes
			\begin{equation}
				\tau^2\, \ddot{\omega} + \tau \, \dot{\omega} - k_0 \, \omega = 0 \, , \qquad k_0 \equiv 1 + \frac{k}{f_0} \, .
			\end{equation}
A straightforward calculation shows that this equation admits a solution if $k_0 > 0$, and two independent ones are
			\begin{equation} \label{omegues}
				\omega_1(\tau) = \tau^m , \quad \omega_2(\tau) = \tau^{-m} , \qquad m = \sqrt{k_0} > 0 \, .
			\end{equation}
Then, the metric line element of the generalised MWH T-models takes the form (\ref{metric-T-2}), where $\omega_i(\tau)$ are given in (\ref{omegues}) and
			\begin{equation} \label{varphi}
				\varphi(\tau) = f_0 \, \tau^2 > 0 \, .
			\end{equation}
Note that, if $k = 1$, then $m > 1$, and we recover the MWH solution, if $k = -1$, then $m < 1$, and if $k = 0$, then $m = 1$.
			It is worth remarking that the choice $\varphi(\tau) = f_0 \, \tau^2$ is equivalent to $\dot{\varphi}^2 = 4 f_0 \, \varphi$ after a redefinition of the coordinate time origin. Consequently, the case (ii) presented in Section \ref{sec-chi-pi} corresponds with the generalised MWH metrics. 
			
			\subsubsection{Pressure and energy density. Study of the energy conditions} 
			Now, we particularise the expressions of the pressure (\ref{pressure-T-2}) and the energy density (\ref{density-T-2}) for the generalised MWH T-models and we obtain
			\begin{equation} \label{pressure-density-McV}
				p = - \frac{m^2}{\tau^2} \, , \qquad \rho = \frac{m^2}{\tau^2} \left[1 + \frac2m \, \frac{\tau^{2m} - Q(r)}{\tau^{2m} + Q(r)} \right] \, .
			\end{equation}
Note that the pressure is negative and this fact disqualifies these solutions as ideal gas models. Nevertheless, it is known that continuous media with negative pressures exist and it is suitable to analyse the energy conditions for these models. \\
			From expressions (\ref{pressure-density-McV}), it follows that the first inequality of the energy conditions E given in (\ref{E}), $-\rho < p$, is equivalent to $(\tau^{2m} - Q)(\tau^{2m} + Q)^{-1} > 0$. Both factors of this expression cannot be simultaneously negative, and both are positive if, and only if,
			\begin{equation} \label{e-cMcV}
				|Q(r)| < \tau^{2m} \, .
			\end{equation}
On the other hand, the second energy inequality, $p \leq \rho$, holds if $\rho > 0$, that is, if $(\tau^{2m} - Q)(\tau^{2m} + Q)^{-1} > - m/2$, which is a consequence of the above condition (\ref{e-cMcV}). Moreover, in this case the expression (\ref{expansion-T-2}) of the expansion becomes
			\begin{equation}
				\theta = \frac{2}{\tau} \left[1 + \frac{m}{2} \, \frac{\tau^{2m} - Q(r)}{\tau^{2m} + Q(r)} \right] \, .
			\end{equation}
Note that if the energy conditions hold, then the sign of the expansion depends on the sign of the time coordinate. \\ \\
			Thus, it follows that the generalised MWH solution fulfils the energy conditions in the spacetime domain defined by (\ref{e-cMcV}). Moreover, for a given time $\tau_1$, we can always choose the inhomogeneity function $Q(r)$ such that (\ref{e-cMcV}) holds. Then, for expanding models the energy conditions hold for $\tau > \tau_1 > 0$ (in the future); and, for contracting models ($\tau < 0$) the energy conditions hold for $\tau < \tau_1 < 0$ (in the past). 
			
			\subsubsection{Speed of sound. Compressibility conditions} 
			We can obtain the indicatrix function $\chi(\rho,p)$, which gives the square of the speed of sound, by specifying the general expression (\ref{chi-Tmodels}) of the T-models for this case. Indeed, from the expressions (\ref{omegues}$-$\ref{varphi}) of the metric functions, and taking into account (\ref{abcd}$-$\ref{rho-punt}), we can determine the functions (\ref{ABCcal}). Then, by substituting in (\ref{chi-Tmodels}) we obtain:
			\begin{equation} \label{chi-McV}
				c_s^2 = \chi(\pi) \equiv \frac{4 \, \pi^2}{(\pi + 1) [(4 - m^2) \pi - m^2]} \, , \quad \pi \equiv \frac{p}{\rho} \, .
			\end{equation}
Note that the indicatrix function is of the ideal gas type, in accordance with case (ii) of Section \ref{sec-chi-pi}. Then, we can analyse the compressibility conditions H$^{\rm G}_1$ in the regions where the energy conditions meet. In this case we have $-1 < \pi < 0$, and then the first inequality in (\ref{H1G}), $0 < \chi$, implies $m^2 < (4 - m^2) \pi = (m^2 - 4)(-\pi) < m^2 - 4$. This contradiction shows that compressibility conditions are not satisfied anywhere.
			In summary, the McVittie-Wiltshire-Herlt solution is not a good model to represent a perfect fluid in local thermal equilibrium.
		
		\subsection{Why there are no solutions that model a classical \mbox{ideal gas?}} \label{sec-CIG}
		Classical ideal gases are the ideal gases that also fulfil the classical dependence of the specific internal energy on the temperature, $\epsilon = c_v \Theta$ (see Section \ref{sec-classic-ideal-gas}). For them, the indicatrix function takes the form (\ref{chi-gas-ideal-classic}). The study undertaken in Section \ref{sec-chi-pi} on the T-models compatible with the EoS (\ref{eq. estat gas ideal}) of a generic ideal gas leads to an indicatrix function of the form (\ref{chi-T-ideal}), which is incompatible with (\ref{chi-gas-ideal-classic}). Thus, no T-solution that models a classical ideal gas exists. \\ \\
		It is worth noting that a similar negative result is found in analysing the ideal gas models belonging to the family of the Szekeres-Szafron solutions of class II, in both singular \cite{C-F-S_SzSz_Singular} and regular \cite{C-F-S_SzSz_Regular} models. In \cite{CFS-CIG}, Coll \textit{et al.} also search for classical ideal gas solutions in the family of the R-models in geodesic motion, and the result is also negative: the only solutions are the homogeneous ones (classical ideal gas FLRW models \cite{CFS-CIG}). \\ \\
		A question naturally arises: are these negative results a consequence of a more general basic result? The answer is affirmative and is given in Proposition \ref{prop-kinematic-CIG}: a geodesic and expanding timelike unit vector $u$ is the unit velocity of a classical ideal gas if, and only if, $u$ is vorticity-free and its expansion is homogeneous. \\ \\
		Note that the Szekeres-Szafron solutions have a geodesic and expanding fluid flow. Consequently, only those with homogeneous expansion can be a candidate to model a classical ideal gas. But, for these metrics, homogeneous expansion is tantamount to barotropic evolution. Thus \cite{Inhomogeneous_Cosmological_Models, FS-SS}, for class II (and consequently in the limit admitting a G$_3$, the T-models) the metric is either a FLRW model or a KCKS solution; while for class I (and consequently in the limit admitting a G$_3$, the geodesic R-models), the metric is necessarily a FLRW model. \\ \\
		Note that the constraints on the kinematics of a classical ideal gas studied in \cite{CFS-CIG} are a consequence of the sole hydrodynamic equations and they do not depend on the field equations. This means that there are also no test solutions modelling a classical ideal gas that is comoving with the prefect fluid flow of the non-homogeneous solutions quoted above.

	\section{T-model field equations: The general solution} \label{T-model-general-sol}
	In the first half of this chapter, we have proposed a thermodynamic interpretation of the T-models. We have obtained the thermodynamic schemes associated with a specific T-model and we have determined the solutions that can model a generic ideal gas. On the other hand, we have generalised and analysed from a thermodynamic point of view the McVittie-Wiltshire-Herlt solution. This T-model can be obtained by applying the Herlt algorithm to the homogeneous T-model presented by McVittie and Wiltshire \cite{McVittie}. \\ \\
	However, there are very few T-models for which we know the explicit analytic expression of the metric functions, and it would be suitable to know more solutions for a better understanding of the physical and geometric properties of the T-models. In this second half of the chapter, we analyse the field equations for the T-models, we revisit the Herlt integration algorithm, and we propose new ones that provide the general solution without making any indefinite integral.

		\subsection{Revisiting the field equations for the T-models} \label{sec-Tmodels r}
		In the previous section, we have shown that the field equations for the T-models can be written as the second order differential equation (\ref{eq-T-1}), which is linear for a specific choice of the metric functions. For such election, the metric line element is given by (\ref{metric-T-1}), the unit velocity of the fluid $u = \sqrt{v} \, \partial_t$ is geodesic and its expansion, {\em pressure} $p$ and the {\em energy density} $\rho$ are then given by (\ref{expansion-T-1}), (\ref{pressure-T-1}) and (\ref{density-T-1}) respectively. \\ \\
		The known T-models have usually been obtained by considering the functions $v(t)$, $\omega(t,r)$ and $Y(t)$ as unknown metric functions. The field equations are linear in the functions $v(t)$ and $\omega(t,r)$, and this fact plays an important role in the integration process. Recall that our choice of the metric function $\varphi = Y^2$ as an unknown of the field equations, led us to equation (\ref{eq-T-1}), which is also a linear equation for $\varphi$. Thus, this equation is linear for the three involved metric functions, a significant quality that will help us in our approach. \\ \\
		Moreover, since (\ref{eq-T-1}) is a homogeneous linear second order differential equation for the function $\omega(t,r)$ when $v(t)$ and $\varphi(t)$ are given, we chose the coordinate $r$ so that $\omega(t,r)$ was written as in (\ref{w-w1-w2}), where $Q(r)$ is an arbitrary real function, and $\omega_i(t)$ being two particular solutions to equation (\ref{eq-T-1}). \\ \\
		The spacetime metric does not change with a redefinition of the time coordinate, $t = t(T)$. Every choice of $t$ can be realised by imposing a constraint on the time-dependent functions $v(t)$, $\varphi(t)$ and $\omega_i(t)$. This coordinate condition, and equation (\ref{eq-T-1}) imposed on each of the functions $\omega_i$, constitute a set of three constraints for the four metric functions $\{\varphi(t), \omega_i(t), v(t)\}$. Consequently, the space of solutions depends on an arbitrary real function depending on time, and another real function, $Q(r)$, depending on $r$. \\ \\
		It is quite usual in literature (see, for example, \cite{ Inhomogeneous_Cosmological_Models, Kramer}) to choose the time coordinate such that $t = Y = \sqrt{\varphi}$. Then, the functions $\omega_i(t)$ are determined by equation (\ref{eq-T-1}) if we give the function $v(t)$. In this case, the space of solutions is controlled by the functions $\{v(t), Q(r)\}$.
		Alternatively, we can give as input one of the functions $\omega_i$, say $\omega_2$, and then equation (\ref{eq-T-1}) becomes a first order linear differential equation for the function $v(t)$; once this equation is solved, we can proceed to determine $\omega_1$ by once again using (\ref{eq-T-1}) with the $v(t)$ previously obtained. This procedure by Herlt \cite{Herlt} shows that the field equation can be solved by quadratures, and the space of solutions is controlled by the functions $\{\omega_2(t), Q(r)\}$. \\ \\ 
		In the previous section, we have taken as time coordinate the proper time $\tau$ of the Lagrangian observer associated with the fluid. This means that $v = 1$, and then, for every choice of the function $\varphi(\tau)$, equation (\ref{eq-T-1}) determines two particular solutions $\omega_i(\tau)$. Thus, with this choice, the space of solutions is controlled by the functions $\{\varphi(\tau), Q(r)\}$.
	
			\subsubsection{Some new solutions with $k \neq 0$} \label{subsec-v=1-knot=0}
			When $v(\tau) = 1$, the field equation (\ref{eq-T-1}) becomes
			\begin{equation} \label{eq-T-v=1}
				2 \varphi \, \ddot{\omega} + \dot{\varphi}\, \dot{\omega} - (\ddot{\varphi} + 2k) \, \omega = 0 \, .
			\end{equation}
If we consider the case $\tilde{\gamma} = 4/3$ in the family of the ideal T-models given in (\ref{metric-T-ideal}$-$\ref{phi-alpha}), we have $\ddot{\varphi} = 0$ and $2 \varphi \, \ddot{\omega} + \dot{\varphi}\, \dot{\omega} = 0$. We can extend this solution to non-plane symmetry, $k = \pm 1$, by imposing on $\omega(\tau)$ this last equation and by considering $\varphi(\tau)$ such that $\ddot{\varphi} + 2k = 0$. Then, we can introduce the change of time $\tau = \kappa \, t$, with $\kappa > 0$ an arbitrary positive constant, so that the solution to this equation can be expressed as
			\begin{equation}
				\varphi(t) = \kappa^2(\varepsilon - k t^2) \, , \label{phi-knot0}
			\end{equation}
where $\varepsilon = \pm 1$ if $k = -1$, and $\varepsilon = + 1$ if $k = 1$. Moreover, $\omega_1(t) = 1$ is a particular solution to equation (\ref{eq-T-v=1}), and another one is
			\begin{align}
				\omega_2(t) &= \arcsin t & {\rm if} & \quad k = 1, &\label{omega-k=1} \\  
				\omega_2(t) &= \textrm{arcsinh} \, t & {\rm if} & \quad k = -1 \quad {\rm and} \quad \varepsilon = +1, &\label{omega-k=-1+} \\
				\omega_2(t) &= \textrm{arcosh} \, t & {\rm if} & \quad k = -1 \quad {\rm and} \quad \varepsilon = -1. &\label{omega-k=-1-}
			\end{align}
Then, the pressure and the energy density take the expressions
			\begin{eqnarray} \label{pressure-T-v=1}
				p = \frac{k \, \varepsilon}{\kappa^2 (\varepsilon - k t^2)^2} \, , \\[1mm]
				\rho = p \left[1 - \frac{2 t \, \sqrt{\varepsilon - k t^2} Q }{\varepsilon [1 + \omega_2(t) Q]} \right] \, . \label{density-T-v=1}
			\end{eqnarray}
Moreover, from the expression (\ref{chi-Tmodels}, \ref{ABCcal}) of the indicatrix function of a T-model, we obtain that the square of the speed of sound is given by
			\begin{equation} \label{chi-Tmodels-v=1}
				\chi(\rho,p) = \frac{8 p^2 (1 - \frac{\varepsilon}{\kappa} \sqrt{\frac{k \varepsilon}{p}})}{\rho^2 + 4 \rho \, p (1 - \frac{\varepsilon}{\kappa} \sqrt{\frac{k \varepsilon}{p}})+ p^2 (3 - 4 \frac{\varepsilon}{\kappa} \sqrt{\frac{k \varepsilon}{p}})} .
			\end{equation}
\\ \\
		\begin{figure}[]
			\centerline{
			\parbox[c]{0.5\textwidth}{\includegraphics[width=0.49\textwidth]{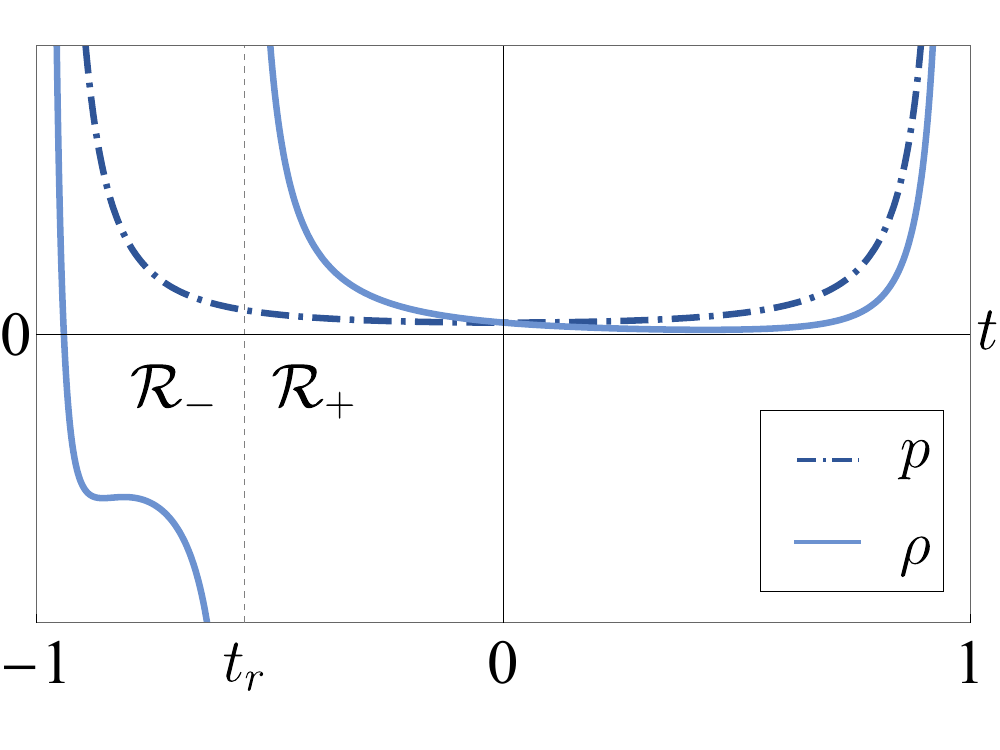}\\[-2mm] \centering{(a)}}
			\hspace{10pt}
			\parbox[c]{0.5\textwidth}{\includegraphics[width=0.49\textwidth]{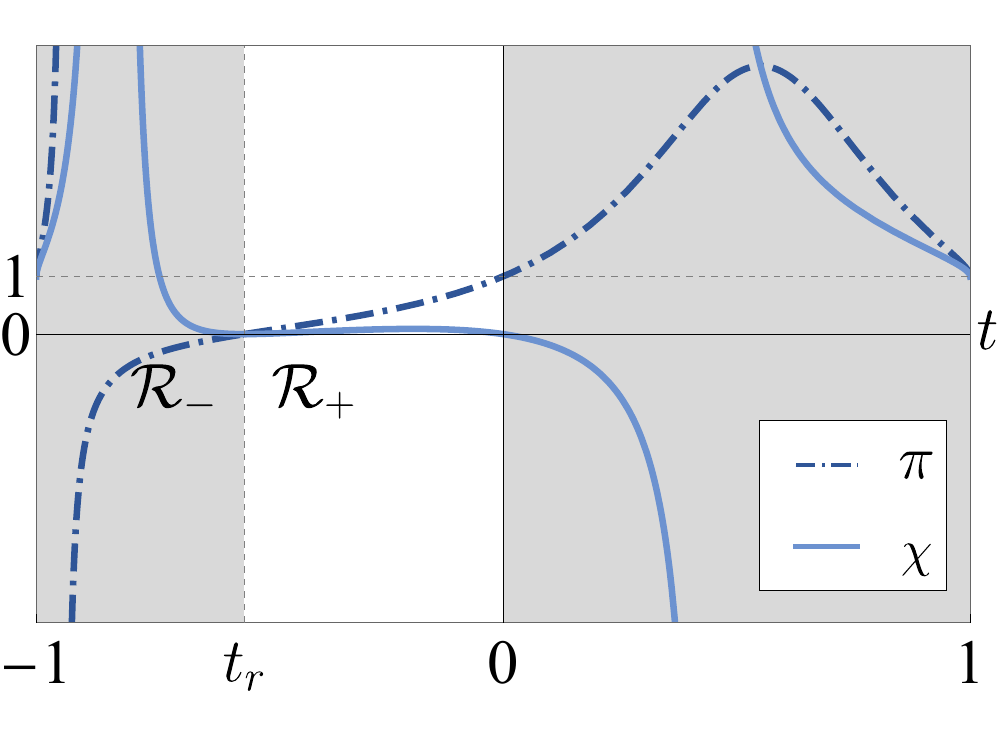}\\[-2mm] \centering{(b)}}}
			\vspace{-2mm}
			\caption{This figure shows the behaviour of the hydrodynamic quantities of the spherically symmetric T-model defined by the functions (\ref{phi-knot0}, \ref{omega-k=1}). We have considered the case $Q(r) = Q_r > 0$. The case $Q_r < 0$ follows by changing $t$ by $-t$. On (a), we have plotted the dependence on time of the energy density $\rho$ and the pressure $p$. Note that $\rho$ is positive in the whole region ${\cal R}_+$. On (b), we have plotted the quotient $\pi = p/\rho$, and the square of the speed of sound $\chi = u(p)/u(\rho)$. Notice that the energy conditions ($0 < \pi < 1$), and the causal sonic condition ($0 < \chi < 1$) only hold in the subregion $]t_r, 0[$ of the region ${\cal R}_+$ (unshaded interval).}
			\label{Fig-3}
		\end{figure}
			$\hspace{-3mm}$The analysis of these solutions shows that their good physical behaviour is constrained to limited spacetime domains:
			\begin{itemize}
\item[(i)]
The spherically symmetric case, $\omega_2(t) = \arcsin t$, $t \in[-1, 1]$, leads to a positive pressure everywhere. The metric has a curvature singularity at $t = t_r \equiv - \sin [1/Q(r)]$ that disconnects two spacetime regions ${\cal R}_-$ ($t < t_r$) and ${\cal R}_+$ ($t > t_r$). In the spacetime domain where $Q(r) > 0$ (respectively, $Q(r) < 0$) the energy density is positive in the region ${\cal R}_+$, (respectively, ${\cal R}_-$), as Figure \ref{Fig-3}(a) shows. Moreover, there is always a spacetime domain in which the macroscopic conditions for physical reality hold (see Figure \ref{Fig-3}(b)). 
\item[(ii)]
The case $\omega_2(t) = \textrm{arcsinh} \, t$ leads to a negative pressure everywhere. Moreover, whatever the values of $Q(r)$, the energy conditions and the compressibility conditions do not hold simultaneously for any value of $t$.
\item[(iii)]
The case $\omega_2(t) = \textrm{arcosh} \, t$ leads to a positive pressure everywhere. Moreover, in the domain where $Q(r) < 0$, the energy conditions and the compressibility conditions hold simultaneously in an interval of time $t \in [1, t_1[$, $t_1 < t_r \equiv \cosh [1/Q(r)]$. The metric has a curvature singularity at $t = t_r$. 
			\end{itemize}
		
		\subsection{Integration algorithms} \label{sec-integracio-1}
	
			\subsubsection{The Herlt algorithm} \label{subsec-Herlt}
			The field equation (\ref{eq-T-1}) is a first order linear differential equation for the metric function $v(t)$ that can be written as
			\begin{equation} \label{eq-Herlt}
			A \dot{v} + 2 \dot{A} v - 2 k \omega = 0 \, , \qquad A \equiv \varphi \dot{\omega} - \frac12 \dot{\varphi} \omega \, .
			\end{equation}
			Herlt \cite{Herlt} proposed an integration algorithm based on this fact. He considers the spherically symmetric case, chooses the time coordinate as $t = Y = \sqrt{\varphi}$ and he establishes the following steps (that we report with our notation):
			\begin{itemize}
\item[h1]
Choose an arbitrary function $\omega_2(t)$.
\item[h2]
Set equation (\ref{eq-Herlt}) for $v(t)$ by taking $\omega = \omega_2(t)$ and $\varphi = t^2$, and obtain the general solution $v(t)$. 
\item[h3]
Set equation (\ref{eq-T-1}) for the function $\omega(t)$, by taking $\varphi = t^2$ and $v(t)$ the function obtained in step h2, and obtain a particular solution $\omega_1(t)$. 
			\end{itemize}
			Herlt \cite{Herlt} remarked that steps h1 and h2 of his algorithm determine a homogeneous KCKS T-model, and step h3 completes a non-homogeneous solution. He applies this algorithm to obtain a non-homogeneous T-model from the homogeneous one presented by McVittie and Wiltshire \cite{McVittie} in which $\omega_2 = t^m$. \\ \\
			It is worth remarking that the Herlt algorithm provides the solution by quadratures. Indeed, two integrals determine the solution $v(t)$ of the non-homogeneous linear first order differential equation (\ref{eq-Herlt}). And, if we know a particular solution $\omega_2(t)$ of the homogeneous linear second order differential equation (\ref{eq-T-1}), then we can obtain another solution $\omega_1(t)$ with two indefinite integrals. \\ \\
			Now we revisit the Herlt algorithm and we show that: (i) it can be generalised to the plane and hyperbolic symmetries, (ii) it can be implemented without any specific choice of the time coordinate $t$, and (iii) it is only necessary to obtain two indefinite integrals to get the solution. \\ \\
			Let us take two arbitrary functions $\{\varphi(t), \omega_2(t)\}$, which fix the time coordinate $t$ and a solution of the field equations (for every $Q(r)$). Note that the general solution of the homogeneous equation associated with equation (\ref{eq-Herlt}) can be obtained without any integral, and it is $v_0(t) = C A^{-2}$. Then, the function $C(t) \equiv v(t) A^2(t)$ satisfies the differential equation $\dot{C} = 2 k \omega_2 A$, and therefore
			\begin{equation}
				C(t) = K_0 + 2k \! \int \! \omega_2(t) A(t) \textrm{d} t \, , \quad K_0 = constant \, .
			\end{equation}
Consequently, we have obtained $v(t)$ by performing a single quadrature. Furthermore, $\omega_1 = \omega_2 \zeta$ is an independent solution to the homogeneous linear equation (\ref{eq-T-1}) if, and only if, function $\zeta(t)$ is non-constant and fulfils the second order differential equation
			\begin{equation}
				2 \frac{\ddot{\zeta}}{\dot{\zeta}} + 4 \frac{\dot{\omega}_2}{\omega_2} + \frac{(\varphi v)^{\cdot}}{\varphi v} = 0 \, .
			\end{equation}
This equation is equivalent to $\dot{\zeta}^2 \omega_2^4 \varphi v = K_1^2$, where $K_1$ is a non-vanishing constant. Consequently, we obtain $\zeta(t)$ (and then $\omega_1(t)$) by taking a single quadrature:
			\begin{equation}
				\zeta(t) = \pm K_1 \! \int \! \frac{\textrm{d} t}{\omega_2^2(t) \sqrt{\varphi(t) v(t)}} \, .
			\end{equation}
Note that, being $Q(r)$ an arbitrary function, $\zeta(t)$ can be redefined by an arbitrary additive constant. Following this line of reasoning, we arrive to the following way of applying the Herlt algorithm: 
			\begin{itemize}
\item[H1]
Choose two arbitrary functions $\{\varphi(t), \omega_2(t)\}$, and obtain the function $A(t) \equiv \varphi(t) \, \dot{\omega}_2(t) \, - \, \frac12 \dot{\varphi}(t) \, \omega_2(t)$. 
\item[H2]
Determine the indefinite integral
			\begin{equation}
				H(t) = \! \! \int \! \! \omega_2(t) A(t)\, \textrm{d} t \, , 
			\end{equation}
and obtain the metric function
			\begin{equation}
				v(t) = \frac{1}{A^2(t)} [K_0 + 2k H(t)] \, .
			\end{equation}
\item[H3]
Determine the indefinite integral
			\begin{equation}
				\zeta(t) = \! \int \! \frac{A(t) \textrm{d} t}{\omega_2^2(t) \sqrt{\varphi(t)}\sqrt{K_0 + 2 k H(t)}} \, ,
			\end{equation}
and obtain the metric function
			\begin{equation}
				\omega(t,r) = \omega_2(t)[\zeta(t) + Q(r)] \, ,
			\end{equation}
			\end{itemize}
where $Q(r)$ is an arbitrary real function. Then, the metric functions $\{\varphi(t), \omega(t,r), v(t)\}$ define a T-model (\ref{metric-T-1}) that is a solution of the field equation (\ref{eq-T-1}). \\ \\
			Note that this algorithm allows us to solve the field equation by quadratures. Nevertheless, only in a few cases the indefinite integrals can be calculated to obtain an explicit expression of the solution. For example, as explained in Section \ref{sec-McVittie}, Herlt \cite{Herlt} considered $\varphi = t^2$ and $\omega_2 = t^m$ in the spherically symmetric case $k = 1$. The second step in the above algorithm determines the metric function $v(t)$ (\ref{v-MWH}), which corresponds to the homogeneous solution by McVittie and Wiltshire \cite{McVittie}. The third step, which determines the function $\zeta(t)$, cannot be explicitly achieved for an arbitrary value of the constant $K_0$. When $K_0 = 0$ we obtain an inhomogeneous solution with $\zeta(t) = t^{-2m}$. This is the MWH solution analysed in Section \ref{sec-McVittie}. \\ \\
			In the following sections, we look for other algorithms alternative to the Herlt one that may allow us to obtain new T-model solutions.
		
			\subsubsection{Field equations for the variables $(\varphi, \alpha, v)$}
			Let us consider the function $\alpha(t,r)$ defined by the condition $\omega = \alpha \sqrt{\varphi}$. Then, in terms of the metric functions $\{\varphi, \alpha, v\}$, the metric tensor (\ref{metric-T-1}) becomes
			\begin{equation} \label{metric-Tsol-2}
				ds^2 = -\frac{1}{v(t)}dt^2 + \varphi(t)[\alpha^2(t,r) dr^2 + C^2 (dx^2+dy^2)],
			\end{equation}
where $C$ is given in (\ref{metric-ss-3}). Moreover, the field equation (\ref{eq-T-1}) takes the expression
			\begin{equation} \label{eq-Tsol-2}
				2 v \varphi \, \ddot{\alpha} + (\dot{v} \varphi + 3 v \dot{\varphi})\, \dot{\alpha} - 2k \alpha = 0 \, .
			\end{equation}
			On the other hand, the pressure keeps the expression (\ref{pressure-T-1}), and the expansion (\ref{expansion-T-1}) and the energy density (\ref{density-T-1}) become
			\begin{equation} \label{expansion-Tsol-2}
				\theta = \sqrt{v} \left(\frac32 \frac{\dot{\varphi}}{\varphi} + \frac{\dot{\alpha}}{\alpha}\right) = \sqrt{v} \, \partial_t [\ln(\varphi^{3/2} \alpha)] \, , 
\\[1mm]
			\end{equation}
			\begin{equation} \label{density-Tsol-2}
				\rho = v \left[\frac34 \frac{\dot{\varphi}^2}{\varphi^2} + \frac{\dot{\varphi}}{\varphi} \frac{\dot{\alpha}}{\alpha}\right] + \frac{k}{\varphi} \, .
			\end{equation}
			Note that (\ref{eq-Tsol-2}) is a non-homogeneous linear first order differential equation for both $v(t)$ and $\varphi(t)$, and a homogeneous linear second order differential equation for the function $\alpha(t,r)$. We then have
			\begin{equation} \label{alpha-1-2}
				\alpha(t,r) = \alpha_1(t) + \alpha_2(t) \, Q(r)\, ,
			\end{equation}
where $Q(r)$ is an arbitrary real function, and $\alpha_i(t)$ being two particular solutions to equation (\ref{eq-Tsol-2}). Notice that the two metric functions $\alpha_i(t)$ cannot both be constant. If one of them is, a redefinition of the arbitrary function $Q(r)$ and a coordinate rescaling allow us to relabel the functions so that $\alpha_2(t)$ is always non-constant. \\ \\ 
All things considered, we have that the four metric functions $\{\varphi(t), \alpha_i(t), v(t)\}$ are submitted to two differential equations and a constraint that fixes the time coordinate. Consequently, the space of solutions depends on an arbitrary real function depending on time, and another real function, $Q(r)$, depending on $r$.
		
			\subsubsection{The modified Herlt algorithm}
			Given two arbitrary functions $\{\varphi(t), \alpha_2(t)\}$, the general solution of the homogeneous equation associated with equation (\ref{eq-Tsol-2}) for $v(t)$ is $v_0(t) = D \varphi^{-3} \dot{\alpha}_2^{-2}$, $D$ being a constant. As a result, the function $D(t) \equiv v(t) \varphi^3(t) \dot{\alpha}_2^2(t)$ satisfies the differential equation $\dot{D} = 2 k \alpha_2 \dot{\alpha}_2 \varphi^2$ and, consequently, we can obtain $v(t)$ by performing only one quadrature. \\ \\
			Furthermore, $\alpha_1 = \alpha_2 \zeta$ is an independent solution to the homogeneous linear equation (\ref{eq-Tsol-2}) if, and only if, the function $\zeta(t)$ is non-constant and fulfils the same second order differential equation than in the Herlt algorithm, which now leads to $\dot{\zeta}^2 \alpha_2^4 \varphi^3 v = K_1^2$. Then, we obtain $\zeta(t)$ (and then $\alpha_1(t)$) through a single quadrature. \\
			The factor $\dot{\alpha}_2 \neq 0$ appears in the two functions that we must integrate to obtain the solution. Thus, it is now suitable to choose the time coordinate $t$ such that $\alpha_2(t) = t$. Then, following a similar line of reasoning to that in the Herlt algorithm, we arrive at the following integration algorithm: 
			\begin{itemize}
\item[A1]
Choose two arbitrary real functions $\{\varphi(t), Q(r)\}$. 
\item[A2]
Determine the indefinite integral
			\begin{equation}
				D(t) = \! \int \! t \, \varphi^2(t) \textrm{d} t \, , 
			\end{equation}
and obtain the metric function
			\begin{equation} \label{A-v}
				v(t) = \frac{1}{\varphi^3(t)} [K_0 + 2k D(t)] \, .
			\end{equation}
\item[A3]
Determine the indefinite integral
			\begin{equation} \label{A-gamma}
				\zeta(t) = \! \int \! \frac{\textrm{d} t}{t^2 \, \sqrt{K_0 + 2 k D(t)}} \, ,
			\end{equation}
and obtain the metric function
			\begin{equation}
				\alpha(t,r) = t[\zeta(t) + Q(r)] \, .
			\end{equation}
			\end{itemize}
			Then, the metric functions $\{\varphi(t), \alpha(t,r), v(t)\}$ define a T-model (\ref{metric-Tsol-2}) that is a solution of the field equation (\ref{eq-Tsol-2}). 
			Note that the steps A1 and A2 provide a particular homogeneous KCKS T-model, and step A3 completes the non-homogeneous solution, for which two indefinite integrals are necessary. \\ \\
			In the next section, we will determine the general solution for $k = 0$ making use of this algorithm. The case $k \neq 0$ requires further analysis in order to obtain the general solution without the need for any integration (see Section \ref{sec-solucio-knot=0} below). 
		
		\subsection[The general solution for $k = 0$]{The general solution for \boldmath$k = 0$} \label{sec-solucio-k=0}
	
			\subsubsection{Metric and hydrodynamic quantities}
			The explicit general solution for the plane symmetry can be obtained by using both the Herlt algorithm and the modified Herlt algorithm. The latter provides a more direct reasoning. Indeed, notice that, when $k = 0$, (\ref{A-v}) and (\ref{A-gamma}) imply, respectively, $v = K_0 \varphi^{-3}$ and $\zeta = - (\sqrt{K_0}\, t)^{-1}$. Then the arbitrary functions $\varphi(t)$ and $Q(r)$, and the spatial coordinates $\{r, x ,y\}$, can be redefined by a factor in such a way that the metric line element (\ref{metric-T-2}) becomes
			\begin{equation} \label{metric-T-k=0}
				ds^2 = -\varphi^3(t) dt^2 + \varphi(t)([t\, Q(r) + 1]^2 dr^2 + dx^2 + dy^2) \, .
			\end{equation}
The unit velocity of the fluid $u = \varphi^{-3/2} \, \partial_t$ has an expansion given by
			\begin{equation} \label{expansion-T-k=0}
				\theta = \frac{1}{\varphi^{3/2}} \left(\frac32 \frac{\dot{\varphi}}{\varphi} \!+\! \frac{Q}{t \, Q + 1}\right) \! = \! \frac{1}{\varphi^{3/2}} \partial_t [\ln(\varphi^{3/2} (t \, Q + 1)] \, .
			\end{equation}
And the pressure $p$ and the energy density $\rho$ are then given by 
			\begin{eqnarray} \label{pressure-T-k=0}
				p = \frac{1}{\varphi^{3}} \left[\frac74 \frac{\dot{\varphi}^2}{\varphi^2} - \frac{\ddot{\varphi}}{\varphi}\right] \, , \\[1mm]
				\rho = \frac{1}{\varphi^{3}} \left[\frac34 \frac{\dot{\varphi}^2}{\varphi^2} + \frac{\dot{\varphi} \, Q}{\varphi(t \, Q + 1)}\right] \, .
\label{density-T-k=0}
			\end{eqnarray}
			On the other hand, we can specify the indicatrix function (\ref{chi-Tmodels}) in this case by calculating the implicit functions of $p$ given in (\ref{ABCcal}) in terms of $\varphi(t)$ and its derivatives: 
			\begin{subequations} \label{ABCcal-02}
				\begin{align}
					{\cal A}(p) & \equiv \frac{4 \varphi^{10}}{\dot{\varphi}(35 \dot{\varphi}^3 + 4 \varphi^2 \dddot{\varphi}- 30\varphi\dot{\varphi} \ddot{\varphi})} , \\[1mm] 
					{\cal B}(p) & \equiv {\cal A}(p) \left[p + \frac{3 \dot{\varphi}^2}{4 \varphi^5}\right] , \qquad 
					{\cal C}(p) \equiv {\cal A}(p) \, p \, \frac{3 \dot{\varphi}^2}{4 \varphi^5} .
				\end{align} 
			\end{subequations}
\\ \\ \\ \\
			It is worth remarking that we can recover previously known T-models with plane symmetry by giving specific expressions of the function $\varphi(t)$:
			\begin{itemize}
\item[(i)]
If we take $\varphi(t) = t^{-4/3}$ we obtain the dust solution. This T-model was considered by Vajk and Eltgroth \cite{Vajk} for the homogeneous case ($Q = constant$). The proper time of the fluid is $\tau = -1/t$.
\item[(ii)]
If we take $\varphi(t) = t^m$, $m > -4/3$, $m \neq 0 $, we obtain the ideal T-models with $\tilde{\gamma} = \frac{4}{3m} + 2 \neq 2$ studied in Section \ref{sec-chi-pi}. The proper time of the fluid is $\tau = \frac{1}{\bar{m}} t^{\bar{m}}, \ \bar{m} \equiv 1 + \frac{3m}{2}$.
\item[(iii)]
If we take $\varphi(t) = e^{2t/3}$ we obtain the ideal T-models with $\tilde{\gamma} = 2$ studied in Section \ref{sec-chi-pi}. The proper time of the fluid is $\tau = e^t$. 
\end{itemize}

			\subsubsection{A new solution with $k = 0$}
			\begin{figure}[]
			\centerline{
			\parbox[c]{0.5\textwidth}{\includegraphics[width=0.49\textwidth]{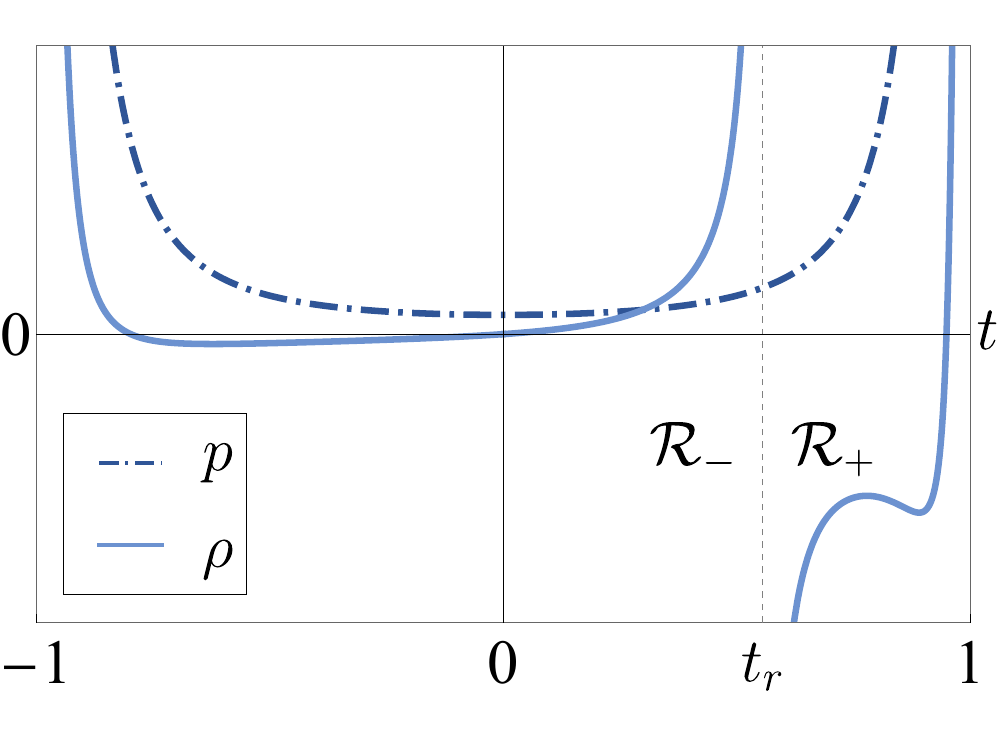}\\[-2mm] \centering{(a)}}
			\hspace{10pt}
			\parbox[c]{0.5\textwidth}{\includegraphics[width=0.49\textwidth]{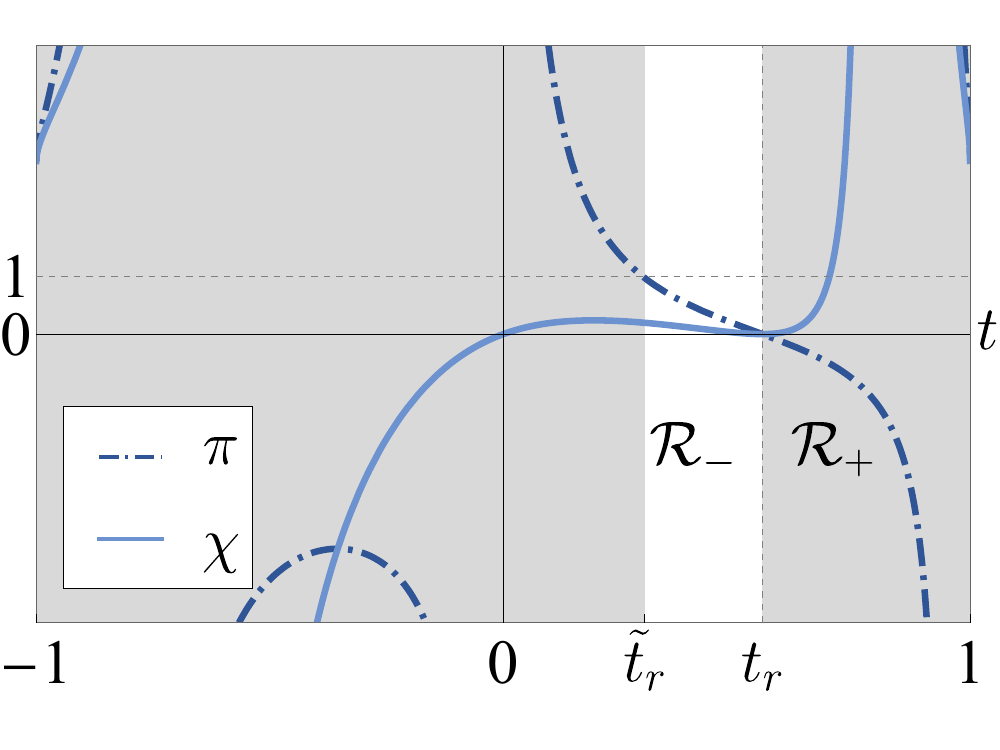}\\[-2mm] \centering{(b)}}}
			\vspace{-2mm}
			\caption{This figure shows the behaviour of the hydrodynamic quantities of the T-model with plane symmetry defined by the function (\ref{phi-k=0new}). We have considered the case $Q(r) = Q_r < 0$. The case $Q_r > 0$ follows by changing $t$ by $-t$. On (a), we have plotted the dependence on time of the energy density $\rho$ and the pressure $p$. Note that $\rho$ is positive in subregion $] \, 0, t_r[$ of the region ${\cal R}_-$. On (b), we have plotted the quotient $\pi = p/\rho$, and the square of the speed of sound $\chi = u(p)/u(\rho)$. Notice that the energy conditions ($0 < \pi < 1$), and the causal sonic condition ($0 < \chi < 1$) only hold in the subregion $]\tilde{t}_r, t_r[$ of the region ${\cal R}_-$, where $\tilde{t}_r$ is defined by the condition $\pi(\tilde{t}_r) = 1$ (unshaded interval).}
			\label{Fig-4}
			\end{figure}
			As an example illustrating how the above method to obtain the general solution for $k = 0$ works, we now obtain a new solution. We take
			\begin{equation} \label{phi-k=0new}
				\varphi(T) = \cos^{2/3}(\kappa T) \, .
			\end{equation}
			Then, we obtain a T-model with $k = 0$ if we replace this expression of $\varphi(T)$ in the metric line element (\ref{metric-T-k=0}). Moreover, we can analyse the physical behaviour of the solution taking into account the expressions (\ref{pressure-T-k=0}) and (\ref{density-T-k=0}) of the pressure and energy density, and the expression (\ref{chi-Tmodels}, \ref{ABCcal-02}) of the indicatrix function $\chi(\rho, p)$. Nevertheless, in this case we can easily obtain the proper time $\tau$ of the fluid. Indeed, we have $d \tau = \varphi^{3/2} d T = \cos(\kappa T) d T$ and, consequently, $\tau = \frac{1}{\kappa} \sin(\kappa T)$. Then, if we introduce the time
			\begin{equation}
				t = \kappa \tau = \sin(\kappa T) \in ]-1, 1[ \, ,
			\end{equation}
and we make use of the general expressions (\ref{pressure-T-1}, \ref{density-T-1}) for the hydrodynamic quantities, we obtain
			\begin{eqnarray} \label{pressure-T-k=0new}
				p = \frac{\kappa^2(2 + t^2)}{3(1 - t^2)^2} \, , \\[1mm]
				\rho = \frac{\kappa^2 t^2}{3(1 - t^2)^2}\left[1 - \frac{2\sqrt{1 - t^2} \, Q}{t(Q \, \textrm{arcsin} \, t + \kappa)}\right] \, .
\label{density-T-k=0new}
			\end{eqnarray}
			Note that this solution has a positive pressure everywhere, and the metric has a curvature singularity at $t = t_r \equiv \sin[-\kappa / Q(r)]$, which disconnects two spacetime regions ${\cal R}_-$($t < t_r$) and ${\cal R}_+$ ($t > t_r$). In the spacetime domain $Q(r) < 0$ (respectively, $Q(r) > 0$) the energy density is positive in the subregion $t \in \, ] \, 0,t_r[$ of ${\cal R}_-$ (respectively, $t \in \, ]\, t_r,0[$ of ${\cal R}_+$), as Figure \ref{Fig-4}(a) shows. Moreover, there is always a spacetime domain in which the macroscopic conditions for physical reality holds if, and only if, $|\frac{Q}{\kappa}| > \frac{2}{\pi}$ (see Figure \ref{Fig-4}(b)).
		
		\subsection[The general solution for $k \neq 0$]{The general solution for \boldmath$k \neq 0$} \label{sec-solucio-knot=0}
	
			\subsubsection{The field equation for the variables $(\varphi, \alpha, \beta)$}
			Now we introduce a new function $\beta(t)$ as unknown metric function. Let us define 
			\begin{equation} \label{beta}
				\beta(t) = v(t) \, \varphi^3(t) > 0 \, .
			\end{equation}
Then, the field equation (\ref{eq-Tsol-2}) becomes
			\begin{equation} \label{eq-T-beta}
				2 \beta \, \ddot{\alpha} + \dot{\beta}\, \dot{\alpha} - 2k\, \alpha \, \varphi^2 = 0 \, .
			\end{equation}
The solution $\alpha(t)$ to this equation is of the form (\ref{alpha-1-2}), where $Q(r)$ is an arbitrary real function, and $\alpha_i(t)$ being two particular solutions to equation (\ref{eq-T-beta}). A straightforward calculation shows that, if $\alpha_1(t)$ fulfils (\ref{eq-T-beta}), then another independent solution can be written as $\alpha_2(t) = \gamma(t) \alpha_1(t)$ where $\gamma(t)$ meets the equation
			\begin{equation} \label{eq-T-gamma}
				\dot{\gamma}^2 \, \alpha_1^4 \, \beta = 1 \, .
			\end{equation}
			It is worth remarking that (\ref{eq-T-beta}) is an algebraic equation for the function $\varphi(t)$. Consequently, $\varphi(t)$ can be obtained without quadratures in terms of $\beta(t)$ and $\alpha_1(t)$. This fact and equation (\ref{eq-T-gamma}) allow us to obtain the general solution for $k \neq 0$ without needing to calculate any integral. In the following, we develop two algorithms that determine this solution in terms of an arbitrary function of time.
		
			\subsubsection{The \boldmath$\gamma$-algorithm}
			Note that any solution $\alpha(t)$ to equation (\ref{eq-T-beta}) is a non-constant function when $k \neq 0$. Thus, we can take the time coordinate $t$ such that
			\begin{equation} \label{alpha1=t}
				\alpha_1(t) = t \, .
			\end{equation}
Then, equations (\ref{eq-T-beta}) and (\ref{eq-T-gamma}) become, respectively,
			\begin{equation} \label{eq-T-beta-gamma}
				\dot{\beta} = 2k \, t \, \varphi^2 \, , \qquad \dot{\gamma}^2 \, t^4 \, \beta = 1 \, .
			\end{equation}
From these expressions we can perform the following algorithm to obtain the general solution of the field equations:
			\begin{itemize}
\item[G1]
Choose two arbitrary real functions $\{\gamma(t), Q(r)\}$. 
\item[G2]
Determine the function 
			\begin{equation}
				\beta(t) = \frac{1}{t^4 \, \dot{\gamma}^2(t)} \, .
			\end{equation}
\item[G3]
Determine the metric functions 
			\begin{subequations}
				\begin{align}
					v(t) & = \frac{\beta(t)}{\varphi^3(t)} \, , \qquad 
\varphi(t) = \sqrt{\frac{|\dot{\beta}(t)|}{2 t}} \, , \\[1mm]
					\alpha(t,r) & = t \,[1 + \gamma(t) Q(r)]\, .
				\end{align}
			\end{subequations}
			\end{itemize}
Then, the triad $\{\varphi(t), \alpha(t,r), v(t)\}$ defines a T-model (\ref{metric-Tsol-2}) which is a solution of the field equation (\ref{eq-Tsol-2}) for
			\begin{itemize}
\item[-]
spherical symmetry, $k \! = \! + 1$, in the spacetime domain where $\dot{\beta}(t) > 0$, 
\item[-]
hyperbolic symmetry, $k \! = \! - 1$, in the spacetime domain where $\dot{\beta}(t) < 0$. 
			\end{itemize}
Moreover, if $\dot{\beta}(t_1) = 0$ the metric is singular at $t_1$. \\ \\
			We can recover previously known T-models with non-plane symmetry by giving specific expressions of the function $\gamma(t)$:
			\begin{itemize}
\item[(i)]
If we take $\gamma(t) = t^{-\frac{2m}{m - 1}}$, we obtain the McVittie-Wilshire-Herlt solution studied in Section \ref{sec-McVittie}. The proper time of the fluid is $\tau = \frac{[k(m - 1)]^{3/4}}{\sqrt{2m}}t^{\frac{1}{m - 1}}$.
\item[(ii)]
If we take $\gamma(t) \! = \! \arcsin \! \sqrt{1 \! - \! \frac{1}{\kappa^2 t^2}}$, we obtain the spherically symmetric model (\ref{phi-knot0}, \ref{omega-k=1}) obtained in Section \ref{subsec-v=1-knot=0}. The proper time of the fluid is $\tau = \sqrt{\kappa^2 \! - \! \frac{1}{t^2}}$.
\item[(iii)]
If we take $\gamma(t) \! = \! \textrm{arcsinh} \sqrt{\frac{1}{\kappa^2 t^2} \! - \! \ 1}$, we obtain the hyperbolically symmetric model (\ref{phi-knot0}, \ref{omega-k=-1+}) obtained in Section \ref{subsec-v=1-knot=0}. The proper time of the fluid is $\tau = \sqrt{\frac{1}{t^2} \! - \! \kappa^2} $.
\item[(iv)]
If we take $\gamma(t) \! = \! \textrm{arccosh} \sqrt{\frac{1}{\kappa^2 t^2} \! + \! \ 1}$, we obtain the hyperbolically symmetric model (\ref{phi-knot0}, \ref{omega-k=-1-}) obtained in Section \ref{subsec-v=1-knot=0}. The proper time of the fluid is $\tau = \sqrt{\frac{1}{t^2} \! + \! \kappa^2}$.
\end{itemize}

			\subsubsection{The \boldmath$\xi$-algorithm}
			If $\alpha_i(t)$ are two independent solutions to equation (\ref{eq-T-beta}), then $\alpha_2 (t) = \alpha_1(t) \gamma(t)$, with $\dot{\gamma} \neq 0$. Thus, we can take the time coordinate $t$ such that
			\begin{equation} \label{gamma=t}
				\gamma(t) = t \, .
			\end{equation}
Then, if $\xi (t) = 1/\alpha_1(t)$, equations (\ref{eq-T-beta}) and (\ref{eq-T-gamma}) become, respectively,
			\begin{equation} \label{eq-T-beta-xi}
				\beta = \xi^4 \, , \qquad \xi^3 \ddot{\xi} + k \varphi^2 = 0 \, .
			\end{equation}
From these expressions we can perform the following algorithm to obtain the general solution of the field equations:
			\begin{itemize}
\item[X1]
Choose two arbitrary real functions $\{\xi(t), Q(r)\}$. 
\item[X2]
Determine the metric functions 
			\begin{equation}
					v(t) \! = \! \frac{1}{\sqrt{\xi(t)\, |\ddot{\xi}(t)|^3}} , \quad \; \varphi(t) \! = \! \sqrt{\xi^3(t)\, |\ddot{\xi}(t)|} , \quad \; \alpha(t,r) \! = \! \frac{1 + t\, Q(r)}{\xi(t)}\, .
			\end{equation}
			\end{itemize}
Then, the triad $\{\varphi(t), \alpha(t,r), v(t)\}$ defines a T-model (\ref{metric-Tsol-2}) which is a solution of the field equation (\ref{eq-Tsol-2}) for
			\begin{itemize}
\item[-]
spherical symmetry, $k \! = \! + 1$, in the spacetime domain where $\xi(t) \, \ddot{\xi}(t) < 0$, 
\item[-]
hyperbolic symmetry, $k \! = \! - 1$, in the spacetime domain where $\xi(t) \, \ddot{\xi}(t) > 0$. 
			\end{itemize}
Moreover, if $\ddot{\xi}(t_1) = 0$ the metric is singular at $t_1$. \\ \\
			We can recover previously known T-models with non-plane symmetry by giving specific expressions of the function $\xi(t)$:
			\begin{itemize}
\item[(i)]
If we take $\xi(t) = t^{\frac{m - 1}{2m}}$, we obtain the McVittie-Wilshire-Herlt solution studied in Section \ref{sec-McVittie}. The proper time of the fluid is $\tau = -\frac{[k(m - 1)]^{3/4}}{\sqrt{2m}}t^{-\frac{1}{2m}}$.
\item[(ii)]
If we take $\xi(t) = \kappa \cos t$, we obtain the spherically symmetric model (\ref{phi-knot0}, \ref{omega-k=1}) obtained in Section \ref{subsec-v=1-knot=0}. The proper time of the fluid is $\tau = \kappa \sin t$.
\item[(iii)]
If we take $\xi(t) = \kappa \cosh t$, we obtain the hyperbolically symmetric model (\ref{phi-knot0}, \ref{omega-k=-1+}) obtained in Section \ref{subsec-v=1-knot=0}. The proper time of the fluid is $\tau = \kappa \sinh t$.
\item[(iv)]
If we take $\xi(t) = \kappa \sinh t$, we obtain the hyperbolically symmetric model (\ref{phi-knot0}, \ref{omega-k=-1-}) obtained in Section \ref{subsec-v=1-knot=0}. The proper time of the fluid is $\tau = \kappa \cosh t$.
			\end{itemize}

			\subsubsection{A new spherically symmetric solution}
\begin{figure}[t]
			\centerline{
			\parbox[c]{0.5\textwidth}{\includegraphics[width=0.49\textwidth]{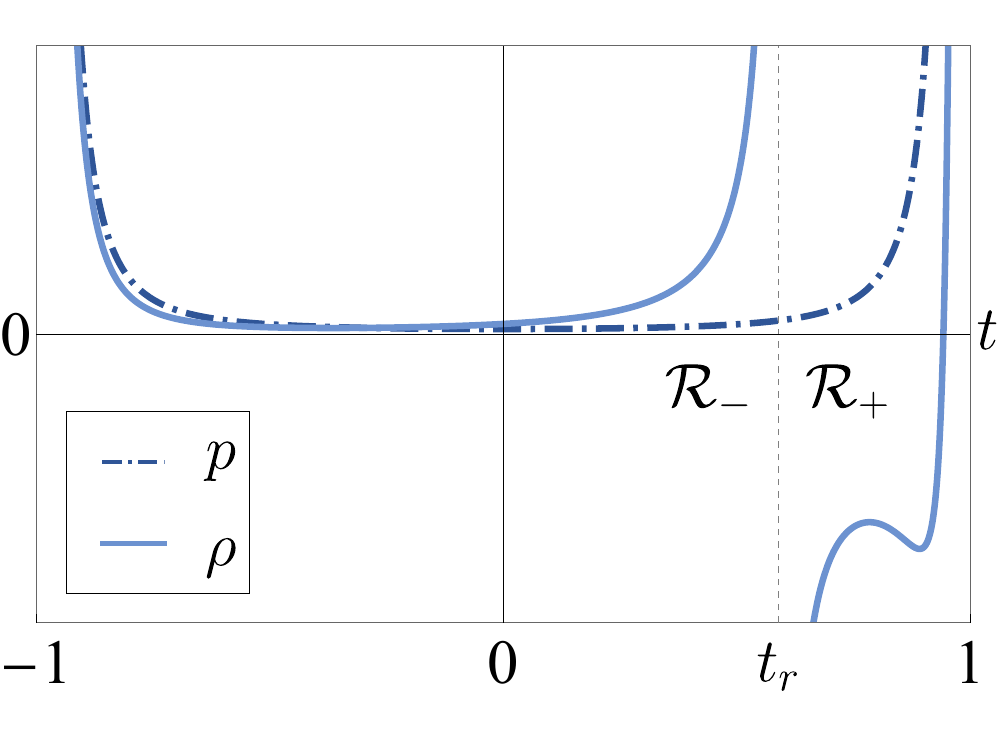}\\[-2mm] \centering{(a)}}
			\hspace{18pt}
			\parbox[c]{0.5\textwidth}{\includegraphics[width=0.49\textwidth]{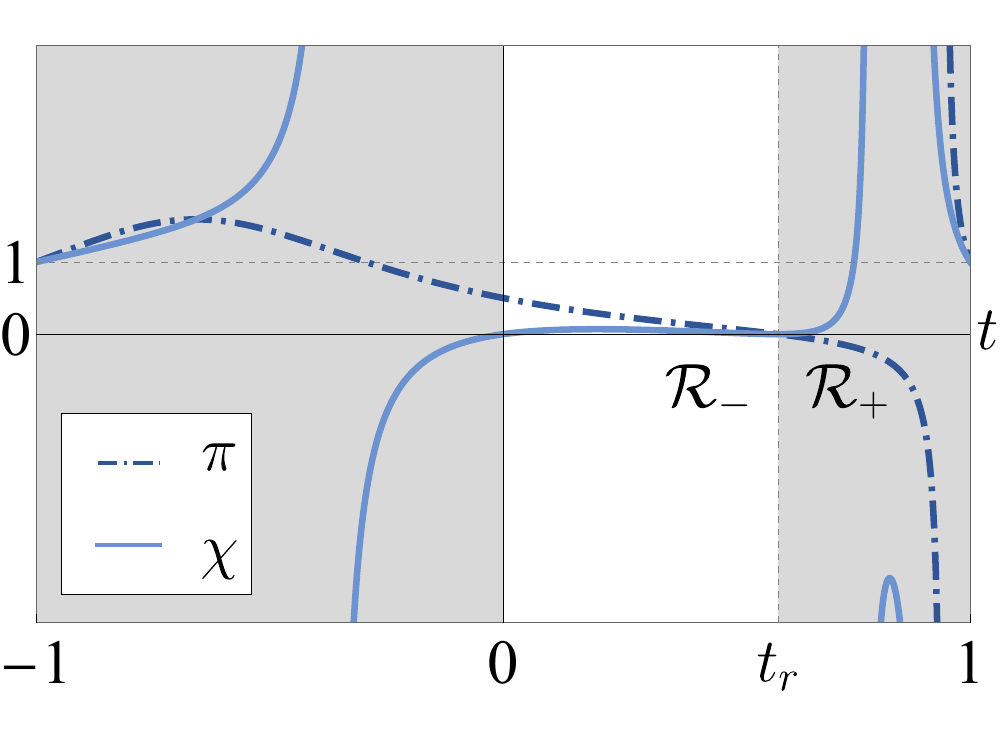}\\[-2mm] \centering{(b)}}}
			\vspace{-2mm}
			\caption{This figure shows the behaviour of the hydrodynamic quantities of the T-model with spherical symmetry determined by the $\xi$-algorithm with $\xi = 1 - t^2$. We have considered the case $Q(r) = Q_r < 0$. The case $Q_r > 0$ follows by changing $t$ by $-t$. On (a), we have plotted the dependence on time of the energy density $\rho$ and the pressure $p$. Note that $\rho$ is positive in the whole region ${\cal R}_-$ and in a part of the region ${\cal R}_+$. On (b), we have plotted the quotient $\pi = p/\rho$, and the square of the speed of sound $\chi = u(p)/u(\rho)$. Notice that the energy conditions ($0 < \pi < 1$), and the causal sonic condition ($0 < \chi < 1$) only hold in the subregions $]0,t_r[$ (unshaded interval).}
			\label{Fig-5}
			\end{figure}

			Now we consider an example to see how the above algorithms to obtain the general solution for $k \neq 0$ work. We take
			\begin{equation} \label{xi-knot0new}
				\xi(t) = 1 - t^m \, .
			\end{equation}
Then, we get a T-model with $k \neq 0$ if we apply the $\xi$-algorithm. We have that for any $m$ out of the range $[0,1]$ the solution is spherically symmetric ($\xi(t) \, \ddot{\xi}(t) < 0$) in a spacetime domain. Moreover, we can analyse the physical behaviour of the solutions taking into account the general expressions (\ref{pressure-T-1}) for the pressure and (\ref{density-Tsol-2}) for the energy density, and the expression (\ref{chi-Tmodels}, \ref{ABCcal}) of the indicatrix function $\chi(\rho, p)$. The metric has a curvature singularity at $t = t_r \equiv -1/Q(r)$, which disconnects two spacetime regions ${\cal R}_-$ ($t < t_r$) and ${\cal R}_+$ ($t > t_r$). \\ \\
			For sake of simplicity, we now focus on the case $m = 2$. Then, the solution is spherically symmetric in the interval $t \in \, ]-1,1[$, and the pressure and the energy density take the expressions
			\begin{eqnarray} \label{pressure-T-knot=0new}
				p = \frac{4 - t^2}{8 \sqrt{2}(1 - t^2)^{5/2}} \, , \quad \\[1mm]
				\rho = \frac{7Q \, t^3 - 5t^2 - 4Q \, t + 8}{8 \sqrt{2}(1 - t^2)^{5/2}(1 + Q \, t)} \, . \quad
\label{density-T-knot0new}
			\end{eqnarray}
Note that this solution has a positive pressure everywhere, and when $Q(r) < 0$ (respectively, $Q(r) > 0$) the energy density is positive in the region ${\cal R}_-$ (respectively, in ${\cal R}_+$) and in a part of the region ${\cal R}_+$ (respectively, in ${\cal R}_-$), as Figure \ref{Fig-5}(a) shows. Moreover, there is always a spacetime domain in which the macroscopic conditions for physical reality hold (see Figure \ref{Fig-5}(b)).

		\subsection{Some remarks} \label{subsec-remarks-T-models-solution}
		The Herlt and our modified Herlt algorithms provide a (particular) homogeneous solution with a quadrature, and the general solution of the non-homogeneous case follows by obtaining another indefinite integral. The general solution of the homogeneous T-models corresponds with the non-homogeneous one for the case $Q(r) = constant$. Hence, our study also provides the general solutions of the KCKS T-models. \\ \\
		The Szekeres-Szafron solutions of class II \cite{Krasinski-Plebanski, Inhomogeneous_Cosmological_Models, Szekeres, Szafron, FS-SS} are a generalisation without symmetries of the T-models. As already mentioned in Section \ref{sec-Tmodels}, a thermodynamic analysis of these solutions shows \cite{C-F-S_SzSz_Singular} that three subfamilies in l.t.e. can be considered: the singular models, the regular models and the T-models. The latter are the object of the present chapter. The Szekeres-Szafron singular and regular models have been studied in \cite{C-F-S_SzSz_Singular} and \cite{C-F-S_SzSz_Regular}, respectively. In both cases, the metric line element and the field equation are similar to those of the T-models, with the function $Q(r)$ replaced by 
		\begin{equation}
			\tilde{Q}(r,x,y) = \frac12 U(r) (x^2 + y^2) + V_1 (r) x + V_2 (r) y + 2\,W(r) ,
		\end{equation}
where $V_1 (r) , V_2 (r), W(r) $ are arbitrary real functions, and $U(r) + k W(r) = 1$ for the regular models, and $U(r) = 0$ and $k = 0$ for the singular models. Consequently, all the integration algorithms obtained in this section for the T-models also apply for the thermodynamic class II Szekeres-Szafron solutions (singular and regular models).

	\section[Physical interpretation of the Kompaneets$\,$-Chernov$\,$-Kantowski$\,$- \\ Sachs solution]{Physical interpretation of the Kompaneets$\,$- \\ Chernov$\,$-Kantowski$\,$-Sachs solution} \label{chap-physical interpretation KS}
	Kompaneets-Chernov-Kantowski-Sachs (KCKS) solutions are homogeneous but anisotropic cosmological solutions to the Einstein equations. Some physical properties of these solutions where investigated by Kompaneets and Chernov \cite{Kompaneets-Chernov} and they were studied with more detail by Kantowski and Sachs for the case of dust \cite{Kantowski-Sachs}. \\ \\ 
	The subfamily of spherically symmetric solutions are spatially homogeneous metrics that do not permit a simply transitive group of motions, and the subspaces of constant time coordinate do not contain the centres of symmetry. However, the counterparts with plane and hyperbolic symmetry are of Bianchi types I and III, respectively, so they do have simply transitive three-dimensional subgroups \cite{Inhomogeneous_Cosmological_Models, Krasinski-Plebanski, Ellis-Maartens-MacCallum, Kramer}. \\ \\
As mentioned in Section \ref{Termo_T-models}, the KCKS metrics are the spatially homogeneous limit of the T-models (\ref{metric-T-1}$-$\ref{density-T-1}). They can also be characterised by one of the following three equivalent conditions: (i) the metric function $\omega(t,r)$ factorises, and then one can take the coordinate $r$ so that $\omega = \omega(t)$, (ii) the energy density is homogeneous, $\rho = \rho(t)$, and (iii) the fluid expansion is homogeneous, $\theta = \theta(t)$. \\ \\
Many works have been devoted to studying the behaviour of KCKS metrics and have used it to describe the Universe before a period of rapid expansion (see \cite{Oliveira-Neto} and references therein). However, its interpretation as a physically realistic fluid is still an open problem and several authors have pointed out the difficulties in associating a realistic equation of state to these solutions and their non-homogeneous generalisations, the T-models and the class II Szekeres-Szafron solutions \cite{Lima-Tiomno-a, Bolejko}. Nevertheless, although many of the solutions already studied have no clear physical meaning \cite{Vajk, McVittie, Herlt}, our results of previous sections and those in \cite{C-F-S_SzSz_Singular, C-F-S_SzSz_Regular} show that these families can admit meaningful thermodynamic interpretations. \\ \\ \\ \\
	Indeed, in the previous section we mentioned that the KCKS solutions may be interpreted as representing an isentropic evolution of a thermodynamic T-model. Here, we will study in detail such interpretation of the KCKS solutions to Einstein's equations in Section \ref{sec-KCKS-T-models}. Then, in Section \ref{sec-KCKS-other} we show how to study other potential interpretations.

		\subsection{Isentropic limit of a T-model} \label{sec-KCKS-T-models}
\begin{figure}[]
				\centerline{
				\parbox[c]{0.33\textwidth}{\includegraphics[width=0.33\textwidth]{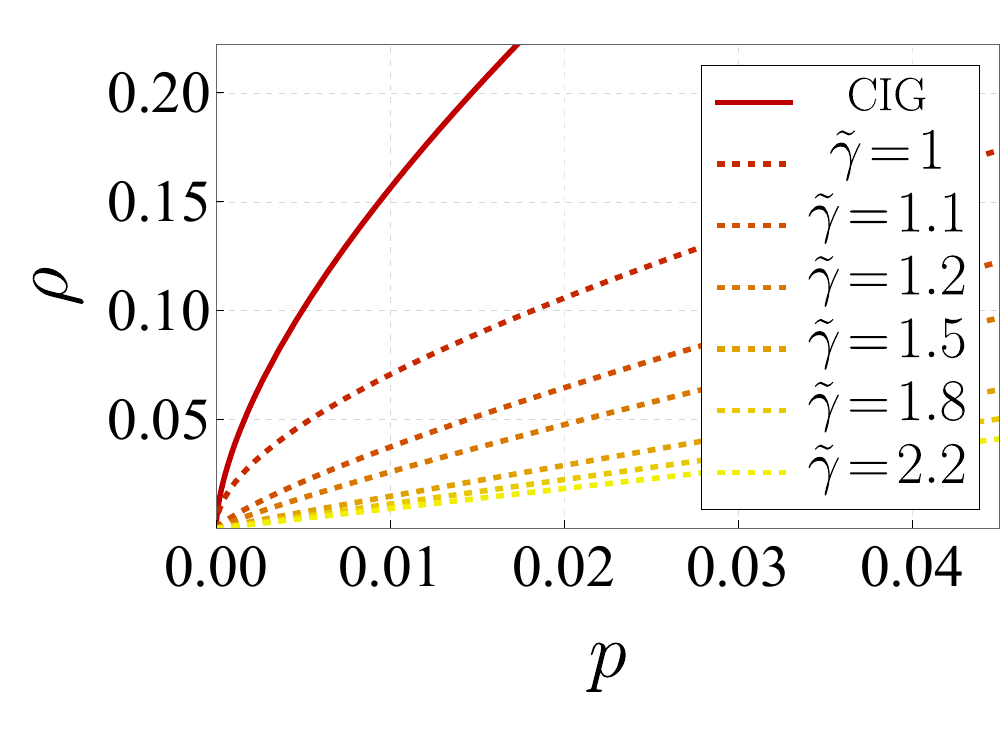}\\[-1mm] \centering{(a)}}
				\hspace{2pt}
				\parbox[c]{0.33\textwidth}{\includegraphics[width=0.33\textwidth]{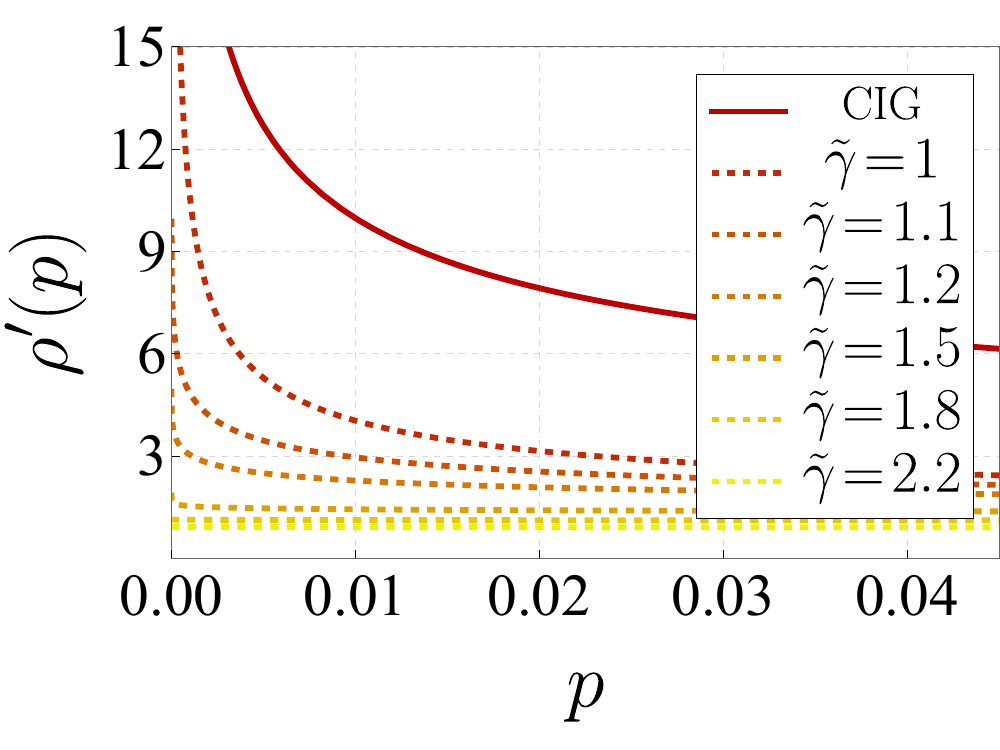}\\[-1mm] \centering{(b)}}
				\hspace{1pt}
				\parbox[c]{0.33\textwidth}{\includegraphics[width=0.33\textwidth]{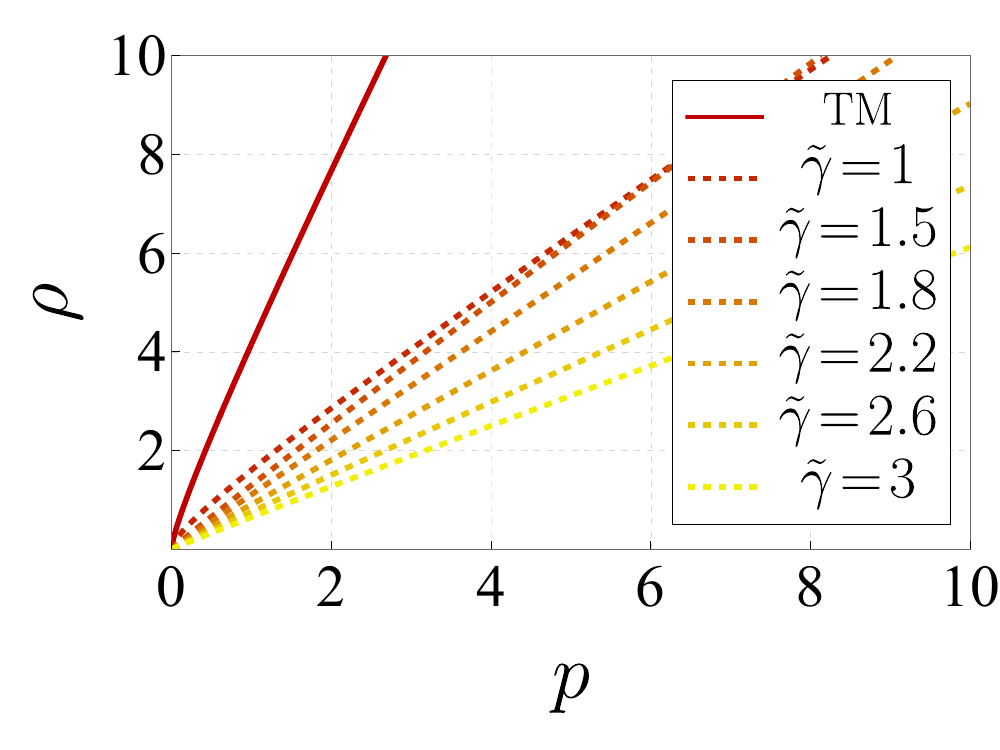}\\[-1mm] \centering{(c)}}}
				\vspace{-2mm}
				\caption{Comparative of (a) the barotropic relation $\rho = \rho (p)$ and (b) its first derivative, for a classical ideal gas with $\gamma = 5/3$ and for an ideal T-model in the isentropic KCKS limit for different values of $\tilde{\gamma}$. (c) Comparative of the barotropic relation $\rho = \rho (p)$ of an isentropic Taub-Matheus (TM) ultra-relativistic fluid and of an ideal T-model in the isentropic KCKS limit for different values of $\tilde{\gamma}$.}
				\label{Fig-KCKS1}
			\end{figure}

		In Section \ref{sec-thermo-Tmodels}, we have seen that each election of the arbitrary metric function $Q(r)$ defines a different particular T-model. Each of these particular solutions, in turn, has an associated set of thermodynamic schemes, given by (\ref{Q(prho)}$-$\ref{s-n-Tmodels}) and (\ref{T-Tmodels}$-$\ref{ell-m}). The different thermodynamic schemes correspond to the different elections of the arbitrary functions $s(Q)$ and $N(Q)$. \\ \\
		Now, the isentropic limit of these thermodynamic schemes is achieved by making $Q = Q_0$ constant, which therefore corresponds to the spatially homogeneous KCKS limit. Consequently, the KCKS solutions can be interpreted as the isentropic limit of the T-models. Each of these isentropic limits is determined by a different value of the constant $Q_0$, and their thermodynamic interpretation is given by families of thermodynamic schemes with four parameters $\lbrace N_0$, $s_0$, $N'_0$, $s'_0 \rbrace$, where $N_0 \equiv N(Q_0)$, $s_0 \equiv s(Q_0)$, $N'_0 \equiv N'(Q_0)$ and $s'_0 \equiv s'(Q_0)$. \\ \\
	In order to exemplify this discussion, let us consider the subfamily of the T-models compatible with the equation of state of a generic ideal gas studied in Section \ref{sec-chi-pi}. The first row of Table \ref{table-3} gives the general expression of the thermodynamic scheme for this case, which can be written as
		\begin{eqnarray} \label{n-Theta-ideal}
			n(\rho, p) \!\!\!\!\! &=& \!\!\!\!\! \rho \, (1 - \pi)^{\frac{\tilde{\gamma}}{2 - \tilde{\gamma}}} \left(\frac{\tilde{\gamma} - 1}{\tilde{\gamma} -1 - \pi}\right)^{\frac{2(\tilde{\gamma} - 1)}{2 - \tilde{\gamma}}} , \qquad \Theta(\rho, p) = \frac{p}{\tilde{k} \, n(\rho, p)} \, ; \\[-1mm]
			\label{s-ideal}
			s(\rho, p) \!\!\!\!\! &=& \!\!\!\!\! \bar{s}_0 + \tilde{k} \ln \left[\frac1p \left(\frac{p - \rho \, (\tilde{\gamma} - 1)}{\rho - p}\right)^{\frac{2\tilde{\gamma}}{2 - \tilde{\gamma}}}\right] \, ,
		\end{eqnarray}
where we have set $e_0$ so that $e(0) = 1$. Therefore, the spatially homogeneous KCKS limit, achievable by setting $s(\rho, p) = s_0$, yields in this case the barotropic relation
		\begin{equation} \label{rel-barotropia-T-model-ideal-isentropic}
			\rho = p \frac{1 + \tilde{B} \, p^{\frac{2 - \tilde{\gamma}}{2\tilde{\gamma}}}}{(\tilde{\gamma} - 1) + \tilde{B} \, p^{\frac{2 - \tilde{\gamma}}{2\tilde{\gamma}}}} \equiv \rho (p) \, ,
		\end{equation}
with $\tilde{B} \equiv \exp\left\lbrace\frac{s_0 - \bar{s}_0}{\tilde{k}} \frac{2 - \tilde{\gamma}}{2\tilde{\gamma}}\right\rbrace$. \\ \\
		This barotropic relation behaves, at low temperatures, as that of a classical ideal gas (\ref{CIG-isentropic}) at zero order (see Figure \ref{Fig-KCKS1}(a)). At first order, it approaches better that of a classical ideal gas with $\gamma = 5/3$ the closer $\tilde{\gamma}$ is to $1$, but never reaches its behaviour (see Figure \ref{Fig-KCKS1}(b)). At high temperatures however, it does not behave at all as the barotropic relation of an isentropic Taub-Matheus ultra-relativistic fluid, given in (\ref{TM-barotrop}) (see Figure \ref{Fig-KCKS1}(c)).

		\subsection{Other interpretations} \label{sec-KCKS-other}
		In the previous section, we have seen that a KCKS solution can represent the isentropic evolution of a T-model. However, it might also be the case that these solutions represent the isentropic limit of other families of solutions. \\ \\ \\ \\
		The isentropic evolutions of thermodynamic solutions always follow a barotropic relation $p = p(\rho, s_0)$. Therefore, by imposing specific barotropic relations, we can obtain the KCKS subfamilies representing the isentropic evolution of more general non-barotropic families of solutions fulfilling such barotropic relations along the evolution.
	
			\subsubsection{Isentropic evolution of a Classical Ideal Gas}
\begin{figure}[H]
			\centerline{
			\parbox[c]{0.5\textwidth}{\includegraphics[width=0.49\textwidth]{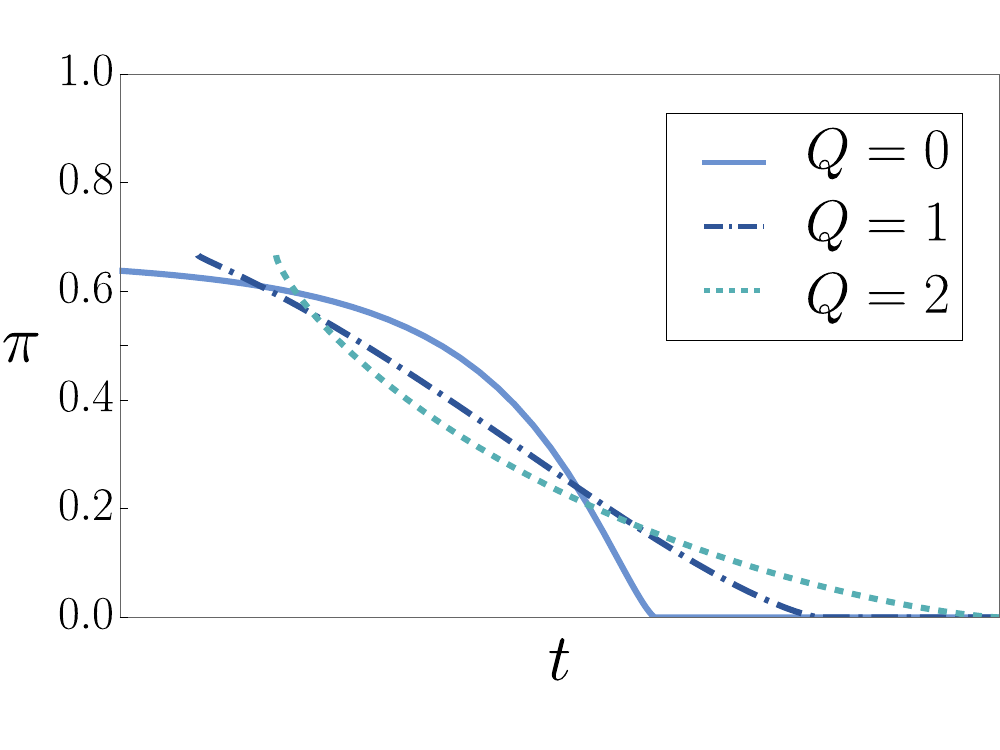}\\[-2mm] \centering{(a)}}
			\hspace{8pt}
			\parbox[c]{0.5\textwidth}{\includegraphics[width=0.49\textwidth]{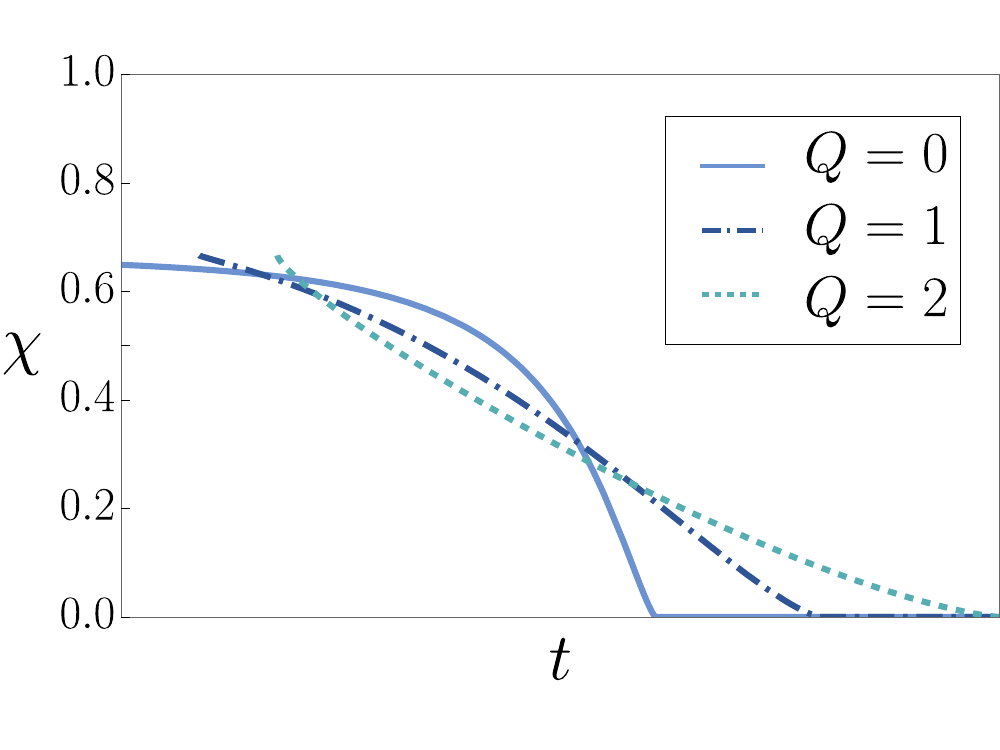}\\[-2mm] \centering{(b)}}}
			\vspace{-2mm}
			\caption{Comparative of (a) $\pi(t)$ and (b) $\chi(t)$ for different values of $Q$. $\qquad \qquad \hspace{1cm}$}
			\label{Fig-KCKS3}
			\end{figure}

\begin{figure}[H]
			\centering
			\hspace{-6mm} \includegraphics[width=0.82\textwidth]{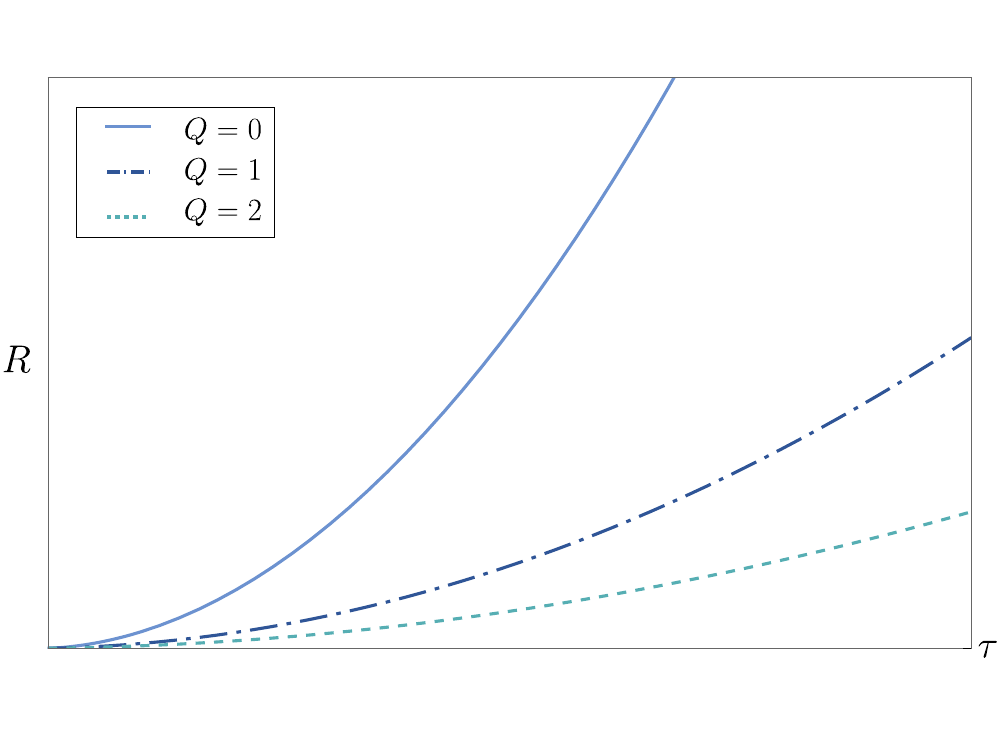}
			\vspace{-8mm}
			\caption{Comparative of $R(\tau) \equiv [\varphi(\tau) \, \omega(\tau)]^{1/3} = \varphi(\tau)^{3/2}[1 + t(\tau)\, Q]$ for different values of $Q$.}
		\label{Fig-KCKS4}
		\end{figure}

			A CIG in isentropic evolution fulfils the barotropic relation (\ref{CIG-isentropic}). Integrating the conservation of matter equation (\ref{matter-conservation}) with $n$ given in (\ref{n-CIG-isentr}) and $\theta$ in (\ref{expansion-T-1}) we get $n = n_0 \left(\frac{R_0}{R}\right)^3$, with $n_0$ constant and $R \equiv (\varphi \omega)^{1/3}$. With that, we can write the pressure as a function of the metric functions for this case as
			\begin{equation} \label{p-GIC-isentropic}
				p = p_0 (\varphi \, \omega)^{-\gamma}, \qquad p_0 = constant.
			\end{equation}
The energy density, $\rho$, is given by (\ref{density-T-1}) with $\omega = \omega_1 (t) + \omega_2 (t) \, Q$ and $Q$ constant. Thus, plugging that and (\ref{p-GIC-isentropic}) into the barotropic relation (\ref{CIG-isentropic}), we get the following differential equation for the metric functions of a KCKS solution representing the isentropic evolution of a non-barotropic family of solutions fulfilling the CIG barotropic relation along the evolution:
			\begin{equation} \label{rel-barotropia-KCKS-GIC-isentropic}
				\bar{n}_0 [\varphi (\omega_1 + \omega_2 \, Q)]^{-1} + \frac{\bar{p}_0}{\gamma - 1} [\varphi (\omega_1 + \omega_2 \, Q)]^{-\gamma} = v \left( \frac14 \frac{\dot{\varphi}^2}{\varphi^2} + \frac{\dot{\varphi}}{\varphi} \frac{\dot{\omega}_1 + \dot{\omega}_2 \, Q}{\omega_1 + \omega_2 \, Q} \right) + \frac{k}{\varphi} \, .
			\end{equation}
\\
			It is worth remarking that in (\ref{rel-barotropia-KCKS-GIC-isentropic}) the metric functions still need to fulfil the perfect fluid field equations (\ref{eq-T-1}). Thus, equation (\ref{rel-barotropia-KCKS-GIC-isentropic}) together with the field equations obtained by imposing (\ref{eq-T-1}) on each of the functions $\omega_i$ constitute a set of three constraints on four metric functions. We can now use the freedom to choose the time coordinate to impose a fourth equation that closes the system. \\ \\ 
			In Section \ref{T-model-general-sol}, we have already used three of these four restrictions to write three of the metric functions in terms of the fourth one. Thus, we can now use those results to write (\ref{rel-barotropia-KCKS-GIC-isentropic}) as a differential equation for the only remaining metric function. \\ \\
			\begin{figure}[t]
				\centerline{
				\parbox[c]{0.33\textwidth}{\includegraphics[width=0.33\textwidth]{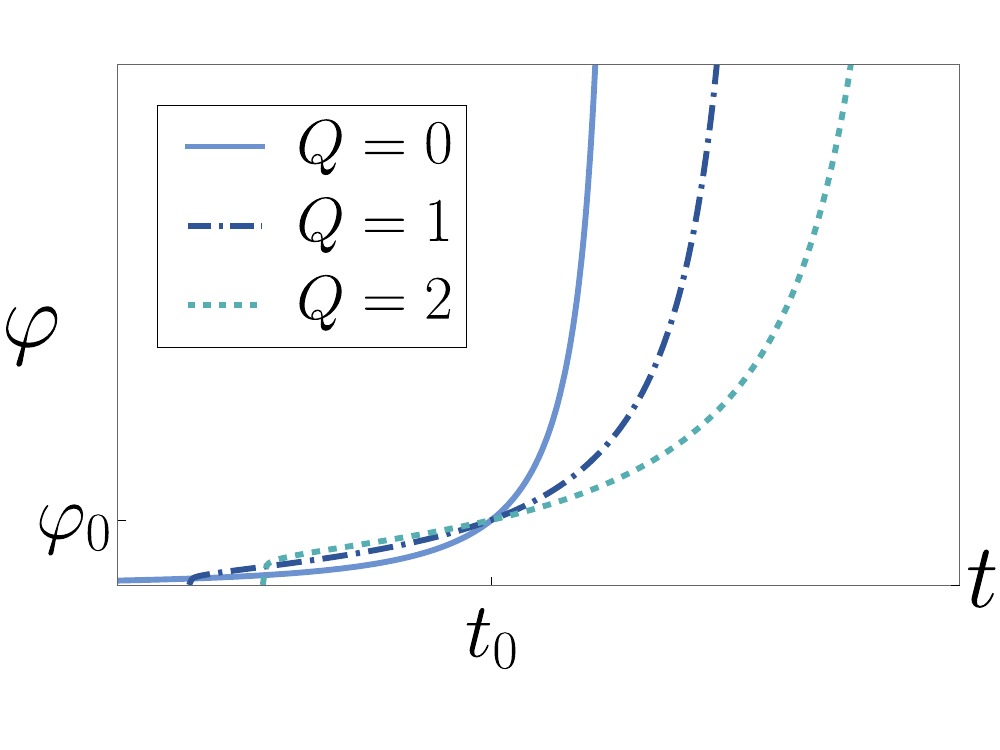}\\[-1mm] \centering{(a)}}
				\hspace{2pt}
				\parbox[c]{0.33\textwidth}{\includegraphics[width=0.33\textwidth]{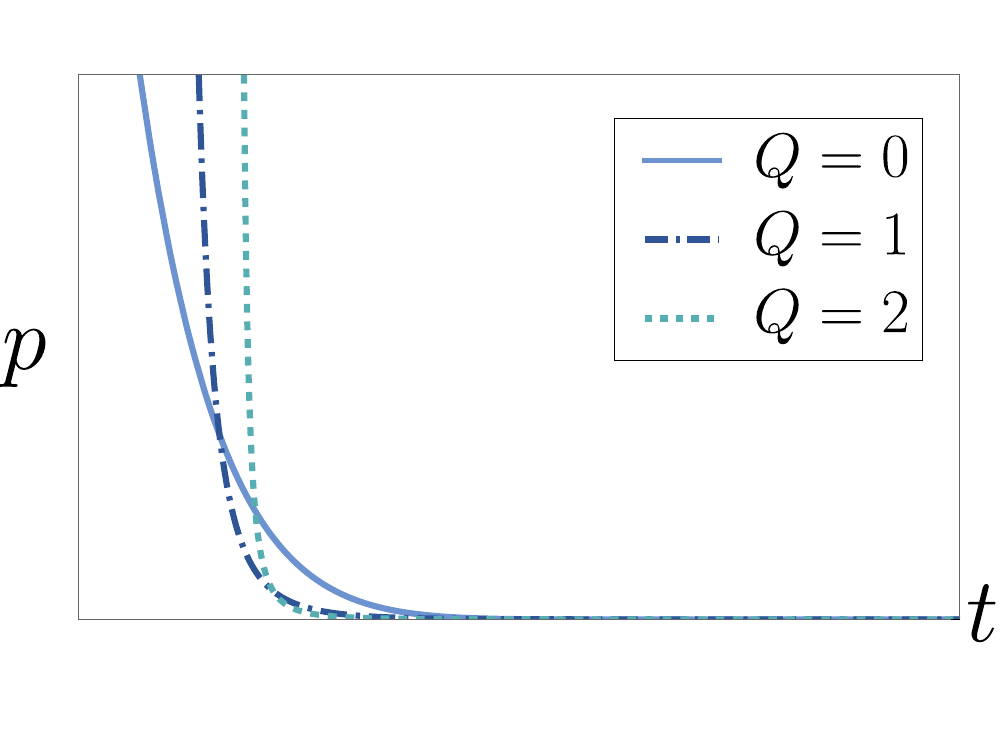}\\[-1mm] \centering{(b)}}
				\hspace{1pt}
				\parbox[c]{0.33\textwidth}{\includegraphics[width=0.33\textwidth]{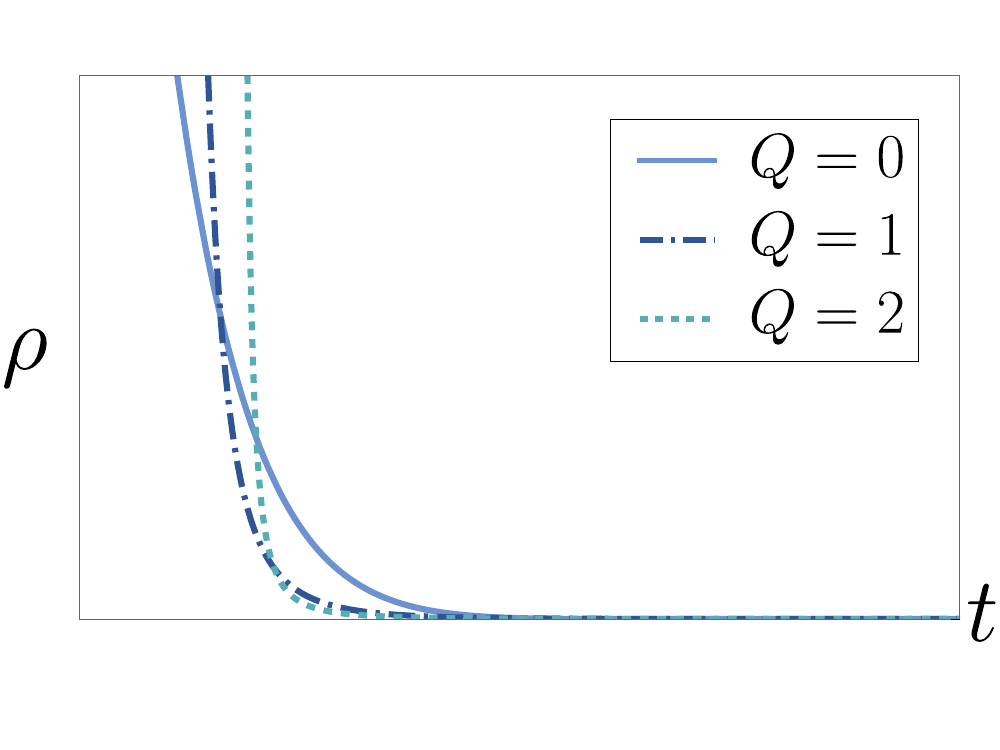}\\[-1mm] \centering{(c)}}}
				\vspace{-2mm}
				\caption{Comparative of (a) the the solution $\varphi(t)$ of the differential equation (\ref{rel-barotropia-KCKS-GIC-isentropic-k_0}) with $\gamma = 5/3$ and initial condition $\varphi(t_0) = \varphi_0$, (b) the pressure $p(t)$ and (c) the energy density $\rho(t)$ obtained from it for different values of $Q$.}
				\label{Fig-KCKS2}
			\end{figure}
\\[-7mm]
			$\hspace{-2mm}$For the case $k = 0$, the modified Herlt algorithm uses the field equations to give $v = \varphi^{-3}$ and $\omega = \sqrt{\varphi} (1 + t \, Q)$. By substituting this in (\ref{rel-barotropia-KCKS-GIC-isentropic}), we get the following first order differential equation for the metric function $\varphi$:
			\begin{equation} \label{rel-barotropia-KCKS-GIC-isentropic-k_0}
				\bar{n}_0 [\varphi^{3/2} (1 + t \, Q)]^{-1} + \frac{\bar{p}_0}{\gamma - 1} [\varphi^{3/2} (1 + t \, Q)]^{-\gamma} = \frac{1}{\varphi^3} \left( \frac34 \frac{\dot{\varphi}^2}{\varphi^2} + \frac{\dot{\varphi}}{\varphi} \frac{Q}{1 + t \, Q} \right) \, .
			\end{equation}
This differential equation needs to be solved numerically. In Figure \ref{Fig-KCKS2}(a) we show its solution for the case $\gamma = 5/3$ for different values of the constant parameter $Q$, while in Figures \ref{Fig-KCKS2}(b) and \ref{Fig-KCKS2}(c) we show the corresponding pressures and energy densities, respectively. In Figures \ref{Fig-KCKS3}(a) and \ref{Fig-KCKS3}(b) we also show the behaviour of their variable $\pi$ and indicatrix function. \\
			$\hspace{-5mm}$It is worth remarking that we recover the FLRW solutions in the limit $Q = 0$. Recall, however, that here we are taking $\alpha_2$ as the time coordinate. Thus, this limit is not evident in the plots and they are not easy to interpret. If we want to compare these solutions with the well known results for the FLRW metrics, we need to plot these quantities as functions of the proper time $\tau = \int \varphi(t)^{3/2} \textrm{d}t = \tau(t)$. Figure \ref{Fig-KCKS4} shows the function $R(\tau) \equiv [\varphi(\tau) \, \omega(\tau)]^{1/3} = \varphi(\tau)^{3/2}[1 + t(\tau)\, Q]$ that results of doing so for different values of $Q$. We can see that the effect of an increase of the constant $Q$ is to slow the growth of the function $R$ compared with the FLRW case $Q = 0$.

			\subsubsection{Relativistic \boldmath$\gamma$-law barotropic relation}
			Another barotropic relation of interest is the relativistic $\gamma$-law, $p = (\gamma - 1)\rho$ (see \cite{Harrison, Vajk, Assad-Lima, Kramer}). If we choose the proper time as the time coordinate ($v = 1$) and plug (\ref{pressure-T-1}) and (\ref{density-T-1}) into the $\gamma$-law relation, we get a differential equation that can be easily integrated in the case $k = 0$. Then, using our results in Section \ref{T-model-general-sol} we can write it as a differential equation for the sole metric function $\varphi(\tau)$. In \cite{Vajk}, the same result is obtained through a different integration method and solutions for $k \neq 0$ are also investigated.

\chapter{R-models admitting an orthogonal flat synchronisation} \label{chap-R-models-plans}
The main goal of the second part of this thesis is to investigate the physical interpretation of the perfect fluid solutions admitting a G$_3/$S$_2$. In the previous chapter, such a study was carried out in detail for the T-models, one of the two subfamilies of perfect fluid solutions with a G$_3/$S$_2$. This chapter, which is based on the results in \cite{FM-R-models-plans}, constitutes the first step towards undertaking a similar study for the other subfamily, the R-models (whose orbit curvature has a gradient that is not tangent to the fluid flow). \\ \\
The Lemaître-Tolman (LT) model \cite{Lemaitre,Tolman} is a remarkable spherically symmetric dust solution for modelling both gravitational collapse and cosmological inhomogeneities (see also references \cite{Bondi, Kramer, Inhomogeneous_Cosmological_Models, Krasinski-Plebanski, Ellis-Maartens-MacCallum}), and its generalisation for non-vanishing pressure conforms the geodesic limit of the R-models. \\ \\
In the last three decades, a large number of studies have been devoted to analysing the cosmic censorship conjecture by using the LT models (see \cite{Inhomogeneous_Cosmological_Models, Krasinski-Plebanski, Lapi-Morales, Mosani-J} and references therein). In cosmology, the LT solutions provide exact inhomogeneous models for studying the formation of structures \cite{Inhomogeneous_Cosmological_Models, Krasinski-Plebanski, Ellis-Maartens-MacCallum, Krasinski-Hellaby} and for analysing the effect of the non-linear inhomogeneities on the cosmic microwave background radiation \cite{Inhomogeneous_Cosmological_Models, Krasinski-Plebanski, SDiego1, SDiego2}. \\ \\
Although Lemaître \cite{Lemaitre} also considered a non-null pressure in his pioneering paper, the dust model is the only one contemplated in the above cited references and in most approaches where the LT metric is considered. Nevertheless, in some papers the role of pressure is analysed \cite{Lasky-Lun, Lynden-Bicak} (see also references therein), and some models with anisotropic pressure have been considered \cite{Sussman-98, Sussman-Pavon}. However, something is still lacking in the study of these LT metrics with pressure: their interpretation as a thermodynamic perfect fluid in l.t.e. \\ \\
The R-models are the perfect fluid solutions whose metric has the form (\ref{metric-ss-1}$-$\ref{metric-ss-3}) with $Y'(t, r) \neq 0$, where a prime represents partial derivative with respect to the coordinate $r$. For the case of the R-models with geodesic motion, the metric line element can be written as \cite{Krasinski-Plebanski}:
\begin{equation} \label{metric-Y-E}
	ds^2 = - d\tau^2 + \frac{[Y'(\tau,r)]^2}{1 + 2 E(r)} \, dr^2 + Y^2(\tau,r) \, d\tilde{\Omega}^2,
\end{equation}
where $d\tilde{\Omega}^2$ is a two-dimensional metric of constant curvature. 
The commonly considered LT models are the metrics of the form (\ref{metric-Y-E}) ({\em LT metrics}) with zero pressure and cosmological constant. Although very few explicit solutions with non-constant pressure are known, the perfect fluid general solution can be obtained by quadratures for the spatially flat case $E(r) = 0$ \cite{Bona-Stela-a}. This result made it possible to construct ‘‘Swiss cheese’’ cosmological models with pressure \cite{Bona-Stela-c}. \\ \\
The procedure described in Sections \ref{sec-other-phys-re-conds} and \ref{sec-generic-ideal-gas} to analyse the physical meaning of the perfect fluid solutions could be applied to the full set of the perfect fluid solutions of the form (\ref{metric-Y-E}). In this case, the obtained constraints would be simply formal and of little practical interest. Nevertheless, when we apply our procedure to a family of explicit solutions, we can go further in our analysis of the physical meaning of the solutions. Therefore, we will limit ourselves here to studying the case in which explicit solutions are known, that is, when $E(r) = 0$. \\ \\
Two comments about this spatially flat case. Firstly, only the spherical symmetry is compatible \cite{Kramer}, and consequently $d\tilde{\Omega}^2 = d \Omega^2$ is the metric of a two-sphere. Secondly, the spherically symmetric perfect fluid solutions that admit a flat slice orthogonal to the fluid flow have a geodesic motion \cite{Bona-Stela-a}, and thus they coincide with the perfect fluid solutions for the LT metrics (\ref{metric-Y-E}) with $E(r) = 0$. \\ \\ \\
Accordingly, in this chapter we study the thermodynamics and analyse the macroscopic conditions for physical reality of the spherically symmetric perfect fluid solutions admitting a flat synchronisation orthogonal to the fluid flow.
In Section \ref{sec-metric} we present the metric line element and remark on the linearity of the field equations. Several approaches that can be considered in solving them are also sketched. As an example we obtain the general solution of the flat dust LT models and we comment on the previously known results. \\ \\
In Section \ref{sec-termo} we undertake the general study of the thermodynamics of the solutions. On the one hand, we obtain the kinematic and hydrodynamic quantities (expansion, pressure and energy density of the fluid) as well as the indicatrix function that gives the square of the speed of sound. On the other hand, we determine the thermodynamic schemes that are compatible with each model. \\ \\
As in the previous chapter, if we want to go further in our analysis of the physical meaning of the solutions we must specify the general expressions by considering particular solutions or by adding some physical properties. In Section \ref{sec-R-chi-pi} we impose a significant physical constraint on the models; their compatibility with the EoS of a generic ideal gas. The general equations that characterise these {\em ideal models} are obtained. \\ \\
In Section \ref{sec-t^q} we analyse when the spherically symmetric limit of the Szafron solution \cite{Szafron} fulfils the ideal constraints previously studied. The behaviour of the subsequent {\em ideal Szafron models} is accurately analysed, and the spacetime domains where the macroscopic constraints for physical reality hold are obtained. A similar study is undertaken in Section \ref{sec-t^1/2} for another ideal model, which can be considered as a limit of the ideal Szafron models. \\ \\
In Section \ref{sec-T(t)} we study the conditions that characterise the models consistent with a homogeneous temperature, namely, those compatible with a non-vanishing thermal conductivity coefficient. 

	\section{Metric and general solution} \label{sec-metric}
	In synchronous comoving coordinates, the metric of the spherically symmetric perfect fluid solutions that admit a flat slice orthogonal to de fluid flow is (\ref{metric-Y-E}) with $E(r) = 0$:
	\begin{equation} \label{metric-Y}
		ds^2 = - d\tau^2 + [Y'(\tau,r)]^2 \, dr^2 + Y^2(\tau,r) \, d\Omega^2.
	\end{equation}
	The unit velocity of the fluid is $u = \partial_\tau$, and its expansion and the non-vanishing components of the shear tensor are, respectively
	\begin{equation} \label{expansion-Y}
		\theta = 2 \, \frac{\dot{Y}}{Y} + \frac{\dot{Y}'}{Y'} = \partial_\tau [\ln(Y^2 \, Y')] ,\\[1mm]
	\end{equation}
	\begin{equation} \label{shear-Y}
		\sigma^1_1 = \sigma^2_2 = -\frac{1}{2}\sigma^3_3 = \bar{\sigma} \equiv \frac{1}{3}\left( \frac{\dot{Y}}{Y} - \frac{\dot{Y}'}{Y'} \right) ,
	\end{equation}
where a dot represents partial derivative with respect to the time coordinate $\tau$. Note that, $u$ being geodesic, we have $u(\phi) \equiv u^{\alpha} \partial_{\alpha} \phi = \dot{\phi}$, for any spacetime function $\phi(x^{\alpha})$. \\ \\
	The Einstein equations for a perfect fluid source reduce to the following expressions for the \textit{pressure} $p$ and the \textit{energy density} $\rho$ \cite{Kramer}
	\begin{equation} \label{pressure and density-Y}
		p = -2 \, \frac{\ddot{Y}}{Y} - \frac{\dot{Y}^2}{Y^2} \, , \quad \, \quad\, \rho = 2 \, \frac{\dot{Y}}{Y}\frac{\dot{Y}'}{Y'} + \frac{\dot{Y}^2}{Y^2} \, ,
	\end{equation}
where $p = p(\tau)$ as a consequence of the conservation of the energy tensor, $\nabla \cdot T = 0$. If we do not add any additional physical requirement, only the first equation in (\ref{pressure and density-Y}) constrains the metric function $Y(\tau,r)$, and the second one gives the energy density for a given solution.
	
		\subsection{General solution of the field equations} \label{subsec-metric-solucio}
		By performing the substitution
		\begin{equation} \label{def. Z}
			Y = Z^{2/3} \, ,
		\end{equation}
expressions (\ref{expansion-Y}), (\ref{shear-Y}), and (\ref{pressure and density-Y}) for the expansion, shear, pressure, and energy density become
		\begin{equation} \label{expansion-shear-Z}
			\theta = \frac{\dot{Z}}{Z} + \frac{\dot{Z}'}{Z'} \, , \qquad \bar{\sigma} = \frac{1}{3}\left( \frac{\dot{Z}}{Z} - \frac{\dot{Z}'}{Z'}\right) , \\
		\end{equation}
		\begin{equation} \label{pressure and density-Z}
			p = -\frac{4}{3} \, \frac{\ddot{Z}}{Z} \, ,\qquad \quad \rho = \frac{4}{3}\, \frac{\dot{Z}}{Z}\, \frac{\dot{Z}'}{Z'} \, .
		\end{equation}
Then, from the expression for the pressure we obtain the following equation \cite{Bona-Stela-a}
		\begin{equation} \label{eq. Z}
			\ddot{Z} + \frac{3}{4} p(\tau)Z = 0 \, ,
		\end{equation}
which is linear in Z. Thus, the general solution is of the form \cite{Bona-Stela-a}
		\begin{equation} \label{sol. general eq. Z}
			Z = a(r) \, f(\tau) + b(r) \, g(\tau) \, ,
		\end{equation}
where $a(r)$ and $b(r)$ are arbitrary functions of the radial coordinate $r$, and $f(\tau)$ and $g(\tau)$ are two independent particular solutions of (\ref{eq. Z}). \\ \\
		Therefore, the problem of finding particular solutions of the field equations can be approached in different ways:
		\begin{itemize}
\item[(i)] 
On the one hand, we can give an arbitrary pressure $p(\tau)$ and look for the general solution to (\ref{eq. Z}). Note that the particular solutions $f$ and $g$ to equation (\ref{eq. Z}) are related by 
		\begin{equation} \label{gtt(f)}
			\ddot{f}/f = \ddot{g}/g \, ,
		\end{equation}
which can be integrated to give $\dot{g} f = \dot{f} g + C$, where $C$ is an arbitrary integration constant. However, since we only need $g$ to be any nontrivial particular solution of (\ref{eq. Z}) independent to $f$, we can set $C = 1$ and therefore obtain 
		\begin{equation} \label{gt(f)}
			\dot{g} f = \dot{f} g + 1 \, .
		\end{equation}
This allows us to obtain, from a known particular solution $f(\tau)$, another particular solution $g(\tau)$ as \cite{Bona-Stela-a, Kramer}
		\begin{equation} \label{g(f)}
			g(\tau) = f(\tau) \! \int \! f^{-2}(\tau) \, d\tau \, .
		\end{equation}
For instance, for $p = 0 $ equation (\ref{eq. Z}) can easily be solved to obtain $Z(\tau,r) = a(r) \, \tau + b(r)$, which corresponds to the parabolic subset ($E(r) = 0$) of the Lema\^itre-Tolmann dust models \cite{Bona-Stela-a, Krasinski-Plebanski}.
\item[(ii)]
On the other hand, we can give an arbitrary function $f(\tau)$ as input, and then determine the pressure as $3 p = -4 \ddot{f}/f$. The solution of (\ref{eq. Z}) is then completed by using (\ref{g(f)}). Thus, we can obtain the solution to the field equations by quadratures. For instance, by choosing $f(\tau) = \tau^q$, $q \neq 1/2$, then $p \sim \tau^{-2}$ and $g(\tau) = \tau^{1 - q}/(1 - 2q)$. These solutions correspond to the spherically symmetric subset of a wider family of solutions considered by Szafron \cite{Szafron}.
\item[(iii)]
However, one could also start by making a particular election of an arbitrary function $\varphi(\tau)$ such that $\dot{\varphi}(\tau) > 0$, and then obtain $f(\tau)$ and $g(\tau)$ as \cite{Bona-Stela-b}
		\begin{equation}
			f(\tau) = [\dot{\varphi}(\tau)]^{-1/2} \, , \qquad g(\tau) = f(\tau) \varphi(\tau) \, .
		\end{equation}
Note that this approach allows us to obtain the general solutions of the field equations without having to compute any integrals.
		\end{itemize}
		It is worth remarking that the same procedures can be carried out for the flat class I Szekeres-Szafron solutions since equation (\ref{eq. Z}) is also fulfilled in that case \cite{Bona-Stela-b, Szafron}. However, although our study of the thermodynamic properties is based on the solution of equation (\ref{eq. Z}), our results only apply in spherical symmetry since the class I Szekeres-Szafron solutions only admit a thermodynamic interpretation in this case \cite{Krasinski-et-al}. \\ \\
		The degrees of freedom of the spherically symmetric perfect fluid solutions admitting a flat synchronisation are given by the election of two arbitrary functions of $r$, $a(r)$ and $b(r)$, and an arbitrary function of time, either $f(\tau)$, $p(\tau)$ or $\varphi(\tau)$. Therefore, the gravitational field is determined by a pair $\{f(\tau), a(r), b(r)\}$. \\ \\
		The spatially homogeneous limit of these solutions are the flat FLRW metrics, which can be characterised by one of the following five equivalent conditions: (i) the metric function $Y(\tau,r)$ factorises, and then the coordinate $r$ can be taken so that $Y(\tau,r) = r R(\tau)$; (ii) $\alpha(r) = a(r)/b(r)$ is a constant function; (iii) the fluid expansion is homogeneous, $\theta = \theta(\tau)$; (iv) the fluid flow is shear-free, $\bar{\sigma} = 0$; and (v) the energy density is homogeneous $\rho = \rho(\tau)$, and then the fluid evolution is barotropic, $\textrm{d} \rho \wedge \textrm{d} p = 0$. 
		
		\subsection{On the dust flat LT models} \label{subsec-pressure constant}
		As an example, in this subsection we will consider the dust flat LT models, that is, the solutions in which the pressure $p$ takes a constant value, $p = -\Lambda$ \cite{Krasinski-Plebanski, barrow}. Now, equation (\ref{eq. Z}) becomes 
		\begin{equation} \label{eq:Z-Lambda}
			\ddot{Z} - \frac34 \Lambda \; Z = 0 \, ,
		\end{equation}
and the energy density is $\rho = \Lambda + \rho_H$, where $\rho_H$ is the hydrodynamic energy density ($\equiv$ matter density). \\ \\
The case $\Lambda = 0$ is the most frequently considered in the literature \cite{Inhomogeneous_Cosmological_Models, Krasinski-Plebanski}, and the solution can be written as
		\begin{equation} \label{Z-Lambda=0}
			Z = Z_0(r) [\tau - \tau_0(r)] \, , 
		\end{equation}
where $\tau_0(r)$ is the non-simultaneous big bang time, and $M(r) = (2/9) Z_0^2(r)$ is the effective gravitational mass. Moreover, the matter density takes the expression
		\begin{equation}
			\rho_H = \frac{4}{3(\tau \! - \! \tau_0)^2 \left[1 - \frac{Z_0\, \tau_0'}{Z_0' (\tau - \tau_0)}\right]} \, .
		\end{equation}
		These expressions can be generalised for $\Lambda \neq 0$. Indeed, the general solution to equation (\ref{eq:Z-Lambda}) is
		\begin{equation} 
\hspace*{-2.0mm}
		Z(\tau,r) \! = \!	
			\begin{cases}
 				a(r) \sinh \, \omega \tau + b(r) \cosh \, \omega \tau , \quad \textrm{if} \; \; \Lambda > 0 \cr 
 				a(r) \sin \, \omega \tau + b(r) \cos \, \omega \tau , \quad \quad \, \textrm{if} \; \; \Lambda < 0 
 			\end{cases}
		\end{equation}
where $\omega^2 \equiv \frac34 |\Lambda|$. It is worth remarking that, for the $\Lambda > 0$ case, if $a(r) > b(r)$ (respectively, $b(r) > a(r)$) these arbitrary functions can be written as $a(r) = Z_0(r) \cosh[\omega \tau_0(r)]$ and $b(r) = -Z_0(r) \sinh[\omega \tau_0(r)]$ (respectively, $b(r) = Z_0(r) \cosh[\omega \tau_0(r)]$ and $a(r) = -Z_0(r) \, \sinh[\omega \tau_0(r)]$). Consequently, we obtain that the general solution $Z(\tau,r)$ of equation (\ref{eq:Z-Lambda}) for $\Lambda > 0$ leads to two models,
		\begin{equation} \label{Z-Lambda>0-a}
			Z(\tau,r) = Z_0(r) \sinh (\omega [\tau \! -\! \tau_0(r)]) , \\[1mm]
		\end{equation}
		\begin{equation} \label{Z-Lambda>0-b}
			Z(\tau,r) = Z_0(r) \cosh (\omega [\tau \! -\! \tau_0(r)]) .
		\end{equation}
For the model (\ref{Z-Lambda>0-a}), $M(r) = (\Lambda/6) Z_0^2(r)$ and the matter density is
		\begin{equation} \label{hydro-rho}
			\rho_H = \frac{\Lambda}{\sinh [\omega(\tau \! - \! \tau_0)]\left[1 \! - \! \frac{\omega Z_0 \, \tau_0'}{Z_0' \coth [\omega(\tau - \tau_0)]}\right]} .
		\end{equation}
We obtain a similar expression for the case (\ref{Z-Lambda>0-b}), which follows by changing $\Lambda \rightarrow - \Lambda$ and $\sinh \rightarrow \cosh$. \\ \\
On the other hand, for the case $\Lambda < 0$ the arbitrary functions can always be written as $a(r) = Z_0(r) \cos[\omega \tau_0(r)]$ and $b(r) = - Z_0(r) \sin[\omega \tau_0(r)]$. Consequently, the general solution $Z(\tau,r)$ of equation (\ref{eq:Z-Lambda}) for $\Lambda < 0$ becomes
		\begin{equation} \label{Z-Lambda<0}
			Z(\tau,r) = Z_0(r) \sin(\omega[\tau - \tau_0(r)]).
		\end{equation}
Now $M(r) = (|\Lambda|/6) Z_0^2(r)$, and the expression of the hydrodynamic energy density is like (\ref{hydro-rho}) with the changes $\Lambda \rightarrow |\Lambda|$ and $(\sinh, \coth) \rightarrow (\sin, \cot)$. \\ \\
		Note that the FLRW homogeneous limit follows by considering $\tau_0(r) = constant$. In this homogeneous case the matter density is positive everywhere for $\Lambda = 0$, for model (\ref{Z-Lambda>0-a}), and for model (\ref{Z-Lambda<0}), and negative for model (\ref{Z-Lambda>0-b}). Nevertheless, shell-crossing singularities \cite{Krasinski-Plebanski} could exist in the inhomogeneous models, which disconnect spacetime domains with positive and negative matter density. \\ \\
		It is worth remarking that the flat dust LT models with $\Lambda = 0$ given in (\ref{Z-Lambda=0}) are the most commonly considered in the basic cosmology books \cite{Inhomogeneous_Cosmological_Models, Krasinski-Plebanski, Ellis-Maartens-MacCallum}. On the other hand, the homogeneous limit of solution (\ref{Z-Lambda>0-a}) is the background universe in the standard $\Lambda$CDM cosmological models. This analytical expression is little used by observational and numerical cosmologists, although it was already considered for the first time by Lema\^itre \cite{Lemaitre}, and its generalisation to a $\gamma$-law is also known \cite{Kramer, Harrison}. The inhomogeneous model (\ref{Z-Lambda>0-a}) has also been considered previously \cite{barrow}. \\ \\
		
	\section{Thermodynamics of the solutions} \label{sec-termo}
	Now, we analyse when the R-models (\ref{metric-Y}) represent the evolution in local thermal equilibrium of a fluid that meets the suitable macroscopic physical constraints stated in Sections \ref{sec-lte} and \ref{sec-other-phys-re-conds}. Note that, due to the symmetries of the metric (\ref{metric-Y}), all scalar invariants depend on two functions at most. Then, the hydrodynamic sonic condition S is fulfilled automatically and step 1 in the procedure presented in Section \ref{sec-other-phys-re-conds} is achieved for the full set. Thus, we proceed to analyse step 2.
	
		\subsection{Hydrodynamic quantities: energy density, pressure and speed \mbox{of sound}} \label{subsec-R-Termo-hydrodynamic}
		Let us consider an inhomogeneous solution of the field equations $Z = a(r) f(\tau) + b(r) g(\tau)$, $\alpha'(r) \neq 0$, $\alpha \equiv a(r)/b(r)$. If $\beta = \beta(r) \equiv a'(r)/b'(r)$, the fluid expansion (\ref{expansion-shear-Z}) can be written as
		\begin{equation} \label{expansion-Zb}
			\theta = \partial_\tau (\ln [(\alpha f \! + \! g)(\beta f \! + \! g)]) =
\frac{\alpha \dot{f} + \dot{g}}{\alpha f + g} + \frac{\beta \dot{f} + \dot{g}}{\beta f + g} ,
		\end{equation}
and the pressure and energy density (\ref{pressure and density-Z}) take the following expressions
		\begin{equation} \label{pressure and density-Zb}
			p = -\frac{4}{3}\frac{\ddot{f}}{f} \, , \quad \rho = \frac{4}{3} \, \frac{\alpha \beta \dot{f}^2 + (\alpha \! + \! \beta) \dot{f}\dot{g} + \dot{g}^2} {\alpha \beta f^2 + (\alpha \! + \! \beta) f g + g^2} \, .
		\end{equation}
		The square of the speed of sound can be obtained using the definition (\ref{indicatrix def}) of the indicatrix function and the energy conservation condition $\dot{\rho} + (\rho + p) \theta = 0$,
		\begin{equation} \label{chi-general}
			c_s^2 = \frac{\dot{p}}{\dot{\rho}} = -\frac{\dot{p}}{\theta (\rho + p)} \equiv \chi (\tau,r) \, ,
		\end{equation}
where $\theta(\tau,r)$, $p(\tau)$, and $\rho(\tau,r)$ are given in (\ref{expansion-Zb}) and (\ref{pressure and density-Zb}). \\ \\
		If we consider a specific solution of the field equations, we can know the expansion $\theta(\tau,r)$, the pressure $p(\tau)$, the energy density $\rho(\tau,r)$ and the indicatrix function $\chi(\tau,r)$ as spacetime functions, explicitly fulfilling step 2 of our approach presented in Section \ref{sec-other-phys-re-conds}. Then, we could analyse the physical behaviour of the solutions. In particular, we could study the spacetime regions where the energy conditions or the relativistic compressibility conditions hold (step 3). It is worth remarking that each solution represents a specific evolution of a family of fluids. If we are interested in the thermodynamic properties of these fluids regardless of the evolution, we should obtain the explicit dependence on the energetic variables of the indicatrix function and thus determine the equation of state $c_s^2 = \chi(\rho,p)$. 
		
		\subsection{Thermodynamic scheme: Entropy, matter density \\ and temperature} \label{subsec-R-Termo-scheme}
		Before considering any specific solution of the field equations, we proceed here with step 4 of the procedure proposed in Section \ref{sec-other-phys-re-conds} for the general case, which corresponds to solving the inverse problem for our metrics (\ref{metric-Y}) by obtaining the associated thermodynamic schemes. Recall that each associated thermodynamic scheme provides a physical interpretation of the solution. \\ \\ 
		In Section \ref{sec-lte} we have shown that the specific entropies $s$ and the matter densities $n$ associated with $T$ are of the form $s = s(\bar{s})$ and $n = \bar{n}/N(\bar{s})$, where $s(\bar{s})$ and $N(\bar{s})$ are arbitrary real functions of a particular solution $\bar{s} = \bar{s}(\rho, p)$ to the specific entropy conservation equation, $u(s) = 0$, and $\bar{n} = \bar{n}(\rho,p)$ is a particular solution to the matter conservation equation, (\ref{matter-conservation}). \\ \\
		From expression (\ref{expansion-Zb}) of the expansion, it is easy to see that $\bar{n} = [(\alpha f \! + \! g)(\beta f \! + \! g)]^{-1}$ is a particular solution to equation (\ref{matter-conservation}), and any function of the radial coordinate $\bar{s} = \bar{s}(r)$ fulfils $u(s) = \dot{s} = 0$. Then, we have that the thermodynamic schemes associated with the perfect fluid solutions of the form (\ref{metric-Y}) are determined by a specific entropy $s$ given by $s(\rho, p) = s(r)$, and a matter density $n$ of the form
		\begin{equation} \label{n-general-R}
			n(\rho,p) = \frac{1}{N(r)[\alpha \beta f^2 + (\alpha \! + \! \beta) f g + g^2]} \, ,
		\end{equation}
where $s = s(r)$ and $N = N(r)$ are two arbitrary real functions. \\ \\ \\ \\ \\
		The temperature $\Theta$ of each thermodynamic scheme determined by a pair $\lbrace s, \, n \rbrace$ can be obtained from (\ref{thermo-first-law-h}) as $\Theta = \left(\frac{\partial h}{\partial s}\right)_p = \frac{1}{s'(r)} \left(\frac{\partial h}{\partial r}\right)_\tau$. Using (\ref{pressure and density-Zb}) and (\ref{n-general-R}) we obtain that the specific enthalpy takes the following expression:
		\begin{equation} \label{h-general-R}
			h \! = \! \frac43 N \! \left[\dot{g}^2 \! - \! g^2 \ddot{f}/f \! - \! \alpha \beta f^2 (\dot{f}/f\dot{)} \! \! - \! (\alpha \! + \! \beta) g^2 (\dot{f}/g\dot{)} \right] \! .
		\end{equation}
Then, we obtain that the temperature $\Theta$ associated with the thermodynamic scheme defined by the functions $\lbrace s(r), N(r) \rbrace$ has the following expression,
		\begin{subequations} \label{temperatura-general-R}
			\begin{eqnarray} 
				\Theta = \tau_1 (\tau)\, r_1 (r) + \tau_2 (\tau) \, r_2 (r) + \tau_3 (\tau)\, r_3 (r) , \quad \quad \qquad \label{temperatura-general-R-a} \\[2mm]
	 \tau_1(\tau) \equiv \dot{g}^2 \! - \! g^2 \ddot{f}/f , \qquad r_1(r) \equiv \frac34 N'/s' , \qquad \qquad \qquad \\[2mm]
	 \tau_2(\tau) \equiv - f^2 [\dot{f}/f\dot{]\;} , \quad \quad \, r_2(r) \equiv \frac34 (N \alpha \beta)'/s' , \qquad \qquad \\[2mm]
	 \tau_3(\tau) \equiv - g^2 [\dot{f}/g\dot{]\;} , \quad \quad \ r_3(r) \equiv \frac34 [N (\alpha + \beta)]'/s' . \quad \quad \ 
			\end{eqnarray}
		\end{subequations}

		\subsection{Imposing additional physical properties}
		If we want to go further in our analysis of the physical meaning of the solutions, we must impose complementary physical properties on the solutions, and thus specify the general expressions presented in this section. Afterwards, we will be able to impose the macroscopic constraints for physical reality. Among others, we can consider the following approaches:

		\begin{itemize}
\item[(i)] 
We can specify a solution by providing the function $f(\tau)$ that determines the time evolution of the model. For example, we can consider the Szafron solution \cite{Szafron} by taking $f(\tau) = \tau^q$ (see Section \ref{sec-t^q}).
\item[(ii)]
We can impose physical constraints on the indicatrix function that fix the (hydrodynamic) equation of state $c_s^2 = \chi(\rho, p)$. For example, we can establish that $\chi(\rho,p)$ is that of a generic ideal gas (see next section). 
\item[(iii)] 
We can impose physical constraints on the thermodynamic scheme. For example, we can demand a homogeneous temperature (see Section \ref{sec-T(t)}).
		\end{itemize}

	\section{Ideal models} \label{sec-R-chi-pi}
	Now we will analyse when the models considered above are compatible with (\ref{eq. estat gas ideal}), the EoS of a {\em generic ideal gas}.
	As we mention in Section \ref{sec-generic-ideal-gas}, equation (\ref{eq. estat gas ideal}) restricts the functional dependence of the indicatrix function $c_s^2 = \chi(\rho,p)$. More precisely, a perfect energy tensor $T \equiv \{u, \rho, p\}$ represents the evolution of a generic ideal gas in l.t.e. if, and only if, it fulfils the ideal gas sonic condition (\ref{cond. sonica ideal}). In the following sections we study the restriction that (\ref{cond. sonica ideal}) imposes on the metric functions.
	
		\subsection{Study of the ideal sonic condition $\textrm{S}^\textrm{G}$}
		We want to impose the constraint $\textrm{d} \chi \wedge \textrm{d} \pi = 0$, where $\chi = \chi(\tau,r)$ is given in (\ref{chi-general}) and where, from (\ref{pressure and density-Zb}), $\pi$ takes the expression
		\begin{equation}
			\pi = \frac{p}{\rho} = \pi(\tau,r) \equiv - \frac{\ddot{f}[\alpha \beta f^2 + (\alpha \! + \! \beta) f g + g^2]}{f[\alpha \beta \dot{f}^2 + (\alpha \! + \! \beta) \dot{f}\dot{g} + \dot{g}^2]} \, .
		\end{equation}
Then, using (\ref{gtt(f)}) together with (\ref{gt(f)}) to replace the derivatives of $g$, a long but straightforward calculation shows that the ideal sonic condition S$^{\rm G}$ is equivalent to
		\begin{equation} \label{ISC}
			\sum_{i = 1}^8 R_i(r) T_i(\tau) = 0 \, ,
		\end{equation}
where the functions $R_i = R_i(r)$ are given by
		\begin{subequations} \label{eq:R}
			\begin{align}
				R_1 &\equiv 1 + \beta'(\alpha) , & R_2 \equiv \alpha + \beta \beta'(\alpha), \qquad \ \ R_3 &\equiv \beta + \alpha \beta'(\alpha) ,
\\[2mm] 
				R_4 &\equiv \alpha \beta \, [1 + \beta'(\alpha)] , & R_5 \equiv \beta^2 + \alpha^2 \beta'(\alpha) , \qquad R_6 &\equiv \beta^3 + \alpha^3 \beta'(\alpha) ,
\\[2mm]
				R_7 &\equiv \alpha \beta \, [\beta + \alpha \beta'(\alpha)] , & R_8 \equiv \alpha\beta \, [\beta^2 + \alpha^2 \beta'(\alpha)] , \quad \, & \qquad \qquad
			\end{align}
		\end{subequations}
and where the functions $T_i = T_i(\tau)$ are given by 
		\allowdisplaybreaks
		\begin{subequations} \label{eq:T}
			\begin{align}
				T_1 &\equiv - g^2 (1 + g \dot{f})^2 \{\dot{f}^2 \ddot{f}^2 + 2 f\dot{f} \ddot{f}\, \dddot{f} + f(-2 \ddot{f}^3 - 3 f\dddot{f}^2 + 2 f\ddot{f}\, \ddddot{f})\} , \label{subeq:T1} \\[2mm] 
				T_2 &\equiv f \{ -g^3 \dot{f}^4 \ddot{f}^2 - g^2 \dot{f}^3 \ddot{f} (\ddot{f} + 2 g f\dddot{f}) \nonumber \\ 
					& \quad + g \dot{f}^2 [\ddot{f}^2 + 2 g^2 f\ddot{f}^3 + 3 g^2 f^2 \dddot{f}^2 - g f\ddot{f} (3 \dddot{f} + 2 g f\ddddot{f})] \nonumber \\ 
					& \quad + \dot{f} [\ddot{f}^2 + 2 g^2 f\ddot{f}^3 + 4 g^2 f^2 \dddot{f}^2 - g f\ddot{f} (2 \dddot{f} + 3 g f\ddddot{f})] \nonumber \\ 
					& \quad + f[-\ddot{f}\, \dddot{f} + g^2 f\ddot{f}^2 \dddot{f} + g (\ddot{f}^3 + f\dddot{f}^2 - f\ddot{f}\, \ddddot{f})]\}, \label{subeq:T2} \\[2mm]
				T_3 &\equiv f\{ -3 g^3 \dot{f}^4 \ddot{f}^2 - g^2 \dot{f}^3 \ddot{f} (5 \ddot{f} + 6 g f\dddot{f}) \nonumber \\ 
					& \quad + 3 g \dot{f}^2 [-\ddot{f}^2 + 2 g^2 f\ddot{f}^3 + 3 g^2 f^2 \dddot{f}^2 - g f\ddot{f} (3 \dddot{f} + 2 g f\ddddot{f})] \nonumber \\ 
					& \quad + \dot{f} [-\ddot{f}^2 + 10 g^2 f\ddot{f}^3 + 14 g^2 f^2 \dddot{f}^2 - g f\ddot{f} (2 \dddot{f} + 9 g f\ddddot{f})] \nonumber \\ 
					& \quad + f[\ddot{f} \,\dddot{f} - g^2 f\ddot{f}^2 \dddot{f} + g (3 \ddot{f}^3 \!+\! 5 f\dddot{f}^2 - 3 f\ddot{f}\, \ddddot{f})]\}, \label{subeq:T3} \\[2mm] 
				T_4 &\equiv f^{2}\{ -3 g^2 \dot{f}^4 \ddot{f}^2 - 2 g \dot{f}^3 \ddot{f} (\ddot{f} \! + \! 3 g f\dddot{f}) + f(\ddot{f}^3 \! + \! 2 g f\ddot{f}^2 \dddot{f} \! + \! f\dddot{f}^2 \! \! - \! f\ddot{f}\, \ddddot{f}) \nonumber \\ 
    				& \quad + \dot{f}^2 [\ddot{f}^2 + 6 g^2 f\ddot{f}^3 + 9 g^2 f^2 \dddot{f}^2 - 6 g f\ddot{f} (\dddot{f} + g f\ddddot{f})] \nonumber \\ 
    				& \quad - 2 f\dot{f} [\ddot{f}\, \dddot{f} + g (-2 \ddot{f}^3 - 4 f\dddot{f}^2 + 3 f\ddot{f}\, \ddddot{f})]\} , \label{subeq:T4} \\[2mm] 
				T_5 &\equiv f^{2}\{-3 g^2 \dot{f}^4 \ddot{f}^2 - 2 g \dot{f}^3 \ddot{f} (2 \ddot{f} + 3 g f\dddot{f}) \nonumber \\ 
					& \quad + 2 g f\dot{f} (4 \ddot{f}^3 \! + \! 5 f\dddot{f}^2 \! \! - \! 3 f\ddot{f}\, \ddddot{f}) + f(\ddot{f}^3 \! - 2 g f\ddot{f}^2 \dddot{f} \! + 2 f\dddot{f}^2 \! \! - \! f\ddot{f}\, \ddddot{f}) \nonumber \\
					& \quad + \dot{f}^2 [-2 \ddot{f}^2 + 6 g^2 f\ddot{f}^3 + 9 g^2 f^2 \dddot{f}^2 - 6 g f\ddot{f} (\dddot{f} + g f\ddddot{f})]\} , \label{subeq:T5} \\[2mm] 
				T_6 &\equiv f^{3}\{-g \dot{f}^4 \ddot{f}^2 \! - \! f^2 \ddot{f}^2 \dddot{f} \! \! - \! \dot{f}^3 \ddot{f} (\ddot{f} \! + \! 2 g f\dddot{f}) + f\dot{f} (2 \ddot{f}^3 \! + 2 f\dddot{f}^2 \! \! - \! f\ddot{f}\, \ddddot{f}) \nonumber \\
					& \quad + f\dot{f}^2 [-\ddot{f}\, \dddot{f} + g (2 \ddot{f}^3 + 3 f\dddot{f}^2 - 2 f\ddot{f}\, \ddddot{f})]\} , \label{subeq:T6} \\[2mm] 
				T_7 &\equiv f^{3}\{-3 g \dot{f}^4 \ddot{f}^2 \! + \! f^2 \ddot{f}^2 \dddot{f} \! \! - \! \dot{f}^3 \ddot{f} (\ddot{f} \! + \! 6 g f\dddot{f}) \! + \! f\dot{f} (2 \ddot{f}^3 \! + \! 4 f\dddot{f}^2 \! \! - 3 f\ddot{f}\, \ddddot{f}) \nonumber \\ 
					& \quad + 3 f\dot{f}^2 [-\ddot{f}\, \dddot{f} + g (2 \ddot{f}^3 + 3 f\dddot{f}^2 - 2 f\ddot{f}\, \ddddot{f})]\} , \label{subeq:T7} \\[2mm] 
				T_8 &\equiv - f^{4} \dot{f}^2 \{\dot{f}^2 \ddot{f}^2 + 2 f\dot{f} \ddot{f}\, \dddot{f} + f(-2 \ddot{f}^3 - 3 f\dddot{f}^2 + 2 f\ddot{f} \, \ddddot{f})\} . \label{subeq:T8}              
\end{align}
\end{subequations}

		\subsection{Analysing the ideal model equation} \label{subsec-analysys-ideal-eq}
		The above analysis of the ideal sonic condition S$^\textrm{G}$ leads to equation (\ref{ISC}). This constraint and (\ref{gt(f)}) constitute a differential system for the functions $\{f(\tau), g(\tau); \beta(\alpha)\}$. The study of the general solution to this system is a complex task that requires the use of numerical methods and that falls outside the scope of this thesis. \\ \\
		Alternatively, we can use an analytical approach in looking for some particular solutions. We can choose a particular function $\beta = \beta(\alpha)$, then determine the functions $R_i(r)$, and finally solve the system of equations that (\ref{ISC}) and (\ref{gt(f)}) impose on $\{f(\tau), g(\tau)\}$. For example, let us consider $\beta = -\alpha$. Then, $R_1 = R_4 = R_5 = R_8 = 0$, $R_2 = -R_3 = 2 \alpha$, $R_7 = -R_6 = 2 \alpha^3$. Consequently, equation (\ref{ISC}) is equivalent to
		\begin{equation} \label{R1=0}
				T_2 - T_3 \equiv E_1(\ddddot{f}, \dddot{f}, \ddot{f},\dot{f}, f, g) = 0 \, , \quad \;
				T_6 - T_7 \equiv E_2(\ddddot{f}, \dddot{f}, \ddot{f},\dot{f}, f, g) = 0 \, .  
		\end{equation}
Then, we can eliminate the fourth derivative from (\ref{R1=0}) and obtain (considering a non-constant pressure): 
		\begin{equation} \label{R1=0-2}
			E_3 \equiv \frac{\dddot{f}}{\ddot{f}} \frac{\dot{g}}{g} 
- \frac{\ddot{f}}{\dot{f}} \left[\frac{\dot{f}}{f} + \frac{\dot{g}}{g}\right] + \frac{\dot{g}^2}{g^2} = 0 \, .
		\end{equation}
It is easy to prove that this equation implies (\ref{R1=0}). Thus, the functions $\{f(\tau), g(\tau)\}$ must fulfil the third-order differential system (\ref{gt(f)}, \ref{R1=0-2}), and then a solution for each initial condition $\{f(\tau_0), \dot{f}(\tau_0), \ddot{f}(\tau_0), g(\tau_0)\}$ exists. It is worth remarking that the ideal Szafron models with $c = -1$ (see next section) are the solution to these equations for specific initial conditions. The study of the solutions corresponding to other initial conditions requires a numerical approach that is beyond the scope of this work. \\ \\
		Finally, we can also consider a family of solutions to the field equations, and then analyse whether a subfamily fulfils the ideal sonic condition (\ref{ISC}). This is the approach that we follow in the next section for the spherically symmetric limit of the Szafron solution \cite{Szafron}.
		
	\section[The ideal Szafron model $f(\tau) = \tau^q$]{The ideal Szafron model \boldmath$f(\tau) = \tau^q$} \label{sec-t^q}
	Now we shall study whether the Szafron solution \cite{Szafron}, which is defined by the choice $f(\tau) = \tau^q$, $q \neq 1/2$, is compatible with the ideal sonic condition (\ref{ISC}). From (\ref{g(f)}), we obtain that the general solution of the field equations takes the form (\ref{sol. general eq. Z}) with
	\begin{equation} \label{g-ideal}
		f(\tau) = \tau^q , \qquad g(\tau) = - \sigma^{-1} \tau^{1 - q}, \qquad \sigma \equiv 2q \! - \! 1 \neq 0 \, .
	\end{equation}
From these expressions, we obtain that the functions $T_i =T_i(\tau)$ given in (\ref{eq:T}) become
	\begin{equation} \label{eq:T-Szafron}
			T_1 = T_2 = T_3 = T_6 = T_7 = T_8 = 0 , \qquad T_4 = -T_5 = \frac18 \sigma (1 - \sigma^2)^2 \tau^{-6} . 
	\end{equation}
Thus, we have that all the functions $T_i$ identically vanish if, and only if, $\sigma = \pm 1$, which corresponds to the dust LT-model with $\Lambda = 0$. Otherwise, the ideal sonic condition (\ref{ISC}) holds when $R_4 = R_5$. Expressions in (\ref{eq:R}) for these functions imply that the functions $\alpha(r) \equiv a(r)/b(r)$ and $\beta(r) \equiv a'(r)/b'(r)$ fulfil the following relation:
	\begin{equation} \label{beta(alpha)-ideal}
		\beta(\alpha) = c \alpha \, ,  
	\end{equation}
where $c \neq 1$ is a constant (note that $c = 1$ leads to the FLRW limit). This equation also holds if we change the functions $\alpha$ and $\beta$ by a factor, since it can be absorbed into $b(r)$ if we compensate it in $g(\tau)$. Thus, we can choose the multiplicative factor of $g(\tau)$ and take $g(\tau) = \tau^{1 - q}$. \\ \\
	Then, taking into account that (\ref{beta(alpha)-ideal}) is a differential equation relating $a(r)$ and $b(r)$, we obtain a solution to the perfect fluid Einstein equations, that is compatible with the ideal sonic condition {\rm S}$^\textrm{G}$ (\ref{cond. sonica ideal}), given by the metric {\rm (\ref{metric-Y})} with the following election of the metric function $Y(\tau,r)$,
	\begin{equation} \label{Y-ideal}
		Y = Z^{2/3} , \qquad Z(t,r) = \tau^{\frac{1 - \sigma}{2}} b(r) [1 + \alpha(r) \tau^{\sigma}] , 
	\end{equation}
where $b(r)$ is given by
	\begin{equation} \label{b-ideal}
		b(r) = |\alpha(r)|^{1/(c - 1)} \, , \qquad c \neq 1 \, .
	\end{equation}
	Regarding the expansion of the fluid flow, from (\ref{expansion-Zb}) we have 
	\begin{equation} \label{expansion-ideal}
		\theta = \frac{1}{\tau}\left[1 \! + \! \sigma - \frac{\sigma}{1 + \alpha \tau^{\sigma}} - \frac{\sigma}{1 + c \, \alpha \tau^{\sigma}}\right] \, . 
	\end{equation} 
	It is worth noting that we are considering, as Szafron did, expanding models with $\tau > 0$. Nevertheless, the change $\tau \leftrightarrow -\tau$, with $\tau < 0$, leads to contracting models, and our subsequent analysis of the physical properties of the solutions is also valid in this case.
	
		\subsection[Hydrodynamic quantities: Energy density, pressure, and speed of sound for $f(\tau) = \tau^q$]{Hydrodynamic quantities: Energy density, pressure, and speed of sound for \boldmath$f(\tau) = \tau^q$} \label{subsec-hydro-ideal}
		Now we can obtain expressions for the hydrodynamic quantities $\rho$, $p$, and the indicatrix function $c_s^2 = \chi(\pi)$. The time dependence of the pressure and the energy density can easily be obtained from the expressions in (\ref{pressure and density-Zb}). The pressure is
		\begin{equation} \label{p-ideal}
			p = \frac{1 - \sigma^2}{3 \tau^2} \, ,
		\end{equation}
and the energy density is
		\begin{equation} \label{rho-ideal-1}
 			\rho = \frac{1}{3\tau^2} \frac{[(1 \! - \! \sigma) \! + \! (1 \! + \! \sigma) \alpha \tau^{\sigma} ] [(1 \! - \! \sigma) \! + \! (1 \! + \! \sigma) c \, \alpha \tau^{\sigma}]}{(1 + \alpha \tau^{\sigma})(1 + c \, \alpha \tau^{\sigma})} . 
		\end{equation}
		Now, solving the equation $\rho = \rho(\tau,\alpha)$ for $\alpha$ and using (\ref{p-ideal}) to eliminate $\tau$, the following function of state can be obtained when $c \neq 0$,
		\begin{subequations} \label{alpha(rho,p)-1}
			\begin{equation}
				\alpha = \alpha (\rho, p) \equiv \kappa_0 \sqrt{(3p)^{\sigma}} \, \frac{(c \! + \! 1)(1 - \pi) + \varepsilon \sigma F(\pi)}{(1 \! - \! \sigma) - (1 \! + \! \sigma) \pi} , \qquad \qquad
			\end{equation}
			\begin{equation} \label{F(pi)-1}
				F(\pi) = \sqrt{\hat{c}^2 (1 - \pi)^2 + 16 c \pi/(1 \! - \! \sigma^2)} \, , \qquad \qquad
			\end{equation}
\vspace{-2mm}
			\begin{equation} \label{F(pi)-1-kappa}
				\kappa_0 \equiv -(1 - \sigma^2)^{1 - \frac{\sigma}{2}}/[2 (1 + \sigma) c] \, , \qquad  \varepsilon = \pm 1, \qquad \hat{c} = (1 - c)/\sigma \, . 
			\end{equation}
		\end{subequations}
		Finally, from expression (\ref{chi-general}) with (\ref{g-ideal}), (\ref{beta(alpha)-ideal}), and (\ref{expansion-ideal}), and taking into account (\ref{p-ideal}) and (\ref{alpha(rho,p)-1}) to eliminate $\tau$ and $\alpha$, the indicatrix function $\chi(\pi)$ can be determined,
		\begin{equation} \label{chi(pi)-1}
			c_s^2 = \chi(\pi) \! \equiv \! \frac{4\pi^2[\hat{c}^2 (1 \! + \! \pi) + (1 \! + \! c) \varepsilon F(\pi)]}{(1 \! + \! \pi) [\hat{c}^2(1 \! - \! \sigma^2) (1 \! + \! \pi)^2 \! + \! 4(1 \! + \! c)^2 \pi]} . 
		\end{equation}
When $c = 0$, the above expression for $\chi(\pi)$ remains valid by taking $\varepsilon = +1$. \\ \\
		Here we are interested in non-barotropic ($\alpha \! \neq$ constant) solutions with a non-vanishing pressure ($\sigma^2 \! \neq \! 1$). Besides, only a positive pressure ($\sigma^2 \! < \! 1$) is compatible with the ideal gas EoS (\ref{eq. estat gas ideal}). On the other hand, the changes $(\sigma, c, \alpha) \leftrightarrow (-\sigma, c^{-1}, \alpha^{-1})$ leave the metric unchanged. Thus, $\alpha(r)$ being a non-constant arbitrary function, we can analyse all the ideal Szafron models by considering $\sigma^2 < 1$ and $-1 \leq c < 1$. \\
		Notice that, as Szafron already pointed out \cite{Szafron}, the solutions approach a FLRW model with a relativistic $\gamma$-law, $p = (\gamma - 1) \rho$, when $\tau \rightarrow 0$ or $\tau \rightarrow \infty$. Indeed, from (\ref{p-ideal}) and (\ref{rho-ideal-1}), we obtain:
		\begin{equation} \label{limits}
			\rho (\tau \rightarrow 0) = \frac{1 \! - \! |\sigma|}{1 \! + \! |\sigma|}\, p , \qquad \rho (\tau \rightarrow \infty) = \frac{1 \! + \! |\sigma|}{1 \! - \! |\sigma|} \, p .
		\end{equation}

		\subsection[Fluid properties: Compressibility conditions $\textrm{H}^\textrm{G}_1$ for $f(\tau) = \tau^q$]{Fluid properties: Compressibility conditions $\textrm{H}^\textrm{G}_1$ \\ for \boldmath$f(\tau) = \tau^q$} \label{subsec-compress-ideal}
\begin{figure}[]
			\centerline{
			\parbox[c]{0.33\textwidth}{\includegraphics[width=0.32\textwidth]{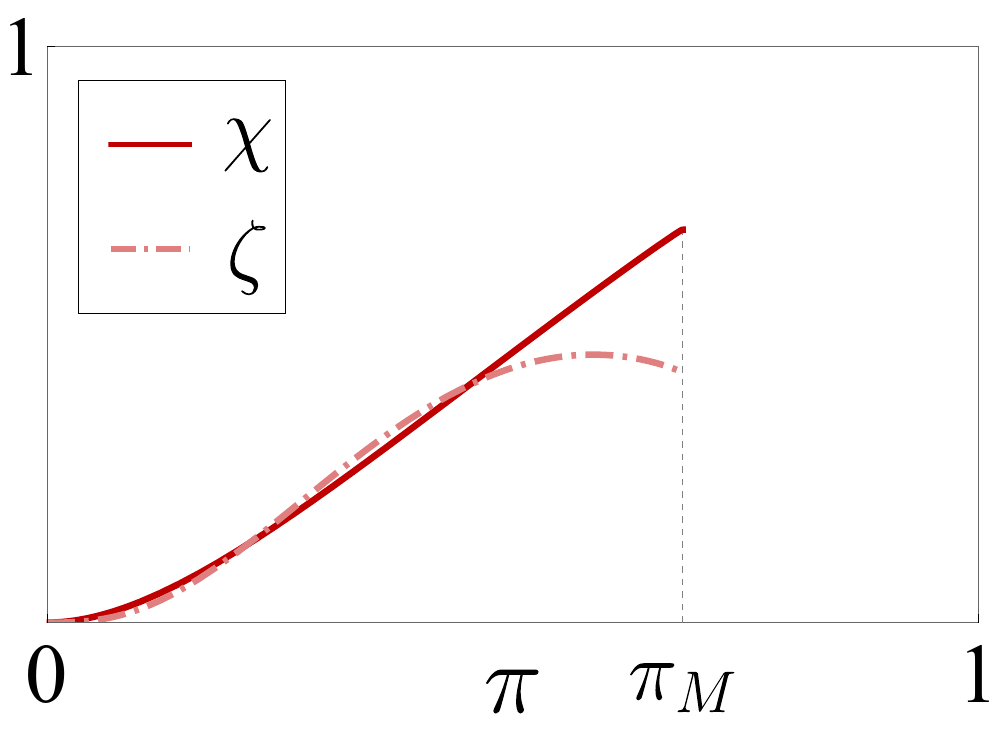}\\[-1mm] \centering{(a)}}
			\hspace{5pt}
			\raisebox{-3pt}{\parbox[c]{0.33\textwidth}{\includegraphics[width=0.32\textwidth]{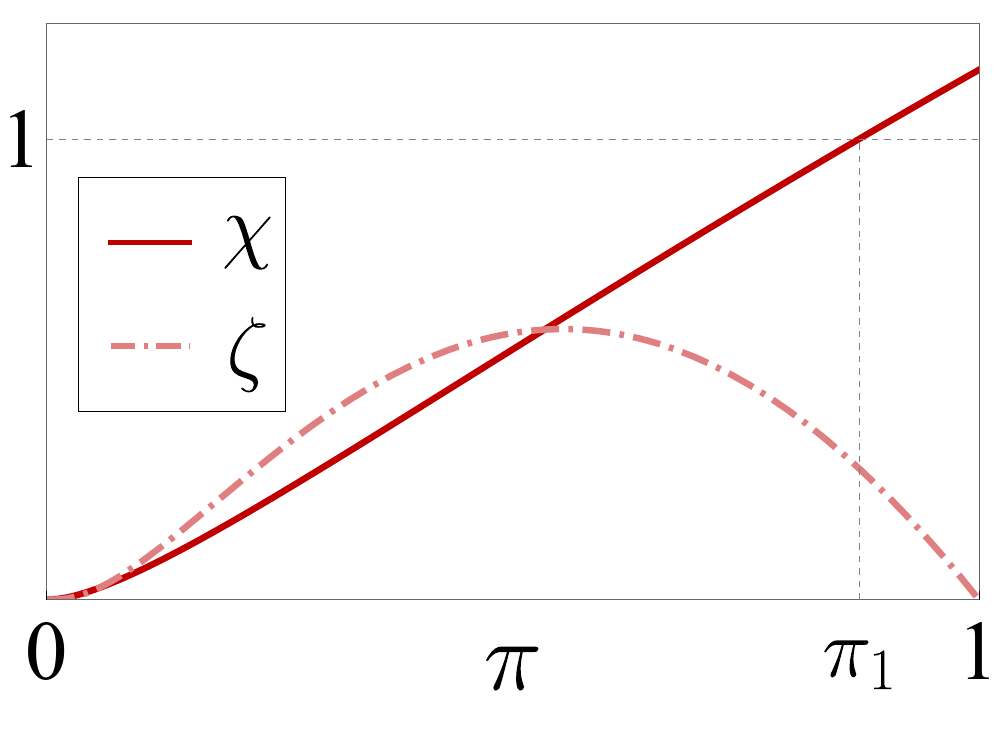}\\[-1mm] \centering{(b)}}}
			\hspace{5pt}
			\parbox[c]{0.33\textwidth}{\includegraphics[width=0.32\textwidth]{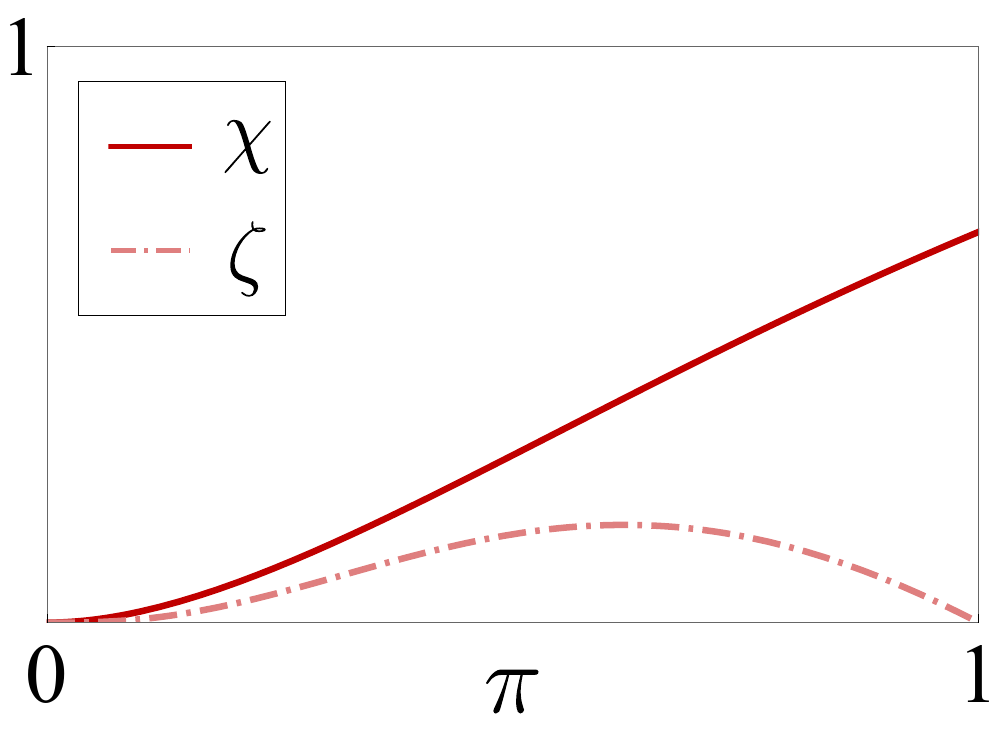}\\[-1mm] \centering{(c)}}}
			\vspace{-2mm}
			\caption{The red solid line shows the behaviour of the indicatrix function $\chi = \chi(\pi)$ for the ideal Szafron models. (a) Case $-1 \leq c \leq 0$; (b) case $c > 0$, $\epsilon = +1$; (c) case $c > 0$, $\epsilon = -1$, $|\sigma| < \sigma_0$. We also plot the function $\zeta(\pi)$ given in (\ref{H1G}) (pink dashed line), which is positive in the interval where $\chi(\pi) \in \, ]0,1[$.}
			\label{Fig-6}
			\end{figure}

		Expression (\ref{chi(pi)-1}) of the indicatrix function $\chi(\pi)$ defines a function of state that characterises a family of fluids. We can analyse the physical reality of these fluids regardless of the particular evolution that the ideal Szafron models represent. More specifically, now we study the compressibility conditions H$_1^{\rm G}$ given in (\ref{H1G}). 
		We must analyse the behaviour of the function $\chi(\pi)$ in the interval $0 < \pi < 1$ where the energy conditions E$^\textrm{G}$ hold. Note that $\chi(\pi)$ depends on the parameters $c$ and $\sigma^2$, and the sign $\varepsilon$. \\ \\
		Firstly, we analyse the first constraint in (\ref{H1G}), the causal condition $0 < \chi(\pi) < 1$. When $-1 \leq c \leq 0$, $\chi(\pi)$ is an increasing function, it is well defined and fulfils this compressibility condition in an interval $]0, \pi_M[$, where $\pi_M \equiv 1 \! + \! \hat{\kappa} \! - \! \sqrt{\hat{\kappa} (\hat{\kappa} \! + \! 2)}$, with $\hat{\kappa} \equiv -8 c /[\hat{c}^2 (1 \! - \! \sigma^2)] > 0$ (see Figure \ref{Fig-6}(a)). Note that $\pi_M$ is close to $1$ (respectively, is close to $0$) when $c$ or $\sigma$ are close to zero (respectively, $\sigma^2$ is close to $1$). \\ \\
		When $1 > c > 0$, the behaviour of the indicatrix function depends on $\varepsilon$. If $\varepsilon = +1$, then $\chi(\pi)$ is a positive increasing function in the whole interval $]0, 1[$, and fulfils the causal constraint in the interval $\,0 < \pi < \pi_1 < 1$, with $\pi_1$ defined by the condition $\chi(\pi_1) = 1$ (see Figure \ref{Fig-6}(b)). Moreover, $\pi_1$ is close to $1$ (respectively, to $0$) when $c$ or $\sigma$ are close to zero (respectively, $c$ is close to $1$). \\ \\
		If $c > 0$ and $\varepsilon = -1$, we have two possibilities: when $|\sigma| < \sigma_0 \equiv (1 - c)/(1 + c)$, $\chi(\pi)$ is a positive increasing function and fulfils the causal constraint in the whole interval $]0, 1[$ (see Figure \ref{Fig-6}(c)); whereas when $|\sigma| > \sigma_0$, $\chi(\pi)$ is a negative function and thus does not satisfy the causal constraint at any point. \\ \\
		Regarding the second of the compressibility conditions H$_1^{\rm G}$, $\zeta(\pi) > 0$, it holds throughout the interval where $\chi(\pi)$ is well defined and fulfils the causal condition in each of the cases considered above (see Figure \ref{Fig-6}). \\ \\
		It is worth remarking that, as it was also the case for the T-models (see Section \ref{sec-CIG}), the ideal Szafron models do not represent the evolution of a classical ideal gas. The equation of state (\ref{chi(pi)-1}) is not compatible with the one of a classical ideal gas (\ref{chi-gas-ideal-classic}). Again, this result agrees with the fact that a geodesic and expanding timelike unit vector is the unit velocity of a classical ideal gas if, and only if, it is vorticity-free and its expansion is homogeneous. \\ \\
		Note that the function $\chi(\pi)$ (\ref{chi(pi)-1}) verifies $\chi(0) = \chi'(0) = 0$. Thus, according to the results in Section \ref{sec-approximations}, it approaches that of a classical ideal gas (or a monoatomic Synge gas) at zero-order (but not at first-order) for small values of $\pi$. For every value of $\sigma$, a value of $c$ exists for which the indicatrix function (\ref{chi(pi)-1}) approaches that of the Synge gas at zero-order in the ultrarelativistic regime, $\chi(1/3) = 1/3$. 

		\subsection[Curvature singularities and spacetime domains for $f(\tau) = \tau^q$]{Curvature singularities and spacetime domains \\ for \boldmath$f(\tau) = \tau^q$} \label{subsec-singularities}
		The expressions for the metric line element, expansion, energy density and pressure of the ideal Szafron models given at the beginning of this section show that our models can have up to three different singularities. The first one takes place at $\tau = 0$. At this singularity, the line element of the 3-spaces $\tau = constant$ vanishes and the energy density and pressure become infinite. Thus, this is a big bang singularity. \\ \\
		Secondly, when $\tau_1^{\sigma}\, \alpha(r) = -1$, the metric line element of the sphere vanishes while the metric distance on the $r$-coordinate lines becomes infinite. Moreover, at this singularity, which is not simultaneous for the comoving observer ($\tau_1 = \tau_1(r)$), the energy density is infinite. 
		\begin{figure}[t]
			\centerline{\;
				\parbox[c]{0.33\textwidth}{\includegraphics[width=0.32\textwidth]{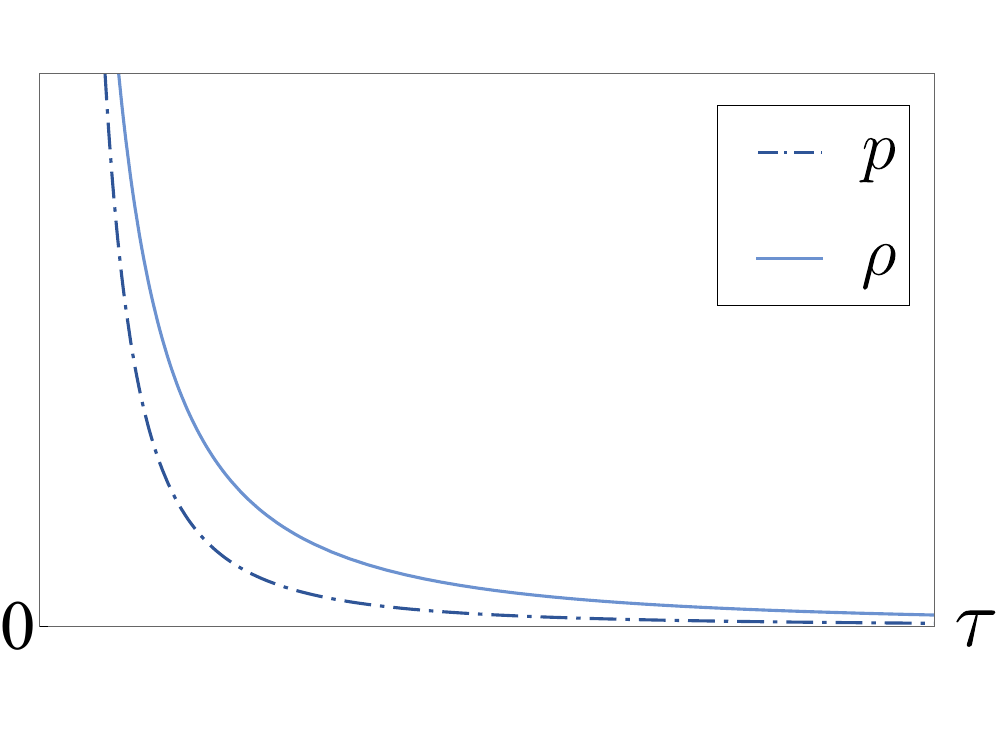}} 
				\raisebox{-3pt}{\parbox[c]{0.33\textwidth}{\includegraphics[width=0.32\textwidth]{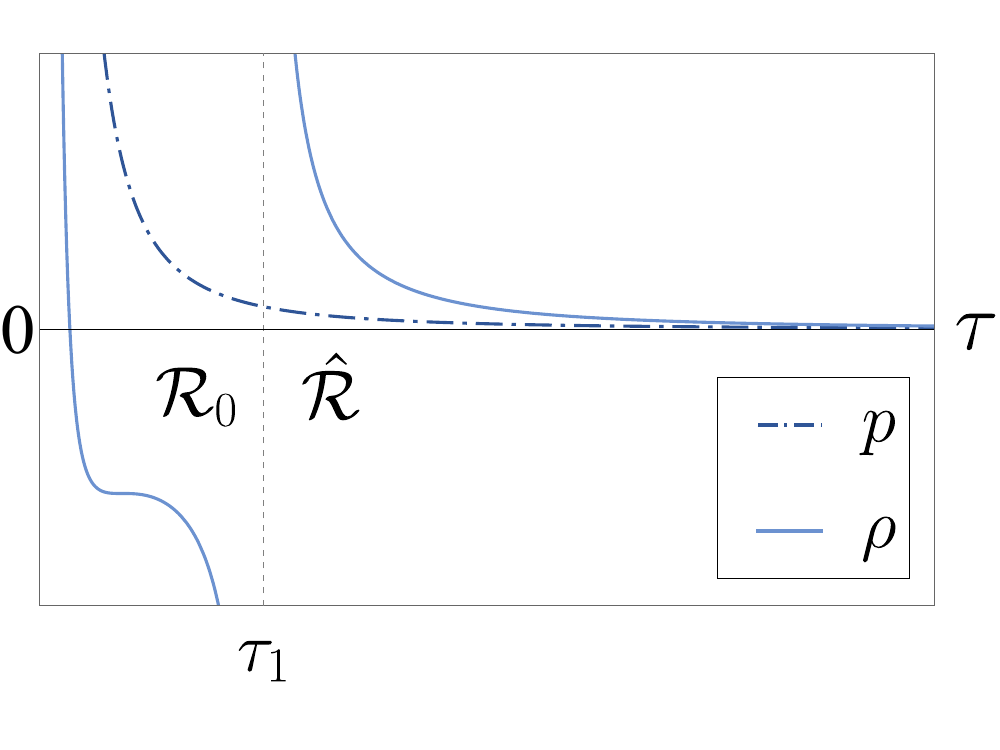}}}
				\raisebox{-3pt}{\parbox[c]{0.33\textwidth}{\includegraphics[width=0.32\textwidth]{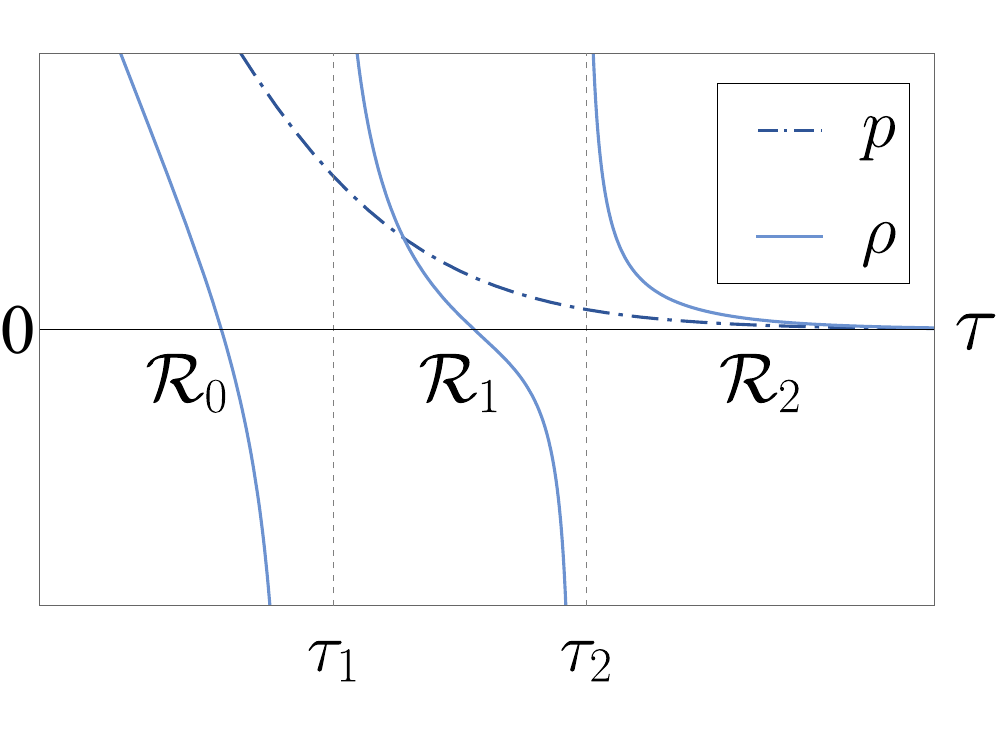}}}}
			\centerline{
				\parbox[c]{0.33\textwidth}{\includegraphics[width=0.32\textwidth]{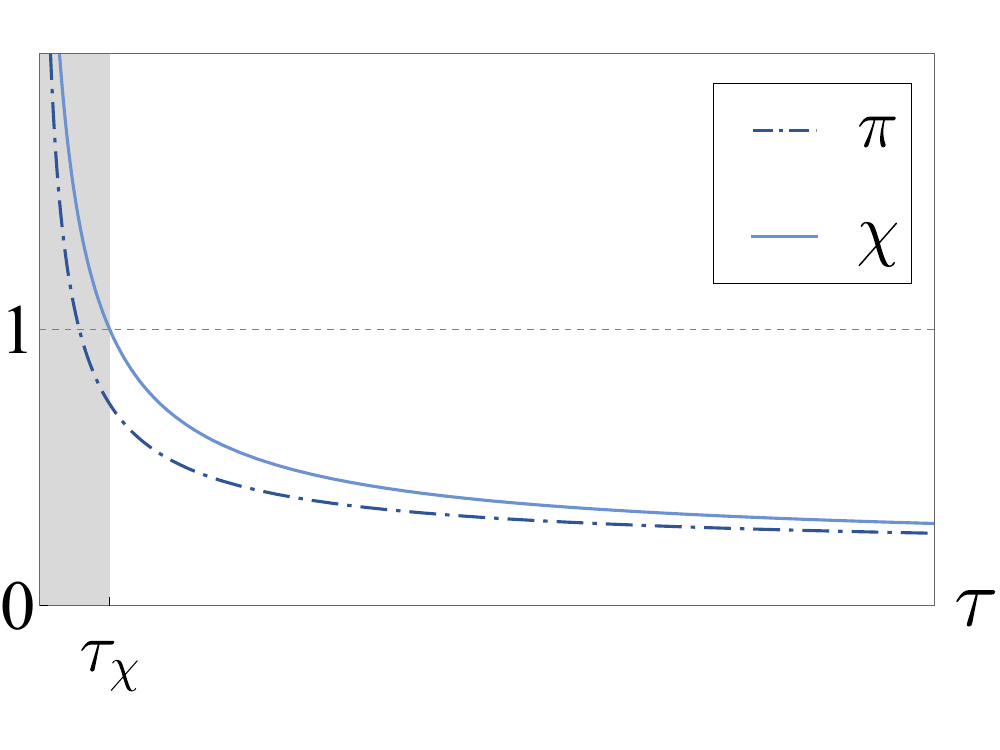}\\[-2mm] \centering{(a)}}
				\parbox[c]{0.33\textwidth}{\includegraphics[width=0.32\textwidth]{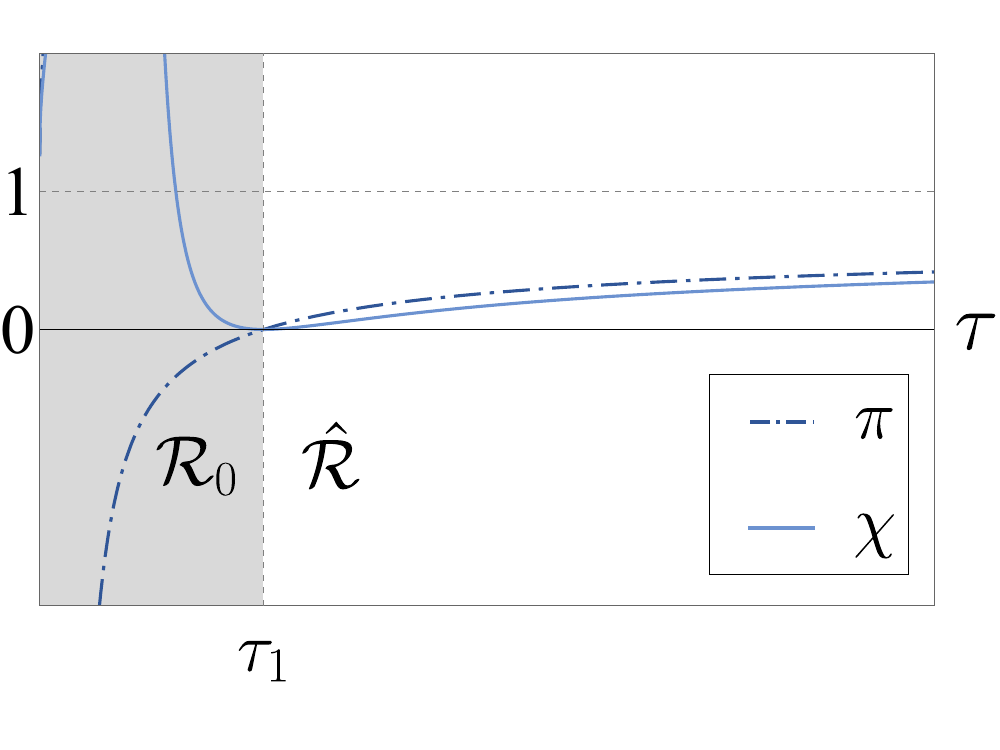}\\[-2mm] \centering{(b)}}
				\parbox[c]{0.33\textwidth}{\includegraphics[width=0.32\textwidth]{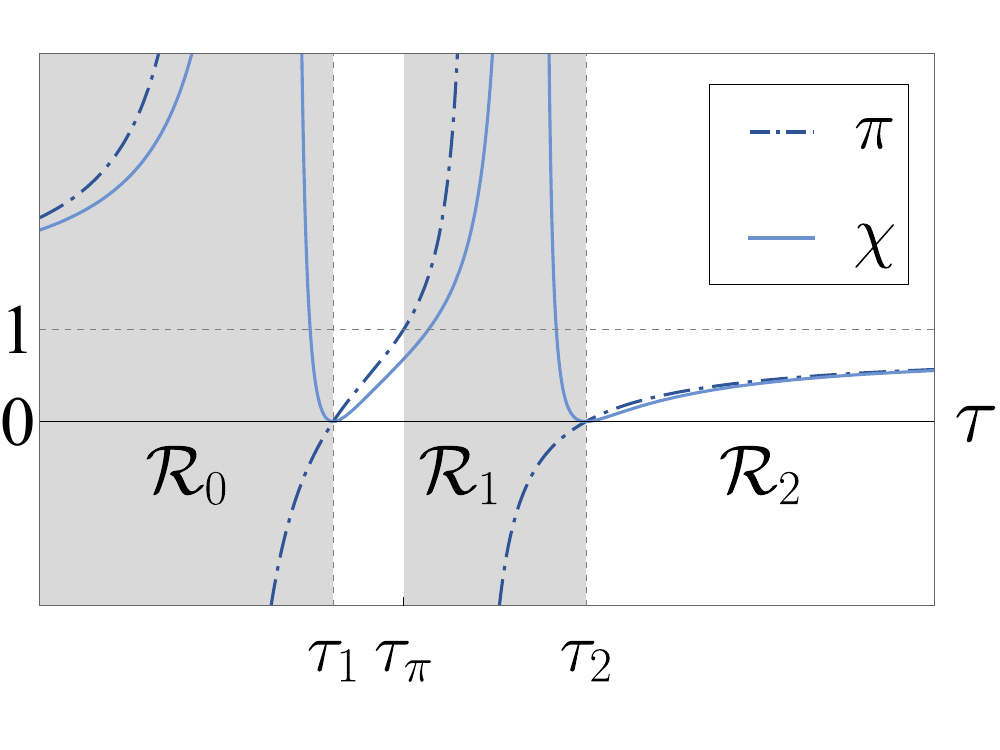}\\[-2mm] \centering{(c)}}}
			\vspace{-2mm}
			\caption{Time evolution of the hydrodynamic quantities of the ideal Szafron models for a fixed $r$. At the top, we plot the energy density $\rho(\tau,r)$ (light blue solid line) and the pressure $p(\tau,r)$ (dark blue dot-dashed line); these figures also indicate the spacetime regions defined by the singularities. At the bottom, we show the hydrodynamic quantities $\chi(\tau,r)$ (light blue solid line) and $\pi(\tau,r)$ (dark blue dot-dashed line), with the spacetime domains where the conditions for physical reality do not hold shaded. (a) Case $c \geq 0$, $\alpha > 0$. (b) Case $c \leq 0$, $\alpha < 0$ (the case $c < 0$, $\alpha>0$ is similar exchanging $\tau_1$ for $\tau_2$). (c) Case $c > 0$, $\alpha < 0$, $\sigma > 0$ and $\tau_1 < \hat{\tau}_{\rho}$; the temporal axis is represented in logarithmic scale. The case $c > 0$, $\alpha < 0$, $\sigma > 0$, and $\tau_1 > \hat{\tau}_{\rho}$ is similar but without the non-shaded domain $]\tau_1, \tau_{\pi}[$. For the same case (c), but with $\sigma < 0$, the situation is analogous, with the exchanges $\tau_\rho \, \leftrightarrow \, \hat{\tau}_\rho$ and $\tau_1 \, \leftrightarrow \, \tau_2$.}
		\label{Fig-7}
		\end{figure}
\\ \\
		Finally, a singularity appears when $\tau_2^{\sigma} \, \alpha(r) = -c^{-1}$. At this singularity, which is not simultaneous either ($\tau_2 = \tau_2(r)$), the metric distance on the coordinate lines of coordinate $r$ vanishes and we have infinite energy density again. \\
		Depending on how many of these singularities appear, three different cases can be distinguished (see Figure \ref{Fig-7}): 
		\begin{itemize}
\item[(i)] If $c \geq 0$ and $\alpha(r) > 0$, then only the singularity at $\tau = 0$ occurs. Consequently, the solution is defined in the full domain $\tau > 0$ (see Figure \ref{Fig-7}(a)).
\item[(ii)] If $c \leq 0$ and $\alpha < 0$, we have a singularity at $\hat{\tau}(r) = \tau_1$, and if $c < 0$ and $\alpha > 0$ we have a singularity at $\hat{\tau}(r) = \tau_2$. In both cases the singularity at $\tau = 0$ occurs. Now, two disconnected spacetimes domains exist: ${\cal R}_0 = \{0 < \tau < \hat{\tau}\}$ and $\hat{\cal R} = \{\hat{\tau} < \tau < \infty\}$ (see Figure \ref{Fig-7}(b)).
\item[(iii)] Finally, if $c > 0$ and $\alpha(r) < 0$, then we have all three singularities, and three disconnected spacetimes domains exist: ${\cal R}_0 = \{0 < \tau < \tau_1\}$, ${\cal R}_1 = \{\tau_1 < \tau < \tau_2\}$ and ${\cal R}_2 = \{\tau_2 < \tau < \infty\}$ (see Figure \ref{Fig-7}(c)).
		\end{itemize}

		\subsection[Analysis of the evolution: Energy and compressibility \\ conditions for $f(\tau) = \tau^q$]{Analysis of the evolution: Energy and compressibility conditions for \boldmath$f(\tau) = \tau^q$} \label{subsec-energy-c}
		Now we study the spacetime domains where the energy conditions E$^\textrm{G}$ given in (\ref{EG}) and the compressibility conditions H$_1^\textrm{G}$ given in (\ref{H1G}) hold. These domains are related to those considered above and defined by the spacetime singularities, but they also depend on the times at which the energy density vanishes or at which $\pi(\tau,r) = 1$. \\ \\ 
		All these times are not simultaneous for the comoving observer and depend on $\alpha(r)$. The last one, $\tau_{\pi}(r)$, is defined by the constraint $c \, \tau_{\pi}^{\sigma} \alpha(r) = \sqrt{(1 \! - \! \sigma)/(1 \! + \! \sigma)}$, which can be obtained from the expression $\pi(\tau, \alpha) = p(\tau)/ \rho(\tau,\alpha)$, where $p(\tau)$ and $\rho(\tau,\alpha)$ are given in (\ref{p-ideal}) and (\ref{rho-ideal-1}), respectively. \\ \\ 
		On the other hand, the energy density vanishes at two times, $\tau_{\rho}(r)$ and $\hat{\tau}_{\rho}(r)$, on the region ${\cal R}_0$ when $c > 0$ and $\alpha < 0$, and it vanishes at one of them when $c < 0$, or when $c = 0$ and $\alpha < 0$. These times are defined by the constraints $\, \tau_{\rho}^{\sigma} \alpha(r) = (1 \! - \! \sigma)/(1 \! + \! \sigma)$ and $c\, {\hat{\tau}}_{\rho}^{\sigma} \alpha(r) = (1 \! - \! \sigma)/(1 \! + \! \sigma)$. Finally, we must also consider the times $\tau_{\chi}(r)$, $\bar{\tau}_{\chi}(r)$ and $\hat{\tau}_{\chi}(r)$, defined by the condition $\chi(\tau, r) = 1$. \\ \\ 
		The role played by the above times in defining the spacetime regions where the energy conditions E$^\textrm{G}$ and the compressibility conditions H$_1^\textrm{G}$ hold depend on the signs of $\sigma$ and $c$. The results for $\sigma > 0$ are summarised in Table \ref{table-5}. For models with $\sigma < 0$, the results are the same, but exchanging $\tau_\rho \, \leftrightarrow \, \hat{\tau}_\rho$ and $\tau_1 \, \leftrightarrow \, \tau_2$ in the $c > 0$ and $\alpha(r) < 0$ case. Figure \ref{Fig-7} also shows the different possible cases.

		\begin{table*}[t]
			\begin{tabular}{ccccc}
				\noalign{\hrule height 1.05pt} \\[-3mm]
  				& & $\rho > 0$ & $0 < \pi < 1$ & $0 < \chi < 1$ \\[1mm]
				\hline \\[-3mm]
   				\! $\!c \geq 0 ,\; \alpha(r) > 0$ & & $[0, \infty[$ & $]\tau_\pi, \infty[$ & $]\tau_\chi,  \infty[ $ \\[0.8mm] 
				\\[-3.5mm]
				\hspace{-1mm} & $ \! \! \! \! \tau_1 > \hat{\tau}_\rho$ & $[0, \tau_\rho[ \, \cup \, ]\hat{\tau}_\rho, \tau_1[ \, \cup \, ]\tau_2, \infty[ $ & $]\tau_\pi, \tau_1[ \, \cup \, ]\tau_2, \infty[ $  & $]\tau_\chi, \infty[ $
				\\[-1mm]
				\! $\!c > 0, \; \alpha(r) < 0$ & & & & 
				\\[-2mm]
				\hspace{-1mm} & $ \! \! \! \! \tau_1 < \hat{\tau}_\rho$ & $[0, \tau_\rho[ \, \cup \, ]\tau_1, \hat{\tau}_\rho[ \, \cup \, ]\tau_2, \infty[ $ & $]\tau_\pi, \hat{\tau}_\rho[ \, \cup \, ]\tau_2, \infty[ $ & $]\hat{\tau}_\chi, \bar{\tau}_\chi[ \, \cup \, ]\tau_\chi, \infty[ $ \\[1mm] 
				\hline \\[-3mm]
				\! $\!c < 0, \; \alpha(r) > 0$ & & $[0, \hat{\tau}_\rho[ \, \cup \, ]\tau_2, \infty[$ & $]\tau_2, \infty[$ & $]\tau_\chi, \infty[$ \\[2mm]
				\! $\!c \leq 0, \; \alpha(r) < 0$ & & $[0, \tau_\rho[ \, \cup \, ]\tau_1, \infty[$ & $]\tau_1, \infty[$ & $]\tau_\chi, \infty[$   
				\\[1mm] \noalign{\hrule height 1.05pt}
			\end{tabular}
\caption{This table provides the space-time regions where the hydrodynamic conditions for physical reality hold for models with $\sigma > 0$, which differ depending on the sign of the parameter $c$. The boundary times $\tau_\pi(r)$, $\tau_\rho(r)$, $\hat{\tau}_\rho(r)$, $\tau_1(r)$, $\tau_2(r)$, $\tau_\chi(r)$, $\bar{\tau}_{\chi}(r)$, and $\hat{\tau}_{\chi}(r)$, are defined in Sections \ref{subsec-singularities} and \ref{subsec-energy-c}. For models with $\sigma < 0$, the results are the same, but exchanging $\tau_\rho \, \leftrightarrow \, \hat{\tau}_\rho$ and $\tau_1 \, \leftrightarrow \, \tau_2$ in the $c > 0$ and $\alpha(r) < 0$ case.}
\label{table-5}
		\end{table*}

		\subsection[Thermodynamic schemes for $f(\tau) = \tau^q$]{Thermodynamic schemes for \boldmath$f(\tau) = \tau^q$} \label{subsec-scheme-ideal}
		Now that we have analysed the hydrodynamic constraints for physical reality for the ideal case, those regarding the hydrodynamic quantities, we can proceed to analyse the thermodynamic ones. To do so, the thermodynamic schemes of this particular subset of solutions must be obtained. This can easily be done by substituting (\ref{g-ideal}) and (\ref{beta(alpha)-ideal}) in (\ref{n-general-R}) and (\ref{temperatura-general-R}), and using (\ref{p-ideal}). It is worth remarking that, in these particular cases, the metric function $\alpha(r)$ is a function of state given by (\ref{alpha(rho,p)-1}). Furthermore, we have that $\dot{\alpha} = 0$ and, therefore, it is a particular solution of $u(s) = 0$. \\ \\
		Taking all this into account, we have that the ideal Szafron models have a specific entropy $s$ which is an arbitrary function of the function of state $\alpha(\rho,p)$ given in (\ref{alpha(rho,p)-1}), $s(\rho,p) = s(\alpha)$. Moreover, they have a matter density given by
		\begin{equation} \label{n(rho,p)-ideal-1}
			n(\rho,p) \! = \! \frac{1}{N(\alpha)[c\alpha^2 \tau^{1 \! + \! \sigma} \! + \! (1 \! + \! c) \alpha \tau \! + \! \tau^{1 - \sigma}]} , 
		\end{equation}
and the temperature $\Theta(\rho,p)$ is given by (\ref{temperatura-general-R-a}), with
		\begin{subequations} \label{temperatura-ideal-R}
			\begin{eqnarray}
				\quad \tau_1(\tau) \equiv \frac{1 \! - \!\sigma}{2} \tau^{-(1 + \sigma)}, \quad \ r_1(r) \equiv \frac34 N'/s' , \qquad \qquad \ \\[-0mm]
				\quad \tau_2(\tau) \equiv \frac{1 \! + \! \sigma}{2} \tau^{-(1 - \sigma)} , \quad \ r_2(r) \equiv \frac34 c (N \alpha^2)'/s' , \qquad \\[0mm]  
				\; \tau_3(\tau) \equiv \frac{1 \! - \! \sigma^2}{2} \tau^{-1}, \quad \quad \, r_3(r) \equiv \frac34 (1 \! + \! c)(N \alpha)'/s' , \, \;
			\end{eqnarray}
		\end{subequations}
where $N(\alpha)$ is an arbitrary function of $\alpha(\rho,p)$, and $\tau = \tau(p) \equiv \sqrt{(1 \! - \! \sigma^2)/(3p)}$. \\ \\
		The set of thermodynamic schemes $\{n, \epsilon, s, \Theta\}$ associated with each ideal Szafron model presented above defines a family of fluids that gives the different interpretations of the solutions, and solves the inverse problem for this case. One of these schemes corresponds to the generic ideal gas, namely, the one that fulfils the ideal gas EoS (\ref{eq. estat gas ideal}). \\ \\
		In Section \ref{sec-generic-ideal-gas}, we have given an algorithm to obtain this ideal gas thermodynamic scheme from the indicatrix function $\chi(\pi)$. This algorithm involves determining (\ref{e(pi) i f(pi)}), two quadratures that cannot be computed for the indicatrix function (\ref{chi(pi)-1}). Nevertheless, we can alternatively look for the functions $s(\alpha)$ and $N(\alpha)$ that lead to this ideal scheme. Indeed, in this case the temperature $\Theta$ and the specific energy density $e = \rho/n$ depend on $\pi$, $\Theta = \Theta(\pi)$ and $e = e(\pi)$. Then, if we impose these conditions on the expressions of $n(\rho,p)$ given in (\ref{n(rho,p)-ideal-1}) and of $\Theta(\rho,p)$ given in (\ref{temperatura-general-R}, \ref{temperatura-ideal-R}), we obtain the ideal gas thermodynamic schemes if
		\begin{equation}
			N(\alpha) = n_1 \alpha^{\frac{1 - \sigma}{\sigma}} , \qquad s(\alpha) = s_0 + s_1 \alpha^2 ,
		\end{equation}
where $s_1 = 9 \tilde{k}/(16 \sigma)$. 
		The positivity conditions P given in (\ref{P}) and the compressibility conditions H$_2$ given in (\ref{H2}) must be required for each thermodynamic scheme to define a physically realistic fluid. For the ideal gas scheme, conditions P hold if we take $n_1 < 0$. Moreover, H$_2$ becomes H$_2^{\rm G}$ given in (\ref{H2G}). Note that $\chi(0) = \chi'(0) = 0$, $\eta(0) = 0$ and $\eta'(0) = 1$. Consequently, condition (\ref{H2G}) does not hold in a neighbourhood of zero. Nevertheless, for the three cases considered in Figure \ref{Fig-6} there exists a $\pi_m$ such that the indicatrix function $\chi(\pi)$ fulfils this constraint if $\pi > \pi_m > 0$. 
		
	\section[The ideal model $f(\tau) = \sqrt{\tau}$]{The ideal model \boldmath$f(\tau) = \sqrt{\tau}$} \label{sec-t^1/2}
	We now consider the case $q = 1/2$, that is, $f(\tau) = \sqrt{\tau}$, and study whether it is compatible with the ideal sonic condition (\ref{ISC}). For this election, the solution of the field equations takes form (\ref{sol. general eq. Z}) with
	\begin{equation} \label{g-ideal-2}
		f(\tau) = \sqrt{\tau} , \qquad g(\tau) = \sqrt{\tau} \ln \tau \, .
	\end{equation}
From these expressions we now obtain that the only functions $T_i = T_i(\tau)$ that do not vanish are
	\begin{equation} \label{eq:T-Szafron-2}
		T_2 = -T_3 = \frac18 \tau^{-6} .
	\end{equation}
Thus, the ideal sonic condition (\ref{ISC}) holds when $R_2 = R_3$. The expressions (\ref{eq:R}) for these functions imply that the functions $\alpha(r)$ and $\beta(r)$ fulfil the following relation:
	\begin{equation} \label{beta(alpha)-ideal-2}
		\beta(\alpha) = \alpha + \tilde{c}\, ,  
	\end{equation}
where $\tilde{c} \neq 0$ is a constant. Then, taking into account that (\ref{beta(alpha)-ideal-2}) is a differential equation that relates $a(r)$ and $b(r)$, we obtain that a solution of the perfect fluid Einstein equations which is compatible with the ideal sonic condition S$^\textrm{G}$ (\ref{cond. sonica ideal}) is given by the metric (\ref{metric-Y}) with the following election of the metric function $Y(\tau,r)$:
	\begin{equation} \label{Y-ideal-2}
		Y = Z^{2/3} , \quad Z(\tau,r) = b(r)\sqrt{\tau} \, [\ln \tau + \alpha(r)] , 
	\end{equation}
where $b(r)$ is given by
	\begin{equation} \label{b-ideal-2}
		b(r) = e^{\alpha(r)/\tilde{c}} \, , \quad \tilde{c} \neq 0 \, .
	\end{equation}
	From (\ref{expansion-Y}), the expansion in this case is 
	\begin{equation} \label{expansion-ideal-2} 
		\theta = \frac{1}{\tau}\left[ 1 + \frac{1}{\alpha + \ln \tau} + \frac{1}{\alpha + \tilde{c} + \ln \tau}\right] \, .
\end{equation} 
Again, the change $\tau \leftrightarrow -\tau$, with $\tau < 0$, leads to contracting models, whose properties are similar to those we study below for the expanding models.

		\subsection[Hydrodynamic quantities: Energy density, pressure, and speed of sound for $f(\tau) = \sqrt{\tau}$]{Hydrodynamic quantities: Energy density, pressure, and speed of sound for \boldmath$f(\tau) = \sqrt{\tau}$} \label{subsec-hydro-ideal-2}
		Now, the pressure also takes the expression (\ref{p-ideal}), and the energy density is
		\begin{equation} \label{rho-ideal-2}
			\rho = \frac{1}{3\tau^2} \frac{(2 + \alpha + \ln \tau)(2 + \alpha + \tilde{c} + \ln \tau)}{(\alpha + \ln \tau)(\alpha + \tilde{c} + \ln \tau)} .
		\end{equation}
From this expression we obtain
		\begin{eqnarray} \label{alpha(rho,p)-2}
			\hspace{-4.0mm} 
			\alpha = \tilde{\alpha} (\rho,p) \equiv \frac{4\pi + \varepsilon \tilde{F}(\pi)}{2(1 \! - \! \pi)} + \frac12 \ln (3p) \! - \! \frac{\tilde{c}}{2} \, , \\[2mm]
			\label{F(pi)b}
			\tilde{F}(\pi) \equiv \sqrt{\tilde{c}^2(1 - \pi )^2 + 16\pi} \, .
		\end{eqnarray}
Then, we can determine the indicatrix function $\chi = \tilde{\chi}(\pi)$,
		\begin{equation} \label{chi(pi)-2}
			c_s^2 = \tilde{\chi}(\pi) \equiv \frac{4\pi^2[ \tilde{c}^2 (1 + \pi) + 2 \varepsilon \tilde{F}(\pi)]}{(1 + \pi) [ \tilde{c}^2 (1 + \pi)^2 + 16\pi]} \, . 
		\end{equation}
		It is worth remarking that this ideal model can be obtained from the ideal Szafron model studied in the previous section by taking the limit $\sigma \rightarrow 0$, $c \rightarrow 1$ and $\hat{c} = (1 - c)/\sigma \rightarrow \tilde{c}$. Consequently, to study this model we can start from the expressions obtained in this subsection, or we could sometimes use the analysis already made for the ideal Szafron models. \\ \\
		Now, when $\tau \rightarrow 0$ or when $\tau \rightarrow \infty$, the solution becomes a shift FLRW model $p = \rho$, as we can deduce by taking $\sigma = 0$ in (\ref{limits}).
		
		\subsection[Fluid properties: Compressibility conditions $\textrm{H}^\textrm{G}_1$ for $f(\tau) = \sqrt{\tau}$]{Fluid properties: Compressibility conditions \boldmath$\textrm{H}^\textrm{G}_1$ \\ for $f(\tau) = \sqrt{\tau}$} \label{subsec-compress-ideal-2}
		We must analyse the behaviour of the equation of state $\tilde{\chi}(\pi)$ in the interval $0 < \pi < 1$ where the energy conditions E$^\textrm{G}$ hold. Now, $\tilde{\chi}(\pi)$ defines a family of fluids depending on the parameter $\tilde{c}^2$ and the sign $\varepsilon$. \\ \\
		The function $\tilde{\chi}(\pi)$ can have two different behaviours depending on the sign $\varepsilon$. If we choose the positive sign, $\varepsilon = +1$, $\tilde{\chi}(\pi)$ is an increasing function, and the first of the compressibility conditions H$_1^{\rm G}$ holds in the whole interval $0 < \pi < 1$, and $\tilde{\chi}(1) = 1$. This case appears as the limit of the ideal Szafron model plotted in Figure \ref{Fig-6}(a). \\ \\
		If $\varepsilon = -1$, $\tilde{\chi}(\pi)$ identically vanishes when $|\tilde{c}| = 2$. When $|\tilde{c}| > 2$, it is an increasing function, and the causal condition holds in the whole interval $0 < \pi < 1$. This case appears as the limit of the ideal Szafron model plotted in Figure \ref{Fig-6}(c). Otherwise, when $|\tilde{c}| < 2$, $\tilde{\chi}(\pi)$ is a negative function and it does not fulfil the causal condition in any interval. \\ \\
		The second compressibility condition H$_1^{\rm G}$ holds in the interval where $\tilde{\chi}(\pi)$ is well defined and fulfils the causal condition in the cases considered above.
		
		\subsection[Curvature singularities and spacetime domains for $f(\tau) = \sqrt{\tau}$]{Curvature singularities and spacetime domains \\ for \boldmath$f(\tau) = \sqrt{\tau}$} \label{subsec-singularities-2}
		For this model we can also have up to three different singularities. First, a big bang singularity at $\tau = 0$. Secondly, the metric line element of the sphere vanishes and the energy density is infinite at $\tilde{\tau}_1 = \tilde{\tau}_1(r) = e^{-\alpha(r)}$. Finally, a singularity could appear at $\tilde{\tau}_2 = \tilde{\tau}_2(r) = e^{-\alpha(r)-\tilde{c}}$, and then the metric distance on the coordinate lines of coordinate $r$ vanishes and we have infinite energy density again. \\ \\
		Now, two different situations can be distinguished:
		\begin{itemize}
\item[(i)] 
If $|\tilde{c}| = 2$, we have the singularity at $\tau = 0$ and also the one at $\hat{\tau}(r)$, where $\hat{\tau} = \tau_1$ if $\tilde{c} = 2$, and $\hat{\tau} = \tau_2$ if $\tilde{c} = -2$. Hence, two disconnected spacetime domains exist: ${\cal R}_0 = \{0 < \tau < \hat{\tau}\}$ and $\hat{\cal R} = \{\hat{\tau} < \tau < \infty\}$.
\item[(ii)] 
If $|\tilde{c}| \neq 2$, then we have all three singularities and three disconnected spacetime domains exist: ${\cal R}_0 = \{0 < \tau < \tau_m\}$, ${\cal R}_m = \{\tau_m < \tau < \tau_M\}$ and ${\cal R}_M = \{\tau_M < t < \infty\}$, where $\tau_m = {\rm min}\{\tau_1,\tau_2\}$ and $\tau_M = {\rm max}\{\tau_1,\tau_2\}$.
		\end{itemize}
Note that these two cases are similar to cases (ii) and (iii) considered in Section \ref{subsec-singularities} for the ideal Szafron models (see Figures \ref{Fig-7}(b) and \ref{Fig-7}(c)).
		
		\subsection[Analysis of the solutions and energy conditions for $f(\tau) = \sqrt{\tau}$]{Analysis of the solutions and energy conditions \\ for \boldmath$f(\tau) = \sqrt{\tau}$} \label{subsec-energy-c-2}
		The spacetime domains where the energy conditions E$^{\rm G}$ and the compressibility conditions H$_1^{\rm G}$ hold depend on the value of $\tilde{c}$. These domains are defined by the times $\tau_1(r)$ and $\tau_2(r)$ that determine the singularities and the times $\tau_\rho = \tau_\rho(r) = e^{-\alpha(r) - 2}$, $\tilde{\tau}_\rho = \tilde{\tau}_\rho(r) = e^{-\alpha(r) - \tilde{c} - 2}$ and $\tau_\pi = \tau_\pi(r) = e^{-(2 + 2\alpha(r) + \tilde{c})/2}$. \\ \\
		We have a behaviour that is similar to some cases of the ideal Szafron models summarised in Table \ref{table-5}. If $|\tilde{c}| = 2$ (respectively, $|\tilde{c}| < 2$ or $|\tilde{c}| > 2$) the behaviour is that of the fifth and sixth rows (respectively, third and fourth rows) in Table \ref{table-5}. Exchanging $\tilde{c} \leftrightarrow -\tilde{c}$ produces the exchange of $\tau_1 \leftrightarrow \tau_2$ and $\tau_\rho \leftrightarrow \tilde{\tau}_\rho$. 
		
		\subsection[Thermodynamic schemes for $f(\tau) = \sqrt{\tau}$]{Thermodynamic schemes for \boldmath$f(\tau) = \sqrt{\tau}$}
		The specific entropy $s$ is again an arbitrary function of the function of state $\alpha(\rho,p)$ (\ref{alpha(rho,p)-2}), $s(\rho,p) = s(\alpha)$. Moreover, the matter density is given by
		\begin{equation} \label{n(rho,p)-ideal}
			n(\rho,p) = \frac{4\sqrt{3p}}{N(\alpha)[2\alpha - \ln(3p)][2\alpha + 2\tilde{c} - \ln(3p)]} \, , 
		\end{equation}
where $N(\alpha)$ is an arbitrary function of $\alpha(\rho,p)$. The temperature $\Theta(\rho,p)$ is given by (\ref{temperatura-general-R-a}), with 
		\begin{subequations} \label{temperatura-ideal-2}
			\begin{eqnarray}
				\ \tau_1(\tau) \equiv \frac{1}{\tau}(1 + \ln \tau) , \qquad \, r_1(r) \equiv \frac34 N'/s' ,   \qquad \qquad \quad \ \ \qquad \\[-0mm]
				\ \tau_2(\tau) \equiv \frac{1}{2\tau} , \qquad \qquad \quad \ \ r_2(r) \equiv \frac34 [N \alpha (\alpha + \tilde{c})]'/s' , \quad \qquad \
	  \\[0mm]  
				\ \tau_3(\tau) \equiv \frac{1}{2t}(1 + \frac12 \ln \tau) , \quad r_3(r) \equiv \frac34 [N (2 \alpha + \tilde{c})]'/s' , \, \ \quad \qquad
			\end{eqnarray}
		\end{subequations}
where $\tau = \tau(p) \equiv 1/\sqrt{3p}$.
		In this case, the ideal thermodynamic scheme, the one fulfilling the EoS (\ref{eq. estat gas ideal}), can be obtained by taking $N(\alpha) = n_0 e^{- \alpha}$ and $s(\alpha) = s_0 + s_1 \alpha$, with $n_0 > 0$ and $s_1 = -9\tilde{k}/8$. \\ \\ \\ \\ \\
		Analogously to the case considered in the previous section, condition (\ref{H2G}) does not hold in a neighbourhood of zero. Nevertheless, for the cases in which the causal condition holds, there exists a $\tilde{\pi}_m$ such that the indicatrix function $\tilde{\chi}(\pi)$ fulfils this constraint if $\pi > \tilde{\pi}_m > 0$.
	
	\section{On the models with homogeneous temperature} \label{sec-T(t)}
	As explained in Section \ref{subsec-intro-thermal-conduc-coeff}, when a non-perfect fluid admits particular evolutions in which the dissipative fluxes vanish, these evolutions are well described by a perfect energy tensor. Moreover, the shear, the expansion and the acceleration of the fluid undergo strong restrictions as a consequence of the constitutive equations. Specifically, if the thermal conductivity coefficient does not vanish, then the fluid acceleration is constrained by (\ref{Fourier}). \\ \\
	After these considerations, we can look for perfect fluid solutions to the Einstein equations that describe both (i) a thermodynamic perfect fluid in local thermal equilibrium, and (ii) an inviscid (with negligible shear and bulk viscosity coefficients) non-perfect fluid in equilibrium. Then, the thermal conductivity coefficient does not vanish and, when the fluid flux is geodesic as the solutions we are considering here, equation (\ref{Fourier}) implies a homogeneous temperature $\Theta = \Theta(\tau)$. \\ \\
	Thus, a forthcoming study we can address is to find the solutions with homogeneous temperature. To do so, we must analyse the compatibility of the expression of the temperature (\ref{temperatura-general-R}), with the constraint $\Theta = \Theta(\tau) \neq 0$. This analysis requires us to consider different cases. \\ \\
	One possibility is to look for the thermodynamic schemes $\{s(\alpha), N(\alpha)\}$ and the metric function $\beta(\alpha)$ that are compatible with a homogeneous temperature for any solution $\{f(\tau), g(\tau)\}$ of the field equations. In this case, the three functions $r_i(r)$ given in (\ref{temperatura-general-R}) are constant and then a straightforward calculation leads to
	\begin{equation} \label{T(t)-beta}
			\beta(\alpha) = \frac{\beta_0 \alpha + \beta_1}{\beta_2 \alpha - \beta_0} , \qquad s(\alpha) = \frac{n_0 \alpha \beta(\alpha) - s_1}{s_0 - n_1 \alpha \beta(\alpha)}, \qquad N(\alpha) = n_0 + n_1 s(\alpha) ,
	\end{equation}
with $\beta_0 = n_1 s_1 - n_0 s_0$. Note that $\beta_1$ and $\beta_2$ cannot be cancelled simultaneously (this leads to $s'(\alpha) = 0$). Thus, the above expression of $\beta(\alpha)$ is compatible with neither (\ref{beta(alpha)-ideal}) nor (\ref{beta(alpha)-ideal-2}). Moreover, the three functions $\tau_i(\tau)$ given in (\ref{temperatura-general-R}) are independent for the ideal models studied in Sections \ref{sec-t^q} and \ref{sec-t^1/2}. Then, these models cannot represent an inviscid fluid with a non-vanishing thermal conductivity coefficient. The compatibility of (\ref{T(t)-beta}) with the general ideal sonic condition (\ref{ISC}) leads to a system of five fourth-order differential equations, which we leave for future work. \\ \\
	When at least one of the functions $r_i(r)$ is non-constant, we can consider different cases that lead to solutions admitting thermodynamic schemes with homogeneous temperature. They will be analysed in the future too.

\chapter{The thermodynamic Stephani universes} \label{chap-Stephani}
We finish the second part of this thesis with the thermodynamic study of the Stephani universes, a work based on the results published in \cite{SFM-Stephani}. The \textit{Stephani universes} are the conformally flat solutions of the Barnes-Stephani metrics \cite{Stephani,Barnes}. They can also be characterised as the spacetimes verifying a weak cosmological principle without any hypothesis on the energy tensor \cite{bc-0, Bona-Coll}. \\ \\
Bona and Coll \cite{Bona-Coll} showed that the necessary and sufficient condition for a Stephani universe to represent the evolution of a fluid in l.t.e. is to admit a three-dimensional isometry group on two-dimensional orbits. This result was later recovered in \cite{Krasinski-et-al}, and spherically symmetric Stephani universes that may be interpreted either as a classical monoatomic ideal gas or as a matter-radiation mixture were considered in \cite{Sussman}. In \cite{CF-Stephani}, Coll and Ferrando studied the Stephani universes that can be interpreted as a generic ideal gas in l.t.e. \\ \\
Despite these results, there are a lot of questions to deal with concerning the physical meaning of the Stephani universes. In Section \ref{sec-TSU}, we obtain the speed of sound, $c_s^2 = \chi(\rho,p)$, for a generic thermodynamic Stephani universe, and we outline several approaches to undertake the field equations. We also determine the corresponding associated thermodynamic schemes $\{n, s , \Theta\}$. \\ \\
In order to understand better the physical meaning of the thermodynamic Stephani universes we must demand complementary significant physical qualities. In Section \ref{sec-qualities} we analyse how some of these constraints restrict the models. Firstly, we impose the {\em ideal sonic condition}, $\chi = \chi(\pi) \neq \pi$, which leads to the ideal gas Stephani universes. These models have already been studied \cite{CF-Stephani, Coll_Ferrando_i_Saez_2020b} and here we summarise some results that we will use later. Secondly, we analyse the compatibility of the solutions with a fluid with non-vanishing thermal conductivity coefficient, and we show that only the FLRW models are possible. Finally, we study the constraints on the models approaching a classical ideal gas at low temperatures. And finally, we determine the restrictions when we demand a good behaviour at high temperatures. \\ \\
Section \ref{sec-Isotropic} is devoted to studying the perfect fluids with an irrotational unit velocity measuring isotropic radiation. Starting from the result of Clarkson and Barrett \cite{Clarkson-1999}, we derive the constraints on the metric line element and we write it for the spherically symmetric case. We show that the {\em Dabrowski metric} \cite{Dabrowski}, which was considered in \cite{Clarkson-1999} as a cosmological model with isotropic radiation, is a solution that does not fulfil the macroscopic constraint for physical reality as a fluid in local thermal equilibrium. We also give some general properties of the ideal gas models with isotropic radiation. \\ \\
Finally, in Section \ref{sec-Model-Nos} we study the Stephani universes modelling an ultrarelativistic gas with the comoving observer measuring isotropic radiation. They approximate a Synge gas at high temperatures and fulfil the compressibility conditions. The so-called singular model is analysed in detail by obtaining the spacetime regions where the energy conditions hold, getting the time evolution and radial profile of the thermodynamic quantities, and studying the generalised Friedmann equation. 

	\section{The thermodynamics of the Stephani universes} \label{sec-TSU}
	As mentioned above, Bona and Coll showed in \cite{Bona-Coll} that the Stephani universes that model the evolution of a fluid in l.t.e. are those admitting a three-dimensional isometry group on two-dimensional orbits, G$_3$/S$_2$. They also showed that the metric line element of the thermodynamic Stephani universes may be written as:
	\begin{subequations} \label{metrica}
		\begin{eqnarray}
			\textrm{d} s^2 = -\alpha^2 \textrm{d} t^2 + \Omega^2 (\textrm{d} x^2 + \textrm{d} y^2 + \textrm{d} z^2) \, ; \qquad \quad \label{the-ste-uni} \\[2mm]
			\alpha \equiv R \partial_R \ln L \, , \quad  
			\Omega \equiv \frac{w}{2z} L\, , \quad L \equiv \frac{R(t)}{1 + b(t) w} \, , \quad \label{eq:termetric-1} \\[0mm] 
			w \equiv \frac{2z}{1 + {\varepsilon \over 4}r^2} \, , \qquad r^2 \equiv x^2 + y^2 + z^2 \, , \qquad \quad \label{eq:termetric-2}
		\end{eqnarray}
	\end{subequations}
$R(t)$ and $b(t)$ being two arbitrary functions of time. Its symmetry group is spherical, plane or hyperbolical depending on whether $\varepsilon$ is $1$, $0$ or $-1$. \\ \\
	Furthermore, the fluid unit velocity is $u = (1/\alpha)\partial_t$, and the energy density, the pressure, the expansion and the 3-space curvature are given by
	\begin{eqnarray}
		\rho = \displaystyle \frac{3}{R^2} (\dot{R}^2 + \varepsilon - 4 b^2) , \qquad p = - \rho - {R \over 3} {\rho'(R) \over \alpha} , \qquad \label{tdp-1} \\[0mm]
		\theta \displaystyle = \frac{3\dot{R}}{R} \neq 0 , \quad \qquad \kappa \displaystyle = \frac{1}{R^2} (\varepsilon - 4 b^2) , \qquad \label{tdp-2}
	\end{eqnarray}
where, for a function $f$ depending on the coordinate $t$, we may write $f = f(R)$ and $f' = f'(R)$. Note that the metric and the invariant quantities depend on two arbitrary functions of time $\{R(t), b(t)\}$. \\ \\
	The FLRW limit occurs when one of the following three equivalent conditions holds: (i) $b(t) = constant$, (ii) the fluid flow is geodesic ($\alpha = 1$), (iii) the pressure is homogeneous, $p = p(t)$.
	
		\subsection[Speed of sound: indicatrix function $\chi(\rho, p)$]{Speed of sound: indicatrix function \boldmath$\chi(\rho, p)$} \label{subsec-Chi-Stephani}
		Due to the symmetries of the metric line element (\ref{metrica}), all scalar invariants depend on two functions $(t, w)$ at most. Then, the hydrodynamic sonic condition S given in (\ref{cond. sonica}) is automatically fulfilled and step 1 in the procedure presented in Section \ref{sec-other-phys-re-conds} is achieved for the full set of solutions. Now, we study the general expression of $\chi(\rho,p)$, which collects all the thermodynamic information that can be expressed using exclusively hydrodynamic quantities. \\ \\
		From the second equation in (\ref{tdp-1}), a direct calculation leads to:
		\begin{eqnarray}
			\label{eq:pi0}
			\pi = \frac{p}{\rho} = \frac{a}{\alpha} -1 , \qquad a = a(R) \equiv - \frac{R\, \rho'(R)}{3 \rho} , \qquad \\
 			\chi \equiv \frac{u(p)}{u(\rho)} = \frac{\partial_R (\rho \pi)}{\partial_R \rho} = \pi - \frac{R}{3a} \partial_R \left(\frac{a}{\alpha}\right) , \qquad
		\end{eqnarray}
and, from the definition of $\alpha$ in (\ref{eq:termetric-1}), it follows that
		\begin{equation} \label{eq:alpha}
  			\alpha = \alpha(R,w) \equiv \frac{1 + (b - R b') w}{1 + b w}.
		\end{equation}
Then, from these expressions, we obtain:
		\begin{eqnarray}
  			\pi \! = \! \pi(R, w) \! \equiv \! \frac{a(1 + bw)}{1 + (b - R b')w} - 1 , \qquad \qquad \qquad \label{eq:pi} \\
			\chi \! = \! \chi(\pi, \! R) \! \equiv \! \pi \! + \! \frac{1}{3} \! + \! \frac{1}{3}(\pi \! + \! 1)[(\pi \! + \! 1)A_1(R) \! + \! A_2(R)], \quad \quad \label{eq:chi} \\[1mm] 
			A_1(R) \equiv -\frac{Rb''}{a^2b'} , \qquad A_2(R) \equiv \frac{Rb''}{ab'} - \frac{a'R}{a^2} - \frac{1}{a}. \qquad \quad \label{Ai(R)}
		\end{eqnarray}
Since $\rho$ is an effective function of $R$, the functions $A_1$ and $A_2$ can be considered as depending on $\rho$. Therefore, we get that the speed of sound, $c_s^2 = \chi(\rho,p)$, of a thermodynamic Stephani universe (\ref{metrica}) is given by
		\begin{equation} \label{eq:chi(rho,p)}
			\chi(\rho, p) \! = \! \pi \! + \! \frac{1}{3} \! + \! \frac{1}{3}(\pi \! + \! 1)[(\pi \! + \! 1) A_1(\rho) \! + \! A_2(\rho)], \quad
		\end{equation}
where $\pi = p/\rho$, and $A_1(\rho)$ and $A_2(\rho)$ are two real functions.
		Every choice of these two functions determines the indicatrix function, which fixes the hydrodynamic properties of a specific thermodynamic Stephani universe.
		
		\subsection{On the generalised Friedmann equations} \label{subsec-eq-Stephani}
		When studying the physical properties of the solutions, we can adopt different approaches. On the one hand, we can give the functions of time $R(t)$ and $b(t)$, which determine a solution, and from (\ref{tdp-1}) and (\ref{Ai(R)}), determine the functions $\rho(R)$, $A_1(R)$ and $A_2(R)$. Then, we can obtain $R(\rho)$, and (\ref{eq:chi(rho,p)}) gives the indicatrix function $\chi(\rho,p)$, which would have to be analysed to know the thermodynamic meaning of this specific solution. \\ \\
		On the other hand, we can prescribe the functions $A_1(\rho)$ and $A_2(\rho)$ so that the indicatrix function $\chi(\rho,p)$ has specific physical properties. This choice defines a differential system for the metric functions $R(t)$ and $b(t)$, that must be solved.
		This second approach is the one Coll and Ferrando take when studying the ideal gas Stephani universes \cite{CF-Stephani}. The ideal gas equation of state requires the indicatrix function to be of the form $\chi = \chi(\pi)$, $\pi = p/\rho$. As a result, $A_1(\rho)$ and $A_2(\rho)$ must be constant functions. The study of the subsequent equations (\ref{Ai(R)}) leads us to distinguish between the regular and the singular models, yielding five possible classes of ideal gas Stephani models \cite{CF-Stephani, Coll_Ferrando_i_Saez_2020b} (see Section \ref{subsec-ideal} below). \\ \\
		When studying the field equations for a given choice of the functions $A_i(\rho)$, it could be suitable to consider all the functions of $t$ as depending on the variable $\rho$. Then, equations (\ref{Ai(R)}) are equivalent to
		\begin{eqnarray}
			\hspace{-8mm} a_1(\rho) R^2 - R''(\rho) R - R'(\rho)[a_2(\rho) R \! - \! 2 R'(\rho)] = 0, \label{R(rho)} \\[1mm]
			\displaystyle b(\rho) = \! \int \! \left[ R^2(\rho) e^{\! - \! \int \! a_2(\rho) \textrm{d} \rho}\right] \textrm{d} \rho , \label{b(rho)} \\[0mm]
			a_1(\rho) \equiv \frac{A_1(\rho)}{9 \rho^2} , \qquad a_2(\rho) \equiv \frac{A_2(\rho) + 3}{3 \rho} .
		\end{eqnarray}
Thus, we obtain the second-order differential equation (\ref{R(rho)}) for $R(\rho)$. Once solved, expression (\ref{b(rho)}) determines $b(\rho)$. Finally, we must solve the generalised Friedmann equation for $\rho(t)$ that follows from (\ref{tdp-1}):
		\begin{equation} \label{gfe}
			\rho R^2(\rho) = \displaystyle 3 \, [R'(\rho)^2 \dot{\rho}^2 + \varepsilon - 4\, b(\rho)^2] .
		\end{equation}

		\subsection{Thermodynamic schemes: entropy, matter density \\ and temperature} \label{subsec-Schemes-Stephani}
		Each of the solutions considered above can be furnished with a family of thermodynamic schemes $\{n, \epsilon, s, \Theta\}$, which offer different thermodynamic interpretations of this solution. In Section \ref{sec-lte}, we have shown that the specific entropies $s$ and the matter densities $n$ associated with $T$ are of the form $s = s(\bar{s})$ and $n = \bar{n}/N(\bar{s})$, where $s(\bar{s})$ and $N(\bar{s})$ are arbitrary real functions of a particular solution $\bar{s} = \bar{s}(\rho, p)$ to the equation $u(s) = 0$, and $\bar{n} = \bar{n}(\rho,p)$ is a particular solution to equation (\ref{matter-conservation}). The metric function $w$ given in (\ref{eq:termetric-2}) is a function of state that plays an important role in obtaining these thermodynamic schemes. From expression (\ref{eq:pi}) we obtain
		\begin{equation} \label{w(rhop)}
			w = \frac{\pi + 1 -a(R)}{(\pi \! + \! 1)[R b'(R) \! - \! b(R)] \! + \! a(R)b(R)} \equiv w(\rho,p) \, .
		\end{equation}
Note that $w = w(\rho,p)$ is a function of state whose dependence on $p$ is explicit, while its dependence on $\rho$ is partially implicit through the function of time $R(\rho)$. Moreover, we have that $w$ fulfils $u(w) = 0$. Consequently, the specific entropy is an arbitrary real function depending on $w$, $s = s(w)$. \\ \\
		On the other hand, from the expression (\ref{tdp-2}) of the expansion, it follows that $\bar{n} = L^{-3}$ is a particular solution of the matter conservation equation (\ref{matter-conservation}). Then, taking into account the expression (\ref{eq:termetric-2}) of $L$, we obtain that the thermodynamic schemes associated with a thermodynamic Stephani universe (\ref{metrica}) are determined by a specific entropy $s$ and a matter density $n$ of the form
		\begin{equation} \label{s-n-Stephani}
			s(\rho, p) \! = \! s(w) , \qquad \quad n(\rho,p) \! = \! \frac{(1 + b w)^3}{R^3 \, N(w)} ,
		\end{equation}
where $s(w)$ and $N(w)$ are two arbitrary real functions of the function $w = w(\rho, p)$ given in (\ref{w(rhop)}), and $b(R)$ and $R$ depend on $\rho$ through the function $R = R(\rho)$. \\ \\
		The temperature of the thermodynamic scheme defined by each pair $\{s, n\}$ given above can be obtained from the thermodynamic relation (\ref{thermo-first-law-h}) as 
		\begin{equation} \label{T-Stephani-1}
			\Theta = - \frac{\rho + p}{n^2}\left[\frac{\partial n}{\partial s}\right]_{\rho} = - \frac{\rho + p}{n^2s'(w)}\left[\frac{\partial n}{\partial w}\right]_{R} . \qquad \quad  
		\end{equation}
Then, taking into account expressions (\ref{eq:pi0}), (\ref{eq:alpha}) and (\ref{s-n-Stephani}), we obtain that, for a thermodynamic Stephani universe (\ref{metrica}), the temperature $\Theta$ of the thermodynamic schemes (\ref{s-n-Stephani}) takes the expression
		\begin{equation} \label{T-Stephani}
			\Theta = \frac{(\rho \! + \! p) R^3[N'(w)(1 \! + \! b w) \! - \! 3 N(w) b]}{s'(w) (1 \! + \! b w)^4} \equiv \Theta(\rho,p) ,
		\end{equation}
where $w = w(\rho, p)$ is given in (\ref{w(rhop)}), and $b(R)$ and $R$ depend on $\rho$ through the function $R = R(\rho)$.
		
		\subsection{Constraints for physical reality} \label{subsec-Phys-reality}
		When studying a specific Stephani universe defined by the functions $\{R(t),b(t)\}$, we must determine the spacetime domain (set of values of the coordinates $\{t, w\}$) where the functions $\{\rho(t), p(t,w)\}$ fulfil the energy conditions E given in (\ref{E}). We must also impose the compressibility conditions H$_1$ given in (\ref{H1}) on the indicatrix function (\ref{eq:chi(rho,p)}) within this domain. This means that the functions $A_i(\rho)$ and their derivatives will be constrained by some inequalities. \\ \\
		On the other hand, the functions $\{s(w), N(w)\}$ defining a thermodynamic scheme will be constrained by the positivity conditions P given in (\ref{P}) and the compressibility condition H$_2$ given in (\ref{H2}). \\ \\ 
		The study of all these constraints for a generic Stephani universe results too formal and useless. We delay this study for specific solutions that can be obtained under the demand of meaningful physical qualities. In the following section we study some of them.
		
	\section{Imposing some significant physical qualities} \label{sec-qualities}
	The general expressions of the hydrodynamic and thermodynamic quantities obtained above can be useful when we particularise them in looking for thermodynamic Stephani universes that model a perfect fluid with specific physical properties. We analyse some of these requirements in this section.
	
		\subsection{Ideal gas Stephani Universes} \label{subsec-ideal}
		A notable physical property that can be required for a perfect fluid solution is that it represents the evolution of a generic ideal gas, which is defined by the EoS (\ref{eq. estat gas ideal}). In this subsection, we summarise the results of \cite{CF-Stephani}, where the compatibility of the thermodynamic Stephani universes with such EoS is analysed, namely, when the indicatrix function (\ref{eq:chi(rho,p)}) satisfies the ideal sonic condition S$^\textrm{G}$, given in (\ref{cond. sonica ideal}). \\ \\
		For the Stephani universes, the ideal condition (\ref{cond. sonica ideal}) implies $A_i(\rho) = c_i = constant$, and the indicatrix function (\ref{eq:chi(rho,p)}) becomes
		\begin{subequations} \label{chi-stephani-ideal}
			\begin{eqnarray} 
				\chi(\pi) = \frac13 c_1 \pi^2 + \lambda \pi + \delta , \qquad \quad \\ 
				\lambda \equiv 1 + \frac13 (2 c_1 + c_2), \quad \delta \equiv \frac13 (1 + c_1 + c_2) .
			\end{eqnarray}
		\end{subequations}
		Depending on the principal constants $c_i$ five classes exist: (C1) $c_1 = c_2 = 0$, (C2) $c_1 = 0, c_2 \neq 0$, (C3) $\Delta \equiv c_2^2 - 4c_1 = 0$, $c_1 \neq 0$, (C4) $\Delta > 0$, $c_1 \neq 0$ and (C5) $\Delta < 0$.
		For every class, the associated ideal thermodynamic scheme (\ref{epsilon i Theta ideals}$\,-$\ref{psi(pi) i phi(pi)}) is determined in \cite{CF-Stephani} by explicitly obtaining the generating functions $e(\pi)$ and $f(\pi)$, given in (\ref{e(pi) i f(pi)}). \\ \\
		On the other hand, the study of equations (\ref{Ai(R)}), with $A_i(\rho) = c_i$, leads to distinguish two families of models. If $a'(R) = 0$, we have the singular models, for which $a(R) = a_0 \neq 0$ and $b(R)$ is given by $b' = b_0 R^{-c_1 a_0^2}$, with $b_0 \neq 0$ an arbitrary constant. These models are compatible with classes C2, C3, and C4. \\ \\ 
		If $a'(R) \neq 0$, we have the regular models, for which $a(R)$ is the solution of $a' R = - a(c_1 a^2 + c_2 a + 1)$ and $b(R) = b_1 + b_2 / a(R)$, with $b_1$ and $b_2 \neq 0$ arbitrary constants. These models are compatible with the five classes Cn \cite{CF-Stephani}. The differential equation for $a(R)$ can be integrated by performing a change of variables. The second equation in (\ref{eq:pi0}) can be written as
		\begin{equation} \label{q(a)}
			\frac{1}{\rho}\frac{\textrm{d}\rho}{\textrm{d}R} = - \frac{3 a}{R} \frac{\textrm{d}R}{\textrm{d}a} = \frac{3}{q(a)} \, , \qquad q(a) \equiv c_1 a^2 + c_2 a + 1 \, .
		\end{equation}
With that, in \cite{CF-Stephani} they get that
		\begin{equation}
			R(a) = \frac{R_0}{a} \sqrt{|q(a) Q^{c_2}(a)|} \, , \qquad Q(a) \equiv \exp \left\lbrace\int \frac{\textrm{d}a}{q(a)} \right\rbrace \, ,  
		\end{equation}
and $a = a(t)$ is a solution to the generalised Friedmann equation
		\begin{equation}
			\rho_0 Q^3(a) = \frac{3 \dot{a}^2}{a^2 q^2(a)} + \frac{3}{R^2(a)}[\varepsilon - b^2(a)] \, .
		\end{equation}
The analytic expression of the function $Q(a)$ depends on the class Cn. Here, we write the expressions for the classes that we will need later: (C1) $Q(a) = \exp (a)$ and (C2) $Q(a) = (c_2 a + 1)^{1/c_2}$. \\ \\
		Now, we obtain the thermodynamic schemes allowing to interpret these solutions as an ideal gas. Note that the ideal thermodynamic scheme (\ref{epsilon i Theta ideals}$\,-$\ref{psi(pi) i phi(pi)}) for a Stephani ideal model must correspond to a specific choice of the functions $s(w)$ and $N(w)$ in (\ref{s-n-Stephani}), where the schemes associated with a thermodynamic Stephani universe are given. According to (\ref{epsilon i Theta ideals}$\,-$\ref{psi(pi) i phi(pi)}), in the ideal thermodynamic schemes, $\rho/n$ and $\rho \exp(s/\tilde{k})$ are functions of $\pi$. Checking the compatibility of this with (\ref{s-n-Stephani}), and using the expressions for $a(R)$ and $b(R)$ particular to each model, we obtain that the functions $s(w)$ and $N(w)$ must fulfil the differential conditions
		\begin{equation} \label{s(w)-N(w)}
				s'(w) = \frac{1}{\sigma_0 + \sigma_1 w + \sigma_2 w^2} , \qquad  
				\frac{N'(w)}{N(w)} = \frac{\mu_0 + \mu_1 w}{\nu_0 + \nu_1 w + \nu_2 w^2} , 
		\end{equation}
where the constants $\sigma_1$, $\mu_i$ and $\nu_i$ depend on the parameters of the specific Stephani ideal model. \\[-6mm]
		
		\subsection{Schemes compatible with thermal conductivity \\ and shear-viscosity} \label{subsec-Schemes-Thermal}
		As explained in Section \ref{subsec-intro-thermal-conduc-coeff}, non-perfect fluids exist admitting particular evolutions in which the dissipative fluxes vanish. Along these evolutions, the shear, the expansion and the acceleration of the fluid undergo strong restrictions (see Section \ref{subsec-intro-thermal-conduc-coeff}). \\ \\[-1mm] 
		The thermodynamic Stephani universes studied here have a shear-free flow and, consequently, are compatible with a non-vanishing shear-viscosity coefficient. Now, we study if thermodynamic schemes, which are compatible with a non-vanishing thermal conductivity coefficient, exist. \\ \\[-1mm]
		A strict Stephani universe has a non-vanishing acceleration $a = \perp \! \textrm{d} \ln \alpha$, where $\alpha$ is given in (\ref{eq:termetric-1}). Then, the constraint (\ref{Fourier}) states $\partial_w(\alpha \Theta) = 0$. If we impose this condition on the temperature (\ref{T-Stephani}), we obtain that no solution exists for a non-constant $b(t)$. Consequently, we have that a thermodynamic Stephani universe (\ref{metrica}) can model a fluid with non-vanishing shear-viscosity. It also models a fluid with non-vanishing thermal conductivity coefficient if, and only if, it is a FLRW universe. \\[-6mm]
		
		\subsection[Good behaviour at low temperatures: $\chi(\rho, 0) = 0$]{Good behaviour at low temperatures: \boldmath$\chi(\rho, 0) = 0$} \label{subsec-Low}
		The indicatrix function of a classical ideal gas (CIG) is of the form (\ref{chi-gas-ideal-classic}). 
		The expression (\ref{chi-stephani-ideal}) for the indicatrix function of the ideal gas Stephani solutions shows that no strict Stephani universe exists modelling the evolution of a CIG. Only the FLRW limit gives a barotropic solution modelling a CIG in isentropic evolution (see \cite{CFS-CIG} for more details). \\
		Regardless, we can analyse the thermodynamic Stephani models that approach a CIG in the vicinity of $p = 0$, which is the region where the classical ideal gas EoS is a good model.
		For the indicatrix function (\ref{chi-gas-ideal-classic}) of a CIG, $\chi_c(\pi) = \tilde{\chi}_c(\rho,p)$, we have
		\begin{eqnarray}
			\chi_c(0) = \tilde{\chi}_c(\rho,0) = 0, \qquad \chi_c'(0) = \gamma , \label{cc-ideal} \\[1mm]
			\partial_p \tilde{\chi}_c(\rho,0) = \frac{\gamma}{\rho} , \qquad \partial_{\rho} \tilde{\chi}_c(\rho,0) = 0 .   
		\end{eqnarray}
And for the indicatrix $\chi(\rho,p)$ of a generic thermodynamic Stephani universe (\ref{eq:chi(rho,p)}) we have
		\begin{subequations} \label{chi-stephani-low}
			\begin{equation}
				\chi(\rho,0) = \frac13[1 + A_1(\rho) + A_2(\rho)],
			\end{equation}
			\begin{equation}   
				\partial_p \chi(\rho,0) = \frac{1}{\rho}\{1 + \frac13[2A_1(\rho) + A_2(\rho)]\} , \qquad
				\partial_{\rho} \chi(\rho,0) = \frac13[A_1'(\rho) + A_2'(\rho)] .   
			\end{equation}
		\end{subequations}
		Consequently, we obtain:
		\begin{itemize}
\item[(i)] 
A thermodynamic Stephani universe approaches a CIG up to zero order at $p = 0$ if, and only if, the functions $A_i(\rho)$ are constrained by the condition:
			\begin{equation}
				1 + A_1(\rho) + A_2(\rho) = 0 . \label{zero-p=0}
			\end{equation}
\item[(ii)] 
A thermodynamic Stephani universe approaches a CIG (with adiabatic index $\gamma$) up to first order at $p = 0$ if, and only if, it models a generic ideal gas ($A_i(\rho) = c_i)$ with an indicatrix function of the form:
			\begin{equation}
				\chi(\pi) = \gamma \pi + (\gamma - 2/3) \pi^2 . \label{first-p=0}
			\end{equation}
		\end{itemize}
In \cite{Coll_Ferrando_i_Saez_2020b}, it has been analysed when an ideal gas Stephani model fulfils the ideal compressibility conditions H$_1^\textrm{G}$ and H$_2^\textrm{G}$. In particular, it is shown that if $\gamma > 1$, the indicatrix function (\ref{first-p=0}) fulfils H$_1^\textrm{G}$ and H$_2^\textrm{G}$ in an interval $[0, \pi_M]$, with $0 < \pi_M \equiv \frac{2}{\gamma + \sqrt{\gamma^2 + 4(\gamma - 2/3)}} < 1$. 
		
		\subsection[Good behaviour at high temperatures: $\chi(\rho, \rho/3) = 1/3$]{Good behaviour at high temperatures: \boldmath$\chi(\rho, \rho/3) = 1/3$} \label{subsec-High}
		As we have seen in Chapter \ref{chap-Synge}, the macroscopic EoS of a relativistic non-degenerate monoatomic gas (Synge gas) can be expressed by means of second kind modified Bessel functions (\ref{synge}), and we have proposed several simpler analytical approaches. The Taub-Mathews EoS (\ref{TM-2}) approximates the Synge gas one at first order at both low and high temperatures, and its indicatrix function is given by (\ref{chi-TM}). At first order in $p = 0$, the Synge gas coincides with a CIG with adiabatic index $5/3$. Thus, the behaviour at low temperatures of the Synge gas has been analysed in the above subsection. Now, we analyse the thermodynamic Stephani models that approach a Synge gas at high temperatures. \\ \\
		The indicatrix function (\ref{chi-TM}), $\chi_{_{T\!M}}(\pi) = \tilde{\chi}_{_{T\!M}}(\rho,p)$, approximates a Synge gas in the interval $[0, 1/3]$. When the temperature increases and tends to infinity, $\pi$ approaches $1/3$ ($\rho = 3p$) and $\rho$ tends to infinity. We have:
		\begin{eqnarray}
			\chi_{_{T\!M}}\!(1/3) = \tilde{\chi}_{_{T\!M}}\!(\rho,\rho/3) \! = \! 1/3, \quad \chi_{_{T\!M}}'(1/3) = 1/2 , \qquad \\[1mm]
			\partial_p \tilde{\chi}_{_{T\!M}}\!(\rho,\rho/3) = \frac{1}{2\rho} , \quad \partial_{\rho} \tilde{\chi}_{_{T\!M}}\!(\rho,\rho/3) = -\frac{1}{6\rho} . \qquad  
		\end{eqnarray}
For the indicatrix, $\chi(\rho,p)$, of a generic thermodynamic Stephani universe (\ref{eq:chi(rho,p)}) we have:
		\begin{subequations} \label{chi-stephani-high}
			\begin{eqnarray}
				\chi(\rho,\rho/3) = \frac23\{1 + \frac23[\frac43 A_1(\rho) + A_2(\rho)]\}, \qquad \ \ \\[1mm]  
				\partial_p \chi(\rho,\rho/3) = \frac{1}{\rho}\{1 + \frac13[\frac83 A_1(\rho) + A_2(\rho)]\} , \qquad \ \ \\
				\partial_{\rho} \chi(\rho,\rho/3) \! = \! -\frac13 \partial_p \chi(\rho,\rho/3) \! + \! \frac49[\frac43 A_1'(\rho) \! + \! A_2'(\rho)] . \qquad \ \
			\end{eqnarray}
		\end{subequations}
Consequently, we obtain:
		\begin{itemize}
\item[(i)] 
A thermodynamic Stephani universe approaches a Synge gas up to zero order at $\rho = 3p$ if, and only if, 
the functions $A_i(\rho)$ are constrained by the condition:
			\begin{equation} \label{zero-3p=rho}
				9 + 16 A_1(\rho) + 12 A_2(\rho) = 0 .
			\end{equation}
\item[(ii)] 
A thermodynamic Stephani universe approaches a Synge gas up to first order at $\rho = 3p$ if, and only if, it models a generic ideal gas ($A_i(\rho) = c_i)$ with an indicatrix function of the form:
			\begin{equation} \label{first-3p=rho}
				\chi(\pi) = \frac{1}{16} [7/3 + 10\pi - 3 \pi^2] . 
			\end{equation}
\end{itemize}
The analysis of the domains in which the indicatrix function (\ref{first-3p=rho}) fulfils the ideal compressibility conditions H$_1^\textrm{G}$ and H$_2^\textrm{G}$ is also performed in \cite{Coll_Ferrando_i_Saez_2020b}. The conclusion is that H$_1^\textrm{G}$ and H$_2^\textrm{G}$ are satisfied in the interval $[0, 1/3]$.
		
	\section{Universes with isotropic radiation} \label{sec-Isotropic}
	The high level of isotropy of the cosmic microwave background radiation is usually considered as a proof that a good cosmological model must be close to a FLRW universe. This conception rests on the the Ehlers-Geren-Sachs (EGS) theorem that states \cite{EGS}: if the cosmological observer of a dust solution measures isotropic radiation, then the spacetime is a FLRW model. This result follows from a previous one by Tauber and Weinberg \cite{Tauber-EGS} on the isotropic solutions of the Liouville equation, which was later generalised for the case when an isotropic collision term exists \cite{TE-EGS}. \\ \\
	The general form of the Einstein equations for spacetimes with isotropic radiation measured by an irrotational observer has been obtained in \cite{FMP-isotropa}. This study shows that the geodesic character of the cosmological observer is a necessary requirement in generalising the EGS result. In fact, Clarkson and Barrett \cite{Clarkson-1999} proved that the perfect fluid solutions with a comoving irrotational observer measuring isotropic radiation are a subclass of the thermodynamic Stephani universes, which only have a geodesic flow in the FLRW limit. \\ \\
	An essential property that generates all the results in the above references is the following: a unit vector $u$ defines an observer measuring isotropic radiation if, and only if, it fulfils
	\begin{equation} \label{u-CKV}
		\sigma = 0 , \qquad \textrm{d} \left[a - \frac13 \theta u \right] = 0 ,
	\end{equation}
where $\sigma$, $a$ and $\theta$ are, respectively, the shear, the acceleration and the expansion of $u$.
Conditions (\ref{u-CKV}) state that $u$ is proportional to a a conformal Killing vector. Consequently, the spacetime is conformally stationary. On the other hand, the energy density, the pressure and the temperature of the radiation fluid are given by \cite{FMP-isotropa}
	\begin{equation} \label{rho-p-radiacio}
		\rho_r \! = \! 3 p_r \! = \! a_R \Theta_r^4, \qquad \Theta_r \! = \! \Theta_0 \beta^{-1}, \qquad \textrm{d} \ln \beta = a\! - \! \frac13 \theta u .
	\end{equation}
	It is known \cite{Clarkson-1999} that any perfect fluid solution with a comoving irrotational observer measuring isotropic radiation is a thermodynamic Stephani universe. In order to determine the subclass with this property, we must impose equations (\ref{u-CKV}) on the cosmological observer. The first condition, $\sigma = 0$, is fulfilled for any Stephani universe. \\ \\
	Now we impose the second one. From expression (\ref{eq:alpha}) of $\alpha$, and taking into account that $a = \partial_w(\ln \alpha)\, \textrm{d} w$, $\theta u = - 3\, \alpha \, \textrm{d} \ln R$, a straightforward calculation shows that the second condition in (\ref{u-CKV}) is equivalent to $b''(R) = 0$. Moreover, the function $\beta = R \alpha$ fulfils equation (\ref{rho-p-radiacio}). Consequently, we can state that the perfect fluid solutions with an irrotational comoving observer measuring isotropic radiation are the thermodynamic Stephani universes (\ref{metrica}) with 
	\begin{equation} \label{b(R)-isotropic}
		b(R) = b_1 + b_2 R ,
	\end{equation}
where $b_i$ are arbitrary constants. The (test) radiation fluid has an energy density, a pressure and a temperature given by:
	\begin{equation} \label{rho-p-radiacio-3}
		\rho_r = 3p_r = a_R \Theta_r^4 , \quad \Theta_r = \Theta_0 \! \left( \! \frac{R_0}{R} \! \right) \! \! \left[1 \! + \! \frac{b_2 w R}{1 \! + \! b_1 w}\right] .
	\end{equation}
Moreover, the indicatrix function takes the expression (\ref{eq:chi}$-$\ref{Ai(R)}), with $A_1(R) = 0$. \\ \\
	Note that this isotropic radiation defines a test fluid that is comoving with the flow of the source of the field equations: a perfect fluid with energy density and pressure given in (\ref{tdp-1}).
	
		\subsection{Spherical symmetry} \label{subsec-esferica}
		So far, all the results apply for spherical, plane and hyperbolic symmetries. From now on, we consider some specific models that could be developed for any symmetry, but that we only analyse for the spherically symmetric case. \\ \\
		The metric of the thermodynamic Stephani universes with spherical symmetry is given by (\ref{metrica}) with $\varepsilon = 1$. If we perform the following change of spatial coordinates:
	\begin{subequations}
	\begin{eqnarray}
		x = \frac{x'}{H} , \quad y = \frac{y'}{H} , \quad z = \frac{z' -1 + \frac12 r'^2}{H} , \ \\
		r'^2 = x'^2 + y'^2 + z'^2 , \quad H = 1 + z' + \frac14 r'^2 , \
	\end{eqnarray}
	\end{subequations}
we define
	\begin{equation} \label{funcions barra}
		\bar{R} \equiv \frac{2 R}{1 - 2 b} \, , \qquad k \equiv 4 \frac{1 + 2b}{1 - 2b} \, ,
	\end{equation}
and perform the change of temporal coordinate from $t$ to $t'$ such that
	\begin{equation}
		\theta (t') = 3 \frac{\dot{\bar{R}}}{\bar{R}} \, ;
	\end{equation}
we get that the metric of the thermodynamic Stephani universes with spherical symmetry can be written as
		\begin{subequations} \label{metrica-ss}
			\begin{eqnarray}
				\textrm{d} s^2 = -\alpha^2 \textrm{d} t^2 + \Omega^2 (\textrm{d} r^2 + r^2 \textrm{d} \tilde{\Omega}^2) \, , \quad \label{the-ssste-uni} \\
				\Omega \equiv \frac{R(t)}{1 + \frac14 k(t) \, r^2} \, , \qquad \alpha \equiv R \, \partial_{R} \ln\Omega \, ; \quad \label{eq:sstermetric-1}  
			\end{eqnarray}
		\end{subequations}
where we have removed the primes and bars to simplify the notation, with $R(t)$ and $k(t)$ two arbitrary functions of time and $\textrm{d} \tilde{\Omega}^2$ the metric of the unitary sphere. The energy density, the pressure, the expansion and the 3-space curvature are given by
		\begin{eqnarray} \label{rho-p_esf}
			\rho = \frac{3}{R^2}(\dot{R}^2 + k) \, , \qquad p = -\rho - \frac{R}{3}\frac{\partial_{R} \rho}{\alpha} \, , \quad \\  
			\theta = 3 \frac{\dot{R}}{R} \neq 0 \, , \qquad \qquad \kappa = \frac{k}{R^2} \, . \quad \label{curvature-ss}
		\end{eqnarray}
		Note that expressions in Sections \ref{subsec-Chi-Stephani}, \ref{subsec-eq-Stephani} and \ref{subsec-Schemes-Stephani} also apply for this case by changing $b \rightarrow k/4$ and $w \rightarrow r^2$. Only in the generalised Friedmann equation (\ref{gfe}) the change must be $\varepsilon \! - \! 4 b^2 \rightarrow k$. As a consequence, the condition for the cosmological observer to measure isotropic radiation in these coordinates is $k''(R) = 0$, namely, $k(R)= k_1 + k_2 R$, accordingly with the result in \cite{Clarkson-1999}. \\ \\
		It is worth remarking that, using the definitions given in (\ref{funcions barra}), it can be seen that this condition is coherent with (\ref{b(R)-isotropic}). Then, the results in (\ref{b(R)-isotropic}) and (\ref{rho-p-radiacio-3}) apply by changing $b_i \rightarrow k_i/4$ and $w \rightarrow r^2$.
		
		\subsection{Analysis of the Dabrowski solution} \label{subsec-roin}
		As explained in Section \ref{subsec-eq-Stephani}, a possible approach to study the physical properties of a solution is to start by prescribing the functions of time $R(t)$ and $k(t)$. Now, we consider one of the solutions considered by Dabrowski \cite{Dabrowski} in studying the general properties of the local isometric embedding of the Stephani universes (hereinafter, the \textit{Dabrowski solution}):
	\begin{equation} \label{R-k_Dabrowski}
		R(t) = D_2 \, t^2 + D_1 , \quad k(t) = - 4 D_2 R + (1 \! - \! D_1^2) ,
	\end{equation}
where $D_i$ are two real parameters. \\ \\
	For this solution, we can obtain $\rho (t)$, $p (t, r)$ and $\pi (t, r)$ from (\ref{rho-p_esf}), and then study the energy conditions (\ref{EG}). This study was done by Barrett and Clarkson in \cite{Barrett-2000}, and they concluded that the energy conditions are fulfilled for a certain range of values of the parameters $D_i$. \\ \\
	Note that the Dabrowski solution (\ref{R-k_Dabrowski}) can represent a universe with a cosmological observer measuring isotropic radiation. This fact was already pointed out in \cite{Barrett-2000}, where the physical, geometrical and observational characteristics of these inhomogeneous models were analysed in detail. \\ \\
	In the conclusions of \cite{Barrett-2000} the authors claim that the Dabrowski solutions (\ref{R-k_Dabrowski}) ``admit a thermodynamic interpretation" although ``there is no equation of state". We want to point out and to clarify these assertions. As explained in Section \ref{sec-intro-physical-interpretation}, the existence of a formal thermodynamic scheme (subject to equations of state) can be characterised in terms of the hydrodynamic quantities by the sonic condition (\ref{cond. sonica}). In fact, the Dabrowski solution fulfils the ideal sonic condition (\ref{cond. sonica ideal}). Indeed, if we use (\ref{R-k_Dabrowski}) to compute $A_1(\rho)$, $A_2(\rho)$ and $\chi(\rho,p)$, we get that $A_1 = 0$, $A_2 = -3/2$ (it is an ideal gas solution) and (\ref{chi-stephani-ideal}) becomes
	\begin{equation} \label{chi-Dabrowski}
		\chi(\rho, p) = \chi(\pi) = (3\pi - 1)/6 . 
	\end{equation}
Then, the Dabrowski solution is compatible with the ideal gas EoS $p = \tilde{k} n \Theta$. However, a question arises: do these models represent the evolution of a realistic perfect fluid? \\ \\
	The indicatrix function (\ref{chi-Dabrowski}) only fulfils the causal compressibility conditions, $0 < \chi < 1$, in the interval $\pi \! \in \, ] \, 1/3 \, , 1 \, [$. Moreover, $\zeta(\pi) = - \frac{17}{36} + \pi - \frac34 \pi^2$ and $\eta(\pi) = - \frac16 - \frac56 \pi + \pi^2$ are negative in the whole domain $]0, 1[$. Therefore, the indicatrix function (\ref{chi-Dabrowski}) does not fulfil the compressibility conditions H$_1^{\rm G}$ and H$_2^{\rm G}$ given in (\ref{H1G}) and (\ref{H2G}). \\ \\
Consequently, the Dabrowski solution fulfils the energy conditions and can be taken into account as a cosmological model, but it cannot be interpreted as a physically admissible thermodynamic perfect fluid.
		
		\subsection{Isotropic radiation with an ideal gas source} \label{subsec-isotropa+gasideal}
		In order to obtain solutions that can be interpreted as an ideal gas source with the comoving observer measuring isotropic radiation, instead of prescribing the functions $R(t)$ and $k(t)$, we will impose these physical properties to our solution and then we will study how they restrict the metric functions. \\ \\
Now, all the results in Section \ref{subsec-ideal} apply with the additional constraint $A_1 = c_1 = 0$, imposed by the isotropic radiation condition (\ref{b(R)-isotropic}).
With the condition $c_1 = 0$, the indicatrix function (\ref{chi-stephani-ideal}) of the ideal models becomes:
		\begin{equation} \label{indicatriu-isotropicradiation}
			\chi (\pi) = \lambda \, \pi + \lambda - \frac23 \, , \qquad \lambda = 1 + \frac13 \, c_2 \, .
		\end{equation}
If $1/3 < \lambda < 5/3$, this indicatrix function verifies the causal condition, $0 < \chi < 1$, for $\pi \in \, ] \, \pi_m , \pi_M \, [ \,$, where $\pi_m \equiv \frac{2}{3\lambda} - 1$ and $\pi_M \equiv \frac{5}{3\lambda} - 1$. If $2/3 \leqslant \lambda \leqslant 5/6$, these conditions are fulfilled in the whole domain $\pi \in \, ] \, 0 , 1 \, [ \, $. \\ \\
		To study the remaining compressibility conditions, we need to analyse the \mbox{domains} in which $\zeta(\pi)$ and $\eta(\pi)$, defined in (\ref{H1G}) and (\ref{H2G}), are positive. Using (\ref{indicatriu-isotropicradiation}), we get $\zeta(\pi) = \! - \lambda(\lambda + 1)\pi^2 + \lambda \, (3 - 2\lambda) \, \pi - \lambda \, (\lambda - 4) - \frac{20}{9}$ and $\eta(\pi) = 2\lambda \, \pi^2 + \\ \left( 3\lambda - \frac73 \right) \, \pi + \lambda - \frac23$. \\ \\
		If $16/29 \approx 0.5517 < \lambda < 10/3$, there exists an interval, $\pi \in \, ] \, \pi_- , \pi_+ \, [, $ in which $\zeta(\pi)$ is positive, where
		\begin{equation}
			\pi_\pm \equiv \frac{3\lambda(3 - 2\lambda) \pm \sqrt{5\lambda(29\lambda - 16)}}{6\lambda(\lambda + 1)} \, .
		\end{equation}
Moreover, if $2/3 < \lambda < 5/6 \approx 0.8333$, $\zeta(\pi)$ is positive in the whole domain $\pi \in \, ] \, 0 , 1 \, [ \, $. \\ \\
		If $\lambda > 1/2$ instead, there exists an interval $\pi \in \, ] \, \bar{\pi}_+ , 1 \, [$ in which $\xi(\pi)$ is positive, and if $\lambda > 2/3$, it is also positive in an interval $\pi \in \, ] \, 0 , \bar{\pi}_- \, [ \, $, where 
		\begin{equation}
			\bar{\pi}_\pm \equiv \frac{7 - 9\lambda \pm \sqrt{9\lambda^2 - 78\lambda + 49}}{12\lambda} \, .
		\end{equation}
Moreover, if $\lambda > (13 - 2\sqrt{30})/3 = \lambda_{\eta} \approx 0.682$, $\eta(\pi)$ is positive in the whole domain $\pi \in \, ] \, 0 , 1 \, [ \, $. \\ \\
		Taking all this analysis into account, we can state that the indicatrix function of the ideal gas models with the comoving observer measuring isotropic radiation takes the expression (\ref{indicatriu-isotropicradiation}), and it fulfils all the compressibility conditions {\rm H}$_1^{\rm G}$ and {\rm H}$_2^{\rm G}$ in the whole domain $\pi \in \, ] \, 0 , 1 \, [ \, $ for $\lambda$ in the interval $]\lambda_{\eta} , \, 5/6 \, [ \, $, where $\lambda_{\eta} \equiv (13 - 2\sqrt{30})/3$. \\ \\
		In Section \ref{subsec-ideal} we have explained how to integrate equations (\ref{Ai(R)}) for ideal gas models. The isotropic radiation condition $c_1 = 0$ is only compatible with regular models of class C1 ($c_2 = 0$) and both singular and regular models of class C2 ($c_2 \neq 0$) (see Section \ref{subsec-ideal}). Taking all this into account, we get that the generalised Friedmann equation (\ref{gfe}) of the ideal gas models with the comoving observer measuring isotropic radiation takes the expression
		\begin{equation} \label{Friedmann-ideal-radiacio}
			\rho(R) = \frac{3}{R^2}(\dot{R}^2 + k_1 + k_2 R) ,
		\end{equation}
where $\rho(R)$ depends on three different models:
		\begin{itemize}
\item[(i)]
C2 singular ($\lambda \neq 1$): $\displaystyle \ \rho(R) = \rho_0 \left(\frac{R_0}{R }\right)^{\frac{1}{1 - \lambda}}$.
\item[(ii)]
C2 regular ($\lambda \neq 1$): $\displaystyle \ \ \rho(R) = \rho _0 \left(1 + \frac{\tilde{R}_0}{R }\right)^{\frac{1}{1 - \lambda}}$.
\item[(iii)]
C1 regular ($\lambda = 1$): $\displaystyle \ \ \rho(R) = \rho _0 \exp \left(\frac{\hat{R}_0}{R}\right)$.
		\end{itemize}
Now, we could analyse the energy conditions (\ref{EG}) for these three cases separately, but we will leave that study for particular cases with extra physical restrictions.
		
	\section{Ultra-relativistic gas with isotropic radiation} \label{sec-Model-Nos}
	Another possible situation of physical interest is that in which the source of the gravitational field is an ultrarelativistic fluid and the cosmological observer measures isotropic radiation. The conditions for a thermodynamic Stephani universe to behave as an ultrarelativistic fluid up to first order are studied in Section \ref{subsec-High}. However, if we also want it to be compatible with isotropic radiation, we can only impose the good behaviour at high temperatures up to zero order. \\ \\ 
	By imposing (\ref{zero-3p=rho}) and the isotropic radiation condition $k(R) = k_1 + k_2 R$, we get that $A_1(R) = c_1 = 0$ and $A_2(R) = c_2 = -3/4$. Thus, this situation is a particular case of the one studied in Section \ref{subsec-isotropa+gasideal}, with $c_2 = -3/4$. Now, $\lambda = 3/4 \in \, ]\lambda_{\xi}, 5/6[ \,$, and therefore the indicatrix function fulfils the compressibility conditions. \\ \\
	After considering all the above, we can state that the models approximating the Synge gas at high temperatures and with the comoving observer measuring isotropic radiation have an indicatrix function of the form
	\begin{equation} \label{indicatriu-Synge-radiation}
		\chi(\rho,p) = \chi(\pi) = (9 \pi + 1)/12 .
	\end{equation}
This indicatrix function fulfils all the compressibility conditions H$_1^{\rm G}$ and H$_2^{\rm G}$ on the spacetime domain where the energy conditions, $0 < \pi <1$, hold.
	The associated ideal thermodynamic scheme $\{n, \epsilon, s, \Theta\}$ is defined by (\ref{epsilon i Theta ideals}$-$\ref{n i s ideals}), where the generating functions $f(\pi)$ and $e(\pi)$ now take the expressions:
	\begin{equation} \label{f-e-Synge-radiacio}
		f(\pi) = \frac{f_0}{(1 \! - \! 3 \pi )^4} , \qquad e(\pi) = \frac{e_0}{|1 \! - \!3 \pi |(\pi \! + \! 1)^3} .
	\end{equation}
	On the other hand, if we make $\lambda = 3/4$ in (\ref{indicatriu-isotropicradiation}) we have that the \mbox{generalised} Friedmann equation of the models approximating the Synge gas at high temperatures and with the comoving observer measuring isotropic radiation takes the expression (\ref{Friedmann-ideal-radiacio}), where $\rho(R)$ depends on two different models:
	\begin{eqnarray} \label{singular-rho}
		\hspace{-32mm} {\rm (i)} & \! {\rm Singular \ models:} \qquad \ \rho(R) & \!\!\!\! = \rho_0 \left(\frac{R_0}{R}\right)^4, \quad \, \\
		\hspace{-32mm} {\rm (ii)} & \! {\rm Regular \ models:} \qquad \ \rho(R) & \!\!\!\! = \rho _2 \left(1 + \frac{\tilde{R}_0}{R}\right)^4. \label{regular-rho}
	\end{eqnarray}
Some remarks:
	\begin{itemize}	
\item[(i)] 
The hydrodynamic properties of the thermodynamics are given by the indicatrix function (\ref{indicatriu-Synge-radiation}), which takes the same expression for both the singular and regular models, and which fulfils the compressibility conditions in $]0,1[$. Nevertheless, this indicatrix function can only approximate the Synge one, $\chi_{Synge}$, in the interval $]0, 1/3[$ where this last one is defined (see Figure \ref{Fig-9}(a)). Note that this approximation gets worse the closer we get to $\pi = 0$. 
	\begin{figure}
		\centerline{
		\parbox[c]{0.5\textwidth}{\includegraphics[width=0.49\textwidth]{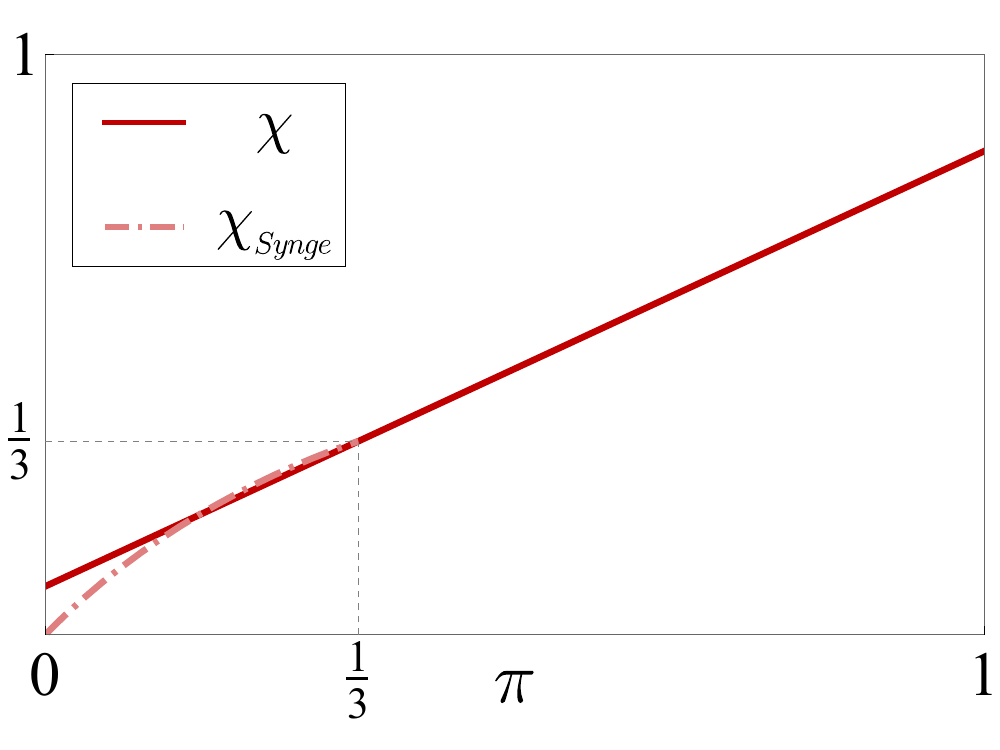}\\[-2mm] \centering{(a)}}
		\raisebox{-2pt}{\parbox[c]{0.5\textwidth}{\includegraphics[width=0.48\textwidth]{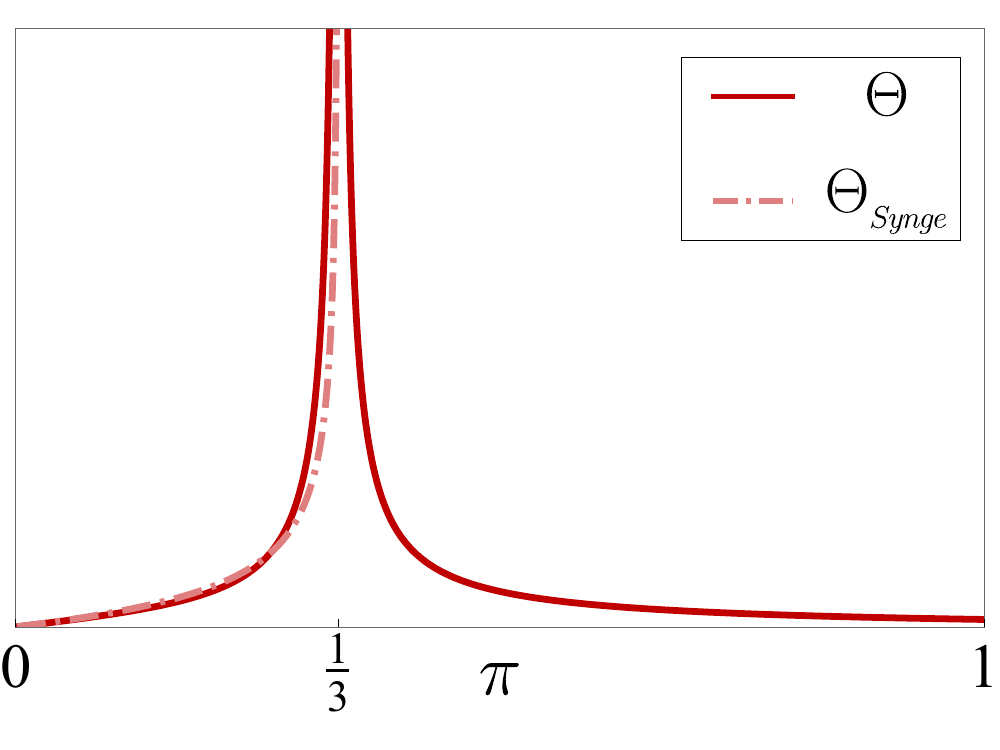}\\[-2mm] \centering{(b)}}}}
		\vspace{-2mm}
		\caption{(a) Here, we show the behaviour of the indicatrix function $\chi(\pi)$ of our model defined in the whole interval $]0,1[$ (red solid line), and the indicatrix function $\chi_{Synge}(\pi)$ of the Synge gas defined in the interval $]0,1/3[$ (pink dashed line). (b) Here, we show a similar situation for the associated temperatures.}
	\label{Fig-9}
	\end{figure}
\item[(ii)] 
Similarly, the ideal thermodynamic scheme determined by functions (\ref{f-e-Synge-radiacio}) is defined in the interval $]0,1[$, but it only approximates the Synge thermodynamic quantities in $]0, 1/3[$. Figure \ref{Fig-9}(b) shows the behaviour of the temperatures.  
\item[(iii)] 
In any case, the indicatrix function (\ref{indicatriu-Synge-radiation}) could be furnished with another (non-ideal) thermodynamic scheme, determined by a pair of functions $\{s(r^2), n(r^2)\}$ (see Section \ref{subsec-Schemes-Stephani}), different from (\ref{s(w)-N(w)}), which can be defined in the interval $]0,1[$. In this case, we could be modelling a physically realistic fluid but that does not satisfy the ideal gas equation of state.
\item[(iv)] 
The energy density $\rho_r(R,r)$, the pressure $p_r(R,r)$ and the temperature $\Theta_r(R,r)$ of the (test) radiation fluid take expressions (\ref{rho-p-radiacio-3}) (with the change $b_i \rightarrow k_i/4$) for both the singular and the regular models. Of course, the coordinate function $R(t)$ does depend on the model.
\item[(v)] 
The ideal thermodynamic scheme defined by the functions (\ref{f-e-Synge-radiacio}) depends on three parameters, $f_0$, $e_0$ and $\tilde{k}$. The first one, $f_0$, fixes the origin of entropy, and we can consider that the different values correspond to a sole ideal gas. The second parameter, $e_0$, modifies the specific energy in a constant factor and, consequently, the temperature and the specific volume $1/n$ change in the same factor. Note that $e_0$ settles the origin of internal energy. If we impose $\epsilon = 0$ at zero pressure, we must take $e_0 = 1$. Finally, the third one, $\tilde{k} = k_B/m$ is determined by the mass of the gas particles. 
\item[(vi)] 
Singular models depend on four parameters $\{k_1, k_2, \rho_0, R_0\}$. The $k_i$ determine the function $k(R)$; $R_0$ is an initial condition for the generalised Friedmann equation (\ref{Friedmann-ideal-radiacio}), $R(t_0) = R_0$; and $\rho_0$ is the energy density at this initial time, $\rho_0 = \rho(t_0)$. However, regular models also depend on a fifth parameter $\tilde{R}_0$. The constant $\rho_2$ takes the expression $\rho_2 \equiv \rho_0 (1 + \tilde{R}_0/R_0)^{-4}$.
	\end{itemize}

		\subsection{Singular models: spacetime domains and energy conditions} \label{subsec-Model-Nos-condition}
\begin{figure}[h]
			\vspace{1mm}
			\qquad \qquad \qquad \qquad \qquad \quad $k_2 < 0$  
			\qquad \qquad \quad \qquad \qquad \qquad \qquad \; $k_2 >0$ \\[-2mm]
			\parbox[c]{0.10\textwidth}{$k_1 \geq 0$} \ 
			\parbox[c]{0.43\textwidth}{\includegraphics[width=0.43\textwidth]{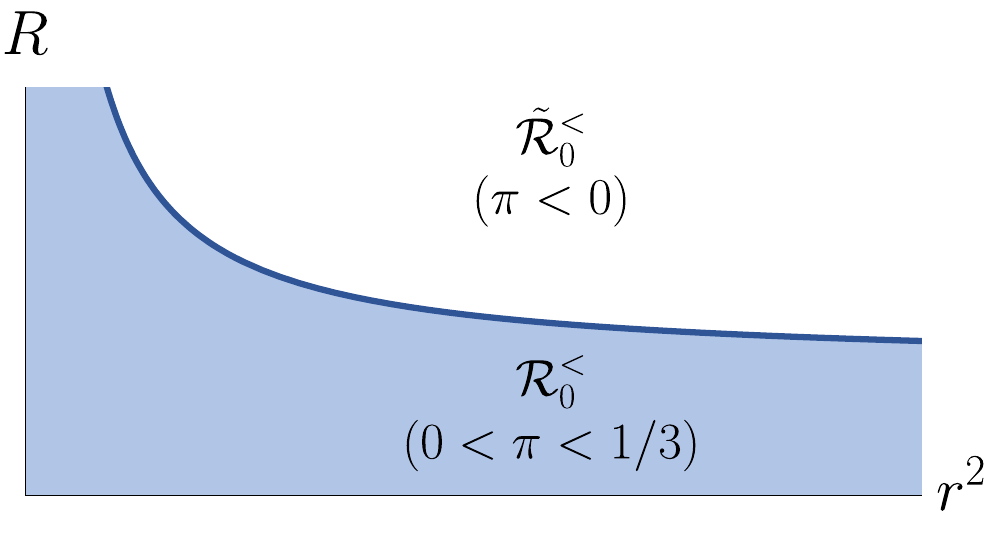}} \ \ 
			\parbox[c]{0.43\textwidth}{\includegraphics[width=0.43\textwidth]{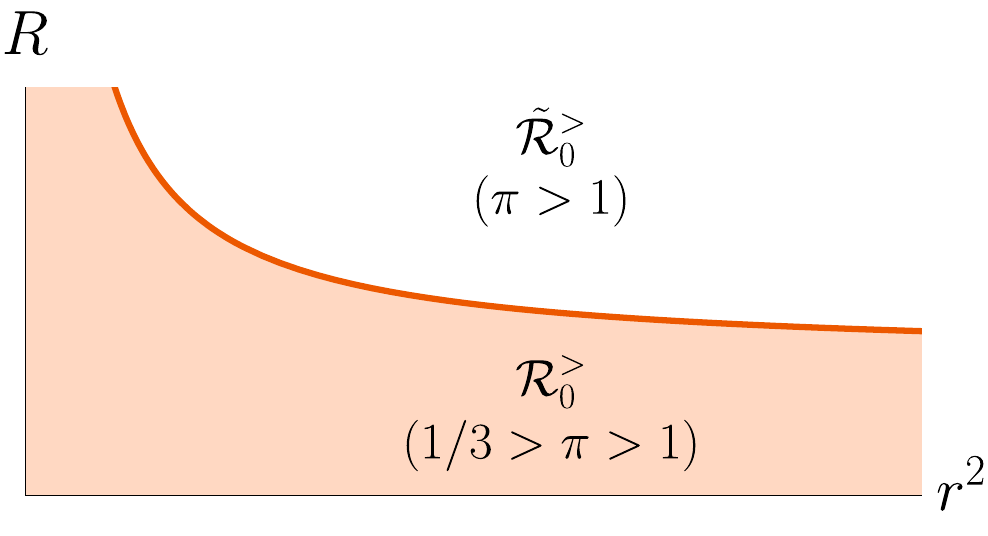}} \\
			\parbox[c]{0.10\textwidth}{$k_1 < 0$} \ 
			\parbox[c]{0.43\textwidth}{\includegraphics[width=0.43\textwidth]{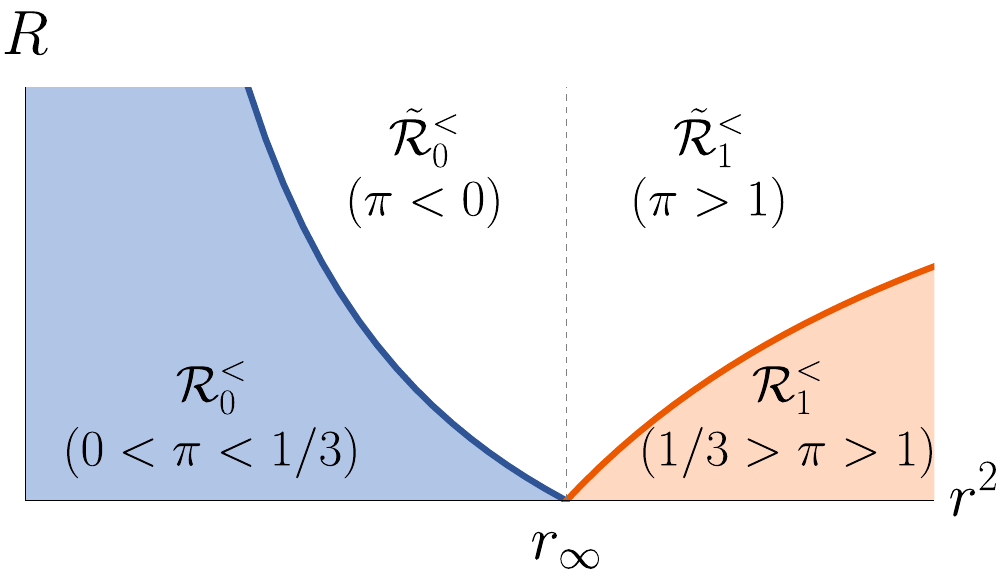}} \ \
			\parbox[c]{0.43\textwidth}{\includegraphics[width=0.43\textwidth]{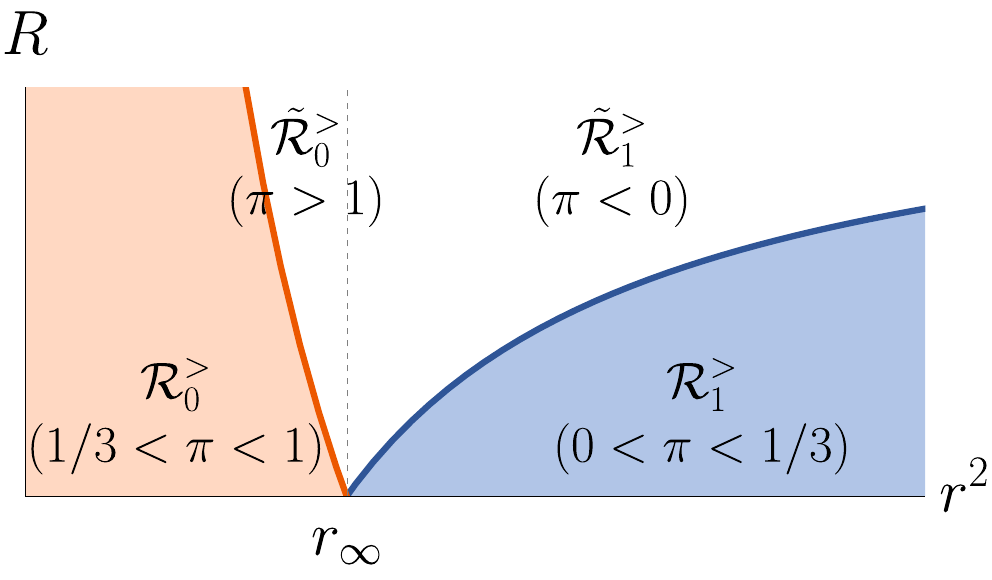}}
			\caption{Spacetime coordinate domains and their physically realistic regions depending on the values of the parameters $k_i$. Upper panels: the case $k_1 \geq 0$ has a single coordinate domain ${\cal D}_0^+$. Lower panels: the case $k_1 < 0$ has two coordinate domains, ${\cal D}_0^-$ and ${\cal D}_1^-$, separated by the straight line $r = r_{\infty}$. The dark blue lines are the hyperbolas $\pi = 0$, and the orange lines are the hyperbolas $\pi = 1$. The energy conditions hold in the shaded regions, and the model approximates a Synge gas in the dark blue regions.}
		\label{Fig-10}
		\end{figure}

		For the singular models, the energy density is given in (\ref{singular-rho}) and the pressure takes the expression
		\begin{equation}
			p(R,r) = \frac13 \rho_0 \left(\frac{R_0}{R}\right)^4 \left[1 + \frac{k_2 R \, r^2}{1 + \frac14 k_1 r^2}\right] .
		\end{equation}
Consequently, the hydrodynamic quantity $\pi = p/\rho$ is
		\begin{equation} \label{pi-singular}
			\pi(R,r) = \frac13 \left[1 + \frac{k_2 R \, r^2}{1 + \frac14 k_1 r^2}\right] .
		\end{equation}
		Note that the entire line element of the 3-spaces $t = $ constant vanishes at $R = 0$, and we have a big bang singularity, where both energy density and pressure diverge. However, $\pi$ takes the value $1/3$. On the other hand, when $k_1 < 0$ we have another curvature singularity at $r_{\infty} = 2\sqrt{-1/k_1}$, where the pressure (and $\pi$) diverges. \\ \\
		Then, when $k_1 \geq 0$ we have a single coordinate domain:
		\begin{equation} \label{domain}
			{\cal D}_0^+ = \{(R,r), \quad R>0, \ r \geq 0\} ,
		\end{equation}
and when $k_1 < 0$ we have two coordinate domains:
		\begin{subequations} \label{domains}
			\begin{eqnarray}
				{\cal D}_0^- = \{(R,r), \quad R > 0, \ 0 \leq r < r_{\infty}\}, \ \ \\[2mm]  
				{\cal D}_1^- = \{(R,r), \quad R > 0, \ r > r_{\infty}\}. \ \  
			\end{eqnarray}
		\end{subequations}
		Now, expression (\ref{pi-singular}) enables us to analyse the regions of the different domains where the energy conditions ($0 < \pi < 1$) hold, and the regions where the model approximates a Synge gas ($0 < \pi < 1/3$). We represent all these regions in a $\{r^2, R\}$ diagram (see Figure \ref{Fig-10}).
Note that $\pi = 1/3$ if $r = 0$. The spacetime events where $\pi = 0$ or $\pi = 1$ are defined, respectively, by the hyperbolas 
		\begin{equation}
			R = - \frac{1}{k_2}\left[\frac{k_1}{4} + \frac{1}{r^2}\right], \qquad R = \frac{2}{k_2}\left[\frac{k_1}{4} + \frac{1}{r^2}\right] .
		\end{equation}
Each of the domains (\ref{domain}, \ref{domains}) contains one of these hyperbolas that divides it into two regions, and the energy conditions only meet in the region next to the coordinate axes (see Figure \ref{Fig-10}):
		\begin{itemize}
\item[(i)]
Case $k_2 < 0$. Domains ${\cal D}_0^+$ and ${\cal D}_0^-$ contain a region ${\cal R}_0^{<}$ where the model approximates a Synge gas, $0 < \pi < 1/3$, and a region $\tilde{\cal R}_0^{<}$ where $\pi < 0$. Domain ${\cal D}_1^-$ contains a region ${\cal R}_1^{<}$ where the model meets the energy conditions but does not approximate a Synge gas, $1/3 < \pi < 1$, and a region $\tilde{\cal R}_1^{<}$ where $\pi > 1$ (see the left panels of Figure \ref{Fig-10}).
\item[(ii)] 
Case $k_2 > 0$. Domains ${\cal D}_0^+$ and ${\cal D}_0^-$ contain a region ${\cal R}_0^{>}$ where the model meets the energy conditions but it does not approximate a Synge gas, $1/3 < \pi < 1$, and a region $\tilde{\cal R}_0^{>}$ where $\pi > 1$. Domain ${\cal D}_1^-$ contains a region ${\cal R}_1^{>}$ where the model approximates a Synge gas, $0 < \pi < 1/3$, and a region $\tilde{\cal R}_1^{>}$ where $\pi <0$ (see the right panels of Figure \ref{Fig-10}).
		\end{itemize}

		\subsection{Singular models: \textit{R}-dependence and radial profiles} \label{subsec-Model-Nos-Evolution}
\begin{figure}[t]
			\centerline{
				\parbox[c]{0.33\textwidth}{\includegraphics[width=0.32\textwidth]{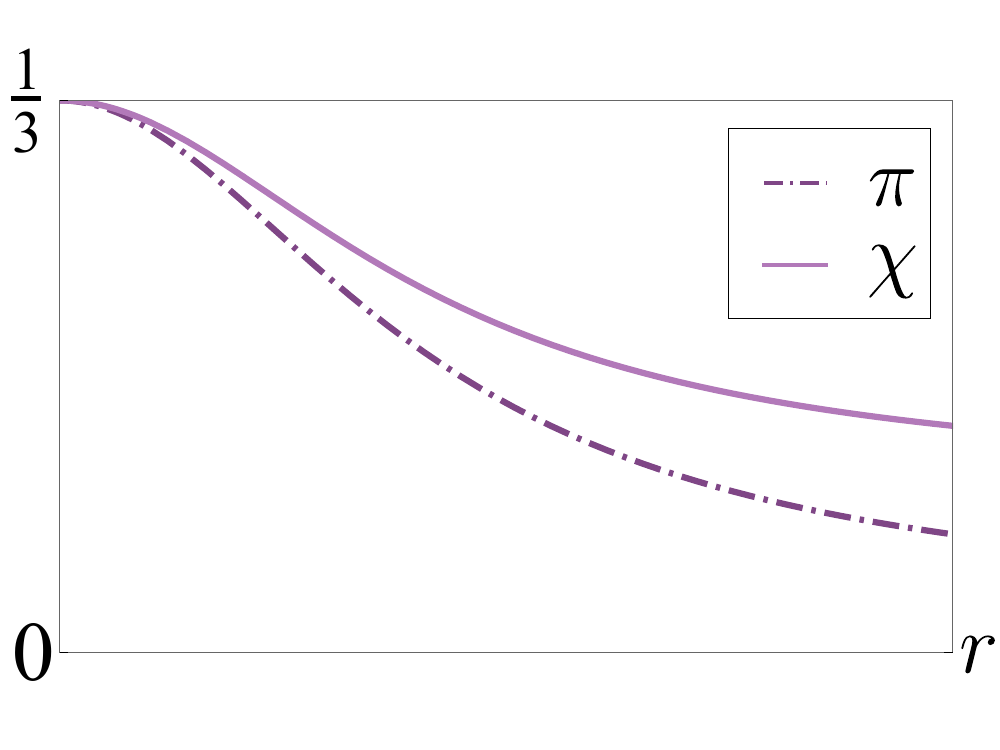}\\[-2mm] \centering{(a)}}
				\raisebox{-2pt}{\parbox[c]{0.33\textwidth}{\includegraphics[width=0.32\textwidth]{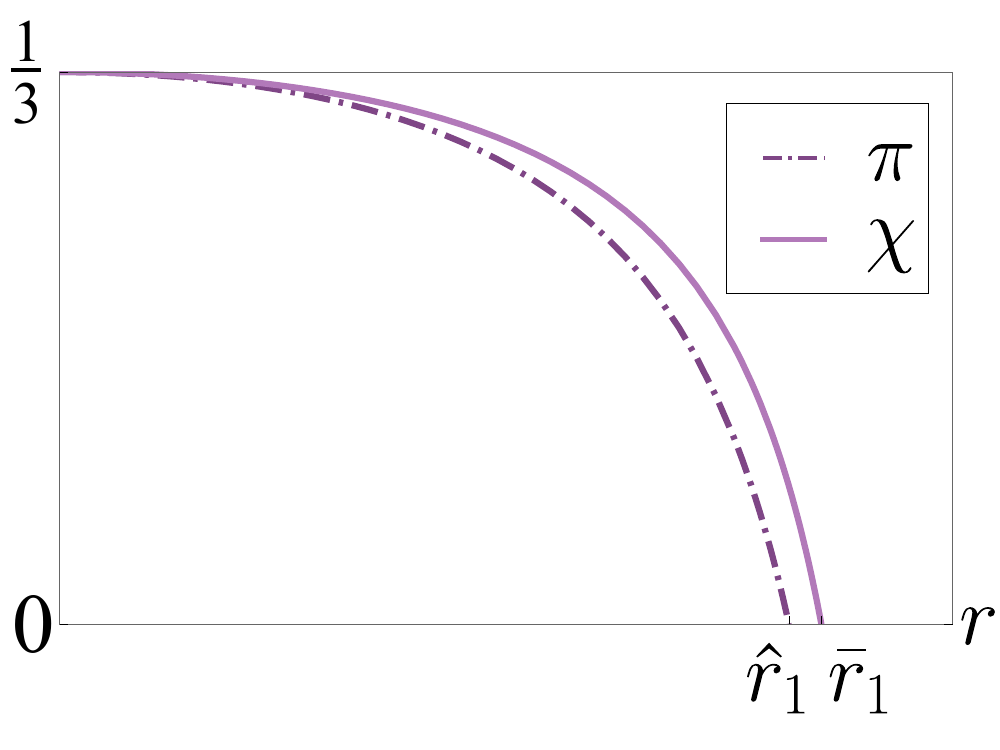}\\[-3mm] \centering{(b)}}}
				\raisebox{-2pt}{\parbox[c]{0.33\textwidth}{\includegraphics[width=0.32\textwidth]{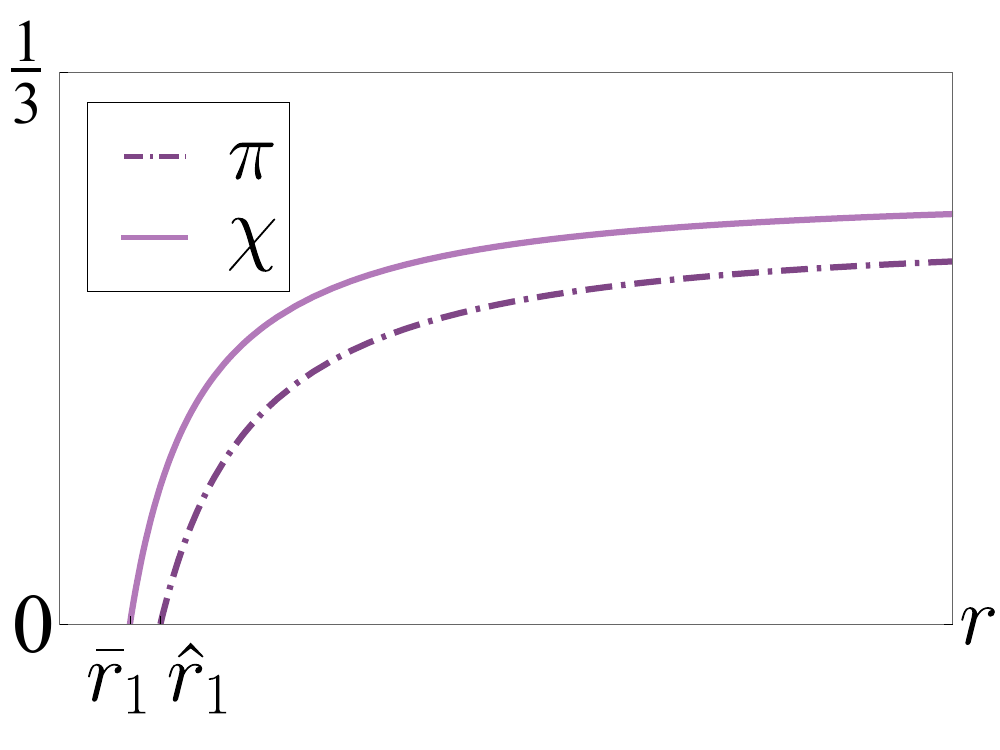}\\[-3mm] \centering{(c)}}}}
\vspace{-2mm}
			\caption{The radial profile of the hydrodynamic quantities $\pi$ and $\chi$ depending on the values of the parameters $k_i$ for a fixed $R_1$. (a) Case I with $R_1 \leq - k_1/(4 k_2)$. (b) Case II and case I with $R_1 > - k_1/(4 k_2)$. (c) Case III.}
		\label{Fig-12}
		\end{figure}

\begin{figure}[t]
			\centerline{
				\parbox[c]{0.33\textwidth}{\includegraphics[width=0.32\textwidth]{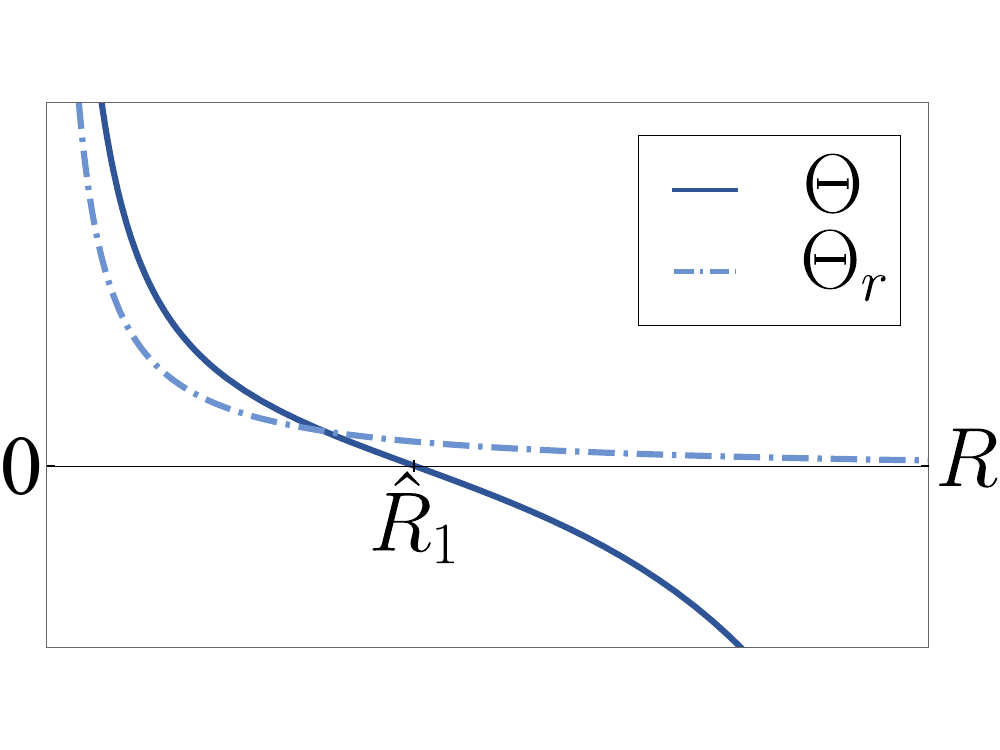}\\[-3mm] \centering{(a)}}
				\parbox[c]{0.33\textwidth}{\includegraphics[width=0.32\textwidth]{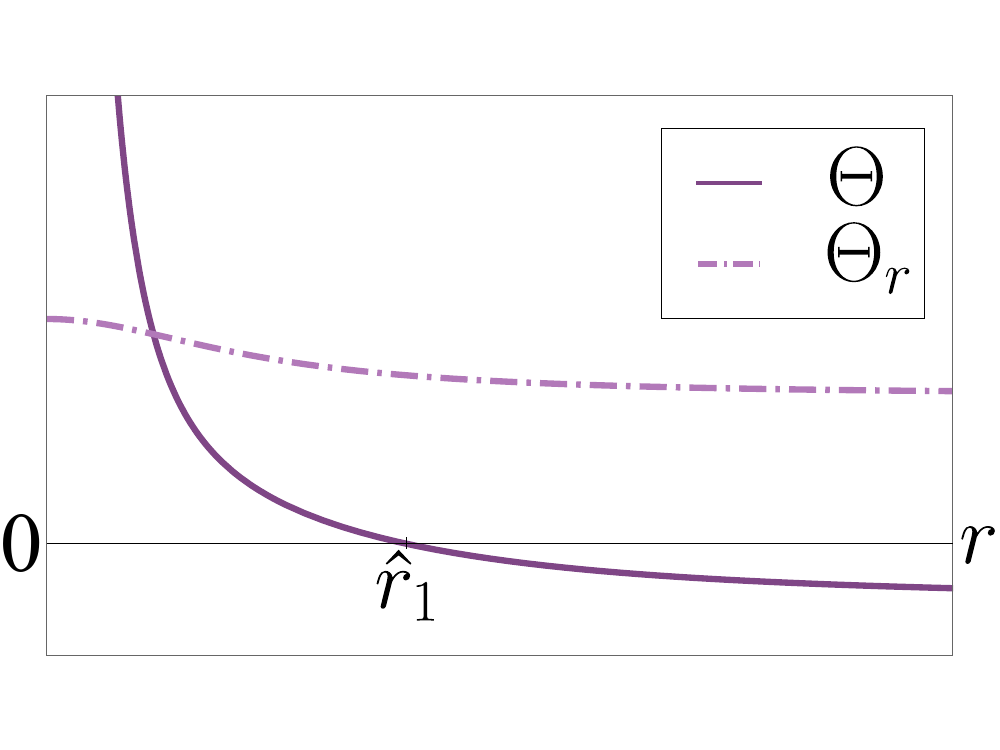}\\[-3mm] \centering{(b)}}
				\parbox[c]{0.33\textwidth}{\includegraphics[width=0.32\textwidth]{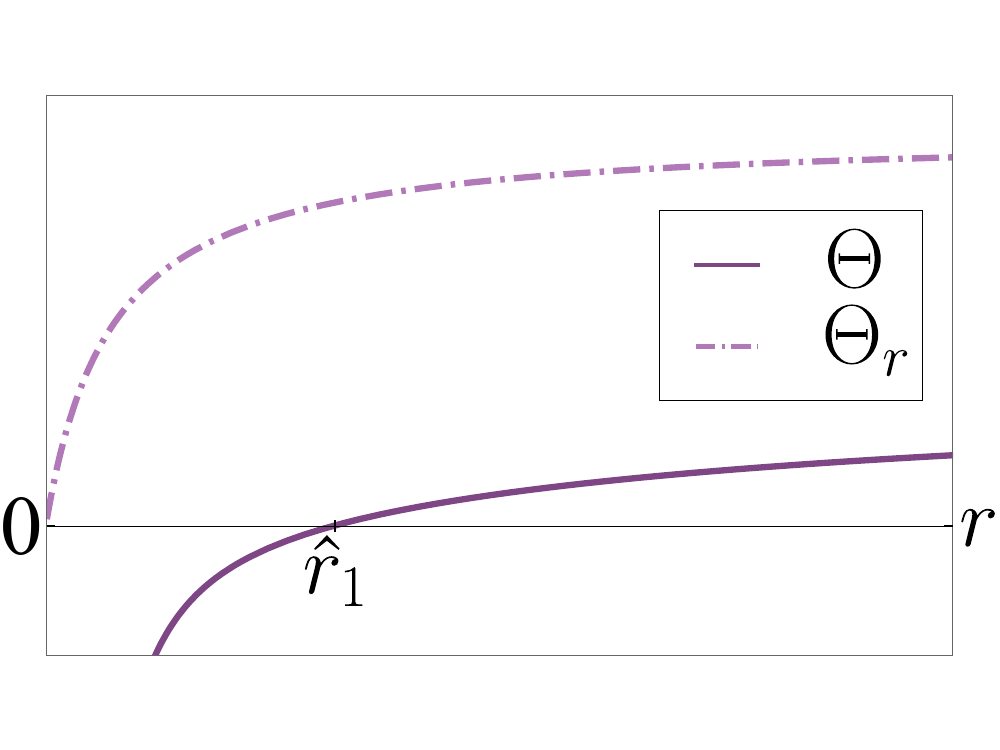}\\[-3mm] \centering{(c)}}}
\vspace{-2mm}
			\caption{(a) $R$-dependence of the temperatures $\Theta$, of the source ideal gas, and $\Theta_r$, of the test radiation fluid, for a fixed $r_1$. The radial profile of these temperatures for a fixed $R_1$ are plotted in (b) (case I with $R_1 > - k_1/(4 k_2)$ and case II) and (c) (case III).}
		\label{Fig-14}
		\end{figure}

		From now on, we only consider the regions where the singular models approach a Synge gas (dark blue regions in Figure \ref{Fig-10}). Then, we have three cases: I) ${\cal R}_0^{<}$ ($k_2 < 0$) with $k_1 \geq 0$, II) ${\cal R}_0^{<}$ ($k_2 < 0$) with $k_1 < 0$, and III) ${\cal R}_1^{>}$ ($k_2 > 0$ and $k_1 < 0$). 
		\begin{figure}[]
			\centerline{
\parbox[c]{0.33\textwidth}{\includegraphics[width=0.32\textwidth]{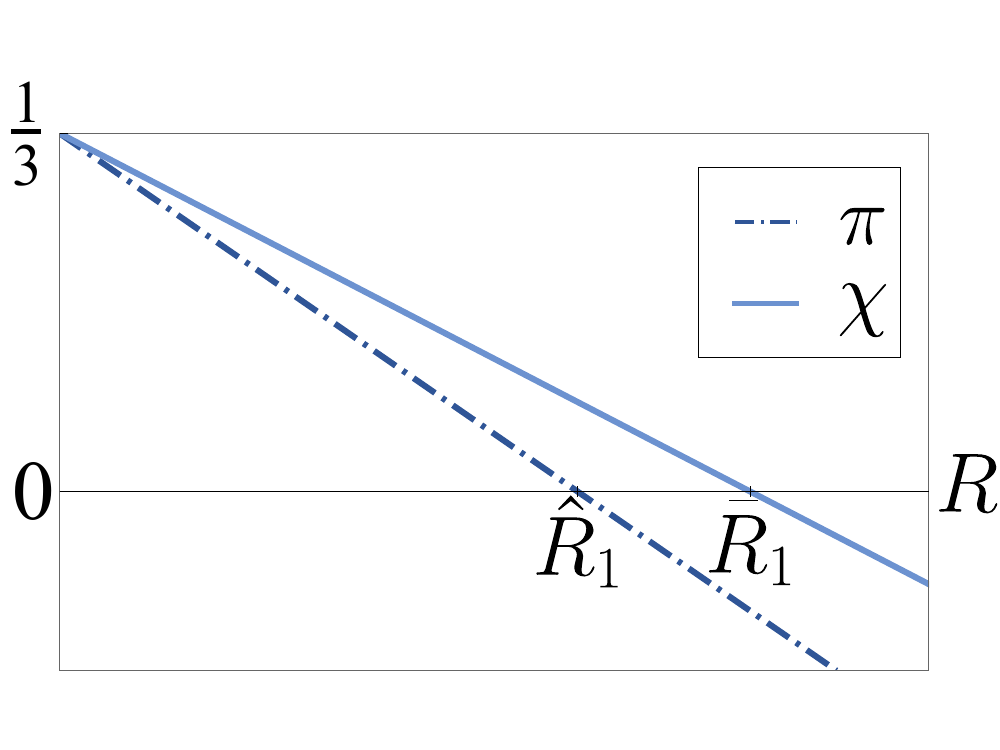}\\[-2mm] \centering{(a)}}
\raisebox{-2pt}{\parbox[c]{0.33\textwidth}{\includegraphics[width=0.32\textwidth]{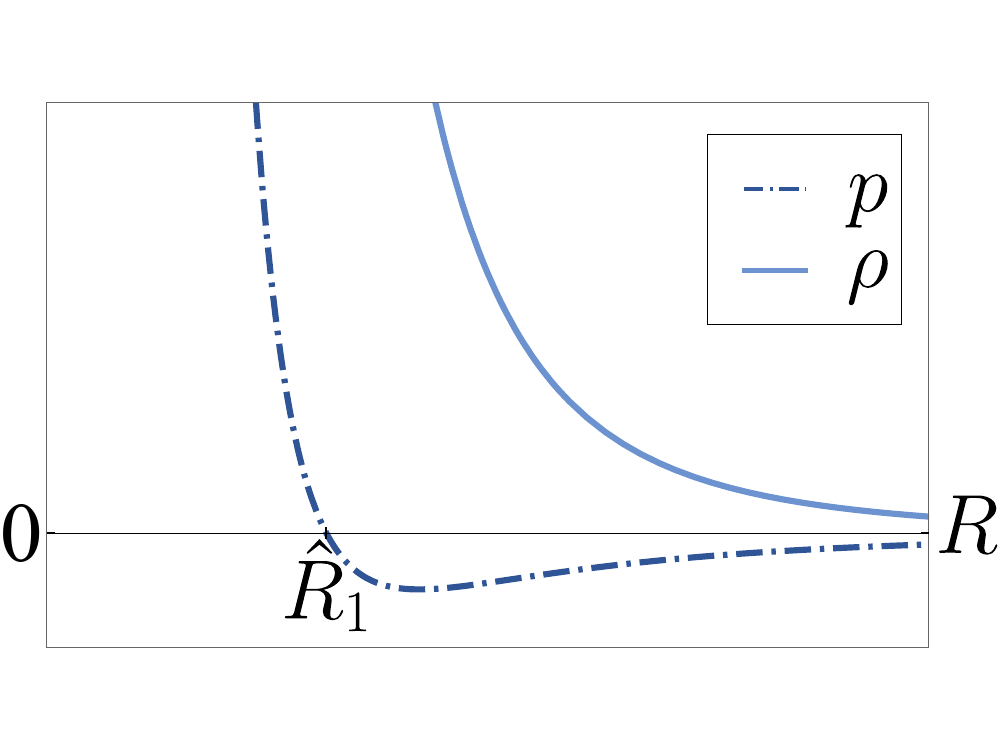}\\[-2mm] \centering{(b)}}}
\parbox[c]{0.33\textwidth}{\includegraphics[width=0.32\textwidth]{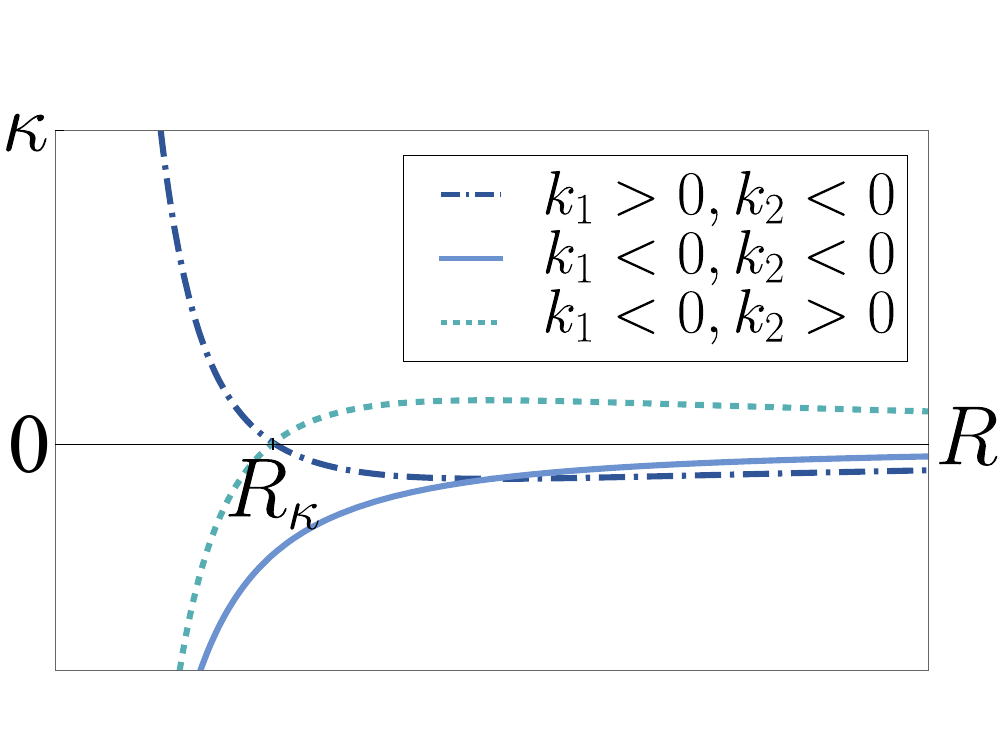}\\[-2mm] \centering{(c)}}}
\vspace{-2mm}
			\caption{The behaviour of the models for a fixed $r_1$. (a) Evolution of the hydrodynamic quantities $\pi$ and $\chi$. (b) Evolution of the energy density and pressure. (c) Evolution of the three-space curvature.}
		\label{Fig-11}
		\end{figure}
\\ \\
		In any case, for a suitable fixed value of the radial coordinate $r_1$, the hyperbola $\pi = 0$ contains a point $(r_1^2, \hat{R}_1)$. In fact, the real function $\pi_1(R) \equiv \pi(r_1^2, R)$ decreases in the interval $[0, \hat{R}_1]$ between $1/3$ and zero (see Figure \ref{Fig-11}(a)). Consequently, the pressure $p_1(R) \equiv p(r^2_1, R)$ is a decreasing function that vanishes at $\hat{R}_1$ (see Figure \ref{Fig-11}(b)). \\ \\
		In the Stephani universes, the three-space curvature depends on time, $\kappa = \kappa(R)$, and expression (\ref{curvature-ss}) implies that its sign depends on the sign of the function $k(R) = k_1 + k_2 R$. When $k_1 k_2 \leq 0$, the curvature vanishes at $R_{\kappa} = - k_1/k_2$. In case I, $k_2 < 0$, $k_1 \geq 0$, the curvature is a decreasing function and $R_{\kappa} < \hat{R}_1$ (region ${\cal R}_0^{<}$) for $r$ such that $r^2 < 4/(3 k_1)$. In case II, $k_2 < 0$, $k_1 <0$, the curvature $\kappa$ is always negative. Finally, in case III, $k_2 > 0$, $k_1 <0$, the curvature is an increasing function and $R_{\kappa}> R_1$; consequently it is negative on the physical region ${\cal R}_1^{>}$ (see Figure \ref{Fig-11}(c)). \\ \\
		Given a fixed value $R_1$ of the function $R$, the radial profiles of the thermodynamic quantities also depend on the three different considered cases (see Figure \ref{Fig-12}). In case I (region ${\cal R}^{<}_0$ with $k_1 \geq 0$), if $R_1 \leq - k_1/(4 k_2)$, the hydrodynamic functions $\pi$ and $\chi$ are decreasing functions which take values between $1/3$ and a non-negative real number (Figure \ref{Fig-12}(a)). In case II (region ${\cal R}^{<}_0$ with $k_1 < 0$), or in case I with $R_1 > - k_1/(4 k_2)$, $\pi$ and $\chi$ are also decreasing functions which take the value $1/3$ at $r = 0$ and vanish at a finite $r = \hat{r}_1$ and $r = \bar{r}_1$ (Figure \ref{Fig-12}(b)). Finally, in case III (region ${\cal R}_1^>$), $\pi$ and $\chi$ are increasing functions which are positive for $r>\hat{r}_1$ and $r > \bar{r}_1$ (Figure \ref{Fig-12}(c)). \\ \\
		\begin{figure}[t]
			\centerline{
				\parbox[c]{0.33\textwidth}{\includegraphics[width=0.32\textwidth]{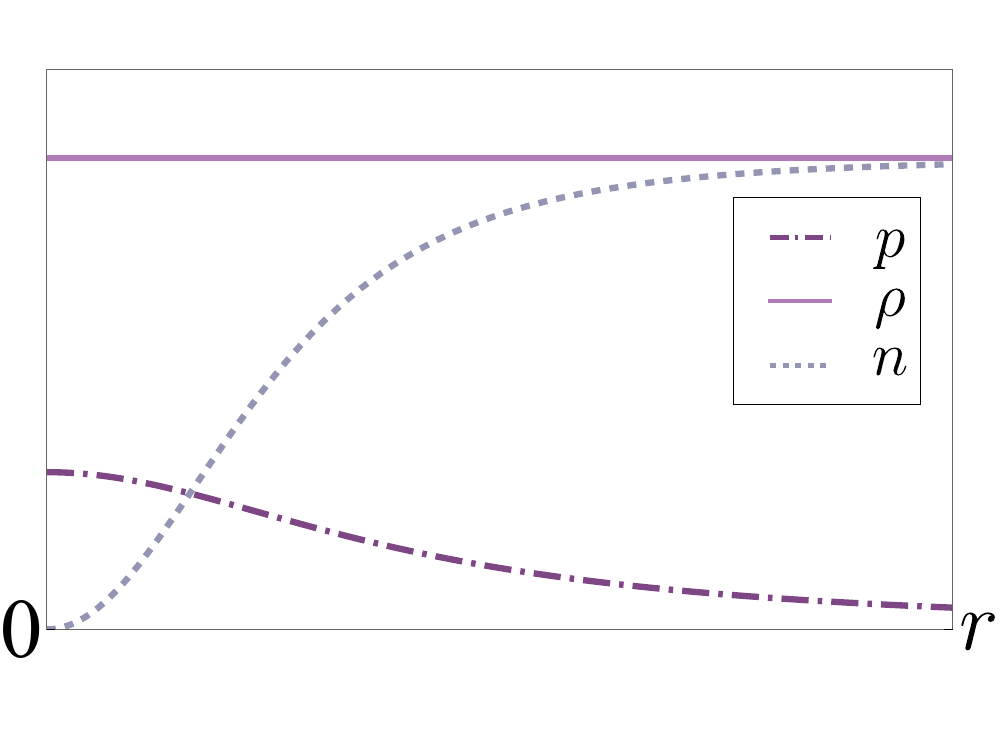}\\[-2mm] \centering{(a)}}
				\raisebox{-2pt}{\parbox[c]{0.33\textwidth}{\includegraphics[width=0.32\textwidth]{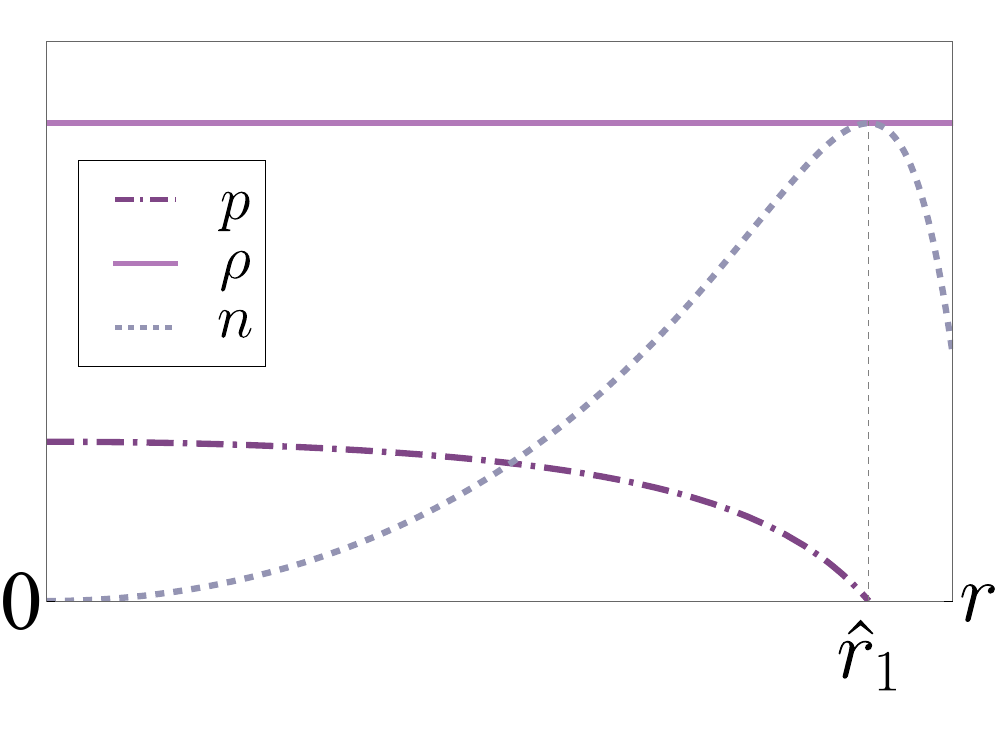}\\[-3mm] \centering{(b)}}}
				\raisebox{-2pt}{\parbox[c]{0.33\textwidth}{\includegraphics[width=0.32\textwidth]{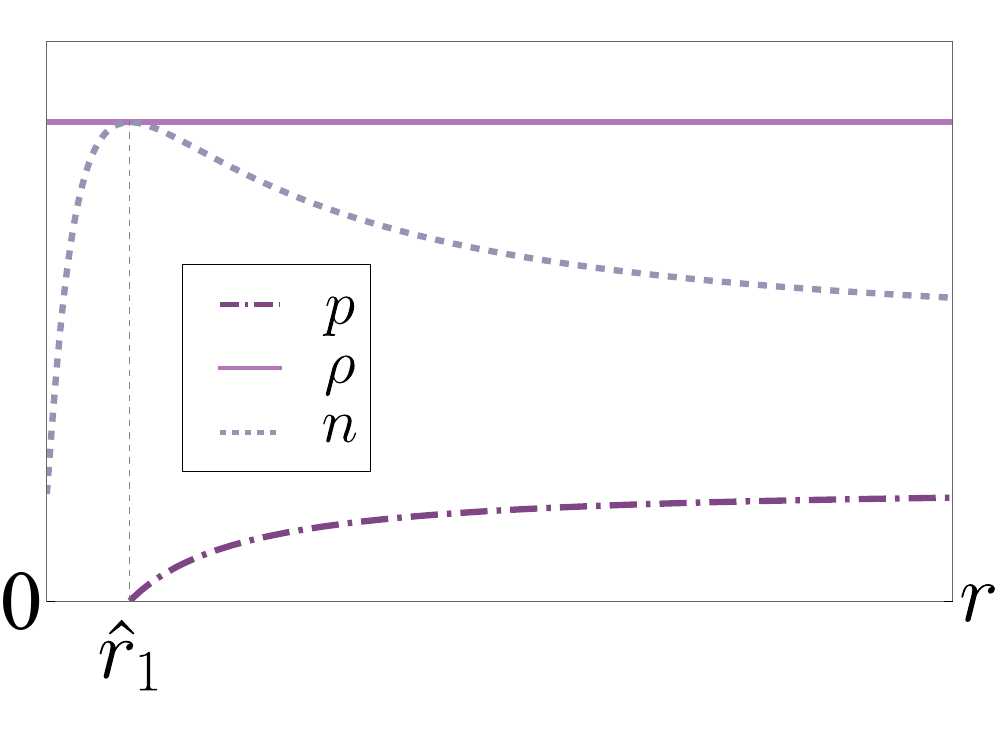}\\[-3mm] \centering{(c)}}}}
\vspace{-2mm}
			\caption{The radial profile of the energy density $\rho$, the pressure $p$ and the matter density $n$ depending on the values of the parameters $k_i$ for a fixed $R_1$. (a) Case I with $R_1 \leq - k_1/(4 k_2)$. (b) Case II and case I with $R_1 > - k_1/(4 k_2)$. (c) Case III.}
		\label{Fig-13}
		\end{figure}
		$\hspace{-2mm}$On the other hand, Figure \ref{Fig-13} shows, also for a fixed $R_1$, the radial profile of the energy density $\rho$ (constant), pressure $p$ and matter density $n$. Again, the behaviour is different for the three aforementioned cases. Figure \ref{Fig-13}(a): for $r > 0$, $p$ is decreasing and $n$ increasing, both positive. Figure \ref{Fig-13}(b): for $r < \hat{r}_1$, $p$ and $n$ have the same behaviour, and $p(\hat{r}_1) = 0$ and $n = \rho$ at $r = \hat{r}_1$. Figure \ref{Fig-13}(c): in this case $p$ is increasing and $n$ decreasing for $r > \hat{r}_1$. \\ \\ \\
		Figure \ref{Fig-14} describes the behaviour of both the temperature $\Theta$ of the source ideal gas and the temperature $\Theta_r$ of the test radiation fluid. For a fixed $r_1$, both temperatures decrease with $R$, and $\Theta$ can vanish at $\hat{R}_1$ (see Figure \ref{Fig-14}(a)). For a fixed $R_1$, the radial profile depends on the model. In Figure \ref{Fig-14}(b) we have plotted the cases I with $R_1 > - k_1/(4 k_2)$ and II, where both temperatures decrease and $\Theta$ vanishes at $r = \hat{r}_1$. Figure \ref{Fig-14}(c) shows case III, where both temperatures increase and $\Theta$ is positive for $r > \hat{r}_1$.

		\subsection{Singular models: the generalised Friedmann equation} \label{subsec-Model-Nos-Friedmann}
\begin{figure}[t]
			\vspace{1mm}
			\centerline{
				\parbox[c]{0.5\textwidth}{\centering{$k_2 < 0$}\\[-2mm] \includegraphics[width=0.49\textwidth]{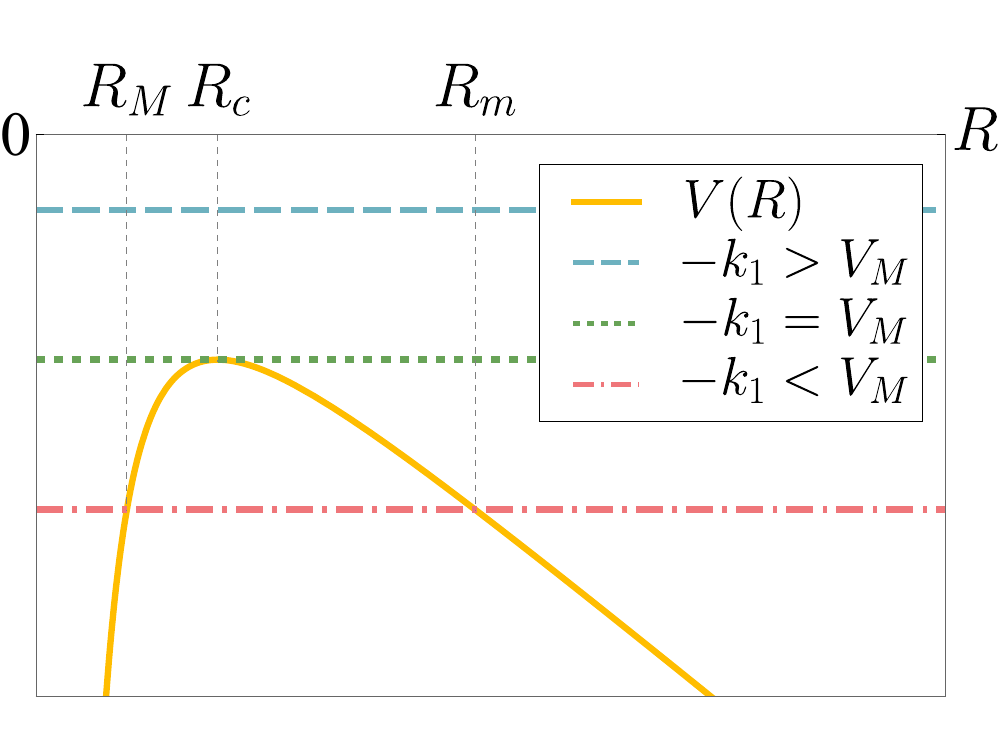} \\[-5mm]}
				\parbox[c]{0.5\textwidth}{\centering{$k_2 > 0$}\\[-2mm] \includegraphics[width=0.49\textwidth]{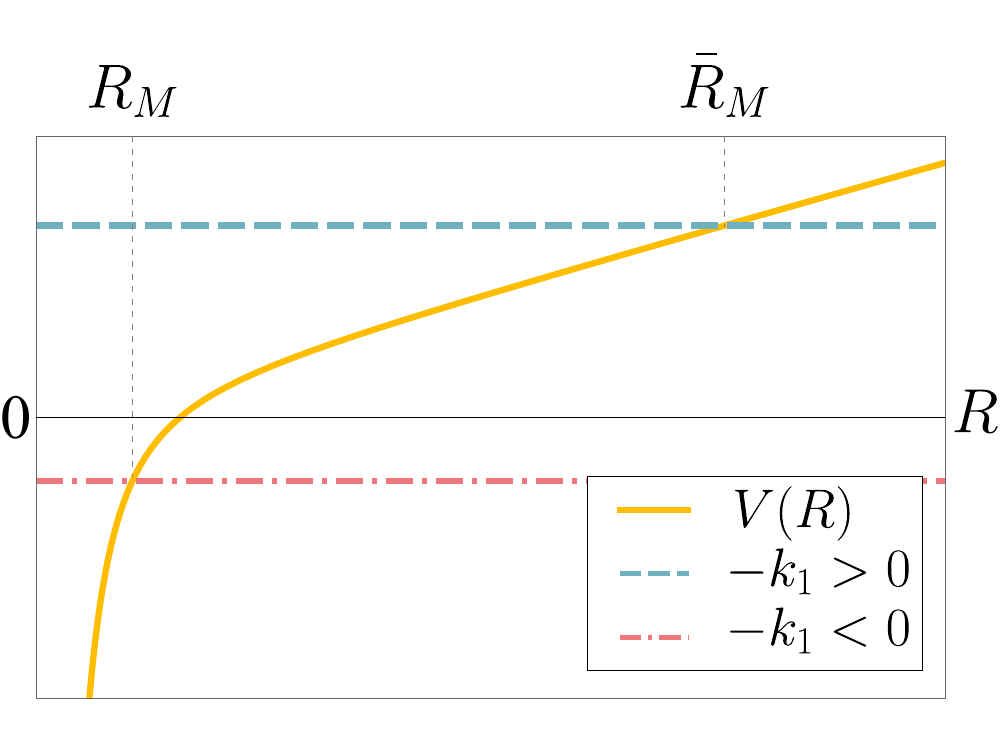} \\[-5mm]}
			}
			\centerline{
				\parbox[c]{0.5\textwidth}{\includegraphics[width=0.49\textwidth]{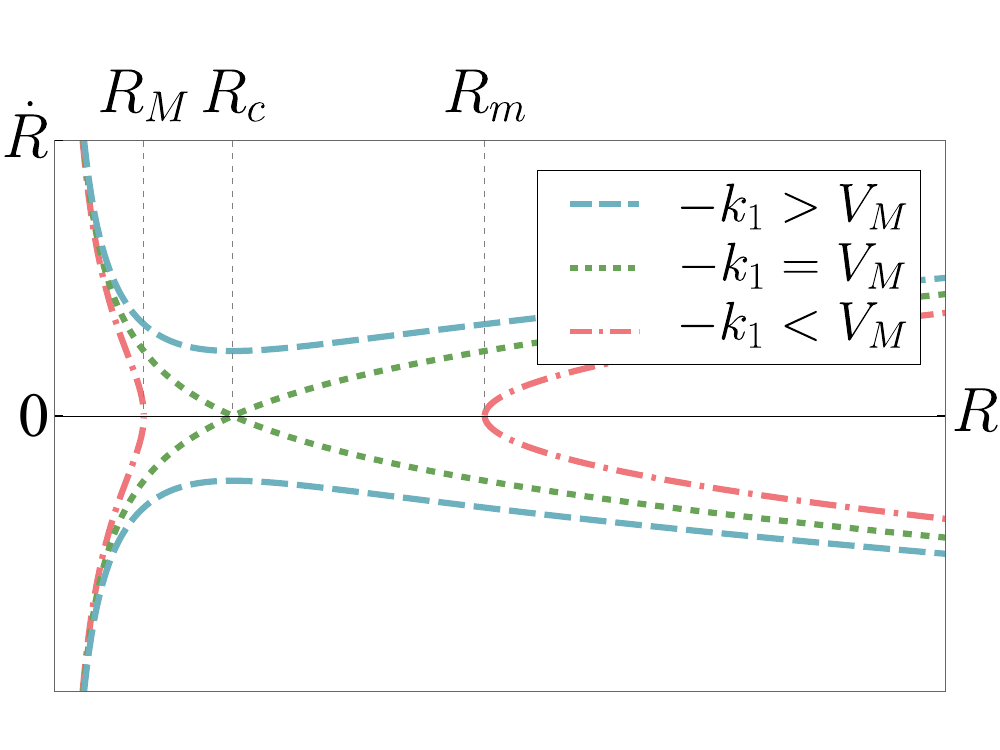}\\[-3mm] \centering{(a)}}
				\parbox[c]{0.5\textwidth}{\includegraphics[width=0.49\textwidth]{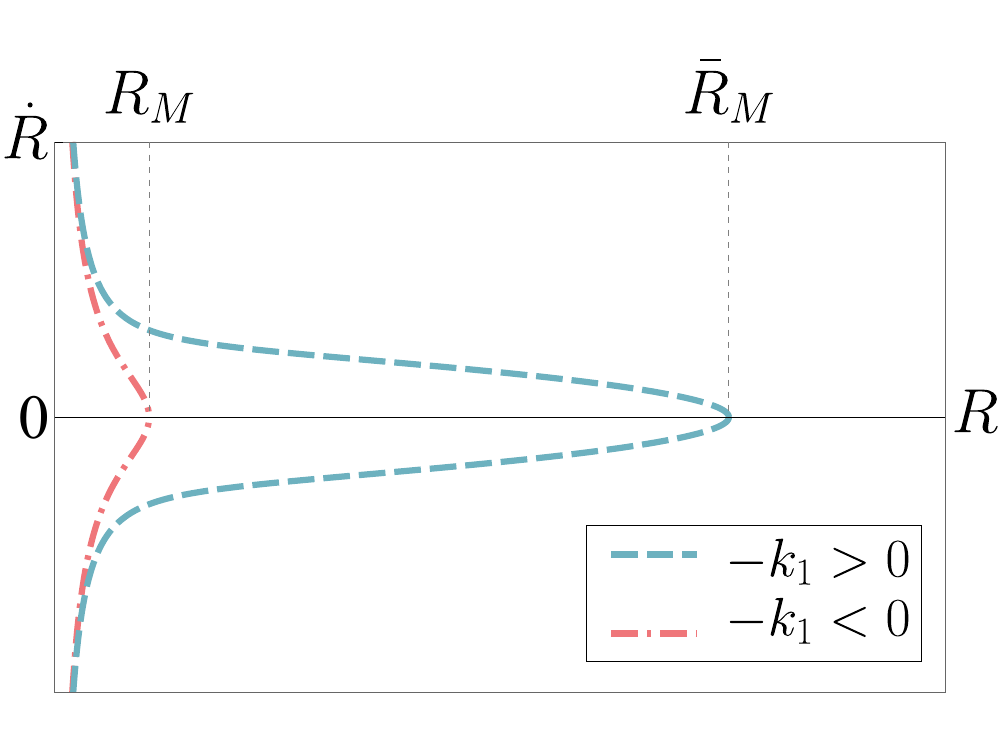}\\[-3mm] \centering{(b)}}
			}
			\vspace{-2mm}
			\caption{Trajectories in the phase space $\{R, \dot{R}\}$, which determine the behaviour of the metric function $R(t)$. (a) $k_2 < 0$, we can have closed, asymptotic and open models. (b) $k_2 > 0$, there are only closed models.}
		\label{Fig-15}
		\end{figure}

		For the singular models (\ref{singular-rho}), the generalised Friedmann equation (\ref{Friedmann-ideal-radiacio}) can be written as
		\begin{equation}
			\dot{R} = \sqrt{\! - k_1 \! - \! V(R)}, \qquad V(R) \equiv k_2 R - \frac{\tilde{\rho}_0}{R^2} , 
		\end{equation}
where $\tilde{\rho}_0 = \frac13 \rho_0 R_0^4$. Then, we can study the qualitative behaviour of the function $R(t)$ by drawing the effective potential $V(R)$ and analysing the trajectories in the phase plane $\{R, \dot{R}\}$ (see Figure \ref{Fig-15}).
\\ \\
		Depending on the sign of $k_2$, we have two qualitatively different effective potentials (see Figure \ref{Fig-15}). If $k_2 < 0$, it has a maximum at $R_c = (-2 \tilde{\rho}_0 / k_2)^{1/3}$, $V_M \equiv V(R_c) = k_2 R_c - \tilde{\rho}_0 R_c^2$. If $k_2 > 0$, it is a growing function.
		Now, we analyse the three cases I, II and III considered in the previous subsection. \\ \\ \\ \\
		In case II (region ${\cal R}_0^{<}$ with $k_1 < 0$), we have $k_2 < 0$ (see Figure \ref{Fig-15}(a)). Moreover, $-k_1 > V_M$, and the solution is valid for $r < r_{\infty}$. Then, we obtain an accelerated expanding model for values of $R$ larger than the critic value $R_c$, and for any $r_1 < r_{\infty}$ the pressure vanishes at a finite time $t_1$ ($R(t_1)) = R_1$). \\ \\
		In case I (region ${\cal R}_0^{<}$ with $k_1 \geq 0$), we have $k_2 < 0$ (see Figure \ref{Fig-15}(a)), and three different models can occur. If $-k_1 > V_M$, we obtain a model similar to that of case II but now valid for any $r > 0$. If $-k_1 = V_M$, we obtain an asymptotic expanding model with $R \rightarrow R_c$; for small values of $r$, the pressure never becomes zero, but for large values of $r$, the pressure vanishes at $R_1 < R_c$. Finally, if $-k_1 < V_M$, we have closed models, with a maximum value of $R$, $R_M < R_{\kappa}$. Generically, a $\bar{r}$ exists such that $R_M < \hat{R}_1$ if $r < \bar{r}$ and $R_M > \hat{R}_1$ if $r > \bar{r}$ ($\bar{r} = \infty$ for large values of $k_1$). \\ \\
		In case III (region ${\cal R}_1^{>}$ with $k_1 < 0$), we have $k_2 > 0$ (see Figure \ref{Fig-15}(b)), and we also obtain closed models, but the pressure vanishes before the contracting era, $\hat{R}_1 < R_M$. Moreover, $\hat{R}_1 < R_{\kappa}$.
		
		\subsection{Singular models: physical interpretation} \label{subsec-Model-Nos-phys-int}
		To sum up, the solutions considered in this section model a spherically symmetric spacetime inhomogeneity caused by an ultrarelativistic gas with homogeneous energy density, and inhomogeneous pressure, matter density and temperature. This inhomogeneity is compatible with a decoupled test inhomogeneous radiation fluid, which is isotropic as measured by the observer comoving with the matter fluid. \\ \\
		Wide ranges of parameters lead to physically realistic models. All of them start with a hot ultrarelativistic fluid that cools down with time, in most cases even becoming dust. Their radial profiles, however, depend on the considered model. Most of them are only physically admissible up to or from a certain value of the radial coordinate, where the pressure vanishes. In some cases we have a void of hot matter (cases I and II), and in other cases the matter density decreases, and the temperature increases, with $r$ (case III). \\ \\
		Thus, the models are useful to describe local inhomogeneities. Nevertheless, in order to be useful globally, they must be matched with other dust models beyond the hypersurface $p = 0$.

\part{IDEAL characterisation of spacetime and fluid \mbox{properties} with \textit{xAct}}

\chapter{IDEAL characterisations and their implementation} \label{chap-IDEAL}

	\section{IDEAL characterisations and determinations} \label{sec-IDEAL}
	Usually, a set of $r$ quantities $\phi_i$ is subject to a set of (differential) equations involving another set of $q$ quantities $\psi_j$, $\mathcal{D}_n(\phi_1, ..., \phi_r, \psi_1, ..., \psi_q) = 0$. Then, one question that one might raise is whether there exists a conditional system $\hat{\mathcal{D}}_m(\phi_1, ..., \phi_r) = 0$, namely an equivalent set of (differential) equations involving only the first set of quantities $\phi_i$. \\ \\ 
	For instance, an exact one-form $f$ is characterised by $f = \textrm{d}\alpha$, where $\alpha$ is a scalar function. Verifying whether a given one-form is exact using this characterisation requires checking the existence of a potential function whose differential yields the considered one-form. However, according to the Poincaré Lemma we have that a one-form $f$ is exact if, and only if, it is closed, $\textrm{d}f = 0$. This provides a more convenient characterisation of exact one-forms. \\ \\
	The sonic condition (\ref{cond. sonica}) offers another instance where a system of differential equations involving several quantities admits a conditional system involving only some of them. In particular, we have that the local thermal equilibrium is characterised by ${\cal D}_n(u, \rho, p, n, \epsilon, s, \Theta) = 0$ given in (\ref{matter-conservation}$-$\ref{thermo-first-law}). This system involves two sets of quantities, the hydrodynamic quantities $\{u, \rho, p\}$ appearing in the perfect fluid energy tensor and the thermodynamic quantities $\{n, \epsilon, s, \Theta\}$, and admits a conditional system involving only one of them, $\hat{\cal D}_m(u, \rho, p) = 0$. \\ \\ 
	Solving the direct problem is equivalent to obtaining this conditional system, the sonic condition in this case. The inverse problem, in turn, is equivalent to obtaining the whole set of quantities involved in the system of (differential) equations. Therefore, this terminology can be extended to other cases: the problem of finding the conditional system $\hat{\mathcal{D}}_m(\phi_1, ..., \phi_r) = 0$ will be called the \textit{direct problem} and that of finding the solution of the whole system $\mathcal{D}_n(\phi_1, ..., \phi_r, \psi_1, ..., \psi_q) = 0$ will be called the \textit{inverse problem}. \\ \\
	The invariant characterisation of spacetimes can also be formulated in these terms. For example, Birkhoff's theorem \cite{Birkhoff} states that ``the only spherically symmetric vacuum solution of the Einstein equations is the Schwarzschild metric". This gives a complete characterisation of the Schwarzschild spacetime by means of two conditions on the metric tensor $g$. The first one implies that its Ricci tensor is zero, ${\rm Ric}(g) = 0$, while the second one states that $g$ admits an isometry group G$_3$ acting on spacelike two-dimensional orbits with positive curvature. \\ \\ 
	The first condition is intrinsic, only involves metric concomitants (other tensorial quantities that can be built out of the metric and its covariant derivatives), and explicit. The second one can also be written explicitly, but it involves elements other than the metric tensor, such as the Killing vectors $X_i \; (i = 1, 2, 3)$ of the spherical symmetry. \\ \\ 
	Then, we have that the Schwarzschild spacetime is characterised by a system $\mathcal{D}_n(g, X_i) = 0$. In \cite{FS-Schwarzschild} (see also Section \ref{IDEAL-Sch}), Ferrando and Sáez give an equivalent characterisation of the Schwarzschild spacetime expressed solely in terms of the metric tensor and its differential concomitants, $\hat{\mathcal{D}}_m(g) = 0$. \\ \\
	This result solves, for a very particular case, a long-standing problem in General Relativity: determining whether a newly discovered solution to Einstein's equations is, in fact, an already known one expressed in a different coordinate system. Accordingly, the question of whether two explicit solutions represent (locally) the same spacetime, differing only by a coordinate transformation, known as the \textit{equivalence problem}, has been extensively studied (see Chapter 9 of \cite{Kramer} and references therein). \\ \\ \\
	One of the earliest approaches to solving the equivalence problem was the so-called \textit{Cartan-Karlhede algorithm} \cite{Karlhede-1980, Karlhede-2006}. This technique is based on a more general method, due to Cartan \cite{Cartan}, applicable to the equivalence of sets of differential forms on manifolds under appropriate transformation groups. It consists of assigning a normalised frame, specially adapted to the algebraic structures of the Weyl tensor and Ricci tensor, to the spacetime and then obtaining the so-called \textit{Cartan invariants} uniquely characterising the spacetime locally. \\ \\
	In this context, the intrinsic characterisation of spherically symmetric spacetimes was established and partially solved by Takeno \cite{Takeno-1952} (see also \cite{Takeno-1966}). However, he claimed that his characterisation was not “of the ideal form” since it did not “contain no tensor other than the metric tensor, the metric volume element and the curvature tensor”. In other words, for Takeno, the ideal way of characterising a spacetime was through a set of equations covariantly constructed out of the spacetime metric and its concomitants, as Ferrando and Sáez would later do in \cite{FS-Schwarzschild} for the Schwarzschild spacetime. \\ \\
	In \cite{FS-Spherical-IDEAL}, Ferrando and Sáez started using the acronym IDEAL (although Coll already used it before) to refer to spacetime characterisations that are Intrinsic (depending solely on the metric tensor $g$), Deductive (requiring no additional inference process), Explicit (with all involved concomitants expressed explicitly in terms of $g$) and ALgorithmic (admitting a flow chart with a finite number of steps). Some historical examples of what we now call \textit{IDEAL characterisations} include those of locally flat Riemann spaces \cite{Riemann-1867}, Riemann spaces with a maximal group of isometries \cite{Bianchi-1902} and locally conformally flat Riemann spaces \cite{Cotton-1899, Weyl-1918, Schouten-1921}. \\ \\ 
	More recently, this approach has gained renewed interest in the literature \cite{FS-Schwarzschild, FS-TypeD, FS-Kerr, FS-Spherical-IDEAL, FS-SS, Igor-2018, Igor-2019, Igor-2024}. Moreover, it can also be employed to algorithmically derive expressions for intrinsic quantities and properties of the spacetime under consideration, using only metric concomitants. We will refer to such procedures as \textit{IDEAL determinations}. While they typically appear as part of IDEAL characterisations, they can also be valuable in broader contexts. \\ \\ \\
	Another illustrative example of an IDEAL characterisation is provided by the Rainich conditions \cite{Rainich}. As already stated in Section \ref{subsec-Rainich}, Rainich formulated the first theory that characterises the non-vacuum solutions of the Einstein field equations corresponding to a specific energy content. He gave the necessary and sufficient conditions for a metric tensor $g$ to be an Einstein-Maxwell solution for a non-null electromagnetic field. \\ \\ 
	An Einstein-Maxwell solution consists of a pair $\lbrace g, F \rbrace$, where $g$ is a Lorentzian metric and $F$ the electromagnetic 2-form solution to Maxwell's equations, satisfying Einstein's field equations (\ref{Einstein-equations}) where the source is the energy tensor of an electromagnetic field, $T = \frac12 [F^2 + (*F)^2]$ with $*F$ the Hodge dual of $F$ (see Appendix \ref{AppendixB} for a summary of the notation). Therefore, the metric is subject to a set of differential equations, $\mathcal{D}(g, F) = 0$, that also involve the electromagnetic tensor. Rainich conditions are a set of conditions, $\hat{\mathcal{D}}(g) = 0$, on the metric alone ensuring that it is an Einstein-Maxwell solution, i.e. that there exists a two-form $F$ such that the pair $\lbrace g, F \rbrace$ verifies $\mathcal{D}(g, F) = 0$. \\ \\
	As already discussed in Section \ref{subsec-Rainich}, the results presented there offer a Rainich-like theory (and therefore an IDEAL characterisation) of the Einstein-Synge solutions. This follows the same philosophy as in \cite{Coll-Ferrando-termo, Hydro-LTE, CFS-CIG}, where the IDEAL characterisations of the perfect fluid, thermodynamic perfect fluid, generic ideal gas and classical ideal gas solutions are obtained. \\ \\
	Besides solving the problem of the metric equivalence, spacetime IDEAL characterisations are particularly useful in many fields. Firstly, they are of interest in achieving a fully algorithmic characterisation of the initial data associated with a given solution. For instance, the IDEAL approaches to the Schwarzschild, Kerr and type D solutions \cite{FS-Schwarzschild, FS-Kerr, FS-EM-sym-D, FS-typeD-vacuum} have served as the starting point in subsequent works \cite{Alfonso-Kerr-ID, Alfonso-TypeD-ID}. \\ \\ 
	Secondly, in numerical relativity, the near-satisfaction of the tensor equations associated to a particular IDEAL characterisation may be indicative of the local proximity of a numerical spacetime to the corresponding geometry \cite{Bhagwat, Okounkova}. In addition, the approach to zero of these equations could be used to study either linear or nonlinear stability of reference geometries, in an unambiguous and gauge independent way. \\ \\ 
	Thirdly, IDEAL characterisations are closely related to linearised gauge-invariant observables \cite{Igor-2017, Frob-Hack-Igor}.
	Finally, they have been proposed as a fundamental tool in epistemic relativity for making gravimetry by using relativistic positioning systems and relativistic stereometric systems \cite{Coll-2016}. \\ \\
	It is worth noting that, so far, we have focused exclusively on spacetime IDEAL characterisations and determinations. However, the concept can be extended beyond this context. For example, the Poincaré Lemma discussed at the beginning of this section provides an IDEAL characterisation of exact one-forms.
	
	\section{Computer algebra programs in GR} \label{sec-computer-algebra}
	The mathematical complexity of computations in General Relativity, particularly when expressions are specified for a given spacetime and coordinate system, was a key motivation behind the early development of computer algebra systems. The term “computer algebra” covers the theory and implementation of computer programs to perform the symbolic manipulations and calculations usual in mathematics, in particular those arising in algebra and calculus \cite{MacCallum-2018}. These symbolic computation systems offer a fast and accurate way of handling more complicated calculations than can be performed by hand.\\ \\
	The aim of this section is not to provide a complete description of the design, implementation, features, or history of computer algebra systems, but rather to offer a glimpse of the context in which this third part of the thesis is framed (see \cite{MacCallum-2018} for an extensive review). \\ \\
	 One of the many uses of computer algebra systems in General Relativity is to solve the equivalence problem. The Cartan-Karlhede algorithm was first implemented in a formal calculation program by $\mathring{\rm{A}}$man, specifically in the CLASSI package \cite{Aman-Karlhede}, built on SHEEP \cite{SHEEP}. However, due to its lack of simplification methods and other limitations, others incorporated the same algorithm in more modern computer algebra programs. For instance, in \cite{PollneyI, PollneyII, PollneyIII}, the authors implemented it in GRTensor for \textit{Maple} using the spinor formalism. \\ \\ 
	 These algorithms need to obtain a canonical frame for the Weyl spinor as a first step, which involves solving a quartic equation. In some cases, the expression of this standard frame is too complicated or cannot even be found \cite{PollneyIII}. This is one of the reasons why alternative methods in which the frame is left completely arbitrary, such as the IDEAL algorithms, are necessary. This does not mean that free-frame methods will always provide a simpler solution faster than fixed-frame methods. Both approaches should be considered as complementary; depending on the spacetime considered and the frame in which it is expressed, one approach (or none) may be preferable over the other.\\ \\
	 Among the main general purpose computer algebra languages with available packages for General Relativity and/or differential geometry at the time of writing this thesis are the previously mentioned \textit{Maple}, as well as \textit{Macsyma}, \textit{Mathematica}, and \textit{Reduce}. A list with a brief description of such packages and links to their corresponding websites can be found at \cite{MacCallum-2018}. Here, we will focus on \textit{xAct}.\\ \\
	 \textit{xAct}, “Efficient tensor computer algebra for the Wolfram Language” \cite{xAct}, is a suite of \textit{Mathematica} packages for tensorial computer algebra, providing tools for both the manipulation of general expressions involving abstract indices and the evaluation of components with specific indices in a chosen coordinate system and metric. Its usefulness and popularity are reflected in the extensive bibliography of articles and theses that have employed it (see the articles section in \cite{xAct}). \\ \\
	 Four of these packages (\textit{xCore}, \textit{xPerm}, \textit{xTensor} and \textit{xCoba}) act as the kernel for the rest, while a number of contributed modules have been developed by various authors to extend the core functionalities for different purposes. One such extension, introduced in the following subsection, constitutes the main subject of this third part of the thesis.
	
	\subsection{xIdeal} \label{subsec-xIdeal}
	The algorithmic structure of the IDEAL approach described in Section \ref{sec-IDEAL} makes it especially well-suited for implementation in some of the computer algebra programs mentioned in the previous section. In particular, in October 2023, I started developing a package for \textit{xAct} in collaboration with A. García-Parrado, which we have named \textit{xIdeal}. The idea behind this package is to implement all these IDEAL characterisations and determinations. As new IDEAL algorithms continue to be developed, this task remains open-ended. Nevertheless, a functional version is already available at \cite{xIdeal}. \\ \\
	\textit{xIdeal} includes both IDEAL characterisations and determinations of properties related to perfect fluids, the main topic of this thesis, as well as of other spacetime geometric properties and quantities. In the following chapters, we present all of these characterisations and determinations without proof, providing references where the corresponding demonstrations can be found. Then, we briefly describe the \textit{xIdeal} functions that implement them and show some examples where they have been used to recover previously known results. Most of these examples can be found in the documentation of their corresponding functions, the remaining more lengthy ones can be found in the form of \textit{Mathematica} notebooks in \cite{S-GitHub}. \\ \\
	In addition, \textit{xIdeal} also contains a database of various metrics called \texttt{xActSolData}. The idea behind this database is to collect these metrics expressed in different coordinate systems allowing the user to load them directly without having to enter them manually. It also includes a function that enables users to add new metrics. Moreover, each stored metric is accompanied by a list of known properties, which can be accessed as needed.

\chapter{Perfect fluid properties} \label{chap-PerFlu-xIdeal}

	\section{Thermodynamic perfect fluid. The Rainich-\mbox{like theory}} \label{sec-thermo-perflu-Rainich}
	As mentioned in the previous chapter, \textit{xIdeal} includes functions that implement IDEAL characterisations related to perfect fluids. In this section, we describe those that characterise properties of the source, namely when a spacetime is compatible with a (thermodynamic) perfect fluid source and the IDEAL determination of the corresponding perfect energy tensor quantities. \\ \\
	In \cite{Coll-Ferrando-termo}, Coll and Ferrando developed the Rainich-like theory for the thermodynamic perfect fluid solutions. As a first step, they used the IDEAL characterisation of a perfect fluid solution \cite{BCM-1992}. In the next step, they determined the additional IDEAL condition for this perfect energy tensor to be compatible with l.t.e. We present it here using the more recent notation adopted in \cite{SFM-simetries-Bianchi} (see Appendix \ref{AppendixB} for further details): \\ \\
	\textbf{Perfect fluid.} \textit{Consider the following concomitants of the Ricci tensor} $R$\textit{:}
		\begin{equation} \label{fluper-definitions}
			\hspace{-10mm} r \equiv \textrm{tr} R , \qquad S \equiv R - \frac14 r \, g , \qquad q \equiv - 2 \frac{\textrm{tr} S^3}{\textrm{tr} S^2} , \qquad Q \equiv S - \frac14 q \, g . 
		\end{equation}
\textit{A spacetime is a perfect fluid solution if, and only if, the Ricci tensor} $R$ \textit{satisfies:}
		\begin{equation} \label{fluper-conditions-A2}
			Q^2 + q Q = 0 , \qquad \qquad q \, Q(v,v) > 0 ,
		\end{equation}
\textit{where} $v$ \textit{is any timelike vector. Moreover, the energy density} $\rho$\textit{, the pressure} $p$ \textit{and the unit velocity} $u$ \textit{of the fluid are given by:}
		\begin{equation} \label{fluper-hydro-variables}
			\hspace{-10mm} \rho = \frac14 (3 q + r) , \qquad p = \frac14 (q - r) , \qquad u = \frac{P(v)}{\sqrt{P(v,v)}} , \qquad P \equiv \frac{1}{q} Q .
		\end{equation}
	\textbf{Thermodynamic perfect fluid.} \textit{Consider the concomitants of the Ricci tensor} $R$ \textit{given in} (\ref{fluper-definitions}) \textit{and} (\ref{fluper-hydro-variables})\textit{. A spacetime is a thermodynamic perfect fluid solution in isoenergetic evolution if, and only if, the Ricci tensor} $R$ \textit{satisfies} (\ref{fluper-conditions-A2}) \textit{and} $u(\rho) = 0$\textit{. Moreover, a spacetime is a thermodynamic perfect fluid solution in non-isoenergetic evolution if, and only if, the Ricci tensor} $R$ \textit{satisfies} (\ref{fluper-conditions-A2}) \textit{and the concomitant} $\chi \equiv u(p)/u(\rho)$ \textit{fulfils:}
		\begin{equation} \label{cond. sonica. IDEAL}
			\textrm{d}\chi \wedge \textrm{d}\rho \wedge \textrm{d}p = 0 \, .
		\end{equation}
Notice that this IDEAL characterisation of the l.t.e. was presented in Theorem \ref{thm-hydro-LTE} of Section \ref{sec-intro-physical-interpretation}, where the goal was to characterise this property using only hydrodynamic quantities. Similarly, Theorem \ref{Theorem-ideal-direct-problem} of the same section gives the IDEAL condition enabling us to build a Rainich-like theory for a generic ideal gas: \\ \\
	\textbf{Generic ideal gas.} \textit{Consider the concomitants of the Ricci tensor} $R$ \textit{given in} (\ref{fluper-definitions}) \textit{and} (\ref{fluper-hydro-variables})\textit{. A spacetime is a thermodynamic perfect fluid solution in non-isoenergetic evolution compatible with the EoS of a generic ideal gas if, and only if, the Ricci tensor} $R$ \textit{satisfies} (\ref{fluper-conditions-A2}) \textit{and the concomitants} $\chi \equiv u(p)/u(\rho)$ \textit{and} $\pi \equiv p / \rho$ \textit{fulfil:}
		\begin{equation} \label{cond. sonica ideal IDEAL}
			\emph{d}\chi\wedge \emph{d}\pi = 0\, ,\qquad \chi \neq \pi\, .
		\end{equation}
	The perfect fluid characterisation (\ref{fluper-definitions}$-$\ref{fluper-conditions-A2}) is implemented in \textit{xIdeal} through the function \texttt{PerfectFluidQ}, which returns \texttt{True}, \texttt{False} or \texttt{Unknown} depending on whether \texttt{metric} fulfils conditions (\ref{fluper-conditions-A2}). To do so, it needs to be given the considered metric and an arbitrary timelike vector as inputs. The function returns \texttt{Unknown} if it is not able to determine whether the second condition is fulfilled, usually because not enough \texttt{Assumptions} about the input metric are given. \\ \\
	The IDEAL determination of the perfect energy tensor quantities (\ref{fluper-hydro-variables}) is implemented through the function \texttt{PerfectFluidVariables}, which checks whether the input metric is compatible with a perfect fluid source and, if so, returns a list with the energy density $\rho$, the pressure $p$ and the unit velocity $u$. This function also needs to be given an arbitrary timelike vector to work. \\ \\
	The \textit{xIdeal} commands implementing the IDEAL characterisations (\ref{cond. sonica. IDEAL}) and (\ref{cond. sonica ideal IDEAL}) are \texttt{ThermodynamicPerfectFluidQ} and \texttt{GenericIdealGasQ}, respectively. They both check if the input metric is compatible with a perfect fluid source and, if so, return \texttt{True} or \texttt{False} depending on whether the respective conditions are fulfilled. \\ \\
	To test these functions, we applied them to two examples: the ideal T-models (\ref{metric-T-ideal}$-$\ref{phi-alpha}) with $\tilde{\gamma} \neq 2$ and the spherically symmetric Stephani Universes (\ref{metrica-ss}). In both cases, \texttt{PerfectFluidQ} and \texttt{ThermodynamicPerfectFluidQ} return \texttt{True}, and \texttt{PerfectFluidVariables} returns the corresponding expressions for the energy density, pressure and fluid flow, (\ref{pressure-density-T-ideal}) and (\ref{tdp-1}). Moreover, \texttt{GenericIdealGasQ} correctly returns \texttt{True} for the ideal T-models and \texttt{False} for the spherically symmetric Stephani Universes, since only the particular case with $A_i(R) = c_i = constant$ in (\ref{Ai(R)}) is compatible with the generic ideal gas EoS (see Sections \ref{subsec-Chi-Stephani} and \ref{subsec-ideal} for details).
	
	\section{IDEAL characterisation of perfect fluid solutions} \label{sec-perflu-sols-characterisation}
	In the previous section, we presented the \textit{xIdeal} functions implementing IDEAL algorithms for characterising properties of the perfect fluid source in a given spacetime. However, as explained in Chapter \ref{chap-IDEAL}, the IDEAL approach can also be used to address the metric equivalence problem through invariant characterisations. In this section, we introduce the \textit{xIdeal} functions that implement IDEAL characterisations of perfect fluid solutions. \\ \\
	Ferrando and Sáez studied the IDEAL characterisation of spherically symmetric spacetimes in \cite{FS-Simetria-Esferica-R}. Among other results, they obtained the IDEAL characterisation of the Stephani universes using that they are the expanding conformally flat perfect fluid solutions: \\ \\ \\ \\
	\textbf{Stephani universes.} \textit{Consider the concomitants of the Ricci tensor} $R$ \textit{given in} (\ref{fluper-definitions})\textit{. The Stephani universes are characterised by conditions} (\ref{fluper-conditions-A2}) \textit{together with}
		\begin{equation} \label{Stephani-uni-IDEAL-charac}
			W = 0 \, , \qquad \qquad 3 \, \textrm{d}q + \textrm{d}r \neq 0 \, ,
		\end{equation}
\textit{where} $W$ \textit{is the Weyl tensor.} \\ \\
	Then, as a corollary, they also obtained the IDEAL characterisation of the FLRW universes: \\ \\ 
	\textbf{FLRW universes.} \textit{Consider the concomitants of the Ricci tensor} $R$ \textit{given in} (\ref{fluper-definitions})\textit{. The FLRW universes are characterised by conditions} (\ref{fluper-conditions-A2}) and (\ref{Stephani-uni-IDEAL-charac}) \textit{together with} $\textrm{d}q \wedge \textrm{d}r = 0$\textit{.} \\ \\
	Now, although it is not done in \cite{FS-Simetria-Esferica-R}, we can also obtain an IDEAL characterisation of the Kustaanheimo-Qvist spacetimes. Recall that the Stephani-Barnes solutions are the irrotational and shear-free perfect fluid solutions with non-zero expansion, and they are either conformally flat (Stephani Universes) or of Petrov-Bel type D (Kustaanheimo-Qvist spacetimes) \cite{Barnes}. \\ \\ 
	Then, we have that the Kustaanheimo-Qvist spacetimes are characterised by (\ref{fluper-conditions-A2}), $\omega = \sigma = 0 \neq \theta$ and $W \neq 0$, where $\omega$ and $\sigma$ are called the vorticity and shear tensors \cite{Rezzolla}, respectively. They are the antisymmetric and trace-free symmetric parts of $\nabla u$ computed in the space orthogonal to $u$, $V \equiv (\nabla u)_\perp$. Hence, they both vanish if, and only if, $V$ is proportional to the induced metric on the orthogonal space $\gamma$. In that case moreover, $V \neq 0$ if, and only if, $\theta \neq 0$. Therefore, we get the following result: \\ \\
	\textbf{Kustaanheimo-Qvist spacetimes}. \textit{Consider the concomitants of the Ricci tensor} $R$ \textit{given in} (\ref{fluper-definitions})\textit{. The Kustaanheimo-Qvist spacetimes are characterised by conditions} (\ref{fluper-conditions-A2}) \textit{together with}
		\begin{equation} \label{KQ-IDEAL-charac}
			W \neq 0 \, , \qquad \qquad V = \frac13 (\textrm{tr}V) \gamma \neq 0 \, ,
		\end{equation}
\textit{where} $V \equiv (\nabla u)_\perp$ \textit{is the part of} $\nabla u$ \textit{orthogonal to} $u$, $\gamma = g + u \otimes u$ \textit{is the induced metric on the orthogonal space and} $u$ \textit{can be obtained from} (\ref{fluper-hydro-variables}) \textit{.} \\
	These three IDEAL characterisations are implemented in \textit{xIdeal} via the functions \texttt{StephaniUniverseQ}, \texttt{FriedmannQ} and \texttt{KustaanheimoQvistQ}, respectively, which return \texttt{True}, \texttt{False} or \texttt{Unknown} depending on whether the metric they are given belongs to the corresponding families. They all need to be given an arbitrary timelike vector to work. \\ \\
	The \textit{xIdeal} function \texttt{StephaniUniverseQ} was tested with the spherically symmetric Stephani universes (\ref{metrica-ss}) and the ideal T-models (\ref{metric-T-ideal}$-$\ref{phi-alpha}) with $\tilde{\gamma} \neq 2$. It correctly returns \texttt{True} in the first case and \texttt{False} in the second one. \texttt{FriedmannQ} was tested with the FLRW metric and the metric for the spherically symmetric case in spherical coordinates. It returns \texttt{True} in both cases. Finally, to test the function \texttt{KustaanheimoQvistQ} we used a particular case of the Kustaanheimo-Qvist metrics (see the function's documentation for more details) and the spherically symmetric Stephani universes (\ref{metrica-ss}). It returns \texttt{True} in the first case and \texttt{False} in the second one as expected.
	
	\section{IDEAL characterisation of perfect fluid flows} \label{velocitats}
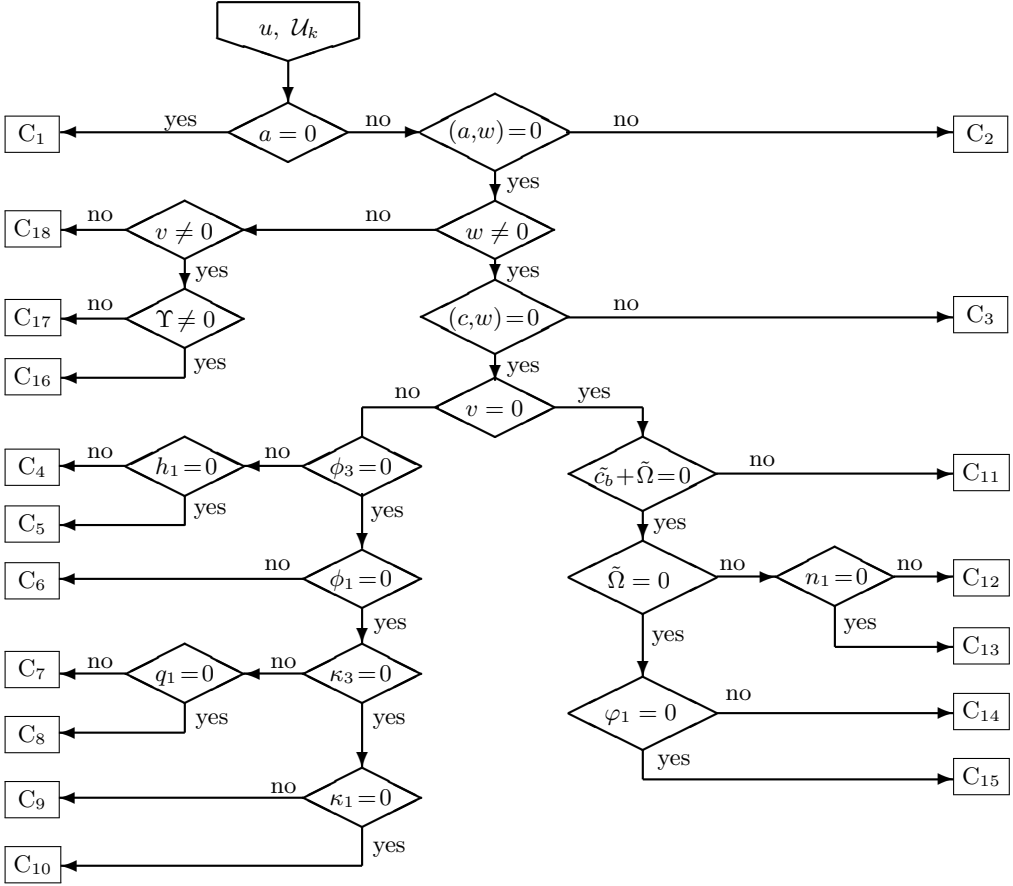
\begin{figure}[H]
	\hspace*{7mm} \setlength{\unitlength}{0.78cm} {\small \noindent
		\begin{picture}(16,16)
			\thicklines
			\put(4.7,15){\line(-3,-1){1.2}}
			\put(2.3,15){\line(3,-1){1.2}}
			\put(4.7,15){\line(0,1){0.6}} \put(2.3,15.6){\line(1,0){2.4}}
			\put(2.3,15.6){\line(0,-1){0.6}}

			\put(3,15.05){$ u, \ {\cal U}_k$}

			\put(3.5,14.60){\vector(0,-1){0.7}}

			\put(3.5,13.9){\line(-2,-1){1}} \put(3.5,13.9){\line(2,-1){1}}
			\put(3.5,12.9){\line(2,1){1}} \put(3.5,12.9){\line(-2,1){1}}
			\put(3,13.25){$a = 0$}

			\put(2.5,13.4){\vector(-1,0){2.86}}
			\put(-1.3,13.25){\framebox{\,C$_1$\,}}
			\put(4.5,13.4){\vector(1,0){1.25}}

			\put(7,14.05){\line(-2,-1){1.25}} \put(7,14.05){\line(2,-1){1.25}}
			\put(7,12.75){\line(2,1){1.25}} \put(7,12.75){\line(-2,1){1.25}}
			\put(6.2,13.3){$(a ,\! w) \! = \! 0$}

			\put(7,12.25){\line(-2,-1){1}} \put(7,12.25){\line(2,-1){1}}
			\put(7,11.25){\line(2,1){1}} \put(7,11.25){\line(-2,1){1}}
			\put(6.5,11.6){$w \neq 0$}

			\put(1.75,12.25){\line(-2,-1){1}} \put(1.75,12.25){\line(2,-1){1}} 
			\put(1.75,11.25){\line(2,1){1}} \put(1.75,11.25){\line(-2,1){1}} 
			\put(1.25,11.6){$v \neq 0$}

			\put(0.75,11.75){\vector(-1,0){1.1}}
			\put(0.75,10.25){\vector(-1,0){1.1}}
			\put(-1.3,11.65){\framebox{C$_{18}$}}
			\put(-1.3,10.15){\framebox{C$_{17}$}}

			\put(6,11.75){\vector(-1,0){3.3}}

			\put(1.75,9.75){\line(0,-1){0.5}}
			\put(1.75,9.25){\vector(-1,0){2.1}}
			\put(1.75,11.25){\vector(0,-1){0.5}}
			\put(-1.3,9.15){\framebox{C$_{16}$}}

			\put(1.75,10.75){\line(-2,-1){1}} \put(1.75,10.75){\line(2,-1){1}} 
			\put(1.75,9.75){\line(2,1){1}} \put(1.75,9.75){\line(-2,1){1}} 
			\put(1.25,10.1){$\Upsilon \! \neq 0$}

			\put(7,10.9 ){\line(-2,-1){1.25}} \put(7,10.9){\line(2,-1){1.25}} 
			\put(7,9.65){\line(2,1){1.25}} \put(7,9.65){\line(-2,1){1.25}} 
			\put(6.2,10.15){$(c ,\! w) \! = \! 0$}

			\put(7,9.25){\line(-2,-1){1}} \put(7,9.25){\line(2,-1){1}}
			\put(7,8.25){\line(2,1){1}} \put(7,8.25){\line(-2,1){1}}
			\put(6.5,8.6){$v = 0$}
			\put(9.5,8.25){\line(-2,-1){1.25}} \put(9.5,8.25){\line(2,-1){1.25}} 
			\put(9.5,7){\line(2,1){1.25}} \put(9.5,7){\line(-2,1){1.25}}
			\put(8.67,7.475){$\tilde{c_b} \! + \! \tilde{\Omega} \! = \! 0$}

			\put(9.5,6.5){\line(-2,-1){1.25}} \put(9.5,6.5){\line(2,-1){1.25}}
			\put(9.5,5.25){\line(2,1){1.25}} \put(9.5,5.25){\line(-2,1){1.25}}
			\put(8.9,5.7){$\tilde{\Omega} = 0$}

			\put(9.5,4.2){\line(-2,-1){1.25}} \put(9.5,4.2){\line(2,-1){1.25}}
			\put(9.5,2.95){\line(2,1){1.25}} \put(9.5,2.95){\line(-2,1){1.25}}
			\put(8.85,3.47){$\varphi_1 = 0$}

			\put(10.72,3.58){\vector(1,0){4.04}}

			\put(14.75,3.49){\framebox{C$_{14}$}}

			\put(14.75,2.39){\framebox{C$_{15}$}}

			\put(12.75,6.39){\line(-2,-1){1}} \put(12.75,6.39){\line(2,-1){1}} 
			\put(12.75,5.39){\line(2,1){1}} \put(12.75,5.39){\line(-2,1){1}} 
			\put(12.25,5.77){$n_1 \! = \! 0$}

			\put(14.75,5.75){\framebox{C$_{12}$}}

			\put(14.75,4.6){\framebox{C$_{13}$}}

			\put(9.5,7){\vector(0,-1){0.5}}

			\put(10.75,7.63){\vector(1,0){4}}

			\put(14.75,7.53){\framebox{C$_{11}$}}

			\put(10.75,5.89){\vector(1,0){1}}

			\put(13.75,5.89){\vector(1,0){1}} \put(12.75,4.7){\vector(1,0){2}}

			\put(12.75,5.4){\line(0,-1){0.7}}

			\put(9.5,5.27){\vector(0,-1){1.08}}

			\put(9.5,2.96){\line(0,-1){0.5}}

			\put(9.5,2.46){\vector(1,0){5.26}}

			\put(14.75,13.25){\framebox{\,C$_2$\,}}
			\put(8.2,13.4){\vector(1,0){6.55}}
			\put(14.75,10.2){\framebox{\,C$_3$\,}}
			\put(8.2,10.27){\vector(1,0){6.55}}
			\put(3.75,7.76){\vector(-1,0){1}}

			\put(0.75,7.76){\vector(-1,0){1.13}}

			\put(-1.3,7.62){\framebox{\,C$_4\,$}}

			\put(1.75,7.25){\line(0,-1){0.5}}
			\put(1.75,6.76){\vector(-1,0){2.12}}

			\put(-1.3,6.66){\framebox{\,C$_5$\,}}
			\put(4.75,8.25){\line(-2,-1){1}} \put(4.75,8.25){\line(2,-1){1}} 
			\put(4.75,7.25){\line(2,1){1}} \put(4.75,7.25){\line(-2,1){1}}
			\put(4.2,7.63){$\phi_3 \! = \! 0$}

			\put(1.75,8.25){\line(-2,-1){1}} \put(1.75,8.25){\line(2,-1){1}} 
			\put(1.75,7.25){\line(2,1){1 }} \put(1.75,7.25){\line(-2,1){1}}
			\put(1.25,7.63){$h_1 \! = \! 0$}

			\put(4.75,8.75){\line(0,-1){0.5}}

			\put(4.75,8.75){\line(1,0){1.3}}

			\put(8,8.75){\line(1,0){1.5}}

			\put(9.5,8.75){\vector(0,-1){0.5}}

			\put(3.8,4.25){\vector(-1,0){1.05}}

			\put(3.8,5.85){\vector(-1,0){4.18}}

			\put(4.75,6.35){\line(-2,-1){1}} \put(4.75,6.35){\line(2,-1){1}}
			\put(4.75,5.35){\line(2,1){1 }} \put(4.75,5.35){\line(-2,1){1 }}
			\put(4.2,5.73){$\phi_1 \! = \! 0$}

			\put(-1.3,5.7){\framebox{\,C$_6$\,}}
			\put(4.75,4.75){\line(-2,-1){1}} \put(4.75,4.75){\line(2,-1){1}}
			\put(4.75,3.75){\line(2,1){1}} \put(4.75,3.75){\line(-2,1){1}}
			\put(4.2,4.13){$\kappa_3 \! = \! 0$}

			\put(1.75,4.75){\line(-2,-1){1}} \put(1.75,4.75){\line(2,-1){1}}
			\put(1.75,3.75){\line(2,1){1}} \put(1.75,3.75){\line(-2,1){1}}
			\put(1.25,4.13){$q_1 \! = \! 0$}

			\put(1.75,3.75){\line(0,-1){0.5}}

			\put(1.75,3.25){\vector(-1,0){2.13}}

			\put(-1.3,4.2){\framebox{\,C$_7$\,}}

			\put(-1.3,3.1){\framebox{\,C$_8$\,}}

			\put(0.75,4.25){\vector(-1,0){1.12}}

			\put(-1.3,2){\framebox{\,C$_9$\,}}

			\put(-1.3,0.9){\framebox{C$_{10}$}}

			\put(3.75,2.15){\vector(-1,0){4.13}}
			\put(4.75,1){\vector(-1,0){5.1}}

			\put(4.75,1.65){\line(0,-1){0.65}}

			\put(4.75,2.65){\line(-2,-1){1}} \put(4.75,2.65){\line(2,-1){1}}
			\put(4.75,1.65){\line(2,1){1}} \put(4.75,1.65){\line(-2,1){1}}
			\put(4.2,2.03){$\kappa_1 \! = \! 0$}

			\put(4.75,3.75){\vector(0,-1){1.1}}

			\put(4.75,5.35){\vector(0,-1){0.6}}

			\put(4.75,7.3){\vector(0,-1){0.94}}

			\put(7,9.65){\vector(0,-1){0.45}}

			\put(7,11.25){\vector(0,-1){0.35}}

			\put(7,12.75){\vector(0,-1){0.5}}
			\put(1.4,13.5){yes}
			\put(4.8,13.5){no} 
			\put(9,13.5){no}
			\put(9,10.4){no}
			\put(7.2,12.45){yes}
			\put(4.8,11.9){no}
			\put(0.1,11.9){no}
			\put(0.1,10.4){no}
			\put(1.9,9.4){yes}
			\put(7.2,10.95){yes}
			\put(7.2,9.35){yes}
			\put(8.4,8.9){yes}
			\put(5.3,8.9){no}
			\put(0.1,7.9){no}
			\put(3.1,7.9){no}
			\put(11.3,7.75){no}
			\put(1.9,6.95){yes}
			\put(4.9,6.9){yes}
			\put(9.68,6.7){yes}
			\put(3.1,6){no}
			\put(10.8,6){no}
			\put(13.8,6){no}
			\put(12.9,5){yes}
			\put(9.65,4.8){yes} 
			\put(4.95,5){yes}
			\put(3.2,4.35){no}
			\put(0.1,4.35){no}
			\put(1.93,3.4){yes}
			\put(1.93,10.95){yes}
			\put(4.93,3.4){yes}
			\put(4.93,1.23){yes} 
			\put(3.2,2.23){no}
			\put(10.9,3.8){no}
			\put(9.75,2.7){yes}
	\end{picture} }
	\vspace{-7mm}
	\caption{Algorithm to distinguish the different classes of flows of perfect energy tensors.}
	\label{Fig-u}
	\end{figure}

	We conclude this chapter by presenting an IDEAL algorithm that we developed in \cite{SFM-velocitats}. In this section, we summarise the main results obtained there. Although this algorithm is not yet implemented in \textit{xIdeal}, we plan to incorporate it in the near future. \\ \\
	Every perfect fluid energy tensor $T \equiv \{u, \rho, p\}$ defines a point of the space ${\bf U} \times {\bf F} \times {\bf F}$, where ${\bf U}$ denotes the set of the timelike unit vector fields, and ${\bf F}$ the set of functions over a given domain of the spacetime. 
	For isolated media, we must impose the conservation of the energy tensor, 
	\begin{equation} \label{conservation-1}
		\nabla \cdot T = 0 \, ,
	\end{equation}
a first order differential system of four equations for the {\em hydrodynamic quantities} $\{u, \rho, p\}$. The set of solutions of (\ref{conservation-1}), ${\cal S} = \{T \equiv (u, \rho, p) \, | \, \nabla \cdot T = 0 \}$, defines a parallelepiped ${\cal S} = {\bf U}_c \times {\bf F}_{\rho} \times {\bf F}_p \subset {\bf U} \times {\bf F} \times {\bf F}$. \\ \\ \\
	${\bf U}_c$ is a proper subset of ${\bf U}$, ${\bf U}_c \neq {\bf U}$. Consequently, in general, a conservative perfect energy tensor having an arbitrary flow does not exist. Thus, it is natural to ask the following question: is it possible to intrinsically define ${\bf U}_c$? Or, more precisely, is it possible to express, solely in terms of $u$ and its derivatives, the necessary and sufficient conditions for $u$ to be the unit velocity of a conservative perfect energy tensor? In \cite{SFM-velocitats}, we show that the answer is affirmative and obtain these conditions. \\ \\
	Our results in \cite{SFM-velocitats} mean that the system (\ref{conservation-1}) for the hydrodynamic quantities, ${\cal D}_n(u, \rho, p) = 0$, admits a {\em conditional system} for the sole \mbox{{\em kinematic quantity} $u$}:
	\begin{equation} \label{D(u)}
		\hat{{\cal D}}_m(u) = 0 \, .
	\end{equation}
This means that (\ref{D(u)}) is a consequence of (\ref{conservation-1}), and conversely, for any solution $u$ of (\ref{D(u)}), a solution $\{u, \rho, p\}$ of (\ref{conservation-1}) exists. In other words, (\ref{D(u)}) is the integrability condition for the system (\ref{conservation-1}) to admit a solution $\{\rho,p\}$. \\ \\
	The search for the conditional system (\ref{D(u)}) leads to a classification of the time-like unit vectors in eighteen classes. For each class, we obtain the necessary and sufficient conditions in $u$ to ensure that it belongs to ${\bf U}_c$ (direct problem). Furthermore, for each class we solve the inverse problem by obtaining the pairs $\{\rho, p\}$ that complete a solution to the system (\ref{conservation-1}). \\ \\
	As we saw in Section \ref{sec-intro-physical-interpretation}, in terms of the hydrodynamic quantities $\{u, \rho, p\}$ the conservation equation (\ref{conservation-1}) takes the expression:
	\begin{equation} \label{conservation-2}
		\textrm{d} p + \dot{p} u + (\rho + p) a = 0 \, , \qquad \dot{\rho} + (\rho + p) \theta = 0 \, ,
	\end{equation}
where $a$ and $\theta$ are, respectively, the acceleration and the expansion of $u$, 
	\begin{equation} \label{acceleracio}
		a \equiv i(u) \nabla u \, , \qquad \quad \theta \equiv \nabla \cdot u \, , \qquad
	\end{equation}
and where a dot denotes the directional derivative, with respect to $u$, of a quantity $q$, $\dot{q} = u(q) = u^{\alpha} \partial_{\alpha} q$, a notation that we will keep throughout this section. \\ \\ \\
	Moreover, we denote $w$ and $v$ the rotation vectors associated with the unit velocity $u$ and the acceleration vector, respectively,
	\begin{equation} \label{rotations}
		w \equiv *(u \wedge \textrm{d} u) \, , \qquad v \equiv *(a \wedge \textrm{d} a) \, .
	\end{equation}
	Note that the change of the scalar quantities $\rho$ and $p$ by a constant factor, and by an additive constant without changing $\rho + p$, leaves equations (\ref{conservation-2}) invariant. Therefore, if $\{u, \rho, p\}$ is a solution of the conservation equations (\ref{conservation-2}), then $\{u, \bar{\rho},\bar{p}\}$ is also a solution, where
	\begin{equation} \label{kappa_i}
		\bar{p} = c_1 p + c_2 \, , \qquad \bar{\rho} = c_1 \rho - c_2 \, ,
	\end{equation}
$c_1, c_2$ being two arbitrary constants. \\ \\
	As previously commented, obtaining the conditional system (\ref{D(u)}) associated to the differential system (\ref{conservation-2}) requires considering different cases, C$_n$, that lead to a classification of the perfect fluid flows. Let us start with class of geodesic flows, \mbox{$a = 0$ (C$_1$).} \\ \\ 
	If $u$ defines a geodesic flow, then the first equation in (\ref{conservation-2}) becomes $\textrm{d} p = - \dot{p} u$. Then, either $u$ is an integrable 1-form or $p$ is a constant. In the first case, we have an irrotational flow, $u$ must be closed, and a function $\tau$ exists such that $u = - \textrm{d} \tau$. Hence, we can take $p \equiv p(\tau)$ as an arbitrary function, and the second equation in (\ref{conservation-2}) becomes a linear differential equation for $\rho$, which can be integrated. The general solution depends on an arbitrary $u$-invariant function, $\psi = \psi(x^1, x^2, x^3)$ (with $(x^i)$ defining a coordinate system in the hypersurface $\tau = constant$), and it is given by 
	\begin{equation} \label{rho-a=0-1} 
		\rho = \left[ \psi(x^i) - \int \! \theta \, p(\tau) \, e^{\int \! \theta \textrm{d} \tau} \ \textrm{d} \tau\right] e^{- \int \! {\theta \textrm{d} \tau}} \, .
	\end{equation}
If we have a rotating flow, then necessarily $p = p_0$. Therefore, the solution of the second equation in (\ref{conservation-2}) is of the form
	\begin{equation} \label{rho-a=0-2} 
		\rho = - p_0 + \psi(x^i) e^{- \int \! {\theta \textrm{d} \tau}} \, ,
	\end{equation}
where $\psi = \psi(x^1, x^2, x^3)$ is an arbitrary $u$-invariant function, $(\tau, x^i)$ being a coordinate system adapted to the vector field $u$, $u = \partial_{\tau}$. Consequently, a geodesic time-like unit vector $u$ is always the velocity of a perfect energy tensor. If $u$ is irrotational ($w = 0$), then $u = -\textrm{d} \tau$, and the pressure is given by an arbitrary real function $p = p(\tau)$ and the energy density $\rho$ is given by (\ref{rho-a=0-1}). If $u$ defines a rotating flow ($w \neq 0$) the pressure is an arbitrary constant $p = p_0$ and the energy density $\rho$ is given by (\ref{rho-a=0-2}). In both cases $\psi(x^i)$ is an arbitrary $u$-invariant function. \\ \\
	Now, let us suppose that $u$ defines a non-geodesic flow. Then, using arguments similar to those for the geodesic case, it is shown in \cite{SFM-velocitats} that a non-geodesic timelike unit vector $u$ is the velocity of a perfect energy tensor if, and only if, a function $\Psi$ exists such that the pair $\{u, \Psi\}$ fulfils
	\begin{equation} \label{eq-uchi}
		\textrm{d} s = q \wedge s\, , \quad \textrm{d} q = 0 \, , \quad s = a + \Psi u \, , \quad \Psi q = (\Psi - \theta) s + i(u) \textrm{d} s \, .
	\end{equation}
Moreover, the solutions $\{u, \bar{\rho}, \bar{p}\}$ of the conservation equations (\ref{conservation-2}) are defined by (\ref{kappa_i}), where $\{\rho, p\}$ are given by
	\begin{align}
		& p = - \lambda, \ \quad \rho = -p + e^{-\mu} & {\rm if} \quad \Psi \neq 0 \, , \hspace{1cm} \label{rhop-not0} \\
		& p = p(\lambda), \quad \rho = -p(\lambda) - p'(\lambda) e^{-\mu} & {\rm if} \quad \Psi = 0 \, , \hspace{1cm} \label{rhop-0}
	\end{align}
with $\{\lambda, \mu\}$ two functions fulfilling $s = e^{\mu} \, \textrm{d} \lambda$ and $q = \textrm{d} \mu$. \\ \\
	The case $\Psi = 0$ corresponds to an isobaric evolution ($\dot{p} = 0$). Then, $s = a$, and equations (\ref{eq-uchi}) imply
	\begin{equation} \label{eq-chi=0}
		\textrm{d} a \wedge a = 0 \, , \qquad \quad i(u) \textrm{d} a = \theta \, a \, .
	\end{equation}
Note that the first equation above means that $a = e^{\mu} \textrm{d} \lambda$, and then $\textrm{d}a = q \wedge a$ with $q = \textrm{d} \mu$, and equations (\ref{eq-uchi}) hold. Thus, a non-geodesic timelike unit vector $u$ is the velocity of a perfect energy tensor with $\dot{p} = 0$ if, and only if, it fulfils equations (\ref{eq-chi=0}). Then, $a = e^{\mu} \textrm{d} \lambda$, and the isobaric solutions $\{u, \bar{\rho}, \bar{p}\}$ of the conservation equations (\ref{conservation-2}) are defined by (\ref{kappa_i}), where $\{\rho, p\}$ are given in (\ref{rhop-0}). \\ \\
	It is worth remarking that the above result states that a unit timelike vector $u$ fulfilling (\ref{eq-chi=0}) is the velocity of an isobaric solution. Nevertheless, this $u$ can also be the velocity of a non-isobaric evolution when another pair $\{u, \Psi\}$, with $\Psi \neq 0$, is also a solution of equations (\ref{eq-uchi}). Some of the classes with a non-geodesic flow considered below are compatible with equation (\ref{eq-chi=0}), and this constraint defines a subset of flows in each of these classes. \\ \\
	Having established these general results, we now summarise the remaining ones. Consider the following $u$-concomitants needed to define the classes C$_n$ which we classify in terms of their derivative order:
	\begin{itemize}
\item 1st order
		\begin{equation} \label{Cn-1st-order}
			b \equiv - u + \frac{1}{a^2} i(a) \textrm{d} u \, , \qquad \ell_3 \equiv b - u \, .
		\end{equation}
\item 2nd order
		\begin{subequations} \label{Cn-2nd-order}
			\begin{equation}
				c \equiv \frac{1}{a^2} i(a) \textrm{d} a \, , \qquad \Omega \equiv \theta - c_u \, , \qquad \Upsilon \equiv \frac{1}{v^2} *(\textrm{d} \theta \wedge u \wedge a \wedge v) \, ,
			\end{equation}
			\begin{equation}
				\nu \equiv \Omega + 2 c_b \, , \qquad \ell_1 \equiv \nu u - 2a \, , \qquad e_4 \equiv \Omega u + 2c \, , \qquad e_5 \equiv 2b + z \, .
			\end{equation}
		\end{subequations}
\item 3rd order
		\begin{equation} \label{Cn-3rd-order}
			\phi_1 \equiv \tilde{\nu} - 5 \gamma + \pi - \gamma z_b \, , \quad \phi_3 \equiv \tilde{\gamma} + 2 \gamma c_b \, , \quad \varphi_1 \equiv \nu^* \! + 2(\theta + c_b - 2c_u) \, ,
		\end{equation}
		\begin{equation}
				f_1 \! \equiv -\gamma^{-1}[\nu^* \! + 2(\theta + c_b \! - \! 2 c_u)] \, a \! - \! 2 b - \! z \, , \quad \bar{\varphi}_1 \! \equiv \varphi_1/\tilde{\Omega} \, , \quad \bar{\phi}_1 \! \equiv \phi_1 / \phi_3 \, , \\[-1mm]
			\end{equation}
			\begin{equation}			 
				f_3 \equiv \Omega u \! - \gamma^{-1}[\gamma^* \! + 3 \gamma \! + \! \pi] \, a \! + \! 2c \, , \quad \pi \equiv \gamma \! + \! \gamma^* \! + \! \tilde{\Omega} - \gamma z_b \, , \quad y \equiv -\frac{i(a)\textrm{d}c}{a^2} \, .
			\end{equation}
\item 4th order
		\begin{subequations} \label{Cn-4th-order}
			\begin{equation}
				h_1 \equiv \textrm{d} \bar{\phi}_1 + \bar{\phi}_1 [\ell_1 - \bar{\phi}_1 \gamma u] + \bar{\phi}_2 [\ell_3 - \bar{\phi}_1 a] + f_1 - \bar{\phi}_1 f_3 \, , \\[-3mm]
			\end{equation}
			\begin{equation}
				\kappa_1 \equiv i(u) i(a)K_1 \, , \qquad \kappa_3 \equiv i(u) i(a)K_3 \, , \\[-1mm]
			\end{equation}
			\begin{equation}
				n_1 \equiv \textrm{d} \bar{\varphi}_1 + \bar{\varphi}_2 \ell_3 + \bar{\varphi}_1 \, (\ell_1 - \bar{\varphi}_1 \gamma u) + (2 \bar{\varphi}_1 - \bar{\varphi}_1^*)a - \bar{\varphi}_1 e_4 - e_5 \, , \\[1mm]
			\end{equation}
			\begin{equation}
				K_1 \equiv \textrm{d}f_1 + \ell_1 \wedge f_1 + \ell_3 \wedge f_2 + f_1 \wedge f_3 \, , \quad K_3 \equiv \textrm{d}f_3 + \gamma u \wedge f_1 + a \wedge f_2 \, , \\[-1mm]
			\end{equation}
			\begin{equation}
				\varphi_2 \equiv \tilde{\Omega}^* \! + \dot{\Omega} + 2(\Omega c_b + \tilde{\Omega}) \, , \qquad \bar{\varphi}_2 \equiv \varphi_2/\tilde{\Omega} \, , \qquad \bar{\kappa}_1 \equiv \kappa_1 / \kappa_3 \, , \\[-1mm]
			\end{equation}
			\begin{equation}
				f_2 \equiv \Omega^* u - \gamma^{-1}[\pi^* \! + \! \dot{\Omega} + \gamma z_u + 2 (\Omega c_b \! + \! \pi)] \, a + 2 \Omega \, b \! - c - y \, .
			\end{equation}
		\end{subequations}
\item 5th order
		\begin{equation} \label{Cn-5th-order}
			q_1 \equiv \textrm{d} \bar{\kappa}_1 + \bar{\kappa}_1 [\ell_1 - \bar{\kappa}_1 \gamma u] + \bar{\kappa}_2 [\ell_3 - \bar{\kappa}_1 a] + f_1 - \bar{\kappa}_1 f_3 \, ,
		\end{equation}
	\end{itemize}
with $a$, $\theta$, $w$ and $v$ ($u$-concomitants of first and second derivative order) given in (\ref{acceleracio}) and (\ref{rotations}), respectively, and where, for a vector $x$ and for a function $\varphi$ we have considered the following notation:
	\begin{equation} \label{x-base}
		x_u = - (x, u) \, , \quad x_a = \frac{1}{a^2}(x, a) \, , \quad x_b = \frac{1}{b^2}(x, b) \, , \quad x_w = \frac{1}{w^2}(x, w) \, ,
	\end{equation}
	\begin{equation} \label{dphi-base}
		\dot{\varphi} = -(\textrm{d} \varphi)_u \, , \quad \varphi^* = (\textrm{d} \varphi)_a \, , \quad \tilde{\varphi} = (\textrm{d} \varphi)_b \, , \quad \hat{\varphi} = (\textrm{d} \varphi)_w \, .
	\end{equation}
Consider also the following $u$-concomitants needed to write the necessary and sufficient conditions for a unit vector of class C$_n$ to belong to ${\bf U}_c$, which we classify in terms of their derivative order:
	\begin{itemize}
\item 2nd order 
		\begin{subequations} \label{Sn-2nd-order}
			\begin{equation}
				\gamma \equiv \frac{(v, w)}{w^2} \, , \qquad z \equiv -\frac{i(a) \textrm{d}b}{a^2} \, , \qquad \Psi_2 \equiv \frac{(u, v)}{(a, w)} \, ,
			\end{equation}
			\begin{equation}
				\ell_4 \equiv \Omega u - a + c \, .
			\end{equation}
		\end{subequations}
\item 3rd order 
		\begin{subequations} \label{Sn-3rd-order}
			\begin{equation}
				\Psi_3 \equiv \frac{\gamma z_w - \hat{\Omega}}{2 \, c_w} \, , \quad \Psi_{14} \equiv - \frac{2 \Omega c_b + \dot{\Omega}}{\varphi_1} \, ,
			\end{equation} 
			\begin{equation}
				\Psi_{16} \equiv \Upsilon^{-1} i(u)i(a)[\textrm{d} \theta \wedge u + \textrm{d}(\Upsilon a)] \, ,
			\end{equation}
			\begin{equation}
				\ell_2 \equiv \pi u + \Omega a - \gamma b \, , \qquad e_6 \equiv c + y - \Omega^* u - 2 \Omega \, b \, .
			\end{equation}
		\end{subequations}
\item 4th order 
		\begin{subequations} \label{Sn-4th-order}
			\begin{equation}
				\phi_2 \equiv \tilde{\pi} - \dot{\gamma} - \gamma(c_u \! + \! 4 c_b \! - \! 3 \theta \! + \! y_b) + \pi c_b \, , \quad \bar{\phi}_2 \equiv \phi_2/\tilde{\Omega} \, ,
			\end{equation}
			\begin{equation}
				\Psi_6 \equiv - \frac{\phi_2}{\phi_1} \, , \qquad \Psi_{11} \equiv - \frac{\tilde{\tilde{\Omega}} + \tilde{\Omega} \, c_b}{2(\tilde{c_b} + \tilde{\Omega})} \, .
			\end{equation}
		\end{subequations}
\item 5th order 
		\begin{subequations} \label{Sn-5th-order}
			\begin{equation}
				h_2 \equiv \textrm{d} \bar{\phi}_2 + \bar{\phi}_1 [\ell_2 - \bar{\phi}_2 \gamma u] + \bar{\phi}_2[\ell_4 - \bar{\phi}_2 a] + f_2 - \bar{\phi}_2 f_3 \, , \quad \kappa_2 \equiv i(u) i(a)K_2 \, ,
			\end{equation}
			\begin{equation}
				n_2 \equiv \textrm{d} \bar{\varphi}_2 + \bar{\varphi}_1 \ell_2 + \bar{\varphi}_2 \, (\ell_4 - \bar{\varphi}_1 \gamma u) + (\bar{\varphi}_2 - \Omega \bar{\varphi}_1 - \bar{\varphi}_2^*)a - \bar{\varphi}_2 e_4 - e_6 \, ,
			\end{equation}
			\begin{equation}
				\Psi_4 \equiv - \frac{\bar{g}(h_1, h_2)}{\bar{g}(h_1, h_1)} \, , \qquad \Psi_9 \equiv - \frac{\kappa_2}{\kappa_1} \, , \qquad \Psi_{12} \equiv - \frac{\bar{g}(n_1, n_2)}{\bar{g}(n_1, n_1)} \, .
			\end{equation}
		\end{subequations}
\item 6th order 
		\begin{subequations} \label{Sn-6th-order}
			\begin{equation}
				q_2 \equiv \textrm{d} \bar{\kappa}_2 + \bar{\kappa}_1[\ell_2 - \bar{\kappa}_2 \gamma u] + \bar{\kappa}_2 [\ell_4 - \bar{\kappa}_2 a] + f_2 - \bar{\kappa}_2 f_3 \, , \quad \Psi_7 \equiv - \frac{\bar{g}(q_1, q_2)}{\bar{g}(q_1, q_1)} \, ,
			\end{equation}
		\end{subequations}
	\end{itemize}
where $\bar{g}$ is the Riemannian metric $\bar{g} \equiv g + 2 u \otimes u$. \\ \\
	\begin{table}[t]
	\noindent
	\normalsize{
		\begin{tabular}{ll}
			\noalign{\hrule height 1.05pt}
			Classes & Definition relations \phantom{\large $\frac{I}{I}$} \\
			\hline
			C$_1$ & $a \! = \! 0$ \phantom{\large $\frac{I}{I}$} \\
			C$_2$ & $a \! \neq \! 0$, \quad $(a, \! w) \! \neq \! 0$ \phantom{\large $\frac{I}{I}$} \\
			C$_3$ & $a \! \neq \! 0$, \quad $(a,\! w) \! = \! 0$, \quad $(c, \! w) \! \neq \! 0$ \phantom{\large $\frac{I}{I}$} \\
			C$_4$ & $a \! \neq \! 0$, \quad $w \! \neq \! 0$, \quad  $(a, \! w) \! = \! (c, \! w) \! = \! 0$, \quad $v \! \neq \! 0$, \quad $\phi_3 \! \neq 0$, \quad  $\! h_1 \! \neq \!0$ \phantom{\large $\frac{I}{I}$} \\ 
			C$_5$ & $a\! \neq \! 0$, \quad $w \! \neq \! 0$, \quad  $(a, \! w) \! = \! (c, \! w) \! = \! 0$, \quad $v \! \neq \! 0$, \quad $\phi_3 \! \neq 0$, \quad  $\! h_1 \! = \!0$ \phantom{\large $\frac{I}{I}$} \\
			C$_6$ & $a \! \neq \! 0$, \quad $w \! \neq \! 0$, \quad $(a, \! w) \! = \! (c, \! w) \! = \! 0$, \quad $v \! \neq \! 0$, \quad $\phi_3 \! = 0$, \quad  $\! \phi_1 \! \neq \! 0$ \phantom{\large $\frac{I}{I}$} \\ 
			C$_7$ & $a\! \neq \! 0$, \quad $w \! \neq \! 0$, \quad $(a, \! w) \! = \! (c, \! w) \! = \! 0$, \quad $v \! \neq \! 0$, \quad $\phi_3\! = \! \phi_1 \! = \! 0$, \quad $\kappa_3 \! \neq \! 0$, \quad $q_1 \! \neq \! 0 \! \! \! \! \! \!$ \phantom{\large $\frac{I}{I}$} \\
			C$_8$ & $a \! \neq \! 0$, \quad $w \! \neq \! 0$, \quad $(a, \! w) \! = \! (c, \! w) \! = \! 0$, \quad $v \! \neq \! 0$, \quad $\phi_3 \! = \! \phi_1 \! = \! 0$, \quad $\kappa_3 \! \neq \! 0$, \quad $q_1 \! = \! 0 \! \! \! \! \! \!$ \phantom{\large$\frac{I}{I}$} \\
			C$_9$ & $a \! \neq \! 0$, \quad $w \! \neq \! 0$, \quad $(a, \! w) \! = \! (c, \! w) \! = \! 0$, \quad $v \! \neq \! 0$, \quad $\phi_3 \! = \! \phi_1 \! = \! 0$, \quad $\kappa_3 \! = \!0$, \quad $\kappa_1 \! \neq \! 0 \! \! \! \! \! \!$ \phantom{\large $\frac{I}{I}$} \\
			C$_{10}$ & $a \! \neq \! 0$, \quad $w \! \neq \! 0$, \quad  $(a, \! w) \! = \! (c, \! w) \! = \! 0$, \quad $v \! \neq \! 0$, \quad $\phi_3 \! = \! \phi_1 \! = \! 0$, \quad $\kappa_3 \! = \! 0$, \quad $\kappa_1 \! = \! 0 \! \! \! \! \! \!$ \phantom{\large $\frac{I}{I}$} \\
			C$_{11}$ & $a \! \neq \! 0$, \quad $w \! \neq \!0$, \quad $(a, \! w) \! = \! (c, \! w) \! = \! 0$, \quad $v \! = \! 0$, \quad 
$\tilde{c_b} \! + \! \tilde{\Omega} \! \neq \!0$ \phantom{\large $\frac{I}{I}$} \\ 
			C$_{12}$ & $a \! \neq \! 0$, \quad $w \! \neq \! 0$, \quad $(a, \! w) \! = \! (c, \! w) \! = \! 0$, \quad $v\! =\! 0$, \quad $\tilde{c_b} \! = \! -\tilde{\Omega} \! \neq \! 0$, \quad $n_1 \! \neq \! 0$ \phantom{\large $\frac{I}{I}$} \\
			C$_{13}$ & $a \! \neq \! 0$, \quad $w \! \neq \! 0$, \quad $(a, \! w) \! = \! (c, \! w) \! = \! 0$, \quad $v \! = \! 0$, \quad $\tilde{c_b}\! = \! -\tilde{\Omega} \! \neq \!0$, \quad  $n_1 \! = \! 0$  \phantom{\large $\frac{I}{I}$}\\ 
			C$_{14}$ & $a \! \neq \! 0$, \quad $w \! \neq \! 0$, \quad $(a, \! w) \! = \! (c, \! w) \! = \! 0$, \quad $v \! = \! 0$, \quad $\tilde{c_b}\! = \! \tilde{\Omega} \! = \! 0$, \quad \ \ $\varphi_1 \neq 0$ \phantom{\large $\frac{I}{I}$}  \\
			C$_{15}$ & $a \! \neq \! 0$, \quad $w \! \neq \! 0$, \quad $(a, \! w) \! = \! (c, \! w) \! = \! 0$, \quad $v \! = \! 0$, \quad $\tilde{c_b} \! = \! \tilde{\Omega} \! = \! 0$, \quad \ \ $\varphi_1 = 0$ \phantom{\large $\frac{I}{I}$} \\
			C$_{16}$ & $a \! \neq \! 0$, \quad $w \! = \! 0$, \quad $v \! \neq \! 0$, \quad 
$\Upsilon \! \neq \! 0$  \phantom{\large $\frac{I}{I}$} \\
			C$_{17}$ & $a \! \neq \! 0$, \quad $w \! = \! 0$, \quad $v \! \neq \! 0$, \quad 
$\Upsilon \! = \! 0$ \phantom{\large $\frac{I}{I}$} \\ 
			C$_{18}$ & $a \! \neq \! 0$, \quad $w \! = \! 0$, \quad $v \! = \! 0$ \phantom{\large $\frac{I}{I}$} \\
			\noalign{\hrule height 1.05pt}
		\end{tabular}
}
	\caption{The timelike unit vectors $u$ can be classified in eighteen classes C$_n$ ($n = 1, ..., 18$) defined by relations imposed on the $u$-concomitants given in (\ref{Cn-1st-order}$-$\ref{Cn-5th-order}).} \label{table-7}
	\end{table}
	$\hspace{-2mm}$With all these $u$-concomitants at hand, we can now introduce the following definition: a timelike unit vector $u$ is said to be of class C$_n$ ($n = 1, ..., 18$) if it satisfies the relations given in Table \ref{table-7}. Then, a timelike unit vector $u$ of class C$_n$ is the velocity of a perfect energy tensor if, and only if, it fulfils the differential system S$_n$ given in the second column of Table \ref{table-8}. Moreover, the pairs $\{\rho, p\}$ of hydrodynamic quantities associated with a velocity of class C$_n$ that fulfils the condition S$_n$ are determined by the expressions H$_n$ given in the third column of Table \ref{table-8}. For classes C$_5$, C$_8$, C$_{10}$, C$_{13}$, C$_{15}$, C$_{17}$ and C$_{18}$ these expressions are given by:
	\begin{table}[t]
	\noindent
	\normalsize{
		\begin{tabular}{lll}
			\noalign{\hrule height 1.05pt}
			C$_n$ \ \,\ & S$_n$: nec.$\, $\&  suf.$\, $conditions&H$_n$: invers problem $\{\rho, p\}$ \phantom{\large $\frac{I^*}{I}$}  \\
			\hline \\[-3mm]
			& \qquad \qquad \qquad & [$u = \textrm{d} \tau$] \qquad $p \! = \! p(\tau)$, \quad $\rho \ \ $ given in (\ref{rho-a=0-1}) \\[-2mm]
			C$_1$ & $\emptyset$  & \\[-2mm]
 			& \qquad \qquad \qquad & [$u = \partial_\tau$] \qquad $p \! = \! p_0$, \quad \, \, $\rho \ \ $ given in (\ref{rho-a=0-2}) \\[1mm]
			\hline
			C$_2$ & (\ref{eq-uchi}), \quad with $\ \Psi = \! \Psi_2$ \phantom{\large $\frac{I^2}{I}$} & \\
			C$_3$ & (\ref{eq-uchi}), \quad with $\ \Psi = \! \Psi_3$\phantom{\large $\frac{I^2}{I}$} & \\
			C$_4$ & (\ref{eq-uchi}), \quad with $\ \Psi = \! \Psi_4$ \phantom{\large $\frac{I^2}{I}$} & [$ \ s = a + \Psi u, \quad \Psi q = (\Psi - \theta ) s + i(u) \textrm{d} s\ $] \\ 
			C$_6$ & (\ref{eq-uchi}), \quad with $\ \Psi = \! \Psi_6$ \phantom{\large $\frac{I^2}{I}$} & [$ \ s = e^{\mu} \textrm{d} \lambda, \qquad \ q = \textrm{d} \mu \ $] \\ 
			C$_7$ & (\ref{eq-uchi}), \quad with $\ \Psi = \! \Psi_7$ \phantom{\large $\frac{I^2}{I}$} & \\
			C$_9$ & (\ref{eq-uchi}), \quad with $\ \Psi = \! \Psi_9$ \phantom{\large $\frac{I^2}{I}$} &
$p = - \lambda, \ \ \rho = -p + e^{-\mu}$ \qquad \, \quad if \ \, $\Psi \neq 0$ \\
			C$_{11}$ & (\ref{eq-uchi}), \quad with $\ \Psi = \! \Psi_{11}$ \phantom{\large $\frac{I^2}{I}$}& $p \! = \! p(\lambda), \ \, \rho \! = \! -p(\lambda) \! - \! p'(\lambda) e^{-\mu}$\ \, if \ \, $\Psi = 0$ \\ 
			C$_{12}$ & (\ref{eq-uchi}), \quad with $\ \Psi = \! \Psi_{12}$ \phantom{\large $\frac{I^2}{I}$} & \\
			C$_{14}$ & (\ref{eq-uchi}), \quad with $\ \Psi = \! \Psi_{14}$ \phantom{\large $\frac{I}{I}$} & \\
			C$_{16}$ & (\ref{eq-uchi}), \quad with $\ \Psi = \! \Psi_{16}$ \phantom{\large $\frac{I^2}{I}$} & \\
			\hline
			C$_5$ & $(u, v) = \hat{\Omega} \! - \! \gamma z_w \! = \! h_2 \! = \! 0$ \phantom{\large $\frac{I^2}{I}$} & \\
			C$_8$ & $(u, v) = \hat{\Omega} \! - \! \gamma z_w \! = \! \phi_2 \! = \! q_2 = 0$  \phantom{\large$\frac{I^2}{I}$} & $\{\rho, p\}$ given in (\ref{ce-eq_wnot=0}$-$\ref{dpdot-dxi}) \\
			C$_{13}$ & $\hat{\Omega} = \tilde{\tilde{\Omega}} + \tilde{\Omega}\, c_b = n_2 = 0$  \phantom{\large $\frac{I^2}{I}$} & \\
			\hline
			C$_{10}$ & $(u, v) = \hat{\Omega} \! - \! \gamma z_w \! = \! \phi_2 \! = \! \kappa_2 \! = \!0$ \phantom{\large $\frac{I^2}{I}$} & $\{\rho, p\}$ given in (\ref{ce-eq_wnot=0}) and (\ref{dpdot-dxi-drho*}) \\
			C$_{15}$ & $\hat{\Omega} = \dot{\Omega} + 2 \Omega c_b = 0$ \phantom{\large $\frac{I^2}{I}$} & $\{\rho, p\}$ given in (\ref{rhode-p-C}) \\
			C$_{17}$ & $(u, v) = \textrm{d} \theta \wedge u = 0$ \phantom{\large $\frac{I^2}{I}$} & $\{\rho, p\}$ given in (\ref{rhop-thetanot=0}$-$\ref{rhop-theta=0}) \\
			C$_{18}$ & $\textrm{d} \theta \wedge u \wedge a = 0$ \phantom{\large $\frac{I^2}{I}$} & $\{\rho, p\}$ given in (\ref{rhode-p}) \\
			\noalign{\hrule height 1.05pt}
		\end{tabular}
}
		\caption{The differential system S$_n$ in the second column gives the necessary and sufficient conditions for a unit vector of class C$_n$ to belong to {\bf U}$_c$. The third column gives the expression H$_n$ of the pairs $\{\rho,p\}$ associated with a velocity in the set C$_n \cap {\bf U}_c$.} \label{table-8}
	\end{table}
	\begin{itemize}
		\item C$_5$, C$_{8}$ and C$_{13}$: $\{\rho, p\}$ are determined by
			\begin{equation} \label{ce-eq_wnot=0}
				\textrm{d} p = - \dot{p} \, u - \xi \, a \, , \qquad \xi \equiv \rho + p \, , \qquad
			\end{equation}
		where $\{\dot{p}, \xi\}$ is the general solution of the exterior system
			\begin{equation} \label{dpdot-dxi}
				\textrm{d} \dot{p} = \dot{p} \, g_1 + \xi \, g_2 \, , \qquad \textrm{d} \xi = \dot{p} \, g_3 + \xi \, g_4 \, , \qquad
			\end{equation}
with the expressions of $g_i$ depending on the class.
					\begin{subequations} \label{g_i-C5}
						\begin{equation}
							\hspace{-30mm} \textrm{- C}_5 \! : \hspace{20mm} g_1 \equiv \ell_1 - \gamma \bar{\phi}_1 u \, , \quad \ g_2 \equiv \ell_2 - \gamma \bar{\phi}_2 u \, , \\[-3mm]
						\end{equation}
						\begin{equation}
							\ g_3 \equiv \ell_3 - \bar{\phi}_1 a \, , \quad \ g_4 \equiv \ell_4 -  \bar{\phi}_2 a \, . \\[-3mm]
						\end{equation}
					\end{subequations}
					\begin{subequations} \label{g_i-C8}
					\begin{equation}
							\hspace{-30mm} \textrm{- C}_8 \! : \hspace{20mm} g_1 \equiv \ell_1 - \gamma \bar{\kappa}_1 u \, , \quad \ g_2 \equiv \ell_2 - \gamma \bar{\kappa}_2 u \, , \\[-3mm]
						\end{equation}
						\begin{equation}
							\ g_3 \equiv \ell_3 - \bar{\kappa}_1 a \, , \quad \ g_4 \equiv \ell_4 - \bar{\kappa}_2 a \, . \\[-3mm]
						\end{equation}
					\end{subequations}
					\begin{subequations} \label{g_i-C13}
					\begin{equation}
							\hspace{-41mm} \textrm{- C}_{13} \! : \hspace{31mm} g_1 \equiv \ell_1 \, , \quad \ g_2 \equiv \ell_2 \, , \\[-3mm]
						\end{equation}
						\begin{equation}
							g_3 \equiv \ell_3 - \bar{\varphi}_1 a \, , \quad g_4 \equiv \ell_4 - \bar{\varphi}_2 a \, .
						\end{equation}
					\end{subequations}
		\item C$_{10}$: $\{\rho, p\}$ are determined by (\ref{ce-eq_wnot=0}) where $\{\dot{p}, \xi, \rho^*\}$ is the general solution of the exterior system
					\begin{subequations} \label{dpdot-dxi-drho*}
						\begin{eqnarray}
							\textrm{d} \dot{p} = \dot{p} \, \ell_1 + \xi \ell_2 + \rho^* \gamma u \, , \\[-1mm]
							\textrm{d} \xi = \dot{p} \, \ell_3 + \xi \ell_4 + \rho^* a \, , \\[-1mm]
							\textrm{d} \rho^* = \dot{p} \, f_1 + \xi \, f_2 +  \rho^* f_3 \, .
						\end{eqnarray}
					\end{subequations}
		\item C$_{15}$: $\{\rho, p\}$ are given by
			\begin{equation} \label{rhode-p-C}
				\rho = - p - \beta^{-1}[\mu \, p_{,\tau} + p_{,\alpha}] \, , \qquad p = p(\tau, \alpha) \, , 
			\end{equation}
where $\tau$, $\alpha$ and $\mu$ are three independent functions such that $\beta = \beta(\tau, \alpha, \mu)$, $u \! = \! -\textrm{d} \tau \! + \! \mu \, \textrm{d} \alpha$ and $a \! = \! \beta \, \textrm{d} \alpha$.
		\item C$_{17}$: $\{\rho, p\}$ are given by
			\begin{eqnarray}
				\rho = \rho(\tau) , \quad & p = p(\tau, \alpha) = - \rho(\tau) - \frac{\rho'(\tau)}{\theta(\tau)}e^{-\alpha} , \quad & {\rm if} \quad \theta \neq 0 \, , \quad \label{rhop-thetanot=0} \\
				\rho = \rho_0 , \qquad & p = p(\tau, \alpha) = - \rho_0 - \varphi(\tau)e^{-\alpha} , \quad \quad & {\rm if} \quad \theta = 0 \, , \quad \label{rhop-theta=0}
			\end{eqnarray}
where $\tau$ and $\alpha$ are two independent functions such that $u = - e^{\alpha} \textrm{d} \tau$, and $\rho(\tau)$ and $\varphi(\tau)$ are arbitrary real functions.
		\item C$_{18}$: $\{\rho, p\}$ are given by
			\begin{equation} \label{rhode-p}
				\rho = - p - \frac{p_{,\beta}}{\alpha_{,\beta}} \, , \qquad p = p(\tau, \beta) \, , 
			\end{equation}
where $\tau$ and $\beta$ are two independent functions such that $u = - e^{\alpha} \textrm{d} \tau$ and $a = \alpha_{,\beta} \, \textrm{d} \beta$, and $p(\tau, \beta)$ is any solution of
			\begin{equation} \label{p-taubeta}
				p_{,\tau \beta} = [(\ln \alpha_{,\beta})_{, \tau} - \theta\, e^{\alpha}]\, p_{,\beta} - \alpha_{,\beta} \, p_{,\tau}  .
			\end{equation}
	\end{itemize}
	It is worth remarking that these results offer an IDEAL characterisation of the velocities of a perfect energy tensor. This means that an algorithm can be built that enables us to distinguish every class C$_n$ and to test the labelling conditions S$_n$. We present this algorithm in Figure \ref{Fig-u}. The input data are $u$ itself and the differential $u$-concomitants ${\cal U}_k$ presented in (\ref{Cn-1st-order}$-$\ref{Cn-5th-order}). If a velocity belongs to class C$_n$, it must fulfil the necessary and sufficient conditions S$_n$ in order to be the velocity of a perfect energy tensor.

		\subsection{Some examples}
		Here, we want to highlight the interest of the theoretical results presented above by analysing some examples with our approach of both test solutions and self-gravitating systems. We only stress some first outcomes that we will study in more detail in further work.
		
			\subsubsection{Perfect fluids with a stationary flow}
			Let $u$ be a unit velocity such that $\xi = |\xi| u$ is a Killing vector. Then, $\theta = 0$, $\sigma = 0$, and $\textrm{d} a = 0$, where $\sigma$ is the shear. Moreover $a = \textrm{d} \alpha$, where $\alpha = \ln |\xi|$. Then, the study of the compatibility of these kinematic restrictions with our classification shows that only classes C$_1$, C$_2$, C$_{15}$ and C$_{18}$ are compatible. Considering now our above study of
$\hspace{-5mm}$these classes, we get that given a time-like Killing vector $\xi$, then $u = \xi/|\xi|$ is always the unit velocity of a perfect energy tensor.
			\begin{itemize}
				\item Case $a = 0$ [C$_1$]. (i) if $w \neq 0$, then the pressure is an arbitrary constant, $p = p_0$; (ii) if $w = 0$, then $u = - \textrm{d} \tau$ and $p = p(\tau)$. In both cases the energy density is an arbitrary $u$-invariant function, $\rho = \rho(\varphi_i)$, $\dot{\varphi_i} = 0$, $i = 1, 2, 3$. 
				\item Case $(w, a) \neq 0$ [C$_2$]. Then, the pressure is an arbitrary function of the norm of $\xi$, $p = p(\alpha)$, $\alpha = \ln |\xi|$, and the energy density is given by $\rho = \rho(\alpha) \equiv - p(\alpha) - p'(\alpha)$.
				\item Case $a \neq 0, \, w \neq 0, \, (w, a) =0 $ [C$_{15}$]. Then, $u = - \textrm{d} \tau + \mu \textrm{d} \alpha$, and the pressure and the energy density are given, respectively, by $p = c_1 \tau e^{-2\alpha} + p_1(\alpha)$ and $\rho = c_1 (\tau + \mu) e^{-2\alpha} - p_1(\alpha) - p_1'(\alpha)$.
				\item Case $a \neq 0, \, w = 0$ [C$_{18}$]. Then, $u = - e^{-\alpha} \textrm{d} \tau$, and the pressure and the energy density are given, respectively, by $p = p_0(\tau) e^{-\alpha} + p_1(\alpha)$ and $\rho = -p_1(\alpha) - p_1'(\alpha)$.
			\end{itemize}
			If we have a stationary flow, then $u$ fulfils equations (\ref{eq-chi=0}) and the conservation equations admit isobaric ($\dot{p} = 0$) solutions. This agrees with the above statement. Moreover, in this case we have a barotropic evolution, $\textrm{d} \rho \wedge \textrm{d} p = 0$. However, in cases C$_{1}$, C$_{15}$ and C$_{18}$ there are also non-isobaric solutions. \\ \\
			Note that the non-geodesic observer at rest with respect to a static gravitational field defines a flow of type C$_{18}$, and the associated energy tensor is stationary when $p_0(\tau) = constant$. In this case, each choice of the function $p_1(\alpha)$ defines a specific barotropic relation $p = p(\rho)$. A test fluid at rest in the Schwarzschild spacetime or a self gravitating sphere in equilibrium belongs to this class. \\ \\
			However, the perfect fluids at rest in the Kerr solution or in any stationary axisymmetric gravitational field are, generically, of class C$_2$. The determination of the subset of stationary axisymmetric solutions with a flow of class C$_{15}$ is an open problem that will be considered in the future.
			
			\subsubsection{Radial flow in a spherically symmetric spacetime}
			Test fluids with a radial flow can model different astrophysical and cosmological scenarios. For example, the radial accretion of a fluid onto a spherically symmetric central object has been used as a simple but non-trivial accretion model to study the effects of back-reaction (see \cite{schn-baug-shap, mach-malec} and references therein). \\ \\
			A spherically symmetric metric can be written in the general form $\textrm{d} s^2 = -e^{2\nu (t, r)} \textrm{d} t^2 + e^{2\lambda(t,r)} \textrm{d} r^2 + R^2(t,r) (\textrm{d} \vartheta^2 + \sin^2 \vartheta \textrm{d} \varphi^2)$, while a fluid radially moving has a unit velocity of the form:
			\begin{equation}
				u = - e^{\nu} \cosh \psi \, \textrm{d} t + e^{\lambda} \sinh \psi \, \textrm{d} r \, , \qquad \psi = \psi(t,r) \, .
			\end{equation}
Then, the fluid acceleration takes the expression: 
			\begin{equation}
				a = A(t,r)(-e^{-\lambda} \sinh \psi \, \textrm{d} t + e^{-\nu} \cosh \psi \, \textrm{d} r) \, ,
			\end{equation}
with $A(t,r) \equiv (e^{\lambda} \sinh \psi)_{,t} + (e^{\nu} \cosh \psi)_{,r}$. \\ \\[-1mm]
			If the flow is geodesic (class C$_1$), the function $\psi(t,r)$ fulfils the first-order \mbox{partial} differential equation $A(t,r) = 0$, which always admits solution. Otherwise, if $a \neq 0$, we have $a \wedge \textrm{d} a = u \wedge \textrm{d} u = 0$, and it necessarily belongs to class C$_{18}$. In both cases, such a $u$ defines the flow of a solution to the conservation equation (\ref{conservation-1}). \\ \\[-1mm]
			The solution $\{\rho, p\}$ to the inverse problem depends on the specific metric functions $\nu(t,r)$ and $\lambda(t,r)$. For example, in the Minkowski spacetime, $\nu = \lambda = 0$, and for a geodesic flow, equation $A(t,r) = 0$ has the general solution $t \! - \! r \coth \psi \! = \! f(\psi)$, where $f(\psi)$ is an arbitrary real function. The particular solution $f(\psi) = t_0$ leads to the Milne's flow, and then $u = - \textrm{d} \tau$, $\tau = [1 - z^2]^{-1/2}$, $z \equiv r / (t - t_0)$. In this case, $\textrm{d} \theta \wedge u = 0$, and according to a result in \cite{CFS-CIG}, $u$ is the flow of a classical ideal gas. It is also the flow of the perfect energy tensors defined by the solution (\ref{rho-a=0-1}) of the inverse problem. \\[-5mm]
			
			\subsubsection{On the Von Zeipel theorem and its extensions}
			Let us consider a stationary axisymmetric spacetime with a metric line element of the form $\textrm{d} s^2 = g_{tt} \textrm{d} t^2 + 2 g_{t \varphi} \textrm{d} t \textrm{d} \varphi + g_{rr} \textrm{d} r^2 + g_{\vartheta \vartheta} \textrm{d} \vartheta^2 + g_{\varphi \varphi} \textrm{d} \varphi^2$, $g_{\alpha \beta} = g_{\alpha \beta}(r, \vartheta)$, where $(t, r, \vartheta, \varphi)$ are spherical-like coordinates and $X = \partial_t$ and $Y = \partial_{\varphi}$ are Killing vectors. A stationary axisymmetric perfect fluid with purely circular flow has a unit velocity of the form \cite{Rezzolla}:
			\begin{equation} \label{von-zeipel-a}
				u = (u^{\alpha}) = u^t (\partial_t + \Omega \partial_{\varphi}) \, , \quad (u_{\alpha}) = u_t (\textrm{d} t - \ell \textrm{d} \varphi) \, , \quad u^t u_t = [\Omega \ell -1]^{-1} ,
			\end{equation}
where $\Omega$ is the {\em coordinate angular velocity} and $\ell$ the {\em specific angular momentum}. \\ \\[-1mm]
			The so-called relativistic Von Zeipel theorem \cite{Rezzolla, Abramowicz} states that the perfect fluid is barotropic if, and only if, the hypersurfaces of constant $\Omega$ coincide with the hypersurfaces of constant $\ell$. Note that the first is a hydrodynamic property, $\textrm{d} \rho \wedge \textrm{d} p = 0$, while the second one is a purely kinematic condition, $\textrm{d} \Omega \wedge \textrm{d} \ell = 0$. \\ \\
			This fact suggests that our kinematic characterisation could bring new insights into the Von Zeipel theorem. A straightforward calculation leads to:
			\begin{equation} \label{von-zeipel-b}
				u \wedge \textrm{d} u = u_t^2 \, \textrm{d} t \wedge \textrm{d} \varphi \wedge \textrm{d} \ell \, , \qquad \textrm{d} a = \ell [\Omega \ell -1]^{-2} \textrm{d} \ell \wedge \textrm{d} \Omega \, .
			\end{equation}
Thus, the kinematic constraint $\textrm{d} \Omega \wedge \textrm{d} \ell = 0$ is equivalent to the intrinsic kinematic condition $\textrm{d} a = 0$. \\ \\ 
			Note that a stationary fluid implies $\dot{p} = 0$. Then, the flow is geodesic (class C$_1$) if, and only if, $p = constant$. Moreover, the first equation in (\ref{von-zeipel-b}) implies that the flow is irrotational ($w = 0$) if, and only if, the specific angular momentum is constant ($\textrm{d} \ell = 0$), and then $u$ belongs to class C$_{18}$. Otherwise, when $w \neq 0$, the flow belongs to class C$_{15}$. In these two cases, the inverse problem (see Table \ref{table-8}) leads to a pressure and an energy density given by:
			\begin{equation} \label{von-zeipel-c}
				p = p(\alpha) \, , \qquad \rho (\alpha) = - p(\alpha) - p'(\alpha) \, ,
			\end{equation}
where $p(\alpha)$ is an arbitrary real function of the acceleration potential $\alpha$, $\textrm{d} \alpha = a$. Then, we have $\textrm{d} \rho \wedge \textrm{d} p = 0$, and we recover the Von Zeipel theorem. \\ \\ 
			The comprehensive study of the unit velocities of the form (\ref{von-zeipel-a}) taking into account the results of \cite{SFM-velocitats} summarised in this section shows that all these vector fields define the flow of a conservative perfect fluid, and that, in addition to the classes quoted above, only classes C$_2$ and C$_3$ are possible. The analysis of the inverse problem for these two classes and the study of the compatible non-barotropic equations of state could lead to potential extensions of the Von Zeipel theorem.
			
			\subsubsection{The flow of the Stephani universes}			
			 In the three previous examples, we have considered test fluids in a given gravitational field. The analysis of the kinematics of a self-gravitating perfect fluid taking into account our approach can also be of interest. The divergence-free condition holds as a consequence of the field equations but, depending on the class C$_n$, the inverse problem can provide test perfect fluids that are comoving with the self-gravitating system. As an example, we analyse the Stephani universes. \\
			 Recall that the Stephani universes are the conformally flat perfect fluid solutions to Einstein equations with non-zero expansion, and they are irrotational and shear-free. The metric line element takes the expression $\textrm{d} s^2 = -\alpha^2 \textrm{d}t^2 + \Omega^2 \delta_{ij} \textrm{d}x^i \textrm{d}x^j$, where $\alpha \equiv R \, \partial_R \ln \Omega$ and $\Omega \equiv R(t)[1 + 2 \vec{b}(t) \cdot \vec{{\rm r}} + \frac{1}{4} K(t) r^2]^{-1}$, and the unit velocity of the fluid is $u = \alpha^{-1} \partial_t$. Moreover, the energy density, pressure and expansion are:
			\begin{equation}
				\rho = \frac{3}{R^2}(\dot{R}^2 + K - 4b^2), \qquad p = - \rho - {R \over 3} {\partial_R \rho \over \alpha}, \qquad \theta(t) = {3 \dot{R} \over R} \neq 0. \label{eq:dp}
			\end{equation}
The geodesic case (class C$_1$) leads to the LFRW limit. The strict Stephani universes ($a \neq 0$) have an irrotational flow ($w = 0$) with homogeneous expansion ($\textrm{d} \theta \wedge u = 0$). Consequently, only classes C$_{17}$ and C$_{18}$ are compatible. \\ \\
			In class C$_{17}$ ($v \neq 0$), the inverse problem (see Table \ref{table-8}) leads to an arbitrary homogeneous energy density, $\bar{\rho} = \bar{\rho}(t)$ (which is not given by the first expression in (\ref{eq:dp})), and a pressure $\bar{p}$ given by the second expression in (\ref{eq:dp}) replacing $\rho$ by $\bar{\rho}$. Besides, it can be shown that condition $v = 0$ (class C$_{18}$) leads to the Stephani universes admitting a G$_3$, that is, the thermodynamic Stephani universes studied in Chapter \ref{chap-Stephani}.
	
\chapter{Geometric properties} \label{chap-geometric-xIdeal}
In this chapter, we present the remaining \textit{xIdeal} functions implemented up to the date of this thesis, which are not necessarily related to perfect fluids. We also describe some IDEAL determinations that, although not yet implemented, have been successfully tested using \textit{xIdeal} and are planned for future development.

	\section[Petrov-Bel type and multiple Debever null directions]{Petrov-Bel type and multiple Debever \\ null directions}
	We will consider ${\cal W} = \frac{1}{2} (W - {\rm i} *W)$ the self-dual Weyl tensor and ${\cal G} = \frac{1}{2} (G - {\rm i} \eta)$ the canonical metric on the space of self-dual 2-forms (bivectors), where $G = \frac{1}{2} \, g \wedge g$, and $\eta$ is the metric volume element. The algebraic classification of the Weyl tensor $W$ can be obtained \cite{Petrov, bel-3} by studying the traceless linear map defined by the self-dual Weyl tensor on the space of bivectors. An alternative approach consists of studying the relative positions between the `null cones' determined by the canonical metric and the Weyl tensor as a quadratic form on the bivectors \cite{Debever-56}. These `null cones' cut, generically, on four null bivectors ${\cal H}$ that are the solutions of the equations ${\cal G}({\cal H}, {\cal H}) = 0$, ${\cal W}({\cal H}, {\cal H}) = 0$. We call them {\em Debever null bivectors} \cite{FMS-Weyl}. The fundamental vector $\ell$ of each one defines a null direction on the spacetime, which is usually called {\em Debever null direction} \cite{Debever}. \\ \\
	Petrov-Bel types N, III and II admit a single multiple Debever null direction which is quadruple, triple and double, respectively. These three types correspond to the Bel radiative gravitational fields, and this multiple Debever direction is called the {\em fundamental direction} of the gravitational field \cite{bel-3}. Petrov-Bel type D admits two double Debever directions, and in type I there is no multiple Debever direction. \\ \\ \\
	The underlying geometry associated with each Petrov-Bel type was thoroughly analysed in \cite{FMS-Weyl} by Ferrando \textit{et al.}, where they also provided an algorithm to determine all these geometric elements defined by the Weyl tensor. In particular, they presented IDEAL algorithms for identifying the Petrov type of a given metric and for determining the corresponding Debever null directions. In \cite{SMF-Debever}, we offer an alternative approach based on the Bel-Robinson tensor \cite{bel-3} to obtain the same elements. \\ \\
	In this section, we present the four algorithms derived from both approaches and describe their implementation within the corresponding \textit{xIdeal} functions. We also provide examples where they have been used to reproduce established results from the literature. \\ \\
	\textbf{Petrov-Bel type with self-dual Weyl tensor.} \textit{Consider the self-dual Weyl tensor} ${\cal W} \neq 0$ (${\cal W} = 0$ \textit{leads to type O})\textit{, the canonical metric on the space of bivectors} ${\cal G}$ \textit{and the scalars} $a \equiv {\rm Tr} {\cal W}^2$ \textit{and} $b \equiv {\rm Tr} {\cal W}^3$\textit{. The Petrov-Bel type of the Weyl Tensor is given by the following algorithm:}
	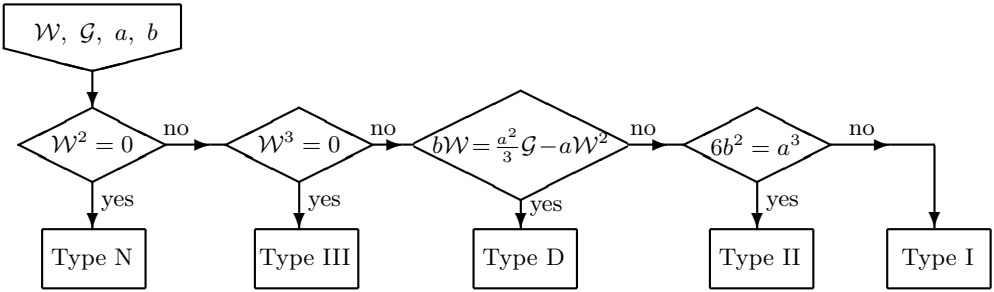
\begin{figure}[H]
	\hspace*{6mm} \setlength{\unitlength}{0.78cm} {\small \noindent
	\begin{picture}(16,6)
		\thicklines 
		\put(2,5){\line(-4,-1){1.5}}
		\put(-1,5){\line(4,-1){1.5}}
		\put(-1,5.75){\line(0,-1){0.75}} \put(2,5.75){\line(-1,0){3}}
		\put(2,5.75){\line(0,-1){0.75}} \put(-0.9,5.15){$\quad {\cal W} , \ {\cal G}, \ a , \ b$}
		\put(0.5,4.65){\vector(0,-1){0.65}}
		\put(0.5,4){\line(-2,-1){1.25}} \put(0.5,4){\line(2,-1){1.25}}
		\put(0.5,2.75){\line(2,1){1.25}} \put(0.5 ,2.75){\line(-2,1){1.25}}
		\put(-0.20,3.25){${\cal W}^2 = 0$}
		\put(4,4){\line(-2,-1){1.25}} \put(4,4){\line(2,-1){1.25}}
		\put(4,2.75){\line(2,1){1.25}} \put(4,2.75){\line(-2,1){1.25}}
		\put(3.3,3.25){${\cal W}^3 = 0$}
		\put(7.75,4.3){\line(-2,-1){1.8}}
		\put(7.75,4.3){\line(2,-1){1.8}} \put(7.75,2.45){\line(2,1){1.8}}
		\put(7.75,2.45){\line(-2,1){1.8}} \put(6.25,3.23){$b {\cal W} \! = \!\frac{a^2}{3} {\cal G} \! - \! a {\cal W}^2$}
		\put(11.75,4){\line(-2,-1){1.25}} \put(11.75,4){\line(2,-1){1.25}}
		\put(11.75,2.75){\line(2,1){1.25}}
		\put(11.75,2.75){\line(-2,1){1.25}} \put(10.96,3.2){$6 b^2 = a^3 $}
		\put(4,2.75){\vector(0,-1){0.8}} \put(0.5,2.75){\vector(0,-1){0.8}}
		\put(11.75,2.75){\vector(0,-1){0.8}}
		\put(7.75,2.45){\vector(0,-1){0.50}}
		\put(1.75,3.38){\vector(1,0){0.75}}\put(2.5,3.38){\line(1,0){0.25}}
		\put(5.25,3.38){\vector(1,0){0.5}}\put(5.75,3.38){\line(1,0){0.20}}
		\put(9.55,3.38){\vector(1,0){0.70}}
		\put(10.25,3.38){\line(1,0){0.25}}
		\put(13 ,3.38){\vector(1,0){1}}\put(14 ,3.38){\line(1,0){0.76}}
		\put(14.75,3.37){\vector(0,-1){1.4}}
		\put(3.25,0.95){\line(1,0){1.75}} \put(3.25,0.95 ){\line(0,1){1}}
		\put(5,1.95 ){\line(-1,0){1.75}} \put(5,1.95 ){\line(0,-1){1}}
		\put(3.32,1.35){Type III}
		\put(-0.35,0.95){\line(1,0){1.75}} \put(-0.35,0.95){\line(0,1){1}}
		\put(1.4,1.95 ){\line(-1,0){1.75}} \put(1.4,1.95 ){\line(0,-1){1}}
		\put(-0.2,1.35){Type N}
		\put(6.95,0.95){\line(1,0){1.75}} \put(6.95,0.95 ){\line(0,1){1}}
		\put(8.7,1.95 ){\line(-1,0){1.75}} \put(8.7,1.95 ){\line(0,-1){1}}
		\put(7.1,1.35){Type D}
		\put(10.95,0.95){\line(1,0){1.75}} \put(10.95,0.95){\line(0,1){1}}
		\put(12.7,1.95 ){\line(-1,0){1.75}} \put(12.7,1.95){\line(0,-1){1}}
		\put(11.1,1.35){Type II}
		\put(13.95,0.95){\line(1,0){1.75}} \put(13.95,0.95){\line(0,1){1}}
		\put(15.7,1.95 ){\line(-1,0){1.75}} \put(15.7,1.95){\line(0,-1){1}}
		\put(14.19,1.35){Type I}
		\put(11.9,2.35){yes} 
		\put(7.9,2.25){yes}
		\put(4.15,2.35){yes} 
		\put(0.65,2.35){yes}
		\put(13.3,3.55){no} 
		\put(9.6,3.55){no} 
		\put(5.2,3.55){no}
		\put(1.7,3.55){no}
	\end{picture} }
	\vspace{-8mm}
	\caption{Algorithm to determine the Petrov-Bel type with the self-dual Weyl tensor. $\qquad$}
	\label{Fig-16}
	\end{figure}
	$\hspace{-5mm}$If $E$ and $H$ are the electric and magnetic parts of the Weyl tensor with respect to an observer $u$, the {\em Petrov matrix} relative to $u$ is ${\cal Q} = E - \textrm{i} H$, and it can be obtained as:
\begin{equation}\label{Q-W}
{\cal Q}_{\beta \nu} = 2 u^{\alpha} u^{\mu} \, {\cal W}_{\alpha \beta \mu \nu} \, .
\end{equation}
In \cite{SMF-Debever}, we offer an alternative version of the algorithm in Figure \ref{Fig-16} to determine the Petrov-Bel type of a given metric in terms of the Petrov matrix: \\ \\
	\textbf{Petrov-Bel type with Petrov matrix.} \textit{Consider the Petrov matrix} ${\cal Q} \neq 0$ (${\cal Q} = 0$ \textit{leads to type O})\textit{, the spatial metric} $\gamma = g + u \otimes u$ \textit{and the scalars} $\hat{a} \equiv {\rm tr} {\cal Q}^2$ \textit{and} $\hat{b} \equiv - {\rm tr} {\cal Q}^3$\textit{. The Petrov-Bel type of the Weyl Tensor is given by the following algorithm:}
	\begin{figure}[H]
	\hspace*{6mm} \setlength{\unitlength}{0.78cm} {\small \noindent
	\begin{picture}(16,6)
		\thicklines 
		\put(2,5){\line(-4,-1){1.5}}
		\put(-1,5){\line(4,-1){1.5}}
		\put(-1,5.75){\line(0,-1){0.75}} \put(2,5.75){\line(-1,0){3}}
		\put(2,5.75){\line(0,-1){0.75}} \put(-0.9,5.15){$\quad {\cal Q} , \ \gamma, \ \hat{a} , \ \hat{b}$}
		\put(0.5,4.65){\vector(0,-1){0.65}}
		\put(0.5,4){\line(-2,-1){1.25}} \put(0.5,4){\line(2,-1){1.25}}
		\put(0.5,2.75){\line(2,1){1.25}} \put(0.5,2.75){\line(-2,1){1.25}}
		\put(-0.20,3.25){${\cal Q}^2 = 0$}
		\put(4,4){\line(-2,-1){1.25}} \put(4,4){\line(2,-1){1.25}}
		\put(4,2.75){\line(2,1){1.25}} \put(4,2.75){\line(-2,1){1.25}}
		\put(3.3,3.25){${\cal Q}^3 = 0$}
		\put(7.75,4.25){\line(-2,-1){1.75}}
		\put(7.75,4.25){\line(2,-1){1.75}} \put(7.75,2.5){\line(2,1){1.75}}
		\put(7.75,2.5){\line(-2,1){1.75}} \put(6.37,3.2){$\hat{b} {\cal Q}\! = \!\frac{\hat{a}^2}{3} \! \gamma \! - \! \hat{a} {\cal Q}^2$}
		\put(11.75,4){\line(-2,-1){1.25}} \put(11.75,4){\line(2,-1){1.25}}
		\put(11.75,2.75){\line(2,1){1.25}}
		\put(11.75,2.75){\line(-2,1){1.25}} \put(10.96,3.2){$6 \hat{b}^2 = \hat{a}^3 $}
		\put(4,2.75){\vector(0,-1){0.8}} \put(0.5,2.75){\vector(0,-1){0.8}}
		\put(11.75,2.75){\vector(0,-1){0.8}}
		\put(7.75,2.5){\vector(0,-1){0.55}}
		\put(1.75,3.38){\vector(1,0){0.75}}\put(2.5,3.38){\line(1,0){0.25}}
		\put(5.25,3.38){\vector(1,0){0.5}}\put(5.75,3.38){\line(1,0){0.25}}
		\put(9.5,3.38){\vector(1,0){0.75}}
		\put(10.25,3.38){\line(1,0){0.25}}
		\put(13 ,3.38){\vector(1,0){1}}\put(14 ,3.38){\line(1,0){0.76}}
		\put(14.75,3.37){\vector(0,-1){1.4}}
		\put(3.25,0.95){\line(1,0){1.75}} \put(3.25,0.95 ){\line(0,1){1}}
		\put(5,1.95 ){\line(-1,0){1.75}} \put(5,1.95 ){\line(0,-1){1}}
		\put(3.32,1.35){Type III}
		\put(-0.35,0.95){\line(1,0){1.75}} \put(-0.35,0.95){\line(0,1){1}}
		\put(1.4,1.95 ){\line(-1,0){1.75}} \put(1.4,1.95 ){\line(0,-1){1}}
		\put(-0.2,1.35){Type N}
		\put(6.95,0.95){\line(1,0){1.75}} \put(6.95,0.95 ){\line(0,1){1}}
		\put(8.7,1.95 ){\line(-1,0){1.75}} \put(8.7,1.95 ){\line(0,-1){1}}
		\put(7.1,1.35){Type D}
		\put(10.95,0.95){\line(1,0){1.75}} \put(10.95,0.95){\line(0,1){1}}
		\put(12.7,1.95 ){\line(-1,0){1.75}} \put(12.7,1.95){\line(0,-1){1}}
		\put(11.1,1.35){Type II}
		\put(13.95,0.95){\line(1,0){1.75}} \put(13.95,0.95){\line(0,1){1}}
		\put(15.7,1.95 ){\line(-1,0){1.75}} \put(15.7,1.95){\line(0,-1){1}}
		\put(14.19,1.35){Type I}
		\put(11.9,2.35){yes} 
		\put(7.9,2.25){yes}
		\put(4.15,2.35){yes} 
		\put(0.65,2.35){yes}
		\put(13.3,3.55){no} 
		\put(9.6,3.55){no} 
		\put(5.2,3.55){no}
		\put(1.7,3.55){no}
	\end{picture} }
	\vspace{-8mm}
	\caption{Algorithm to determine the Petrov-Bel type with the Petrov matrix.$\qquad \qquad \qquad$}
	\label{Fig-17}
	\end{figure}
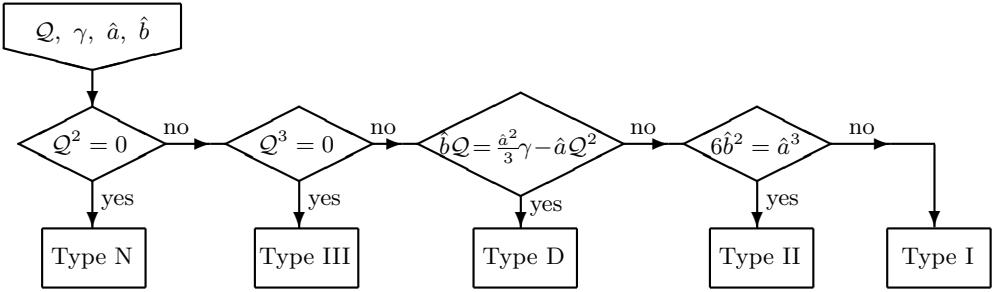
	$\hspace{-5mm}$\textbf{Multiple Debever null directions with self-dual Weyl tensor.} \textit{In terms of the self-dual Weyl tensor} ${\cal W}$\textit{, the fundamental direction} $\ell$ \textit{of a Bel radiative gravitational field can be obtained as}
		\begin{equation} \label{ell-a}
			\ell = \frac{H^2(v)}{\sqrt{-H^2(v, v)}} \, , \qquad H = \frac{1}{\sqrt{2}}({\cal H} + \bar{{\cal H}}) \, , \qquad {\cal H} = \frac{{\cal P}({\cal X})}{\sqrt{{\cal P}({\cal X}, {\cal X})}}
		\end{equation}
\textit{where a bar denotes complex conjugation and}
		\begin{itemize}
\item
${\cal P} \equiv {\cal W}$ \textit{if the Weyl tensor is of type N.}
\item
${\cal P} \equiv - {\cal W}^2$ \textit{if the Weyl tensor is of type III.}
\item
${\cal P} \equiv \frac{1}{3\psi}({\cal W} - \psi {\cal G})({\cal W} + 2 \psi {\cal G})$\textit{, with} $\psi = - \frac{\textrm{Tr} {\cal W}^3}{\textrm{Tr} {\cal W}^2}$\textit{, if the Weyl tensor is of type II.}
		\end{itemize}
\textit{For a type D Weyl tensor, the double Debever null directions} $\ell_\pm$ \textit{can be obtained as}
		\begin{equation}
			\ell_\pm \propto \left[ U^2 \pm U \right] (v) \, ,
		\end{equation}
\textit{where} $v$ \textit{is an arbitrary timelike future-pointing vector and} $U$ \textit{can be obtained as}
		\begin{equation} \label{TypeD-directions}
			U = \frac{1}{\sqrt{2}} ({\cal U} + \bar{\cal U}) \, , \qquad {\cal U} \equiv \frac{{\cal A} ({\cal X})}{\sqrt{- {\cal A} ^2 ({\cal X} , {\cal X})}} \, , \qquad {\cal A} \equiv {\cal W} - \psi \, {\cal G} \, ,
		\end{equation}
\textit{with} ${\cal X}$ \textit{an arbitrary bivector such that} ${\cal A} ({\cal X}) \neq 0$. \\ \\
	\textbf{Multiple Debever null directions with Petrov matrix.} \textit{In terms of the observer} $u$ \textit{and the Petrov matrix} ${\cal Q} = E - \textrm{i} H$\textit{, the fundamental direction} $\ell$ \textit{of a Bel radiative gravitational field can be obtained as}
		\begin{equation} \label{ell-b}
			\ell = \textrm{tr}({\bar{\cal P}} \cdot {\cal P}) u + 2 \, \textrm{i} \, *\!({\bar{\cal P}} \cdot {\cal P})(u) ,
		\end{equation}
\textit{where}
		\begin{itemize}
\item
${\cal P} \equiv {\cal Q}$ \textit{if the Weyl tensor is of type N,}
\item
${\cal P} \equiv - {\cal Q}^2$ \textit{if the Weyl tensor is of type III,}
\item
${\cal P} \equiv \hat{\psi} {\cal Q} + 2 \hat{\psi}^2 \gamma - {\cal Q}^2$\textit{ if the Weyl tensor is of type II,}
		\end{itemize}
\textit{with} $\gamma = g + u \otimes u$\textit{,} $\hat{\psi} = \frac{\textrm{tr} {\cal Q}^3}{\textrm{tr} {\cal Q}^2}$\textit{. For a type D Weyl tensor, the double Debever null directions} $\ell_\pm$ \textit{can be obtained as}
		\begin{equation} \label{ell+-}
			\ell_{\pm} = v_0 \pm v_1 \, ,
		\end{equation}
\textit{where} 
		\begin{equation} \label{e0e1}
			v_0 = \cosh^2 \phi\, u - \frac{\textrm{i}}{\sqrt{\zeta}} \, *({\bar{\cal R}} \cdot {\cal R})(u) \, , \qquad v_1 = \frac{S(w)}{\sqrt{S(w,w)}} \, ,
		\end{equation}
$w$ \textit{being an arbitrary vector such that} $S(w) \neq 0$\textit{, and with}
		\begin{eqnarray} \label{P-S}
			{\cal R} \equiv \frac{1}{3 \hat{\psi}}{\cal Q} \, , \qquad \cosh^2 \phi \equiv \frac12 (1 \! + \! \sqrt{\zeta}) \, , \qquad \zeta \equiv \textrm{tr}[{\cal R} \cdot \bar{\cal R}] + \frac13 \, , \quad \\[2mm] 
			S \equiv \frac{1}{4}(1 \! + \! \frac{2}{3\sqrt{\zeta}})({\cal R} + \bar{\cal R}) + \frac{1}{4 \sqrt{\zeta}} ({\cal R} \cdot \bar{\cal R} + \bar{\cal R} \cdot {\cal R}) + \frac16 (1 \! + \! \frac{1}{3\sqrt{\zeta}}) \gamma \, . \label{alpha-z-a}
		\end{eqnarray}
These four algorithms are implemented in \textit{xIdeal} by means of two functions: \texttt{PetrovType} and \texttt{DebeverNullDirections}. In the following subsections, we show two examples in which they have been tested and some considerations for their use.

		\subsection{An example of a radiative gravitational field}
		The following metric is a particular case of a type II perfect fluid solution with a geodesic, shearfree, non-expanding multiple Debever null direction \cite{Kramer}:
		\begin{equation}
			\hspace{-1cm} ds^2 = 2 F^{-2} (dx^2 \! + \! dy^2) - 2(dv \! + \! 2L \, dy)[dr \! + \! 2B \, dy + A(dv \! + \! 2L \, dy)] \, ,
		\end{equation}
with $F^2 = \frac43 \kappa_0 x^3$, $A = -\kappa_0 x/2$, $B = \frac34 \sqrt{2}/x$ and $L = \frac38 \sqrt{2}/(\kappa_0 x^2)$. It is a good example to check our functions since its Petrov-Bel type and null direction are already known. \\ \\ 
		First, we check that, indeed, it is of Petrov-Bel type II with the $\texttt{PetrovType}$ function. This function can determine the Petrov-Bel type of a given metric following the two algorithms presented above. With both, we get that the considered metric is of type II. Then, we use the $\texttt{DebeverNullDirections}$ function to obtain its multiple Debever null directions. Again, this function can follow the algorithm based on the self-dual Weyl tensor (by default) or the one based on the Petrov matrix. Regardless of the method, we get that the multiple Debever null vector has the direction of $\partial_r$, as reported in the literature.
		
		\subsection{An example of type D} \label{subsec-TypeD-example}
		A similar approach can be followed with Kerr-NUT metric. Kerr-NUT metric generalises both Kerr and NUT vacuum solutions, and can be written as \cite{Kerr-Weir}
		\begin{equation}
			ds^2 = \frac{- \, \alpha^2}{x^2 \! + \! y^2}(y^2 dt \! + \! dz)^2 + \! \frac{x^2 \! + \! y^2}{\alpha^2} dx^2 \! + \frac{\beta^2}{x^2 \! + \! y^2}(-x^2 dt + dz)^2 \! + \frac{x^2 \! + \! y^2}{\beta^2}dy^2 ,
		\end{equation}
where
		\begin{equation} \label{alpha i beta}
			\hspace{-1cm} \alpha^2 = p x^2 + \frac{k(3 - k^2)}{(1 + k^2)^3} x + s \, , \qquad \quad \beta^2 = -p y^2 + \frac{3k^2 - 1}{(1 + k^2)^3}y + s \, .
		\end{equation}
By setting $3k^2 = 1$ we recover the Kerr metric. \\ \\
		Using the $\texttt{PetrovType}$ function, we get that the Kerr-NUT metric is of type D. It is worth remarking that, in our computations, we added the assumption that the coordinates satisfy $x^2 + y^2 > 0$. This helps the program to simplify some expressions and this, in turn, improves the computation speed. We explain this in more detail in the next subsection. For the same reason, we did not specify the expressions of the scalar functions $\alpha$ and $\beta$. Thus, the results we get here are valid for any $\alpha(x)$ and $\beta(y)$. However, only expressions (\ref{alpha i beta}) are solution of the vacuum Einstein equations. \\ \\
		To get the multiple Debever null directions of Kerr-NUT metric, we use the $\texttt{DebeverNullDirections}$ function with the following extra assumptions: $y^4 \alpha^2 + x^4 \beta^2 > 0$ and $y^4 \alpha^2 - x^4 \beta^2 > 0$. These make sure that the function computes the quantity $\zeta$ defined in (\ref{alpha-z-a}) with the proper sign taking into account that we are considering the region outside the horizons. With that, we get that the two multiple Debever null directions of a Kerr-NUT metric are
		\begin{equation}
			\ell_\pm^\mu = ( y^2 , \, \pm \, y^2 \, \alpha^2 , \, 0 \, , \, x^2 \, y^2 ) \, ,
		\end{equation}
with $\alpha$ defined in (\ref{alpha i beta}). 
		
		\subsection{Other algorithms and discussion} \label{subsec-other-algorithms}
		There exist other algorithms to obtain the geometric spacetime quantities discussed in this section besides the ones presented here (see \cite{MacCallum-2018} for an extensive review). One of the first such algorithms was the Cartan-Karlhede algorithm \cite{Karlhede-2006} outlined in Section \ref{sec-computer-algebra}. Recall that these algorithms are tetrad based and should be considered as complementary. \\ \\
		Traditionally, tetrad based algorithms were thought to be more efficient than classical coordinate methods. However, as shown in \cite{Pollney}, a good simplification strategy is crucial to minimise the computing time. In the previous subsection, for example, we did not specify the expressions of $\alpha (x)$ and $\beta (y)$ for the Kerr-NUT metric, leaving them as arbitrary functions of the coordinates. We also added the assumption that the coordinates satisfy $x^2 + y^2 > 0$. This significantly improved the time spent simplifying some expressions. There is no universal rule to know which or how many assumptions are needed, but in some cases they are necessary to even get a result. In the previous subsection again, if we do not add the assumptions that $y^4 \alpha^2 + x^4 \beta^2 > 0$ and $y^4 \alpha^2 - x^4 \beta^2 > 0$, the multiple Debever null directions are not properly obtained. \\ 
		It is also worth mentioning that our functions require arbitrary vectors to compute some projections. The only theoretical restrictions on these arbitrary vectors is the non-nullity of some contractions as explained above. However, vectors that take simple expressions in the frame considered are obviously preferable. \\ \\
		The algorithm based on the Petrov matrix may be interesting because it makes use of the electric and magnetic parts of the Weyl tensor relative to an observer, which in some contexts are the natural working variables. Regardless, this new algorithm can also be advantageous if it improves the computing time. Indeed, in the first example of the previous subsections, the Petrov matrix method requires less computing time than the algorithm based on the self-dual Weyl tensor, whereas in the second example, only the Petrov matrix method provides a result in a reasonable time.
	
	\section{Dimension of the isometry group} \label{sec-simetries}
	The Cartan-Karlhede algorithm can also be used to obtain the dimension of the orbits and the number of Killing fields admitted by a given spacetime \cite{Kramer, Karlhede-MacCallum, tomoda}. Although it is theoretically clearly established, the number of scalars that have to be computed is formidable (see \cite{tomoda} and references therein for a recent review on this subject). \\ \\ 
	In this section, we present a set of algorithms for determining the dimension of the isometry group by taking the successive derivatives of the so called {\it connection tensor} up to the fourth order in two situations: when an invariant frame can be determined \cite{SFM-simetries} (Section \ref{subsec-simetries-RFrame}) and for type N vacuum spacetimes \cite{SMF-DimensioTN} (Section \ref{subsec-dim-TipusN}). A similar approach has been performed in obtaining the dimension of the isometry group in three-dimensional Riemannian spaces \cite{FS-K3}, and in studying the homogeneous three-dimensional spaces, both Riemannian \cite{FS-G3} and Lorentzian \cite{FS-L3}. Essentially, these algorithms compute the Cartan invariants in the invariant frame without the explicit determination of the frame itself in some of the cases.
	
		\subsection{Spacetimes with an invariant frame} \label{subsec-simetries-RFrame}
\begin{figure}[H]
		\vspace{-1mm} \hspace*{6mm} \setlength{\unitlength}{0.78cm} {\small \noindent
		\begin{picture}(16,19)
		\thicklines
			\put(1.7,18){\line(-3,-1){1.2}}
			\put(-0.7,18){\line(3,-1){1.2}}
			\put(1.7,18){\line(0,1){0.6}} \put(-0.7,18.6){\line(1,0){2.4}}
			\put(-0.7,18.6){\line(0,-1){0.6}}
			\put(-0.1,18.05 ){$H, \; C^{[q]}$}
			\put(2.1,16.7){yes}
			\put(2.1,15){yes}
			\put(6.1,15){yes}
			\put(6.1,13.3){yes}
			\put(10.1,13.3){yes}
			\put(10.1,11.5){yes}
			\put(14.1,11.5){yes}
			\put(2.1,6.5){yes}
			\put(6.1,6.5){yes}
			\put(6.1,4.75){yes}
			\put(10.1,4.75){yes}
			\put(6.1,2.3){yes}
			\put(2.1,2.3){yes}
			\put(6.1,9){yes}
			\put(10.1,9){yes}
			\put(0.76,15.56){no}
			\put(0.76,13.86){no}
			\put(4.76,13.86){no}
			\put(4.76,12.1){no}
			\put(8.76,12.1){no}
			\put(8.76,10.45){no}
			\put(12.76,10.45){no}
			\put(8.76,7.9){no}
			\put(4.76,7.9){no}
			\put(4.76,5.4){no}
			\put(0.76,5.4){no}
			\put(4.76,3.7){no} 
			\put(8.76,3.7){no}
			\put(0.76,1.25){no}
			\put(4.76,1.25){no}
			\put(0.5,17.60){\vector(0,-1){0.6 }}
			\put(0.5,15.9){\vector(0,-1){0.6 }}
			\put(0.5,14.2){\vector(0,-1){7.35 }}
			\put(0.5,5.75){\vector(0,-1){3.15 }}
			\put(4.5,14.2){\vector(0,-1){0.6 }}
			\put(4.5,12.5){\vector(0,-1){3.15 }}
			\put(4.5,5.75){\vector(0,-1){0.65 }}
			\put(8.5,12.5){\vector(0,-1){0.65 }}
			\put(2.1,16.45){\vector(1,0){12.8}}
			\put(2.1,14.75){\vector(1,0){0.8}}
			\put(6.1,14.75){\vector(1,0){8.8}}
			\put(6.1,13.05){\vector(1,0){0.8}}
			\put(10.1,13.05){\vector(1,0){4.8}}
			\put(10.1,11.3){\vector(1,0){0.8}}
			\put(14.1,11.3){\vector(1,0){0.8}}
			\put(6.1,8.8){\vector(1,0){0.8}}
			\put(10.1,8.8){\vector(1,0){4.8}}
			\put(2.1,6.3){\vector(1,0){0.8}}
			\put(6.1,6.3){\vector(1,0){8.8}}
			\put(6.1,4.55){\vector(1,0){0.8}}
			\put(10.1,4.55){\vector(1,0){4.8}}
			\put(2.1,2.05){\vector(1,0){0.8}}
			\put(6.1,2.05){\vector(1,0){8.8}}
			\put(0.5,17){\line(-3,-1){1.62}}  \put(0.5,17){\line(3,-1){1.62}}
			\put(0.5,15.9){\line(3,1){1.62}} \put(0.5,15.9){\line(-3,1){1.62}}
			\put(-0.34,16.31){$C^{[1]} = 0$}
			\put(14.9,16.4){\framebox{\,G$_4$\,\hspace*{-0.4cm}$\phantom{{A^B_C}}\!\!$}}
			\put(0.5,15.3){\line(-3,-1){1.62}} \put(0.5,15.3){\line(3,-1){1.62}}
			\put(0.5,14.2){\line(3,1){1.62}} \put(0.5,14.2){\line(-3,1){1.62}}
			\put(-0.24,14.65){$[11] = 0$}
			\put(4.5,15.3){\line(-3,-1){1.62}} \put(4.5,15.3){\line(3,-1){1.62}}
			\put(4.5,14.2){\line(3,1){1.62}} \put(4.5,14.2){\line(-3,1){1.62}}
			\put(3.76,14.65){$[12] = 0$}
			\put(14.9,14.7){\framebox{\,G$_3$\,\hspace*{-0.4cm}$\phantom{{A^B_C}}\!\!$}}
			\put(4.5,13.6){\line(-3,-1){1.62}} \put(4.5,13.6){\line(3,-1){1.62}}
			\put(4.5,12.5){\line(3,1){1.62}} \put(4.5,12.5){\line(-3,1){1.62}}
			\put(3.7,12.95){$[122] \! = \! 0$}
			\put(8.5,13.6){\line(-3,-1){1.62}} \put(8.5,13.6){\line(3,-1){1.62}}
			\put(8.5,12.5){\line(3,1){1.62}} \put(8.5,12.5){\line(-3,1){1.62}}
			\put(7.7,12.95){$[123] \! = \! 0$}
			\put(14.9,13){\framebox{\,\hspace*{-0.06cm}G$_{2b}$\,\hspace*{-0.45cm}$\phantom{{A^B_C}}\!\!$}}
			\put(8.5,11.85){\line(-3,-1){1.62}}
			\put(8.5,11.85){\line(3,-1){1.62}} \put(8.5,10.75){\line(3,1){1.62}}
			\put(8.5,10.75){\line(-3,1){1.62}} \put(7.68,11.17){$[1233] \! = \! 0$}
			\put(8.4,9.75){{\large $\nexists$}}
			\put(8.48,10.75){\vector(0,-1){0.45}}
			\put(12.5,11.85){\line(-3,-1){1.62}}
			\put(12.5,11.85){\line(3,-1){1.62}}
			\put(12.5,10.75){\line(3,1){1.62}}
			\put(12.5,10.75){\line(-3,1){1.62}}
			\put(11.68,11.17){$[1234] \! = \! 0$}
			\put(12.4,9.75){{\large $\nexists$}}
			\put(12.48,10.75){\vector(0,-1){0.45}}
			\put(14.9,11.22){\framebox{\,\hspace*{-0.06cm}G$_{1d}$\,\hspace*{-0.45cm}$\phantom{{A^B_C}}\!\!$}}
			\put(4.5,9.35){\line(-3,-1){1.62}} \put(4.5,9.35){\line(3,-1){1.62}}
			\put(4.5,8.25){\line(3,1){1.62}} \put(4.5,8.25){\line(-3,1){1.62}}
			\put(3.68,8.7){$[1222] \! = \! 0$}
			\put(8.5,9.35){\line(-3,-1){1.62}} \put(8.5,9.35){\line(3,-1){1.62}}
			\put(8.5,8.25){\line(3,1){1.62}} \put(8.5,8.25){\line(-3,1){1.62}}
			\put(7.68,8.7){$[1223] \! = \! 0$}
			\put(14.9,8.75){\framebox{\,\hspace*{-0.06cm}G$_{1c}$\,\hspace*{-0.45cm}$\phantom{{A^B_C}}\!\!$}}
			\put(4.5,8.25){\vector(0,-1){0.45}} \put(4.4,7.3){{\large $\nexists$}}
			\put(8.4,7.3){{\large $\nexists$}}
			\put(8.5,8.25){\vector(0,-1){0.45}}
			\put(0.5,6.85){\line(-3,-1){1.62}} \put(0.5,6.85){\line(3,-1){1.62}}
			\put(0.5,5.75){\line(3,1){1.62}} \put(0.5,5.75){\line(-3,1){1.62}}
			\put(-0.36,6.2){$[111] \! = \! 0$}
			\put(4.5,6.85){\line(-3,-1){1.62}} \put(4.5,6.85){\line(3,-1){1.62}}
			\put(4.5,5.75){\line(3,1){1.62}} \put(4.5,5.75){\line(-3,1){1.62}}
			\put(3.7,6.2){$[112] \! = \! 0$}
			\put(14.9,6.25){\framebox{\,\hspace*{-0.06cm}G$_{2a}$\,\hspace*{-0.45cm}$\phantom{{A^B_C}}\!\!$}}
			\put(4.5,5.1){\line(-3,-1){1.62}} \put(4.5,5.1){\line(3,-1){1.62}}
			\put(4.5,4){\line(3,1){1.62}} \put(4.5,4){\line(-3,1){1.62}}
			\put(3.64,4.45){$[1122] \! = \! 0$}
			\put(8.5,5.1){\line(-3,-1){1.62}} \put(8.5,5.1){\line(3,-1){1.62}}
			\put(8.5,4){\line(3,1){1.62}} \put(8.5,4){\line(-3,1){1.62}}
			\put(7.68,4.45){$[1123] \! = \! 0$}
			\put(14.9,4.45){\framebox{\,\hspace*{-0.06cm}G$_{1b}$\,\hspace*{-0.45cm}$\phantom{{A^B_C}}\!\!$}}
			\put(8.4,3.05){{\large $\nexists$}}
			\put(8.5,4){\vector(0,-1){0.45}}
			\put(4.4,3.05){{\large $\nexists$}}
			\put(4.5,4){\vector(0,-1){0.45}}
			\put(0.5,2.6){\line(-3,-1){1.62}}  \put(0.5,2.6){\line(3,-1){1.62}}
			\put(0.5,1.5){\line(3,1){1.62}} \put(0.5,1.5){\line(-3,1){1.62}}
			\put(-0.34,1.95){$[1111] \! = \! 0$}
			\put(4.5,2.6){\line(-3,-1){1.62}}  \put(4.5,2.6){\line(3,-1){1.62}}
			\put(4.5,1.5){\line(3,1){1.62}} \put(4.5,1.5){\line(-3,1){1.62}}
			\put(3.68,1.95){$[1112] \! = \! 0$}
			\put(4.4,0.5){{\large $\nexists$}}
			\put(4.5,1.5){\vector(0,-1){0.45}}
			\put(0.4,0.5){{\large $\nexists$}}
			\put(0.5,1.5){\vector(0,-1){0.45}}
			\put(14.9,1.9){\framebox{\,\hspace*{-0.06cm}G$_{1a}$\,\hspace*{-0.45cm}$\phantom{{A^B_C}}\!\!$}}
		\end{picture} }
		\vspace{-7mm}
		\caption{Algorithm to determine the dimension of the groups of isometries of a spacetime admitting a Riemann-frame.}
		\label{Fig-18}
		\end{figure}
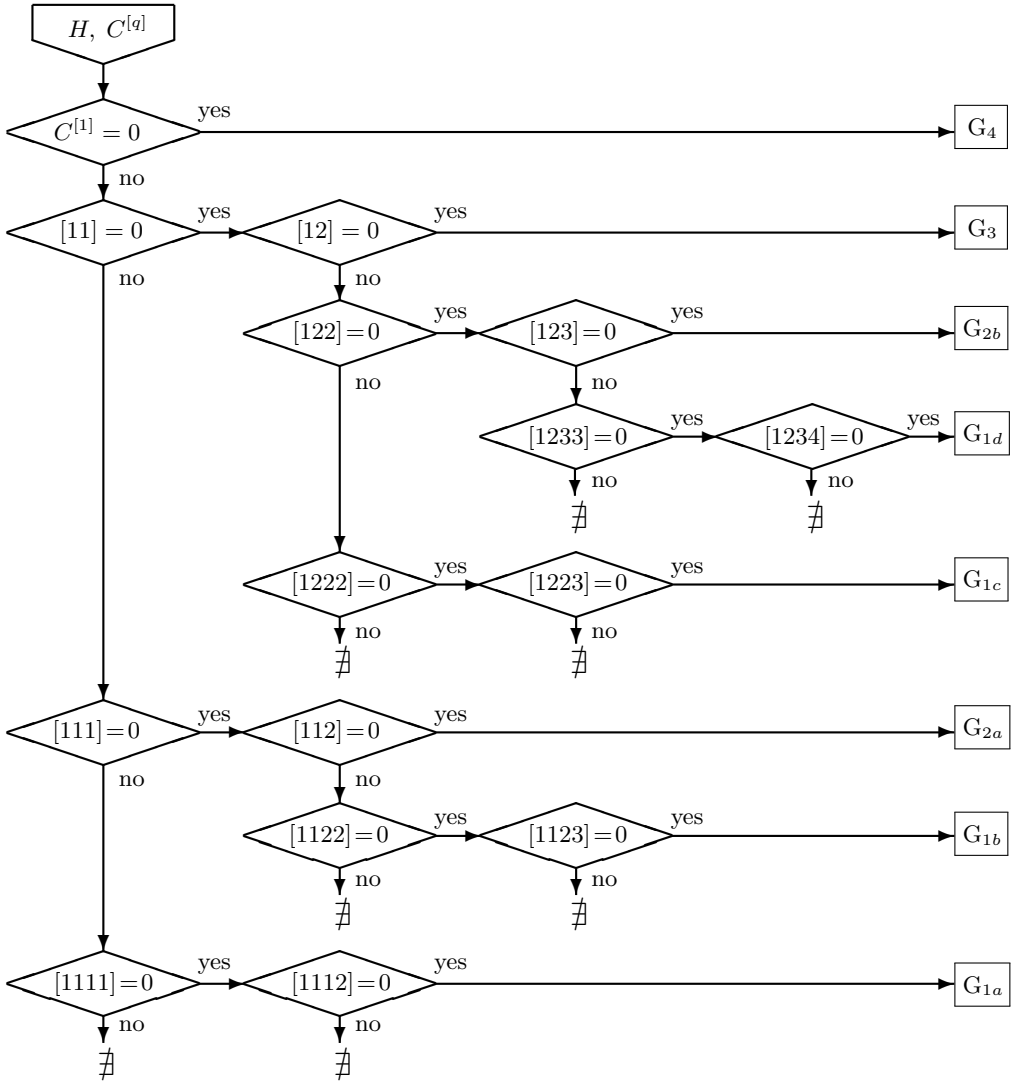

		An invariant frame that can be built from the Riemann tensor $R$ and its covariant derivatives will be called a Riemann-frame ($R$-frame). If a $R$-frame exists, we can always orthonormalise it and then work with such oriented orthonormal $R$-frame $\{ e_{a} \}$, $\eta(e_0, e_1, e_2, e_3) = 1$, where Latin indices count the vectors of the frame and indicate the components of a tensor in this non-holonomic frame. Let $\{ \theta^{a} \}$ be its algebraic dual basis. We can collect the connection coefficients in the connection tensor $H$ defined as
		\begin{equation} \label{defH}
			H = \gamma^{c}_{a b} \, \theta^{b} \otimes \theta^a \otimes e_{c} .
		\end{equation}
We shall denote with the same symbol a tensor and the metric equivalent tensors that follow by raising and lowering indices with $g$. In that sense, the tensor that results from $H$ when raising the second index can be obtained from the frame $\{ e_{a} \}$ as
		\begin{equation} \label{H-wedge}
			H = - \frac12 \tilde{\eta}^{a b}\nabla e_{a} \bar{\wedge} e_b \, , 
		\end{equation}
where $\tilde{\eta}^{ab}$ is the signature symbol, $\tilde{\eta}^{ab} = diag(-1,1,1,1)$. \\ \\
		\textbf{Isometry group dimension.} \textit{Consider the following concomitants of the connection tensor} $H$ \textit{of a given spacetime:}
		\begin{align} \label{cq}
			C^{[q]}_{\alpha_q \cdots \, \alpha_1 \mu \nu \sigma} &= \nabla_{\alpha_q} C^{[q - \! 1]}_{\alpha_{q - \! 1} \cdots \, \alpha_1 \mu \nu \sigma} + {H_{\alpha_{q} \alpha_{q - \! 1}}}^{\! \rho} C^{[q - \! 1]}_{\rho \cdots \, \alpha_1 \mu \nu \sigma} + \cdots \\ 
			&\quad + {H_{\alpha_q \sigma}}^{\! \rho} C^{[q- \! 1]}_{\alpha_{q - \! 1} \cdots\, \alpha_1 \mu \nu \rho} \, , \notag
		\end{align}
\textit{The dimension} $r$ \textit{of the admitted isometry group} G$_r$ \textit{is given by the algorithm in Figure} \ref{Fig-18}\textit{, where the notation} $[qr] \equiv \eta(C^{[q]}\!, C^{[r]})$\textit{,} $[p q r] \equiv \eta(C^{[p]}\!, C^{[q]}\!, C^{[r]})$, $[p q r s] \equiv \eta(C^{[p]}\!, C^{[q]}\!, C^{[r]}\!,  C^{[s]})$ \textit{indicates the contraction of one, two, three or four indices of the volume element} $\eta$ \textit{with the first index of the tensors} $C^{[q]}$. \\ \\ \\ \\ \\ \\ \\ \\
		$\hspace{-5mm}$In spacetimes of Petrov-Bel types I, II or III, the Weyl tensor defines a principal frame. Using this, in \cite{SFM-simetries} an IDEAL determination of the associated connection tensor is obtained without explicitly obtaining the frame: \\ \\
		\textbf{Type I, II and III connection tensor.} \textit{The connection tensor} $H$ \textit{can be expressed in terms of the self-dual connection tensor} ${\cal H} \equiv \frac{1}{\sqrt{2}} (H - \textrm{i} * H)$ {as}
		\begin{equation} \label{H-Hautodual}
			H = \frac{1}{\sqrt{2}}({\cal H} + \bar{\cal H}) \, .
		\end{equation}
		\textit{The self-dual connection tensor} ${\cal H}$ \textit{associated with the Weyl principal frame of a Petrov-Bel type I, II or III spacetime can be obtained as follows:}
		\begin{itemize}
		\item \textit{Type I:}
			\begin{subequations}
				\begin{equation}
					{\cal H}_{\alpha \mu \nu} = \frac{1}{\sqrt{2}} {\cal X}_{\alpha \lambda \rho} {{\cal Y}^{\lambda \rho}}_{\mu \nu} \, ,
				\end{equation}
				\begin{equation}
					{\cal X}_{\lambda \alpha \rho} \equiv \frac{1}{2} \nabla_{\lambda} {\cal W}_{\alpha \beta \mu \nu} {{\cal W}^{\mu \nu \beta}}_{\rho}\, , \qquad {\cal Y} \equiv \frac{1}{\Delta} (3 a {\cal W}^2 + 6 b {\cal W} + \frac{1}{2} a^2 {\cal G}) \, ,
				\end{equation}
			\end{subequations}
\textit{where} $a \equiv \textrm{Tr} {\cal W}^2$\textit{,} $b \equiv \textrm{Tr} {\cal W}^3$ \textit{and} $\Delta \equiv a^3 - 6 b^2 \neq 0$.
		\item \textit{Type II:}
			\begin{subequations}
				\begin{equation} \label{TII-self-dual-H}
					{\cal H} \equiv - \frac{1}{\sqrt{2}} \left( \nabla {\cal U} \cdot {\cal U} + \nabla {\cal L}_+ \! \cdot \! {\cal L}_- + \nabla {\cal L}_- \! \cdot \! {\cal L}_+ \right) ,
				\end{equation}
				\begin{equation}
					{\cal U} = \frac{{\cal A}({\cal Y})}{\sqrt{-{\cal A}^2({\cal Y},{\cal Y})}}, \qquad {\cal L}_+ = \frac{{\cal P}({\cal X})}{\sqrt{{\cal P}({\cal X},{\cal X})}},
				\end{equation}
				\begin{equation} \label{L_-}
					{\cal L}_- = \frac{1}{2 ({\cal L}_+ , {\cal X})^2} \left[{\cal S}({\cal X}, {\cal X}) {\cal L}_+ - 2({\cal L}_+ , {\cal X}) {\cal S}({\cal X}) \right] , \qquad {\cal S} \equiv {\cal G} + {\cal U} \otimes {\cal U} \, ,
				\end{equation}
			\end{subequations}
\textit{with} ${\cal A} \equiv ({\cal W} - \psi {\cal G})^2$\textit{,} ${\cal P} \equiv \frac{1}{3 \psi} ({\cal W} - \psi {\cal G})({\cal W} + 2 \psi {\cal G})$\textit{,} $\psi \equiv -b/a$\textit{,} $a \equiv \textrm{Tr} {\cal W}^2$ \textit{and} $b \equiv \textrm{Tr} {\cal W}^3$\textit{, and where} ${\cal Y}$ \textit{and} ${\cal X}$ \textit{are arbitrary bivectors such that} ${\cal A}({\cal Y}) \neq 0$ \textit{and} ${\cal P}({\cal X}) \neq 0$\textit{, respectively.}
		\item \textit{Type III:} ${\cal H}$ \textit{in} (\ref{TII-self-dual-H}) \textit{with}
			\begin{equation} \label{L_+-U-TIII}
				{\cal L}_+ = \frac{- \, {\cal W}^2({\cal X})}{\sqrt{- {\cal W}^2({\cal X}, {\cal X})}}, \qquad {\cal U} = \frac{2 ({\cal L}_+ , {\cal X}) {\cal W}({\cal X}) - {\cal W}({\cal X}, {\cal X}) {\cal L}_+}{2 ({\cal L}_+ ,{\cal X})^2} ,
			\end{equation}
${\cal L}_-$ \textit{given in} (\ref{L_-})\textit{, and where} ${\cal X}$ \textit{is an arbitrary bivector such that} ${\cal W}^2({\cal X}) \neq 0$.
		\end{itemize}
		The \textit{xIdeal} functions that implement these two algorithms are called \texttt{ConnectionTensor} and \texttt{IsometryGroupDimension}. \texttt{ConnectionTensor} needs to be given a metric and a previously obtained $R$-frame. Then it obtains the connection tensor $H$ as in (\ref{H-wedge}). If no $R$-frame is given but the metric is of Petrov-Bel type I, II or III, it still obtains the connection tensor as in (\ref{H-Hautodual}$-$\ref{L_+-U-TIII}). Regarding \texttt{IsometryGroupDimension}, the input data are a metric and a connection tensor. Then, it determines the dimension of the group of isometries, if there is any, by following the algorithm in Figure \ref{Fig-18}. \\ \\
		Now, we consider two examples in which these two functions have been tested: the Petrov homogeneous vacuum solution, which is of Petrov-Bel type I and therefore the connection tensor can be obtained without explicitly constructing the $R$-frame, and the Wils conformally flat pure radiation solution, where we need to consider Riemann derivatives to obtain the $R$-frame.
		
			\subsubsection{The Petrov homogeneous vacuum solution}
			The Petrov homogeneous vacuum solution \cite{Petrov-sol, Kramer} can be characterised as the only type I vacuum solution with constant Weyl eigenvalues \cite{FS-typeI-a}. It admits a group of isometries G$_4$, and only one Killing vector has its Killing two-form aligned with a Weyl principal bivector \cite{FS-typeI-b}. We use the \textit{xIdeal} function \texttt{ConnectionTensor} to obtain its associated connection tensor without explicitly constructing the $R$-frame and we pass the output to the function \texttt{IsometryGroupDimension}, which returns G$_4$ as expected.
			
			\subsubsection{The Wils conformally flat pure radiation solution}
			The conformally flat pure radiation solution \cite{Wils} is a metric with no symmetries or polynomial invariants \cite{Koutras-Mc}. The metric line element is given by
			\begin{subequations} \label{Wils-metric}
				\begin{eqnarray}
					\textrm{d} s^2 \! = \! - 2 x \, \textrm{d} u  \textrm{d} w + 2 w \, \textrm{d} u \textrm{d} x - F \textrm{d} u^2 + \textrm{d} x^2 + \textrm{d} y^2, \\[2mm]
 					F \equiv \frac12 \phi^2(x^2 + y^2) - w^2 , \qquad \phi \equiv 2 \sqrt{|x f(u)|} \, , \quad
				\end{eqnarray}
			\end{subequations}
where $f(u)$ is an arbitrary real function. Let us consider the null frame $\{ \ell, k, e_2, e_3\}$, 
			\begin{equation} \label{Wils-frame}
				\hspace{-3mm} \ell = \frac{1}{2 \phi}[F \textrm{d} u + 2 x \textrm{d} w - 2 w \textrm{d} x] , \qquad k = \phi \, \textrm{d} u , \qquad e_2 = \textrm{d} x , \qquad e_3 = \textrm{d} y .
			\end{equation}
The Weyl tensor vanishes, and the study of the Ricci tensor $R$ and its derivatives leads to the following relations:
			\begin{equation} \label{Wils-relations}
				\begin{array}{c}
					\hspace{-3mm} R = k \! \otimes \! k , \qquad x^2 Q = R \! \otimes \! R, \qquad S^2 = e_2 \! \otimes \! e_3, \qquad 2 k \otimes \ell = U - U^2 , \\[2mm]
					\hspace{-3mm} Q_{\alpha \beta \lambda \mu} \equiv \nabla_{\alpha} R_{\beta \rho} \nabla_\lambda R_{\mu}{}^{\rho} , \qquad S \equiv \frac12 \nabla \textrm{d} x^2 \! - \! g , \qquad U \equiv *(e_2 \wedge e_3) .
				\end{array}
			\end{equation}
From these expressions, it follows that the tetrad (\ref{Wils-frame}) defines a $R$-frame. Then, we can obtain the connection tensor $H$ given in (\ref{H-wedge}), with $\sqrt{2} \, e_0 = \ell + k$ and $\sqrt{2} \, e_1 = \ell - k$. We give this $R$-frame to the \textit{xIdeal} function \texttt{ConnectionTensor} and pass its output to the function \texttt{IsometryGroupDimension}. As a result, we obtain that no isometry group is admitted by this solution.
		
		\subsection{Type N vacuum solutions} \label{subsec-dim-TipusN}
\begin{figure}[t]
		\setlength{\unitlength}{0.79cm} {\small \noindent
		\begin{picture}(18,14)
		\thicklines
			\put(9.5,11.5){\footnotesize{yes}}
			\put(13,11.5){\footnotesize{yes}}
			\put(8.45,10.35){\footnotesize{no}}
			\put(11.7,10.35){\footnotesize{no}}
			\put(9.65,9.45){\footnotesize{yes}}
			\put(6.55,9.45){\footnotesize{no}}
			\put(12.25,8.15){\footnotesize{no}}
			\put(3.85,8.15){\footnotesize{no}}
			\put(12.25,6.2){\footnotesize{no}} 
			\put(3.85,6.2){\footnotesize{no}}
			\put(12.4,4.2){\footnotesize{no}} 
			\put(3.7,4.2){\footnotesize{no}}
			\put(12.4,2.2){\footnotesize{no}}
			\put(11.03,7.15){\footnotesize{yes}}
			\put(5.93,7.15){\footnotesize{yes}}
			\put(11.03,5.15){\footnotesize{yes}}
			\put(5.93,5.15){\footnotesize{yes}}
			\put(11.03,3){\footnotesize{yes}} 
			\put(5.93,3){\footnotesize{yes}}
			\put(11.03,1.2){\footnotesize{yes}}
			\put(8.3,12.5){\vector(0,-1){0.5}}
			\put(8.3,10.75){\vector(0,-1){0.75}}
			\put(9.5,9.35){\line(1,0){1.35}} \put(7.1,9.35){\line(-1,0){1.35}}
			\put(10.85,9.35){\vector(0,-1){0.7}}
			\put(5.75,9.35){\vector(0,-1){0.7}}
			\put(4.5,8.03){\vector(-1,0){1.6}} \put(1.7,8.05){\line(-1,0){0.6}}
			\put(5.75,7.4){\vector(0,-1){0.7}}
			\put(10.85,7.4){\vector(0,-1){0.7}}
			\put(5.75,5.47){\vector(0,-1){0.67}}
			\put(10.85,5.47){\vector(0,-1){0.67}}
			\put(5.75,3.3){\line(0,-1){0.7}}
			\put(10.85,3.3){\vector(0,-1){0.59}}
			\put(10.85,1.45){\line(0,-1){0.65}}
			\put(10.85,0.8){\vector(1,0){2.75}}
			\put(14.8,0.8){\line(1,0){0.6}}
			\put(12.1,2.08){\vector(1,0){1.51}} \put(14.8,2.08){\line(1,0){0.6}}
			\put(12.3,4.05){\vector(1,0){1.31}} \put(14.8,4.05){\line(1,0){0.6}}
			\put(4.3,4.05){\vector(-1,0){1.4}}
			\put(1.7,4.05){\line(-1,0){0.6}}
			\put(5.75,2.6){\vector(-1,0){2.85}}
			\put(1.68,2.6){\line(-1,0){0.55}}
			\put(12.1,6.08){\vector(1,0){1.5}}
			\put(4.5,6.08){\vector(-1,0){1.6}} \put(1.7,6.08){\line(-1,0){0.6}}
			\put(8.3,12.9){{\oval(5,0.8)}} \put(6.1,12.75){$\ \widehat{C7}$ ($\varphi_n = 0 $) ;  $ \ \, 6 Q = \Lambda g$}

			\put(8.3,11.98){\line(-2,-1){1.25}}
			\put(8.3,11.98){\line(2,-1){1.25}} \put(8.3,10.73){\line(2,1){1.25}}
			\put(8.3,10.73){\line(-2,1){1.25}} \put(7.8,11.25 ){$\Lambda = 0$}
			\put(11.55,12.1){\line(-2,-1){1.5}}
			\put(11.55,12.1){\line(2,-1){1.5}} \put(11.55,10.6){\line(2,1){1.5}}
			\put(11.55,10.6){\line(-2,1){1.5}} \put(10.44,11.2){$\textrm{d} b_\ell \! = \! \textrm{d} c_\ell \! = \! 0$}
			\put(13,11.35){\vector(1,0){0.6}}
			\put(14.78,11.35){\line(1,0){0.63}}
			\put(14.2,11.4){{\oval(1.2,0.8)}} \put(13.85,11.25){OS1}
			\put(15.4,11.25){\framebox{$G_4/N_3$}}
			\put(9.5,11.35){\vector(1,0){0.54}}
			\put(14.2,10.1){{\oval(1.2,0.8)}} \put(13.85,9.95){OS2}
			\put(15.4,9.95){\framebox{$G_6/O_4$}}
			\put(14.78,10){\line(1,0){0.63}}
			\put(11.56,10){\vector(1,0){2.05}}
			\put(11.56,10.6){\line(0,-1){0.6}}
			\put(8.3,9.98){\line(-2,-1){1.25}} \put(8.3,9.98){\line(2,-1){1.25}}
			\put(8.3,8.73){\line(2,1){1.2}} \put(8.3,8.73){\line(-2,1){1.25}}
			\put(7.54,9.23 ){$\ell \wedge c = 0$}
			\put(10.85,8.65){\line(-2,-1){1.25}}
			\put(10.85,8.65){\line(2,-1){1.25}}
			\put(10.85,7.4){\line(2,1){1.25}} \put(10.85,7.4){\line(-2,1){1.25}}
			\put(10.22,7.9){$\textrm{d} b = 0$}
			\put(12.1,8.03){\vector(1,0){1.5}}
			\put(14.2,8.1){{\oval(1.2,0.8)}} \put(13.85,7.95){BR1}
			\put(15.4,7.9){\framebox{$H$}}
			\put(14.8,8.05){\line(1,0){0.6}}
			\put(5.75,8.65){\line(-2,-1){1.25}}
			\put(5.75,8.65){\line(2,-1){1.25}} \put(5.75,7.4){\line(2,1){1.25}}
			\put(5.75,7.4){\line(-2,1){1.25}} \put(5.12,7.9){$\textrm{d} b = 0$}
			\put(2.3,8.1){{\oval(1.2,0.8)}} \put(1.9,7.95){AR1}
			\put(0.45,7.9){\framebox{$H$}}
			\put(5.75,6.7){\line(-2,-1){1.25}} \put(5.75,6.7){\line(2,-1){1.25}}
			\put(5.75,5.45){\line(2,1){1.25}} \put(5.75,5.45){\line(-2,1){1.25}}
			\put(5.15,5.97){$\Psi_\ell = 0$}
			\put(10.85,6.7){\line(-2,-1){1.25}}
			\put(10.85,6.7){\line(2,-1){1.25}}
			\put(10.85,5.45){\line(2,1){1.25}}
			\put(10.85,5.45){\line(-2,1){1.25}} \put(10.1,5.97){$\alpha \alpha^* = 0$}
			\put(2.3,6.1){{\oval(1.2,0.8)}} \put(1.9,5.95){AR2}
			\put(0.45,5.9){\framebox{$H$}}
			\put(14.2,6.1){{\oval(1.2,0.8)}} \put(13.85,5.95){BR2}
			\put(15.4,5.9){\framebox{$H$}} \put(14.8,6.05){\line(1,0){0.6}}
			\put(10.85,4.8){\line(-2,-1){1.5}} \put(10.85,4.8){\line(2,-1){1.5}}
			\put(10.85,3.3){\line(2,1){1.5}} \put(10.85,3.3){\line(-2,1){1.5}}
			\put(9.82,3.92){$\textrm{d} \phi \! \wedge \! \textrm{d} \phi_\ell \! = \! 0$}
			\put(14.2,4.1){{\oval(1.2,0.8)}} \put(13.85,3.95){BS1}
			\put(15.4,3.95){\framebox{$G_3/N_2$}}
			\put(14.2,2.1){{\oval(1.2,0.8)}} \put(13.85,1.95){BS2}
			\put(15.4,1.95){\framebox{$G_4/T_3$}}
			\put(14.2,0.8){{\oval(1.2,0.8)}} \put(13.85,0.65){BS3}
			\put(15.4,0.65){\framebox{$G_5/O_4$}}
			\put(5.75,4.8){\line(-2,-1){1.5}} \put(5.75,4.8){\line(2,-1){1.5}}
			\put(5.75,3.3){\line(2,1){1.5}} \put(5.75,3.3){\line(-2,1){1.5}}
			\put(4.7,3.9){$3\chi c^2 \! = \! 5 \gamma \gamma^*$}
			\put(2.3,4.1){{\oval(1.2,0.8)}} \put(1.89,3.95){AR3}
			\put(0.45,3.9){\framebox{$H$}}
			\put(2.3,2.6){{\oval(1.2,0.8)}} \put(2,2.45){AS}
			\put(-0.3,2.5){\framebox{$G_4/T_3$}}
			\put(10.85,2.7){\line(-2,-1){1.25}}
			\put(10.85,2.7){\line(2,-1){1.25}}
			\put(10.85,1.45){\line(2,1){1.25}}
			\put(10.85,1.45){\line(-2,1){1.25}} \put(10.2,1.97){$\textrm{d} \phi = 0$}
		\end{picture} }
		\vspace{-7mm}
		\caption{Algorithm to distinguish the different classes in the family $\widehat{\rm C7}$ of the type N vacuum solutions. It also gives the dimension of the group and orbits for the singular ones.}
		\label{Fig-20}
		\end{figure}
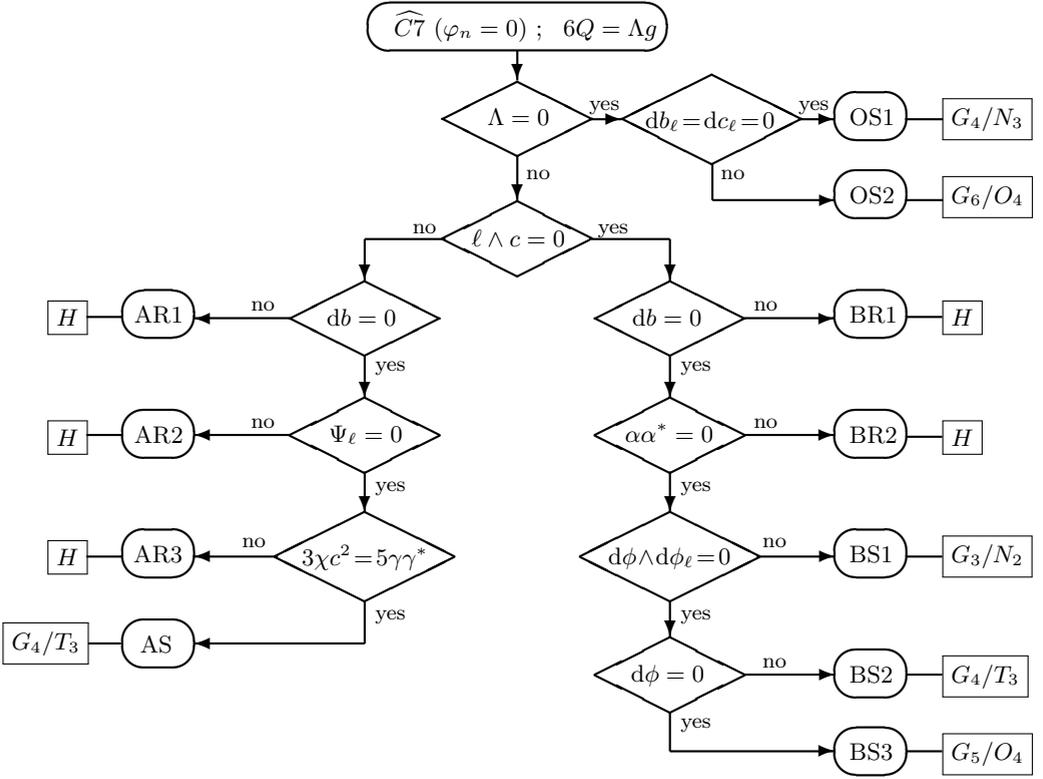

		For solutions of Petrov-Bel types D, N or O, the existence of a $R$-frame depends on the algebraic properties of the Ricci tensor or on the successive covariant derivatives of both the Ricci and Weyl tensors. Consequently, the determination of the number of isometries for these solutions requires a wider analysis. In \cite{SMF-DimensioTN}, we present a new step in solving this question: the study of this problem for the type N vacuum solutions with cosmological constant $\Lambda$. Here, we summarise the results in \cite{SMF-DimensioTN} and show how \textit{xIdeal} assists in applying them. \\ \\
		Note that, in this case, the Ricci tensor does not define any scalar or tensorial invariant, and consequently, all the characterisation conditions depend, necessarily, on the Weyl tensor and its covariant derivatives. Our study requires considering an invariant classification of the type N vacuum solutions. The {\em regular classes} are defined by a condition that ensures the existence of a $R$-frame. For them, we can apply the results in the previous subsection. For the {\em singular classes} (no $R$-frame exists), we need to know the scalars, directions and 2-forms that can be obtained from the Weyl tensor, and we make use of the Eisenhart \cite{Eisenhart-1933} and Defrise \cite{Defrise} results to determine the dimension of the isometry group and of its orbits. Now, we introduce the definition and notation of each class, which is useful to summarise the results. \\ \\ \\
		Let ${\cal L}$ and $\ell$ be the fundamental bivector and the fundamental vector of a type N Weyl tensor (${\cal W} = {\cal L} \otimes {\cal L}$, $\ell \wedge {\cal L} = 0$), which can be obtained as
		\begin{equation} \label{L-ell}
			{\cal L} \equiv \frac{{\cal W}({\cal X})}{\sqrt{{\cal W}({\cal X},{\cal X})}} \ , \quad \quad \ell \equiv \frac{L^2(x)}{\sqrt{-L^2(x, x)}}\, , \quad \quad L \equiv \frac{1}{\sqrt{2}}({\cal L} + \bar{{\cal L}}) \, ,
		\end{equation}
where ${\cal X}$ is an arbitrary bivector such that ${\cal W}({\cal X}) \neq 0$ and $x$ an arbitrary timelike vector. Note that both ${\cal L}$ and $\ell$ are determined up to sign, but (\ref{L-ell}) gives a future-pointing fundamental vector $\ell$ if $x$ is also future-pointing. \\ \\
		Now, let us define the following Weyl concomitants: \\[-7mm]
		\begin{subequations} \label{varphi_n}
			\begin{align}
				&\varphi_1 \equiv ({\cal F}_1,{\cal F}_1), \quad \varphi_2 \equiv |({\cal F}_2,{\cal F}_2)| + |({\cal F}_2^*,{\cal F}_2^*)|, \quad \varphi_3 \equiv (F_3, F_3), \\[1mm]    
				&\varphi_4 \equiv (F_4, F_4,), \quad \varphi_5 \equiv (F_5, F_5), \quad \varphi_6 \equiv *(\ell \wedge f \wedge f^*) , \quad \varphi_7 \equiv *(\ell \wedge b \wedge c) ,
			\end{align}
		\end{subequations}
where \\[-7mm]
		\begin{subequations}
			\begin{align}
				&{\cal F}_1 \equiv \nabla_{\ell}{\cal L} \, , \quad {\cal F}_2 \equiv \nabla_{a}{\cal L} \, , \quad {\cal F}_2^* \equiv \nabla_{a^*}{\cal L} \, , \\[1mm] 
				&F_3 \equiv \textrm{d} \ell \, , \quad F_4 \equiv \ell \wedge c \, , \quad F_5 \equiv \textrm{d} b \, ,
			\end{align}
		\end{subequations}
and 
		\begin{equation} \label{a_a*}
			a \equiv -\frac{1}{(\ell,v)}(\nabla L \cdot \ell)(v) , \qquad a^* \equiv \frac{1}{(\ell,v)}(\nabla \! * \! L \cdot \ell)(v) ,	
		\end{equation}
		\begin{equation} \label{alpha}
			\alpha \equiv \frac{(a,v)}{(\ell,v)} , \qquad  \alpha^* \equiv \frac{(a^*,v)}{(\ell,v)} ,
		\end{equation}
		\begin{equation} \label{B-b}
			B \equiv \nabla \ell + ^t \! \nabla \ell \, , \qquad b \equiv \frac{1}{(\ell,v)} \left[B(v) - \frac{B(v,v)}{2 (\ell, v)} \ell \right] ,
		\end{equation}
		\begin{equation} \label{C-c}
			C \equiv - Sym(\nabla \! * \! L \cdot L) \, , \qquad c \equiv \frac{1}{(\ell,v)^3}[3(\ell, v) C(v,v) \! - \! 2 C(v,v,v) \ell],
		\end{equation}
		\begin{equation} \label{f_f*}
			f \equiv \textrm{d} \alpha - \alpha^* c, \qquad f^* \equiv \textrm{d} \alpha^* + \alpha c,
		\end{equation}
with $v$ an arbitrary timelike vector. 
		\begin{itemize}
			\item[$\circ$] We say that a type N spacetime is of class {\rm C}n if $\varphi_i = 0$ $\forall i < n$ and $\varphi_n \neq 0$.
			\item[$\circ$] We denote $\widehat{{\rm C}n}$ the family of type N spacetimes such that $\varphi_i = 0$ $\forall i \leqslant n$.
		\end{itemize}
		The type N vacuum solutions with $\Lambda = 0$ in the family $\widehat{\rm C7}$ are the plane waves. They admit a non-trivial isotropy group of dimension two, and we can consider two classes:
		\begin{itemize}
			\item[$\circ$] OS1: $\textrm{d} b_\ell \neq 0$ or $\textrm{d} c_\ell \neq 0$. The metric admits a G$_5$ on $3$-dimensional null orbits.
			\item[$\circ$] OS2: $\textrm{d} b_\ell = \textrm{d} c_\ell = 0$. The metric admits a G$_6$ on $4$-dimensional orbits. In this last class we can consider two subclasses: OS2a when $b_\ell \neq 0$, and OS2b when $b_\ell = 0$.
		\end{itemize}
		In the family $\widehat{\rm C7}$ ($\varphi_n = 0, n \leqslant 7$) of type N vacuum solutions with non-vanishing cosmological constant, we consider the following classes defined by invariant conditions:
		\[\begin{array}{lllll}
			\hspace{-6mm} \circ \;\, {\rm AR1:} & \quad \ell \wedge c \neq 0 & \quad \textrm{d} b \neq 0 & \quad & \\
			\hspace{-6mm} \circ \;\, {\rm AR2:} & \quad \ell \wedge c \neq 0 & \quad \textrm{d} b = 0 & \quad \psi_\ell \neq 0 & \\
			\hspace{-6mm} \circ \;\, {\rm AR3:} & \quad \ell \wedge c \neq 0 & \quad \textrm{d} b = 0 & \quad \psi_\ell = 0 & \quad 3 \chi c^2 \neq 5 \gamma \gamma^* \\
			\hspace{-6mm} \circ \;\, {\rm AS:} & \quad \ell \wedge c \neq 0 & \quad \textrm{d} b = 0 & \quad \psi_\ell = 0 & \quad 3 \chi c^2 = 5 \gamma \gamma^* \\
			\hspace{-6mm} \circ \;\, {\rm BR1:} & \quad \ell \wedge c = 0 & \quad \textrm{d} b \neq 0 & \quad & \\
			\hspace{-6mm} \circ \;\, {\rm BR2:} & \quad \ell \wedge c = 0 &\quad \textrm{d} b = 0 & \quad \alpha \alpha^* \neq 0 & \\
			\hspace{-6mm} \circ \;\, {\rm BS1:} & \quad \ell \wedge c = 0 & \quad \textrm{d} b = 0 & \quad \alpha \alpha^* = 0 & \quad \textrm{d} \phi \wedge \textrm{d} \phi_\ell \neq 0 \\
			\hspace{-6mm} \circ \;\, {\rm BS2:} & \quad \ell \wedge c = 0 & \quad \textrm{d} b = 0 & \quad \alpha \alpha^* = 0 & \quad \textrm{d} \phi \wedge \textrm{d} \phi_\ell = 0 \qquad \textrm{d} \phi \neq 0 \\
			\hspace{-6mm} \circ \;\, {\rm BS3:} & \quad \ell \wedge c = 0 & \quad \textrm{d} b = 0 & \quad \alpha \alpha^* = 0 & \quad \textrm{d} \phi = 0 
		\end{array}\]
where the involved Weyl concomitants can be obtained as follows:
		\begin{equation} \label{gamma-psi_ell_c}
			\gamma \equiv -\frac{L(c,v)}{(\ell,v)} , \quad \gamma^* \equiv \frac{*L(c,v)}{(\ell,v)} , \quad \psi_c \equiv \frac{(b,c)}{c^2}, \quad \psi_{\ell} \equiv \frac{1}{(\ell,v)}(b - \psi_c c,v) ,	
		\end{equation}
		\begin{equation} \label{chi}
			\chi \equiv \frac{P(v, v)}{(\ell, v)^2} , \qquad P \equiv \nabla c + (\gamma \alpha + \gamma^* \alpha^*) g - \frac45 \psi_c c \otimes c ,
		\end{equation}
		\begin{equation} \label{phi-q}
			\phi \equiv \frac{(q,v)}{(\ell,v)} , \qquad q \equiv \frac{1}{(\ell,v)^2} [(\ell, v) S(v) - \frac12 S(v,v) \ell],
		\end{equation}
		\begin{equation} \label{S}
			S \equiv \nabla b + ^t \! \nabla b + 5 (\alpha^2 + \alpha^{*2}) g - \frac45 b \otimes b,
		\end{equation}
		\begin{equation} \label{phi_ell_b}
			\phi_{\ell} \equiv \frac{1}{(\ell,v)}(\textrm{d} \phi - \phi_b b,v), \qquad \phi_b \equiv -2(\varepsilon + \frac15 \phi) ,
		\end{equation}
where $v$ is an arbitrary timelike vector, and $\varepsilon = 1$ if $\alpha^* = 0$ and $\varepsilon = -1$ if $\alpha = 0$. \\ \\
		With all this, we have the following algorithm to classify and determine the dimension of the isometry group of a type N vacuum solution: \\ \\
		\textbf{Isometry group dimension of type N vacuum solutions. } \textit{Consider the Weyl concomitants} $\varphi_n$ \textit{defined in} (\ref{varphi_n}$-$\ref{f_f*})\textit{. The following algorithm gives the first part of the classification of the type N vacuum solutions:}
		\begin{figure}[H]
		\setlength{\unitlength}{1cm} {\normalsize \noindent
		\begin{picture}(16,8)
		\thicklines
			\put(1.4,4.9){\vector(1,0){0.6}}
			\put(4,4.9){\vector(1,0){0.8}} \put(6.7,4.9){\vector(1,0){0.8}}
			\put(9.5,4.9){\vector(1,0){0.8}}
			\put(2,3.4){\vector(-1,0){0.46}}
			\put(4.8,3.4){\vector(-1,0){0.8}} \put(7.5,3.4){\vector(-1,0){0.8}}
			\put(9.5,4.9){\vector(1,0){0.8}}
			\put(3,2.9){\vector(0,-1){0.5}} \put(3,1.6){\line(0,-1){0.55}}
			\put(5.75,2.9){\vector(0,-1){0.5}} \put(5.75,1.6){\line(0,-1){0.55}}
			\put(8.5,2.9){\vector(0,-1){0.5}} \put(8.5,1.6){\line(0,-1){0.55}}
			\put(3,5.4){\vector(0,1){0.5}} \put(3,6.7){\line(0,1){0.54}}
			\put(5.75,5.4){\vector(0,1){0.5}} \put(5.75,6.7){\line(0,1){0.54}}
			\put(8.5,5.4){\vector(0,1){0.5}} \put(8.5,6.7){\line(0,1){0.54}}
			\put(11.25,5.4){\vector(0,1){0.5}} \put(11.25,6.7){\line(0,1){0.54}}
			\put(11.25,4.4){\line(0,-1){1}} \put(11.25,3.4){\vector(-1,0){1.8}}
			\put(0.5,4.5){\line(0,1){0.8}} \put(0.5,5.3){\line(1,0){0.5}}
			\put(0.5,4.5){\line(1,0){0.5}} \put(1,5.3){\line(1,-1){0.4}}
			\put(1,4.5){\line(1,1){0.4}}
			\put(0.6,4.8){$\varphi_n$}
			\put(1.4,4.9){\line(1,0){0.6}}
			\put(2.5,4.8){$\varphi_1 \! = \! 0$}
			\put(3,5.4){\line(-2,-1){1}} \put(3,5.4){\line(2,-1){1}}
			\put(3,4.4){\line(2,1){1 }} \put(3,4.4 ){\line(-2,1){1 }}
			\put(5.2,4.8){$\varphi_2 \! = \! 0$}
			\put(5.75,5.4){\line(-2,-1){1}} \put(5.75,5.4){\line(2,-1){1}}
			\put(5.75,4.4){\line(2,1){1 }} \put(5.75,4.4 ){\line(-2,1){1}}
			\put(8.0,4.8){$\varphi_3 \! = \! 0$} \put(8.5,5.4){\line(-2,-1){1}}
			\put(8.5,5.4){\line(2,-1){1}} \put(8.5,4.4){\line(2,1){1 }}
			\put(8.5,4.4 ){\line(-2,1){1 }}
			\put(10.75,4.8){$\varphi_4 \! = \! 0$} \put(11.25,5.4){\line(-2,-1){1}}
			\put(11.25,5.4){\line(2,-1){1}} \put(11.25,4.4){\line(2,1){1 }}
			\put(11.25,4.4 ){\line(-2,1){1 }}
			\put(2.5,3.3){$\varphi_7 \! = \! 0$}
			\put(3,3.9){\line(-2,-1){1}} \put(3,3.9){\line(2,-1){1}}
			\put(3,2.9){\line(2,1){1 }} \put(3,2.9 ){\line(-2,1){1 }}
			\put(5.25,3.3){$\varphi_6 \! = \! 0$}
			\put(5.75,3.9){\line(-2,-1){1}} \put(5.75,3.9){\line(2,-1){1}}
			\put(5.75,2.9){\line(2,1){1 }} \put(5.75,2.9 ){\line(-2,1){1 }}
			\put(8.0,3.3){$\varphi_5 \! = \! 0$} \put(8.5,3.9){\line(-2,-1){1}}
			\put(8.5,3.9){\line(2,-1){1}} \put(8.5,2.9){\line(2,1){1 }}
			\put(8.5,2.9 ){\line(-2,1){1 }}
			\put(0.95,3.5){{\oval(1.2,0.8)}}
			\put(0.7,3.3){$\widehat{\textrm{C}7}$}
			\put(3,6.3){{\oval(1.2,0.8)}} \put(2.75,6.15){C$1$}
			\put(2.2,7.4){\framebox{$H = H_1$}}
			\put(5.75,6.3){{\oval(1.2,0.8)}} \put(5.5,6.15){C$2$}
			\put(4.95,7.4){\framebox{$H = H_2$}}
			\put(8.5,6.3){{\oval(1.2,0.8)}} \put(8.25,6.15){C$3$}
			\put(7.7,7.4){\framebox{$H = H_3$}}
			\put(11.25,6.3){{\oval(1.2,0.8)}} \put(11,6.15){C$4$}
			\put(10.45,7.4){\framebox{$H = H_4$}}
			\put(3,2){{\oval(1.2,0.8)}} \put(2.75,1.85){C$7$}
			\put(2.2,0.7){\framebox{$H = H_7$}}
			\put(5.75,2){{\oval(1.2,0.8)}} \put(5.5,1.85){C$6$}
			\put(4.95,0.7){\framebox{$H = H_6$}}
			\put(8.5,2){{\oval(1.2,0.8)}} \put(8.25,1.85){C$5$}
			\put(7.7,0.7){\framebox{$H = H_6$}}
			\put(4.1,5){\footnotesize{yes}}
			\put(6.8,5){\footnotesize{yes}}
			\put(9.6,5){\footnotesize{yes}}
			\put(9.9,3.5){\footnotesize{yes}}
			\put(7,3.5){\footnotesize{yes}}
			\put(4.3,3.5){\footnotesize{yes}}
			\put(1.66,3.55){\footnotesize{yes}}
			\put(3.1,5.5){\footnotesize{no}}
			\put(5.9,5.5){\footnotesize{no}}
			\put(8.6,5.5){\footnotesize{no}}
			\put(11.4,5.5){\footnotesize{no}}
			\put(8.6,2.65){\footnotesize{no}} 
			\put(5.9,2.65){\footnotesize{no}}
			\put(3.1,2.65){\footnotesize{no}}
		\end{picture} }
		\vspace{-7mm}
		\caption{Algorithm to distinguish the regular classes C$n$, $n \leqslant 7$ and the complementary family $\widehat{\rm C7}$ of the type N vacuum solutions.}
		\label{Fig-19}
		\end{figure}
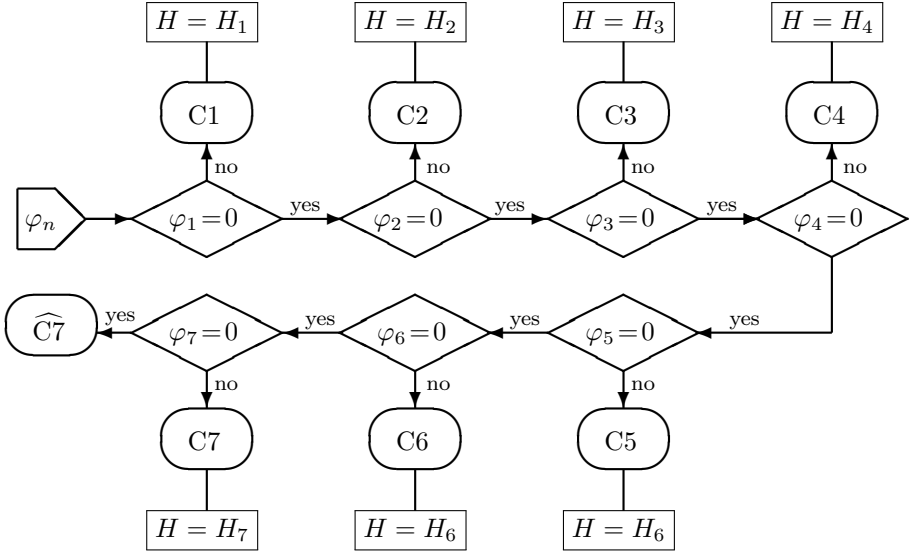
$\hspace{-5mm}$\textit{All the type} N \textit{spacetimes of class} C\textit{n,} $n = 1,...,7$\textit{, are regular, that is, they admit a} $R$\textit{-frame} $\{\ell, k, e_2, e_3\}$\textit{, which can be obtained as follows:}
		\begin{itemize}
			\item \textit{Classes} C1\textit{,} C2\textit{,} C3\textit{,} C4 \textit{and} C5
				\begin{equation} \label{null-frame-1}
					k = -\frac{(U^2 + \epsilon \, U)(v)}{2 (v, \ell)} , \quad \epsilon = \frac{U(v,\ell)}{(v, \ell)}, \quad e_2 = L(k) , \quad e_3 = - *\!L(k) ,
				\end{equation}
			\textit{where} $v$ \textit{is an arbitrary timelike vector and the t-volume} $U$ \textit{is given by}
				\begin{itemize}
					\item[$\circ$] $U = \frac{1}{\sqrt{2}}({\cal U} + \bar{{\cal U}})$\textit{, } $\, {\cal U} = \frac{{\cal F}}{\sqrt{-({\cal F}, {\cal F})}}$ \textit{with} ${\cal F} = {\cal F}_1$ \textit{for class} C1 \textit{and} ${\cal F} = {\cal F}_2$ \textit{if} $({\cal F}_2, {\cal F}_2) \neq 0$ \textit{or} ${\cal F} = {\cal F}_2^*$ \textit{if} $({\cal F}_2^*, {\cal F}_2^*) \neq 0$\textit{, for class} C2.
					\item[$\circ$] $\displaystyle{U = \frac{F_n}{\sqrt{- \, (F_n, F_n)}}}$ \textit{for classes} C3\textit{,} C4 \textit{and} C5. \\
				\end{itemize}
			\item \textit{Classes} C6 \textit{and} C7
				\begin{subequations} \label{null-frame-2}
					\begin{align}
						&k = -\frac{(U^2 + \epsilon U)(v)}{2 (v, \ell)} , \qquad \epsilon = \frac{U(v,\ell)}{(v, \ell)}, \qquad U = *(e_2 \wedge e_3) ,\\  
						&e_2 = \frac{b}{\sqrt{b^2}} , \qquad e_3 = \frac{\hat{c}}{\sqrt{\hat{c}^2}}, \qquad \hat{c} = c - \frac{(c,b)}{b^2}b ,
					\end{align}
				\end{subequations}
				\textit{with} $\{b, c\} = \{f, f^*\}$ \textit{for class} C6 \textit{and} $v$ \textit{an arbitrary timelike vector.}
		\end{itemize}
		\textit{The null frame} $\{\ell, k, e_2, e_3\}$ \textit{obtained in} (\ref{null-frame-1}) \textit{is oriented, and the one obtained in} (\ref{null-frame-2}) \textit{becomes oriented with the change} $e_3 \hookrightarrow \epsilon e_3$\textit{. From these null frames, we can determine an orthonormal} $R$\textit{-frame} $\{e_0, e_1, e_2, e_3\}$\textit{,} $\sqrt{2}\, e_0 = \ell + k , \ \sqrt{2}\, e_1 = \ell - k$\textit{, and obtain the associated connection tensor using} (\ref{defH})\textit{. Then, we can use the algorithm in Figure} \ref{Fig-18} \textit{to obtain the dimension of the corresponding isometry group.} \\ \\  
		\textit{The last end arrow in Figure} \ref{Fig-19} \textit{leads to the algorithm in Figure} \ref{Fig-20} \textit{for the family} $\widehat{\rm C7}$\textit{.}
\textit{All the type }N\textit{ vacuum solutions with non-vanishing cosmological constant in the classes} AR\textit{n,} $n = 1,2,3$\textit{, and} BR\textit{p, }$p = 1, 2$\textit{, are regular, that is, they admit a }$R$\textit{-frame }$\{\ell, k, e_2, e_3\}$\textit{, which can be obtained as in} (\ref{null-frame-2}) \textit{with:}
		\begin{itemize}
			\item \textit{Class} AR1\textit{,} $\{b, c\} = \{c, \textrm{d} \mu\}$ \textit{if} $\mu \neq 0$ \textit{or} $\{b, c\} = \{c, \textrm{d} \mu^*\}$ \textit{if} $\mu^* \neq 0$.
			\item \textit{Class} AR2, $\{b, c\} = \{c, \textrm{d} \psi_\ell\}$.
			\item \textit{Class} AR3, $\{b, c\} = \{c, \textrm{d} \chi\}$.
			\item \textit{Class} BR1, $\{b, c\} = \{b,q\}$.
			\item \textit{Class} BR2, $\{b, c\} = \{b, \textrm{d} \phi\}$.
		\end{itemize}
\textit{From these null frames we can determine an orthonormal} $R$\textit{-frame} $\{e_0, e_1, e_2, e_3\}$\textit{,} $\sqrt{2}\, e_0 = \ell + k , \ \sqrt{2}\, e_1 = \ell - k$\textit{, and obtain the associated connection tensor using} (\ref{defH})\textit{. Then, we can use the algorithm in the previous subsection to obtain the dimension of the corresponding isometry group.} \\ \\
		\textit{On the other hand, all the type N vacuum solutions with non-vanishing cosmological constant in the classes} AS\textit{, and }BS\textit{p,} $p = 1, 2, 3$\textit{, are singular, that is, they admit a non-trivial isotropy group.}
		\begin{itemize}
			\item \textit{Class} AS\textit{: the metric admits a} G$_4$ \textit{on} $3$\textit{-dimensional timelike orbits.}
			\item \textit{Class} BS1\textit{: the metric admits a} G$_3$ \textit{on} $2$\textit{-dimensional null orbits.}
			\item \textit{Class} BS2\textit{: the metric admits a} G$_4$ \textit{on} $3$\textit{-dimensional timelike orbits.}
			\item \textit{Class} BS3\textit{: the metric admits a} G$_5$ \textit{on} $4$\textit{-dimensional orbits.}
		\end{itemize}
		The implementation of the above algorithms in \textit{xIdeal} functions that automatise their application is a work in progress. Still, we now use \textit{xIdeal} to help us apply them to meaningful families of solutions as examples: the pp-waves, the Kundt waves and the Siklos metrics. For the three of them, we will obtain which class or classes Cn they belong to. Once the class is determined, for the pp-waves and the Kundt waves we will determine a $R$-frame and obtain the corresponding connection tensors. After that, we could apply the \textit{xIdeal} function \texttt{IsometryGroupDimension} to determine the dimension of the isometry group for the most general cases. However, such \textit{xIdeal} function works faster for expressions where the dependence of the metric functions on the space-time coordinates is more explicit. Thus, for the pp-waves we will apply it to known particular subfamilies, while for the Kundt waves we will simply prove, with the help of \textit{xIdeal}, that they admit, at most, a G$_2$ as it is already known. For the Siklos solutions, instead, we will apply the algorithm in Figure \ref{Fig-20} to the particular solutions belonging to the $\widehat{\rm C7}$ family. All these computations can be found in a \textit{Mathematica} notebook in \cite{S-GitHub}.
		
			\subsubsection{The pp-waves with trivial isotropy group}
			For vacuum and Maxwell null or radiation fields, the metric of the pp-waves takes the following expression \cite{Kramer}:
			\begin{equation} \label{metric-pp}
				\textrm{d} s^2 = \textrm{d} x^2 + \textrm{d} y^2 - 2 \textrm{d} u \textrm{d} v - 2h(u,x,y) \textrm{d} u^2.
			\end{equation}
In the vacuum case, $h(u,x,y) = f(u, \zeta) + \bar{f}(u, \bar{\zeta})$, where $f$ is an analytical complex function and $\zeta = x + \textrm{i} y$, that is, $h_{xx} + h_{yy} = 0$. The symmetries of the vacuum pp-waves can be found in the Ehlers and Kundt review \cite{Ehlers-Kundt} (see also \cite{Kramer}), and the study for the non-vacuum case was achieved in \cite{Sippel-Goenner-1986}. \\ \\
			Invariant analyses of the vacuum pp-waves have been addressed with different methods. It is worth mentioning the broad Cartan-Karlhede approach by Milson {\it et al.} \cite{Coley-JMP}, the scalar differential invariants procedure \cite{Kruglikov}, or the recent IDEAL characterisation by Khavkine {\it et al.} \cite{Igor-2024}. Here, we apply our algorithms and compare our results with the already known ones. \\ \\ \\
			If we apply the algorithm in Figure \ref{Fig-19} to the metric (\ref{metric-pp}), we obtain that, generically, it belongs to class C7. Only in a degenerate subclass we have $\varphi_7 = 0$, which leads to the plane waves. As already stated above, the class OS1 and the two subclasses OS2a and OS2b match with the three classes of plane waves considered in the literature \cite{Kramer, Ehlers-Kundt}. A Cartan-Karlhede approach to the plane waves can be found in \cite{Coley-2012}. \\ \\ 
			Now, we focus on the pp-waves in class C7, that is, those with a trivial isotropy group. We can determine a $R$-frame as in (\ref{null-frame-2}) and obtain the corresponding connection tensor $H_7$. The \textit{xIdeal} function \texttt{IsometryGroupDimension} enables us to determine the symmetries of a specific pp-wave solution. Note that the pp-waves fulfil $\textrm{d} b = 0$, and thus there is a Killing vector collinear with $\ell$ (see \cite{SMF-DimensioTN} for more details). Consequently, at least a G$_1$ exists and there are no space-like orbits. \\ \\
			The expression of the function $f(u, \zeta)$ corresponding to the different isometry groups is known \cite{Kramer, Ehlers-Kundt}. We apply our approach to each of these metrics and the results are summarised in Table \ref{table-9}. It is worth mentioning that, for the metrics in rows 6 and 7, we actually applied our approach to two particular cases. In the column `Invariant conditions' we find, on the left, the condition in Figure \ref{Fig-18} determining the dimension of the group that each metric fulfils, where we have denoted by ${\cal G}_r(H)$ the set of conditions in Figure \ref{Fig-18} leading to the group G$_r$. On the right, we find the invariant conditions that complete the properties of the group. \\ \\
			\begin{table*}[t]
				$\begin{array}{llcll}
					\noalign{\hrule height 1.05pt} \\[-3mm]
					f(u, \zeta) & \quad {\rm Invariant} & \!\!\! {\rm conditions} & \quad {\rm Group} \ {\rm G}_r & \ {\rm O}_r \ {\rm Orbits} \\ 
					\hline \\[-2mm]
					e^{\textrm{i} \alpha} \zeta^{2 \textrm{i} k} & \quad \; \, {\cal G}_3(H) & Z = 0 & \quad {\rm G_3} \quad {\rm BI} & \ {\rm T}_3 \ \ [\tilde{m}^2 \! > \! 0] \\[0.5mm]
					e^{2 k \zeta} & \quad \; \, {\cal G}_3(H) & {\cal B}_1(Z) & \quad {\rm G_3} \quad {\rm BIII} & \ {\rm T}_3 \ \ [\tilde{m}^2 \! > \! 0] \\[0.5mm] 
					\ln \zeta & \quad \; \, {\cal G}_3(H) & {\cal B}_2(Z) & \quad {\rm G_3} \quad {\rm BVI_{0}} & \ {\rm T}_3 \ \ [\tilde{m}^2 \! > \! 0] \\[0.5mm]
					a u^{-2} \ln \zeta & \quad \; \, {\cal G}_3(H) & {\cal C}_2(Z) & \quad {\rm G_3} \quad {\rm BVI_{0}} & \ {\rm T}_3 \ \ [\tilde{m}^2 \! > \! 0] \\[0.5mm]
					A(u) \ln \zeta & \quad \; \, {\cal G}_{2a}(H) & (A, H) \! = \! 0 & \quad {\rm G_2} \quad \textsf{Abelian} & \ {\rm N}_2 \ \ [(\! A, \! A \!) \! = \! 0] \\[0.5mm]
					f(\zeta e^{\textrm{i} k u}) & \quad \; \, {\cal G}_{2a}(H) & (A, H) \! = \! 0 & \quad {\rm G_2} \quad \textsf{Abelian} & \ {\rm T}_2 \ \ [(\! A, \! A \!) \! < \! 0] \\[0.5mm] 
					u^{-2}A(\zeta u^{\textrm{i} k}) & \quad \; \, {\cal G}_{2a}(H) & (A, H) \! \neq \! 0 & \quad {\rm G_2} \quad \textsf{non-Abelian} & \ {\rm T}_2 \ \ [(\! A, \! A \!) \! < \! 0] \\[0.5mm]
					f(\zeta,u) & \quad & \quad & \quad {\rm G_1} \quad & \ {\rm N}_1 \\
					\noalign{\hrule height 1.05pt}
				\end{array}$
				\caption{This Table summarises the analysis of the regular pp-waves by using our IDEAL approach.}
				\label{table-9}
				\vspace{-4mm}
			\end{table*}
$\hspace{-2mm}$For the G$_3$, we can determine the Bianchi type as pointed out in \cite{SFM-simetries}. For any vector $v$ and any 2-form $V$, the vector $\tilde{m}_\lambda = C^{[1]}_{\lambda \rho \mu \nu}v^\rho V^{\mu \nu} $, defines the (spacelike) direction orthogonal to the orbits. Then, the induced metric on the orbits is $\tilde{\gamma} = g - n \otimes n$, with $n = |\tilde{m}^2|^{-1/2} \tilde{m}$. Moreover, 
			\begin{equation} \label{Z}
				Z_{\alpha \beta} = - \frac{1}{2}\tilde{\gamma}^{\lambda}_{\alpha} \, {H_{\lambda}}^{\mu \nu} \eta_{\mu \nu \beta \rho} \, n^{\rho} .
			\end{equation}
is the structure tensor associated to the isometry group, and we can apply the algorithm to determine the Bianchi type presented in \cite{FS-L3}. Actually, this algorithm is the same one as in Figure \ref{Fig-24} but changing $N$ by $\hat{N} \equiv \hat{\gamma} \cdot N$ in equations (\ref{nu}$-$\ref{nu23}), with $\hat{\gamma} \equiv \tilde{\gamma} + 2 \, \hat{v} \otimes \hat{v}$ a subsidiary Riemannian metric and $\hat{v}$ an arbitrary, unitary, timelike vector. In Table \ref{table-9}, ${\cal B}_1(Z)$ and ${\cal B}_2(Z)$ denote the invariant conditions in Figure \ref{Fig-24} that lead to Bianchi types BIII and BVI$_0$, respectively. The two classes with this last Bianchi type can be discriminated by the invariant condition $b^2 = 0$ ($b^2 \neq 0$), where $b$ is given in (\ref{B-b}). The invariant $\tilde{m}$ is a space-like vector and, therefore, the orbits are time-like. \\ \\
			For the G$_2$, the connection tensor fulfils conditions ${\cal G}_{2a}(H)$ leading to G$_{2a}$ in Figure \ref{Fig-18}. Then the commutative character can be expressed with the invariant condition $(A, H) = 0$, where $(A,H)$ denotes the contraction of the 2-form $A$ with the two first indices of $H$, and $A$ is $A_{\alpha \beta} = \eta(C^{[1]}, C^{[1]})_{\alpha \beta \lambda \mu \nu \rho \sigma \tau} v_1^\lambda V_1^{\mu \nu} v_2^\rho V_2^{\sigma \tau}$, $v_1, v_2$ being two arbitrary vectors and $V_1, V_2$ two arbitrary 2-forms \cite{SFM-simetries}. Moreover, the causal character of the orbits depends on the sign of $(A,A)$. \\ \\
			For any other function $f(\zeta, u)$, we know that, at least, a G$_1$ exists. Thus, necessarily one of the sets conditions ${\cal G}_{1a}(H)$, ${\cal G}_{1b}(H)$, ${\cal G}_{1c}(H)$ or ${\cal G}_{1d}(H)$ leading to G$_{1a}$, G$_{1b}$, G$_{1c}$ or G$_{1d}$ in Figure \ref{Fig-18} holds. This fact can be tested by considering specific choices of the function $f$ not belonging to the previously considered subfamilies.
			
			\subsubsection{On the Kundt waves}
			In his original paper, Kundt \cite{Kundt} considers a family of vacuum and Maxwell null or radiation field solutions whose metric line element takes the expression \cite{Coley-2013, Kruglikov}:
			\begin{equation} \label{metric-Kundtwaves}
				\textrm{d} s^2 = \textrm{d} x^2 + \textrm{d} y^2 - \textrm{d} u \textrm{d} v + \frac{2v}{x} \textrm{d} u \textrm{d} x - [8 x h(u, x, y) - \frac{v^2}{4x^2}] \textrm{d} u^2.
			\end{equation}
In the vacuum case, $h(u, x, y)$ fulfils the partial differential equation $h_{xx} + h_{yy} = 0$. These solutions have been analysed using the Cartan-Karlhede and the scalar differential invariant approaches \cite{Coley-2013, Kruglikov}. Moreover, it is known that they can admit two, one or no symmetries \cite{Coley-2013}. \\ \\
			Here, we do not follow our approach to studying the symmetries in detail. However, we want to comment on two direct results. On the one hand, if we apply our algorithm given in Figure \ref{Fig-19} to the metric (\ref{metric-Kundtwaves}), we obtain $\varphi_i = 0$ for $i < 5$ and  $\varphi_5 = - (4 x^4)^{-1} \neq 0$, and therefore it belongs to class C5. Consequently, all the Kundt waves have a trivial isotropy group, and the study of their isometry group should be addressed using the connection tensor $H$ associated with the $R$-frame obtained as in (\ref{null-frame-1}). \\ \\
			On the other hand, a simple reasoning shows that the Kundt waves admit, at most, a G$_2$. Indeed, $\varphi_5$ is a non-constant scalar invariant. If either $\alpha$ or $\alpha^*$ (given in (\ref{alpha})) determine a new independent invariant scalar, then we have, at most, a $G_2$. Otherwise, we have that $\textrm{d} \alpha \wedge \textrm{d} \varphi_5 = \textrm{d} \alpha^* \wedge \textrm{d} \varphi_5 = 0$. But these conditions lead to $h = \alpha_1(u) e^{k_1 y} \sin(k_1 x + k_2) + \alpha_2(u) x + \alpha_3(u) y + \alpha_4(u)$, with $k_i$ arbitrary constants. \\ \\ 
			If we apply the \textit{xIdeal} function \texttt{IsometryGroupDimension} to such metric, we obtain that, generically, it has no symmetries. Moreover, for this metric $\eta(C^{[1]}, C^{[1]}) \neq 0$, and consequently a G$_3$ cannot exist. If we take $\alpha_2(u) = \alpha_3(u) = \alpha_4(u) = 0$ and $\alpha_1(u) = k_0$, \texttt{IsometryGroupDimension} determines that a G$_2$ exists, in agreement with the results in \cite{Kruglikov}.
			
			\subsubsection{On the Siklos solutions}
			The Siklos metrics \cite{Siklos} are vacuum and Maxwell null or radiation field solutions with negative cosmological constant $\Lambda$. Their metric line element takes the expression \cite{Siklos, Podolsky-98}:
			\begin{equation} \label{metric-Siklos}
				\textrm{d} s^2 = -\frac{\Lambda}{x^2}[\textrm{d} x^2 + \textrm{d} y^2 - 2 \textrm{d} u \textrm{d} v - 2 h(u,x,y) \textrm{d} u^2].
			\end{equation}
In the vacuum case $h(u, x, y)$ fulfils the partial differential equation $h_{xx} + h_{yy} = 2 h_x/x$. These solutions were analysed in \cite{Podolsky-98}, and their symmetries were studied in the Siklos paper \cite{Siklos} and recently revisited in \cite{Calvaruso}. These metrics turn out to be a subclass of the family of Kundt metrics obtained in \cite{Garcia-Plebanski} (see also \cite{Ortaggio}). \\ \\ 
			If we apply the algorithm given in Figure \ref{Fig-19} to the metric (\ref{metric-Siklos}), we obtain that, in general, it belongs to class C7. However, some particular subfamilies belong to the family $\widehat{\rm C7}$. For the general case, belonging to class C7, we can determine a $R$-frame as in (\ref{null-frame-2}) and obtain the corresponding connection tensor $H_7$. The \textit{xIdeal} function \texttt{IsometryGroupDimension} enables us to determine the symmetries of a specific Siklos solution. Note that in this case $\textrm{d} b = 0$, and thus there is a Killing vector collinear with $\ell$ \cite{SMF-DimensioTN}. Consequently, at least a G$_1$ exists and there are no space-like orbits. A detailed analysis, such as that done for pp-waves in the previous subsection, requires further work that goes beyond the scope of this thesis. Now, we only point out that the cases 0, 1, 2, 3, 5 and 6 considered in \cite{Calvaruso} have $\varphi_7 \neq 0$ and belong to class C7. \\ \\
			The application of the algorithm given in Figure \ref{Fig-20} to the cases in the family $\widehat{\rm C7}$ shows that only our singular classes in the branches A and B are possible: cases 4, 7 ($\equiv$ 11), 9 and 10 in \cite{Calvaruso} correspond to our classes BS1, BS3, BS2 and AS, respectively.
	
	\section{IDEAL labelling of the spatially-homogeneous \mbox{cosmologies}} \label{sec-sim-Bianchi}
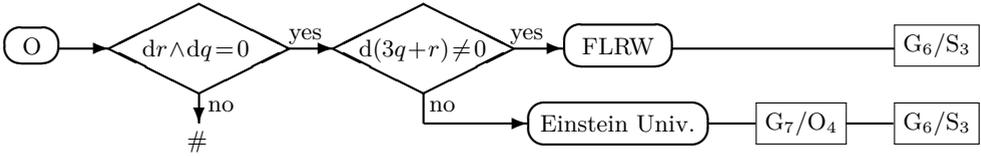
\begin{figure}[H]
		\setlength{\unitlength}{0.79cm} {\small \noindent
		\begin{picture}(18,3.2)
		\thicklines
			\put(0.4,1.9){{\oval(0.9,0.6)}} \put(0.25,1.75){O}
			\put(0.85,1.85){\vector(1,0){0.85}}
			\put(3.2,2.6){\line(-2,-1){1.5}} \put(3.2,2.6){\line(2,-1){1.5}}
			\put(3.2,1.1){\line(2,1){1.5}} \put(3.2,1.1){\line(-2,1){1.5}}
			\put(2.25,1.72){$\textrm{d} r \! \wedge \! \textrm{d} q \! = \! 0$}
			\put(4.65,1.85){\vector(1,0){0.8}}
			\put(6.95,2.6){\line(-2,-1){1.5}} \put(6.95,2.6){\line(2,-1){1.5}}
			\put(6.95,1.1){\line(2,1){1.5}} \put(6.95,1.1){\line(-2,1){1.5}}
			\put(5.87,1.72){$\textrm{d} (3q \! + \! r ) \! \neq \! 0$}
			\put(8.45,1.85){\vector(1,0){0.83}}
			\put(10.2,0.6){\oval(3,0.7)} \put(8.9,0.46){Einstein Univ.}
			\put(10.2,1.9){\oval(1.8,0.7)}\put(9.6,1.75){FLRW}
			\put(12.5,0.5){\framebox{G$_7$/O$_4$}}
			\put(14.81,0.5){\framebox{G$_6$/S$_3$}}
			\put(14.81,1.8){\framebox{G$_6$/S$_3$}}
			\put(6.95,1.1 ){\line(0,-1){0.5}}
			\put(6.95,0.6){\vector(1,0){1.74}}
			\put(3.2,1.1){\vector(0,-1){0.5}} \put(3, 0.2){\#}
			\put(11.1,1.85){\line(1,0){3.72}}
			\put(11.7,0.6){\line(1,0){0.8}} \put(14,0.6){\line(1,0){0.8}}
			\put(3.35,0.8){no}
			\put(7.05,0.8){no}
			\put(8.4,2){yes}
			\put(4.7,2){yes}
		\end{picture} }
		\vspace{-5mm}
		\caption{Algorithm characterising the spatially-homogeneous cosmologies of Petrov-Bel type O.}
		\label{Fig-21}
		\end{figure}

\begin{figure}
		\setlength{\unitlength}{0.79cm} {\small \noindent
		\begin{picture}(18,15)
		\thicklines
			\put(3,16.15){{\oval(1.2,0.8) }}\put(2.86,16){{\bf{D}}}
			\put(3.6,16.12){\vector(1,0){1.2}}
			\put(6,16.75){\line(-2,-1){1.25}} \put(6,16.75){\line(2,-1){1.25}}
			\put(6,15.5){\line(2,1){1.25}} \put(6,15.5){\line(-2,1){1.25}}
			\put(5.4,16 ){$\delta_1 = 0$}
			\put(6,15){\line(-3,-1){2.15}} \put(6,15){\line(3,-1){2.15}}
			\put(6,13.55){\line(3,1){2.15}} \put(6,13.55){\line(-3,1){2.15}}
			\put(4.52,14.14){$\textrm{d} r \! = \textrm{d} q \! = \textrm{d} \psi \! = \! 0$}
			\put(6,13){\line(-2,-1){1.25}} \put(6,13){\line(2,-1){1.25}}
			\put(6,11.75){\line(2,1){1.25}} \put(6,11.75){\line(-2,1){1.25}}
			\put(5.3,12.25 ){$E_2 = 0$}
			\put(1.5,12.3){\framebox{$H \! = \! H_2$}} \put(-0.4,12.3){Fig.$\!$
			\ref{Fig-23}}
			\put(6,11.2){\line(-2,-1){1.25}} \put(6,11.2){\line(2,-1){1.25}}
			\put(6,9.95){\line(2,1){1.25}} \put(6,9.95){\line(-2,1){1.25}}
			\put(5.4,10.45 ){$\delta_\sigma = 0$}
			\put(1.5,10.5){\framebox{$H \! = \! H_\sigma$}} \put(-0.4,10.5){Fig.$\!$
			\ref{Fig-23}}
			\put(6,9.4){\line(-2,-1){1.25}} \put(6,9.4){\line(2,-1){1.25}}
			\put(6,8.15){\line(2,1){1.25}} \put(6,8.15){\line(-2,1){1.25}}
			\put(5.32,8.65 ){$E_3 = 0$}
			\put(1.5,8.7){\framebox{$H \! = \! H_3$}} \put(-0.4,8.7){Fig.$\!$
			\ref{Fig-23}}
			\put(5.2,7.2){\framebox{G$_5$/O$_4$}}
			\put(6,6.65){\line(-2,-1){1.25}} \put(6,6.65){\line(2,-1){1.25}}
			\put(6,5.4 ){\line(2,1){1.25}} \put(6,5.4){\line(-2,1){1.25}}
			\put(5.5,5.88 ){$s = 0$}
			\put(2.5,5.85 ){\framebox{Gödel}}
			\put(-0.4,5.85){\framebox{B III/S$_3$}}
			\put(1.7,4.7 ){\framebox{\,KCKS, \,$k = \! 0 \! = \theta \,$\phantom{$a^b$\hspace*{-3.7mm}}}}
			\put(-0.4,4.7){\framebox{B I/S$_3$}}
			\put(10,14.87){\line(-2,-1){1.25}} \put(10,14.87){\line(2,-1){1.25}}
			\put(10,13.62){\line(2,1){1.25}} \put(10,13.62){\line(-2,1){1.25}}
			\put(9.15,14.14 ){Eq. $\!$(\ref{mu})}
			\put(10,13){\line(-2,-1){1.25}} \put(10,13){\line(2,-1){1.25}}
			\put(10,11.75){\line(2,1){1.25}} \put(10,11.75){\line(-2,1){1.25}}
			\put(9.45,12.25 ){$\delta_4 = 0$}
			\put(12.8,12.3){\framebox{$H = H_4$}} \put(15.5,12.3){Fig.$\!$
			\ref{Fig-23}}
			\put(12.8,16){\framebox{$H = H_1$}} \put(15.5,16){Fig.$\!$ \ref{Fig-23}}
			\put(16.34,14.1){\#}
			\put(16.34,10.5){\#}
			\put(12.8,8.7){\framebox{$H = H_{\tilde{\sigma}}$}}
			\put(15.5,8.7){Fig.$\!$ \ref{Fig-23}}
			\put(12.8,6.86){\framebox{$H = H_5$}} \put(15.5,6.86){Fig.$\!$
			\ref{Fig-23}}
			\put(10,11.2){\line(-2,-1){1.25}} \put(10,11.2){\line(2,-1){1.25}}
			\put(10,9.95){\line(2,1){1.25}} \put(10,9.95){\line(-2,1){1.25}}
			\put(9.4,10.45 ){$\Upsilon = 0$}
			\put(10,9.4){\line(-2,-1){1.25}} \put(10,9.4){\line(2,-1){1.25}}
			\put(10,8.15){\line(2,1){1.25}} \put(10,8.15){\line(-2,1){1.25}}
			\put(9.4,8.6 ){$\tilde{\delta}_{\sigma} = 0$}
			\put(10,7.6){\line(-2,-1){1.25}} \put(10,7.6){\line(2,-1){1.25}}
			\put(10,6.35){\line(2,1){1.25}} \put(10,6.35){\line(-2,1){1.25}}
			\put(9.4,6.85 ){$E_5 = 0$}
			\put(10,5.75){\line(-2,-1){1.25}} \put(10,5.75){\line(2,-1){1.25}}
			\put(10,4.5){\line(2,1){1.25}} \put(10,4.5){\line(-2,1){1.25}}
			\put(9.15,5.02 ){Eq.$\!$ (\ref{lambda})}
			\put(11.2,5.13){\vector(1,0){4.8}} \put(16.34, 5){\#}
			\put(9.3,3.5){\framebox{G$_4$/S$_3$}}
			\put(10,2.75){\line(-2,-1){1.25}} \put(10,2.75){\line(2,-1){1.25}}
			\put(10,1.5){\line(2,1){1.25}} \put(10,1.5){\line(-2,1){1.25}}
			\put(9.4,2){$w_1 = 0$}
			\put(12.2,2.13){\line(1,0){2.8}} \put(11.2,2.13){\vector(1,0){1}}
			\put(12.2,0.95){\line(1,0){2.8}}
			\put(12.2,3.35){\line(1,0){2.5}}
			\put(12.2,0.95){\line(0,1){2.4}}
			\put(15,2.05){\framebox{\ B II/S$_3$}}
			\put(15,0.85){\framebox{B IX/S$_3$}}
			\put(14.7, 3.25){\framebox{B VIII/S$_3$}}
			\put(13.2,1.15){$\zeta > 0$}
			\put(13.2,2.35){$\zeta = 0$}
			\put(13.2,3.55){$\zeta < 0$}
			\put(-0.4,2.65){\framebox{B III/S$_3$}}
			\put(-0.4,1.45){\framebox{B VII$_0$/S$_3$}}
			\put(3.2,2.15){\line(1,0){0.45}}
			\put(5.77,2.13){\vector(-1,0){0.65}}
			\put(8.77,2.13){\vector(-1,0){0.55}}
			\put(3.65,2.05){\framebox{KCKS}}
			\put(3.2,1.45){\line(0,1){1.4}}
			\put(3.2,2.85){\line(-1,0){1.76}}
			\put(3.2,1.45){\line(-1,0){1.42}}
			\put(2,3.01){$\zeta < 0$}
			\put(2,1.65){$\zeta = 0$}
			\put(7,2.75){\line(-2,-1){1.25}} \put(7,2.75){\line(2,-1){1.25}}
			\put(7,1.5){\line(2,1){1.25}} \put(7,1.5){\line(-2,1){1.25}}
			\put(6.45,2.0){$s_1 = 0$}
			\put(7,1.5){\vector(0,-1){0.6}}
			\put(6.2,0.45){\framebox{B V/S$_3$}}
			\put(6,15.5){\vector(0,-1){0.5}}
			\put(6,13.55){\vector(0,-1){0.55}}
			\put(6,11.78){\vector(0,-1){0.6}}
			\put(6,9.95 ){\vector(0,-1){0.55}}
			\put(6,8.16 ){\vector(0,-1){0.5}}
			\put(6,6.98 ){\line(0,-1){0.33}}
			\put(6,5.4 ){\line(0,-1){0.6}}
			\put(6,4.8){\vector(-1,0){0.85}} \put(1.7,4.8){\line(-1,0){0.6}}
			\put(4.8,6.02){\vector(-1,0){0.94}} \put(2.5,6){\line(-1,0){1.06}}
			\put(7.2,16.12){\vector(1,0){5.6}}
			\put(14.55,16.12){\line(1,0){0.6}}
			\put(8.1,14.27){\vector(1,0){0.68}}
			\put(11.23,14.24){\vector(1,0){4.8}}
			\put(11.2,12.38){\vector(1,0){1.6}}
			\put(14.55,12.38){\line(1,0){0.6}}
			\put(4.75,12.38){\vector(-1,0){1.6}}
			\put(1.5,12.38){\line(-1,0){0.4}}
			\put(4.75,10.58){\vector(-1,0){1.6}}
			\put(1.5,10.58){\line(-1,0){0.4}}
			\put(11.23,10.58){\vector(1,0){4.8}}
			\put(4.75,8.78){\vector(-1,0){1.6}} \put(1.5,8.78){\line(-1,0){0.4}}
			\put(11.2,8.78){\vector(1,0){1.6}} \put(14.6,8.78){\line(1,0){0.6}}
			\put(11.2,6.98){\vector(1,0){1.6}} \put(14.55,6.98){\line(1,0){0.65}}
			\put(10,13.62){\vector(0,-1){0.6}}
			\put(10,11.78){\vector(0,-1){0.6}}
			\put(10,9.95 ){\vector(0,-1){0.55}}
			\put(10,8.16 ){\vector(0,-1){0.57}}
			\put(10,6.38 ){\vector(0,-1){0.63}}
			\put(10,4.5 ){\vector(0,-1){0.55}}
			\put(10,3.27){\line(0,-1){0.51}}
			\put(8.1,16.3){no} 
			\put(8.1,14.5){no}
			\put(11.5,14.4){no} 
			\put(11.5,12.6){no}
			\put(11.5,10.8){no} 
			\put(11.5,9){no}
			\put(11.5,7.2){no}
			\put(11.3,2.35){no} 
			\put(8.3,2.35){yes}
			\put(5.3,2.35){yes} 
			\put(7.2,1.1){no}
			\put(4.2,12.6){no} 
			\put(4.2,10.8){no} 
			\put(4.2,9){no}
			\put(11.5,7.2){no}
			\put(4.2,6.2){no}
			\put(6.2,15.2){yes}
			\put(6.2,13.2){yes} 
			\put(10.2,13.2){yes}
			\put(6.2,11.4){yes} 
			\put(10.2,11.4){yes}
			\put(6.2,9.6){yes} 
			\put(10.2,9.6){yes}
			\put(6.2,7.85){yes} 
			\put(6.2,5){yes}
			\put(10.2,7.8){yes}
			\put(10.2,6){yes}
			\put(11.5,5.3){no}
			\put(10.2,4.2){yes}
		\end{picture} }
		\vspace{-4mm}
		\caption{Algorithm that determines the connection tensor for the regular type D cosmologies. Moreover, for singular type D cosmologies, it determines the dimension of the isometry group and distinguishes the Bianchi type of the cases when a G$_3$ acts on three-dimensional spacelike orbits.}
		\label{Fig-22}
		\end{figure}
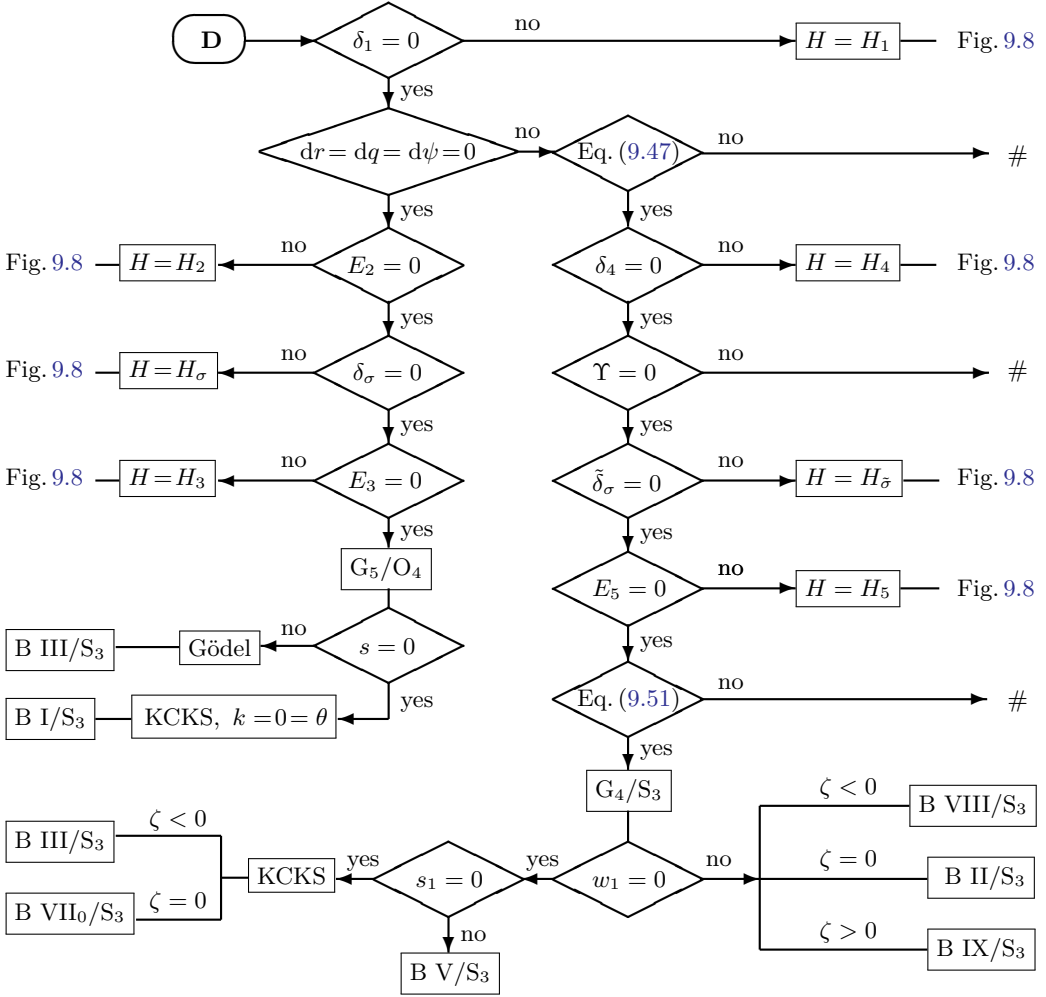

\begin{figure}
		\setlength{\unitlength}{0.79cm} {\small \noindent
		\begin{picture}(18,8)
		\thicklines
			\put(1.35,7){$H$} \put(1.5,6.65){\vector(0,-1){0.54}}
			\put(2.1,6.95){\line(-2,-1){0.6}} \put(0.9,6.95){\line(2,-1){0.6}}
			\put(2.1,6.95){\line(0,1){0.6}} \put(0.9,6.95){\line(0,1){0.6}}
			\put(0.9,7.55){\line(1,0){1.2}}
			\put(1.5,6.1){\line(-2,-1){1.25}} \put(1.5,6.1){\line(2,-1){1.25}}
			\put(1.5,4.85){\line(2,1){1.25}} \put(1.5,4.85){\line(-2,1){1.25}}
			\put(0.95,5.35){${\cal G}_4(H)$} \put(1.5,4.85){\vector(0,-1){1.5}}
			\put(4.15,5.4){{\framebox{G$_4$/O$_4$}}}
			\put(7.5,6.1){\line(-2,-1){1.25}} \put(7.5,6.1){\line(2,-1){1.25}}
			\put(7.5,4.85){\line(2,1){1.25}} \put(7.5,4.85){\line(-2,1){1.25}}
			\put(6.7,5.35){$\, \textrm{tr} H \! = \! 0$} \put(7.5,4.85){\vector(0,-1){0.35}}
			\put(11.25,6.24){\line(-2,-1){1.5}}
			\put(11.25,6.24){\line(2,-1){1.5}}
			\put(11.25,4.74){\line(2,1){1.5}}
			\put(11.25,4.74){\line(-2,1){1.5}} \put(10.2,5.4){$r \! > \! 0, \; q \! > \! 0$}
			\put(6.05,3.76){\vector(-1,0){0.75}} \put(4.73,3.65){\#}
			\put(8.7,5.48){\vector(1,0){1.07}}
			\put(12.7,5.48){\vector(1,0){1.7}}
			\put(11.25,6.88){\vector(1,0){2}}
			\put(11.25,6.25){\line(0,1){0.64}}
			\put(13.4,6.8){\#}
			\put(9,3.75){\vector(1,0){5.3}}
			\put(7.5,4.5){\line(-2,-1){1.5}} \put(7.5,4.5){\line(2,-1){1.5}}
			\put(7.5,3){\line(2,1){1.5}} \put(7.5,3){\line(-2,1){1.5}}
			\put(6.6,3.63){$\tilde{\beta} \! \geq \! 0, q \! > \! 0$}
			\put(14.3,3.66){\framebox{B VI$_h$/S$_3$}}
			\put(12.8,5.7){yes} 
			\put(11.4,6.36){no}
			\put(9.2,4){yes}
			\put(5.67,4){no}
			\put(14.4,5.4){{\framebox{B IX\,/S$_3$}}}
			\put(1.5,3.35){\line(-2,-1){1.75}} \put(1.5,3.35){\line(2,-1){1.75}}
			\put(1.5,1.6){\line(2,1){1.75}} \put(1.5,1.6){\line(-2,1){1.75}}
			\put(0.3,2.35){${\cal G}_3(H) , \, m^2 \! \! < \! 0$}
			\put(1.5,1.6){\vector(0,-1){0.4}} \put(1.33,0.75){\#}
			\put(4.15,2.4){{\framebox{G$_3$/S$_3$}}}
			\put(5.56,2.5){\line(1,0){0.4}}
			\put(6.1,2.35){Fig.$\!$ \ref{Fig-24}}
			\put(3.25,2.48){\vector(1,0){0.9}}
			\put(5.66,5.48){\vector(1,0){0.64}}
			\put(2.69,5.48){\vector(1,0){1.48}}
			\put(2.8,5.7){yes}
			\put(8.8,5.7){yes}
			\put(1.65,4.1){no} 
			\put(1.7,1.4){no} 
			\put(3.3,2.75){yes}
		\end{picture} }
		\vspace{-7mm}
		\caption{Algorithm to characterise the regular spatially-homogeneous cosmologies once the main connection tensor $H$ is known.}
		\label{Fig-23}
		\end{figure}
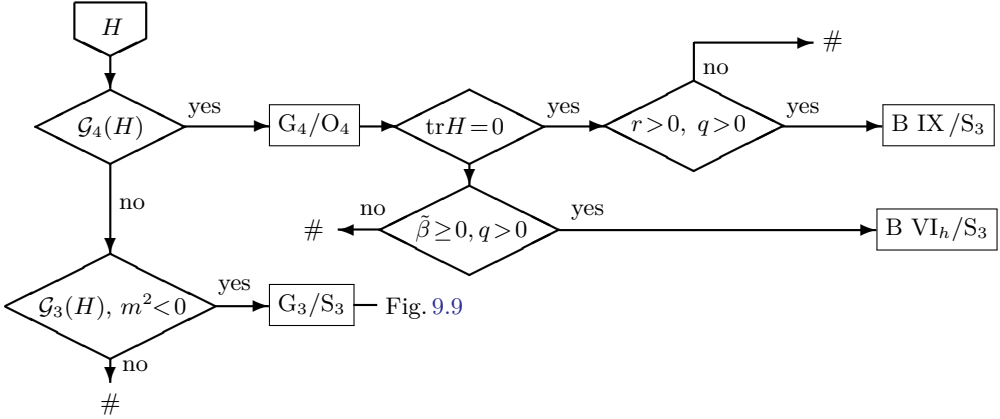

	In this section, we deal with the IDEAL characterisation of the Bianchi cosmological models and the other spatially-homogeneous cosmologies achieved in \cite{SFM-simetries-Bianchi}. This study requires considering different classes of perfect fluid solutions and, for each class, the necessary and sufficient conditions for a solution to define a spatially-homogeneous cosmology. \\ \\[-1mm]
	In the previous section, we have presented a first necessary step for implementing this approach: the IDEAL labelling of the spacetimes admitting a $R$-frame with a maximal group of isometries of dimensions three or four. These results are the starting point of the work carried out in \cite{SFM-simetries-Bianchi}, which we summarise here, enabling us to study the spatially-homogeneous cosmologies admitting a $R$-frame. \\ \\[-1mm]
	Once we know that the spacetime is a spatially-homogeneous perfect fluid solution, a second step consists in determining the Bianchi type when a G$_3$ acting on a S$_3$ exists \cite{Bianchi}. For this task, we make use of the previous results obtained in \cite{FS-G3}, where the different three-dimensional homogeneous Riemannian spaces have been intrinsically characterised. \\ \\[-1mm]
	On the one hand, if we want to label the spatially-homogeneous cosmologies, we must use the invariant characterisation of the perfect fluid solutions (\ref{fluper-definitions}$-$\ref{fluper-hydro-variables}). On the other hand, in spacetimes of Petrov-Bel types I, II or III, the Weyl tensor algebraically defines a principal frame. In (\ref{H-Hautodual}$-$\ref{L_+-U-TIII}), we have presented the connection tensor $H$ for each of those Petrov-Bel types from some specific concomitants of the Weyl tensor. Now, the cosmological observer $u$ can be used to obtain $H$ for a Petrov-Bel type N spacetime, and to get an alternative method for types II and III: \\ \\[-2mm]
	\textbf{Perfect fluid type II, III and N connection tensor.} \textit{The connection tensor} $H$ \textit{of a perfect fluid type II, III or N spacetime can be obtained as follows: \\[-3mm]} 
	\begin{subequations} \label{H-perflu-TII-TIII-TN}
		\begin{equation} \label{H-perflu-TII-TIII-TN-1}
			H = I - J , \qquad I_{\alpha \gamma \lambda} \equiv - \nabla_{\alpha} {P^{\rho}}_{[\gamma} P_{\lambda] \rho} \, ,
		\end{equation}
		\begin{equation} \label{H-perflu-TII-TIII-TN-2}
			J_{\alpha \gamma \lambda} \equiv \nabla_\alpha E^{\mu}{}_{\rho} \, E^{\rho \nu} \, \eta_{\mu \nu \pi \delta} K^{\pi \sigma} \eta_{\sigma \gamma \lambda \beta} P^{\delta \beta} ,
		\end{equation}
		\begin{equation} \label{H-perflu-TII-TIII-TN-3}
			K \equiv \frac{1}{\delta} [ 3 \alpha E^2 + 6 \beta E + \frac{1}{2} \alpha^2 (g \! + \! P)] \, , \quad \alpha \! \equiv \! \textrm{tr} E^2, \ \ \beta \! \equiv \! \textrm{tr} E^3 , \ \ \delta \! \equiv \! \alpha^3 - 6 \beta^2 ,
		\end{equation}
	\end{subequations}
\textit{with} $P$ \textit{given in} (\ref{fluper-hydro-variables}) \textit{and} $E = ({\cal N} + \bar{\cal N})[P]$\textit{, where:}
		\begin{itemize}
\item
${\cal N} \equiv {\cal W}$ \textit{if the Weyl tensor is of type N,}
\item
${\cal N} \equiv {\cal W}^2$ \textit{if the Weyl tensor is of type III,}
\item
${\cal N} \equiv ({\cal W} - \psi {\cal G})({\cal W} + 2 \psi {\cal G})$, with $\psi = - \frac{\textrm{Tr} {\cal W}^3}{\textrm{Tr} {\cal W}^2}$\textit{, if the Weyl tensor is of type II.}
		\end{itemize}
The study of the spatially-homogeneous cosmologies of Petrov-Bel type D leads to two situations: the \textit{regular type D solutions} (a $R$-frame exists) and the \textit{singular type D solutions} (no $R$-frame exists). On the other hand, no conformally flat spatially-homogeneous cosmology admits a $R$-frame. Now, we summarise the results in \cite{SFM-simetries-Bianchi} about these families. \\ \\[-1mm]
	Let $W$ be the Weyl tensor of a perfect fluid solution of Petrov-Bel type D and $R$, its Ricci tensor. Consider now the Riemann concomitants 
	\begin{equation} \label{delta1}
		\delta_1 \equiv (\textrm{tr} E_1^2)^3 - 6(\textrm{tr} E_1^3)^2 \, , \qquad E_1 \equiv W[P] \, ,
	\end{equation}
with $P$ given in (\ref{fluper-hydro-variables}). If $\delta_1 \neq 0$, the connection tensor $H$ associated with a Riemann-frame can be obtained as in (\ref{H-perflu-TII-TIII-TN}), with $E = E_1$. \\ \\[-1mm]
	If $\delta_1 = 0$, we need to distinguish two cases. If the two Ricci algebraic scalar invariants, namely $r$ and $q$ defined in (\ref{fluper-hydro-variables}), and the complex Weyl eigenvalue $\psi \equiv - \frac{\textrm{Tr} {\cal W}^3}{\textrm{Tr} {\cal W}^2} = \psi_R + \textrm{i} \psi_I$ are constant, we have that\\[-3mm]
	\begin{equation} \label{drqpsi}
		\textrm{d} r = \textrm{d} q = \textrm{d} \psi = 0 \, .
	\end{equation}
In that case, the connection tensor $H$ associated with a Riemann-frame can be obtained as in (\ref{H-perflu-TII-TIII-TN}), with:
	\begin{itemize}
\item 
$E = E_2$ if $E_2 \neq 0$;
\item 
$E = \sigma$ if $E_2 = 0$ and $\delta_\sigma \neq 0$;
\item 
$E = E_3$ if $E_2 = 0$, $\sigma \neq 0$, $\delta_\sigma = 0$ and $E_3 \neq 0$;
	\end{itemize}
where
	\begin{subequations}
		\begin{eqnarray} \label{E2-sigma-E3}
			E_2 \equiv \omega^2 \, P_1 - \omega \otimes \omega \, , \quad \sigma \equiv [P(v, v)]^{-1/2} \Sigma(v) \, , \quad \delta_\sigma \equiv (\textrm{tr} \sigma^2)^3 - 6(\textrm{tr} \sigma^3)^2 \, , \quad \, \\[1mm]
			E_3 \equiv P_1 - \frac{1}{s} \sigma - \frac{1}{3} \gamma \, , \quad P_1 \equiv \frac{2}{3} \Big[ \frac{1}{\psi} {\cal W} - {\cal G} \Big] [P] \, , \quad \omega^\alpha \equiv - \eta^{\alpha \beta \lambda \mu} \nabla_{\beta} P_{\lambda \nu} P^{\nu}{}_{\mu} \, , \quad \\
			\Sigma_{\alpha \beta \lambda} \equiv - \frac{1}{2} [ \nabla_{\alpha} P_{\beta \mu} + \nabla_{\beta} P_{\alpha \mu} ] \, P^{\mu}{}_{\lambda} \, , \qquad s \equiv 3 \frac{ \textrm{tr} \sigma^3}{\textrm{tr} \sigma^2} \, , \qquad \gamma = g + u \otimes u \, , \quad \; \;
		\end{eqnarray}
	\end{subequations}
with $P$ given in (\ref{fluper-hydro-variables}) and $v$ an arbitrary timelike vector. \\ \\
	On the other hand, if at least one of the algebraic invariants is a non-constant scalar whose gradient defines a time-like direction we have that:
	\begin{equation} \label{mu}
		\exists \ \mu \in \{r, q, \psi_R, \psi_I\}, \qquad (\textrm{d} \mu)^2 < 0 .
	\end{equation}
Then, the connection tensor $H$ associated with a $R$-frame can be obtained as
	\begin{subequations}
		\begin{eqnarray}
			H = \nabla n \bar{\wedge} n - J , \quad n \equiv [-(\textrm{d} \mu)^2]^{-1/2} \textrm{d} \mu , \quad J_{\alpha \gamma \lambda} \equiv \nabla_\alpha E^{\mu}{}_{\rho} \, E^{\rho \nu} \, \epsilon_{\mu \nu \pi} K^{\pi \sigma} \epsilon_{\sigma \gamma \lambda} , \qquad \\[1mm]
			\epsilon_{\sigma \gamma \lambda} \equiv \eta_{\sigma \gamma \lambda \delta} n^{\delta} , \qquad K \equiv \frac{1}{\delta} [ 3 \alpha E^2 + 6 \beta E + \frac{1}{2} \alpha^2 \tilde{\gamma}] \, , \qquad \qquad \qquad \\[1mm]
			\alpha \equiv \textrm{tr} E^2, \qquad \beta \equiv \textrm{tr} E^3 , \qquad \delta \equiv \alpha^3 - 6 \beta^2 , \qquad \tilde{\gamma} \equiv g + n \otimes n \, , \qquad \qquad
		\end{eqnarray}
	\end{subequations}
with
	\begin{itemize}
\item 
$E = E_4$ if $\delta_4 \neq 0$;
\item 
$E = \tilde{\sigma}$ if $\delta_4 = 0$, $\Upsilon = 0$ and $\tilde{\delta}_\sigma \neq 0$;
\item 
$E = E_5$ if $\delta_4 = 0$, $\Upsilon = 0$, $\tilde{\delta}_\sigma \neq 0$ and $E_5 \neq 0$;
	\end{itemize}
where
	\begin{subequations}
		\begin{eqnarray} \label{E4-sigmatilde-E5}
			\delta_4 \equiv (\textrm{tr} E_4^2)^3 - 6(\textrm{tr} E_4^3)^2 \, , \qquad E_4 \equiv W[n \otimes n] \, , \qquad \Upsilon \equiv \textrm{d}[(\textrm{d} \mu)^2]\wedge n \, , \\[1mm]
			\tilde{\sigma} \equiv \nabla n - \frac{1}{3} \tilde{\theta} \, \tilde{\gamma} \, , \qquad \tilde{\theta} \equiv \nabla \cdot n \, , \qquad \tilde{\delta}_\sigma \equiv (\textrm{tr} \tilde{\sigma}^2)^3 - 6(\textrm{tr} \tilde{\sigma}^3)^2 \, , \qquad \\
			E_5 \equiv \tilde{P}_1 - \frac{1}{\tilde{s}} \tilde{\sigma} - \frac{1}{3} \tilde{\gamma} \, , \qquad \tilde{P}_1 \equiv \frac{2}{3} \Big[ \frac{1}{\psi} {\cal W} - {\cal G} \Big] [n \otimes n] \, , \qquad \tilde{s} \equiv 3\frac{\textrm{tr} \tilde{\sigma}^3}{\textrm{tr} \tilde{\sigma}^2} \, . \,
		\end{eqnarray}
	\end{subequations}
In the time-like principal plane, we have two invariant orthonormal frames, $\{u, e_1\}$ and $\{n, \tilde{e}_1\}$, connected by a boost defined by a hyperbolic angle $\phi$, namely, 
	\begin{equation} \label{phi}
		u = \cosh \phi \, n + \sinh \phi \, \tilde{e}_1 , \qquad \cosh \phi = \Gamma \equiv \sqrt{P(n,n)} \, .
	\end{equation}
A necessary condition for the existence of a G$_3$ is that all the scalar invariants have a collinear gradient:
	\begin{equation} \label{lambda}
		\textrm{d} \lambda \wedge n = 0, \qquad \forall \lambda \in \{ r, q, \Gamma, \psi, \tilde{\theta}, \tilde{s} \} \, .
	\end{equation}
	Based on these considerations, the spatially-homogeneous cosmologies admitting a $R$-frame are characterised in \cite{SFM-simetries-Bianchi}. For the cases in which the spacetime does not admit a $R$-frame, it is shown that a multiply transitive group of isometries exists and a further analysis is carried out to determine whether the solution is spatially-homogeneous. The results are summarised in the following algorithm: \\ \\
	\textbf{IDEAL labelling of the spatially-homogeneous cosmologies.} \textit{Consider the Ricci and Weyl tensors of a given metric. The first step in the characterisation algorithm is to impose that the metric tensor defines a perfect fluid solution, which can be done as stated in} (\ref{fluper-definitions}$-$\ref{fluper-hydro-variables})\textit{. Then, we need to distinguish the Petrov-Bel type of the metric. This can be done by applying either of the algorithms in Figures} \ref{Fig-16} \textit{and} \ref{Fig-17}\textit{. For type O, Figure} \ref{Fig-21} \textit{characterises the spatially-homogeneous solutions, where the number sign} $\#$ \textit{indicates that the metric is not spatially-homogeneous.}
$\hspace{-5mm}$\textit{The Petrov-Bel type D cases can be characterised following the algorithm in Figure} \ref{Fig-22}. \textit{Again, we use the number sign} $\#$ \textit{to indicate that the metric is not spatially-homogeneous. For the type D regular cases, the different connection tensors can be obtained as explained above, while the type D singular cases are characterised by the following concomitants:}
	\begin{subequations}
		\begin{eqnarray}
			4 s_1^2 \equiv \tilde{s}_1^2 , \qquad (\tilde{s}_1)^{\alpha} = \tilde{\gamma}^{\lambda \mu} \nabla_{\lambda} (\tilde{P}_1)_{\mu \nu} (\tilde{P}_1)^{\nu \alpha}, \qquad \\[2mm] 
			4 w_1^2 \equiv \tilde{\omega}_1^2 , \qquad (\tilde{\omega}_1)^{\alpha} = - \eta^{\alpha \beta \lambda \mu} \nabla_{\beta} (\tilde{P}_1)_{\lambda \nu} (\tilde{P}_1)^{\nu}{}_{\mu} \, , \\
			\zeta \equiv 4 w_1^2 + \frac{1}{3}\tilde{s} (\tilde{\theta} - \tilde{s}) - 3 \psi_R \, . \qquad \qquad \;
		\end{eqnarray}
	\end{subequations}
\\ \\
$\hspace{-2mm}$\textit{Perfect fluid solutions of Petrov-Bel types I, II, III and N are all regular and their associated connection tensors can be obtained as in} (\ref{H-Hautodual}$-$\ref{L_+-U-TIII}) \textit{or} (\ref{H-perflu-TII-TIII-TN})\textit{. Figure} \ref{Fig-23} \textit{summarises the results of the regular spatially-homogeneous characterisation in an algorithm where the connection tensor is the only input data and}
	\begin{equation} \label{invaB}
		\hat{\beta} = 1 - \frac{(3q - 5r)(3q - r)(q + r)}{4 (q - r)^3} , \qquad m_\lambda = C^{[1]}_{\lambda \rho \mu \nu}v^\rho V^{\mu \nu} ,
	\end{equation}
\textit{with} $r$ \textit{and} $q$ \textit{given in} (\ref{fluper-definitions}) \textit{and} $C^{[1]}$ \textit{given in} (\ref{cq})\textit{, and where} $v$ \textit{and} $V$ \textit{are, respectively, a vector and a two-form such that} $m \neq 0$\textit{. We denote by} ${\cal G}_r(H)$ \textit{the set of conditions in Figure} \ref{Fig-18} \textit{leading to} G$_r$. \\ \\
$\hspace{-2mm}$\textit{When the maximal group is a} G$_3$\textit{, we must determine the corresponding Bianchi type. This can be done by using the algorithm given in} \cite{FS-G3}\textit{. We present such algorithm in Figure} \ref{Fig-24}\textit{, where}
	\begin{subequations} \label{defs-Fig-24}
		\begin{equation} \label{Z-N}
			Z_{\alpha \beta} = - \frac{1}{2}\tilde{\gamma}^{\lambda}_{\alpha} \,
{H_{\lambda}}^{\mu \nu} \eta_{\mu \nu \beta \rho} \, n^{\rho} , \quad n \equiv |m^2|^{-1/2} m \, , \quad N_{\alpha \beta} = Z_{(\alpha \beta)} - (\textrm{tr} Z) \, \tilde{\gamma}_{\alpha \beta} \, ,
		\end{equation}
		\begin{equation} \label{nu}
			a \equiv - * Z \, , \qquad \nu \equiv \textrm{tr} N \, , \qquad \psi_1 \equiv \textrm{tr} N^2 \, , \qquad \psi_2 \equiv \textrm{tr} N^3 \, ,
		\end{equation}
		\begin{equation} \label{nu23}
			\nu_2 \equiv \frac12 (\nu - \psi_1) \, , \quad \nu_3 \equiv \frac13 \psi_2 - \frac12 \nu \psi_1 + \frac16 \nu^3 \, ,
		\end{equation}
	\end{subequations}
\textit{with} $m$ \textit{given in} (\ref{invaB}) \textit{and} $\tilde{\gamma} = g + n \otimes n$\textit{. The parameter} $h$ \textit{of the Bianchi types VI}$_h$ \textit{and VII}$_h$ \textit{is the invariant} $h = a^2/\nu_2$. \\
		\begin{figure}[H]
		\setlength{\unitlength}{0.85cm} {\small \noindent
		\begin{picture}(15,13.5)
		\thicklines
			\put(8.5,12){\line(-5,-1){1}}
			\put(6.5,12){\line(5,-1){1}}
			\put(6.5,12){\line(0,1){1}} \put(8.5,13){\line(-1,0){2}}
			\put(8.5,13){\line(0,-1){1}} \put(6.62,12.4){$\nu, \nu_{2}, \nu_3 , \ a $}
			\put(7.5,11.8){\vector(0,-1){0.55}}
			\put(9.5,10.58){\vector(1,0){2.78}} \put(5.5,10.58){\vector(-1,0){2.78}}
			\put(7.5,11.25){\line(-3,-1){2}} \put(7.5,11.25){\line(3,-1){2}}
			\put(7.5,9.9){\line(3,1){2}} \put(7.5,9.9){\line(-3,1){2}}
			\put(6.2,10.51){$\nu = \! \nu_2 = \! \nu_3 = 0$}
			\put(3.6,11){$a \! = \! 0$} \put(4.75,10.7){yes}
			\put(4,11.1){\oval(1.1,0.5)} \put(4,10.84){\line(0,-1){0.25}}
			\put(10.7,11){$a\!\neq\!0$} \put(9.75,10.7){yes}
			\put(11.1,11.1){\oval(1.1,0.5)} \put(11.1,10.84){\line(0,-1){0.25}}
			\put(12.3,10.2){\line(1,0){1}} \put(12.3,10.2){\line(0,1){1}}
			\put(13.3,11.2){\line(-1,0){1}} \put(13.3,11.2){\line(0,-1){1}}
			\put(12.65,10.56){V}
			\put(1.7,10.2){\line(1,0){1}} \put(1.7,10.2){\line(0,1){1}}
			\put(2.7,11.2){\line(-1,0){1}} \put(2.7,11.2){\line(0,-1){1}}
			\put(2.15,10.56){I}
			\put(7.5,9.9){\vector(0,-1){0.65}}
			\put(7.7,9.6){no}
			\put(7.5,9.25){\line(-3,-1){2}} \put(7.5,9.25){\line(3,-1){2}}
			\put(7.5,7.9){\line(3,1){2}} \put(7.5,7.9){\line(-3,1){2}}
			\put(6.08,8.45){$\nu\neq\! 0, \nu_2 = \! \nu_3\! =\!0$}
			\put(3.6,9){$a \! = \! 0$} \put(4.75,8.7){yes}
			\put(4,9.1){\oval(1.1,0.5)} \put(4,8.84){\line(0,-1){0.25}}
			\put(10.7,9){$a \! \neq \! 0$} \put(9.75,8.7){yes}
			\put(11.1,9.1){\oval(1.1,0.5)} \put(11.1,8.84){\line(0,-1){0.25}}
			\put(9.5,8.58){\vector(1,0){2.78}} \put(5.5,8.58){\vector(-1,0){2.78}}
			\put(12.3,8.2){\line(1,0){1}} \put(12.3,8.2){\line(0,1){1}}
			\put(13.3,9.2){\line(-1,0){1}} \put(13.3,9.2){\line(0,-1){1}}
			\put(12.65,8.56){IV}
			\put(1.7,8.2){\line(1,0){1}} \put(1.7,8.2){\line(0,1){1}}
			\put(2.7,9.2){\line(-1,0){1}} \put(2.7,9.2){\line(0,-1){1}}
			\put(2.1,8.56){II}
			\put(7.5,7.25){\line(-3,-1){2}}  \put(7.5,7.25){\line(3,-1){2}}
			\put(7.5,5.9){\line(3,1){2}} \put(7.5,5.9){\line(-3,1){2}}
			\put(6.2,6.45){$ \nu_2 < 0, \ \nu_3 = 0$}
			\put(7.5,7.9){\vector(0,-1){0.65}}
			\put(7.7,7.6){no}
			\put(3.6,7){$a \! = \! 0$} \put(4.75,6.7){yes}
			\put(4,7.1){\oval(1.1,0.5)} \put(4,6.84){\line(0,-1){0.25}}
			\put(10.7,7){$a \! \neq \! 0$} \put(9.75,6.7){yes}
			\put(11.1,7.1){\oval(1.1,0.5)} \put(11.1,6.84){\line(0,-1){0.25}}
			\put(9.5,6.58){\vector(1,0){2.78}} \put(5.5,6.58){\vector(-1,0){2.78}}
			\put(12.3,6.17){\line(1,0){1}} \put(12.3,6.15){\line(0,1){1}}
			\put(13.3,7.15){\line(-1,0){1}} \put(13.3,7.17){\line(0,-1){1}}
			\put(12.5,6.75){VI$_h\!$ } \put(12.45,6.35){(III) }
			\put(1.7,6.2){\line(1,0){1}} \put(1.7,6.2){\line(0,1){1}}
			\put(2.7,7.2){\line(-1,0){1}} \put(2.7,7.2){\line(0,-1){1}}
			\put(1.93,6.56){VI$_0$}
			\put(7.5,5.25){\line(-3,-1){2}} \put(7.5,5.25){\line(3,-1){2}}
			\put(7.5,3.9){\line(3,1){2}} \put(7.5,3.9){\line(-3,1){2}}
			\put(6.2,4.45){$\nu_2 > 0, \ \nu_3 = 0$}
			\put(7.5,5.9){\vector(0,-1){0.65}}
			\put(7.7,5.6){no}
			\put(3.6,5){$a \! = \! 0$} \put(4.75,4.7){yes}
			\put(4,5.1){\oval(1.1,0.5)} \put(4,4.84){\line(0,-1){0.25}}
			\put(10.7,5){$a \! \neq \! 0$} \put(9.75,4.7){yes}
			\put(11.1,5.1){\oval(1.1,0.5)} \put(11.1,4.84){\line(0,-1){0.25}}
			\put(9.5,4.58){\vector(1,0){2.78}} \put(5.5,4.58){\vector(-1,0){2.78}}
			\put(12.3,4.2){\line(1,0){1}} \put(12.3,4.2){\line(0,1){1}}
			\put(13.3,5.2){\line(-1,0){1}} \put(13.3,5.2){\line(0,-1){1}}
			\put(12.42,4.56){VII$_h$}
			\put(1.7,4.2){\line(1,0){1}} \put(1.7,4.2){\line(0,1){1}}
			\put(2.7,5.2){\line(-1,0){1}} \put(2.7,5.2){\line(0,-1){1}}
			\put(1.85,4.56){VII$_0$}
			\put(7.5,3.9){\vector(0,-1){0.65}}
			\put(7.7,3.6){no}
			\put(7.5,3.25){\line(-3,-1){2}} \put(7.5,3.25){\line(3,-1){2}}
			\put(7.5,1.9){\line(3,1){2}} \put(7.5,1.9){\line(-3,1){2}}
			\put(6.2,2.45){$ \nu \, \nu_3 > 0, \ \nu_2 > 0$}
			\put(7.5,1.9){\vector(0,-1){0.7}}
			\put(7.7,1.6){no}
			\put(4.75,2.7){yes}
			\put(5.5,2.58){\vector(-1,0){2.78}}
			\put(1.7,2.2){\line(1,0){1}} \put(1.7,2.2){\line(0,1){1}}
			\put(2.7,3.2){\line(-1,0){1}} \put(2.7,3.2){\line(0,-1){1}}
			\put(2,2.56){IX}
			\put(1.7,0.7){\line(1,0){1}} \put(1.7,0.7){\line(0,1){1}}
			\put(2.7,1.7){\line(-1,0){1}} \put(2.7,1.7){\line(0,-1){1}}
			\put(1.83,1.06){VIII}
			\put(7.5,1.2){\vector(-1,0){4.78}}
		\end{picture} }
		\vspace{-4mm}
		\caption{Algorithm to distinguish the Bianchi-Behr types of a G$_3$. $\qquad \qquad \qquad \qquad \qquad$}
		\label{Fig-24}
		\end{figure}
$\hspace{-5mm}$Now, we will use \textit{xIdeal} to apply the above algorithms to particular solutions as examples.

		\subsection{Bianchi type I cosmologies}
		Let us consider now the general expression of the metrics admitting a three-dimensional group of isometries of Bianchi type I \cite{Ellis-Maartens-MacCallum}:
			\begin{equation}
				\textrm{d} s^2 \, = \, - \textrm{d} t^2 + \ell_1^2(t) \, \textrm{d} x^2 + \ell_2^2(t) \, \textrm{d} y^2 + \ell_3^2(t) \, \textrm{d} z^2 \, .
			\end{equation}
This example can serve as a test for our algorithm. Conditions in (\ref{fluper-conditions-A2}) give us the field equations for $\ell_i(t)$ that should be solved in order to obtain perfect fluid solutions, which are always non-tilted ($u = - \textrm{d} t$). Regardless of that, using the \textit{xIdeal} function \texttt{PetrovType}, we get that these metrics are, generically, of Petrov Type I. Hence, we can use the function \texttt{ConnectionTensor} to obtain its connection tensor $H$ as in (\ref{H-Hautodual}$-$\ref{H-perflu-TII-TIII-TN-1}). With that, we can follow the algorithm in Figure \ref{Fig-24}. We get that conditions ${\cal G}_4(H)$ are not fulfilled but equations ${\cal G}_3(H)$ and $m^2 < 0$ are. Therefore, we check that, indeed, the dimension of the isometry group is three. To determine the Bianchi type of the group, we use \textit{xIdeal} to obtain the structure tensor $Z$ given in (\ref{Z-N}) and we get that it vanishes. Thus, the $G_3$ is indeed of Bianchi type I.

		\subsection{The homogeneous Ozsváth solutions of classes II and III}
		Let us consider the Ozsváth solutions of classes II and III as given in \cite{Kramer}, 
		\begin{equation} \label{FK-Kramer}
			\textrm{d} s^2 \, = \, \hat{a}^2 [ (1 - \hat{s}) (\omega^1)^2 + (1 + \hat{s}) (\omega^2)^2 + ( \textrm{d} u + \lambda \omega^3)^2 - 2(\omega^3)^2 ] \, , 
		\end{equation}
with $\lambda = 0$ for class III, and $\lambda^2 = 1 - 2 \hat{s}^2$ for class II. The parameters $\hat{a}$ and $\hat{s}$ are \mbox{constants}, and $\{\omega^1, \omega^2, \omega^3\}$ are the reciprocal group of a Bianchi type VIII, namely \cite{Kramer}, $\omega^1 = \textrm{d} x - \sinh{y} \, \textrm{d} z$, $\omega^2 = \cos{x} \, \textrm{d} y - \sin{x} \cosh{y} \, \textrm{d} z$, and $\omega^3 = \, \sin{x} \, \mathrm{d} y \\ + \cos{x} \cosh{y} \, \textrm{d} z$. \\ \\
		If we now use the \textit{xIdeal} function \texttt{PerfectFluidQ}, we get that: (i) For $\lambda = 0$, the output is \texttt{False}. Indeed, for this case $q \, Q(\omega^3, \omega^3) = 0$. Thus, metric (\ref{FK-Kramer}) is not a perfect fluid solution. In fact, the Ricci tensor has a triple eigenvalue and the first condition in (\ref{fluper-conditions-A2}) holds, but the eigenvector associated with the simple eigenvalue is space-like. (ii) For $\lambda \neq 0$, the output is also \texttt{False}. The first condition in (\ref{fluper-conditions-A2}) does not hold, that is, the Ricci tensor does not have a triple eigenvalue and, consequently, metric (\ref{FK-Kramer}) is not a perfect fluid solution. \\ \\
		Therefore, we conclude that there is a mistake in the transcription made in \cite{Kramer} of the results by Ozsváth. Indeed, in the notation of \cite{Kramer}, the solutions of classes II and III take actually the expression \cite{Ozsvath_b, Farnsworth-Kerr}: 
		\begin{equation} \label{Ozsvath-II-III}
			\textrm{d} s^2 \, = \, \hat{a}^2 [- 2(\omega^1)^2 + (1 + \hat{s}) (\omega^2)^2 + ( \textrm{d} u + \lambda \omega^1)^2 + (1 - \hat{s})(\omega^3)^2 ] \, , 
		\end{equation}
with $\lambda$, $\hat{a}$, $\hat{s}$ and $\{\omega^1, \omega^2, \omega^3\}$ as given after (\ref{FK-Kramer}). Now, we obtain that both conditions in (\ref{fluper-conditions-A2}) are fulfilled, and thus, these metrics are perfect fluid solutions. Moreover, as stated in \cite{Ozsvath_b, Farnsworth-Kerr}, they define dust solutions with cosmological constant and positive matter density if: $\hat{s}^2 < 1$ when $\lambda = 0$ (class III) and $1/4 < \hat{s}^2 < 1/2$ when $\lambda \neq 0$ (class II). \\ \\
		On the other hand, using the \textit{xIdeal} function \texttt{PetrovType} we get that metrics (\ref{Ozsvath-II-III}) are of Petrov Type I when $\lambda \neq 0$, and of type D when $\lambda = 0$. For class III Ozsváth metrics, we follow the algorithm in Figure \ref{Fig-22} and we get that it belongs to the family whose connection tensor is given by $H = H_{\sigma}$. For class II Ozsváth metrics, we could obtain the connection tensor $H$ using the \textit{xIdeal} function \texttt{ConnectionTensor} without giving a $R$-frame. Alternatively, we can give the Weyl principal frame $\{e_{0}, e_{1}, e_{+}, e_{-}\}$, which can be determined as indicated in \cite{FMS-Weyl}:
		\begin{equation}
			e_0 = \hat{a}\sqrt{2}\, \omega_1 , \quad e_1 = \hat{a}\, (\textrm{d} u + \lambda \omega_1), \quad e_{\pm} = \frac{\hat{a}}{\sqrt{2}}\, [\sqrt{1 + \hat{s}}\, \omega_2 \pm \sqrt{1 - \hat{s}}\, \omega_3]. 
		\end{equation}
In the two cases, once we get the connection tensor $H$, we apply the algorithm in Figure \ref{Fig-24}. From it, we get that the dimension of the isometry group is four and that $\textrm{tr} H = 0$, that is, the metric admits an unimodular $G_4$. We also have $q > 0$, $r < 0$. All these invariant conditions characterise the Ozsváth solutions of classes II and III. Nevertheless, they are not spatially-homogeneous solutions, contrary to what Ozsváth \cite{Ozsvath_b} and \cite{Kramer} claim (see \cite{SFM-simetries-Bianchi} for more details). \\ \\ \\ \\ \\ \\
		It is worth remarking that solutions (\ref{Ozsvath-II-III}) were obtained under the hypothesis of a positive matter density. The determination of the solution without this condition is still an open problem that will lead to new solutions. For example, the metric given by
		\begin{equation} \label{FK}
			\textrm{d} s^2 \, = \, \hat{a}^2 [ 2 (\omega^1)^2 + (1 + \hat{s}) (\omega^2)^2 + 2 \textrm{d} u^2 - (1 - \hat{s})(\omega^3)^2 ] \, , 
		\end{equation}
is a dust solution with cosmological constant, which has similar properties as the class III Ozsváth metric, but with a negative matter density.
	
	\section{IDEAL characterisation of the Schwarzschild metric and type D static vacuum solutions} \label{IDEAL-Sch}
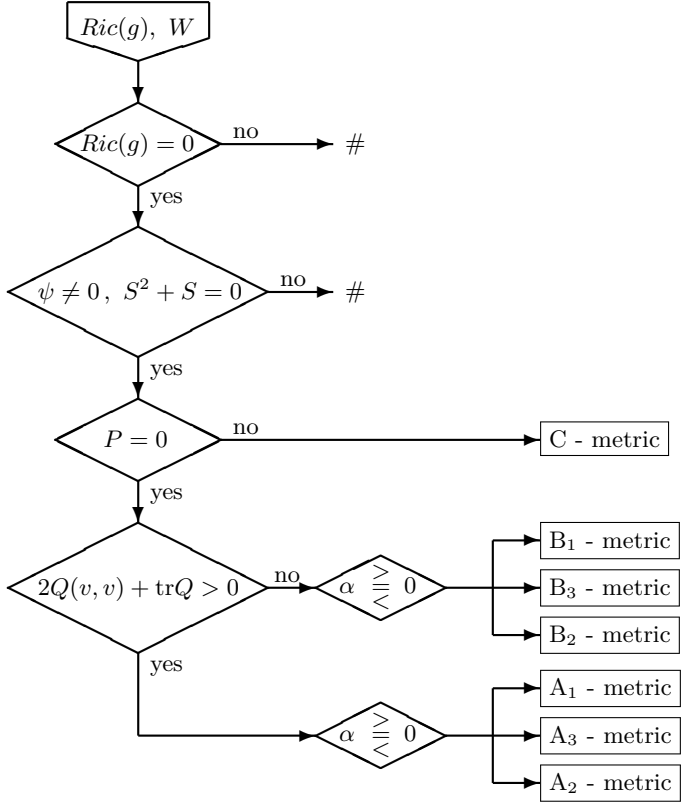
\begin{figure}[h]
	\hspace*{7mm} \setlength{\unitlength}{0.78cm} {\small \noindent
		\begin{picture}(16,14.5)
			\thicklines
			\put(4.7,13.5){\line(-3,-1){1.2}} \put(2.3,13.5){\line(3,-1){1.2}}
			\put(4.7,13.5){\line(0,1){0.6}} \put(2.3,14.1){\line(1,0){2.4}}
			\put(2.3,14.1){\line(0,-1){0.6}}
			\put(2.51,13.55){$Ric(g), \ W$}

			\put(3.5,13.1){\vector(0,-1){0.7}}

			\put(3.5,12.4){\line(-2,-1){1.4}} \put(3.5,12.4){\line(2,-1){1.4}}
			\put(3.5,11){\line(2,1){1.4}} \put(3.5,11){\line(-2,1){1.4}}
			\put(2.51,11.58){$Ric(g) = 0$}
			
			\put(4.9,11.7){\vector(1,0){1.9}}
			\put(5.1,11.8){no}
			\put(7,11.6){\#}
			
			\put(3.5,11){\vector(0,-1){0.7}}
			\put(3.7,10.7){yes}
			
			\put(3.5,10.3){\line(-2,-1){2.2}} \put(3.5,10.3){\line(2,-1){2.2}}
			\put(3.5,8.1){\line(2,1){2.2}} \put(3.5,8.1){\line(-2,1){2.2}}
			\put(1.8,9.08){$\psi \neq 0 \, , \ S^2 + S = 0$}
			
			\put(5.7,9.2){\vector(1,0){1.1}}
			\put(5.9,9.3){no}
			\put(7,9.1){\#}
			
			\put(3.5,8.1){\vector(0,-1){0.7}}
			\put(3.7,7.8){yes}
			
			\put(3.5,7.4){\line(-2,-1){1.4}} \put(3.5,7.4){\line(2,-1){1.4}}
			\put(3.5,6){\line(2,1){1.4}} \put(3.5,6){\line(-2,1){1.4}}
			\put(2.91,6.58){$P = 0$}
			
			\put(4.9,6.7){\vector(1,0){5.4}}
			\put(5.1,6.8){no}
			\put(10.3,6.58){{\framebox{C - metric}}}
			
			\put(3.5,6){\vector(0,-1){0.7}}
			\put(3.7,5.7){yes}
			
			\put(3.5,5.3){\line(-2,-1){2.2}} \put(3.5,5.3){\line(2,-1){2.2}}
			\put(3.5,3.1){\line(2,1){2.2}} \put(3.5,3.1){\line(-2,1){2.2}}
			\put(1.8,4.08){$2Q(v, v) + \textrm{tr}Q > 0$}
			
			\put(5.7,4.2){\vector(1,0){0.8}}
			\put(5.8,4.3){no}
			
			\put(6.5,4.2){\line(2,1){1.1}} \put(7.61,4.75){\line(2,-1){1.1}}
			\put(6.5,4.2){\line(2,-1){1.1}} \put(7.61,3.65){\line(2,1){1.1}}
			\put(6.9,4.11){$\alpha \qquad 0$}
			\put(7.45,4.11){$=$}
			\put(7.45,4.36){$>$}
			\put(7.45,3.86){$<$}
			
			\put(9.5,5){\vector(1,0){0.8}}
			\put(8.7,4.2){\vector(1,0){1.6}}
			\put(9.5,3.4){\vector(1,0){0.8}}
			\put(9.5,5){\line(0,-1){1.6}}
			
			\put(10.3,4.88){{\framebox{B$_1$ - metric}}}
			\put(10.3,4.08){{\framebox{B$_3$ - metric}}}
			\put(10.3,3.28){{\framebox{B$_2$ - metric}}}
			
			\put(3.5,3.1){\line(0,-1){1.4}}
			\put(3.7,2.8){yes}
			\put(3.5,1.7){\vector(1,0){3}}
			
			\put(6.5,1.7){\line(2,1){1.1}} \put(7.61,2.25){\line(2,-1){1.1}}
			\put(6.5,1.7){\line(2,-1){1.1}} \put(7.61,1.15){\line(2,1){1.1}}
			\put(6.9,1.61){$\alpha \qquad 0$}
			\put(7.45,1.61){$=$}
			\put(7.45,1.86){$>$}
			\put(7.45,1.36){$<$}
			
			\put(9.5,2.5){\vector(1,0){0.8}}
			\put(8.7,1.7){\vector(1,0){1.6}}
			\put(9.5,0.9){\vector(1,0){0.8}}
			\put(9.5,2.5){\line(0,-1){1.6}}
			
			\put(10.3,2.38){{\framebox{A$_1$ - metric}}}
			\put(10.3,1.58){{\framebox{A$_3$ - metric}}}
			\put(10.3,0.78){{\framebox{A$_2$ - metric}}}
	\end{picture} }
	\vspace{-7mm}
	\caption{Algorithm to distinguish the different classes of type D static vacuum solutions. Class A$_1$ corresponds to the Schwarzschild metric.}
	\label{Fig-25}
	\end{figure}

	As already mentioned in Section \ref{sec-IDEAL}, Ferrando and Sáez obtained an IDEAL characterisation of the Schwarzschild spacetime \cite{FS-Schwarzschild}. Their result is based on the classification of the type D static vacuum solutions on seven classes (A$_i$, B$_i$, and C with $i = 1, 2, 3$) by Ehlers and Kundt \cite{Ehlers-Kundt}. The Schwarzschild metric corresponds to the class A$_1$. Therefore, by obtaining an IDEAL algorithm to label the type D static vacuum solutions according to this classification, they obtained an IDEAL characterisation of the Schwarzschild spacetime: \\ \\
	\textbf{Type D static vacuum solutions.} \textit{Consider the following concomitants of the Weyl tensor:}
	\begin{subequations}
		\begin{equation}
			\psi \equiv - \frac{\textrm{tr}W^3}{\textrm{tr}W^2} \, , \quad \alpha \equiv \frac19 (\textrm{d} \ln \psi ,\textrm{d} \ln \psi) - 2 \psi \, , \quad P \equiv * W (\textrm{d} \psi; \textrm{d} \psi) \, ,
		\end{equation}
		\begin{equation}
			S \equiv \frac{1}{3 \psi} (W - \psi G) \, , \qquad Q \equiv S (\textrm{d} \psi; \textrm{d} \psi) \, ,
		\end{equation}
	\end{subequations}
\textit{with} $G = \frac12 g \wedge g$\textit{. The algorithm in Figure} \ref{Fig-25} \textit{gives the classification of the type D static vacuum solutions, where} $v$ \textit{is an arbitrary unitary timelike vector. Class} A$_1$ \textit{corresponds to the Scwarzschild metric.}\\ \\
	$\hspace{-3mm}$This algorithm is implemented in \textit{xIdeal} by means of a function called \texttt{StaticVacuumTypeDClassify}. We applied it to the Schwarzschild metric and it correctly returns that the input is an A$_1$-metric.
	
	\section{IDEAL characterisation of the Kerr metric} \label{IDEAL-Kerr}
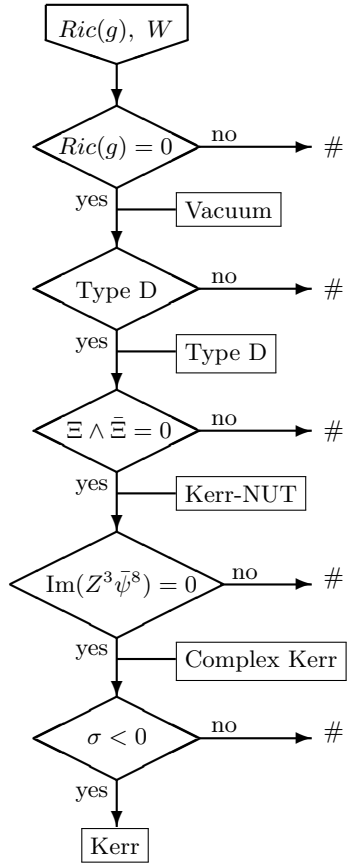
\begin{figure}[H]
	\vspace{-12mm} \hspace*{20mm} \setlength{\unitlength}{0.78cm} {\small \noindent
		\begin{picture}(16,14.5)
			\thicklines
			\put(4.7,13.5){\line(-3,-1){1.2}} \put(2.3,13.5){\line(3,-1){1.2}}
			\put(4.7,13.5){\line(0,1){0.6}} \put(2.3,14.1){\line(1,0){2.4}}
			\put(2.3,14.1){\line(0,-1){0.6}}
			\put(2.51,13.55){$Ric(g), \ W$}

			\put(3.5,13.1){\vector(0,-1){0.7}}

			\put(3.5,12.4){\line(-2,-1){1.4}} \put(3.5,12.4){\line(2,-1){1.4}}
			\put(3.5,11){\line(2,1){1.4}} \put(3.5,11){\line(-2,1){1.4}}
			\put(2.51,11.58){$Ric(g) = 0$}
			
			\put(4.9,11.7){\vector(1,0){1.9}}
			\put(5.1,11.8){no}
			\put(7,11.6){\#}
			
			\put(3.5,11){\vector(0,-1){1}}
			\put(2.8,10.7){yes}
			
			\put(3.5,10.64){\line(1,0){1}}
			\put(4.5,10.5){{\framebox{Vacuum}}}
			
			\put(3.5,10){\line(-2,-1){1.4}} \put(3.5,10){\line(2,-1){1.4}}
			\put(3.5,8.6){\line(2,1){1.4}} \put(3.5,8.6){\line(-2,1){1.4}}
			\put(2.8,9.15){Type D}
			
			\put(4.9,9.3){\vector(1,0){1.9}}
			\put(5.1,9.4){no}
			\put(7,9.2){\#}
			
			\put(3.5,8.6){\vector(0,-1){1}}
			\put(2.8,8.3){yes}
			
			\put(3.5,8.24){\line(1,0){1}}
			\put(4.5,8.1){{\framebox{Type D}}}
			
			\put(3.5,7.6){\line(-2,-1){1.4}} \put(3.5,7.6){\line(2,-1){1.4}}
			\put(3.5,6.2){\line(2,1){1.4}} \put(3.5,6.2){\line(-2,1){1.4}}
			\put(2.65,6.77){$\Xi \wedge \bar{\Xi} = 0$}
			
			\put(4.9,6.9){\vector(1,0){1.9}}
			\put(5.1,7){no}
			\put(7,6.8){\#}
			
			\put(3.5,6.2){\vector(0,-1){1}}
			\put(2.8,5.9){yes}
			
			\put(3.5,5.84){\line(1,0){1}}
			\put(4.5,5.7){{\framebox{Kerr-NUT}}}
			
			\put(3.5,5.2){\line(-2,-1){1.8}} \put(3.5,5.2){\line(2,-1){1.8}}
			\put(3.5,3.4){\line(2,1){1.8}} \put(3.5,3.4){\line(-2,1){1.8}}
			\put(2.3,4.17){$\textrm{Im}(Z^3 \bar{\psi}^8) = 0$}
			
			\put(5.25,4.3){\vector(1,0){1.55}}
			\put(5.45,4.4){no}
			\put(7,4.3){\#}
			
			\put(3.5,3.4){\vector(0,-1){1}}
			\put(2.8,3.1){yes}
			
			\put(3.5,3.04){\line(1,0){1}}
			\put(4.5,2.9){{\framebox{Complex Kerr}}}
			
			\put(3.5,2.4){\line(-2,-1){1.4}} \put(3.5,2.4){\line(2,-1){1.4}}
			\put(3.5,1){\line(2,1){1.4}} \put(3.5,1){\line(-2,1){1.4}}
			\put(3,1.6){$\sigma < 0$}
			
			\put(4.9,1.7){\vector(1,0){1.9}}
			\put(5.1,1.8){no}
			\put(7,1.7){\#}
			
			\put(3.5,1){\vector(0,-1){0.8}}
			\put(2.8,0.7){yes}
			\put(2.9,-0.25){{\framebox{Kerr}}}
	\end{picture} }
	\vspace{1mm}
	\caption{Algorithm to distinguish a vacuum type D solution, identify if it belongs to the Kerr-NUT family and, ultimately, single out the Kerr metric.}
	\label{Fig-26}
	\end{figure}

	The first IDEAL characterisation of the Kerr spacetime was also obtained by Ferrando and Sáez \cite{FS-Kerr}. Later, García-Parrado obtained in \cite{Alfonso-TypeD-ID} an alternative characterisation, which is the one we have implemented in \textit{xIdeal}:\\ \\
	\textbf{Kerr solution.} \textit{Consider the following concomitants of the self-dual Weyl tensor:}
	\begin{subequations}
		\begin{equation}
			\psi \equiv - \frac{\textrm{Tr}{\cal W}^3}{\textrm{Tr}{\cal W}^2} \, , \qquad {\cal S} \equiv {\cal W} - \psi {\cal G} \, , \qquad Z \equiv (\textrm{d}\psi, \textrm{d}\psi) \, ,
		\end{equation}
		\begin{equation}
			\Xi \equiv {\cal S} (\textrm{d}\psi; \textrm{d}\psi) \, , \qquad \sigma \equiv \frac{-\textrm{Re}(Z^3 \bar{\psi}^8)}{[\textrm{Re}(\psi^3 \bar{Z}) - |Z|^2]^3} \, ,
		\end{equation}
	\end{subequations}
\textit{where an overbar denotes complex conjugation. The algorithm in Figure} \ref{Fig-26} \textit{first characterises whether the input metric is a vacuum type D solution. Then, it identifies the Kerr-NUT metrics within this class of solutions. Finally, it singles out the Kerr metric from among them.} \\ \\
	$\hspace{-5mm}$This algorithm is implemented in \textit{xIdeal} by means of a function called \texttt{KerrSolutionQ}. This function prints out the intermediate steps of the algorithm (\texttt{Vacuum}, \texttt{Type D}, \texttt{Kerr-Nut}, \texttt{Complex Kerr} and \texttt{Kerr}) and returns \texttt{True} if the input metric is the Kerr solution. The Petrov-Bel type D condition is checked using the \texttt{PetrovType} \textit{xIdeal} function, which implements the algorithm in Figure \ref{Fig-16}. We applied \texttt{KerrSolutionQ} to the Kerr metric and it correctly returns \texttt{True}.

\chapter[Conclusions and future work]{Conclusions and \\ future work}
This thesis brings together the main results that we have achieved during the doctoral studies. Two different major topics have been addressed, both of them falling within the field of General Relativity, which have given rise to the three parts in which this thesis has been divided. \\ \\ 
In the first two parts, we tried to understand further, from a physical point of view, the jungle of perfect fluid solutions to the Einstein equations. To do so, we used the hydrodynamic approach previously developed by our research group and our own contribution to the topic and then we focused on the solutions with a G$_3/$S$_2$. The objective of the third part was to implement, in a computer algebra program, some of the numerous algorithms that characterise and determine spacetime properties in an IDEAL way.

	\subsection*{Physical interpretation of a perfect energy \mbox{tensor}}
	In Chapter \ref{chap-Synge}, we expand the previous work on the interpretation of perfect energy tensors by presenting a hydrodynamic approach to the Synge gas. This point of view can be useful to look for test solutions or self-gravitating systems that model high-energy scenarios, as the subsequent chapters prove. Our study is also of conceptual interest and it allowed us to formulate the Rainich-like theory for the Einstein-Synge solutions. It is worth remarking that the characterisation theorems presented there can be slightly changed in order to obtain the characterisation of the perfect fluid solutions corresponding to the media with the other EoS considered in the chapter. For example, for the TM EoS the conditions characterising the Synge EoS should be changed to the expression of the indicatrix function of a TM gas. \\ \\ 
	Several analytical approximations to the Synge EoS have shown their usefulness in numerical codes to model a relativistic gas. In this chapter we also carried out a hydrodynamic approach to recover and to analyse the Taub-Mathews EoS. We established that the square of the speed of sound of the TM EoS and the Synge EoS differ at most by 2.36$\%$. Furthermore, our method enabled us to obtain other analytical EoS that approximate the Synge EoS to the same order of approximation than the TM one. However, they are less accurate than the TM EoS and do not fulfil Taub's inequality. We were able to find EoS with better accuracy than that of TM by reaching higher orders in the approximation, and all of them fulfil Taub's inequality. Therefore, the TM equation of state acts as the limiting case.

	\subsection*{Physical interpretation of perfect fluid solutions with \\ spherical symmetry}
	In Chapter \ref{chap-T-models}, we started by studying the T-models, the subfamily of the Szekeres-Szafron solutions of class II admitting a G$_3/$S$_2$. First, we analysed their potential thermodynamic interpretations. To do so, we showed that the field equation can be written as a differential equation that is linear for an adequate choice of the three unknown metric functions, we obtained the coordinate expressions of the hydrodynamic quantities and we proved that the whole set of the T-models can represent the evolution of a fluid in l.t.e. Then, we obtained the general expression of the square of the speed of sound of such fluids and of the associated thermodynamic schemes. Each of these thermodynamic schemes offers a different thermodynamic interpretation of the considered solution. However, complementary macroscopic physical requirement (energy and compressibility conditions and positivity of some thermodynamic quantities) must be imposed on the thermodynamic solutions in order to obtain physically realistic models. \\ \\ 
	First, we obtained the subfamily of the T-models representing an evolution compatible with the EoS of a generic ideal gas and all its associated thermodynamic schemes. Then, we found that there exist several combinations of parameters with wide spacetime domains in which the macroscopic conditions for physical reality hold. In particular, we showed that there exist an ideal gas thermodynamic scheme that is a good approximation to an ultrarelativistic gas. We also analysed from a thermodynamic perspective the previously known MWH solution and we showed that it does not fulfil the necessary conditions to be interpreted as a perfect fluid in l.t.e. Finally, we proved that no T-model can represent the evolution of a classical ideal gas. \\ \\
	We also analysed the T-model field equation with the objective of finding its general solution. Herlt \cite{Herlt} proposed an integration algorithm proving that this field equation can be solved by quadratures. First, we revisited the Herlt approach and proposed a modified procedure. In both algorithms the solution is obtained by calculating two indefinite integrals. Then, by redefining the unknown metric functions, we established some algorithms that solve the equation without calculating any integral. Hence, we gave the explicit expression of the general solution considering separately the case of plane symmetry and that of spherical and hyperbolical symmetry. All of them depend on an arbitrary function of time and an arbitrary function $Q(r)$ of the spatial coordinate $r$. We recovered some known solutions and we obtained new ones by applying the above algorithms. \\ \\ 
	The physical meaning of these T-models can be analysed {\it a posteriori} by using our hydrodynamic approach to the perfect fluid solutions. Nevertheless, it would be appropriate to be able to impose specific physical or geometrical properties established {\it a priori}, as we did with the ideal T-models. This justifies the presentation of different integration methods here, allowing us to choose the most suitable one for the restrictions we impose. Finally, we also argued that our study also provides the general solutions of the KCKS T-models and the thermodynamic class II Szekeres-Szafron solutions (singular and regular models). \\ \\
	We finish this chapter by studying the physical interpretations of the KCKS subfamily of the T-models. First, we showed that these solutions can be interpreted as the isentropic evolution of a fluid whose non-isentropic evolution is represented by a T-model and we obtained their corresponding barotropic relation. Then, we studied their interpretation as representing isentropic evolutions fulfilling the CIG and the relativistic $\gamma$-law barotropic relations. We showed that the CIG evolution generalises the FLRW CIG solution, giving solutions whose scaling function $R$ grows at different speeds. \\ \\
	In Chapter \ref{chap-R-models-plans}, we turned to the study of the R-models that admit an orthogonal flat synchronisation, which constitute a subfamily of the Lemaître-Tolman metrics. The analysis began by noting that the whole subfamily can represent the evolution of a fluid in l.t.e. Then, we obtained general expressions for the hydrodynamic quantities and all the associated thermodynamic schemes. \\ \\ 
	As the next step, we derived the differential equation that must be satisfied by solutions representing an evolution compatible with the ideal gas EoS. This equation involves two functions of the coordinate time, $f(\tau)$ and $g(\tau)$, as well as a function $\beta(\alpha)$, where $\alpha$ is an arbitrary function of the spatial coordinate $r$ and $g(\tau)$ is determined by $f(\tau)$ through the field equations. Two particular solutions were examined: the ideal Szafron model $f(\tau) = \tau^q$ with $q \neq 1/2$ and the solution with $f(\tau) = \sqrt{\tau}$. For both cases, we obtained explicit coordinate expressions for the hydrodynamic quantities and the associated thermodynamic schemes, including the thermodynamic schemes allowing to interpret both solutions as ideal gases. A detailed analysis of their behaviour revealed that there exist wide spacetime domains where the necessary conditions for physical reality are fulfilled. The chapter concluded by proving that none of these two ideal solutions can represent an inviscid fluid with a non-vanishing thermal coefficient, although other ideal solutions might. \\ \\
	The second part of the thesis finishes with Chapter \ref{chap-Stephani}, in which we studied general properties of the thermodynamic Stephani universes and analysed the constraints that some specific physical requirements impose on the models. We focused on the solutions where the observer comoving with the fluid flow can measure a state of isotropic radiation. The models considered highlight a long-known fact (see \cite{FMP-isotropa, Clarkson-1999} and references therein): an inhomogeneous perfect fluid solution can be compatible with an observed inhomogeneous and isotropic radiation. Although our purpose there was not to look for cosmological models compatible with particular observational data, our study shows that some of our models, or other similar ones that could be obtained with an analogous approach, could model local nonlinear inhomogeneities of the real Universe. \\ \\
	After these detailed studies, we can conclude that the hydrodynamic approach introduced in the first part of the thesis offers an effective framework for achieving our objective, enabling us to identify the subfamilies of the three considered models that represent physically realistic fluids with various properties of interest. However, this represents only the first step towards the broader goal of providing a physical interpretation of the extensive body of exact solutions available in the literature. \\ \\
	Besides deepening our understanding of the theory of General Relativity, which is already a significant accomplishment, we believe that our study may also be of interest to the Numerical Relativity community. While it is true that numerical solutions may be better suited to modelling determined physical scenarios, the analytic solutions obtained through our approach can serve as a catalogue of reference spacetimes, potentially useful for testing numerical codes.
	
	\subsection*{Implementation of IDEAL algorithms in xAct}
	In Chapter \ref{chap-PerFlu-xIdeal}, we present IDEAL characterisations related to fluid properties and the \textit{xIdeal} functions that implement them:
	\begin{itemize}
		\item \texttt{PerfectFluidQ}
		\item \texttt{PerfectFluidVariables}
		\item \texttt{ThermodynamicPerfectFluidQ}
		\item \texttt{GenericIdealGasQ}
		\item \texttt{StephaniUniverseQ}
		\item \texttt{FriedmannQ}
		\item \texttt{KustaanheimoQvistQ}
	\end{itemize}
	In this chapter, we also present the IDEAL characterisation of a perfect fluid flow that we developed recently. Although this algorithm has not been implemented in \textit{xIdeal} yet, we applied it to some examples. These preliminary applications yielded promising results, highlighting the potential value of being able to apply the algorithm to a broader range of scenarios. \\ \\
	In Chapter \ref{chap-geometric-xIdeal}, we introduce the other IDEAL characterisations and determinations not necessarily related to fluids that are currently implemented in \textit{xIdeal} and we explained and applied to several examples their corresponding functions:
	\begin{itemize}
		\item \texttt{PetrovType}
		\item \texttt{DebeverNullDirections}
		\item \texttt{ConnectionTensor}
		\item \texttt{IsometryGroupDimension}
		\item \texttt{StaticVacuumTypeDClassify}
		\item \texttt{KerrSolutionQ}
	\end{itemize}
	Here, we also present some recently developed IDEAL algorithms, such as the IDEAL labelling of the spatially-homogeneous cosmologies, which, although not implemented yet, have been successfully applied to different cases with the help of \textit{xIdeal}. \\ \\
	Beyond the more theoretical applications explored in this thesis, these IDEAL algorithms may also prove useful in the field of Numerical Relativity, both for characterising the initial data associated with a given solution and for assessing the local proximity of the numerical solution to a considered spacetime. \\ \\
	Regarding their implementation in a computer algebra program, although there is a huge amount of work yet to be done, as we discuss below, our hope is that \textit{xIdeal} becomes a useful resource for the scientific community working with exact solutions to the Einstein equations. Among its main advantages are the possibility to discern whether a potential new solution is actually an existing one expressed in an unusual coordinate system, to obtain the explicit expressions of intrinsic quantities and properties of the considered spacetime in a matter of seconds and providing a database of metrics with their most important properties. Ultimately, by automating and systematising key aspects of the IDEAL approach, we hope that our package will contribute to advancing research in General Relativity.
	
	\subsection*{Future work}
	In this thesis, we have given the first steps towards the physical interpretations of the perfect fluid solutions to the Einstein equations admitting a G$_3$/S$_2$. However, several open problems have been left unexplored and their completion could expand our current results towards promising directions. \\ \\
	To begin with, as discussed above, our results in Section \ref{T-model-general-sol} allow us to impose physical or geometrical properties \textit{a priory}. Therefore, in a future work we can investigate particular T-model solutions obtained after imposing such physically relevant conditions. \\ \\
	In Section \ref{sec-R-chi-pi}, we obtained the differential equation that the ideal R-model metric functions must fulfil. A preliminary study of this constraint is presented in Section \ref{subsec-analysys-ideal-eq} and two particular solutions are studied in the subsequent sections. However, a complete study would provide other spatially flat R-model solutions compatible with the generic ideal gas EoS. \\ \\ 
	In the analysis of the particular solutions mentioned above, we have that each choice of the functions $\alpha(r)$ leads, by using expressions (\ref{rho-ideal-1}) and (\ref{rho-ideal-2}), to different energy density profiles that may model inhomogeneities. It would be of interest to investigate which specific elections result in physically realistic inhomogeneities and to study their evolution. In fact, a similar analysis was performed for certain subclasses of the general solution (\ref{def. Z}$-$\ref{sol. general eq. Z}) in the context of the so-called ``Swiss cheese” models \cite{Bona-Stela-c}. Moreover, one of these subclasses appears to admit solutions compatible with the ideal sonic condition (\ref{ISC}). Accordingly, exploring the thermodynamics of such solutions and assesing their potential interpretation as a generic ideal gas constitute open problems that deserve being considered for future work. \\ \\
	Chapter \ref{chap-R-models-plans} ends by formulating the problem of interpreting the solutions as describing the evolution of an inviscid non-perfect fluid in equilibrium. Some initial results are given there, but the complete analysis is left for a forthcoming study. \\ \\ \\
	The results on the models studied in Chapter \ref{chap-Stephani} also suggest some open problems whose study goes beyond the scope of this thesis. Firstly, to build a complete model covering the entire spacetime, the singular models considered in Section \ref{sec-Model-Nos} should be matched with a dust model through the junction surface $\pi(R,r) = 0$. Then, it would also be interesting to make an accurate analysis of the parameters of the models to achieve the more suitable values for physically realistic models. Finally, the regular models should be investigated in detail as we have done with the singular ones. \\ \\
	Moreover, some relevant questions arise from the findings of this chapter as well. In the inhomogeneities observed in the real Universe, matter moves with respect to the cosmological observer who observes an almost isotropic background radiation. To study the radial profiles and the evolution of such nonlinear inhomogeneities we are interested in obtaining solutions with test isotropic radiation for a cosmological observer and a perfect fluid source with a non-comoving flow. \\ \\ 
	On the other hand, a further study to be made consists in analysing in depth the flow of the thermodynamic Stephani universes taking into account the kinematic approaches presented in \cite{FS-KCIG} and \cite{SFM-velocitats} (summarised here in Section \ref{velocitats}). This study will enable us to determine other test fluids which are comoving with the Stephani cosmological observer. \\ \\
	At the end of Section \ref{velocitats} we show some first outcomes that also deserve further consideration. For example, we have that the perfect fluids at rest in the Kerr solution or in any stationary axisymmetric gravitational field are, generically, of class C$_2$. The determination of the subset of stationary axisymmetric solutions with a flow of class C$_{15}$ is an open problem that is left for the future. Additional work is also required to study a similar question for the flows defined by a conformal Killing vector field. Moreover, the results obtained there on the Von Zeipel theorem suggest that our study could lead to a generalisation of such theorem. \\ \\
	All these open problems have emerged throughout the various studies presented in this thesis. Nevertheless, substantial work remains to be done in furthering the physical interpretation of perfect fluid solutions. For instance, several R-models have yet to be analysed from a physical perspective: the geodesic R-models that do not admit flat synchronisations, the Kustaanheimo-Qvist solutions and the non-geodesic R-models that fall outside the Stpehani-Barnes family of solutions. Furthermore, it would be worthwhile to explore axisymmetric solutions, not only to study their physical interpretation, but also to apply our hydrodynamic and kinematic approaches to analyse astrophysical scenarios, such as a test fluid accreting onto a central object. \\ \\
	We cannot conclude this section without addressing the future of \textit{xIdeal}. Although the number of implemented IDEAL characterisations and determinations continues to grow, there remains much work to be done. The algorithms presented in Sections \ref{velocitats}, \ref{subsec-dim-TipusN} and \ref{sec-sim-Bianchi} have not yet been implemented. Furthermore, several other IDEAL characterisations from the literature should be incorporated, not to mention that new ones are likely to be developed in the coming years. \\ \\ 
	On the other hand, some IDEAL determinations of geometric spacetime elements necessary for other IDEAL characterisations are already implemented but cannot be accessed through any public function. Making them available for the user is still a pending task. \\ \\
	Further refinement is also required even for the currently implemented functions to ensure that they are user-friendly and robust. For instance, as already commented, in some cases the results heavily depend on the given \texttt{Assumptions}, and the user should be able to know when extra \texttt{Assumptions} are needed. The simplest example of this situation is given by the Schwarzschild metric in standard coordinates. If we do not specify that the Schwarzschild mass $m$ is positive and that $2m < r$ through \texttt{Assumptions}, it is not classified correctly by the \textit{xIdeal} function \texttt{StaticVacuumTypeDClassify}. \\ \\
	Finally, when the input metric corresponds to a family of solutions, the \textit{xIdeal} classification functions always assign it to the most generic case possible. However, it would be interesting if they could also give the conditions that potential subfamilies should fulfil to belong to  particular subclasses.

\appendix

\chapter{Index-free notation} \label{AppendixB}
(i) Products and other expressions involving 2-tensors $A$ and $B$.
	\begin{itemize}
	\item Composition as an endomorphism: \\[-3mm]
		\begin{equation} \vspace{-2mm}
			(A \cdot B)^\alpha{}_\beta = A^\alpha{}_\mu B^\mu{}_\beta \, .
		\end{equation}
	\item Square and cube as endomorphisms: \\[-3mm]
		\begin{equation} \vspace{-2mm}
			A^2 = A \cdot A \, , \qquad\qquad A^3 = A^2 \cdot A \, .
		\end{equation}
	\item Trace as an endomorphism: \\[-3mm]
		\begin{equation} \vspace{-2mm}
			\textrm{tr} A = A^\alpha{}_\alpha \, .
		\end{equation}
	\item Action on a vector $x$ as an endomorphism and as a quadratic form: \\[-3mm]
		\begin{equation} \vspace{-2mm}
			[A(x)]^\alpha = A^\alpha{}_\mu x^\mu \, , \qquad\qquad A(x,x) = A_{\mu\nu} x^\mu x^\nu \, .
		\end{equation}
	\item Antisymmetrised product with a vector $x$: \\[-3mm]
		\begin{equation}
			(A \bar{\wedge} x)_{\alpha \beta \gamma} = A_{\alpha \beta} x_\gamma \! - \! A_{\alpha \gamma} x_\beta \, .
		\end{equation}
	\item Transpose: \\[-3mm]
		\begin{equation}
			(^t \! A)_{\alpha \beta} = A_{\beta \alpha} \, .
		\end{equation}
	\item Exterior product: \\[-3mm]
		\begin{equation}
			(A \wedge B)_{\alpha \beta \mu \nu} = A_{\alpha \mu} B_{\beta \nu} - A_{\alpha \nu} B_{\beta \mu} + A_{\beta \nu} B_{\alpha \mu} - A_{\beta \mu} B_{\alpha \nu} \, .
		\end{equation}
	\item Hodge dual: \\[-3mm]
		\begin{equation}
			*A_{\alpha \beta} = \frac12 \eta _{\alpha \beta \mu \nu} A^{\mu \nu} \, .
		\end{equation}
	\end{itemize}
(ii) Products and other expressions involving double 2-forms ${\cal A}$ and ${\cal B}$. \\ $\phantom{(ii)}$ A double 2-form is a 4-tensor that is antisymmetric in both pairs of indices.
	\begin{itemize}
		\item Action as an endomorphism and as a quadratic form on the space of 2-forms ${\cal X}$: \\[-3mm]
		\begin{equation} \vspace{-2mm}
			[{\cal A}({\cal X})]_{\alpha \beta} = \frac12 {\cal A}_{\alpha \beta \mu \nu} {\cal X}^{\mu \nu} \, , \qquad {\cal A}({\cal X},{\cal X}) = \frac14 {\cal A}_{\mu \nu \lambda \sigma} {\cal X}^{\mu \nu} {\cal X}^{\lambda \sigma} \, .
		\end{equation}
	\item Composition as an endomorphism: \\[-3mm]
		\begin{equation} \vspace{-2mm}
			({\cal A} \cdot {\cal B})^{\alpha \beta}{}_{\kappa \rho} = \frac12 {\cal A}^{\alpha \beta}{}_{\mu \nu} {\cal B}^{\mu \nu}{}_{\kappa \rho} \, .
		\end{equation}
	\item Square and cube as endomorphisms: \\[-3mm]
		\begin{equation} \vspace{-2mm}
			{\cal A}^2 = {\cal A} \cdot {\cal A} \, , \qquad\qquad {\cal A}^3 = {\cal A}^2 \cdot {\cal A} \, .
		\end{equation}
	\item Trace as an endomorphism: \\[-3mm]
		\begin{equation} \vspace{-2mm}
			\textrm{Tr} {\cal A} = \frac12 {\cal A}^{\mu \nu}{}_{\mu \nu} \, .
		\end{equation}
	\item Action on two vectors $x$ and $y$: \\[-3mm]
		\begin{equation}
			{\cal A}(x; y)_{\alpha \beta} = {\cal A}_{\alpha \mu \beta \nu} x^\mu y^\nu \, .
		\end{equation}
	\item Action on a symmetric two-tensor $B$: \\[-3mm]
		\begin{equation}
			{\cal A}[B]_{\alpha \beta} = {\cal A}_{\alpha \mu \beta \nu} \, B^{\mu \nu} \, .
		\end{equation}
	\item Hodge dual: \\[-3mm]
		\begin{equation}
			*{\cal A}_{\alpha \beta \mu \nu} = \frac12 \eta _{\alpha \beta \kappa \rho} {\cal A}^{\kappa \rho}{}_{\mu \nu} \, .
		\end{equation}
	\end{itemize}

\newpage

\addcontentsline{toc}{chapter}{Bibliography}

\renewcommand{\ChapterNumber}{\ifnum\value{chapter}>0 \thechapter \fi}
\renewcommand{\thechapter}{}  

\bibliographystyle{bib-style.bst}
\providecommand{\noopsort}[1]{}\providecommand{\singleletter}[1]{#1}%
\providecommand{\href}[2]{#2}\begingroup\raggedright\endgroup

\newpage
\thispagestyle{empty}

\AddToShipoutPictureBG*{%
  \includegraphics[width=\paperwidth,height=\paperheight]{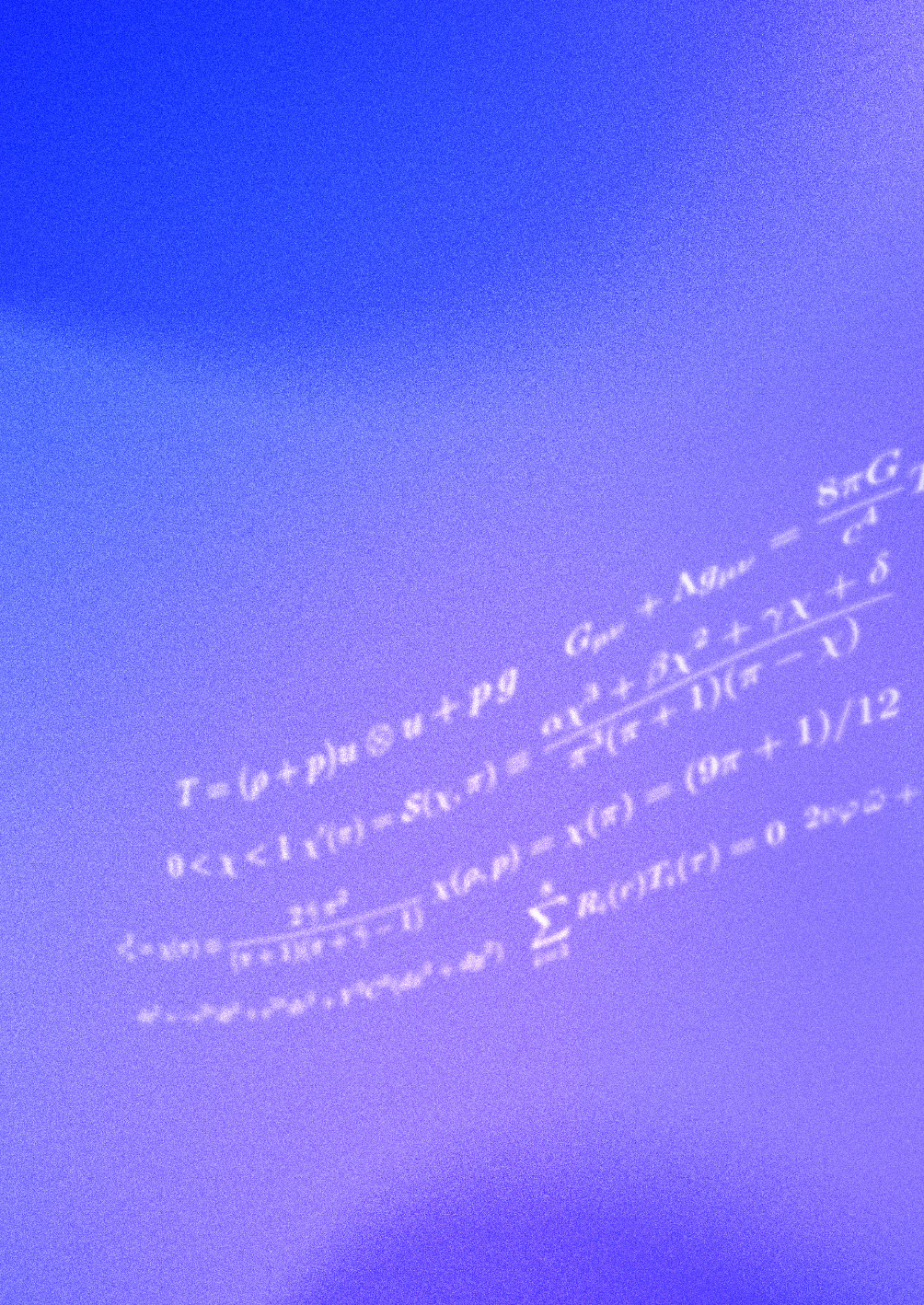}%
}

\begin{flushleft}
\begin{minipage}{0.5\textwidth}
\vspace{1cm}
\color{white}
“The most important step a man can take. It's not the first one, is it? \\
It's the next one. Always the next step.” \\[1em]
\raggedright ― Dalinar Kholin, \\ \textit{Oathbringer} by Brandon Sanderson
\end{minipage}
\end{flushleft}

\end{document}